\newcommand{\draftfinal}{%
  final%
}

\newif\ifdblspc\dblspctrue
\newif\ifdropcaps\dropcapstrue
\newif\iflulu\lulufalse
\newif\ifluluhbk\luluhbkfalse

\newcommand*{\hyperoptions}{colorlinks=true,}
\newcommand*{\myspacing}{onehalfspacing}
\newcommand*{\mysiding}{oneside}

\newcommand{\classoptions}{letterpaper,\mysiding,nodisplayskipstretch,12pt,\myspacing}

\newcommand{\geomoptions}{left=1.5in,right=1in,top=1in,bottom=1.1in,headheight=15pt,includehead,heightrounded,reversemp,marginparwidth=87pt}

\makeatletter
\newcommand{\headfoot}{%
    \fancyhead[L]{\leftmark}%
    \fancyhead[R]{\thepage}%
    \renewcommand{\headrulewidth}{0pt} 
    \renewcommand{\chaptermark}[1]{\markboth {\scuspaced{\if@mainmatter\@chapapp\ \thechapter. \ \fi ##1}}{}}
    \renewcommand{\tocetcmark} [1]{\markboth {\scuspaced{##1}}{}}
}
\makeatother
\newdimen\bleed \bleed=0in

\newcommand*{\xoff}{+0.25in }
\newcommand*{\xtrim}{0 }

\makeatletter
\newcommand{\luluheadfoot}{%
    \newcommand{\sepp}{\hskip 2em}
    \fancyhead[RO]{\rightmark{\sepp\thepage\rblob}}
    \fancyhead[LE]{{\lblob\thepage}\sepp\leftmark}
    \renewcommand{\chaptermark}[1]{\markboth {{\sc\if@mainmatter\@chapapp\ \thechapter. \ \fi ##1}}{{}}}
    \renewcommand{\sectionmark}[1]{\markright{{\ifnum \c@secnumdepth >\z@ \thesection. \ \fi \sc ##1}}}
    \renewcommand{\tocetcmark} [1]{\markboth {{\sc ##1}}{{\sc ##1}}}
    \fancypagestyle{plain}{%
        \fancyhf{}%
        \fancyhead[RO]{{\rblob}}
        \fancyhead[LE]{{\lblob}}
        \renewcommand{\headrulewidth}{0pt}%
        \fancyfoot[C]{\thepage}}%
    \fancypagestyle{empty}{
        \fancyhf{}%
        \fancyhead[RO]{{\rblob}}%
        \fancyhead[LE]{{\lblob}}%
        \renewcommand{\headrulewidth}{0pt}}%
    \fancypagestyle{overview}{%
        \fancyhf{}%
        \fancyhead[RO]{{\overview}}%
        \renewcommand{\headrulewidth}{0pt}}
}
\makeatother

\newcommand{\pngRes}{_LR} 

\documentclass[russian,USenglish,\classoptions,\draftfinal]{book} 
\usepackage{ifdraft}
\usepackage[USenglish=usenglishmax,russian=nohyphenation]{hyphsubst}
\usepackage[OT2,T1]{fontenc} 

\usepackage{graphicx}
\usepackage{pifont}     

\usepackage{fourier-orns} 

\usepackage{babel}      
\usepackage{color}
\usepackage{amsthm}
\usepackage{amsmath}
\usepackage{eqparbox}
\usepackage{textcomp}

\usepackage{nomenlev}\makenomen 
\usepackage{url}

\usepackage{bibentry}   
\makeatletter
\let\BRatbibitem\BR@bibitem
\makeatother

\usepackage[final, \hyperoptions
hyperfootnotes=false,linktoc=all,pagebackref]{hyperref}

\usepackage[sort&compress,mcite]{natbib}
\renewcommand*{\bibnumfmt}[1]{\eqparbox[t]{bblnm}{\hfill#1.}}

\usepackage{mathtools}      
\usepackage{bibunitsLSB}    
\usepackage[final]{pdfpages}

\usepackage{amssymb}
\usepackage{longtable}
\usepackage{subfigure}
\usepackage{footnpag}

\ifdblspc\usepackage{setspace} 
\else
\newenvironment{singlespace}{}{}
\newenvironment{singlespace*}{}{}
\fi
\usepackage[\ifdraft{showframe,}{}\geomoptions]{geometry}
\usepackage{tocbibind} 
\usepackage[perpage,marginal,bottom,symbol,splitrule]{footmisc} 
\ifdblspc
\else\fi

\usepackage{upgreek}
\usepackage{type1cm}
\usepackage{doi}

\usepackage[toc]{appendix}

\usepackage{fnbreak}
\usepackage{sectsty} 
	\sectionfont{\fontfamily{phv}\fontseries{m}\fontsize{15}{17}\selectfont}
	\subsectionfont{\fontfamily{phv}\fontseries{m}\fontsize{13}{15}\selectfont}
\usepackage[Sonny]{fncychap} 
    \ChNameVar{\Large\fontfamily{phv}\selectfont}
    \ChTitleVar{\Large\fontfamily{phv}\selectfont}
\usepackage{emptypage} 
\usepackage[font={small},labelfont={bf},margin=10pt]{caption} 

\usepackage{textcase} 

\usepackage{fancyhdr}

\DeclareRobustCommand*{\scuspaced}[1]{\MakeTextUppercase{#1}}

\newcommand{\doblob}{\rule[-.1\unitlength]{2\unitlength}{.5\unitlength}}
\newcommand{\rblob}{%
    \begin{picture}(0,0)
        \put(1.15,-\value{thumbcount}){\blob}
    \end{picture}}
\newcommand\lblob{%
    \begin{picture}(0,0)
        \put(-3.15,-\value{thumbcount}){\blob}
    \end{picture}
}

\newcounter{thumbcount}
\makeatletter
\newcommand\overview{}
\newcommand*{\listofthumbs}{%
    \cleardoublepage
    \thispagestyle{overview}%
    \@starttoc{thb}
    \mbox{}\newpage}
\makeatother
\newcommand*{\thumbwrite}[2]{\addtocontents{thb}{\string\thumbitem{#1}{#2}}}
\newcommand*{\thumbitem}[2]{\expandafter\gdef\expandafter\overview\expandafter{\overview
    \begin{picture}(0,0)
        \put(0.75,-#2){\llap{\Large{#1}}}
        \put(1.15,-#2){\doblob}
    \end{picture}}}
\newcommand*{\thumb}[1]{%
    \let\blob=\doblob
    \stepcounter{thumbcount}%
    \thumbwrite{#1}{\arabic{thumbcount}}}
\newcommand*{\nothumb}{\let\blob=\relax}
\nothumb

\headfoot
\newcommand{\lofchap}[1]{%
    \addtocontents{lof}{\protect\addvspace{-10pt}}
    \addcontentsline{lof}{lofchapter}{\protect\numberline{\thechapter}#1}%
}

\newcommand{\lotchap}[1]{%
    \addtocontents{lot}{\protect\addvspace{-10pt}}
    \addcontentsline{lot}{lofchapter}{\protect\numberline{\thechapter}#1}%
}

\makeatletter
    \newcommand*\l@lofchapter[2]{%
    \addpenalty{-\@highpenalty}%
    \vskip 1.0em \@plus\p@
    \setlength\@tempdima{1.5em}%
    \begingroup
    \parindent \z@ \rightskip \@pnumwidth
    \leavevmode \bfseries
    \advance\leftskip\@tempdima
    \hskip -\leftskip
    #1\nobreak\par
    \penalty\@highpenalty
    \endgroup}
\makeatother
\ifdropcaps
\usepackage{lettrine}  

\fi
    \providecommand*{\lettrine}[3][blah]{#2#3} 

\usepackage[babel,final]{microtype} 

    \backrefparscanfalse
    \newcommand{\backrefnotcitedstring}{\relax}
    \newcommand{\backrefcitedsinglestring}[1]{Cited on page~#1.}
    \newcommand{\backrefcitedmultistring} [1]{Cited on pages~#1.}
    \AtBeginDocument{
        } 
    \renewcommand*{\backref}[1]{}  
    \renewcommand*{\backrefalt}[4]{
      \ifcase #1\backrefnotcitedstring%
      \or\backrefcitedsinglestring{#2}%
      \else\backrefcitedmultistring{#2}%
      \fi}

\newcommand{\ghz}{~\mathrm{GHz}}
\newcommand{\mhz}{~\mathrm{MHz}}
\newcommand{\khz}{~\mathrm{KHz}}
\newcommand{\hz}{~\mathrm{Hz}}

\newcommand{\MHz}{~\mathrm{MHz}}
\newcommand{\KHz}{~\mathrm{KHz}}

\newcommand{\us}{~\upmu\mathrm{s}}
\newcommand{\ns}{~\mathrm{ns}}
\newcommand{\ms}{~\mathrm{ms}}
\newcommand{\V}{~\mathrm{V}}

\newcommand{\dB}{~\mathrm{dB}}

\newcommand{\ohm}{~\Omega}

\newcommand{\uw}{~\upmu\mathrm{W}}

\newcommand{\ff}{~\mathrm{fF}}
\newcommand{\pF}{~\mathrm{pF}}
\newcommand{\fF}{~\mathrm{fF}}

\newcommand{\nH}{~\mathrm{nH}}

\newcommand{\mA}{~\mathrm{mA}}

\newcommand{\mK}{~\mathrm{mK}}
\newcommand{\K}{~\mathrm{K}}

\newcommand{\nm}{~\mathrm{nm}}
\newcommand{\um}{~\upmu\mathrm{m}}
\newcommand{\mm}{~\mathrm{mm}}

\newcommand{\mycaption}[2]{\caption[#1]{\textbf{\boldmath #1.} #2}}
\newcommand{\capl}[1]{{\textbf{#1}}}

\newcommand*{\eref}[1]{equation~\ref{#1}}
\newcommand*{\Eref}[1]{Equation~\ref{#1}}
\newcommand*{\equref}[1]{Eq.~\ref{#1}}
\newcommand*{\aref}[1]{\hyperref[#1]{appendix~\ref*{#1}}}
\newcommand*{\Aref}[1]{\hyperref[#1]{Appendix~\ref*{#1}}}
\newcommand*{\chref}[1]{\hyperref[#1]{chapter~\ref*{#1}}}
\newcommand*{\Chref}[1]{\hyperref[#1]{Chapter~\ref*{#1}}}
\newcommand*{\sref}[1]{\hyperref[#1]{section~\ref*{#1}}}
\newcommand*{\Sref}[1]{\hyperref[#1]{Section~\ref*{#1}}}

\newcommand*{\figref}[1]{\hyperref[#1]{Fig.~\ref*{#1}}}
\newcommand*{\fref}[1]{\hyperref[#1]{figure~\ref*{#1}}}
\newcommand*{\Fref}[1]{\hyperref[#1]{Figure~\ref*{#1}}}
\newcommand*{\tref}[1]{\hyperref[#1]{table~\ref*{#1}}}
\newcommand*{\Tref}[1]{\hyperref[#1]{Table~\ref*{#1}}}

\newcommand*{\figthanks}[1]{ {(Figure used with permission from~\cite{#1}.  See \nameref{ch:copyperm}.)}}
\newcommand*{\figadapt}[1]{ {(Figure adapted with permission from~\cite{#1}.  See \nameref{ch:copyperm}.)}}
\input{Qcircuit.tex}

\begin{document}
\addtolength{\hoffset}{\bleed}
\addtolength{\voffset}{\bleed}
\addtolength{\pdfpagewidth}{2\bleed}
\addtolength{\pdfpageheight}{2\bleed}

\frontmatter

\iflulu
	\begin{singlespace}
	\listofthumbs
	\end{singlespace}
	\thispagestyle{empty}
    \begin{centering}
    \vspace*{3in}
        \copyright\ 2013 by Matthew David Reed\\
        All rights reserved.\\
\iflulu
		\vfill
		ISBN 978-1-304-08486-6 \\
\fi
    \end{centering}

	\clearpage


\begin{titlepage}
\vspace*{1cm}
\begin{centering}
\begin{Large}                         
    Entanglement and Quantum Error Correction with Superconducting Qubits\\
\end{Large}                           
\vfill
    A Dissertation\\
    Presented to the Faculty of the Graduate School\\
    of\\
    Yale University\\
    in Candidacy for the Degree of\\
    Doctor of Philosophy\\
\vfill
    by\\
    Matthew David Reed\\
\vspace{1cm}
    Dissertation Director: Professor Robert J. Schoelkopf\\
\vspace{1cm}
    December 2013\\
\end{centering}
\end{titlepage}



\thispagestyle{empty}
\begin{singlespace}\begin{center}

\iflulu
	{\sc Abstract} \\
	~\\
	\begin{Large}
		Entanglement and Quantum Error Correction with Superconducting Qubits\\
	\end{Large}
	~\\
	{\large Matthew David Reed}\\
	~\\
\else
	{\large Entanglement and Quantum Error Correction with Superconducting Qubits}\\
	Matthew David Reed
\fi

2013\\
\end{center}\end{singlespace}
\vfill
\noindent

A quantum computer will use the properties of quantum physics to solve certain computational problems much faster than otherwise possible.  One promising potential implementation is to use superconducting quantum bits in the circuit quantum electrodynamics (cQED) architecture.  There, the low energy states of a nonlinear electronic oscillator are isolated and addressed as a qubit.  These qubits are capacitively coupled to the modes of a microwave-frequency transmission line resonator which serves as a quantum communication bus.  Microwave electrical pulses are applied to the resonator to manipulate or measure the qubit state.  State control is calibrated using diagnostic sequences that expose systematic errors.  Hybridization of the resonator with the qubit gives it a nonlinear response when driven strongly, useful for amplifying the measurement signal to enhance accuracy.  Qubits coupled to the same bus may coherently interact with one another via the exchange of virtual photons.  A two-qubit conditional phase gate mediated by this interaction can deterministically entangle its targets, and is used to generate two-qubit Bell states and three-qubit GHZ states.  These three-qubit states are of particular interest because they redundantly encode quantum information.  They are the basis of the quantum repetition code prototypical of more sophisticated schemes required for quantum computation.  Using a three-qubit Toffoli gate, this code is demonstrated to autonomously correct either bit- or phase-flip errors.  Despite observing the expected behavior, the overall fidelity is low because of decoherence.  A superior implementation of cQED replaces the transmission-line resonator with a three-dimensional box mode, increasing lifetimes by an order of magnitude.  In-situ qubit frequency control is enabled with control lines, which are used to fully characterize and control the system Hamiltonian.

\else


\begin{titlepage}
\vspace*{1cm}
\begin{centering}
\begin{Large}                         
    Entanglement and Quantum Error Correction with Superconducting Qubits\\
\end{Large}                           
\vfill
    A Dissertation\\
    Presented to the Faculty of the Graduate School\\
    of\\
    Yale University\\
    in Candidacy for the Degree of\\
    Doctor of Philosophy\\
\vfill
    by\\
    Matthew David Reed\\
\vspace{1cm}
    Dissertation Director: Professor Robert J. Schoelkopf\\
\vspace{1cm}
    December 2013\\
\end{centering}
\end{titlepage}


	\thispagestyle{empty}
    \begin{centering}
    \vspace*{3in}
        \copyright\ 2013 by Matthew David Reed\\
        All rights reserved.\\
\iflulu
		\vfill
		ISBN 978-1-304-08486-6 \\
\fi
    \end{centering}

	\clearpage
	
\fi


\begin{singlespace}
\microtypesetup{protrusion=false} 
\tableofcontents
\listoffigures
\listoftables
\microtypesetup{protrusion=true}
\end{singlespace}

\chapter{Acknowledgements}\newcommand{\tnk}{\textsc}

\lettrine{I}{} have had the great fortune to work with some of the smartest and most talented people in the world during graduate school.  Chief among those I wish to thank is my advisor \tnk{Robert Schoelkopf}, who is both a world-class scientist and a great mentor.  He has taught me how to be a better researcher and to prioritize the questions and decisions that matter now, while disregarding or postponing those that do not.  Rob is extremely patient; he allowed me to work on the quantum error correction experiment for months without tangible results.  His confidence in me during that time not only made possible (eventual!) technical and scientific progress, but gave me valuable experience in how to manage such a large project and better focus my time and energy.

\tnk{Steve Girvin} always made himself available for discussions, despite being in contention for the busiest physicist on the planet.  His ability to step into a complicated technical discussion and immediately contribute original and innovative ideas was amazing; his intellect and intuition are an inspiration to me. 

\tnk{Michel Devoret} has a unique way of approaching problems, distilling even complicated and messy ideas down to their (often beautiful) essence.  His enthusiasm for and confidence in the future of quantum information science is a great source of encouragement not only for the work I did at Yale but also for my career going forward.  I admire his meticulous attention to detail and aesthetic sense, which is refreshing to see in a scientist of his caliber.

\tnk{Liang Jiang} is a much more recent addition to the research effort, but nevertheless made a big impact for me.  His class on quantum information taught me a lot, clarifying things that had been a source of confusion for many years.  He is a gifted researcher and teacher.

The person who I worked closest with in graduate school was \tnk{Leonardo DiCarlo}.  I learned from him not only how to run a complicated experiment but also how to approach scientific problems on a practical level.  There is no one in the field who works harder than Leo, which, combined with the fact that he is an outstandingly brilliant scientist, means that he sets the bar for productivity and knowledge.  He holds himself to a very high standard in everything he does, a trait which I aspire to adopt for myself.  He is a fun and friendly person whom I consider to be a good friend.

More recently, I have had the pleasure of working with \tnk{Kevin Chou}, \tnk{Nissim Ofek}, and \tnk{Jacob Blumoff} on the tunable 3D cQED project.  Kevin's boundless enthusiasm and willingness to work late into the night was a big reason for the project's success.  He is a fast learner and was able to rapidly take over the day-to-day operation of the experiment.  Nissim's practicality and causal confidence were also key, as was his considerable skill in fabrication, assembly, and programming.  Even when other projects were making demands on his time, he was always willing to help with sample assembly or another round of fabrication.  Jacob took the lead in sorting out some of the most difficult outstanding theoretical questions about the device, including how to accurately and robustly model a vertical transmon qubit and calculate its Purcell relaxation.  These tools will be crucial for designing the next generation of experiments, which I know I am leaving in capable hands.

This work was conducted on the fourth floor of Becton Center at Yale University, which has an amazing collaborative environment.  All of the people who have done work on the fourth floor were on some level involved in the work presented in this thesis, but there are a few I would like to thank in particular.  \tnk{Luigi Frunzio} is a staple of the fourth floor, operating and maintaining much of the fabrication and mechanical equipment we all rely on; he is partly responsible for fabricating many of the devices reported on in this thesis.  He is an ebullient and good-humored person, whose personality perpetuates the good culture of Becton.  

I have learned a lot from both \tnk{David Schuster} and \tnk{Andrew Houck}.  Dave is the quintessential physicist who, in addition to having an encyclopedic knowledge of cQED, seems to have a well thought out and compelling opinion about virtually everything from board games to domestic policy.  Andrew had the original idea of my first project of the Purcell filter and taught me many early skills including how to wirebond.  More recently, he has been a great source of advice for planning the next stage of my career.  He is an engaging and friendly person, who genuinely cares about the well-being of those around him.  

\tnk{Blake Johnson} and \tnk{Jerry Chow} were the senior graduate students when I first joined Schoelkopf lab, and were directly involved in training me to run experiments.  Blake included me in his pioneering two-cavity experiment, showing me how to run Clean Sweep and the pulse sequencing program.  Jerry was similarly helpful in my early days, working with me to find my first qubit.  Both Jerry and Blake are fun and friendly people with great senses of humor.

\tnk{Luyan Sun}, \tnk{Gerhard Kirchmair}, and \tnk{Michael Hatridge} are more recent additions to Becton, but are also important in the success of my research.  Luyan is an amazingly hard-working person who is always willing to help debug the experiment or provide advice, and is generous with sharing the credit for his success.  Gerhard is outgoing and personable, and is consistently enthusiastic and ready to help.  He is also incredibly smart; whenever I explained something and he agreed with me, I knew I must be right.  I consider him to be good friend, and enjoyed our frequent games of Munchkin.  Mike has a deep knowledge of all aspects of the experimental equipment, from quantum-limited amplifiers to idiosyncrasies with cryogen-free dilution fridges; he is the guy you go to with a question no one else can answer.  He is a fun and welcoming person, and regularly threw some of the best parties I have been to in recent memory.  \tnk{Hanhee Paik} was also a key member of the floor, whose good sense of humor and thoughtfulness brightened the environment for everyone.

\tnk{Teresa Brecht} and \tnk{Brian Vlastakis} are valued colleagues and friends.  Teresa, whose self-effacing manner belies her remarkable competence and intelligence, holds herself to higher standard than almost anyone I know.  Her enthusiasm and genuine interest in my work has redoubled my own focus.  She is one of my favorite people to hang out with, and we have become close friends.  Brian is an outgoing and friendly person who, like Teresa, does not seem to realize how much cooler he is than everyone else.  He was happy to provide a huge amount of help in setting up the Schr\"{o}dinger cat experiment, and plays a mean game of Munchkin.

\tnk{Eleanor Sarasohn} has become one of my closest friends in my last years of graduate school, and is one of the kindest people I have ever met.  She is a singularly talented poet, whose enthusiasm for learning about all manner of topics is infectious.  I can always rely on her for astute and unvarnished opinions and advice.  Eleanor has made an indelible change to my life -- in particular by helping me to figure out what to do after graduate school -- and knows me better than almost anyone.  She also gave me absurdly detailed feedback and editorial comments on this thesis, reading the entire document cover-to-cover {\it twice}; it is vastly improved as a result.

During my time at Yale, \tnk{Devon Cimini} has become by best friend.  He has a great sense of humor and is generous with his time and ideas.  He is a deeply rational, intelligent, and level-headed person, and is a great source of advice.  Devon was always available to talk about the latest development in politics or technology and offered practical encouragement when things in lab were not going well.  I will always fondly remember board game nights and taking turns playing new indie video games in our living room.

I apparently first met \tnk{Sarah Rusk} in college, though neither of us have any recollection of it; fortunately for me, we reconnected at Yale.  Sarah has an excellent sense of humor, is extremely loyal, caring, and understanding and is one of my favorite people and best friends.  She is also a fantastic cook and has a great appreciation for culture, two properties which I hope to continue to assimilate.  I have learned a lot from her about how to be an adult and enjoy life; together we have become connoisseurs of wine costing less than ten dollars per bottle.  In recent months, she listened to numerous practice job talks and was invaluable to me in helping to sort out what I should do after graduating.

Finally, I would like to thank my family \tnk{Brian Reed}, \tnk{Roberta Reed}, and \tnk{Andrea Wilson}.  Their support was essential to me not only in graduate school, but throughout my life; I would not have gotten the places I have without it.  My father, Brian, got me interested in science from an early age and encouraged that passion and technical thinking.  My mother, Roberta, equipped me with the writing skills that I have relied on heavily.  Both parents were supportive of my decision to enroll at Harvey Mudd College, which was the more expensive option but created many of the the opportunities that I enjoy today.  I am also particularly thankful to my sister, Andrea, for recently contributing her excellent business sense and advice about my future career.

There are of course countless unnamed others who are owed thanks as well, including teachers, mentors, professors, and friends.

\vspace{12 pt} 

{\it Update for arXiv:} Thanks also to \tnk{David DiVincenzo}, who generously accepted the role of outside reader for this thesis.

\chapter{Publication list}
\begin{bibunit}[utphysMDR]
\newcommand{\hiliteLSB}[1]{#1}
\begin{singlespace}
This thesis is based in part on the following publications:
\nocite{Reed2010_bold, Johnson2010_bold, Reed2010b_bold, DiCarlo2010_bold, Sun2012_bold, Reed2012_bold}

\renewcommand{\bibsection}{\relax} 
    \putbib[thesis]
\end{singlespace}
\end{bibunit} 
\chapter{Nomenclature}
\thumb{Nomenclature}
\renewcommand*{\nompreamble}{%
    \begin{list}{}{%
    \renewcommand{\makelabel}[1]{\eqparbox[b]{listlab}{##1}}%
    \setlength{\labelwidth}{\eqboxwidth{listlab}}%
    \setlength{\labelsep}{10pt}%
    \setlength{\parsep}{3pt}%
    \setlength{\leftmargin}{\labelwidth+\labelsep}%
    \setlength{\rightmargin}{0pt}}%
}
\renewcommand*{\nomornmnt}{\decofourright \decofourleft}
\renewcommand*{\nompostamble}{\end{list}}
\renewcommand*{\nomenlist}[1]{\ref{#1}}
\renewcommand*{\nomgroupfont}[1]{{\large\textit{#1:}}}
\renewcommand*{\nomformatlbl}[1]{\color{lightgray}\fbox{#1}} 
\begin{singlespace}
\printnomens
\end{singlespace}  

\mainmatter

\setcounter{chapter}{0}
\chapter{Introduction}
\thumb{Introduction}
\label{ch:intro}


\lettrine{A}{t} the turn of the twentieth century, it was widely believed that physics was complete.  Electricity and magnetism were unified with Maxwell's equations, statistical mechanics accurately predicted the properties of fluids and gases, and optics, acoustics, thermodynamics, and kinetics all seemed to be understood.  This was reflected in the progress of the industrial revolution.  Steam power, transatlantic radio, and the telegraph were direct results of physical understanding, yet several nagging problems remained.  In 1895, Wilhelm R\"{o}ntgen created x-rays and in 1889, Marie Curie discovered radiation, neither of which had an explanation.  In 1902, Philipp Lenard observed that the photoelectric voltage depended on the color of light and not its intensity, to the contrary of Maxwell's predictions.  The Rayleigh law of 1900 absurdly predicted that a black body at thermal equilibrium will emit radiation with infinite power at short wavelengths, a problem known as the ultraviolet catastrophe.  And in 1911, Ernest Rutherford showed that electrons orbit the tiny positively-charged nucleus of the atom, but could not explain why the electrons do not fall in.

Initially, a few postulates were used to rectify these problems.  In 1900, Max Planck suggested that energy was quantized and that light came in integer units of $h \nu$, where $\nu$ is the frequency of the light and $h$ is a number known as Planck's constant.  This conjecture solved the problem of black body radiation,  but its broader implications were unappreciated until much later.  In 1905, Albert Einstein explained the photoelectric effect with this idea of energy quantization.  In 1913, Neils Bohr suggested that electrons orbiting atoms could only occupy certain well-defined orbitals, which explained why electrons did not spiral into an atomic nucleus as well as why atoms emitted only at discrete energy levels.  These theories, despite remaining strictly phenomenological, successfully explained many of the specific experimental difficulties of the age.  However, this ``old quantum theory'' offered no justification for quantization nor underlying structure.

\nomdref{Aepr}{EPR}{Einstein-Podolsky-Rosen}{ch:intro} 

It was not until 1925 that a modern theory of quantum mechanics was developed.   Werner Heisenberg and Erwin Schr\"{o}dinger invented respectively, matrix mechanics and wave mechanics.  Although this modern theory unified the phenomenological postulates, it had bizarre implications.  Particles could be in more than one state at once and properties like position and momentum could not be simultaneously known.  Even determinism -- arguably the deepest postulate of modern science -- would be thrown out.  For this reason, and despite its success at explaining the world, the modern theory had numerous detractors.  Albert Einstein, Boris Podolsky, and Nathan Rosen highlighted a supposed paradox that occurred when two ``entangled'' particles were separated and one measured \cite{Einstein1935}.  The information of this measurement appeared to be instantly transmitted to the unmeasured particle regardless of distance, which seemed to violate the special theory of relativity.  EPR suggested that the only resolution to this problem was that quantum theory was incomplete.

Despite this and other vociferous challenges, quantum theory was simply too effective to repudiate.  It accurately and self-consistently described the world, especially after the development of renormalized quantum field theory that unified special relativity and quantum physics in the early 1950's \cite{Wilczek1999}.  For example, field theory correctly predicts the electron spin $g$-factor to a precision of better than one part per trillion \cite{Gabrielse2006,Gabrielse2007}.  The relativistic objections to quantum mechanics were also dismissed, most famously by John Bell's theorem of 1964 \cite{Bell1964}.  Bell showed that there are physical consequences of quantum entanglement which could not occur if ``local hidden variables'' pre-ordained particle correlations.  The subsequent experimental verification of this theorem by Alain Aspect in 1981 \cite{Aspect1981} and others proved that the universe truly disobeyed local realism.

The development of Bell's theorem greatly strengthened the conceptual foundation of quantum theory, but fundamental questions about the nature of quantum information remained.  For example, could entanglement be used to transmit information faster than the speed of light?  This question\footnotemark ~ led to the development of the No-Cloning Theorem \cite{Wootters1982}, which held that an arbitrary quantum state could not be perfectly copied and thus entanglement could not violate relativity.  Proposals for forgery-proof quantum money \cite{Wiesner1983} and provably secure communication using quantum key distribution \cite{Bennett1984} were made as a direct result.  

\footnotetext{These questions had long been disregarded by the mainstream community, who largely adhered to the ``shut up and calculate'' school of quantum-mechanical thought.  However, a group of counter-culture Bay Area physicists in the 1970's who failed to find jobs following the postwar physics boom formed a cohort to investigate these more philosophical questions.  Their ``Fundamental Fysiks Group'' proposed a method of transmitting information faster than the speed of light, which three groups independently discovered the No-Cloning theorem to resolve \cite{Kaiser2012}.}

These developments raised the question of whether the properties of quantum mechanics could be leveraged for other useful purposes.  In 1982, Richard Feynman suggested that a computer using quantum mechanics might more naturally model the physical world \cite{Feynman1982}.  David Deutsch showed in 1985 that such a ``quantum computer'' could not be efficiently simulated with a classical one, which cemented the supposition that quantum information is fundamentally different from its classical counterpart \cite{Deutsch1985}.  Initially, this was only of theoretical interest since it was unclear how to actually achieve a quantum speed-up.  Though the state of a quantum computer could evolve in a huge parallel superposition, the result of a computation would be randomly chosen from that population when it was measured.  Fortunately, the Deutsch-Jozsa algorithm was discovered in 1992, demonstrating that this was a surmountable problem \cite{Deutsch1992}.  Though that algorithm has little practical use, it runs exponentially faster than any classical solution and proved that, in principle, the computational power of quantum physics could be accessed.  More importantly, Peter Shor discovered an integer factoring algorithm in 1995 which could also realize an exponential speed-up \cite{Shor1995}.  The computational difficulty of factorizing numbers is the basis of many classical encryption algorithms \cite{Rivest1978}, so an efficient algorithm provided significant motivation for further study of quantum information science.

For the same reason that a quantum computer would be powerful, it would also be highly susceptible to errors.  Quantum bits are intrinsically analog devices and are described by continuous variables.  Any spurious interactions with the environment or imprecision in control signals will cause the quantum state to become corrupted.  Moreover, even if each individual error is small, there is nothing to prevent subsequent errors from building up and propagating as an algorithm is run.  Thus, the rate at which errors occur sets a fundamental limit on the duration of a calculation that has any appreciable chance of success, and is quite low for a calculation of any size.  For example, the rate required to run Shor's algorithm on an appreciably large number is ten or more {\it orders of magnitude} lower than could ever feasibly be achieved \cite{Devitt2006}.  Without some means of circumventing this issue, the quantum computer would again be relegated to a mere theoretical curiosity.  Fortunately, in 1995 Peter Shor proposed the first ``quantum error correction'' code, by which a single ``logical'' qubit was redundantly encoded with nine physical qubits \cite{Shor1995b}.  This code makes the effective error rate of a logical qubit much lower than the rates of each constituent qubit.  Error correcting codes requiring 5 or 7 qubits were discovered shortly then after \cite{Steane1996, Laflamme1996}, but merely correcting errors is not enough to compute.  These logical qubits must be usable in algorithms, which means manipulating them in a way that is robust to errors as well.  Peter Shor once again solved this problem\footnotemark, finding in 1996 that an arbitrarily perfect {\it fault tolerant} quantum computer could be built from faulty qubits \cite{Shor1996}.

\footnotetext{It is rather remarkable that the same person discovered the most important algorithm and solved the two biggest theoretical challenges facing quantum information.}

\nomdref{Anmr}{NMR}{nuclear magnetic resonance}{ch:intro}

With a quantum computer shown to be theoretically possible, it turned to experimental groups to attempt to build one.  The earliest efforts used liquid-state nuclear magnetic resonance (NMR).  Due to its applications to medicine and chemistry, NMR already had some of the necessary functionality, such as single-qubit gates and good coherence times.  For that reason, initial progress was rapid with demonstrations of simple two-qubit algorithms \cite{Chuang1998, Chuang1998a, Jones1998} soon followed by a seven-qubit factorization of the number 15 using Shor's algorithm \cite{Vandersypen2001}.  For a variety of reasons including poor measurement signal-to-noise and register initialization, however, liquid-state NMR could not scale much past this point \cite{Warren1997}.

A more promising approach was to use trapped ions \cite{Cirac1995}.  In that system, a linear string of ionized beryllium, calcium, strontium, or another type of atom are confined using electric fields.  Certain electronic transitions of each atom are used as a qubit, with higher transitions used for measurement and initialization \cite{Leibfried2003}.  Ions are coupled to one another by their collective motion, which is essentially a coherent ``phonon bus.''  High-precision lasers are used for both single and multi-qubit manipulations\footnotemark.  The field also enjoyed brisk progress by leveraging the techniques and machinery developed for atomic clocks, and by taking advantage of the long coherence times of atomic transitions \cite{Leibfried2003b, Singer2010, Schindler2011, Monroe2013}.  This system currently holds the record for measurement \cite{Myerson2008} and gate fidelity \cite{Gaebler2012}, as well as for the most qubits simultaneously controlled (fourteen) \cite{Monz2011}.  Despite this, scaling to thousands or millions of trapped ions is an imposing problem.  Correlated noise \cite{Monz2011} harms state fidelity as the system grows, and the experimental apparatus required for these relatively small systems are quite complicated.  Progress has been made toward miniaturizing the trap onto a chip \cite{Kielpinski2002, Stick2006}, but combining the huge current required to trap ions with the constraints of a cryogenic circuit represents a challenge \cite{Wang2010}.

\footnotetext{This fact is amusing in the context of a quote from Erwin Schr\"{o}dinger in 1952, where he said that ``we never experiment with just one electron or atom or (small) molecule. In thought-experiments we sometimes assume that we do; this invariably entails ridiculous consequences... we are not experimenting with single particles any more than we can raise Ichthyosauria in the zoo'' \cite{Schrodinger1952}.  The Ichthyosaur is a dolphin-like marine reptile that has been extinct for 90 million years.}

A variety of other quantum computing systems have recently been introduced.  Some examples of credible architectures are optical lattices of neutral atoms \cite{Briegel2000}, semiconductor quantum dots \cite{Loss1998, Hanson2007, Maune2012, Shulman2012, Awschalom2013}, electrons trapped over liquid helium \cite{Schuster2010}, and diamond nitrogen-vacancy centers \cite{Jelezko2006, Dutt2007, Maurer2012, Awschalom2013}.  One of the most promising new approaches and the subject of this thesis is superconducting circuits \cite{Devoret2013}.  There, the collective motion of Cooper pairs in a nonlinear electronic circuit is quantized and used as a qubit \cite{Bouchiat1998, Makhlin2001, Vion2002}.  The state of this motion can be controlled and detected with microwave signals.  Superconducting circuits have been used to demonstrate a variety of quantum information tasks like single-qubit gates \cite{Chow2009, Chow2012}, two-qubit gates \cite{DiCarlo2010, Ansmann2009, Bialczak2010} and high-fidelity measurement \cite{Wallraff2005, Reed2010b, Vijay2011, Hatridge2013}.  For much of their history, however, there was an open question about whether these circuits could be sufficiently coherent to attain fault tolerance.  Fortunately, in the context of very recent experiments \cite{Paik2011, Chang2013}, it appears that the answer is yes.

\section{Overview of thesis}
\label{sec:overview}

This thesis reports recent results using the circuit quantum electrodynamics (cQED) superconducting architecture.  This system, in which superconducting qubits are coupled to microwave cavities, has proven itself as one of the most promising implementations of superconducting technology and potentially of any known quantum computing architecture.  I begin by introducing the characteristics and experimental implementation of cQED and culminate with the experimental realization of the three-qubit quantum error correcting code.  A variety of other results will also be reported, including a new mechanism for qubit readout and a design for improving the coherence of qubits without sacrificing controllability.

Before examining the details of the system, I provide a brief overview of quantum information science in \chref{ch:concepts}.  I introduce the concept of quantum bits, gates, algorithms, and measurement.  I discuss entanglement and how it can be quantified, and briefly mention the requirements for a quantum computer and a few useful algorithms that can be run on one.  The chapter concludes by emphasizing the need for quantum error correction and listing several approaches to do so.

In \chref{ch:theory}, I summarize the physics of these superconducting systems.  I introduce the transmon qubit, which is the qubit variant used throughout this thesis, and show how it can be coupled to a microwave resonator in cQED.  Control of the resonator enables us to apply single-qubit gates, mediate coupling between qubits, and measure qubit states.  I discuss flux bias lines, which are used to control qubit transition frequencies in-situ, and show how to calculate the expected qubit relaxation.

With the theoretical concepts established, I turn to the details of our experimental implementation in \chref{ch:exptsetup}.  I introduce two approaches to building cQED: the two-dimensional planar design and the three-dimensional cavity design.  Though they are conceptually similar, their designs are quite different.  I show a new variant on 3D cavities that integrates flux bias lines and comes with its own host of design considerations.  I then explain how the devices are cooled in a helium dilution refrigerator, and describe how the fridge is cabled to maximize thermalization, control precision, and measurement fidelity.  Finally, I explain how single-qubit gates are accurately and inexpensively generated at room temperature.

Calibrating these gates in a real experiment is the subject of \chref{ch:singlequbitgates}.  I introduce simple procedures for measuring cavity transmission and qubit spectroscopy which are required to initiate any cQED experiment.  I then show how to progressively tune-up qubit pulses with Rabi and Ramsey oscillations and a sequence called ``AllXY.''  This sequence is more sensitive to a variety of pulse error syndromes than other approaches, and is an archetype for even more sophisticated tune-ups.

\Chref{ch:qubitmeasurement} concerns the details of qubit measurement.  It begins with the conventional dispersive mechanism and calculates the expected signal to noise ratio of such a measurement.  Due to the low signal power and the relatively high noise temperature of the amplifier chain most often used, this SNR and the corresponding measurement fidelity can be low.  Motivated by this fact, I introduce a new element called a ``Purcell filter'' that breaks the relationship between qubit and cavity lifetime.  Apart from enabling the use of low-Q measurement cavities to increase dispersive measurement fidelity, it can also be used to efficiently reset qubits to their ground state.  I then discuss the ``high power'' readout scheme, which exploits the unusual behavior of the cavity when driven very strongly to make a high-fidelity measurement.  This has the advantage of obviating the need for sophisticated amplifiers or design complications to attain good measurement fidelity, but scrambles the qubit state during the measurement, which limits the scope of its application.

I then turn to more sophisticated qubit experiments and discuss how we have generated three-qubit entanglement on demand in \chref{ch:entanglement}.  I start by describing the characteristics of the device we used, which hosts four individually flux-biased transmon qubits.  I show two ways that we can use this flux control to implement two-qubit entangling gates.  Both methods exploit an interaction with higher transmon excited states, but approach it either in the slow (adiabatic) or fast (sudden) limit.  In order to verify that these gates are working as expected and to quantify their fidelity, I discuss how state and process tomography can be efficiently measured with a joint qubit measurement.  Finally, using the sudden two-qubit gate, I explain how we have produced three-qubit entanglement and measured the resulting state with tomography.  We also verified the presence and quality of entanglement with various witnesses.

These techniques lead directly into \chref{ch:qec}, where I discuss our recent demonstration of three-qubit quantum error correction.  The key to this result is an efficient three-qubit Toffoli gate.  This gate leverages our understanding of both sudden and adiabatic gates using higher-excited states to engineer an interaction between a computational state and a third-excited state of one transmon.  I discuss the procedure we have developed to tune the gate up and report the resulting performance as measured by state and process tomography.  Using this gate, we demonstrated both bit- and phase-flip quantum error correction.  We verified that the algorithm worked as expected by measuring the ancilla qubit states after a full bit-flip on a single qubit and confirming the quadratic dependence of fidelity on the effective error rate of all three qubits.  Despite this success, the algorithm never improves the fidelity of a process because the coherence of the qubits in the device was too poor.

Motivated by this result and recent breakthroughs in 3D qubit coherence \cite{Paik2011, Rigetti2012}, I report on preliminary results using a tunable version of the 3D cQED architecture in \chref{ch:tunable}.  I show how the added control lines constitute an unacceptable qubit decay channel as we expected, and that this channel can be effectively turned off with proper filtering.  I then discuss a host of experiments we have performed to measure the system by combining cavity photon-number Fock states with fast flux control.  These techniques enabled us to accurately measure cavity lifetime, coherence, nonlinearity, and dispersive shift as a function of qubit frequency.  These results indicate that, in the absence of qubit hybridization, the cavity can be extremely coherent.  This leads us to study the cavity itself as a quantum resource.  I explain how we have used qubit number splitting to measure the cavity state and the inherited Kerr nonlinearity to produce interesting states to detect.  Finally, combining this with fast flux control, I show how we can ``freeze'' this Kerr evolution, effectively controlling the cavity Hamiltonian on demand.

I conclude this thesis in \chref{ch:conclusion} with an overview of the state of the field and suggestions for future work.  In particular, I list a few straightforward improvements that could be made to the tunable architecture to enable more sophisticated experiments.  A few of these experiments are also proposed.  Finally, I give my perspective on the future of superconducting qubits and their prospects for implementing a quantum computer.

\setcounter{chapter}{1}
\chapter{Concepts of Quantum Information}
\lofchap{Concepts of Quantum Information}
\lotchap{Concepts of Quantum Information}
\thumb{Concepts of Quantum Information}
\label{ch:concepts}


\lettrine{T}{his} chapter will serve as an introduction to the core concepts of quantum information processing.  There will be three sections.  First, we introduce the fundamental building blocks and language of quantum information starting with qubits and single-qubit rotations.  Next, we describe the reasons that quantum information processing is potentially very powerful, introduce some of the potential applications, and describe what constitutes the basic requirements for a ``real'' quantum computer.  We then introduce quantum error correction, which is required for the same reasons that a quantum bit is powerful, and give an example with the simple three-qubit code.  We will conclude with a brief survey of the different kinds of more sophisticated error-correcting codes.

The first section will begin with the idea of a quantum bit or ``qubit,'' which is the fundamental building block of quantum information.  We will introduce a useful geometric picture for their quantum state known as the ``Bloch sphere,'' and single-qubit ``quantum gates'' used to manipulate that state which can be viewed as rotations about some axis of the Bloch sphere.  One convenient way of describing these rotations is with the Pauli matrices, which correspond to full flips about a given axis but can also be applied in small amounts using matrix exponentiation.  Pauli operators are also useful as a language that defines observables of quantum states.  We then consider what happens when you have more than one qubit, both how the quantum state of a register is described and how operators are constructed.  The consequences of having multiple qubits leads to a discussion of the {\it density matrix}.  This accounts for the experimental reality that we are only ever controlling and measuring subsystems that may be coupled with other uncontrolled degrees of freedom.  It also enables us to describe the state of an ensemble of identically-prepared qubits that do not remain the same because of noise, control imprecision, and decay.  We conclude this section by discussing the {\it entanglement} of particles and how it can be generated and detected.

The second section of this chapter will offer an explanation as to why quantum information processing has the potential to be such a powerful tool and what it would take to harness that power.  We start with a general explanation of the properties of quantum information.  Because particles can be in superpositions of states, even a relatively small number of qubits can encode an enormous amount of information.  Moreover, that information can be efficiently manipulated with unitary operations.  This leads into a discussion of {\it quantum algorithms}, which are specialized procedures that utilize this complexity to solve certain problems faster than otherwise possible.  These algorithms operate only with stringent requirements, however, which leads us to the final topic of this section: the DiVincenzo criteria for a quantum computer.  Those criteria constitute the basic hardware requirements that a quantum computer must satisfy.

The final section will explain why quantum bits are much more susceptible to errors for the same reason they are powerful.  There are straightforward methods of correcting errors in classical computers which do not seem to easily generalize to the quantum case.  Fortunately, by taking advantage of another resource unique to quantum mechanics -- entanglement -- we demonstrate how a three-qubit quantum repetition code can be made to correct for arbitrary bit-rotations of any one qubit.  There are two implementations of this code: one relies on measuring error syndromes and classical logic to detect and correct an error; the other implementation combines coherent quantum interactions with non-unitary qubit reset to autonomously correct errors.  Both codes can easily be modified to fix phase-flip errors instead of bit-flips, which may be a more common error for certain qubit implementations.  A qubit is susceptible to both bit- and phase-flips, however, which require a larger code to repair.  One such code which corrects for all possible single-qubit errors is the nine-qubit Shor code, which is a concatenation of the three-qubit bit- and phase-flip repetition codes.  We conclude with a discussion of other kinds of error correcting codes and the concept of fault tolerance.

\section{Fundamental concepts}

The field of quantum information processing typically concerns systems of one or more quantum bits or {\it qubits}.  A quantum bit is a much more sophisticated object than its classical brother.  Whereas a classical bit stores only one bit of information, a qubit state is described by two real numbers -- in some sense, an uncountably infinite amount of information.  As if this enormous multiplication were not enough, when we have multiple qubits, the amount of information needed to describe the overall state grows exponentially as a function of the number of qubits.  This section will introduce the basic tools we use to describe and manipulate this quantum information.

\subsection{Single-qubit states}
	
A single qubit is a quantum object whose allowed states are either $\ket{0}$ or $\ket{1}$.  These ``ground'' and ``excited'' states form what is known as the ``computational basis.''  Being a quantum object, {\it superpositions} of these states are also allowed: the qubit can be in both states simultaneously.  The quantum state of a single qubit is therefore given by $\ket{\psi} = \alpha \ket{0} + \beta \ket{1}$, where $\alpha$ and $\beta$ are complex numbers and $|\alpha|^2 + |\beta|^2 = 1$.  The probability that the qubit is in its ground state is given by $|\alpha|^2$ (or, equivalently, $\alpha^*\alpha$, where the star operator indicates the complex conjugate), and similarly $|\beta|^2$ for the excited state.  Since $\alpha$ and $\beta$ are complex numbers, we can equivalently write $\ket{\psi} = |\alpha| e^{i\phi_\alpha} \ket{0} + |\beta| e^{i\phi_\beta} \ket{1}$.

\subsubsection{The Bloch sphere}

\begin{figure}
\centering
\includegraphics{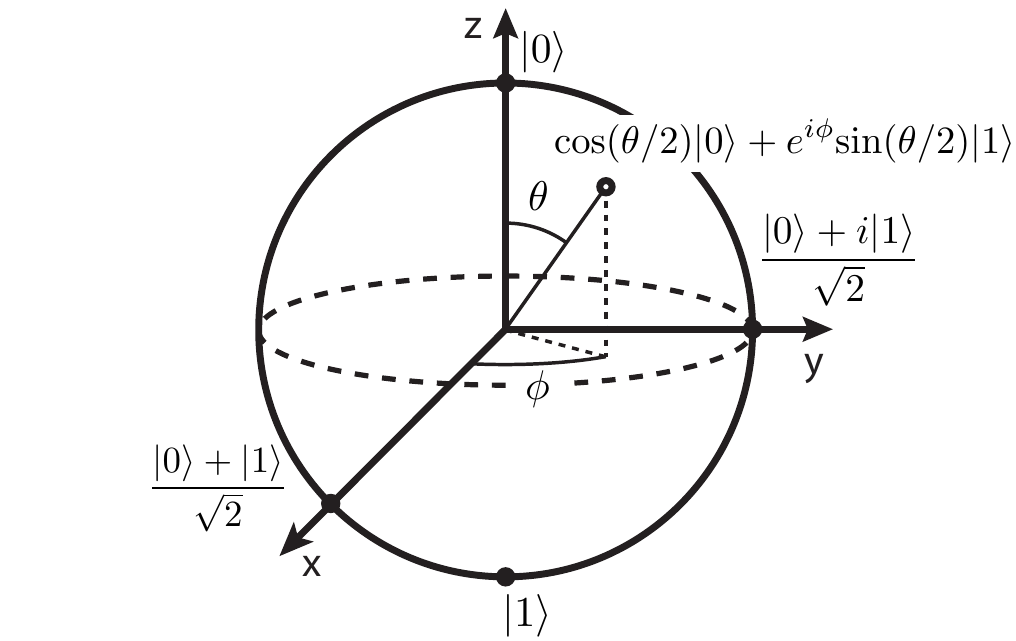}
\mycaption{The Bloch sphere}{A single-qubit wavefunction  $\ket{\psi}=\alpha\ket{0}+\beta\ket{1}$, where $\alpha$ and $\beta$ are complex numbers, is reduced to two real parameters $\ket{\psi} = \mathrm{cos}(\theta/2)\ket{0} + e^{i\phi}\mathrm{sin}(\theta/2)\ket{1}$ with the application of the constraint $|\alpha|^2+|\beta|^2=1$ and that global phases are irrelevant.  The coordinates $\{\theta,\phi\}$ map to a point on the surface of a sphere, known as the Bloch sphere.}
{\label{fig:blochsphere}}
\end{figure}

We can further simplify the representation of $\psi$ by exploiting two facts.  First, we know that the sum of the probability of being found in each state must equal $1$, $|\alpha|^2 + |\beta|^2 = 1$.  Second, the ``global'' phase of a quantum state has no physical meaning.  Then, defining $\phi = \phi_\beta - \phi_\alpha$ and noting that $\mathrm{cos}^2(x) + \mathrm{sin}^2(x) = 1$, we say
\begin{equation}
	\ket{\psi} = \mathrm{cos}\left(\frac{\theta}{2}\right)\ket{0} + e^{i\phi}\mathrm{sin}\left(\frac{\theta}{2}\right)\ket{1}
\end{equation}
 with $\theta \in [0,\pi]$ and $\phi \in [0,2\pi)$.  This representation is chosen to map our quantum state to a point on a sphere, as shown in \figref{fig:blochsphere}.  This is known as the {\it Bloch sphere} representation of a quantum bit.  We can also say that $\psi$ is a two-component vector given by $\left[\mathrm{cos}\left(\frac{\theta}{2}\right),  e^{i\phi}\mathrm{sin}\left(\frac{\theta}{2}\right) \right]$.

\subsection{Single-qubit gates and the Pauli matrices}
\label{subsec:paulimatrices}

\nomdref{Gsigmaz}{$\sigma_z$}{Pauli Z matrix}{subsec:paulimatrices}
\nomdref{Gsigmax}{$\sigma_x$}{Pauli X matrix}{subsec:paulimatrices}
\nomdref{Gsigmay}{$\sigma_y$}{Pauli Y matrix}{subsec:paulimatrices}
\nomdref{Gsigmai}{$\sigma_i$}{Pauli I matrix}{subsec:paulimatrices}
\nomdref{CH}{H}{Hadamard matrix}{subsec:paulimatrices}
\nomdref{CHhat}{$\hat{H}$}{Hamiltonian}{subsec:paulimatrices}
\nomdref{Crntheta}{$R_{\hat{n}}^{\theta}$}{rotation operator about the $\hat{n}$-axis by an angle $\theta$}{subsec:paulimatrices}

Now that we have a single qubit, how do we manipulate its state?  We can describe our state as a two-component complex vector; any possible manipulation of that vector can be represented with a 2 by 2 matrix.  We can further say that this matrix must be Hermitian (e.g. $\hat{O} = \hat{O}^\dagger$) due to the constraints of quantum mechanics.  One convenient way of writing this matrix $\hat{O}$ is as a sum of {\it Pauli matrices}, which are given by:
\begin{equation}
	\sigma_z = \left(\begin{array}{@{}cc@{}}
		1 & 0 \\
		0 & -1 
	\end{array}\right),
	~
	\sigma_x = \left(\begin{array}{@{}cc@{}}
		0 & 1 \\
		1 & 0 
	\end{array}\right),
	~
	\sigma_y = \left(\begin{array}{@{}cc@{}}
		0 & -i \\
		i & 0 
	\end{array}\right),
	~
	\sigma_i = I = \left(\begin{array}{@{}cc@{}}
		1 & 0 \\
		0 & 1 
	\end{array}\right)
\end{equation}
where $\sigma_z$ adds $\pi$ to the relative phase difference between $\ket{0}$ and $\ket{1}$, $\sigma_x$ flips their populations, $\sigma_y$ does both, and $\sigma_i$ does nothing.  They obey the relationship $\sigma_j^2 = I$ and the anti-commutator $\{\sigma_j, \sigma_k\} = 2 \delta_{jk} I$ for all $\{j,k\} = \{x,y,z,i\}$.  From the point of view of the Bloch sphere, the first three of these can also be viewed as a rotation by $\pi$ about the respective axis.  For example, the $\sigma_x$-operation on a qubit rotates the state 180 degrees about the $x$-axis of the Bloch sphere.  These operators also form a complete basis for any Hermitian 2 by 2 matrix, and so any Hamiltonian evolution on a single qubit can be described as a sum of them.  By convention, we say that $\hat{H} = \frac{1}{2} h_0 I  + \frac{1}{2} \left( h_x \sigma_x + h_y \sigma_y + h_z \sigma_z \right) = \frac{1}{2} \vec{h} \cdot \vec{\sigma}$, where in the second step we have omitted the $h_0$ term because it is only an energy offset and is therefore physically irrelevant.  Another common gate is known as the  ``Hadamard,'' which maps $\ket{0}$ to $(\ket{0} + \ket{1})/\sqrt{2}$ and $\ket{1}$ to $(\ket{0} - \ket{1})/\sqrt{2}$, and is defined by the matrix $H=\frac{1}{\sqrt{2}}\left( \begin{smallmatrix} 1&1\\ 1&-1\end{smallmatrix} \right)$.

\subsubsection{Small rotations}

What if we want to apply only a small rotation about one of these axes?  Let us consider the time-dependent Schr\"{o}dinger equation, given by
\begin{equation}
	i \hbar \frac{\partial}{\partial t} \ket{\psi} = \hat{H} \ket{\psi}
\end{equation}
where $\hat{H}$ is the Hamiltonian governing the time evolution.  For time-independent $\hat{H}$, we can solve this equation with $\ket{\psi(t)} = e^{-i\hat{H}t / \hbar} \ket{\psi}$.  If $\hat{H} = \frac{1}{2} \Omega \vec{n} \cdot \vec{\sigma}$, where $\vec{n}$ is an arbitrary vector which defines our rotation axis, the corresponding unitary is given by $\hat{U}=e^{-i \hat{H} t} = e^{-i \frac{\Omega t}{2} \vec{n} \cdot \vec{\sigma}}$.  Taylor-expanding the exponential, $\hat{U} = \sum_{j=0}^{\infty} \frac{1}{j!} \left( \frac{-i\Omega t}{2} \vec{n} \cdot \vec{\sigma}\right)^j$, and utilizing the Pauli operator identities, we are left with $\sum_{j\in\mathrm{evens}}^{\infty} \frac{1}{j!} \left( \frac{-i \Omega t}{2}\right)^j I + \sum_{j\in\mathrm{odds}}^{\infty} \frac{1}{j!} \left( \frac{-i \Omega t}{2}\right)^j \left(\hat{n} \cdot \hat{\sigma}\right)$.  We can identify the two sums as sines and cosines, giving us
\begin{equation}
	\label{eq:unitaryevo}
	\hat{U}(t) = \mathrm{cos}\left(\frac{\Omega t}{2}\right) I - i \mathrm{sin}\left(\frac{\Omega t}{2}\right) \left(\hat{n} \cdot \hat{\sigma}\right)=R_{\hat{n}}^{\Omega t}.
\end{equation}
Thus, we can control the amount of rotation driven by our Hamiltonian by simply changing the period of time for which we apply it.  Equivalently, we could change the parameter $\Omega$, which represents the coupling or drive strength of our rotation.  As we will see in \chref{ch:singlequbitgates}, we control both of these parameters when applying rotations to superconducting qubits using resonant microwave tones.

For example, if we take $\hat{n} = \hat{z}$, then $\hat{U}$ is diagonal in the computational basis as 
\begin{equation}
	\hat{U}(t)  = \left(\begin{array}{@{}cc@{}}
		e^{-i \Omega t / 2} & 0 \\
		0 & e^{+i \Omega t /2}
	\end{array}\right)
	=
	e^{-i \Omega t / 2} \left(\begin{array}{@{}cc@{}}
		1 & 0 \\
		0 & e^{+i \Omega t}
	\end{array}\right)
\end{equation}
wherein the second equation we have factored out the irrelevant global phase.  We can arbitrarily control the phase difference between $\ket{0}$ and $\ket{1}$ by applying the $\sigma_z$ operator.  Geometrically, this corresponds to rotations about the $z$-axis -- as a function of time, our state precesses about $z$.  For arbitrary $\vec{n}$, if we choose our qubit basis as states pointing parallel and anti-parallel to $\vec{n}$, the unitary operation is exactly as written above.

\subsection{Measurement}
\label{subsec:msmtthry}

\nomdref{Aqnd}{QND}{quantum non-demolition}{subsec:msmtthry}

How do we understand measuring the state of a qubit?  Consider the projection operator $\hat{P} = \ket{0} \bra{0}$, which models the act of measuring.  Its expectation value, $\bra{\psi}\hat{P} \ket{\psi}$, gives the probability to be found in the ground state.  (This corresponds to the infinite strength limit of measurement, where the qubit is fully projected.  Finite-strength measurements are extremely subtle \cite{Hatridge2013, Groen2013} and are outside the scope of this introduction.)  Note that measuring a qubit destroys the quantum nature of the qubit -- we have gone from encoding two continuously-valued numbers $\theta$ and $\phi$ in the wavefunction to retrieving a single classical bit of information.  This process is known as ``projecting'' or ``collapsing'' the qubit state.  If the measurement is {\it quantum non-demolition} (QND) to the qubit state, the qubit state will be left in the state in which we measure it.  A non-QND measurement still collapses the wavefunction, but may leave the qubit state in some other (perhaps non-computational) state or states unrelated to the measurement outcome.

We can combine this understanding of measurement with the time-dependent operators we derived in the previous section.  Suppose we apply the Hamiltonian $\hat{H} = \frac{\Omega}{2} \sigma_z$ to the state $\ket{\psi} = 0$ starting at time $t$.  What is the value of $\bra{\psi(t)} \hat{P} \ket{\psi(t)}$?  Using our result from \equref{eq:unitaryevo}, we have 
\begin{equation}
	\hat{U}(t)  = \left(\begin{array}{@{}cc@{}}
		\mathrm{cos}\left(\frac{\Omega t}{2}\right) & -i \mathrm{sin}\left(\frac{\Omega t}{2}\right) \\
		-i \mathrm{sin}\left(\frac{\Omega t}{2}\right) & \mathrm{cos}\left(\frac{\Omega t}{2}\right)
	\end{array}\right).
\end{equation}
Applying this to our state, we have $\ket{\psi(t)} = \left[ \mathrm{cos}\left(\frac{\Omega t}{2}\right), -i \mathrm{sin}\left(\frac{\Omega t}{2}\right) \right]$.  The expected value of the operator $\hat{P}$ is given by $\bra{\psi(t)} \hat{P} \ket{\psi(t)} = \mathrm{cos}^2\left(\frac{\Omega t}{2}\right)$.  This behavior, where the populations of $\ket{0}$ and $\ket{1}$ oscillate as a function of time, is known as a {\it Rabi oscillation} and will be discussed in greater detail in \chref{ch:singlequbitgates}.

How would we measure such a Rabi oscillation?  Our measurement will always either give us 0 or 1, but our prediction is $\mathrm{cos}^2\left(\frac{\Omega t}{2}\right)$, which can take any value between those two extremes.  In order to see this behavior, we need to prepare the qubit many times in the same state, measure it, and average the measurement outcomes.  This is known as an ``ensemble average,'' where  $|\alpha|^2$ fraction of the time we will find the qubit in the ground state and $|\beta|^2$ in the excited state.  The average of these two numbers as a function of $\theta$ will be $\mathrm{cos}^2\left(\theta/2\right)$.  We would repeat the experiment many times for several values of $\theta$ to see the full oscillation.

\subsubsection{Expectation values}

We can also measure the expected value of other projections.  For example, the expected value of a Pauli operator given by $\langle \sigma_j \rangle = \langle \psi | \sigma_j | \psi \rangle$ indicates the projection of our state vector onto that axis of the Bloch sphere.  The projection operator $\hat{P}$ is related to the expected value of $\sigma_z$ by a constant; the expected value of the other Pauli operators can be understood as measuring the qubit along a different axis than the computational basis.  As we will see in \sref{subsec:densitymatrix}, the projections of the qubit along each\footnotemark ~ of the Pauli operators fully specifies the quantum state.

\footnotetext{Only three numbers are required; $\sigma_i$ tells us nothing because $\langle \psi | \sigma_i | \psi \rangle$ is defined to be $1$ by normalization.}

\subsection{Multiple qubits}
\label{subsec:multiplegates}

When we have more than one qubit in our system, the number of computational basis states increases rapidly.  The number of states grows exponentially with the number of qubits $N$, as $2^N$.  For two qubits, we have four basis states: $\ket{00}$, $\ket{10}$, $\ket{01}$, and $\ket{11}$; for three, we have eight states: $\ket{000}$, $\ket{100}$, $\ket{010}$, $\ket{001}$, $\ket{110}$, $\ket{101}$, $\ket{011}$, and $\ket{111}$, and so on.  A quantum state must specify the complex coefficients of all of these basis vectors; this information can no longer be represented as a simple geometrical picture like the Bloch sphere.
	
\subsubsection{Multi-qubit gates}

\nomdref{Acnot}{cNOT}{controlled-NOT gate}{subsec:multiplegates}
\nomdref{Acphase}{cPhase}{controlled-phase gate}{subsec:multiplegates}
\nomdref{Gcnot}{$\Lambda(\sigma_x)$}{controlled-NOT gate}{subsec:multiplegates}
\nomdref{Gcphase}{$\Lambda(\sigma_z)$}{controlled-phase gate}{subsec:multiplegates}

Gates operating on a manifold of multiple qubits must also be realized.  Since the state vector has $2^N$ elements, these operators must be $2^N$ by $2^N$ matrices.  Consider a set of $k=\left(i,x,y,z\right)$ Pauli operators that each act on only the $j$th qubit, $\sigma_k^j$, where the superscript denotes which qubit it addresses.  For example, if we have two qubits, an $X$-operation on the first qubit would be given by the tensor product of $\sigma_x^1$ and $\sigma_i^2$.  A {\it single-qubit gate} is one where all but one of the operators in the tensor product are $I$; having two or more non-identity operations constitutes a {\it multi-qubit gate}.  For example, an $X$-operation on two qubits simultaneously would be given by $\sigma_X^1 \otimes \sigma_X^2$ and is commonly abbreviated as $\sigma_{XX}$ or simply $XX$.  

Some particularly common gates include the ``SWAP gate,'' which maps $\ket{01} \leftrightarrow \ket{10}$ and does nothing to $\ket{00}$ or $\ket{11}$, and is given by the matrix 
\begin{equation}
	\mathrm{SWAP} = \left(\begin{array}{@{}cccc@{}}
		1 & 0 & 0 & 0 \\
		0 & 0 & 1 & 0 \\
		0 & 1 & 0 & 0 \\
		0 & 0 & 0 & 1 
	\end{array}\right).
\end{equation}
There are also {\it controlled NOT gates}, where a target qubit is flipped if and only if a control is excited, and is given by the matrix
\begin{equation}
	\mathrm{cNOT} = \Lambda(\sigma_x) = \left(\begin{array}{@{}cccc@{}}
		1 & 0 & 0 & 0 \\
		0 & 1 & 0 & 0 \\
		0 & 0 & 0 & 1 \\
		0 & 0 & 1 & 0 
	\end{array}\right).
\end{equation}
The cNOT is naturally extendible to being ``controlled'' by more than one qubit; for example, the three-qubit {\it Toffoli gate} flips some qubit if and only if two controls are excited, and is therefore also known as a controlled-controlled-NOT or ccNOT gate \cite{Toffoli1980}.  Another common two-qubit gate is the {\it controlled phase gate}, which flips the phase of only the $\ket{11}$ basis state:
\begin{equation}
	\mathrm{cPhase} = \Lambda(\sigma_z) = \left(\begin{array}{@{}cccc@{}}
		1 & 0 & 0 & 0 \\
		0 & 1 & 0 & 0 \\
		0 & 0 & 1 & 0 \\
		0 & 0 & 0 & -1 
	\end{array}\right).
\end{equation}
The cPhase also has a multi-qubit generalization known as the {\it Toffoli-sign} (or ccPhase) gate which flips the phase of the basis state $\ket{1..1}$.  Many of these multi-qubit gates can be related to one another with single-qubit rotations.  Experimentally, single-qubit gates are implemented essentially the same regardless of the number of qubits.  Multi-qubit gates are much more exotic, however.  Methods of producing these interactions  and using them to entangle qubits will be a major topic of discussion in \chref{ch:entanglement} and \chref{ch:qec}.

\subsubsection{Multi-qubit correlations}

Just as there are multiple-qubit gates, we can also define multi-qubit correlations.  Single-qubit observables are still defined as the expected value of some tensor-product operator that has only one non-identity element.  For example, the expected value of $X$ on the first qubit of a three-qubit register is given by 
\begin{equation}\begin{split}
	\langle X^1\rangle = \langle \psi_1| \langle \psi_2| \langle \psi_3| \sigma_x^1 \otimes \sigma_i^2 \otimes \sigma_i^3 | \psi_3 \rangle| \psi_2 \rangle| \psi_1 \rangle \\
	= \langle \psi_1| \sigma_x^1 | \psi_1\rangle \langle \psi_2| \sigma_i^2 | \psi_2 \rangle \langle \psi_3| \sigma_i^3 | \psi_3 \rangle\\
	= \langle \psi_1| \sigma_x^1 | \psi_1\rangle
	\end{split}\end{equation}
where we have used the fact that operators commute through states that do not share the same labels, and that the states are normalized.

We can also define multi-qubit correlations, where its value is given as the product of the two individual correlations.  For example $\langle Z^1 Z^2 \rangle$ tells us the probability that {\it both} qubits are pointing in the same direction along the $z$-axis.  The state $\ket{\phi^+} = \left(\ket{00} + \ket{11}\right)/\sqrt{2}$ would have a $\langle Z^1 Z^2 \rangle$ value of $+1$, even though both single-qubit $Z$ correlations are zero.  (This is a special state known as a {\it Bell state}, see \sref{subsec:entanglement} below for more.)  Note that this tells us nothing about which direction either one individually is pointing -- merely that they are pointing in the same direction.  As we will see, the fact that these are independent pieces of information will be crucial both in understanding entanglement and for performing quantum error correction.

Just as with the single-qubit case, knowing all of the expected values of the multi-qubit Pauli operators fully specifies a state.  With two qubits there are 15 non-trivial correlations given by $XI$, $YI$, $ZI$, $IX$, $IY$, $IZ$, $XX$, $XY$, $XZ$, $YX$, $YY$, $YZ$, $ZX$, $ZY$, and $ZZ$.  Just as you would expect, the number of these linearly-independent correlations grows exponentially with the number of qubits, as $4^N-1$.  Measuring these multi-qubit correlations can be done by either post-processing individual but simultaneously performed single-qubit measurements, or, as we will see in \chref{ch:entanglement}, by exploiting the properties of more exotic measurement operators.

\subsection{The density matrix}
\label{subsec:densitymatrix}

\nomdref{Grho}{$\rho$}{density matrix}{subsec:densitymatrix}

The previous sections showed how significantly our state description and operators change when we add a single qubit.  But what happens if someone snuck in an extra qubit before we closed the sample box without us noticing?  More generally, how can talk about a subsystem -- our group of qubits -- when there are so many other degrees of freedom around?  Each atom in our sample box, the helium circulating through our fridge, the bench that our control electronics are sitting on, a plane flying overhead -- everything -- should, in principle, be included in the wavefunction describing our system.  We are forced to admit that in our experiments, we are only describing and controlling a subset of the system, and must therefore come up with a new language to describe it.  In particular, this new description must contain more information than just an $2^N$-component vector due to the fact that our state, viewed as a subset of a larger system, can undergo non-unitary evolution.\footnotemark

\footnotetext{Much of this section and the next are adapted from the lecture notes of Liang Jiang from his Spring 2013 course {\it Quantum Information and Computation} at Yale University \cite{Jiang2013}.}

We introduce an object called the {\it density matrix} defined as 
\begin{equation}
	\rho = \sum_i p_i |\psi_i \rangle \langle \psi_i |
\end{equation}
where $\ket{\psi_i}$ are some complete or over-complete set of states of our subsystem that need not be orthogonal to one-another and $p_i\geq 0$ is the {\it probability} that the subsystem is in that state, with $\sum_i p_i = 1$.  (Note that $p_i$ is a real-valued probability, and not a complex probability amplitude.)  The expectation value of some operator $A$ is then given by $\langle A \rangle = \mathrm{tr}\left(\rho A\right) = \sum_i p_i \langle \psi_i | A | \psi_i\rangle$, where the $\mathrm{tr}$ operation is the matrix trace.  The density matrix has three properties worth mentioning right away: it is self-adjoint (e.g. Hermitian) so that $\rho = \rho^\dagger$, it is normalized, so that $\mathrm{tr}\left(\rho\right) = 1$, and it is ``positive'' in the sense that $\langle \psi | \rho | \psi \rangle \geq 0$ for all $\ket{\psi}$.  The last property further implies that all the eigenvalues of the matrix are greater than or equal to $0$.  

We define the {\it state purity} as $\mathrm{tr}\left(\rho^2\right) = \sum_i p_i^2 \leq 1$.  If it is possible to write the density matrix with only one non-zero $p_i$, then the purity is $1$ and the state can be written as a conventional state vector.  For a two-level qubit, the Pauli operators span the space of Hermitian density matrices, so we can write $\rho = \frac{1}{2}\left( \lambda_0 I + \lambda_x \sigma_x + \lambda_y \sigma_y + \lambda_z \sigma_z \right)$.  Since $\mathrm{tr}(\rho)=1$, this simplifies to $\rho = \frac{1}{2} \left( I + \vec{\lambda} \cdot \vec{\sigma}\right)$, with $|\lambda| \leq 1$.  The purity of this state can be shown to be $\frac{1}{2}\left(1 + \vec{\lambda}\cdot\vec{\lambda}\right) \leq 1$.  $\lambda_j = \langle \sigma_j \rangle = \mathrm{tr}(\rho \sigma_j)$, confirming our earlier claim that by knowing the expected values of all the Pauli operators of a single qubit, we know everything that can be known about the state.

We can also introduce a useful property called {\it von Neumann Entropy}.  It is defined as $S(\rho) = -\mathrm{tr}\left[ \rho \mathrm{log}_2\rho\right] = -\sum_j \lambda_j \mathrm{log}_2(\lambda_i)$.  For a pure state like $\rho = |0\rangle\langle 0|$, we have $S(\rho)=0$ and nothing is uncertain; however, for a maximally mixed state like $\rho' = \frac{1}{2} \left( |0\rangle \langle 0| + |1 \rangle\langle 1 |\right)$, $S(\rho') = 1$.  $S$ has several important properties: for any pure state, $S(\rho)=0$; $S(\rho) \geq 0$ for all $\rho$; for a system with a $d$-dimensional Hilbert space, $S(\rho) \leq \mathrm{log}_2 d$; and if systems $A$ and $B$ are in some pure state $\ket{\psi_{AB}}$, $S(\rho_A) = S(\rho_B)$.

\subsubsection{Density matrix examples}

Let us build up some intuition about the density matrix.  Suppose the entire universe contains only two qubits, $A$ and $B$, in some state $\ket{\psi_{AB}}$, and a thief steals away the $B$ qubit while we are distracted by something important.  What can we say about the state of $A$?  We can write the total state $|\psi_{AB}\rangle = \sum_{i,\mu} a_{i,\mu} |i\rangle_A |\mu\rangle_B$ where $|i\rangle$ is a quantum state of system $A$ and $|\mu\rangle$ of system $B$ and $\sum_{i,\mu} |a_{i,\mu}|^2=1$.  The density matrix of qubit $A$ is given by the {\it partial trace} of the full density matrix, where we sum over the degrees of freedom of $B$.  That is, $\rho_A = \mathrm{tr}_B\left(\rho_{AB}\right) = \mathrm{tr}_B\left(|\psi_{AB}\rangle\langle \psi_{AB}|\right) = \sum_{i,j,\mu} a^*_{i,\mu} a_{j,\mu} |j\rangle \langle i |$, where again the indices $i$ and $j$ sum over the $A$ states and $\mu$ over the $B$ states.  Consider a specific case where $|\psi_{AB}\rangle = \alpha |00\rangle + \beta |11\rangle$.  We have $a_{0,0}=\alpha$ and $a_{1,1}=\beta$, so $\rho_A = |\alpha|^2 |0\rangle\langle 0| + |\beta|^2 |1 \rangle \langle 1 | = \left( \begin{smallmatrix} |\alpha|^2&0\\ 0&|\beta|^2 \end{smallmatrix} \right)$.  

Interestingly, this reduced density matrix {\it cannot} be written in terms of a single vector.  The state which you might suppose is most similar to it, $|\psi_{A'}\rangle = \alpha |0\rangle + \beta|1\rangle$ has the density matrix $\rho_{A'} = \left(\alpha |0\rangle + \beta |1 \rangle\right)\left(\alpha^* \langle 0| + \beta^* \langle 1| \right) = |\alpha|^2 |0\rangle\langle 0| + \alpha \beta^* |0\rangle\langle 1| + \beta \alpha^* |1\rangle \langle 0 | + |\beta|^2 |1\rangle\langle 1| = \left( \begin{smallmatrix} |\alpha|^2& \alpha \beta^*\\ \beta \alpha^*&|\beta|^2 \end{smallmatrix} \right)$.  The astute observer will notice that this is not the same matrix: the cross-terms that we picked up when expanding $|\psi\rangle \langle \psi|$ give us off-diagonal terms.  Those terms indicate the {\it coherences} of the state; when we traced away the $B$ qubit above, those coherences were lost.  In the Bloch sphere representation, this is equivalent to forgetting about the $\phi$ information while keeping $\theta$.  We can calculate the purities of both cases; for state $A'$, we find the purity is equal to $(|\alpha|^2 + |\beta|^2)^2 = 1$, while for state $A$ we find a purity of $|\alpha|^4 + |\beta|^4 \leq 1$.  The matrix $\rho_A$ need not have a vector-state equivalence.

So far, we have introduced the density matrix in the context of a subsystem.  However, it also strongly motivated by another experimental reality: the fact that we are always measuring an {\it ensemble} of prepared states.  Suppose that our universe truly contains only one qubit, and we are interested in measuring how well we can prepare that qubit in some state.  Since quantum mechanics only lets us measure a qubit by projecting it along a chosen axis, in order to fully characterize the state we must repeatedly prepare and measure it many times along several distinct axes, in a similar way that we described to measure the Rabi oscillation above.  The question then arises: what happens if we do not actually prepare an identical state every time?  This might happen if some random process affects the qubit, such as dephasing or spontaneous emission, if there is noise in our control machinery, if the environment is changing slowly (e.g. the temperature of the fridge or attenuation of the drive lines), or so on.  It turns out that the density matrix representation is also equipped to deal with this.  An ensemble of imperfect states is described by a density matrix whose purity is reduced as the differences among the states increases.

\subsubsection{Super-operators}
\label{subsubsec:superoperators}

\nomdref{Ct1}{$T_1$}{qubit relaxation time}{subsubsec:superoperators}
\nomdref{Ct2}{$T_2$}{qubit dephasing time}{subsubsec:superoperators}

Quantum mechanics predicts that the evolution of the universe, taken as a whole, is unitary.  However, when we limit ourselves to a small subset of that system, its interactions with degrees of freedom outside of our awareness need not be unitary.  How must we modify our operator formalism to model such evolution and ensemble impurity?  We introduce a new object called a {\it super-operator} which takes a density matrix and maps it to a new one, $\rho_{A'} = \hat{\epsilon}\left(\rho_A\right)$.  $\hat{\epsilon}$ must obey several properties.  First, to be consistent with the properties of classical probability, it must be linear, such that $\hat{\epsilon}(\lambda_1 \rho_1 + \lambda_2 \rho_2) = \lambda_1 \hat{\epsilon}(\rho_1) + \lambda_2 \hat{\epsilon}(\rho_2)$.  Second, it must preserve the hermiticity of $\rho$.  Third, it must be trace preserving, so that $\mathrm{tr}(\rho_{A'})=1$.  Fourth, it must be positive, so that $\langle \psi | \rho_{A'} | \psi \rangle \geq 0$ for all $|\psi\rangle$.  The Kraus representation theorem \cite{Nielsen2000} states that any super-operator $\hat{\epsilon}$ satisfying these criteria has an operator representation 
\begin{equation}
\label{eqn:krausrep}
	\hat{\epsilon}(\rho) = \sum_\mu M_\mu \rho_a M_\mu^\dagger
\end{equation}
for some set of operators $M_\mu$ where $\sum_\mu M_\mu^\dagger M_\mu=1$.  For a pure state $\rho = |\psi\rangle \langle \psi|$ and a single operator $M_0=O$, we see that $\hat{\epsilon}$ reduces to the normal single-matrix operator we are used to seeing with state vectors.  However, if we have more than one $M_i$, the action of $\hat{\epsilon}$ on $\rho$ is no longer unitary and the purity of our state need not be preserved.

This formalism is correct, but quite abstract without a few examples.  Let us first consider a {\it dephasing channel}, where $M_0 = \sqrt{1-p} I$, $M_1 = \sqrt{p} |0\rangle\langle 0|$, and $M_2 = \sqrt{p} |1\rangle\langle 1|$.  In some sense, this is a channel where the qubit is randomly measured, which projects away its phase coherences.  The parameter $p$ can be taken to be the probability that the qubit has been dephased.  Taking $\rho = \left( \begin{smallmatrix} \rho_{00} & \rho_{10}\\ \rho_{01}&\rho_{11} \end{smallmatrix} \right)$, we calculate $\rho' = \hat{\epsilon}(\rho) = (1-p)\rho + p \left( \begin{smallmatrix} \rho_{00} & 0\\ 0&0 \end{smallmatrix} \right) + p \left( \begin{smallmatrix} 0&0\\ 0&\rho_{11} \end{smallmatrix} \right) = \left( \begin{smallmatrix} \rho_{00} & (1-p)\rho_{10}\\ (1-p)\rho_{01}&\rho_{11} \end{smallmatrix} \right)$.  If our state $\rho = \frac{1}{2}\left(1+\vec{\lambda}\cdot\vec{\sigma}\right)$, then the new state $\rho'$ has $\lambda' = \left( \lambda_x (1-p), \lambda_y (1-p), \lambda_z \right)$.  We see that if $p=1$, $\rho$ becomes fully dephased and the off-diagonal terms vanish, just as demonstrated in the above case where our $B$ qubit was stolen.  Note that if the states are separable with $\ket{\psi} = \ket{A}\otimes\ket{B}$, then there are no coherences and dephasing (or tracing over B) has no effect on A.  It is possible to apply this process continuously using the ``Master Equation'' partial differential equation formalism.  Deriving this equation exceeds the scope of this introduction, but it can be shown that if dephasing occurs at a rate $\gamma$, $\rho(t)= \left( \begin{smallmatrix} \rho_{00}(0) & \rho_{10}(0) e^{-\gamma t}\\ \rho_{01}(0)e^{-\gamma t}&\rho_{11}(0) \end{smallmatrix} \right)$.  The dephasing rate $\gamma$ is often specified by a characteristic dephasing time $T_2 = 1/\gamma$.

Another relevant example is the {\it amplitude damping channel} \footnotemark.  There, $M_0 = \left( \begin{smallmatrix} 1&0\\ 0&\sqrt{1-p}\end{smallmatrix} \right)$ and $M_1 = \left( \begin{smallmatrix} 0&\sqrt{p}\\ 0&0\end{smallmatrix} \right)=\sqrt{p} |0\rangle\langle 1|$.  Repeating the calculation above, we find that $\lambda \rightarrow \lambda' = \left( \sqrt{1-p}\lambda_x, \sqrt{1-p} \lambda_y, p + (1-p)\lambda_z\right)$.  Again, if we were to apply this operation continuously, we can define its rate $\gamma$ to be equal to $1/T_1$, where $T_1$ is the characteristic time of the qubit for it to relax from $\ket{1}$ back to $\ket{0}$.  The maximum $T_2$ is given by $2 T_1$.  Interestingly, this decay channel's eigenstate is not a maximally mixed state, but rather the ground state $\ket{0}\bra{0}$.  Finally, the third common process is the {\it depolarizing channel}.  There, $M_0 = \sqrt{1-p} I$, $M_1 = \sqrt{\frac{p}{3}} \sigma_x$, $M_2 = \sqrt{\frac{p}{3}} \sigma_y$, and $M_3 = \sqrt{\frac{p}{3}} \sigma_z$.  This is often considered the most severe damage that you can inflict on a qubit state, since $\lambda \rightarrow \lambda' = \left(1 - \frac{4p}{3}\right) \lambda$, making the qubit completely impure (e.g. $|\lambda'|=0$) for $p=3/4$.  In the context of quantum error correction, undoing a depolarizing error is the most demanding task for a code.

\footnotetext{$M_0$ accounts for the possibility that the channel did not cause the qubit to decay.  Why, then, does $M_0 \ne I$?  Suppose we are performing an experiment where we are measuring the qubit with a photodetector, which clicks when the qubit decays and emits a photon.  As we wait and are attempting to infer the state of the qubit, the longer our photodetector does {\it not} click, the more we have to account for the possibility that the qubit was not in its excited state to begin with.  Similarly, if we fail to detect photons leaking out of a cavity supposedly containing some large coherent state, we must consider the possibility that the cavity actually contains a smaller state that which is less likely to have emitted a photon.  Thus, the $M_0$ operator reduces our estimate of the excited state population.  This phenomenon is known as the ``dog that did not bark''; if our guard dog is silent, it may be that there are no robbers trying to break in, but it also might be that the dog is simply absent (or asleep).}

\subsection{Entanglement}
\label{subsec:entanglement}

The phenomenon of quantum entanglement is one of the most famous and ``spooky'' predictions of quantum theory.  It refers to the fact that the quantum state of one particle can depend on the state of another.  For example, take the Bell state mentioned above where $|\phi^+\rangle = \left(|00\rangle + |11\rangle\right)/\sqrt{2}$.  Measured individually, both qubits are equally likely to point in either direction along the $z$-axis, but if one qubit is found to be pointing up then the other will as well, and vice versa.  That is, their behaviors are random but {\it correlated}.  An entangled state is one where the state $\ket{\psi_{AB}}$ cannot be written as a {\it separable state}, the product of two wave functions individually spanning the constituent subsystems, $\ket{\psi_A}\otimes \ket{\psi_B}$.  Quantifying entanglement requires defining what your subsystems are.  Mathematically, entanglement is a consequence of the fact that each coefficient of a multi-component wavefunction is independent.  Viewed from the point of view of elements in a vector, there is no discrimination between subsystems -- no inherent meaning of the basis vectors.  It is only in the context of delineations between subsystems that entanglement is meaningful.  For example, an atom with four energy levels and two two-level qubits would have the same Hilbert space, but we would only say that the two separate atoms could meaningfully be entangled.

More formally, we define the {\it Schmidt decomposition} for two systems $A$ and $B$ as 
\begin{equation}
	|\psi_{AB}\rangle = \sum_j \sqrt{\lambda_j} |U_j\rangle_A |V_j\rangle_B
\end{equation}
where $\sum_j \lambda_j=1$ and the states $|U\rangle_A$, $|V\rangle_B$ form a complete orthonormal basis for the systems $A$ and $B$, respectively.  Any pure state can be written in this form\footnotemark.  If it is not possible to find a basis in which we can write $|\psi_{AB}\rangle$ using only one non-zero $\lambda_j$, then the state is entangled.  For pure states, we can also say that if the von Neumann entropy of some subsystem is greater than zero, it is entangled with some other subsystem.  If the von Neumann entropy is maximal, the state is also maximally entangled.  Interestingly, since the entropy of the full pure state is zero but the entropy of the subsystems are not, there seems to be some ``negative entropy'' associated with entanglement.

\footnotetext{The proof can be found in Nielsen and Chuang \cite{Nielsen2000} on page 109.}

\subsubsection{Generating entangled states}
\label{subsubsec:generatingentanglement}

How is entanglement generated?  We know immediately that any {\it local} operation cannot change the amount of entanglement.  A local operation is one which leaves invariant the basis vectors of a subsystem that it does not address.  Consider the Hamiltonian $\hat{H}=\hat{H}_A + \hat{H}_B$ where $\hat{H}_A$ leaves invariant system $B$ and vice versa, and $[\hat{H}_A,\hat{H}_B] = 0$.  The unitary evolution driven by this Hamiltonian is also separable, since $\hat{U}=e^{-i \hat{H} t} = e^{-i(\hat{H}_A t + \hat{H}_B t)} = e^{-i \hat{H}_A t} e^{-i \hat{H}_B t} = \hat{U}_A \hat{U}_B$ when their commutator is zero.  If we start with an unentangled state $|\psi_{AB}\rangle = |\psi_A\rangle \otimes |\psi_B\rangle$, applying this unitary to the state also gives us an unentangled state, since $\hat{U} |\psi_{AB}\rangle = \hat{U}_A |\psi_A\rangle \hat{U}_B |\psi_B\rangle = (\hat{U}_A |\psi_A\rangle)\otimes(\hat{U}_B |\psi_B\rangle)$, which is still separable.  Moreover, if we have an entangled state, this local operation will only change the vectors in the Schmidt decomposition and not the number or magnitude of the $\lambda_i$'s, leaving the amount of entanglement unchanged.

Generating entanglement therefore must involve {\it non-local} operations which do not commute with the any of the subsystems we are attempting to entangle.  We can see this with an example.  Suppose $\hat{H}=g \sigma_x^A \otimes \sigma_x^B$, which gives $\hat{U}=e^{-i \hat{H} t} = \mathrm{cos}(gt) I - i \mathrm{sin}(gt) \sigma_x^A \otimes \sigma_x^B$.  Applying this to the state $|00\rangle$ for $t=\frac{\pi}{4g}$ gives us $\frac{1}{\sqrt{2}} \left( |00\rangle - i |11\rangle \right)$, which is a maximally-entangled two-qubit state with $S(\rho_A)=S(\rho_B)=1$ and $S(\rho_{AB})=0$.  Another Hamiltonian that will be useful later is given by $\hat{H} = g \left(\frac{1-\sigma_z^A}{2}\right)\left(\frac{1-\sigma_z^B}{2}\right)$.  This operation, which contains all the Pauli $Z$ operators $\sigma_{ZI}$, $\sigma_{IZ}$, and $\sigma_{ZZ}$, maps to the unitary $\hat{U}=\mathrm{diag}\{ 1,1,1,-1\}$ at $t=\pi/g$, which we identify as a {\it controlled-phase gate} -- it flips only the phase of the basis state $|11\rangle$.  We can also make a controlled-NOT gate, which flips the state of qubit $B$ if and only if qubit $A$ is excited using the Hamiltonian $\hat{H}=g \left(\frac{1-\sigma_z^A}{2}\right) \left(\frac{1-\sigma_x^B}{2} \right)$.

\nomdref{Cghz}{$\ket{GHZ}$}{Greenberger-Horne-Zeilinger state.  For three qubits, $(\ket{000}+\ket{111})/\sqrt{2}$}{subsec:entanglement}
\nomdref{Cw}{$\ket{W}$}{W state.  For three qubits, $(\ket{100}+\ket{010}+\ket{001})/\sqrt{3}$}{subsec:entanglement}

There are certain classes of entanglement that are particularly famous.  For two qubits, we have already mentioned the {\it Bell state} $\ket{\phi^+} = \frac{\ket{00} + \ket{11}}{\sqrt{2}}$.  We can also construct a {\it Bell basis} with the addition of the states $\ket{\phi^-} = \frac{\ket{00} - \ket{11}}{\sqrt{2}}$, $\ket{\psi^+} = \frac{\ket{01} + \ket{10}}{\sqrt{2}}$, and $\ket{\psi^-} = \frac{\ket{01} - \ket{10}}{\sqrt{2}}$.  Each of these states is maximally entangled, mutually orthogonal, and span the two-qubit Hilbert space.  We can write product states in this basis since it is complete, for example $\ket{00} = \frac{\ket{\phi^+} + \ket{\phi^-}}{\sqrt{2}}$.  For three qubits, there are two separate classes of entanglement: GHZ and W-type states \cite{Dur2000}.  The canonical GHZ (Greenberger-Horne-Zeilinger) state \cite{Greenberger1989} is given by $\ket{GHZ} = \left( \ket{000} + \ket{111} \right)/\sqrt{2}$ and the W state by $\ket{W} = \left( \ket{100} + \ket{010} + \ket{001} \right)/\sqrt{3}$.  These two states are not transformable to one another with local operations, so we can say that each of these entanglement classes form a {\it manifold} of states. For example, the state $\ket{GHZ'} = \left(\ket{010} + \ket{101}\right)/\sqrt{2}$ still counts as a GHZ-class state since it is modified from its prototype by an $X$-operation on the second qubit, and no application of single-qubit rotations could change  it into a W state.  As the number of qubits increases, so too does the number of distinct entanglement classes;  $GHZ$ and $W$ types generalize to more qubits in an obvious way, but the additional classes are very complicated and constitute an active research question \cite{Bastin2009}.

\subsubsection{Entanglement of impure states}

The method of detecting entanglement we mentioned earlier does not work if the state under test is impure: states no longer have a defined Schmidt decomposition and their purity can be less than 1 without entanglement.  However, we can extend our language to deal with these states as well.  We previously defined a separable state as one that can be written as $|\psi_{AB}\rangle = |\psi_A\rangle \otimes |\psi_B\rangle$.  When talking about a density matrix, we can distinguish between two types of unentangled states: an {\it uncorrelated} state and a {\it separable} state.  An uncorrelated state can be written as $\rho_{\mathrm{un}} = \rho_A \otimes \rho_B$, while a separable state as $\rho_{\mathrm{sep}} = \sum_k p_k \rho_A^k \otimes \rho_B^k$, with $\sum_k p_k = 1$ and $p_k \geq 0$.  That is, a separable state is an incoherent sum of uncorrelated states.  If our density matrix cannot be written in either of these forms, we can say it is entangled.  Rigorously distinguishing the presence of entanglement with impure states is outside the scope of this introduction.  However, at least for two qubits, the ``positive partial transpose'' (PPT) is definitive \cite{Peres1996, Horodecki1996}.

\section{Computing with qubits}

Why are qubits potentially useful for computing?  Broadly speaking, quantum information science seeks to take advantage of two properties of quantum mechanics: superposition and parallelism.  We have already seen that the amount of information stored in a single qubit is in some sense much greater than a single bit, with its state described by two real numbers.  This fact is compounded when describing the wavefunction of a system of several qubits: for two qubits, there are six real numbers, with three there are fourteen, and so on.  For $N$ qubits, there are $2(2^N-1)$ real numbers that describe a pure quantum state (and more, $2^{N+1}-1$, if the state is impure, though this information is not necessarily useful); for $N=200$, this is more real numbers than there are atoms in the universe.  Even a relatively small number of qubits (by the standard of modern computers), stores an incredible density of information.  Moreover, if we had some unitary that acts on these qubits, it would operate on all the basis states simultaneously.  That is, for some unitary $\hat{U}$, $\ket{00..0} \rightarrow \hat{U}\ket{00..0}$, $\ket{10..0} \rightarrow \hat{U}\ket{10..0}$, and so on through $\ket{11..1} \rightarrow \hat{U}\ket{11..1}$.  Even if our qubit is in some vast superposition of states, a unitary operation acts on each constituent equally rapidly.  Not only can we store a huge amount of information in a register of qubits, then, but we can also {\it manipulate} that information very efficiently as well.

Unfortunately, things are not as favorable as these facts might imply due to the realities of quantum measurement.  As we saw in the previous section, a measurement can only return the value of one observable.  When a qubit is in an eigenstate of that observable (e.g. it is already pointing along the direction you are measuring), the measurement will not destroy any information since the qubit is merely projected to itself.  However, in any other case it will randomly project the qubit along one of the possible eigenstates.  Thus, the best you can do to extract the expected value of some observable is to prepare and measure the state many times in succession.  There is also no way of isolating or measuring a single coefficient of one basis vector without many repetitions; you cannot tell if a coefficient is identically zero, for example, or simply very small.  Thus, it seems that this huge amount of information stored and manipulated in our quantum state is not useful to us because we cannot efficiently retrieve it.

\subsection{Quantum algorithms}

Thankfully, the problem of measurement is surmountable because of quantum interference.  Using it, the answer to a simple (e.g. yes-or-no) question might be encoded in an observable or observables that can be extracted efficiently.  For example, the quantum amplitude of a basis state corresponding to the correct answer might be amplified at the cost of an incorrect state's population.  A {\it quantum algorithm} would take advantage of the qualities of quantum mechanics to solve certain problems faster than is possible classically.  However, because of the difficulty of getting around the issue of measurement, only a few useful quantum algorithms have so far been discovered \cite{Shor2003,Shor2004}.  The most well-known examples include {\it Grover's search algorithm}, which can search an unstructured database quadratically faster than is possible with a classical computer \cite{Grover1996, Grover1997}, {\it Shor's factoring algorithm}, which is exponentially faster than is believed to be possible classically \cite{Shor1995}, and the {\it Deutsch-Jozsa algorithm} for determining whether a function is balanced or constant, which also grants an exponential increase \cite{Deutsch1992}, though many more exist \cite{QuantumZoo}.  The first two algorithms will only return a correct answer with a high probability and may require several repetitions (though it is efficient to check if they succeeded).  This is in contrast to Deutsch-Jozsa, which is deterministic and will always return the correct value if the algorithm was run successfully.  The details of how these algorithms work will not be covered here, but often take advantage of more fundamental quantum subroutines like phase estimation, the quantum discrete Fourier transform, and basis state amplitude amplification \cite{Shor2004}.

\subsection{DiVincenzo criteria}
\label{subsec:divincenzo}

Just as classical algorithms do not require a particular type of computer as long as they implement certain functionality, quantum algorithms are also agnostic about their physical implementation.  Any quantum computer must satisfy certain requirements, however, which were first enumerated by David P. DiVincenzo and are known as the {\it DiVincenzo criteria} \cite{DiVincenzo2000}.  This section will briefly introduce and explain each of these and several additional requirements for error correction.

\subsubsection{1) ``A scalable physical system with well characterized qubits''}

A quantum computer must be made up of many quantum bits, which both exhibit quantum properties (superposition and entanglement) and can be produced in large quantities.  They must be ``well characterized,'' in the sense that their physical parameters are accurately known (for example, the energies of states) as well as their couplings to other qubits and the environment.  The Hamiltonian of the qubit itself must be understood, so that, e.g. if there are higher excited states of the bit we can ensure that qubit population is contained to the computational Hilbert space.

\subsubsection{2) ``The ability to initialize the state of the qubits to a simple fiducial state, such as the ground state''}

We must be able to initialize the computer in some state $|\psi\rangle = |00...\rangle$ for two reasons.  First, any algorithm would require the computational register to be in some known state to begin a computation, since its evolution is unitary and ``garbage in'' would map to ``garbage out.''  Second, as we will see, it is necessary to perform error correction on any quantum computer, which requires a steady stream of qubits in some pure state to extract entropy.  For that reason, this requirement might be expanded to not only include the initialization of the computer in some known state, but also the real-time preparation of extra ``ancilla'' qubits in a known state during computation.

\subsubsection{3) ``Long relevant decoherence times, much longer than the gate operation time''}

In order to successfully run an algorithm, a quantum computer must accurately store the information it is working on.  As discussed above in \sref{subsubsec:superoperators}, any corruption of information is known as {\it decoherence}, and can be understood as the unintentional coupling to the environment, noise in control signals, and so on.  The time it takes to do any part of an algorithm must be much faster than the characteristic time of this information loss.  Note that the {\it total} algorithm run time need not be short compared to the decoherence time because of the existence of quantum error correction.  As long as errors can be quickly identified -- that is, as long as the error rate for a given operation is below some threshold -- some amount of decoherence is tolerable.

\subsubsection{4) ``A `universal' set of quantum gates''}

Any quantum algorithm is a set of deterministic unitary instructions that involve some number of qubits.  For example, an algorithm might call for a unitary $\hat{U}_1$ to be applied, followed by another unitary $\hat{U}_2$, and so on.  Then, to run that algorithm, we would turn on some Hamiltonian $H_1$ for some time, followed by $H_2$, and so on.  In practice, it is very challenging to implement a series of arbitrary Hamiltonians.  It is more convenient to break them down into some set of constituent parts like we did with the Pauli operators for single-qubit rotations.  Just as with a classical computer, where having the NAND gate is sufficient to do any fundamental logic operation (NOT, AND, OR, NOR, XOR, and XNOR operations all have simple circuit diagrams involving only NANDs), many quantum operations can be comprised from the others.  For example, a controlled-phase gate can be turned into a controlled-NOT with the addition of Hadamard gates on the target qubit before and after \cite{Nielsen2000}.  Thus, there are many possible sets of ``universal'' gates.  One set of gates that can be shown to be universal is the combination of the Hadamard, a $Z$ rotation by $\pi/4$ (mysteriously known as the $\pi/8$ gate), and the controlled-NOT gate.  Having such tenuous control is not ideal; it increases the overhead to construct whatever unitary we might need.  More control, like arbitrary rotations of single qubits or additional two or three-qubit gates, is better.

\subsubsection{5) ``A qubit-specific measurement capability''}

We will need to extract information from our computer to retrieve its results, and so some sort of measurement mechanism is required.  The words ``qubit-specific'' are slightly more controversial, however.  A measurement of some group property of qubits, like their projection onto a specific basis state, is insufficient.  While an algorithm could possibly be tailor-made for a different measurement operator, individual qubit measurement is crucial for extracting error syndromes to perform quantum error correction \cite{Fowler2012}.  Also, if a single measurement is not reliable enough, with the ability to measure qubits individually, you could potentially perform controlled-NOT gates between the qubit and several ancillae and then measure those and tabulate the results to improve the fidelity.  For example, we could map $(\alpha\ket{0} + \beta \ket{1}) \otimes \ket{0000} \rightarrow \alpha \ket{00000} + \beta \ket{11111}$ using entangling gates and then do a majority-rules vote on the measurement result of all those qubits.

\subsubsection{Requirements for fault tolerance}

There are also specific requirements for {\it fault tolerant} quantum error correction \cite{Shor1996} (as defined and discussed in \sref{subsec:faulttolerance}), in addition to the well-known error threshold.  We must be able to perform operations on many qubits in parallel, so that we can keep up as we grow the computer and increase the error channels.  Since most physical ``qubits'' are actually larger quantum systems and error correction only addresses the computational Hilbert space, we must be able to ensure that we do not leave that space, or at least have a means of unconditionally pumping population back to it.  We need a continuous source of initialized qubits to absorb entropy; it is not enough to simply initialize all the qubits at the beginning of the algorithm because their purity will drop exponentially as time goes on.  And finally, we require that errors do not scale unfavorably with the size of the computer (e.g. the error rate per qubit does not depend on the size) and that the errors are not strongly correlated between qubits (e.g. many qubits cannot fail simultaneously, or the code will be overwhelmed, as we explain in \sref{subsec:quantumrepetitioncode}).  As a corollary to this, the architecture used to couple qubits together must not cause errors to propagate between qubits \cite{Fowler2012}.

\subsubsection{Additional ``desiderata''}

The five enumerated requirements are necessary to do computation.  There are, however, several other abilities that would be extremely desirable from a practical point of view \cite{Gottesman2002}.  The ability to both convert a stationary quantum bit into a ``flying'' qubit like a photon and then transmit it large distances is crucial for quantum-enabled communication \cite{Turchette1995, Imamoglu1999}.  Access to gates between physically distant qubits can reduce the requirements on error thresholds by one or two orders of magnitude compared to being limited to only nearest-neighbor cNOTs \cite{Svore2005,Szkopek2006}.  Fast and reliable measurement and classical computation, so that we can take advantage of the extremely low error rates of classical computers to implement our error correction, also dramatically reduces overhead \cite{Devitt2009}.  (In principle, a quantum computer can implement the classical logic that would follow a measurement, but since the error rates of quantum gates are much higher and we do not require quantumness to perform these corrections, it is preferable to use a classical computer.)  Some amount of error correlation is tolerable, but the less the better, as is a high degree of parallelism and a large supply of extra ancilla qubits.  Finally, exceeding the error threshold by a large amount also reduces the overhead required by a code; in practice, we would want to exceed it by two or more orders of magnitude so that the number of qubits required is not excessive \cite{Fowler2012}.

\section{Quantum errors and error correction}
\label{sec:errorsandqec}

\nomdref{Aqec}{QEC}{quantum error correction}{sec:errorsandqec}

As we have seen in the previous section, because continuous unitary evolution can occur on a vast multitude of simultaneous basis states, a quantum computer is potentially much more powerful than a classical one.  However, that same property also makes it intrinsically more sensitive to errors.  Quantum algorithms like the quantum discrete Fourier transform often rely on maintaining minute phase differences between states; any source of noise or control error may corrupt these continuous-valued parameters and cause the computation to fail.  Moreover, because evolution is continuous, even tiny errors will inexorably build up as time goes on, foiling our computational aspirations or at least limiting them to trivial problems.

This section will introduce the concept of {\it quantum error correction}, which can theoretically save us from this predicament.  We start by clarifying the difference between a classical and quantum error and will introduce a simple classical error correcting code.  We will then argue that there are three major reasons that we might expect this error correcting code should fail in the quantum case.  However, by being clever enough, we can bypass each issue and propose a quantum bit-flip code.  This code, which requires only three qubits to operate, does not correct for all possible errors but represents a prototype for more sophisticated quantum codes.  We will explain this code in detail and demonstrate that, with a small modification, it can correct for phase errors instead.  We will end by introducing other quantum codes which can correct for arbitrary errors and discuss the requirements for fault tolerance.

\subsection{Classical vs. quantum errors}
	
\begin{figure}
	\centering
	\includegraphics{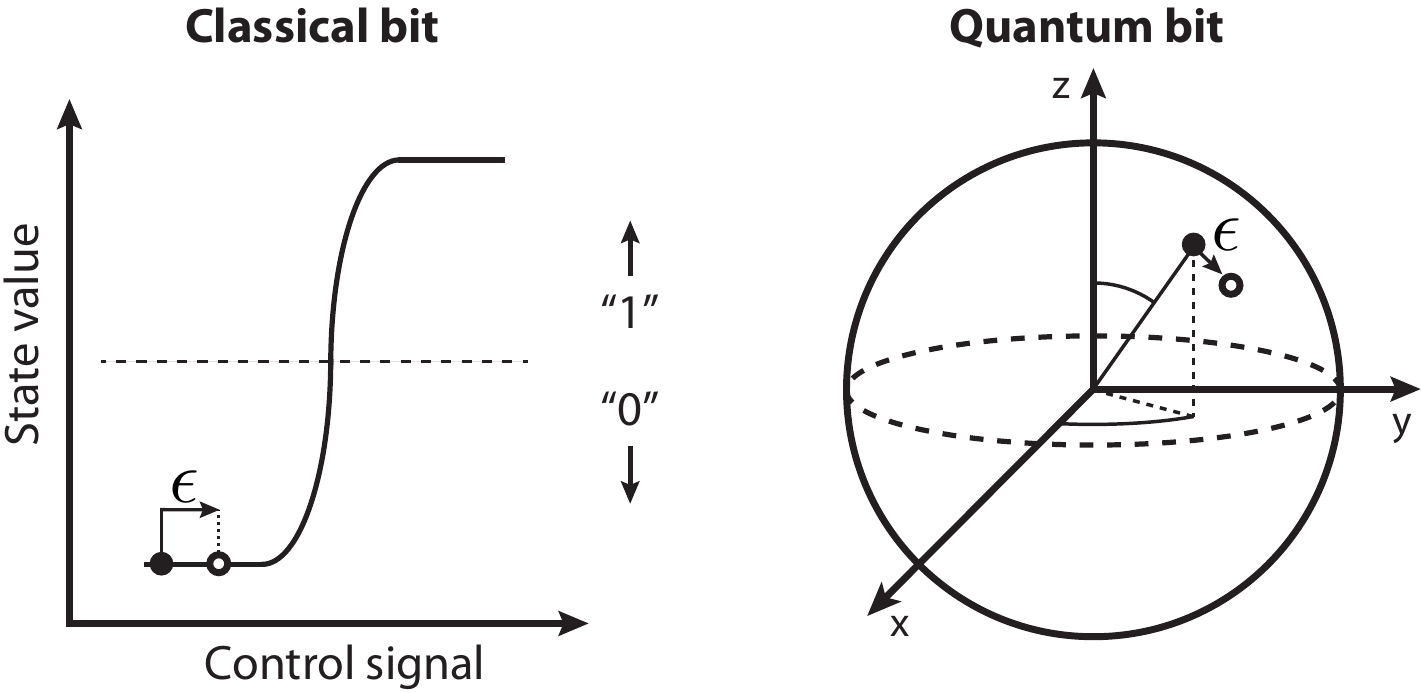}
	\mycaption{Classical vs. quantum errors}{Quantum bits are intrinsically more susceptible to errors than classical bits.  As shown on the left, the state of classical bits is set by some nonlinear mapping of a control signal to a state value.  Any state value below a certain threshold counts as a ``0'' and any value above, a ``1.''  Thus, small amounts of noise in the control parameter will not affect the value stored in the bit.  This fact, where there is not a one-to-one mapping of control to state values is known as ``phase space compression.''  Contrast this to the case of a quantum bit, shown on the right, where the quantum state is continuous-valued and any noise in the control parameters directly change the quantum state.  For this reason, while classical error rates are typically on the order of $10^{-15}$, the best quantum error rates demonstrated to date are on the order of $10^{-2}-10^{-4}$ per gate \cite{Brown2011,Chow2012}.}
	{\label{fig:cvsqerrors}}
\end{figure}

Let us begin by discussing the reasons why a classical computer typically does {\it not} require error correction during normal operation.  Consider the state of a classical switch (e.g. a transistor) as a function of its control parameter (e.g. the applied voltage).  As shown in \figref{fig:cvsqerrors}, because the switch can only take one of two values, small fluctuations in the control parameter will not change its value.  This is called {\it phase space compression}, since there is not a one-to-one mapping of input control states to output switch states.  It is a very good thing for a classical computer; in some sense, the switch is automatically correcting for control errors as it operates.  For this reason, the probability that any given operation will fail (e.g. encode the wrong result) in a modern computer is typically on the order of $\epsilon \sim 10^{-15}$.

Contrast this to the case of a quantum switch.  Its state can take any continuous value, so there is no compression of the mapping of control inputs to state outputs.  Thus, even tiny control errors will change the quantum state in a way that could corrupt the outcome of a calculation.  Moreover, because the state's precise value affects the future evolution of the computer, even small errors will add up as time goes on.  So far, the best single-qubit error rates demonstrated in any system are only on the order of $\epsilon\sim10^{-2}-10^{-4}$ per gate \cite{Brown2011,Chow2012}, with even worse rates for two-qubit gates.  This error rate limits the size of a computation that has any feasible chance of succeeding.  For example, in order to run Shor's algorithm on a number impossible to factor classically (500-1000 digits), we would need an error rate similar to that of a classical computer, $\epsilon \sim 10^{-15}$.  To get that rate, even disregarding the need for our control to be essentially perfect, our qubit coherence would need to be on the order of $T_2 \sim 1$ year; an improvement of 11 orders of magnitude over the current state-of-the-art for superconducting qubits.  We need a more practical way of lowering quantum error rates.

\subsection{The classical repetition code}
\label{subsec:classicalrepetitioncode}

\begin{figure}
	\centering
	\includegraphics{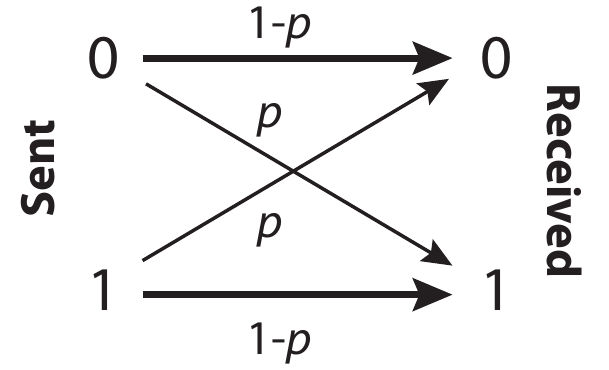}
	\mycaption{Binary symmetric channel}{The cartoon depicts the attempt to send a single bit to a recipient over a noisy channel.  The bit is correctly transmitted with probability $1-p$ but is flipped with probability $p$.  The error probability is independent of which symbol is being transmitted, hence calling the channel {\it symmetric}.
	}
	{\label{fig:symmetricchannel}}
\end{figure}

Though classical error correction is normally unnecessary, there are circumstances where it is useful.  Suppose for example we are attempting to send a bit of information to someone over a noisy communication channel, as shown in \figref{fig:symmetricchannel}.  This channel has a probability $p$ of flipping the bit being sent.  The probability of this error is independent of the information being sent, so this is known as a {\it binary symmetric channel}.  One possible way to reduce our susceptibility to this problem is to encode all the information we send with a {\it repetition code}.  Every time we send a piece of information, we actually transmit it three times (e.g. to send a $1$, we transmit $111$); any message that does not conform to the format of three repeated bits is not allowed and must be due to an error.  The recipient then takes every group of three bits and does a majority-rules vote; in order for the information to be corrupted, two or more errors would have to simultaneously occur.  Of course, we have also gone from having one faulty bit to having three, so the likelihood of an error occurring has substantially increased.  The probability $P_n$ of having $n$ bit-flips is given by $P_0=(1-p)^3$, $P_1=3p(1-p)^2$, $P_2=3p^2(1-p)$ and $P_3=p^3$.  Thus, the probability for the code to fail is given by $p_{\mathrm{eff}}=P_2+P_3=3p^2-2p^3$.  Crucially, this error rate no longer has any linear dependence on $p$; for $p<1/2$, the effective error rate is less than $p$ and adding the redundancy of the extra faulty qubits benefits rather than detracts.  The smaller the value of $p$, the greater the improvement due to performing this code; for example, if $p=10^{-4}$, $p_{\mathrm{eff}}=3 \cdot 10^{-8}$.  Moreover, if we require a higher-order reduction we can employ greater amounts of redundancy (e.g. sending 5 or 7 bits for every 1 we wish to transmit).

\subsection{The challenges of quantum errors}

Implementing such a repetition code with quantum bits seems like it might help our problem.  While we do not necessarily need to transmit a quantum state over a noisy channel to perform quantum computation (though this would be required for things like quantum key distribution), we could think of the ``channel'' as simply the passage of time.  That is, as we let our qubit sit in some state, it will inevitably interact with the environment and decohere, causing an error.  Can we implement the repetition code with quantum bits?  There are actually three good reasons to think the answer is {\it no}, which initially seemed to indicate that quantum error correction was impossible \cite{Nielsen2000}.  First, the no-cloning theorem prohibits the copying of quantum states \cite{Wootters1982}.  Second, measuring the states to see if they are the same does not seem to help us detect the presence of an error and will necessarily destroy their quantum information.  And third, we must correct for continuous rather than discrete errors and so might expect to need an infinite amount of redundancy to distinguish each possible error.  Let us investigate each one of these and introduce why they present a problem.
	
\subsubsection{The no-cloning theorem}
		
The {\it no-cloning theorem} states that it is impossible to create an identical copy of a quantum state.  It is surprisingly easy to prove \cite{Wootters1982}.  Suppose we have a qubit $A$ that is in some state $|\psi\rangle_A$ that we want to copy and a qubit $B$ in some initial state $|i\rangle_B$.  If we have the ability to clone our state, there must be some unitary such that $\hat{U}\ket{\phi}_A\ket{i}_B = \ket{\phi}_A\ket{\phi}_B$.  $\hat{U}$ preserves the inner product, so $\bra{i}_B\bra{\phi}_A \ket{\psi}_A \ket{i}_B = \bra{i}_B\bra{\phi}_A \hat{U}^\dagger \hat{U} \ket{\psi}_A \ket{i}_B = \bra{\phi}_B\bra{\phi}_A \ket{\psi}_A \ket{\psi}_B$, where we have introduced the state $\ket{\phi}_A$ which is some arbitrary state of $A$.  Since $\langle i | i \rangle = 1$, it follows that $\langle \phi | \psi \rangle = \langle \phi | \psi \rangle^2$ and either $\ket{\phi} = \ket{\psi}$ or $\langle \phi | \psi \rangle = 0$.  Since $\phi$ and $\psi$ are arbitrary and need not satisfy either relationship, it follows that $\hat{U}$ cannot clone any quantum state and the no-cloning theorem is proven.  This seems to preclude our application of the repetition code.  We cannot make redundant copes of our qubit to independently transmit, a functionality that seems to be at the core of the code's mechanism.

\subsubsection{The measurement problem}

Even if we bypass the issue of cloning, how might we detect the presence of an error and not destroy the state?  Suppose we had three identical copies of our quantum state through some process (perhaps by running our algorithm three times).  Our only means of extracting information is with a {\it projective} measurement.  Measuring each qubit will give us a $0$ or a $1$, but that does not tell us anything about whether a flip has occurred.  For example, suppose that the state we are trying to protect is $\ket{\psi} = \frac{2}{\sqrt{3}}\ket{0} + \frac{1}{\sqrt{3}}\ket{1}$.  We should expect to get, on average, a $0$ twice and a $1$ once for every three qubits we measure, even in the absence of any error.  In contrast to the classical code, different measurement outcomes do not necessarily indicate that an error occurred because qubits can be in superpositions.  Moreover, once we have measured the qubits, we will have collapsed their wave functions and destroyed the quantum information stored inside.  Our projection maps our two continuous values of $\theta$ and $\phi$ to one classical bit.  Of course, with prior knowledge of $\ket{\psi}$ one could arrange the measurement axis so that the measurement operator is $P=\ket{\psi}\bra{\psi}$.  However, we are trying to come up with a code that can detect errors in {\it unknown} states.
		
\subsubsection{Continuous errors}

In the classical code, if we restrict ourselves to the possibility of zero or one error, we can know exactly which error has occurred.  If we receive for example the codeword $100$, the first bit was corrupted and the intended message was a ``$0$'' (even if more than one error can occur, the most {\it likely} event was the first being flipped only).  We could then flip the first bit back and restore the datum to its unaltered state if we needed to re-transmit those bits.  We would like to be able to do the same thing with our qubits: since their states will need to be manipulated and maintained for the entire length of a computation (which may take months), we must periodically refresh their state by applying the inverse of the error that has occurred\footnotemark.

\footnotetext{In some codes, simply keeping track of exactly which error has occurred to each qubit and incorporating that knowledge into future operations or in interpreting the final result is enough.  You do not necessarily have to correct the error in-situ, but you do have to know exactly what happened.}

For quantum errors, this seems to pose a daunting challenge.  We started this section by arguing that qubits were especially sensitive to errors exactly because there are so many ways that things can go wrong.  While a classical bit can undergo only one discrete error -- being flipped -- a qubit could undergo any spurious evolution in two {\it continuous} dimensions.  If we need to correct a qubit to some high precision, it would seem that we would need to know the precise unitary that caused the error to equal precision.  And because there are in some sense an infinite number of different errors, does that imply that we need an infinite set of resources to distinguish each one?

\subsection{The quantum repetition code}
\label{subsec:quantumrepetitioncode}

Somewhat miraculously, none of these issues are prohibitive.  There is a simple quantum code that can correct for arbitrary bit-rotations of a qubit.  To explain it, we first note that the bit-flip code we described actually uses more information than is necessary to fix an error.  We do not need to know the {\it value} of the bits being sent, we only need to know whether they are {\it the same} as the other bits.  For example, whether our code word is $101$ or $010$ does not change the identification of the error -- the middle bit was flipped in both cases.  In the language we have developed, we need not know $\langle Z \rangle$ for each qubit, but rather  $\langle ZZ\rangle$ for both pairs.  This key observation turns out to save us from all three problems previously identified.

\nomdref{Aghzlike}{GHZ-like}{state of the form $\alpha\ket{000} + \beta\ket{111}$.  GHZ-class when $|\alpha|=|\beta|=1/\sqrt{2}$.}{subsec:quantumrepetitioncode}

The {\it quantum repetition code} encodes a single qubit on a manifold of three qubits, as does its classical analog.  What is this encoding?  We already know that we cannot copy our state three times because of the no-cloning theorem, but we can do something even better using {\it entanglement}.  We map $(\alpha|0\rangle_1+\beta|1\rangle_1) |0\rangle_2|0\rangle_3 \rightarrow \alpha|000\rangle + \beta|111\rangle$, where the first qubit stores some state we would like to protect, the second and third qubits are known as {\it ancillas} and are initialized in the ground state, and the notation $|abc\rangle$ refers to the state of the first, second, and third qubit respectively.  This state, where the three qubits are all in the same state as one another but individually are pointing in no definite direction, is a three-qubit entangled state known as a ``GHZ-like'' state.  When $|\alpha| = |\beta| = \frac{1}{\sqrt{2}}$ it is {\it maximally entangled} and the state is a GHZ state without qualification, as defined in \sref{subsubsec:generatingentanglement}. 

The unitary responsible for mapping $\alpha|0\rangle + \beta|1\rangle$ to this state is two controlled-NOT gates between qubits 1 and 2 and 1 and 3, $\hat{U}=N_1^2N_1^3$.  The operation of a single controlled-NOT flips the target qubit if and only if the control qubit is excited.  Because the operation is coherent, it will act on only the subspace where this is true.  Thus, $N_1^2 (\alpha\ket{0}_1+\beta\ket{1}_1) \ket{0}_2 = \alpha\ket{00} + \beta\ket{11}$ and the extension to three qubits is obvious.  Note that the state is {\it not} being copied; if any one of the qubits were measured, the entire state would be projected.  Interestingly, this kind of state is actually {\it more} susceptible to noise because if any single one of the qubits is dephased, all three states are lost.  For the same reason, an entangled state may be interesting for applications to sensing or detection since its sensitivity to fields that affect all the qubits grows like $N$ instead of $\sqrt{N}$, as found classically\footnotemark.  

\footnotetext{Affecting all the qubits equally would correspond to {\it correlated} noise, which would diminish our ability to correct errors.  For example, if the noise channel never caused a single flip but instead always flipped two bits at a time, the code we have described would fail.  A similar pattern holds true with quantum codes; correlations in noise that make double errors more likely than they would otherwise be must be limited or at least well-understood if we hope to build a quantum computer.}

\subsubsection{GHZ-like states}

These states have the valuable property that they are simultaneous eigenstates of the three Pauli $ZZ$ operators.  For example, $Z_1Z_2 |\phi_0\rangle = Z_1Z_2 \alpha|000\rangle + Z_1Z_2 \beta|111\rangle = (+1)(+1) \alpha|000\rangle + (-1)(-1) \beta|111\rangle = \alpha|000\rangle + \beta|111\rangle$, so the state $|\phi_0\rangle$ has eigenvalue $+1$ for the operator $Z_1Z_2$ (and also for $Z_1Z_3$ and $Z_2Z_3$, though knowing any two of these correlations implies you know the third; we omit $Z_1Z_3$ henceforth).  Something very interesting happens if we suppose that one of the qubits was flipped, so we have $|\phi_1\rangle = \alpha|100\rangle + \beta |011\rangle$.  We calculate that $Z_1Z_2$ now equals $-1$ (and $Z_2Z_3=+1$).  The sign of the eigenvalue has changed in response to the error.  We list the eigenvalues for all four possible cases -- either no error or a bit-flip of exactly one of the three qubits -- in \tref{table:ghzerroreigenvalues}.  Luckily, each of these errors has a distinct set of values of the $ZZ$ eigenvalue -- a unique {\it error syndrome}.  Moreover, we can in principle measure these error syndromes without learning anything about the encoded state -- these syndromes are independent of the values of $\alpha$ and $\beta$.  We have thus made a huge amount of progress.  Not only have we come up with an encoding that does not violate the no-cloning theorem, but there are a set of observables\footnotemark ~ that unambiguously identify which error has occurred on that state (assuming that there was at most one error).  We have thus gotten around two of the three major problems we had identified with correcting quantum errors.  We see how to circumvent the third problem in the next section.

\footnotetext{The operators $Z_1Z_2$ and $Z_2Z_3$ are known as {\it stabilizer operators}, following the formalism of Daniel Gottesman \cite{GottesmanThesis}.  They have the property that they commute both with one another and that our GHZ-like state is an eigenstate of them both.  Our manifold of three qubits is reduced to a single-qubit degree of freedom with the constraint that the three-qubit state must be an eigenstate with eigenvalue +1 for both operations.  The stabilizer operators also commute with the ``logical qubit operators'' $X_{l}=X_1X_2X_3$, $Y_{l}=i X_l Z_l$, and $Z_{l}=Z_1Z_2Z_3$ which perform rotations on the encoded qubit $\alpha|0\rangle_l + \beta|1\rangle_l$, where $|0\rangle_l = |000\rangle$ and $|1\rangle_l=|111\rangle$.}

\begin{table}
	\centering
	\begin{tabular}{|c|c|c|c|}
		\hline
		{\bf State} 										& {\bf Error} 	& {$\mathbf{Z_1Z_2}$} 	& {$\mathbf{Z_2Z_3}$} \\ \hline
		$|\phi_0\rangle=\alpha\ket{000} + \beta\ket{111}$ 	& None 			& $+1$ 					& $+1$ \\ \hline
		$|\phi_1\rangle=\alpha\ket{100} + \beta\ket{011}$ 	& $Q_1$ 		& $-1$ 					& $+1$ \\ \hline
		$|\phi_2\rangle=\alpha\ket{010} + \beta\ket{101}$ 	& $Q_2$ 		& $-1$ 					& $-1$ \\ \hline
		$|\phi_3\rangle=\alpha\ket{001} + \beta\ket{110}$ 	& $Q_3$ 		& $+1$ 					& $-1$ \\ \hline
	\end{tabular}
	\mycaption{Eigenvalues of the GHZ states}{Each of the four error syndromes has a unique signature in the values of $Z_1Z_2$ and $Z_2Z_3$ which can be determined without knowledge of $\alpha$ or $\beta$.  This fact is the basis of the quantum repetition code.}
	\label{table:ghzerroreigenvalues}
\end{table}

\subsubsection{Measurement-based quantum repetition code}
\label{subsubsec:msmtbasedqrc}

\begin{figure}
	\centering
	\includegraphics{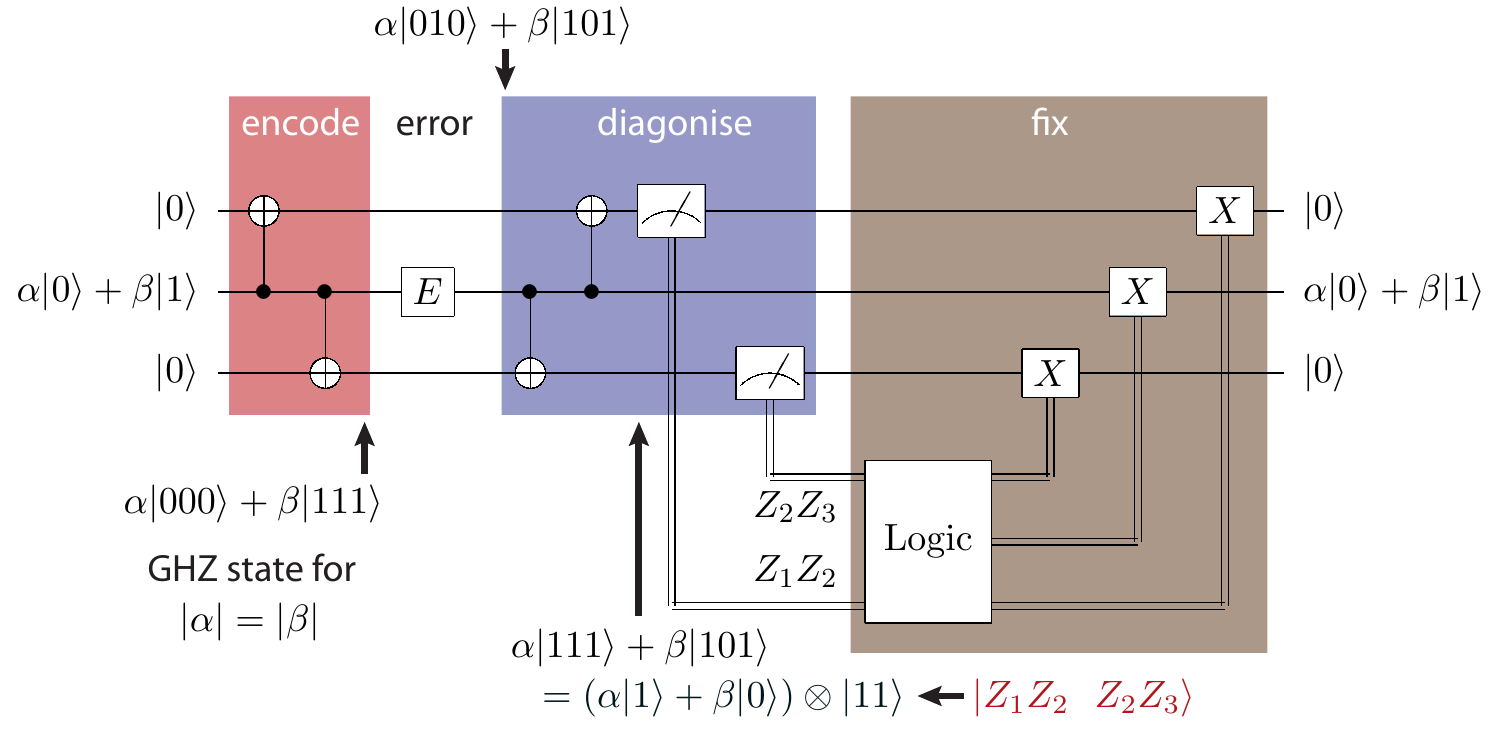}
	\mycaption{Measurement-based QEC circuit}{The code begins by mapping a single-qubit state $\alpha|0\rangle+\beta|1\rangle$ to the three-qubit entangled state $\alpha|000\rangle+\beta|111\rangle$.  A single bit-flip is shown on the middle qubit, making the state $\alpha|010\rangle + \beta|101\rangle$.  The encoding is then reversed, leaving the middle qubit in a quantum state, related to its original state possibly by an error rotation, and the ancilla qubits with the values of the error syndromes.  These qubits are then measured and the results processed with classical logic, which triggers conditional rotations that return the quantum register to its original state.  Small error rotations are projected to full flips or nothing by the measurement process.  This quantum code corrects for a single erroneous rotation of any qubit.}
	{\label{fig:qecmsmtcircuit}}
\end{figure}

We saw in the last section that if we know the values of both $ZZ$ eigenvalues of our GHZ-like state, we can unambiguously detect the presence of a single bit-flip.  But what happens if a finite rotation (that is, a superposition of a bit-flip and no bit-flip) occurs?  It turns out that this is also correctable by the three-qubit quantum repetition code.  To see this, we need to be specific about how the code works.  The code is shown as a ``quantum circuit'' in \figref{fig:qecmsmtcircuit}, where each horizontal line represents the action on a single qubit as a function of ``time.''  The code begins with the situation that we have a qubit in some state that we wish to protect, $|\psi\rangle = \alpha |0\rangle + \beta|1\rangle$, here the second qubit, and two ``ancilla'' qubits -- one and three -- in their ground state.  As described above, the circuit that maps this state $\alpha|0\rangle + \beta|1\rangle \rightarrow \alpha|000\rangle + \beta|111\rangle$ consists of two controlled-NOT gates, depicted by black dots and crossed, open circles connected by a line.  The qubit that has the open circle is flipped if and only if the qubit with the black dot is excited.  After these two gates, the three-qubit manifold is in the encoded GHZ-like state.  Suppose now that one of the qubits is fully flipped; for concreteness, say the second qubit, so the manifold will now be in the state $\alpha|010\rangle + \beta|101\rangle$.

At this point, we wish to detect the presence of an error.  We do this by reversing the encoding step with two more controlled-NOT gates\footnotemark.  For our case of the second qubit being flipped, the qubits will now be in the state $\left( \alpha|1\rangle + \beta|0\rangle \right)\otimes|11\rangle$, where we have first listed the state of the second qubit followed by the state of the first and third qubit.  Notice that the ancilla qubits are no longer in their ground state.  Instead, the ancilla qubits now contain the values of the $ZZ$ eigenvalues between themselves and the middle bit -- the exact error syndromes that we wish to measure.  To see this, consider only two qubits in some state $\alpha|00\rangle + \beta|11\rangle$.  If we perform a cNOT gate on the second qubit, the state will be given by $\alpha|00\rangle + \beta|10\rangle = (\alpha|0\rangle+\beta|1\rangle)\otimes|0\rangle$.  If the first qubit were flipped, the resulting state $\alpha|10\rangle + \beta|01\rangle$ would be mapped to $\alpha|11\rangle + \beta|01\rangle = (\alpha|1\rangle + \beta|0\rangle)\otimes|1\rangle$ by the cNOTs.  If the second qubit were flipped instead, the resulting state would be given by $(\alpha|0\rangle + \beta|1\rangle)\otimes|1\rangle$.  In the first case, the $ZZ$ eigenvalue was $+1$, while in the second and third it was $-1$; the ancilla being left in its excited state indicates that the $ZZ$ eigenvalue was flipped.  

\footnotetext{Reversing the encoding returns our quantum information to a single-qubit state, so we could not detect the presence of additional errors.  In a ``real'' error corrected computer, we are never allowed to store important information in such a risky way.  We must come up with methods of extracting the error syndromes without leaving the safety of our encoded subspace, as discussed below in \sref{subsec:faulttolerance}.}

Immediately after the decoding step, the second qubit is in its original state that may or may not have been flipped and the other two qubits contain the values of the error syndromes $Z_1Z_2$ and $Z_2Z_3$.  Because the values of these correlations contain no information about $\alpha$ and $\beta$, we can proceed by measuring these qubits and using some classical logic to conditionally flip the qubits back as necessary\footnotemark.  In the case of a flipped $Q_2$, the manifold will be $(\alpha|1\rangle + \beta|0\rangle) \otimes |11\rangle$ and we find that we need to flip each qubit to return the manifold to its original state, $(\alpha|0\rangle+\beta|1\rangle)\otimes|00\rangle$.  But what happens if instead of having a full bit-flip, $Q_2$ underwent some small rotation?  Suppose the rotation is by some angle $\theta$; we can define the effective {\it probability} of a full flip to be $p=\mathrm{sin}^2(\theta/2)$.  Immediately after decoding, then, the manifold will be in the state $\sqrt{1-p}(\alpha|0\rangle+\beta|1\rangle)\otimes|00\rangle + \sqrt{p}(\alpha|1\rangle + \beta|0\rangle)\otimes(|11\rangle)$.  That is, the state will be a superposition of $Q_2$ in the correct state with the ancillas indicating no error plus $Q_2$ flipped with the ancillas indicating as such.  If we were to now measure the ancilla qubits, we must either get $00$ or $11$ -- the two-qubit wavefunction will collapse.  If we get $00$, the wavefunction of $Q_2$ will have collapsed to $\alpha|0\rangle + \beta|1\rangle$ -- no error; if we get $11$, it will be $\alpha|1\rangle + \beta|0\rangle$ -- a full flip.  Thus, by measuring the ancilla qubits, we have essentially forced them to {\it decide} whether or not a full error has occurred.  The discreteness of the error syndrome measurement collapses the continuous errors to discrete errors; after the measurement, there are either no errors or there is one bit-flip.  Put another way, the measurement has removed the entropy associated with this error by projecting it into a definite state.  The experimenter need not be aware of the fact that the errors can occur continuously.  Thus error correction still works for superpositions of zero or one full error.  In fact, the code would still work perfectly for superpositions of any of the four possible errors, as long as they never occur simultaneously.

\footnotetext{The fact that these operations are {\it conditional} on a measurement outcome is crucial.  Suppose that our error was not simply a qubit being flipped, but rather that it became entangled with some degree of freedom that we do not control; it is now in an impure state.  Since we do not control those external degrees of freedom, no unitary operation applied to the qubit can reduce this impurity.  The state collapse associated with measurement provides the non-unitarity that we need to unravel this entanglement.  As soon we measure the ancilla, any entanglement with the bath is eliminated.  As we will see later, unconditionally resetting a qubit can serve this purpose as well.}

\subsubsection{Autonomous error correction}
\label{subsubsec:autonomousec}

\begin{figure}
	\centering
	\includegraphics{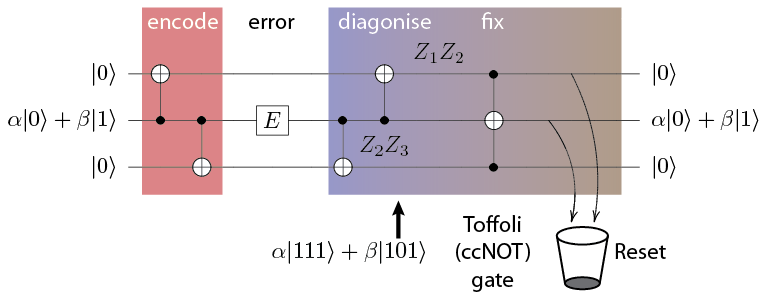}
	\mycaption{Autonomous QEC circuit}{The three-qubit bit-flip code can be modified to eliminate the need for a measurement by implementing the classical logic operation with a coherent controlled-controlled-NOT gate and a non-unitary unconditional reset of the ancilla qubits.}
	{\label{fig:qecautocircuit}}
\end{figure}

\nomdref{Accnot}{ccNOT}{Three-qubit controlled-controlled-NOT gate, also known as Toffoli.}{subsubsec:autonomousec}

The previous section showed how measurement can be used to extract the entropy associated with an error from our quantum state.  In practice, however, it can be challenging to extract information from our quantum state with high fidelity, perform a classical calculation with that information, and manipulate our qubits conditioned on the result, all within a time brief compared to $T_2$.  Fortunately, it is possible to modify the code to eliminate the need for this measurement by using a combination of a coherent multi-qubit operation and an unconditional qubit reset.  As shown in \figref{fig:qecautocircuit}, the first half of the code -- the encoding, error, and decoding steps -- are all the same as shown previously in \figref{fig:qecmsmtcircuit}.  The difference comes after the unencoding and imprinting of the error syndromes in the ancilla qubits.  There, we apply a three-qubit controlled-controlled NOT (ccNOT, also known as Toffoli) gate which flips the target qubit if and only if both control qubits are excited.  This implements the classical logic that would be used in the measurement-based case.  If the ``protected'' qubit were flipped, the ccNOT will flip it back since both $ZZ$ eigenvalues will be $-1$ and the ancillae will be left in their excited state after the unencoding.  This approach will still work with continuous errors because it acts coherently on each subspace.  For example, if we again had $|\psi\rangle = \sqrt{1-p}(\alpha|0\rangle+\beta|1\rangle)\otimes|00\rangle + \sqrt{p}(\alpha|1\rangle + \beta|0\rangle)\otimes(|11\rangle)$ and applied our ccNOT on it, we would be left with the state $|\psi'\rangle = \sqrt{1-p}(\alpha|0\rangle+\beta|1\rangle)\otimes|00\rangle + \sqrt{p}(\alpha|0\rangle + \beta|1\rangle)\otimes(|11\rangle) = (\alpha|0\rangle+\beta|1\rangle)\otimes(\sqrt{1-p}|00\rangle + \sqrt{p}|11\rangle)$.  That is, the protected qubit will be in its correct state and the ancillae will contain the entropy of the error.

At this point, the middle qubit will be in its original state (as long as there were zero or one errors), but we still must return the ancillae to their ground state if we hope to re-encode and correct another error.  Unfortunately, it is not possible to also do this with some multi-qubit unitary gate.  There are two arguments demonstrating this is true: first, a unitary operation cannot reduce the amount of entropy in a system since it would violate the laws of thermodynamics (nor could the amount of entropy change since a unitary is invertible and if entropy increased, its inverse must cause a corresponding decrease).  Since we are treating these errors as random\footnotemark, they increase the entropy of our wavefunction, and therefore cannot be eliminated (only shuffled around, as we have done so far).  Second, if the error we are correcting originates with an interaction with the environment -- perhaps our qubit partially swapped an excitation with some other degree of freedom and is now entangled -- the error will not be correctable because our unitary by convention does not address that subsystem.  Thus, we must introduce some non-unitary operation to extract this entropy.  One method is to  {\it reset} the ancillae to the ground state $|\psi\rangle \rightarrow |0\rangle$  as shown in the final step of the code, by putting them in contact with a cold bath.  (One example of this is described in \sref{subsec:qubitreset}.)

\footnotetext{``Errors'' are not necessarily random.  For example, a systematic under or over-rotation would look to the code like a bit-flip error.  Similarly, if we had some always-on coherent coupling to some other degree of freedom (e.g. the environment), it also would be correctable in principle, even though nothing about it is truly ``random.''  Its preferrable to actually eliminate such errors and reduce the overall error rate, but if that's impossible, error correction could serve as a last line of defense.  Nevertheless, there is no loss of generality of labeling these systematic errors ``entropy'' since they are unknown from the point of view of the code.}

\subsubsection{Phase-flip correction}

The code described thus far corrects only {\it bit-flip} errors, described by the Pauli operator $X$.  That represents only one of the possible rotations we can perform on a qubit.  We can also have $Z$ errors.  ($Y$ errors can be thought of as simultaneous $X$ and $Z$ errors because $iXZ = Y$ and do not count as an additional axis.)  Moreover, $Z$ errors are a much more common type of error for many types of qubits.  For example, a superconducting qubit's frequency can sometimes be changed with applied flux and so noise in that parameter scrambles the phase of a superposition, producing a $Z$ error.  Fortunately, it is possible to modify the bit-flip code to instead correct for phase-flips by changing coordinate systems.  If immediately before (and after) the ``error'' portion of the code we applied a $\pi/2$ ($-\pi/2$) rotation around the $y$-axis (or a Hadamard), the code will correct for phase-flips instead of bit-flips.  This happens because the rotations will map $Z$ errors to $X$ errors and vice versa, since $XY=iZ$, or, more explicitly, $R_y^{\pi/2} R_z^\pi R_y^{-\pi/2} = R_x^{\pi}$ and $R_y^{\pi/2} R_x^\pi R_y^{-\pi/2} = R_z^{\pi}$ (up to a global phase).  These extra rotations can be commuted with the other rotations of the code, so its requirements are no greater than the bit-flip code.

\subsection{The Shor code}

\begin{figure}
	\centering
	\centerline{
	\Qcircuit @C=1em @R=0.8em {
		\lstick{\ket{\psi}} & \ctrl{3} 	& \ctrl{6} 	& \gate{H}	& \ctrl{1}	& \ctrl{2}	& \multigate{8}{E}	& \ctrl{1}	& \ctrl{2} 	& \targ	   	& \gate{H} 	& \ctrl{3} 	& \ctrl{6} 	& \targ 	& \qw & \rstick{\ket{\psi}} \qw \\
		\lstick{\ket{0}} 	& \qw 		& \qw 		& \qw 		& \targ		& \qw		& \ghost{E} 	  	& \targ		& \qw	   	& \ctrl{-1}	& \qw 	   	& \qw 	   	& \qw 	   	& \qw 		& \qw \\
		\lstick{\ket{0}} 	& \qw 		& \qw 		& \qw 		& \qw		& \targ		& \ghost{E} 	  	& \qw		& \targ	   	& \ctrl{-1}	& \qw 	   	& \qw 	   	& \qw 	   	& \qw 		& \qw \\
		\lstick{\ket{0}} 	& \targ 	& \qw 		& \gate{H}	& \ctrl{1}	& \ctrl{2}	& \ghost{E} 	  	& \ctrl{1}	& \ctrl{2} 	& \targ	   	& \gate{H} 	& \targ	   	& \qw	   	& \ctrl{-3}	& \qw \\
		\lstick{\ket{0}} 	& \qw 		& \qw 		& \qw 		& \targ		& \qw		& \ghost{E} 	  	& \targ		& \qw	   	& \ctrl{-1}	& \qw 	   	& \qw 	   	& \qw 	   	& \qw 		& \qw \\
		\lstick{\ket{0}} 	& \qw 		& \qw 		& \qw 		& \qw		& \targ		& \ghost{E} 	  	& \qw		& \targ	   	& \ctrl{-1}	& \qw 	   	& \qw 	   	& \qw 	   	& \qw 		& \qw \\
		\lstick{\ket{0}} 	& \qw 		& \targ 	& \gate{H}	& \ctrl{1}	& \ctrl{2}	& \ghost{E} 	  	& \ctrl{1}	& \ctrl{2} 	& \targ	   	& \gate{H} 	& \qw 	   	& \targ	   	& \ctrl{-3}	& \qw \\
		\lstick{\ket{0}} 	& \qw 		& \qw 		& \qw 		& \targ		& \qw		& \ghost{E} 	  	& \targ		& \qw	   	& \ctrl{-1}	& \qw 	   	& \qw 	   	& \qw 	   	& \qw 		& \qw \\
		\lstick{\ket{0}} 	& \qw 		& \qw 		& \qw 		& \qw		& \targ		& \ghost{E} 	  	& \qw		& \targ	   	& \ctrl{-1}	& \qw 	   	& \qw 	   	& \qw 	   	& \qw 		& \qw
	}
	}                                                                                                                                                                              
	\mycaption{Nine-qubit Shor code}{Arbitrary single-qubit errors can be corrected by concatenating the phase- and bit-flip codes.  This is done by replacing each qubit used in the phase-flip code with three qubits that are bit-flip corrected.  The measurement-free version is shown here; the eight bottom qubits would need to be reset at the end of this operation.}
	{\label{fig:qecshorcode}}
\end{figure}

We have shown how the three-qubit code can correct bit-flip {\it or} phase-flip errors.  But what if we wish to correct both at the same time?  In principle, being able to correct for both $X$ and $Z$ errors (and thus also $Y$ errors) would account for any possible error on a single qubit, including particularly nasty errors like the depolarizing channel.  We can see that this is not possible to achieve with only three qubits by using a counting argument: we need one qubit to store the state we are protecting and two more classical bits to encode the four error syndromes of a single bit or phase flip.  Thus, in order to correct for both errors simultaneously we need more qubits.  One conceptually straightforward way of doing this is the nine-qubit Shor code \cite{Shor1995b}.  The idea of the Shor code is to {\it concatenate} the bit-flip and phase-flip codes together.  As shown in \figref{fig:qecautocircuit}, each ``qubit'' used in the phase-flip code will actually be a composite of three qubits which are themselves bit-flip corrected, so any bit or phase flip of any one of the qubits will be corrected.  The logical basis is then given by $|0\rangle_L = \frac{1}{\sqrt{8}}(|000\rangle + |111\rangle)\otimes(|000\rangle + |111\rangle)\otimes(|000\rangle+|111\rangle)$ and $|1\rangle_L = \frac{1}{\sqrt{8}}(|000\rangle - |111\rangle)\otimes(|000\rangle - |111\rangle)\otimes(|000\rangle - |111\rangle)$, where we would write our encoded state as $\alpha|0\rangle_L + \beta|1\rangle_L$.  We can manipulate this logical qubit with the logical operators $X_L = XXXXXXXXX$ and $Z_L = ZZZZZZZZZ$ (which are {\it not} the same as simultaneous rotations on each qubit individually -- $X_L$ and $Z_L$ are nine-qubit gates).  The error syndromes are given by the operators $S_1 = ZZIIIIIII$, $S_2 = ZIZIIIIII$, $S_3 = IIIZZIIII$, $S_4 = IIIZIZIII$, $S_5 = IIIIIIZZI$, $S_6 = IIIIIIZIZ$, $S_7 = XXXXXXIII$, and $S_8 = XXXIIIXXX$, where the first six discriminate bit-flips of each individual physical qubit and the remaining two detect phase-flips of the three bit-flip corrected composite qubits.  This was one of the first error correcting codes discovered, but is unlikely to be the most practical given its relatively large overhead and complicated logical operations.

\subsubsection{The five-qubit code}

The Shor code is ``inefficient'' in that it can correct for more than one error on a single qubit at a time.  For example, because there is so much redundancy you could have a bit-flip of one of all three ``composite'' qubits and still recover.  It is actually possible to make do with only {\it five} qubits to recover from an arbitrary single-qubit error \cite{Laflamme1996} using the aptly named ``five-qubit code.''  The error syndromes (that is, stabilizer operators) are given by $S_1=XZZXI$, $S_2=IXZZX$, $S_3=XIXZZ$ and $S_4=ZXIXZ$, and the logical operators are $X_L=XXXXX$ and $Z_L=ZZZZZ$.  This is the smallest code possible that can correct any single-qubit error.  Its efficiency can be proven by a simple counting argument.  There are $3 \times 5 + 1 = 16 = 2^4$ errors to detect (a $Z$, $X$, or $Y$ error on each qubit, or no error) and the information of one qubit to encode, thus giving us a total of $4 + 1 = 5$ required qubits.
	
\subsubsection{Other error correcting codes}

There are numerous other error correcting codes that have been discovered over the years.  The 7-qubit Steane code is most similar to those we have previously discussed \cite{Steane1996}. Though requiring two more qubits than minimally necessary for an arbitrary single-qubit error, it has the advantage of being a ``CSS''-type code which eases its analysis and implementation \cite{Calderbank1996, Steane1996, Nielsen2000}.  (CSS codes are named after A. Calderbank, P. Shor, and A. Steane and are a special class of stabilizer codes that are constructed from classical error correcting codes \cite{Nielsen2000}.)  {\it Topological} codes have been developed, which are defined for a two-dimensional (or higher) lattice of qubits \cite{Kitaev1997, Bravyi1998, Fowler2012}.  They have the advantage of being tolerant to relatively high error rates, requiring only nearest-neighbor two-qubit gates for both error detection and logical gate operations, and more conveniently scaling to higher levels of protection than discrete codes, where the only option is concatenation.  More recently, there have been proposals to encode a quantum state into a superposition of coherent states of light \cite{Leghtas2013}, taking advantage of the large Hilbert space and the possibility of fewer types of errors of a harmonic oscillator \cite{Leghtas2012}.

\subsection{Fault tolerance}
\label{subsec:faulttolerance}

It is not enough to be able to correct an error that occurs at some discrete, specified time.  In reality, our computer must be built with the understanding that an error may occur at any point during a calculation, and that the gates we will rely on to correct those errors may themselves be faulty.  A protocol that satisfies these requirements is {\it fault tolerant} \cite{Shor1996, Preskill1998, Gottesman2009}.  To be so, we must not only be able to store a quantum state in a way that can recover from errors, but we must also be able to {\it manipulate} that information using some complete set of logical gate operations without leaving the space of the code.  Put another way, we have to replace all our qubits with {\it logical qubits} that are error corrected.

A more insidious problem that must be dealt with is that our logical gate operations must be performed in a way that do not propagate or magnify errors.  Even if these gates are perfect, they will modify an error that has occurred.  Suppose we have some unitary $\hat{U}$ that acts on a state $|\psi\rangle$ that has suffered an error $E$, so we wish to evaluate $\hat{U} E |\psi\rangle$.  Because of linearity, this is equal to $(\hat{U}E\hat{U}^\dagger)\hat{U}|\psi\rangle$ -- that is, the desired state ($\hat{U}|\psi\rangle$) that has now suffered a new error $\hat{U}E\hat{U}^\dagger$.  If $\hat{U}$ is a single-qubit operation, this may not be such a big deal, since although it will change the error (a Hadamard switches an $X$ error with a $Z$, and so on), our code should be able to mitigate any kind equally well.  A more significant problem arises when $\hat{U}$ is a multi-qubit gate like a cNOT, because it will have the effect of {\it propagating} our error to other qubits.  For example, if the control qubit has suffered an $X$ error, after our cNOT gate, the target qubit will have picked up that error as well -- our single-qubit error has become a two-qubit error that we may not necessarily be able to correct.  In order to be fault-tolerant, then, we must be very careful with our multi-qubit gates so as to avoid this problem.

\begin{figure}
	\centering
	\centerline{
	\Qcircuit @C=1em @R=0.8em {
		 					& \mathrm{~~~~~~encode} & & & & & & & &  \mathrm{correct} & & & & \mathrm{~~~decode}	 \\
		 					&  & & & & & & & & & 	 \\
		\lstick{\ket{\psi}} & \ctrl{1} 	& \ctrl{2}	& \qw	& \multigate{2}{E}	& \qw				& \ctrl{4}	& \qw	   	& \ctrl{5}	& \qw	 	& \qw	 	& \gate{X}		   & \qw 	& \ctrl{2} 	& \ctrl{1} & \rstick{\ket{\psi}} \qw \\
		\lstick{\ket{0}} 	& \targ 	& \qw 	  	& \qw	& \ghost{E} 	  	& \qw				& \qw		& \ctrl{3} 	& \qw		& \qw	  	& \qw 	   	& \gate{X} \cwx	   & \qw	& \qw 		& \targ   & \rstick{\ket{0}} \qw \\
		\lstick{\ket{0}} 	& \qw 		& \targ	  	& \qw	& \ghost{E} 	  	& \qw				& \qw		& \qw	   	& \qw		& \ctrl{3}  & \qw 	   	& \gate{X} \cwx	   & \qw	& \targ		& \qw	  & \rstick{\ket{0}} \qw \\
		 					& 			& 		  	&    	&					&				  	& 			& 		   	& 			& 		 	& 			& \cwx 			   & 		\\
							& 			& 	 	  	&	  	&					& \lstick{\ket{0}}	& \targ		& \targ	   	& \qw		& \qw	   	& \meter 	& \cw \cwx\\
							& 			& 		  	& 		& 					& \lstick{\ket{0}}	& \qw		& \qw	   	& \targ		& \targ	   	& \meter 	& \cw \cwx \\
		 					& & & & & & & & & & & & & & ~
							\gategroup{3}{2}{5}{3}{1.8em}{--}
	                       	\gategroup{3}{12}{9}{7}{0.7em}{--}
							\gategroup{3}{14}{5}{15}{1.8em}{--}                                                                                                                                                   
	}
	}
	\mycaption{Fault-tolerant three-qubit bit-flip code}{Rather than decoding the qubits after the error, one can use two additional ancillae to extract and measure the error syndromes.  This way the quantum information is always protected from bit-flip errors.  Combined with logical gate operations, this constitutes a fault-tolerant implementation of the three-qubit code.}
	{\label{fig:qecthreeft}}
\end{figure}

The solutions to all these problems will not be fully explained here, but we can get the flavor of the necessary changes by looking at how we might modify the three-qubit code.  Since we can never leave the code space, we no longer have the luxury of recycling the ancilla qubits to also extract the error syndromes.  As shown in \figref{fig:qecthreeft}, we instead measure the correlations $Z_1Z_2$ and $Z_2Z_3$ with two additional ancillae through the use of controlled-NOT gates.  In addition to not having our logical qubit leave the subspace, this has the further advantage of potentially allowing us to measure the syndromes several times before making a decision, since the logical qubit is an eigenstate of those operators.  This is helpful if our cNOT or measurement operations are not perfect.  It is also in principle possible to make a fault-tolerant version of autonomous error correction where measurements are not necessary, but it requires a great deal more overhead since the ``classical'' logic can no longer be assumed to be perfect.
	
\section{Conclusions}

This chapter was intended to serve as a brief introduction to quantum information science for people with a background in physics but no particular knowledge of the subject.  We started with a survey of the tools and language used in quantum information: the Bloch sphere, Pauli matrices and operators, the density matrix, and entanglement, to name a few.  This led into a discussion of what makes quantum information processing so powerful but also that harnessing that power to run quantum algorithms has many stringent hardware requirements.  Finally, we showed why quantum information is particularly sensitive to errors, and explained the simplest model of how we might correct them without losing their advantages.  We concluded with a discussion of more sophisticated codes that can correct for arbitrary errors and have different hardware and parameter requirements, and gave an outline of fault tolerant error correction.  With the tools introduced here established, we can now proceed with an explanation of our particular implementation of quantum information processing.  In the next chapter, we discuss superconducting qubits coupled to microwave resonators, known as circuit quantum electrodynamics, with the goal of laying the framework to present our recent experimental achievements.

\setcounter{chapter}{2}
\chapter{Superconducting Qubits and cQED}
\thumb{Superconducting Qubits and cQED}
\lofchap{Superconducting Qubits and cQED}
\label{ch:theory}


\lettrine{T}{he} previous chapter introduced the requirements that a quantum computer must satisfy; however, there was no corresponding statement about the particular physical manifestation of such a machine.  In fact, significant experimental progress has been made using a variety of systems \cite{Ladd2010} including nuclear spins \cite{Vandersypen2005, Ryan2009}, trapped ions \cite{Cirac1995, Leibfried2003, Singer2010, Schindler2011, Monroe2013}, linear optics \cite{Kok2007, Barz2012}, solid state quantum dots \cite{Loss1998, Hanson2007, Maune2012, Shulman2012, Awschalom2013}, diamond color centers \cite{Jelezko2006, Dutt2007, Maurer2012, Awschalom2013}, and superconducting circuits \cite{Devoret2013}.  In addition, there have been dozens of proposals \cite{Dyakonov2012} for additional systems like topologically protected qubits \cite{Nayak2008, Hasan2010, Stern2013}.  Demonstrating the basic building blocks of quantum information science like single-qubit control and on-demand entanglement is a lot different than mastering a system that could be scaled to a fault-tolerant million-qubit system, however.  Thus, the history of quantum computing has seen the rise and fall of several technologies at the experimental forefront.  Though liquid-state NMR systems were always known not to be scalable \cite{Warren1997}, they made rapid initial progress before being superseded by trapped ions.  Recent advances with superconducting circuits\cite{Paik2011, Hatridge2013} and quantum dots \cite{Maune2012, Shulman2012} are now threatening the primacy of ions.  The field is increasingly concerned not only with basic quantum information experiments, but also with which system has a chance to scale to the large number of qubits required to perform calculations or simulations not otherwise possible.

This thesis reports on one of the most promising technologies: superconducting qubits.  Compared to its competitors, superconducting circuits have certain advantages, such as being relatively simple to fabricate using standard lithography techniques \cite{Frunzio2005}, having all-electrical quantum control using microwave light fields \cite{Vion2002, Blais2004}, and large, controllable nonlinearities at small energy scales \cite{Kirchmair2013}.  They require being cooled in a helium dilution refrigerator to milliKelvin temperatures, however, and until recently there were major questions about the limits of their coherence.  There is still a lot of work necessary to build a scalable quantum architecture with these devices, but progress has been very rapid and there is no known reason that will inhibit this technology from advancing further.  Moreover, it has enabled a new class of quantum physics experiments \cite{Hofheinz2009, Johnson2010, Bozyigit2011, Kirchmair2013, Lang2013} that, in addition to being useful for information processing, are intrinsically interesting and beautiful.

This chapter begins with a theoretical introduction to the physics of superconducting qubits and the architecture with which we study them.  These two topics have been discussed in great detail in previous theses \cite{SchusterThesis, BishopThesis, ChowThesis} and papers \cite{Blais2004, Koch2007, Gambetta2007}, and so we summarize rather than exhaustively derive those results.  We focus on the particular kind of qubit, the transmon \cite{Koch2007, Schreier2008}, that we are studying in our lab.  It is conveniently modeled as a nonlinear LC oscillator that has non-degenerate transition frequencies between higher excited states.  These additional levels play an important role in generating entanglement, as will be discussed in \chref{ch:entanglement}.   We couple these qubits to standing-wave modes of microwave resonators in the circuit quantum electrodynamics (cQED) architecture \cite{Wallraff2004, Majer2007}.  The dynamics of cQED are described by the Jaynes-Cummings Hamiltonian \cite{Jaynes1963}.  While being straightforward to derive, it must be approximated and transformed in order to most clearly see the properties we regularly take advantage of for readout, single-qubit gates, and qubit-qubit coupling.  We complete this section by discussing dissipation and the strong dispersive limit, which defines a regime of parameter space where even high-order qubit-cavity coupling terms dominate over decoherence, enabling additional control through the use those effects.

Many of our devices employ flux bias lines (FBLs), which are current-carrying wires used to thread magnetic flux through the qubit SQUID loop.  They allow us to tune qubit transition frequencies in-situ and have been used successfully in the planar cQED architecture \cite{DiCarlo2009, DiCarlo2010}.  The final section of this chapter will discuss our recent extension of FBLs to the 3D cQED architecture.  We will show how to calculate the flux coupling that we can expect for a given geometry, which has been a concern due to the screening effects of the superconducting 3D box, and show that for reasonable geometries it should be possible to thread a flux quantum with a sufficiently small amount of current.  Doing so requires a drastic expansion of the size of the FBL and the SQUID loop that greatly increases the capacitance between the two, creating a new channel for qubit decay.  We show how to calculate the expected qubit lifetime using an effective circuit model and how we can add simple filtering to fix the problem.

\section{Superconducting qubits}
\label{sec:scqubits}

One thing that distinguishes superconducting qubits from virtually all of their competitors is the fact that the quantum degree of freedom is not that of a single or small number of discrete particles.  Instead, superconducting qubits employ the collective motion of a macroscopic number of paired electrons known as Cooper pairs.  These particles have condensed into a single collective ground state.  Electrical circuits built using superconductors will exhibit quantum behavior with energy levels set by their geometric parameters.  They can in principle be lossless if we limit ourselves to dissipationless elements like the capacitor and the inductor.  However, because these are both linear elements, the only Hamiltonian we could engineer with them is the harmonic oscillator.  They would have have precisely evenly spaced levels and the only free parameter is the frequency of that oscillator.  A system must have individually addressable transitions if we intend to control two of its levels as a qubit; linear elements are insufficient.

\nomdref{Cic}{$I_c$}{Josephson junction critical current}{sec:scqubits}
\nomdref{Clj}{$L_J$}{Josephson inductance}{sec:scqubits}
\nomdref{Gwphi0}{$\Phi_0$}{magnetic flux quantum given by $h/2e$}{sec:scqubits}

To our great fortune, superconductors are endowed with the only known simultaneously nonlinear and non-dissipative circuit element\footnotemark, the Josephson junction \cite{Josephson1962, Josephson1974}.  It is the source of anharmonicity used by all types of superconducting qubits.  There are several ways of making a Josephson junction, which is most generally any weak link between two superconductors.  By far the most common method is via two superconducting wires separated by a thin insulating oxide.  Cooper pairs may coherently tunnel across this barrier, which leads to a phase difference $\phi$ between the macroscopic wavefunction on one side of the barrier compared with the other.  The condensates on either side are in the ground state of the BCS Hamiltonian, so this phase difference is the only possible low-energy degree of freedom.  The tunneling supercurrent is given by $I=I_c \mathrm{sin}\phi(t)$, where the critical current $I_c$ is the maximum current that may flow through the junction and is set by parameters like the effective barrier transparency.  The phase $\phi(t)$ evolves in time in the presence of a potential difference $V$ across the junction according to
\begin{equation}
	\hbar \frac{\partial\phi}{\partial t}=2eV.
\end{equation}
Taking the full time derivative of $I$, we have
\begin{equation}
	\dot{I}=(I_c \mathrm{cos}\phi)\frac{\partial\phi}{\partial t} = \frac{2e V I_c}{\hbar}\mathrm{cos}\phi.
\end{equation}
We note that this equation looks like the inductor relationship $V=-L \dot{I}$ with a Josephson inductance given by
\begin{equation}
	L_J = \frac{\hbar}{2 e V I_c \mathrm{cos}\phi}  = \frac{\Phi_0}{2\pi I_c \mathrm{cos}\phi}
\end{equation}
where $\Phi_0=h/2e$ is the magnetic flux quantum.  Thus, the Josephson junction can be thought of as an inductor with an inductance that depends on $\phi$, and thus the applied current, and is therefore nonlinear.

\footnotetext{One possibly exception to this statement is the field of cavity quantum optomechanics, where the low-energy interaction between light and mechanical objects is studied \cite{Gigan2006, Thompson2008, OConnell2010, Teufel2011, Purdy2013}.  The radiation pressure of light couples to the mechanical motion of some oscillator, which can endow the Hamiltonian with a nonlinearity.}

\subsection{The Transmon qubit}
\label{subsec:transmon}

In the early days of superconducting quantum information science, there was a proliferation of qubit circuit topologies that were classified as flux \cite{Mooij1999, Friedman2000, Chiorescu2003}, phase \cite{Yu2002, Martinis2002}, and charge qubits \cite{Bouchiat1998, Nakamura1999, Lehnert2003} based on which degree of freedom was a good quantum number.  As time went on, these distinctions became less and less relevant as ``hybrid'' superconducting qubits were developed that had better properties (e.g. decreased susceptibility to specific sources of noise) and whose eigenstates were no longer number states of any of the three quantities \cite{Vion2002, Koch2007, Schreier2008}.  One such hybrid qubit is known as the {\it transmon}.  Though normally classified as a charge qubit, it enjoys an exponentially suppressed sensitivity to charge noise because its eigenstates are superpositions of several charge states \cite{Koch2007}.  It is often sufficient to consider this qubit as a slightly anharmonic LC oscillator, though as we will see, calculating the detailed level structure requires a more careful treatment.

\subsubsection{The Cooper-pair box}

\nomdref{Acpb}{CPB}{Cooper-pair box}{subsec:transmon}
\nomdref{Ccsum}{$C_\Sigma$}{total qubit capacitance to ground}{subsec:transmon}
\nomdref{Cej}{$E_J$}{Josephson energy}{subsec:transmon}
\nomdref{Cec}{$E_C$}{electrostatic charging energy}{subsec:transmon}
\nomdref{Cnhat}{$\hat{n}$}{Cooper pair number operator}{subsec:transmon}
\nomdref{Cng}{$n_g$}{gate charge}{subsec:transmon}
\nomdref{Gwphihat}{$\hat{\phi}$}{junction phase operator}{subsec:transmon}
\nomdref{Galpha}{$\alpha$}{transmon anharmonicity}{subsec:transmon}

The transmon qubit is a derivative of the Cooper-pair box qubit (CPB).  The CPB is topologically simple, consisting of two superconducting islands connected by a Josephson junction \cite{Bouchiat1998, Nakamura1999}.  Cooper pairs may tunnel across this junction, and since there is only one path for this transit, the number of tunneled pairs is an integer quantity.  This also implies that the junction phase $\phi$ is ``compact'' or periodic in $2\pi$.  (This would not be true if, for example, the junction was shorted out by an inductor as it is with a phase qubit.  There, the potential energy landscape has an overall $\phi^2$ dependence on top of the Josephson sinusoid.  $\phi$ is then non-compact due to the closed-loop topology of the circuit.)  There are two relevant energy scales: $E_C=\frac{e^2}{2 C_\Sigma}$,  the charging energy of a single electron stored on the capacitance, and $E_J=\frac{I_C \Phi_0}{2\pi}$, the energy associated with an electron tunneling across the junction.  $C_\Sigma=C_g+C_J$ is the total capacitance of the CPB to ground, given by the sum of the geometric capacitance and the capacitance of the Josephson junction itself.  The CPB Hamiltonian is then
\begin{equation}
\label{eq:cpbham}
	\hat{H}=4E_C (\hat{n} - n_g)^2 - E_J \mathrm{cos}\hat{\phi}
\end{equation}
where $\hat{n}$ is the integer number of Cooper pairs that have tunneled through the junction, $n_g$ is some offset ``gate'' charge representing an external voltage bias, and $\phi$ is the phase across the junction and is periodic in $2\pi$.  $\hat{n}$ and $\hat{\phi}$ are canonically conjugate variables, where $\hat{n} = \frac{\partial}{\partial \hat{\phi}}$ and $[\hat{n}, e^{\pm i \hat{\phi}}]=\pm e^{\pm i \hat{\phi}}$ \footnotemark.

\footnotetext{If $\hat{\phi}$ is non-compact, we would have $[\hat{n}, \phi]=1$.  When $\phi$ is compact, only operators like $e^{i\hat{\phi}}$ that respect the periodicity of $\hat{\phi}$ are valid.}

We can build some intuition for this Hamiltonian  with a physical analog of a charged quantum rotor, following the argument of Koch {\it et al.} \cite{Koch2007}.  Suppose we have some point-mass $m$ attached to a massless rod of length $l$, which is free to rotate around a fixed pivot.  The potential energy of the mass in a uniform gravitational field is $V=-mgl \mathrm{cos}(\phi)$, where $\phi$ is the angle of the rod relative to its equilibrium position.  Angular momentum is given by $\hat{L}_z = (\vec{r} \times \vec{p}) \cdot \hat{z} = i \hbar \frac{\partial}{\partial \phi}$, so our full Hamiltonian is 
\begin{equation}
	\hat{H}_{\mathrm{rotor}} = \frac{\hat{L}_z^2}{2 m l^2} - m g l \mathrm{cos} \hat{\phi} .
\end{equation}
Since $\hat{L}_z/\hbar$ has integer eigenvalues, we can map $\hat{n}\leftrightarrow \hat{L}_z/\hbar$, $E_J \leftrightarrow mgl$, $E_C \leftrightarrow \hbar^2 / 8 m l^2$, and we find that this is identical to \equref{eq:cpbham} with $n_g=0$.  We can add that term by stipulating that the mass carries some charge and is in a uniform magnetic field parallel to the rotor's pivot, and thus $\vec{p} \rightarrow \vec{p} - q \vec{A}$, giving $L_z \rightarrow L_z + \frac{1}{2} q B_0 l^2$, where the vector potential $A = B_0(-y,x,0)/2$.  This gives $n_g \leftrightarrow qB_0l^2/2\hbar$, completing our one-to-one mapping of the Cooper-pair box Hamiltonian to the charged quantum rotor.

\subsubsection{Transmon regime}

\begin{figure}
	\centering
	\includegraphics{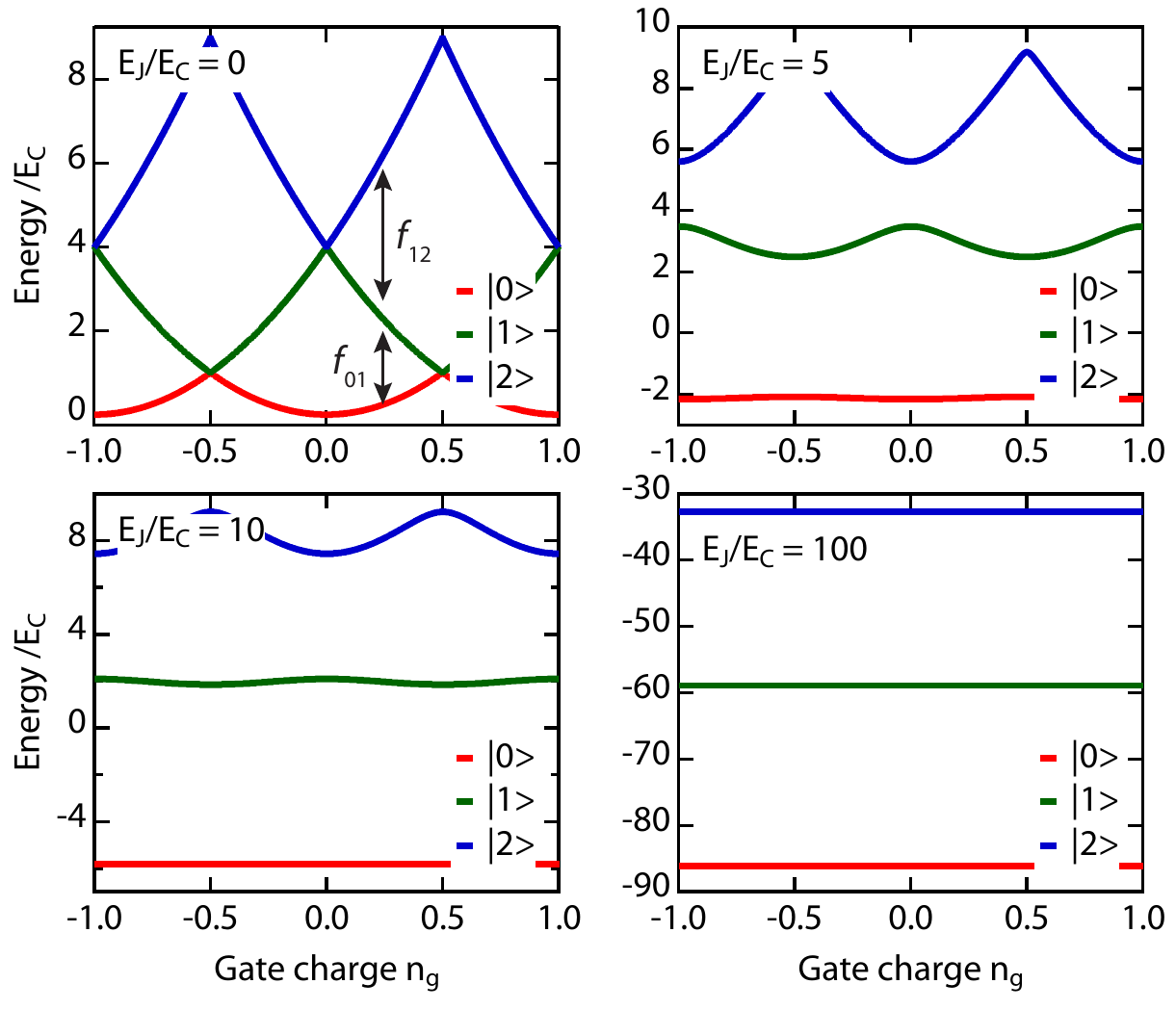}
	\mycaption{Transmon energies as a function of gate charge}{We plot the energy of the solutions to \equref{eq:cpbham} as a function of gate charge $n_g$.  We show the results for various values of $E_J/E_C$, demonstrating the reduction in gate charge sensitivity.}
	{\label{fig:transmonenergy}}
\end{figure}

The gate charge $n_g$ is actually a major nuisance because its value is subject to charge noise.  Electrons drifting around in the environment will cause it to change at random.  If that changes the transition frequency of our qubit, it will induce dephasing.  Fortunately, we can suppress our sensitivity to gate charge by operating our qubit in the ``transmon'' regime where the ratio $E_J/E_C$ is large.  In this limit, variation in transition frequency between two subsequent levels as a function of gate charge\footnotemark ~ scales as $e^{-\sqrt{8 E_J/E_C}}$.  The cost, a small $E_C$, reduces the anharmonicity of the qubit.  However, this effect scales only geometrically with $E_C$ whereas our sensitivity to charge noise scales exponentially -- a huge net win \cite{Koch2007}.  We show numerically-evaluated energies for various values of $E_J/E_C$ as a function of $n_g$ in \figref{fig:transmonenergy}.  This shows that for low values of $E_J/E_C$ the transition energies depend strongly on $n_g$, but for larger values the dependence flattens out.

\footnotetext{\label{foot:matheiu}
In the WKB approximation, the Mathieu characteristic values can be approximated to give the charge dispersion for the {\it m}th level of the transmon as $\epsilon_m \sim (-1)^m E_C \frac{2^{4m+5}}{m!}\sqrt{\frac{2}{\pi}}\left(\frac{E_J}{2 E_C}\right)^{\frac{m}{2}+\frac{3}{4}}e^{-8\sqrt{8E_J/E_C}}$.  Note that the size of charge dispersion increases with the excitation level.}

Large $E_J/E_C$ corresponds to limiting our junction to small phase excursions from $\phi=0$.  For our rotor model, this is imposed when both the gravitational pull and moment of inertia are very large, so the angle $\phi$ only experiences small fluctuations.  In this limit, we can expand the cosine potential as $\hat{H}=\hat{H}_0 + V$, where $\hat{H}_0$ is a linear term given by $4E_C \hat{n}^2 + \frac{E_J}{2}\hat{\phi}^2$ and $V$ is the perturbation
\begin{equation}
	V = E_J \left( -\frac{1}{4!} \hat{\phi}^4 + \frac{1}{6!}\hat{\phi}^6+...\right).
\end{equation}
We identify $\hat{H}_0$ as a simple harmonic oscillator with frequency $\hbar \omega=\sqrt{8E_J E_C}$.  As we mentioned earlier, to zeroth order, the transmon is an LC oscillator with the Josephson junction serving as a large linear inductor.  Importantly, the gate charge $n_g$ does not enter into this calculation.  By limiting ourselves to small values of $\phi$, the boundary condition that $\phi = \phi + 2 \pi$ becomes irrelevant, meaning that $\phi$ can be treated as non-compact.  We are free to impose vanishing boundary conditions on our wavefunction for $\phi\rightarrow\infty$.  This transforms the conjugate momentum $\hat{n}$ from an integer-valued discrete quantity into a continuous quantity and we can therefore neglect the offset term $n_g$.  Limiting ourselves to small phase fluctuations corresponds to large charge fluctuations, which in turn disperse the discreteness of charge \cite{SMG_Singapore}.

We continue our calculation in second quantization with the substitution $\hat{\phi}=\phi_{\mathrm{ZP}}(\hat{a}+\hat{a}^\dagger)$, where $\phi_{\mathrm{ZP}}^2=\frac{\hbar \omega}{2 E_J} = \sqrt{\frac{2 E_C}{E_J}}$.  Plugging this in, we have $\hat{H}_0 = \hbar \omega \hat{a}^\dagger \hat{a}$, the normal harmonic oscillator equation, and $V\approx\frac{-1}{12} E_C (\hat{a}+\hat{a}^\dagger)^4$.  Expanding the term $(\hat{a}+\hat{a}^\dagger)^4$ and dropping rapidly-rotating terms like $\hat{a}^\dagger \hat{a}^\dagger$ (the rotating-wave approximation), we get $V\approx-\frac{E_C}{2}(\hat{a}^\dagger \hat{a}^\dagger \hat{a} \hat{a} + 2 \hat{a}^\dagger \hat{a})$ \cite{SMG_LesHouches}.  Combining terms, we find that the oscillator frequency has been renormalized to 
\begin{equation}
\label{eqn:transmonenergy}
	\hbar \omega' = \sqrt{8 E_J E_C} - E_C
\end{equation}
and has an anharmonicity given by 
\begin{equation}
	\alpha = -E_C.
\end{equation}
Thus we have arrived at our original description of the transmon qubit as a slightly anharmonic LC oscillator with Hamiltonian
\begin{equation}
\label{eqn:anharmonicqubit}
	\hat{H} = \hbar \omega' (\hat{a}^\dagger\hat{a} + 1/2) + \hbar \frac{\alpha}{2} \hat{a}^\dagger \hat{a}^\dagger \hat{a} \hat{a}.
\end{equation}
The transition frequency from ground to first excited state, given by $\hbar \Omega_{01} = \hbar \omega'$, is different from the transition between the first and second excited state $\hbar \Omega_{12} = \hbar \Omega_{01} - \alpha$.  This grants us the ability to control the levels directly.  The anharmonicity $\alpha=-E_C$ is on the order of $\sim 3-5\%$ of $\hbar \omega'$ in the transmon limit, where $E_J / E_C \sim 50-100$.  In typical experiments this is large enough ($\sim 200 \mhz$) to allow for controlling microwave pulses on the order of a few nanoseconds.  The view of the transmon as an anharmonic LC oscillator is particularly relevant with the recent development of the 3D cQED architecture \cite{Paik2011} and black-box quantization \cite{Nigg2011}, where it is useful to numerically calculate parameters in the basis of \equref{eqn:anharmonicqubit}.

\subsubsection{The charge basis}

\begin{figure}
	\centering
	\includegraphics{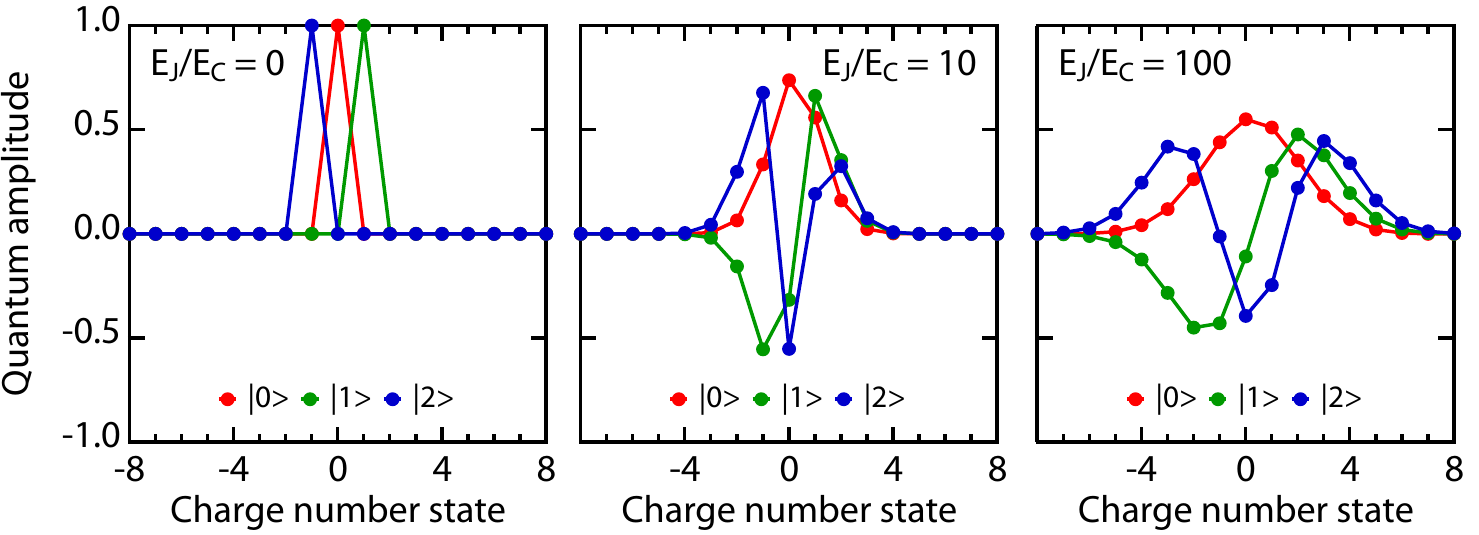}
	\mycaption{Transmon eigenstates in the charge basis}{We plot the solution to \equref{eqn:chargecpb} with $n_g=0.25$ and values of $E_J/E_C$ of 0, 10, and 100.  Notice as this ratio is increased, the amount of hybridization of the eigenstates increases as well.}
	{\label{fig:transmonchargebasis}}
\end{figure}

If we wish to relax our assumptions about the size of $E_J/E_C$, the eigenstates of \equref{eq:cpbham} can also be solved analytically using special ``Mathieu'' functions \cite{BishopThesis}.  In practice, however, these are difficult to evaluate, so it is often more convenient to numerically solve \equref{eq:cpbham} in the charge basis.  Using $\hat{n} = i \frac{\partial}{\partial \hat{\phi}}$ and the commutation relation $[\hat{n}, e^{i\hat{\phi}}] = e^{i\hat{\phi}} \implies e^{i \hat{\phi}}|n\rangle = |n+1\rangle$, we rewrite the Hamiltonian as
\begin{equation}
	\label{eqn:chargecpb}
	\hat{H}=4E_C \sum_{j=-N}^{N}(j-n_g)^2 |j\rangle\langle j| - E_J \sum_{J=-N}^{N-1} |j+1\rangle\langle j| + |j\rangle \langle j+1|
\end{equation}
where we truncate the Hilbert space at some size $N$ large enough to encompass the levels we are interested in evaluating.  In this representation, the underlying tunneling of Cooper pairs is clear, with the first term corresponding to the capacitive energy stored in the charge configuration and the second term to the tunneling across the junction.  We write the solution to this equation in the {\it transmon basis} which diagonalizes it, given by
\begin{equation}
	\hat{H}_q=\hbar \sum_k \omega_k |k\rangle \langle k |
\end{equation}
where $|k\rangle$ are the exact solutions to the CPB Hamiltonian (e.g. in the $\phi$ basis, the Mathieu functions) and $\omega_k=E_k\hbar$ are the eigenenergies.  We will often approximate the $|k\rangle$ with the numerical solutions of \eref{eqn:chargecpb}, with which we can compute much faster.  Several such solutions for various values of $E_J/E_C$ are shown in \figref{fig:transmonchargebasis}.

When the higher excited states are not important, we can approximate the transmon oscillator as a two-level qubit by restricting our sum to only the ground and first excited state.  Ignoring the existence of the higher levels (for example, taking the limit that the anharmonicity goes to infinity), we approximate the qubit Hamiltonian as a spin-$1/2$ particle with 
\begin{equation}
	\label{eq:qubitasaspin}
	\hat{H}_q=\frac{\hbar}{2} \omega_q \hat{\sigma}_z.
\end{equation}
Here, we would assign $\omega_q = \omega'$, the value that we calculated in the previous section.  This approach lends itself more readily to the language of quantum information science, as we discussed in \chref{ch:concepts}.

\subsubsection{Flux sensitivity}
\label{subsubsec:fluxsensitivity}

\nomdref{Asquid}{SQUID}{superconducting quantum interference device}{subsubsec:fluxsensitivity}
\nomdref{Cejmax}{$E_J^{\mathrm{max}}$}{maximum value for the effective $E_J$ under flux tuning}{subsubsec:fluxsensitivity}

Both the Cooper-pair box and the transmon are often designed with a pair of Josephson junctions connecting the islands rather than only one.  This configuration, known as a superconducting quantum interference device or SQUID, allows for the tuning of the effective $E_J$ by changing the magnetic flux $\Phi$ threading through the loop formed by the two junctions.  The Josephson contribution to the CPB Hamiltonian with two junctions is given by
\begin{equation}
	\hat{H}_J = E_{J1} \mathrm{cos}(\theta_1) + E_{J2}\mathrm{cos}(\theta_2)
\end{equation}
where the two junctions have their own Josephson energies $E_{Ji}$ and superconducting phase difference $\theta_i$.  Due to flux quantization (see section 3.2.3 of David Schuster's thesis \cite{SchusterThesis}), the phase difference $\phi = \theta_1 - \theta_2$ can be written in terms of the loop magnetic flux $\Phi$, with $\phi = 2\pi \Phi / \Phi_0$, where $\Phi_0 = h/2e$ is the flux quantum.  Combining this identification with trigonometric substitutions, we write
\begin{equation}
	\label{eqn:splittransmon}
	\hat{H}_J=(E_{J1}+E_{J2})\mathrm{cos}\left( \pi \frac{\Phi}{\Phi_0}\right)\mathrm{cos}\theta + (E_{J2}-E_{J1})\mathrm{sin}\left(\pi \frac{\Phi}{\Phi_0}\right)\mathrm{sin}\theta
\end{equation}
where $\theta = \frac{\theta_1 + \theta_2}{2}$.  This is identical to the term in \equref{eq:cpbham} when $E_{J1}=E_{J2}$ with the effective Josephson energy
\begin{equation}
	\label{eqn:ejeff}
	E_{J}^{\mathrm{eff}} = (E_{J1}+E_{J2})\mathrm{cos}\left( \pi \frac{\Phi}{\Phi_0}\right).
\end{equation}
This ``symmetric junction'' case shows that we can tune the effective Josephson energy of a qubit with the applied magnetic field.  The transition frequency, following \equref{eqn:transmonenergy}, is given by
\begin{equation}
	\label{eqn:transmonfluxtune}
	h f_{01}\approx\sqrt{8 E_C (E_{J1}+E_{J2})|\mathrm{cos}\left(\pi \Phi/\Phi_0\right)|}-E_C.
\end{equation}

\begin{figure}
	\centering
	\includegraphics{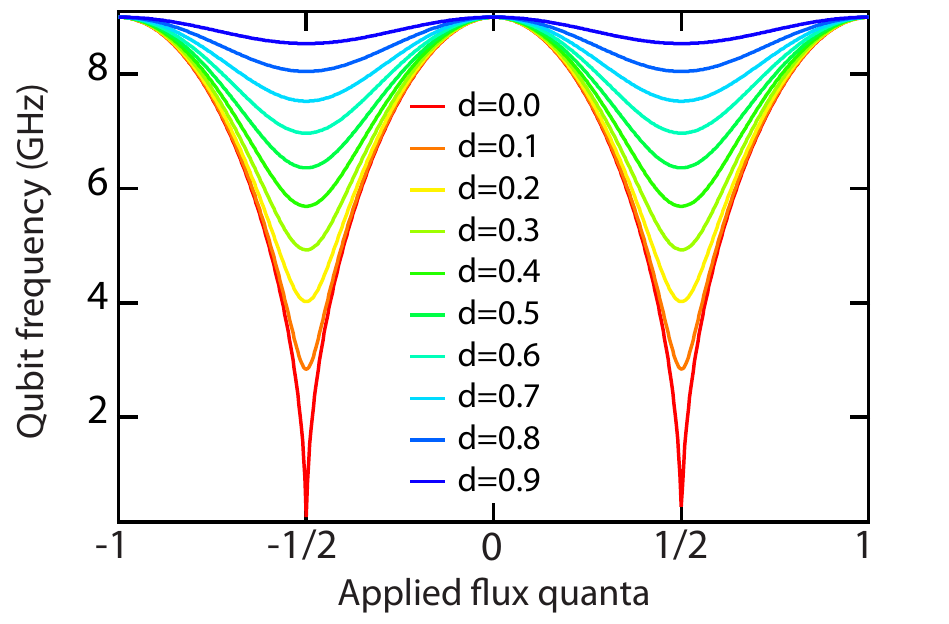}
	\mycaption{Transmon flux tuning as a function of junction asymmetry}{With perfectly symmetric junctions, as a function of applied flux the transmon transition frequency will tune between its maximum frequency down to zero, according to $\omega\sim\sqrt{\left|\mathrm{cos}\left( \pi \Phi / \Phi_0\right)\right|}$.  When the junctions are different sizes, set by the asymmetry parameter $d$, a minimum frequency develops and the qubit is less sensitive to flux.}
	{\label{fig:fluxtune}}
\end{figure}

Things are a bit more complicated when the junctions are not identical.  Defining the asymmetry parameter $d$ as
\begin{equation}
	d=\frac{E_{J1} - E_{J2}}{E_{J1} + E_{J2}}
\end{equation}
we can write \equref{eqn:splittransmon} as
\begin{equation}
	\hat{H}_J = E_{J}^{\mathrm{eff}} \sqrt{1+d^2 \mathrm{tan}^2 \left(\frac{\pi \Phi}{\Phi_0}\right)}\mathrm{cos}(\phi - \phi_0)
\end{equation}
where $E_J^{\mathrm{eff}}$ is given by \equref{eqn:ejeff} and the phase offset $\phi_0$ is given by the transcendental equation $\mathrm{tan}\phi_0 = d \mathrm{tan}(\pi \Phi / \Phi_0)$ and can be eliminated by a variable transformation \cite{Koch2007}.  Thus, the effective $E_J$ now not only has a maximum value $E_{J}^{\mathrm{max}} = E_{J1} + E_{J2}$ but also a minimum value $E_{J}^{\mathrm{min}} = d(E_{J1} + E_{J2}) = E_{J1} - E_{J2}$ in the limit that $\phi\rightarrow \pi/2$.  In practice, $d \lesssim 0.1$, and so this minimum frequency is too low to be experimentally important, but if we intentionally make the junctions quite asymmetric we can engineer a minimum frequency that is relatively high.  As we will see in the next section, this ability may prove interesting if we wish to have more than one ``flux sweet spot.''

\subsection{Flux noise}

Just as charge noise affecting the qubit transition frequency looks like a random $\hat{Z}$ process and causes dephasing (which motivated the development of the transmon), noise in the flux $\Phi$ will also dephase split-junction qubits.  Following Koch {\it et al.}, the dephasing time due to flux noise is given to first order by
\begin{equation}
	\label{eqn:firstorderfluxnoise}
	T_{\phi} = \frac{\hbar}{A} \left| \frac{\partial E_{01}}{\partial \Phi} \right|^{-1} = \frac{\hbar}{A} \frac{\Phi_0}{\pi}\left( 2 E_C E_J^{\mathrm{max}} \left| \mathrm{sin}\frac{\pi\Phi}{\Phi_0} \mathrm{tan} \frac{\pi\Phi}{\Phi_0} \right| \right)^{-1/2}
\end{equation}
where the parameter $A$ is the RMS flux noise\footnotemark, and typically is on the order of $10^{-5} \Phi_0$.  For fixed $A$, the only control we have over this is given by the slope of the frequency curve -- essentially the qubit's susceptibility to this noise.  $T_{\phi}$ goes to infinity for $\Phi = n \Phi_0$, which is known as a {\it flux sweet spot}.  The dephasing time of course does not go to infinity, but rather only becomes second-order sensitive to the flux noise.  We must take the next-order derivative of $\omega(\Phi)$, giving
\begin{equation}
	T_{\phi} = \left| \frac{\pi^2 A^2}{\hbar} \frac{\partial^2 E_{01}}{\partial\Phi^2} \right|_{\Phi=0}^{-1} = \frac{\hbar \Phi_0^2}{A^2 \pi^4 \sqrt{2E_J^{\mathrm{max}} E_C}}.
\end{equation}
At a flux bias of $\Phi=\Phi_0/4$, $A=10^{-5}\Phi_0$, $E_J=30\ghz$, and $E_C = 350\mhz$, \equref{eqn:firstorderfluxnoise} gives $T_\phi \sim 1\us$, while at the flux sweet spot $\Phi=0$, these same parameters give $T_\phi \sim 3~\mathrm{ms}$.  Thus, it is always preferable to operate as close as possible to a flux sweet spot; a lot of effort has gone into engineering $f_{\mathrm{max}}$ for this exact reason.  As we saw in the previous section, any amount of junction asymmetry gives rise to two sweet spots.  In principle, one could choose the junction sizes carefully to set these two frequencies at experimentally relevant values.

\footnotetext{Treating flux noise properly is a bit difficult.  \Eref{eqn:firstorderfluxnoise} is a statement that $T_{\phi}^{-1} = \delta \omega_{\mathrm{RMS}} = \frac{\partial \omega}{\partial \Phi} \Phi_{\mathrm{RMS}}$.  Our job is to define $\Phi_{\mathrm{RMS}}$ in a reasonable way.  Since we are assuming the noise follows a $1/f$ frequency dependence, we have to define some frequency range that we care about.  Thus, $\Phi_{\mathrm{RMS}}^2 = \int_{f_{\mathrm{min}}}^{f_{\mathrm{max}}} \frac{A^2}{f}df = A^2 \mathrm{ln}(\frac{f_{\mathrm{max}}}{f_{\mathrm{min}}})$.  This logarithmic factor is of order unity, so we usually approximate $\Phi_{\mathrm{RMS}} = A$.  Strictly speaking, however, $A$ sets only the scale of noise and not its precise magnitude.}

\section{Circuit quantum electrodynamics}
\label{sec:cqed}

\nomdref{acqed}{cQED}{circuit quantum electrodynamics}{sec:cqed}

The field of circuit quantum electrodynamics (cQED) concerns the study of superconducting qubits that are strongly coupled to a mode of light \cite{Wallraff2004, Blais2004}.  Conventionally, this mode was a standing wave in a 1-dimensional microwave-frequency transmission line \cite{Wallraff2004, Majer2007, DiCarlo2009}, but more recently the resonant modes of a 3D box have supplanted that role \cite{Paik2011}.  This mode, commonly known as the ``cavity'' in reference to the older and recently Nobel prize-winning field of {\it cavity} quantum electrodynamics, fills many important roles.  First, because the frequency of the cavity mode is sensitive to the state of the qubit(s), we can infer those states with a cavity transmission measurement \cite{Wallraff2005,Gambetta2007}.  Second, the cavity gives us an easy way to perform single-qubit rotations with resonant microwave tones \cite{Chow2009}.  Third, as we will see in \sref{subsec:purcelleffect}, it protects the qubit from spontaneous decay by changing the density of photon states \cite{Houck2008}.  Fourth, and finally, it mediates qubit-qubit coupling via a virtual coupling \cite{Majer2007, DiCarlo2009}.  We do not seek to re-derive all the properties of cQED here given the plentiful pre-existing resources that do an excellent job \cite{Blais2004, Koch2007, SMG_LesHouches, SchusterThesis, BishopThesis, ChowThesis}.  We will instead briefly summarize the results of these sources that we will reference in later chapters.

\subsection{Harmonic oscillators}
\label{subsec:harmonicoscillators}

\nomdref{Cd}{$\hat{D}$}{harmonic oscillator displacement operator}{subsec:harmonicoscillators}
\nomdref{Galphaket}{$\ket{\alpha}$}{coherent state}{subsec:harmonicoscillators}
\nomdref{Cadaggera}{$\hat{a}^\dagger$, $\hat{a}$}{creation and annihilation operators for a resonator}{subsec:harmonicoscillators}

Both transmission lines and 3D cavity modes are, in the absence of a qubit, linear harmonic oscillators\footnotemark.  The Hamiltonian of a single harmonic oscillator is given by
\begin{equation}
	\hat{H}_{HO} = \hbar \omega \left( \hat{a}^\dagger \hat{a} + 1/2 \right) = \hbar \omega \left( \hat{N}+ 1/2 \right)
\end{equation}
where $\hat{a}^\dagger$ and $\hat{a}$ are the raising and lowering operators and $\hat{N} = \hat{a}^\dagger \hat{a}$ is the number operator.  The operators satisfy the relation $[\hat{a}, \hat{a}^\dagger]=1$.  We can define a {\it number basis} $|n\rangle$ which is an eigenvector of the number operator with $\hat{N}|n\rangle = n|n\rangle$.  We can also define a conjugate set of variables, charge $\hat{Q}$ and magnetic flux $\hat{\Phi}$ (or equivalently, position and momentum) as $\hat{Q} = -i Q_{\mathrm{ZPF}} (\hat{a} - \hat{a}^\dagger)$ and $\hat{\Phi} = \Phi_{\mathrm{ZPF}} (\hat{a} + \hat{a}^\dagger)$, where $Q_{\mathrm{ZPF}}$ and $\Phi_{\mathrm{ZPF}}$ are the magnitude of the zero-point fluctuations of each variable and can be defined in terms of the physical parameters of the oscillator.  These variables correspond to physical observables of the charge on the capacitor and the flux through the inductor of the effective circuit of the oscillator, and obey the relation $[\hat{\Phi},\hat{Q}]=i \hbar$.

\footnotetext{See, for example, Chapter 2 of Steven Girvin's 2011 Les Houches notes for a modern approach to how circuits are quantized \cite{SMG_LesHouches}.  Other useful resources are Chapter 3 of David Schuster's 2008 thesis \cite{SchusterThesis} and Chapter 2 of Lev Bishop's 2010 thesis \cite{BishopThesis}.  Circuit quantization involves assembling a Lagrangian for the system following an explicit set of instructions and applying a Legendre transformation to get the Hamiltonian.  A quantum LC oscillator is found to have the Lagrangian $L(\phi, \dot{\phi}) = \frac{C \dot{\phi}^2}{2} - \frac{\phi^2}{2L}$, where $\phi$ is the flux in the inductor and $L$ and $C$ are the inductance and capacitance of the circuit, respectively.  From this, the Hamiltonian is found to be the common harmonic oscillator with $\hat{H}=\frac{\hat{q}^2}{2C} + \frac{\hat{\phi}^2}{2L}$, with $\hat{q}$ the operator for charge on the capacitor.  This can be written in terms of raising and lowering operators as $\hat{H}=\hbar \omega\left( \hat{a}^\dagger \hat{a} + \frac{1}{2}\right)$ with $[\hat{a},\hat{a}^\dagger]=1$, $\phi=\sqrt{\frac{\hbar Z}{2}}(\hat{a}+\hat{a}^\dagger)$, $q=-i\sqrt{\frac{\hbar}{2Z}}(\hat{a}-\hat{a}^\dagger)$, $\omega=\frac{1}{\sqrt{LC}}$, and $Z=\sqrt{\frac{L}{C}}$. }

We can calculate the eigenstates of this Hamiltonian in second quantization \cite{Townsend2000}.  First, we note that $\hat{a}^\dagger|n\rangle = \sqrt{n+1}|n+1\rangle$ and $\hat{a}|n\rangle = \sqrt{n}|n-1\rangle$, which implies that $\hat{a}|0\rangle = 0$.  Writing $\hat{a} = \frac{1}{2}\left(\frac{\hat{\Phi}}{\Phi_{\mathrm{ZPF}}} + i \frac{\hat{Q}}{Q_{\mathrm{ZPF}}} \right)$, and noting that $\hat{Q} = \frac{\hbar}{i}\frac{\partial}{\partial \hat{\Phi}}$, we have 
\begin{equation}\begin{split}
	\label{eq:shogs}
	\langle \Phi | a | 0\rangle =&~ \frac{1}{2} \langle \Phi | \left( \frac{\hat{\Phi}}{\Phi_{\mathrm{ZPF}}} + i \frac{\hat{Q}}{Q_{\mathrm{ZPF}}} \right)|0\rangle = \frac{1}{2} \langle \Phi | \left( \frac{\hat{\Phi}}{\Phi_{\mathrm{ZPF}}} +  \frac{i}{Q_{\mathrm{ZPF}}}\frac{\hbar}{i}\frac{\partial}{\partial \hat{\Phi}} \right)|0\rangle  \\
	 =&~ \frac{1}{2} \left( \frac{\Phi}{\Phi_{\mathrm{ZPF}}} \langle \Phi|0\rangle + \frac{\hbar}{Q_{\mathrm{ZPF}}} \frac{\partial}{\partial \hat{\Phi}} \langle \Phi |0\rangle \right) \\
	 =&~ 0
\end{split}\end{equation}
where in the second line we have used $\langle \Phi | \hat{\Phi} | 0\rangle = \Phi \langle \Phi | 0\rangle$ and $\langle \Phi | \frac{\partial}{\partial \hat{\Phi}} | 0\rangle = \frac{\partial}{\partial \Phi} \langle \Phi | 0\rangle$.  Solving, we find the first-order single-variable differential equation 
\begin{equation}
	\label{eq:shode}
	\frac{\partial}{\partial \Phi} \langle \Phi | 0\rangle = \frac{-Q_{\mathrm{ZPF}}}{\hbar \Phi_{\mathrm{ZPF}}} \Phi \langle \Phi | 0 \rangle
\end{equation}
whose solution is given by the gaussian
\begin{equation}
	\langle \Phi | 0\rangle = N e^{\frac{-Q_{\mathrm{ZPF}}}{2\hbar \Phi_{\mathrm{ZPF}}} \Phi^2} = N e^{-\frac{\Phi^2}{2 \Phi_{\mathrm{ZPF}}^2}}
\end{equation}
with the normalization factor $N=\left( \frac{Q_{\mathrm{ZPF}}}{\pi \hbar \Phi_{\mathrm{ZPF}}} \right)^{1/4}$ and noting that $Q_{\mathrm{ZPF}} \Phi_{\mathrm{ZPF}}=\hbar$.  We can also get the higher excited states of the harmonic oscillator by applying the raising operator $\hat{a}^\dagger = \frac{1}{2}\left(\frac{\hat{\Phi}}{\Phi_{\mathrm{ZPF}}} - i \frac{\hat{Q}}{Q_{\mathrm{ZPF}}} \right)$, so that
\begin{equation}\begin{split}
	\langle \Phi | n \rangle =& \frac{1}{\sqrt{n!}} \langle \Phi | (\hat{a}^\dagger)^n |0\rangle \\
	=& \frac{1}{2^n\sqrt{n!}} \left(\frac{\hat{\Phi}}{\Phi_{\mathrm{ZPF}}} - \frac{\hbar}{Q_{\mathrm{ZPF}} } \frac{\partial}{\partial \Phi} \right)^n  \left( \frac{Q_{\mathrm{ZPF}}}{\pi \hbar \Phi_{\mathrm{ZPF}}} \right)^{1/4} e^{\frac{-Q_{\mathrm{ZPF}}}{2\hbar \Phi_{\mathrm{ZPF}}} \Phi^2}.
\end{split}\end{equation}
Omitting normalization, we find for example
\begin{equation}
	\langle \Phi | 1\rangle = \Phi e^{\frac{-Q_{\mathrm{ZPF}}}{2\hbar \Phi_{\mathrm{ZPF}}} \Phi^2}.
\end{equation}

In the absence of some nonlinearity, this quantization of photon number is not physically obvious.  The classical controls that we possess over a cavity are not sources of a definite number of photons, but rather perform what is known as a {\it displacement} on the oscillator.  This is because the oscillator's levels are exactly evenly spaced so we cannot address individual transitions.  For example, if we apply a resonant drive to a cavity in its ground state, we will initially transfer population from $|0\rangle$ to $|1\rangle$.  However, as soon as there is population in $|1\rangle$, transitions to $|2\rangle$ are possible, which then turns on transitions to $|3\rangle$, and so on.  Thus, we can only create some Poisson-distributed superposition of number states.  The unitary displacement operator describing this process is given by
\begin{equation}
	\hat{D}(\alpha)=e^{\alpha \hat{a}^\dagger - \alpha^*\hat{a}}.
\end{equation}
The parameter $\alpha$ is a complex number, with $\alpha = |\alpha|e^{i\phi}$.  Working in the photon number basis, if we apply this displacement operator on the ground state, we will produce a {\it coherent state}
\begin{equation}
	\hat{D}(\alpha)|0\rangle = |\alpha\rangle = e^{-\frac{|\alpha|^2}{2}}\sum_{n=0}^{\infty}\frac{\alpha^n}{\sqrt{n!}} |n\rangle.
\end{equation}
This state is an eigenstate of the lowering operator, with 
\begin{equation}
	\hat{a} |\alpha\rangle = \alpha |\alpha\rangle. 
\end{equation} 
A real-valued displacement corresponds to adding a constant to the flux (position) coordinate of the harmonic oscillator, while an imaginary value corresponds to adding charge (momentum).  For example, a displaced ground state wavefunction is given by
\begin{equation}
	\langle \Phi |\hat{D}(\lambda)|0\rangle = \langle (\Phi - \lambda) |0\rangle = \langle \Phi | \lambda \rangle = \psi_\lambda(\Phi)
\end{equation}
where $\lambda$ is some real-valued displacement along the flux axis.  As a function of time, a displaced state will oscillate between being displaced in flux and in charge, with
\begin{equation}\begin{split}
	\hat{a}(t) |\alpha\rangle =&~ e^{i\omega t} \alpha |\alpha\rangle\\
	\implies |\alpha(t)\rangle =&~ |\alpha e^{i\omega t}\rangle.
\end{split}\end{equation}
It can be shown that any possible linear drive of the form $\hat{H}_d = f(t)(\hat{a}+\hat{a}^\dagger)$ can only ever produce a coherent state, independent of the form of $f(t)$\footnotemark. 

\footnotetext{As a corollary, if we ever want to have some interesting ``non-classical'' state of light, we require some sort of nonlinearity to get out of coherent state purgatory.  A qubit turns out to be a great source of nonlinearity, with several recent results creating and detecting photon number Fock states \cite{Johnson2010, Hofheinz2008} and Schr\"{o}dinger cat states \cite{Kirchmair2013, Vlastakis2013}.}

Both the transmission line and 3D box have higher modes (e.g. the first mode of the transmission line has voltage nodes only at the capacitors, but higher modes will have additional nodes), so we can expand our treatment to explicitly take them into account with
\begin{equation}
	\hat{H}_{nHO} = \hbar \sum_n \omega_n \left( \hat{a}^\dagger_n \hat{a}_n + 1/2 \right)
\end{equation}
where $\omega_n$ is the frequency of the $n$th mode.  Enumerating these modes can be done analytically in the case of a transmission line\footnotemark, but is more complicated with a 3D box and generally requires a numerical simulation of the actual box geometry.  These higher modes are often neglected, but understanding them can be important because they can be a source of decay \cite{Houck2008, Reed2010} and dephasing \cite{Sears2012}, or an additional resource for measurement or storage of quantum information \cite{Eichler2011}.

\footnotetext{For a transmission line resonator of length $d$, capacitance per unit length $c$, and inductance per unit length $l$, the resonant frequencies are approximately $\omega_n = n\pi/d\sqrt{lc}$ for integer $n$.  The approximation comes in from the capacitive loading from the input and output capacitors, which can also be included but would be more complicated.}

\subsection{Qubit-cavity coupling}
\label{subsec:qubitcavitycoupling}

\nomdref{Gbeta}{$\beta$}{voltage division ratio}{subsec:qubitcavitycoupling}
\nomdref{Cgij}{$g_{ij}$}{transmon dipole coupling energy between charge levels $i$ and $j$}{subsec:qubitcavitycoupling}
\nomdref{Cg}{$g$}{vacuum-Rabi coupling frequency, $g=g_{01}$}{subsec:qubitcavitycoupling}
\nomdref{Gzomegar}{$\omega_r$}{resonator transition frequency}{subsec:qubitcavitycoupling}
\nomdref{Gzomegaq}{$\omega_q$}{qubit ground to excited state transition frequency}{subsec:qubitcavitycoupling}
\nomdref{Gdelta}{$\Delta$}{qubit-cavity detuning, $\Delta=\omega_q-\omega_r$}{subsec:qubitcavitycoupling}
\nomdref{Gxchi}{$\chi$}{state-dependent dispersive cavity shift}{subsec:qubitcavitycoupling}
\nomdref{Gxchiij}{$\chi_{ij}$}{state-dependent dispersive cavity shift with respect to the $i$ to $j$ transmon transition}{subsec:qubitcavitycoupling}
\nomdref{Gkappa}{$\kappa$}{cavity linewidth; photon relaxation rate}{subsec:qubitcavitycoupling}
\nomdref{Ajc}{JC}{Jaynes-Cummings}{subsec:qubitcavitycoupling}
\nomdref{Arwa}{RWA}{rotating wave approximation}{subsec:qubitcavitycoupling}
\nomdref{Cbdaggerb}{$b^\dagger$, $b$}{creation and annihilation operators for the qubit in the harmonic approximation}{subsec:qubitcavitycoupling}

\begin{figure}
	\centering
	\includegraphics[scale=1]{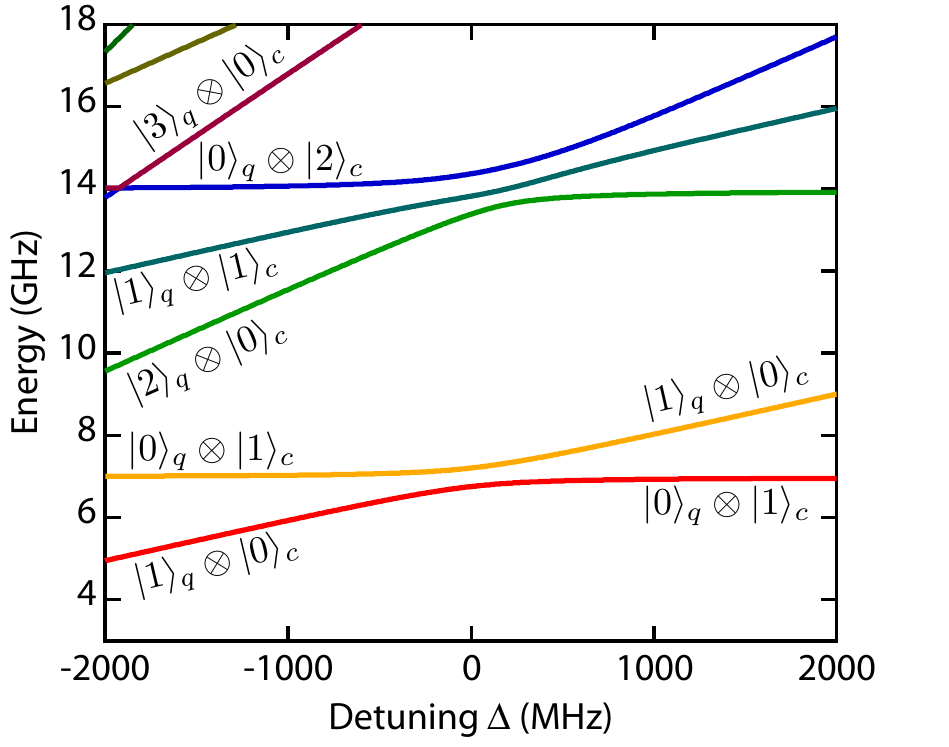}
	\mycaption{cQED eigenenergies}{The solutions to \equref{eq:transmonjc} are plotted as a function of the detuning between the qubit transition frequency between the 0 and 1 state, $\omega_q$, and the bare cavity frequency $\omega_r$.  Eigenstates are labeled according to the excitation levels of the qubit and cavity.  Here, the parameters $\omega_r = 7\ghz, \alpha \approx E_C = 300\mhz, g = 200\mhz,$ and $n_g=0.25$ were used.  The transmon was numerically solved with 21 charge number states and the cavity with 5 photon levels.}
	{\label{fig:cqedsol}}
\end{figure}

In cQED, we are interested in coupling a qubit to the standing-wave mode of a harmonic oscillator.  By placing the qubit in a region where the electric field of the cavity will induce a voltage difference across the two islands, we can attain a coupling between the two degrees of freedom \cite{Koch2007}.  The coupled Hamiltonian is given by
\begin{equation}
	\hat{H} = \hbar \omega_r \hat{a}^\dagger \hat{a} + 4 E_C(\hat{n} - n_g)^2 - E_J \mathrm{cos} \hat{\phi} + 2 \beta e V^0_{\mathrm{rms}} \hat{n}(\hat{a} + \hat{a}^\dagger)
\end{equation}
where $\omega_r$ is the cavity frequency, $\beta= C_g / C_{\Sigma}$ the ratio of the effective capacitance $C_g$ of one side of the transmon to the cavity to total capacitance\footnotemark, and $V^0_{\mathrm{rms}}$ is the root-mean squared voltage at the qubit position per photon of the oscillator.  We can re-write this Hamiltonian in the transmon basis $|i\rangle$ as
\begin{equation}
	\label{eq:nonrwacqed}
	\hat{H} = \hbar \omega_r \hat{a}^\dagger \hat{a} + \hbar \sum_j \omega_j |j\rangle \langle j | + \hbar \sum_{i,j} g_{ij} |i\rangle \langle j | (\hat{a} + \hat{a}^\dagger)
\end{equation}
where the dipole coupling energies are
\begin{equation}
	\hbar g_{ij} = 2 \beta e V^0_{\mathrm{rms}} \langle i | \hat{n} | j\rangle.
\end{equation}
The first term of \equref{eq:nonrwacqed} is the energy of the harmonic oscillator, the second the energy of the transmon, and the third a term that couples the two.  This equation further simplifies if we apply the rotating wave approximation (RWA), where we drop rapidly-rotating terms that do not conserve energy (for example, terms that promote the energy of both the transmon and the photon), giving
\begin{equation}
	\label{eq:transmonjc}
	\hat{H} = \hbar \omega_r \hat{a}^\dagger \hat{a} + \hbar \sum_j \omega_j |j\rangle \langle j | + \left(\hbar \sum_i g_{i,i+1} |i\rangle \langle i | + 1| \hat{a}^\dagger + h.c.\right).
\end{equation}
We have plotted the solutions to this Hamiltonian as a function of qubit frequency in \figref{fig:cqedsol}.

\footnotetext{For the planar architecture, this gate capacitance $C_g$ can be worked out by inverting a five by five capacitance network of the two islands of the transmon and the center pin and two ground planes of the coplanar waveguide cavity.  See, for example, Figure 2.8 in Lev Bishop's 2010 thesis \cite{BishopThesis}.  In the case of the 3D resonator, this geometric picture is no longer correct and it is necessary to numerically simulate the impedances seen by the junctions.}

In the limit of a small number of transmon levels, we can simplify this expression by substituting the qubit with the anharmonic oscillator of \equref{eqn:anharmonicqubit}.  Labeling the cavity degrees of freedom with the operator $\hat{a}$ and the qubit with $\hat{b}$, we have
\begin{equation}
	\hat{H} =  \hbar \omega_r \hat{a}^\dagger \hat{a} + \hbar \omega_q \hat{b}^\dagger \hat{b} - \frac{\alpha}{2} \hat{b}^\dagger \hat{b}^\dagger \hat{b} \hat{b} + \hbar g\left(\hat{a}^\dagger \hat{b} + \hat{a} \hat{b}^\dagger \right).
\end{equation}
We can further simplify this by ignoring the higher levels entirely and, following \equref{eq:qubitasaspin}, treat it like a spin. The result is the {\it Jaynes-Cummings Hamiltonian}
\begin{equation}
	\label{eq:jch}
	\hat{H} = \hbar \omega_r \left(\hat{a}^\dagger \hat{a} + \frac{1}{2} \right) + \frac{\hbar \omega_q}{2} \hat{\sigma}_z + \hbar g \left(\hat{a}^\dagger \sigma_- + \hat{a} \sigma_+\right)
\end{equation}
where $g = g_{0,1}$, $\sigma_\pm$ is the raising and lowering operator of the spin, and where we have again included the vacuum energy of the harmonic oscillator for arbitrary reasons.

Studying these equations is arguably the goal of the remainder of this thesis, but we can immediately highlight some physics by diagonalizing the Jaynes-Cummings Hamiltonian.  Because of our simplifications, the Hamiltonian is a 2 by 2 block diagonal matrix, with dressed eigenstates
\begin{equation}\begin{split}
	|n,+\rangle =&~ \mathrm{cos}(\theta_n)|n-1,e\rangle + \mathrm{sin}(\theta_n)|n,g\rangle \\
	|n,-\rangle =&~ -\mathrm{sin}(\theta_n)|n-1,e\rangle + \mathrm{cos}(\theta_n)|n,g\rangle
\end{split}\end{equation}
where $\mathrm{tan}(2\theta_n) = \frac{2 g\sqrt{n}}{\Delta}$.  This pair of states constitute the {\it Jaynes-Cummings ladder} and parametrized by the total number of excitations $\widehat{n_{tot}} = \hat{a}^\dagger \hat{a} + \sigma_+ \sigma_->0$.  Each pair constitutes an {\it excitation manifold}\footnotemark, though for $n=0$, we have only one state $|0\rangle = |0,g\rangle$.  The energies of these states are given by $E_0 = -\frac{\hbar\Delta}{2}$ and $E_{n,\pm}=n\hbar\omega_r \pm \frac{\hbar}{2}\sqrt{4g^2n+\Delta^2}$ where $\Delta=\omega_q-\omega_r$ is the detuning between the qubit and cavity.  When $\Delta=0$ and the qubit and cavity are in resonance, we are in the {\it vacuum-Rabi} regime, where the eigenstates are symmetrically-weighted odd and even combinations of $|n-1,e\rangle$ and $|n,g\rangle$.  If you were to prepare one of the undressed eigenstates at this point, the excitation would oscillate between the qubit and cavity at the rate $g$ -- a ``vacuum-Rabi oscillation.''

\footnotetext{When talking about multiple-level transmon qubits, the number of states per excitation manifold grows with the excitation number.  As we saw in \figref{fig:cqedsol}, with only one excitation, we have (in the undressed basis) only two states: $|n=1, q=0\rangle$ or $|n=0, q=1\rangle$.  For two excitations, we now can have $|n=2,q=0\rangle$, $|n=1,q=1\rangle$, or $|n=0,q=2\rangle$, and so on.  As you can imagine, with several multi-level qubits at some reasonable level of excitation, the level diagram becomes very complicated (see, for example, \figref{fig:chev003}).  

Incidentally, the reason that the Hamiltonian is block diagonal is that it commutes with the total number of excitations $\hat{N}_{\mathrm{ex}} \equiv \hat{a}^\dagger \hat{a} + \frac{1+\hat{\sigma}_z}{2}$.  This structure gives this ladder of states.  Each block is given by
\begin{equation*}
	\hat{H}_{\mathrm{2x2}}^{(n+1)}/\hbar = \left(\begin{array}{@{}cc@{}}
		n \omega_c + \frac{1}{2} \omega_{01} & g \sqrt{n+1} \\
		g \sqrt{n+1} & (n+1)\omega_c - \frac{1}{2} \omega_{01}
	\end{array}\right).
\end{equation*} 
}


\subsection{Dispersive limit and qubit readout}
\label{subsec:dispersivelimit}

We can further approximate the full transmon Hamiltonian (\equref{eq:jch}) in the {\it dispersive limit}, where $g_{j,j+1} \ll \omega_{j+1,j}-\omega_c$.  Using a unitary transformation \cite{Koch2007}, we can expand the Hamiltonian in powers of $g/\Delta$, giving to second order \cite{BishopThesis}
\begin{equation}\begin{split}
	\label{eq:dispersivetransmoncqed}
	\hat{H}/\hbar =&~ \omega_r \hat{a}^\dagger \hat{a} + \sum_i \left( \omega_i |i\rangle \langle i | + \chi_{i,i+1} |i+1 \rangle \langle i+1|\right) \\
	&~ - \chi_{01} \hat{a}^\dagger \hat{a}|0\rangle \langle 0| + \sum_{i=1}^{\infty}  \left(\chi_{i-1,i}-\chi_{i,i+1}\hat{a}^\dagger \hat{a}\right)|i\rangle\langle i|
\end{split}\end{equation}
where $\chi_{ij} = \frac{g_{ij}^2}{\omega_{ij} - \omega_c}$ are known as dispersive couplings.  Here, the first term is again the cavity and the second the transmon qubit energy.  The third and fourth terms constitute a renormalization of the transmon and cavity energy levels due to their coupling; the fifth term corresponds to the state-selective dispersive shifts.  We can again take the two-level approximation of the qubit, giving the {\it dispersive Jaynes-Cummings} or {\it number splitting} Hamiltonian
\begin{equation}
	\label{eq:dispersivejc}
	\hat{H} = \hbar \left( \omega'_c + \chi \hat{\sigma}_z \right) \hat{a}^\dagger \hat{a} + \frac{\hbar}{2}\omega'_q \hat{\sigma}_z.
\end{equation}
Here, both the qubit transition frequency $\omega'_q = \omega_{01} + \chi_{01}$ and the cavity transition frequency $\omega'_c = \omega_c - \chi_{12}/2$ are renormalized by a quantity known as the {\it Lamb shift}.  The dispersive cavity shift $\chi = \chi_{01} - \frac{\chi_{12}}{2}$ can be approximated for transmons as $\chi \approx \frac{g^2}{\Delta}\frac{E_C}{\hbar\Delta-E_C}$, with $\Delta=\omega_{01}-\omega_C$.

\subsubsection{Number splitting}

\Eref{eq:dispersivejc} deserves some analysis.  We see that the term corresponding to the exchange of excitations between the qubit and cavity is gone, replaced by the term $\chi \hat{\sigma}_z \hat{a}^\dagger\hat{a}$.  There are two ways of interpreting this term.  First, we could group it with the qubit $\sigma_z$ operator, giving $\hat{H}l_q = \hbar \frac{\sigma_z}{2} \left( \omega_q + 2 \chi \hat{a}^\dagger \hat{a} \right)$.  The qubit frequency then depends on the number of photons in the cavity.  Explicitly, the qubit transition frequency will be $\omega_q$ if the cavity is in its ground state, but $\omega_q + 2 \chi$ if there is one photon, $\omega_q + 4 \chi$ for two, and so on.  If there is a superposition of photon number states in the cavity (for example, if there is a coherent or thermal state), the qubit transition frequency will split into many peaks, a phenomenon known as {\it number splitting} \cite{Schuster2007}.  When these peaks are resolved (in the strong dispersive limit, discussed below), it is possible to perform number-selective pulses on the qubit, thus entangling the qubit with the number state of the cavity.  This turns out to be extremely useful for a variety of quantum optics \cite{Johnson2010, Leghtas2013} and quantum information processing applications, as we will discuss in \chref{ch:tunable}.

\subsubsection{Qubit readout}

Alternatively, we could group this extra term with the cavity number operator, giving us $\hat{H}_c = \hbar \left( \omega_c + \chi \sigma_z \right) \hat{a}^\dagger \hat{a}$.  The cavity frequency now depends on the qubit state, moving $2\chi$ lower in frequency when the qubit is excited.  For the case of transmon qubits, both the ground and excited state peaks are moved up in frequency relative to the undressed cavity frequency, but the ground state is shifted more, by what we call $2\chi$.  Thus, in that case, $\omega_c$ does not have physical significance and is instead a convenient frequency with which to write the Hamiltonian.  This dispersive shift is the mechanism that we will use to measure the qubit: by measuring the frequency of the cavity, we can infer the state of the qubit.

Consider the situation where we pulse on a drive of strength $\epsilon_{\mathrm{rf}}$ at some detuning $\delta_{\mathrm{rf}}=\omega_{\mathrm{RF}}-\omega_r$ from the bare cavity frequency.  Ignoring any nonlinearity of the cavity itself, the resulting coherent state in the cavity will depend on the state of the qubit, and will be given by \cite{Gambetta2006, SchusterThesis}
\begin{equation}
	\alpha_{\pm} = \frac{\epsilon_{\mathrm{rf}}}{  \frac{\kappa_{\mathrm{in}} + \kappa_{\mathrm{out}}}{2} + i(\delta_{\mathrm{rf}} \pm \chi)}
\end{equation}
where the subscript denotes the state of the qubit.  This state has an amplitude 
\begin{equation}
	|\alpha_{\pm}|^2 = n_{\pm} = \frac{\epsilon_{\mathrm{rf}}^2}{  \frac{(\kappa_{\mathrm{in}} + \kappa_{\mathrm{out}})^2}{4} + (\delta_{\mathrm{rf}} \pm \chi)^2}.
\end{equation}
The amplitude inside the cavity is given by the product of the drive power with the input coupling strength, $\epsilon_{\mathrm{rf}} = \epsilon_{\mathrm{in}} \sqrt{\kappa_{\mathrm{in}}}$.  We can detect these states by measuring the power leaking out of the cavity in a transmission experiment.  In the simple case where $\chi \gg \kappa$, if we drive on resonance with the ground state cavity peak and get a relatively high transmission, the qubit is in its ground state; if we get a relatively low transmission, the qubit is in its excited state.  It is also possible to drive such that the {\it amplitudes} of $\alpha_\pm$ are the same, but the {\it phase} is different; which choice is optimal depends on system parameters.  The distinguishability of these two states in either case is given by $D=e^{-|\alpha_+ - \alpha_-|^2}$, and corresponds to the overlap of two gaussians on the complex plane.  As we will see in \sref{sec:dispersivereadout}, optimizing this {\it dispersive readout} is a complicated task, especially when accounting for the nonlinearity that the cavity inherits from its qubit coupling.

The measurement operator corresponding to this operation is $|0\rangle\langle 0|$.  What happens if more than one qubit is coupled to the single cavity?  We can generalize the dispersive Hamiltonian as
\begin{equation}
	\label{eq:multidispersivejc}
	\hat{H}/\hbar = \left( \omega'_c + \sum_i\chi^{(i)} \hat{\sigma}_z^{(i)}\right) \hat{a}^\dagger \hat{a} + \frac{1}{2}\sum_i\omega_q^{(i)} \hat{\sigma}_z^{(i)}.
\end{equation}
The cavity frequency will then have a distinct frequency for each eigenstate of the computational state manifold.  For example, with two qubits, the cavity will be found at $\omega'_c$ for $|00\rangle$ but $\omega'_c+\chi^{(1)}$ for $|10\rangle$, $\omega'_c+\chi^{(2)}$ for $|01\rangle$, and $\omega'_c+\chi^{(1)}+\chi^{(2)}$ for $|11\rangle$.  Then, if we measure at $\omega'_c$ and get a high transmission in the limit that the peaks are well-resolved, we will have projected the system with the operator $|00\rangle\langle00|$.  For $N$ qubits the measurement operator can be understood to first order as $|0^{\otimes N}\rangle\langle 0^{\otimes N}|$.  As we will see in \sref{sec:statetomo}, it is somewhat more complicated when we include the fact that the cavity responses have finite overlaps.

\subsection{Single-qubit gates}
\label{subsec:singlequbitgatethry}

\nomdref{Goxi}{$\xi$}{external drive strength}{subsec:singlequbitgatethry}
\nomdref{Gzomega}{$\Omega_R$}{Rabi frequency}{subsec:singlequbitgatethry}

The Jaynes-Cummings Hamiltonian is not the complete story: we must also include the effect of a microwave drive.  Conceptually, a drive can be understood as a coupled second cavity mode that is displaced \cite{BishopThesis} and will leak energy into our primary cavity.  We can approximate its state in the limit that its displacement is large and coupling small as being constant and write
\begin{equation}
	\hat{H}_{d} = (\hat{a}+\hat{a}^\dagger)\left(\xi e^{-i\omega_dt} + \xi^* e^{i \omega_d t} \right) = \hat{a} \xi^* e^{i\omega_d t} + \hat{a}^\dagger \xi e^{-i\omega_d t}
\end{equation}
where $\xi$ is a parameter that defines the strength of the driving (given, in our toy model, by the product of the coupling times the displacement) and in the second part we have applied the rotating wave approximation (assuming $\xi$ is much smaller than any relevant transition frequency).  We can add this term to \eref{eq:transmonjc}, and by applying a unitary transformation to go into the rotating frame of the drive, have
\begin{equation}
	\hat{H}_r / \hbar = (\omega_r - \omega_d) \hat{a}^\dagger \hat{a} + \sum_j (\omega_j - j \omega_d) |j\rangle \langle j | + \left(\sum_i g_{i,i+1} |i\rangle \langle i + 1| \hat{a}^\dagger + h.c.\right) + \hat{a} \xi^* + \hat{a}^\dagger \xi.
\end{equation}

In this form the drive is acting on the cavity state, but we are interested in the dynamics of the {\it transmon}.  To see those instead, we can make another unitary transformation, the Glauber displacement transformation \cite{BishopThesis,Gambetta2006}, producing
\begin{equation}\begin{split}
	\hat{H} / \hbar =&~ \Delta_r \hat{a}^\dagger \hat{a} + \sum_j \Delta_j |j\rangle \langle j | + \left(\sum_i g_{i,i+1} |i\rangle \langle i + 1| \hat{a}^\dagger + |i+1\rangle \langle i| \hat{a}\right) \\
	&+ \frac{1}{2} \sum_j \left(\Omega_R^*(t) |j\rangle\langle j+1| + \Omega_R(t) |j+1 \rangle \langle j |\right)
\end{split}\end{equation}
where $\Delta_r = \omega_r - \omega_d$, and $\Delta_j = \omega_j - j \omega_d$.   $\Omega_R(t)=2g\alpha(t)$ is the Rabi frequency of the drive with $\alpha(t)$ the applied displacement as a function of time; we have $\Omega_R = 2\xi g/\Delta_r$ for the case of a time-independent drive.  Taking the dispersive limit as we did in the previous section, this further simplifies to
\begin{equation}\begin{split}
	\hat{H}/\hbar =&~ \Delta_r \hat{a}^\dagger \hat{a} + \sum_j \Delta_j |j\rangle \langle j | + \sum_j  \chi_{j,j+1} |j+1 \rangle \langle j+1| -  \chi_{01} \hat{a}^\dagger \hat{a}|0\rangle \langle 0| \\
	& + \sum_{j=1} (\chi_{j-1,j}-\chi_{j,j+1}\hat{a}^\dagger \hat{a})|j\rangle\langle j|) + \frac{1}{2} \sum_j  (\Omega_R^*(t) |j\rangle\langle j+1| + \Omega_R(t) |j+1 \rangle \langle j |).
\end{split}\end{equation}
Finally, treating the qubit as a two-level system, we arrive at the {\it driven dispersive Jaynes-Cummings Hamiltonian}
\begin{equation}
	\label{eqn:drivendispersivejc}
	\hat{H} = \frac{\hbar}{2} \Delta_q \hat{\sigma}_z + \hbar(\Delta_r + \chi \hat{\sigma}_z)\hat{a}^\dagger \hat{a} + \hbar(\Omega_R^*(t)\hat{\sigma}_- + \Omega_R(t)\hat{\sigma}_+).
\end{equation}
Choosing $\Omega_R(t) = \Omega_R^x(t)\mathrm{cos}(\omega_d t) + \Omega_R^y(t)\mathrm{sin}(\omega_d t)$,
\begin{equation}
	\hat{H} = \frac{\hbar}{2} \Delta_q \hat{\sigma}_z + \hbar(\Delta_r + \chi \hat{\sigma}_z)\hat{a}^\dagger \hat{a} + \frac{\hbar}{2}(\Omega_R^x(t)\hat{\sigma}_x + \Omega_R^y(t)\hat{\sigma}_y)
\end{equation}
and we see that by choosing the phase and amplitude of the drive, we can perform arbitrary rotations of our qubit about the $x$- and $y$-axes (defined in the rotating frame).

\subsection{Qubit-qubit coupling}
\label{subsec:qubitqubitcoupling}

\nomdref{Gfzeta}{$\xi$}{two-qubit $\sigma_z \otimes \sigma_z$ interaction strength}{subsec:qubitqubitcoupling}
\nomdref{Cj}{$J$}{virtual photon qubit-qubit swap interaction strength}{subsec:qubitqubitcoupling}

We can generalize \equref{eq:dispersivetransmoncqed} to $N$ transmons coupled to a single cavity by writing
\begin{equation}\begin{split}
	\label{eq:multitransmoncqed}
	\hat{H}/\hbar = &  \sum_{n=1}^N \left\{ \sum_i \left( 
	\omega_i^{(n)} |i\rangle_n \langle i |_n + \chi_{i,i+1}^{(n)} |i+1 \rangle_n \langle i+1|_n\right) - \chi_{01}^{(n)} \hat{a}^\dagger \hat{a}|0\rangle_n \langle 0|_n \right. \\
	&~~~~ + \sum_{i=1}^{\infty}  \left(\chi_{i-1,i}^{(n)}-\chi_{i,i+1}^{(n)}\hat{a}^\dagger \hat{a}\right)|i\rangle_n\langle i|_n \\
	&~~~~ + \sum_{m>n} \left( \sum_{ij} \frac{g_{j,j+1}^{(n)} g_{i,i+1}^{(m)} \left( \Delta_j^{(n)} + \Delta_i^{(m)} \right)}{2\Delta_j^{(n)} \Delta_i^{(m)})} \left[ |j\rangle_n\langle j+1|_n \otimes |i+1\rangle_m\langle i|_m + \right.\right. \\	
	&~~~~ \left. + |j+1\rangle_n\langle j|_n \otimes |i\rangle_m\langle i+1|_m \right] \Bigg) \Bigg\} + \omega_r \hat{a}^\dagger \hat{a} 
\end{split}\end{equation}
where $|i\rangle_n$ denotes the $i$th level of the $n$th transmon, and the superscript $^{(n)}$ labels the parameters of the $n$th transmon.   The first two lines are the sum of single transmon energies and cavity couplings and the third and fourth lines control interactions between transmons and the cavity energy.  This equation is a bit intimidating, so let us simplify to the case of $N=2$ and the two-level qubit approximation.  Thus, \eref{eq:multitransmoncqed} simplifies to
\begin{equation}\begin{split}
	\hat{H}/\hbar = & \left( \omega_c + \chi_1 \hat{\sigma}_z^{(1)}+ \chi_2 \hat{\sigma}_z^{(2)}\right)\hat{a}^\dagger \hat{a} + \frac{1}{2}\omega_1\hat{\sigma}_z^{(1)} + \frac{1}{2}\omega_2\hat{\sigma}_z^{(2)} \\
	&~~~~ + \frac{g_1 g_2 \left( \Delta_1 + \Delta_2\right)}{2 \Delta_1 \Delta_2} \left(\hat{\sigma}_+^{(1)}\hat{\sigma}_-^{(2)} + \hat{\sigma}_-^{(1)}\hat{\sigma}_+^{(2)} \right).
\end{split}\end{equation}
We can identify these terms as, respectively, the cavity which may be dispersively shifted by both qubits, the two bare qubit Hamiltonians, and a two-qubit swap interaction that occurs via a {\it virtual} interaction with the cavity.  The direct qubit-qubit interaction strength is known as $J$ splitting\footnotemark, and could potentially be used to generate entanglement, as we will discuss in \sref{subsec:adiabaticcphase}.

\footnotetext{For more about this interaction, see section 4.3.2 of Jerry Chow's 2010 thesis \cite{ChowThesis}.}

The full expression of \eref{eq:multitransmoncqed} indicates that there will be interactions involving the higher excited states of transmons as well.  We have found that the interaction of the computational state $|11\rangle$ with the non-computational state $|02\rangle$ is of particular experimental relevance.  As described in section 4.3.3 of Jerry Chow's thesis \cite{ChowThesis}, we can estimate the size of this interaction with fourth-order perturbation theory and by approximating the transmons as three-level systems \cite{DiCarlo2009}.  Defining the parameter $\xi$ as the magnitude of an effective $\hat{\sigma}_z^{(1)} \otimes \hat{\sigma}_z^{(2)}$ interaction between the qubits, we have
\begin{equation}\begin{split}
	\xi = -2 g_1^2 g_2^2 \left(\frac{1}{\delta_A \Delta_1^2} + \frac{1}{\delta_B \Delta_2^2} + \frac{1}{\Delta_1 \Delta_2^2} + \frac{1}{\Delta_2 \Delta_1^2} \right)
\end{split}\end{equation}
where $\delta_A = \omega_{01}^{(2)} - \omega_{12}^{(1)}$, $\delta_B = \omega_{01}^{(1)} - \omega_{12}^{(2)}$, and $\Delta_i = \omega_{01}^{(i)} - \omega_r$.  This expression is only valid when the $0\leftrightarrow1$ transition of one qubit is far from the $1\leftrightarrow2$ transition of the other, and it diverges when the transitions are in resonance.  A full numerical diagonalization of \equref{eq:multitransmoncqed} accurately predicts the experimentally-measured $\xi$ all the way through the avoided crossing \cite{DiCarlo2009}.  This interaction is the basis of two different implementations of a two-qubit entangling gate, one using the crossing off-resonantly and the other directly in resonance.  We discuss both implementations in \sref{subsec:adiabaticcphase} and \sref{subsec:suddencphase} respectively.  Moreover, the virtual interaction of $|102\rangle$ and $|003\rangle$ in a three-qubit register is the basis of a fast three-qubit gate we use to demonstrate quantum error correction, as we will discuss in \chref{ch:qec}. 

\subsection{Dissipation and the strong dispersive limit}
\label{sec:dissipationanddispersive}

\nomdref{Gcgamma}{$\gamma$}{qubit relaxation rate}{sec:dissipationanddispersive}

So far we have not mentioned that these quantum objects are also coupled to their environment and therefore suffer from dissipation.  In the case of the cavity, we often explicitly control this dissipation with the input and output RF coupling strengths because we want our measurement photons to be transmitted (see \chref{ch:qubitmeasurement}).  This will give the cavity a finite line width $\kappa/2\pi$, which is related to its lifetime by $T_{\mathrm{cav}} = 1/\kappa$.  (If we reduce our coupling enough, $\kappa$ will no longer be dominated by our intentional coupling and instead by, for example, photon absorption on the walls of the cavity.)  The qubit also has a finite lifetime, parametrized by $T_1 = 1/\gamma$.  This is often set by either off-resonant decay through the cavity known as the ``Purcell effect,'' \cite{Purcell1946, Houck2008} (\sref{subsec:purcelleffect}) or some uncontrolled degree of freedom like lossy dielectric.

We define the {\it strong dispersive limit} as occurring when $\chi > \gamma, \kappa$.  There, the cavity will move by more than a linewidth in response to a change of the qubit state, making our measurement highly projective.  More interestingly, in the case of photon number splitting, the qubit transition frequencies will be photon-number resolved \cite{Schuster2007}, enabling photon-number selective qubit operations \cite{Johnson2010, Kirchmair2013}.

\section{Flux bias lines}
\label{sec:fluxbiaslines}

\nomdref{Afbl}{FBL}{flux bias line}{sec:fluxbiaslines}

\begin{figure}
	\centering
	\includegraphics[scale=1]{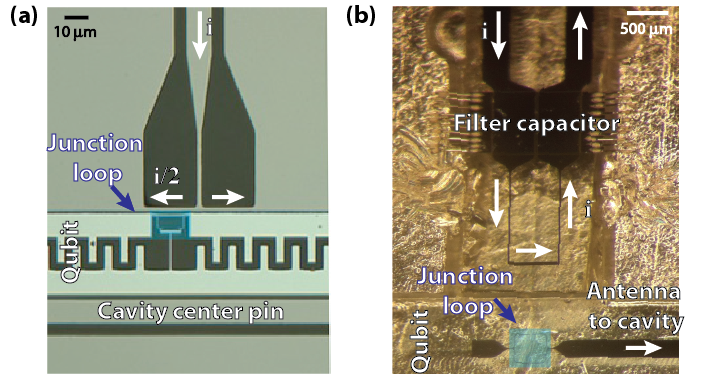}
	\mycaption{Flux bias line designs}{\capl{(a)} Optical micrograph of a typical FBL in the 2D planar architecture.  Current comes in from a single-ended CPW transmission line and is shorted out close to the transmon SQUID loop to create a local magnetic flux.  The current freely returns through the device's ground plane.  \capl{(b)} Optical micrograph of a filtered 3D FBL.  The FBL and qubit are located on separate substrates, and so require much larger geometries.  The resulting increased capacitance between the two objects requires explicit filtering of the FBL to avoid affecting the qubit lifetime. }
	{\label{fig:fbldesigns}}
\end{figure}

As we discussed in \sref{subsubsec:fluxsensitivity}, when using two Josephson junctions in a transmon qubit, an applied magnetic flux through the loop formed by those junctions will change their effective $E_J$.  Initially, this was accomplished by placing a magnetic field coil outside of the sample holder.  This approach has several disadvantages.  First, if there is more than one qubit, it is not possible to independently tune them with only one global field \cite{Majer2007}. Second, because of the large inductance of the coil, the field cannot be changed on timescales comparable to the qubit coherence time.  As we will see in \chref{ch:entanglement}, this is desirable because it would allow for much richer control schemes.  Third, the total magnetic field (and therefore, the total current) needs to be rather large in order to thread a flux quantum through our small loop, which could adversely affect device performance by spawning vortices or heating the fridge.  Finally, it is increasingly common to package devices in superconducting boxes (be it the sample holder or the 3D cavity itself), eliminating the possibility of applying an external field because of the Meissner effect \cite{Meissner1933}.

These issues prompted the development of {\it flux bias lines} or FBLs.  Broadly speaking, FBLs are wires that get very close to the SQUID loop of a single qubit.  Running current through this wire will produce a magnetic field that changes the SQUID flux.  To date, there have been two implementations of these devices: one in the two-dimensional architecture \cite{DiCarlo2009}, and much more recently, a second implementation that works for inline transmon qubits coupled to 3D cavities which we will introduced in \sref{sec:tunable3darchitecture}.  As shown in \figref{fig:fbldesigns}(a), the 2D FBL is implemented with a short-circuited transmission line that is terminated near the loop.  To place the maximum of current close to the loop, the termination is off-center, and is symmetric about the center line in an attempt to control the impedance.  The current returns in an uncontrolled way through the ground plane of the device.  This simplifies things by requiring only one microwave port, though, as we will see in \sref{sec:fourqubitdevice}, it has the disadvantage of having finite DC flux coupling to other qubits.

The design of the 3D FBL is a bit different because the flux line and qubit are on separate substrates, as shown in \figref{fig:fbldesigns}(b).  As with the 2D case, a current-carrying wire comes as close as possible to the qubit SQUID loop, but the size of the loop is significantly larger as a result of the increased distance owing to being on different substrates.  As we will see below, it is necessary to add explicit filtering to these lines to prevent the qubit from decaying due to the substantial capacitance of this geometry.  This FBL design also has two ports so the current flows along a well-defined path in and out of the sample holder.  As a result, there is zero measurable flux cross-coupling between qubits.  

This section will focus primarily on the design considerations for extending FBLs to the 3D architecture.  Specifically, we wish to calculate whether it is possible to thread a flux quantum through the superconducting loop with a reasonable geometry and current magnitude.  Initially, this was a major concern because the loop and FBL are so much farther apart than the 2D case and because of the flux screening of the Meissner effect.  We will show that a flux quantum should be attainable using with an analytic calculation.  We then discuss the consequence of the much larger geometry necessary: an increased capacitive coupling between the qubit and FBL leading to qubit relaxation.  We calculate that the qubit should decay in only a few microseconds due to this effect, but that it can be eliminated with the addition of explicit low-pass filtering on the FBL.  

\subsection{Flux coupling}
\label{subsec:fluxcoupling}

\begin{figure}
	\centering
	\includegraphics[scale=0.75]{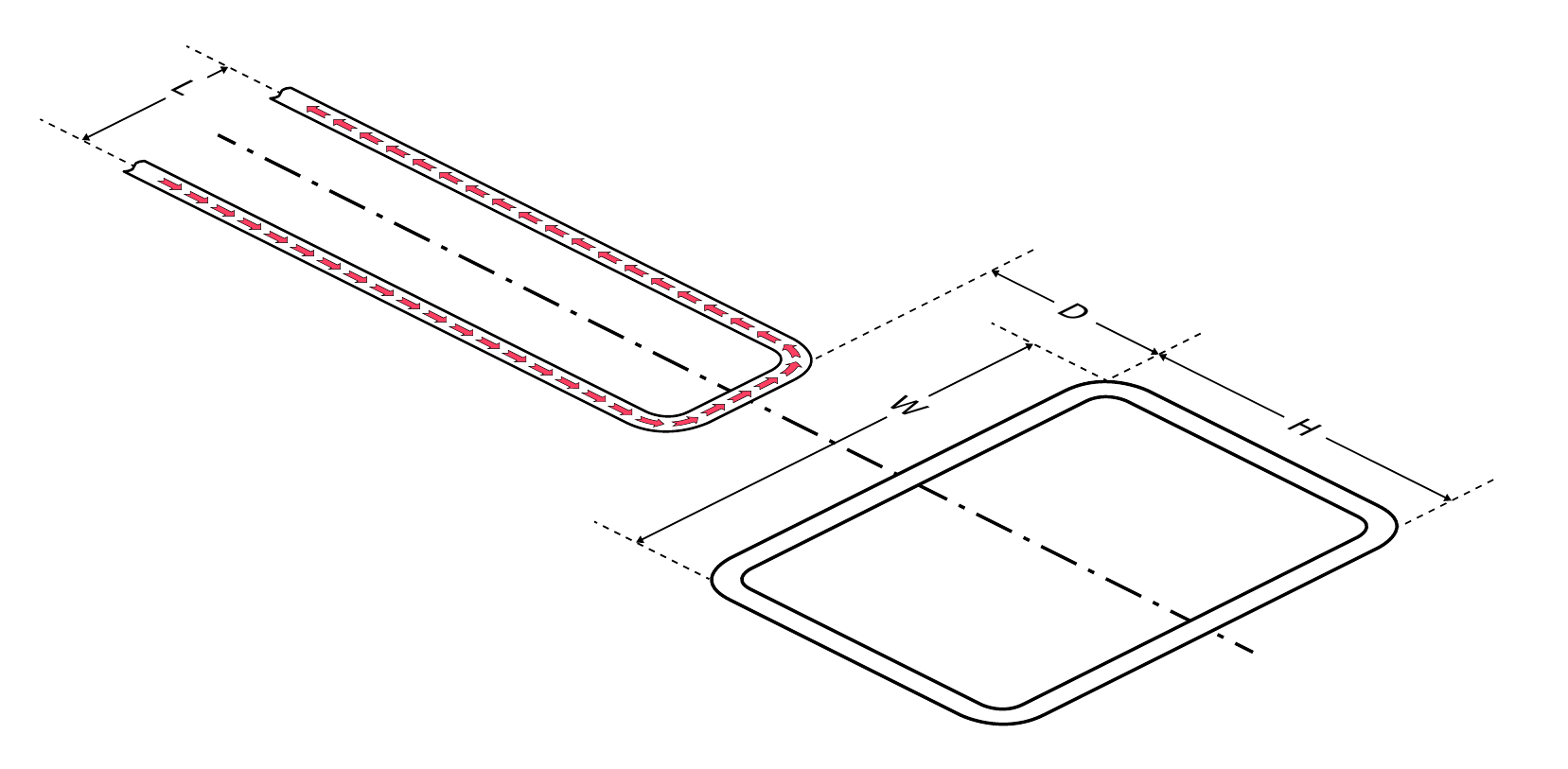}
	\mycaption{Flux bias line geometry}{A current-carrying wire approaches a SQUID loop of width $W$ and height $H$ at a distance $D$ away.  The closest approach of the wire has a length $L$.  The magnetic field of the wire coming to and from this length is cancelled out by symmetry. }
	{\label{fig:fblgeometry}}
\end{figure}

One major question when designing the 3D FBLs was whether it would be possible to couple a flux quantum into the SQUID loop without an excessive amount of current\footnotemark.  Applying more than a few milliamps would constitute an unacceptable heat load on the fridge because of finite cable resistance.  Following the work of Nissim Ofek and Kevin Chou (see \aref{ap:fluxcoupling}), we can estimate the flux coupling analytically using the Biot-Savart Law 
\begin{equation}
	B = \int \frac{\mu_0I}{4\pi}\frac{\vec{dl}\times\vec{r}}{|r|^3}.
\end{equation}
The geometry in question is shown in \figref{fig:fblgeometry}, where we have a SQUID loop of width $w$ and height $h$ positioned a distance $d$ away from a wire of length $L$.  The field at a position $(x,y)$ due to the length of wire is given by
\begin{equation}\begin{split}
	B(x,y) =&~ \frac{\mu_0 I \hat{z}}{4\pi} \int_{-L/2}^{L/2} \frac{y}{((x-l)^2+y^2)^{3/2}}dl \\
	=&~ \frac{\mu_0I\hat{z}}{4\pi y}\left[\frac{\frac{L}2-x}{\sqrt{\left(\frac{L}2-x\right)^2+y^2}}+\frac{\frac{L}2+x}{\sqrt{\left(\frac{L}2+x\right)^2+y^2}}\right].
\end{split}\end{equation}
Integrating to find the total flux in the loop, we have
\begin{equation}\begin{split}
	\label{eqn:fblFlux}
	\Phi=&\mathop{\int}_D^{D+H}dy\mathop{\int}_{-\frac{W}2}^{\frac{W}2}dx\,B(x,y)\\
=&~\frac{\mu_0I}{2\pi}\left.\left.\left[\sqrt{u^2+y^2}+u\log y -u\log\left\{u^2+u\sqrt{u^2+y^2}\right\}\right]\right|_{y=D}^{D+H}\right|_{u=\frac{L-W}{2}}^{\frac{L+W}{2}}.
\end{split}\end{equation}
Note that we can ignore the field due to the wire carrying current to and from the length $L$ because the two sides are equal and opposite, and the SQUID loop is symmetric relative to them.

\footnotetext{A similar calculation for both the flux coupling and capacitive relaxation for the 2D case is found in section 5.3.3 of Jerry Chow's thesis \cite{ChowThesis}.}

Unfortunately, this treatment ignores an important complication.  The flux bias line is surrounded by a superconducting box that initiates surface currents to cancel any magnetic field inside its bulk due to the Meissner effect.  These screening currents tend to reduce the amount of flux looped through the SQUID.  We can impose this boundary condition using the method of images, which involves placing fictitious currents in the bulk of the superconductor.  An image reflected about the wall imposes the $B=0$ condition on that wall, but violates the requirement on the opposite wall.  Another wire placed there similarly solves that side, but again messes up the opposite boundary, albeit to a lesser extent.  Thus, we need an infinite sum of currents placed at integer multiples of $w$ (where the walls are at $y=\pm w/2$), with each ``wire'' carrying $I_n = (-1)^n I$.  We define the ratio $G(d,0) = B(d,0)/B^{0}(d,0)$, where $B^{0}(d,0)$ is the field at $(d,0)$ in the absence of the screening of the walls.  We then have
\begin{equation}\begin{split}
	\label{eqn:fblGfactor}
	G(d,0) \equiv& \frac{B(d,0)}{B^{0}(d,0)} = \sum_{n=-\infty}^{\infty} \frac{1}{n^2 (w/d)^2 + 1}\\
	\approx&~ \frac{1}{\mathrm{cosh\left(\frac{d}{2 w \sigma}\right)^2}}
\end{split}\end{equation}
where the second line is a fit to the numerically-evaluated sum, with $\sigma = 0.955 \frac{\pi^2}{24}$.  A thorough explanation of this process is found in \aref{ap:fluxcoupling}.  Taking the product of \eref{eqn:fblFlux} and \eref{eqn:fblGfactor} and plugging in reasonable parameters of $L=W=325\um$, $D=790\um$, and $L=500\um$, we find that the current required to thread one flux quantum is approximately $1.7\mA$.  Experimentally, this same geometry was found to require about $1.5\mA$ -- very close to the prediction, considering the approximations taken.

\subsection{Relaxation through capacitive coupling to FBLs}
\label{subsec:fblrelaxation}

\begin{figure}
	\centering
	\includegraphics{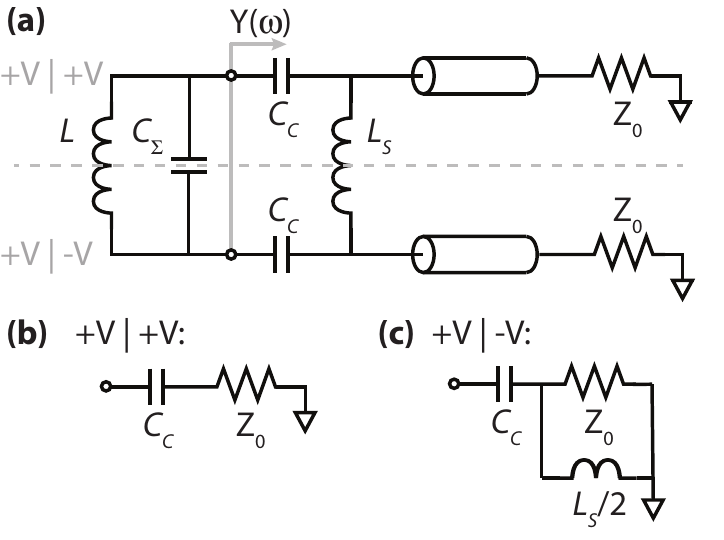}
	\mycaption{Flux bias line circuit}{\capl{(a)} We model the effect of qubit relaxation through the FBL with an equivalent circuit model where the qubit is capacitively coupled (by two capacitors of size $C_c$) to two transmission lines that are shorted together by an inductor $L_s$.  We break this circuit into differential and common modes.  \capl{(b)} The equivalent circuit for the common mode, where $L_s$ has dropped out because there is no voltage drop across it.  \capl{(c)} The equivalent circuit for the differential mode,  where a virtual ground is placed at the mid-point of the inductor $L_s$.}
	{\label{fig:fluxcircuit}}
\end{figure}

An increased capacitive coupling is the cost of the large FBL geometry required for the 3D case.  The coupling of the qubit to the FBL itself will add a relaxation channel, which, especially with the increased qubit lifetimes, can become the dominant source.  We can model the FBL-qubit system with an equivalent circuit, as shown in \figref{fig:fluxcircuit}(a).  The lifetime of the LC oscillator standing in for the qubit will be given by 
\begin{equation}
	\label{eqn:t1fromy}
	T_1 = \frac{C_{\Sigma}}{\mathrm{Re}[Y(\omega_q)]}
\end{equation}
where $C_\Sigma$ is the qubit capacitance and $Y(\omega_q)$ is the admittance of the FBL circuit seen by the qubit \cite{Esteve1986, Neeley2008, Houck2008}.  There are two paths to ground, so we break this circuit into common and differential modes, where the two sides of the qubit junction have either equal or opposite voltages\footnotemark.  The qubit lifetime is then also set by two components, $T_1^{-1} = C_\Sigma^{-1}\left(\mathrm{Re}[Y_{\mathrm{dif}}(\omega_q)]+\mathrm{Re}[Y_{\mathrm{com}}(\omega_q)]\right)$.  The effective circuit diagrams for each of the two cases are shown in \figref{fig:fluxcircuit}(b) and (c).  The impedance for the common case is given by
\begin{equation}
	Z_{\mathrm{com}}=\frac{1}{2}\left(Z_0 + \frac{1}{i \omega C_c}\right)
\end{equation}
where the factor of $1/2$ accounts for the fact that there are two copies of the circuit shown in (b) in parallel with one another.  Taking the real part and the inverse, we then have
\begin{equation}
	\mathrm{Re}[Y_{\mathrm{com}}] = \frac{2}{Z_0} \frac{\omega^2}{\omega^2 + \omega_{c0}^2}
\end{equation}
where $\omega_{c0} = \frac{1}{Z_0 C_c}$.  Doing the equivalent for the differential case, including a factor of $2$ to account for having two of copies of circuit in parallel with the shared ground connecting them, we have
\begin{equation}
	Z_{\mathrm{dif}} = 2 \left(\left(\frac{1}{Z_0} + \frac{1}{i \omega L_s/2} \right)^{-1} + \frac{1}{i \omega C_c} \right)
\end{equation}
\begin{equation}
	\implies \mathrm{Re}[Y_{\mathrm{dif}}] = \frac{Z_0 \omega^4}{2\omega_s^4\left(Z_0^2(\frac{\omega^2}{\omega_{d0}^2} - 1)^2 + \frac{\omega^2}{\omega_{d0}^4 C_c}\right)} \approx Z_0 \frac{\omega^4}{2 \omega_{d0}^4}
\end{equation}
where $\omega_{d0} = \frac{1}{\sqrt{L_s C_c/2}}$ and in the second half we take the low-frequency approximation $\omega \ll \omega_{d0}$.  Using an electrostatic simulation of $C_c \approx 0.25\fF$, a typical qubit capacitance $C_\Sigma = 40\fF$, and estimating $L_s=0.5\nH$, we find that the lifetime of the qubit at $9 \ghz$ would be about $2\us$ -- a number much smaller than typical 3D $T_1$s.  (As we will see in \sref{sec:tunablequbitlifetime}, this is consistent with what we measure experimentally.)  Note that this decay is entirely dominated by the common mode, in which we lose the benefit of the additional filtering of $L_s$.  The differential mode alone would grant a lifetime of about $100\us$, and can be ignored in the limit that the qubit couples to both modes.

\footnotetext{As we will discuss in the next chapter, due to its design, the qubit in practice couples to both the differential and common mode.  In principle, it would be possible to make sure that the two islands have equal capacitance to ground to make the qubit oscillation symmetric, thereby reducing the matrix element to the common mode.  For this treatment, however, we assume that both modes are equally important in the approximation of a highly asymmetric qubit island capacitance.}

\subsection{Filtering}
\label{subsec:fblfiltering}

\begin{figure}
	\centering
	\includegraphics{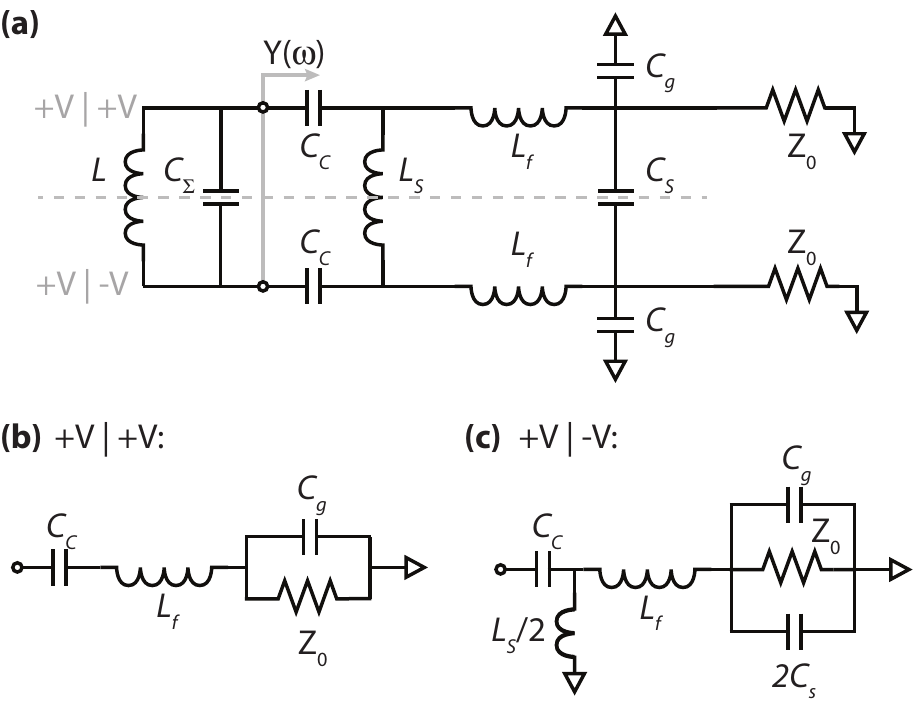}
	\mycaption{Filtered flux bias line circuit}{\capl{(a)} In order to reduce qubit relaxation via capacitive coupling to the outside world, we add a low-pass filter to the FBL.  We are free to do this because the FBL control signals are always below $1\ghz$ while the qubit transition frequencies are typically $5-10 \ghz$.  The filter is implemented with a series inductor and parallel capacitor, and again can be split into common and differential modes, shown in \capl{(b)} and \capl{(c)}.}
	{\label{fig:fluxfilter}}
\end{figure}

\begin{figure}
	\centering
	\includegraphics{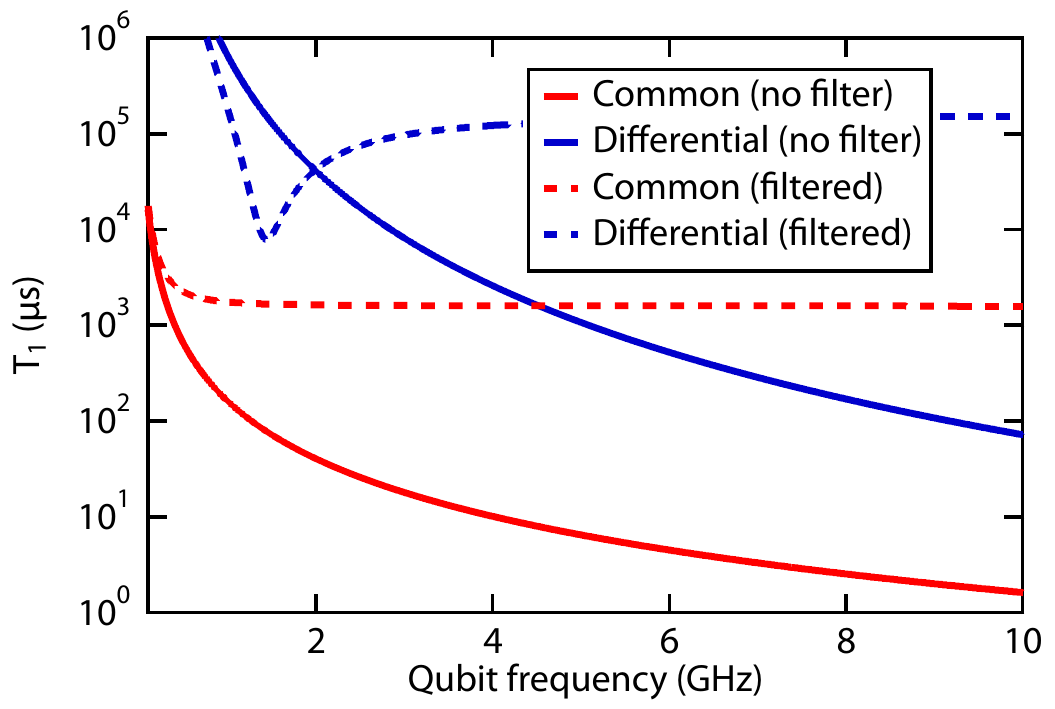}
	\mycaption{Qubit lifetime due to flux bias lines with and without filters}{The predicted qubit lifetime due to coupling to the FBL common and differential modes, with and without filtering, are plotted.  Without filtering, because of the large geometry leading to a capacitance $C_c \approx 0.25\fF$, the qubit is expected to decay in only $2\us$ via the common mode.  This can be ameliorated with the use of a low-pass filter, which both flattens out the frequency dependence above the characteristic cutoff frequency and pushes the lifetime up by several orders of magnitude.  Since these numbers are very sensitive to the particular qubit and FBL parameters ($C_\Sigma$ and $C_c$ in particular), one should not pay so much attention to the numerical values of these predictions but rather the ratio of improvement realized with filtering.  The parameters used in these calculations are $C_c = 0.25\fF, C_\Sigma = 40\fF, L_s = 0.5\nH, C_g = 10\pF, L_f = 1\nH,$ and $C_s = 0$. }
	{\label{fig:fluxlifetime}}
\end{figure}

For the large geometry needed to couple a flux quantum in the 3D case, the capacitive coupling to the flux bias line can be the dominant source of relaxation.  Fortunately, it is possible to ameliorate this decay by adding a low-pass filter, as shown in \figref{fig:fluxfilter}.  There, we have added a series inductor and capacitor both to ground and between the two arms of the bias line.  We  can again calculate the impedance for the common and differential modes of this structure, with
\begin{equation}
	\label{eq:filteredZcom}
	Z_{\mathrm{com}} = \frac{1}{2} \left( \left(\frac{1}{Z_0} + i \omega C_{g} \right)^{-1} +  i \omega L_f + \frac{1}{i \omega C_c} \right)
\end{equation}
\begin{equation}
	\label{eq:filteredZdif}
	Z_{\mathrm{dif}} = 2 \left( \left( \left( \frac{1}{Z_0} + i \omega \left(C_{g} + 2C_{s}\right) \right)^{-1} + i \omega L_f \right)^{-1} + \frac{1}{i \omega L_s/2} \right)^{-1} + \frac{2}{i \omega C_c}.
\end{equation}
As before, inverting these equations and taking their real part gives us the expected qubit lifetime using \equref{eqn:t1fromy}.  The largest geometric inductance that can be made on this scale without being affected by self-resonances is about $L_f^{\mathrm{max}}\approx 1\nH$.  Thus, the job of filtering comes down to the capacitance.  Experimentally, it is possible to get a large capacitance by using a three-layer lithography process, where $C_g$ can be as large as tens of pF.  Evaluating \equref{eq:filteredZcom} and \equref{eq:filteredZdif} with reasonable parameters $(C_c = 0.25\fF, C_\Sigma = 40\fF, L_s = 0.5\nH, C_g = 10\pF, L_f = 1\nH)$, and assuming $C_s=0$ since it is much smaller than $C_g$, we predict a common mode lifetime of $\sim 1.6\ms$ and a differential mode lifetime of $0.15~\mathrm{sec}$ at $9\ghz$.  Thus, this simple filtering should completely turn off decay through the flux line.  The qubit lifetime as a function of frequency for both the filtered and unfiltered case is shown in \figref{fig:fluxlifetime}.

\section{Conclusions}
\label{sec:theoryconclusions}

This chapter introduced the superconducting transmon qubit and several ways of looking at it.  In the most correct case, we exactly diagonalize the Cooper-pair box Hamiltonian in the charge basis.  This often provides more precision than necessary, so we can approximate the transmon as either an anharmonic oscillator or a two-level spin-1/2 particle.  We couple these qubits to a standing-wave mode of a microwave resonator in the cQED architecture, which grants us the ability to perform single-qubit rotations and qubit measurements, and mediates couplings between qubits that will be used to generate entanglement.  The terms responsible for these properties are found in the Jaynes-Cummings Hamiltonian, though some work is required to convert them into a transparent form.  We use flux bias lines to control the frequency of qubits in-situ, and recently extended them to the 3D architecture.  There, the geometry required to couple a flux quantum results in a substantial capacitance between the FBL and qubit and therefore a new channel for qubit decay that can dominate for typical parameters.  We have the capacity to to eliminate this channel with the addition of a low-pass filter to the FBL.  In the next chapter, we will explain the actual experimental realization of both planar and tunable 3D cQED.

\setcounter{chapter}{3}
\chapter{Experimental Design and Setup}
\thumb{Experimental Design and Setup}
\lofchap{Experimental Design and Setup}
\label{ch:exptsetup}


\lettrine{N}{ow} that we have a solid theoretical background in both quantum information processing and superconducting qubits, we can shift our attention to the experiment.  This chapter begins by introducing the planar cQED architecture \cite{Blais2004, Wallraff2004} which has been the primary focus of the Schoelkopf lab for many years.  It uses a coplanar waveguide geometry that is entirely specified with optical lithography.  We will introduce the basic features of this geometry and conclude with how such devices are fabricated and packaged.

More recently, our lab has shifted to a new architecture that replaces the CPW cavity with a three-dimensional superconducting box \cite{Paik2011}.  This change has realized huge improvements in qubit coherence, though it initially came the a cost of reduced control.  We seek to re-introduce the ability to tune the qubits using flux without giving up the lifetime gains of the 3D transmon.  This motivation sets the stage to introduce the {\it tunable 3D cQED architecture}, which integrates flux tunability by utilizing a {\it vertical transmon} whose Josephson junction is located in the bulk aluminum rather than inside the cavity itself.  We can therefore access the junction with a wire placed on a separate piece of sapphire.  We will first introduce the design principles of the tunable architecture and show how it is assembled.  We next explain the design of the cavity, sample box, qubits, and flux bias lines.  

We then discuss more broadly the experimental details of measuring any cQED device.  We use a helium dilution fridge to cool the device to $\sim20\mK$.  The cables going into and out of the fridge must be thoroughly filtered, with attenuation on the incoming lines and circulators on the outgoing ones.  We then examine a typical wiring diagram used both to generate control pulses and measure a cQED device, and explain common areas for variants and recent improvements.  We conclude with a description of how mixers can be used to implement pulse modulation much more cheaply than using a full vector generator, but how doing so requires a series of calibrations.

\section{Planar design}
\label{sec:planardesign}

\nomdref{Acpw}{CPW}{coplanar waveguide}{sec:planardesign}
\nomdref{Cqc}{$Q_C$}{cavity coupling quality factor}{sec:planardesign}
\nomdref{Cqi}{$Q_I$}{cavity internal quality factor}{sec:planardesign}
\nomdref{Cq}{$Q$}{cavity quality factor, also $Q_{\mathrm{tot}}$}{sec:planardesign}
\nomdref{Geeff}{$\epsilon_{\mathrm{eff}}$}{effective dielectric strength}{sec:planardesign}

The planar or ``2D'' geometry used in most of the experiments presented in this thesis has been developed for several years, beginning with the first realization of cQED with Cooper-pair boxes in 2004 \cite{Wallraff2004}.  Other than the transition to transmon qubits and the introduction of flux bias lines, the overall geometry has not changed significantly in the years since.  Therefore, instead of introducing all the details of the architecture which have been thoroughly discussed in previous papers \cite{Wallraff2004, Majer2007, DiCarlo2009} and theses \cite{ChowThesis, JohnsonThesis, SchusterThesis}, we will instead summarize the features of planar cQED.

\begin{figure}
	\centering
	\includegraphics{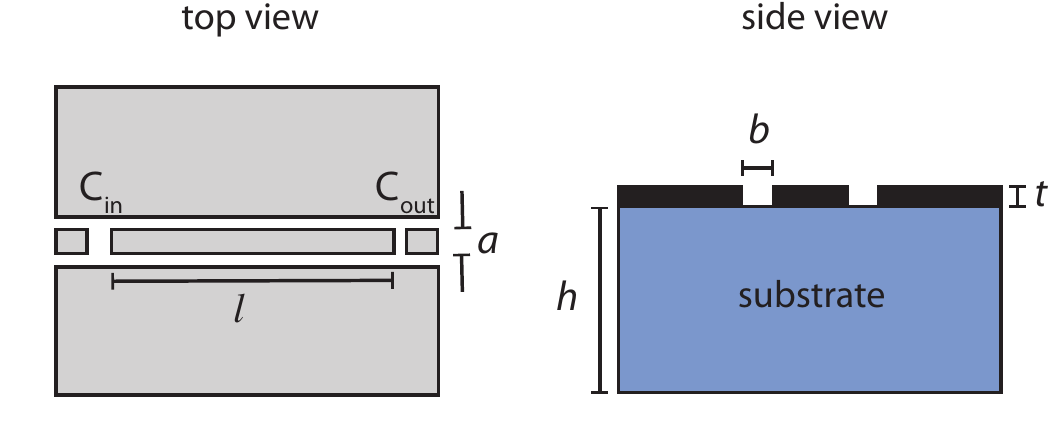}
	\mycaption{Coplanar waveguide geometry}{\capl{(top)} The planar cavity is implemented with a coplanar waveguide terminated with two capacitors.  The length between these capacitors sets the resonance frequency of the cavity, and the size of the capacitors determines its coupling $Q_c$.  \capl{(side)} The waveguide is patterned on top of a sapphire substrate with thickness $h \approx 420\um$.  The width $a$ of the CPW is typically $10\um$, with a spacing from the ground plane $b=4.2\um$.  The thickness of the niobium film $t = 200\nm$.  These parameters will give an impedance $Z\approx50\ohm$.}
	{\label{fig:cpwgeometry}}
\end{figure}

The resonators in planar cQED are implemented with a coplanar waveguide (CPW) geometry, as shown in \figref{fig:cpwgeometry}.  The length of the center pin between two capacitors sets the frequency of the resonator and the magnitude of those capacitances sets the coupling $Q_c$.  The large dielectric strength of the sapphire substrates set an effective $\epsilon_{\mathrm{eff}}\approx 6$ for the wave, giving a cavity resonance frequency of $\omega_r = \frac{c\pi}{l\sqrt{\epsilon_{\mathrm{eff}}}}$, where $l$ is the physical length of the center conductor.  The input and output capacitors $C_{\mathrm{in}}$ and $C_{\mathrm{out}}$ can be different sizes, and making the output capacitor large is especially favorable for maximizing the signal power sent into the amplification chain for dispersive readout (see \sref{sec:dispersivereadout}).  The overall coupling $Q_c$ of the cavity is given by
\begin{equation}
	Q_c = \frac{\pi}{2} \frac{1}{\omega Z_0^2 \left(C_{\mathrm{in}}^2 + C_{\mathrm{out}}^2 \right)}
\end{equation}
where $\omega$ is the frequency of the fundamental $\lambda/2$ mode of the resonator and $Z_0$ is the characteristic impedance of the line (ideally $50\ohm$).  The impedance is a function of the center pin width $a = 10\um$, ground plane gap $b = 4.2\um$, film thickness $t = 200\nm$, and substrate height $h = 420\um$.  It can be calculated analytically, but using a numerical method like AWR Microwave Office's TXLINE is more practical.  The total $Q_{\mathrm{tot}}$ is given by the inverse sum of the coupling $Q_c$ and the non-radiative internal $Q_i$ that takes into account uncontrolled losses through (for example) dielectrics, with $Q_{\mathrm{tot}}^{-1} = Q_c^{-1} + Q_i^{-1}$.

\begin{figure}
	\centering
	\includegraphics{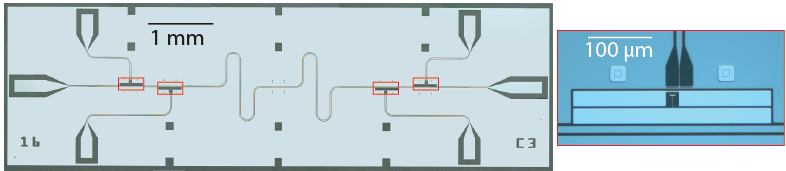}
	\mycaption{Four-qubit planar cQED device}{ \capl{(left)} Optical micrograph of whole chip.  We can see the CPW cavity patterned in the center of the $2\mm$ by $7\mm$ chip, with looping to set the frequency to $\sim 9\ghz$.  There are four transmon pockets highlighted with a red box, each equipped with its own flux bias line.  \capl{(right)} An optical micrograph of a transmon qubit.  This qubit was designed without interdigitation between the two islands in an attempt to reduce dielectric participation.  At the top of the picture, the termination of the FBL is visible.  The two squares above and to the right and left of the qubit are alignment markers used during fabrication.  \figadapt{DiCarlo2010}}
	{\label{fig:fourqubitdevice}}
\end{figure}

A typical device is shown in \figref{fig:fourqubitdevice}.  The devices are $2\mm$ by $7\mm$ by $\sim 500\um$, with the CPW structure patterned with niobium on a sapphire substrate.  There are $300\um$ by $30\um$ pockets alongside the cavity center pin for transmon qubits, highlighted in red in the figure.  These pockets are located near the ends, where the anti-nodes of the standing wave are located, to maximize coupling.  In the case of this four-qubit device, the qubits are offset from one another to minimize direct qubit-qubit capacitance.  (Direct capacitance would add an undesirable constant $ZZ$ interaction between the qubits.)  These devices employ flux bias lines that are similarly implemented with $50\ohm$ CPW but short-circuited near the location of the transmon SQUID as described in \sref{sec:fluxbiaslines}.  The figure shows that each of the transmon pockets is equipped with a FBL, making this a six-port device when the input and output cavity ports are counted.  On the right, a zoom-in of one of the transmon qubits is shown.  Engineering the parameters $E_C$ and $\beta$ of the transmon qubit requires inverting a 5 by 5 capacitance matrix (see Ref.~\citenum{Koch2007}).  However, to first order, $\beta$ can be controlled with the long dimension of the qubit (closest to the center pin) and $E_C$ with spacing between the two islands.  In this picture, the end of the FBL is visible, aligned so that one of the two current paths is centered on the transmon SQUID loop.

\subsection{Planar fabrication}
\label{subsec:planarfab}

All planar resonators in this thesis were produced by a similar process \cite{Frunzio2005}.  It begins with sputtering a $\sim200\nm$ thick niobium film on 2-inch C-plane corundum ($\alpha-\mathrm{Al}_2\mathrm{O}_3$) wafer using a d.c.-magnetron.  This wafer is covered with S1808 photoresist, on which the resonator structures are patterned with contact UV optical lithography.  The niobium is dry etched with a fluorine-based reactive ion etcher, followed by lift-off in acetone.  The wafer is diced into $2\mm$ x $7\mm$ chips.  To create the qubits, each individual chip is covered with two additional layers of photoresist (top $100\nm$ of 950K PMMA A3, bottom 550 nm MMA(8.5)-MAA EL13 copolymer) and patterned with electron-beam lithography.  The bottom layer is more sensitive to electrons than the top, giving a natural $\sim80\nm$ undercut which eases lift-off.  After development, aluminum is deposited with a double-angle electron beam evaporation to create the $\mathrm{Al}-\mathrm{AlO}_x-\mathrm{Al}$ Josephson junctions and qubit capacitors, with a $15\%$ $O_2$ 15 torr 12-minute oxidation step between subsequent depositions.  Finally, the undeveloped photoresist is removed with lift-off, leaving behind only the aluminum that was defined by electron lithography.  Junction resistances (which are related to $E_J^{\mathrm{max}}$) can be tested by either fabricating explicit test junctions alongside the qubits or by directly probing the transmon islands.  As long as appropriate precautions are taken to limit the test current and prevent static shocks, this measurement does not appear to harm the junction.

\subsection{Planar sample holders}
\label{subsec:planarsampleholders}

\nomdref{Apcb}{PCB}{printed circuit board}{subsec:planarsampleholders}

\begin{figure}
	\centering
	\includegraphics{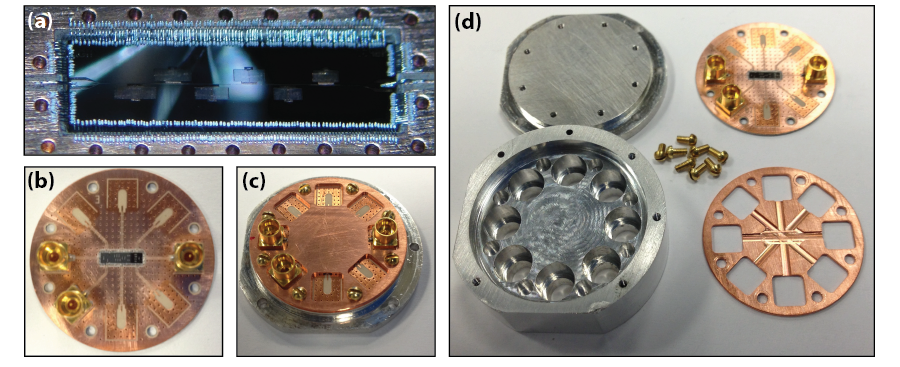}
	\mycaption{Planar sample packaging}{ \capl{(a)} Typical wire bonding of planar device to PC board.  For assembly, the planar chips are placed into a pocket of a PC board and thoroughly wire bonded to join the two ground planes.  Connections between the center pins of the device and the PC board are also made.  To cut down on extraneous on-chip modes, wire bonds are sometimes also placed on the device itself.  Note that this picture is not the same device as shown in the others of this figure.  \capl{(b)} PC board view.  The copper traces of the PCB are connected to Rosenberger connectors which mate with cables connected to the sample holder.  \capl{(c)} Assembled sample bottom.  The PCB is screwed to an ``octobox lid'' and a copper shim designed to control the 3D mode structure of the sample holder is placed over the top.  \capl{(d)} Octobox lid and base, PC board with installed device, and copper shim. }
	{\label{fig:planarsampleholder}}
\end{figure}

The packaging of the planar device is depicted in \figref{fig:planarsampleholder}.  The devices are placed on a copper PC board which has an Arlon dielectric.  As many as eight microwave Rosenberger connectors can be soldered to this board to interface with RF cables.  The board is covered with vias to short the two copper ground planes together.  The devices are wire-bonded hundreds of times around their perimeter to short the PCB ground and sample ground, and the microwave connections (e.g. RF and flux bias) are similarly bonded to the associated printed circuit board (PCB) trace.  Additional wire bonds are often placed on the sample chip itself, to short together adjacent ground planes that are split by control lines\footnotemark.  Over the top of this, a ``flip chip'' is placed to reduce the volume surrounding to the chip in an attempt to control the resonant mode structure.  Finally, the assembly is screwed onto the ``octobox lid'' (made of either copper or aluminum) and placed into the ``octobox bottom'', where the Rosenberger connectors are mated to RF cables.  There are eight RF cables in the octobox sample holder, as referred to by its name.  The whole package is placed inside a magnetic shield and bolted to the base plate of the helium dilution fridge.

\footnotetext{See section 5.3.3 of Jerry Chow's thesis for more on this \cite{ChowThesis}.}

\section{Tunable 3D architecture}
\label{sec:tunable3darchitecture}

\begin{figure}
	\centering
	\includegraphics{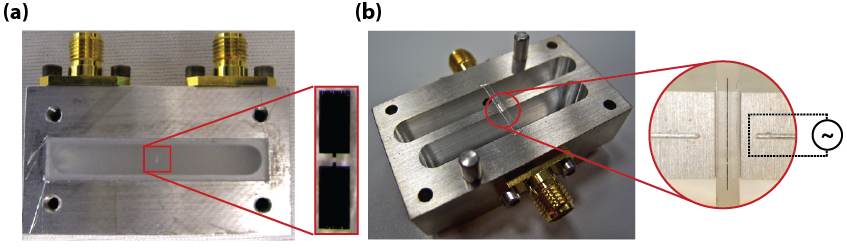}
	\mycaption{One- and two-cavity 3D cQED designs}{ \capl{(a)} The canonical single-cavity 3D cQED architecture.  A superconducting 3D cavity is made by mating two metal pieces with pockets that line up and constitute the walls of the cavity.  A sapphire substrate hosting a transmon qubit is placed between the two cavity halves.  This qubit is much larger in size to couple strongly to the more diffuse electric field of the cavity, but is functionally identical to a planar transmon.  It is isolated in the middle of the cavity, and so is impossible to control directly using this geometry.  The device is controlled exclusively via RF ports, shown in gold at the top of the picture.  \capl{(b)} Two-cavity 3D cQED device with a ``vertical transmon'' coupled to both.  A thin sapphire substrate containing a qubit is placed across the cavities.  Its Josephson junction is located between the cavity walls and in the bulk of the aluminum, potentially giving us the ability to place a wire (shown in a dashed black line) next to it for flux control.}
	{\label{fig:3doctomotivation}}
\end{figure}

Our lab has recently shifted to coupling qubits to the electromagnetic modes of a three-dimensional box rather than a planar CPW, an architecture commonly known as ``3D cQED'' \cite{Paik2011}.  The conventional design is shown in \figref{fig:3doctomotivation}(a), where we see a bisected aluminum 3D cavity with a wafer of sapphire laying across the plane of the cut.  Two of these cavity halves constitute the device (second half not shown), which are screwed together with the sapphire wafer laying between them.  As shown in the inset, a transmon qubit is patterned on this wafer.  It must be a great deal larger than conventional 2D designs in order to attain the same cavity coupling strength because the electromagnetic field is more diffuse, but is otherwise equivalent in behavior to planar devices.  RF coupling antenna are inserted into the top of the cavity (whose gold-colored connectors are seen at the top of the picture), and are also functionally identical to the RF drive ports of the planar devices.  (See Adam Sears's thesis \cite{SearsThesis} for a more detailed description of the geometry of these devices.)

This 3D architecture has realized huge gains in coherence time, with measured $T_1$ and $T_2$ in excess of $100\us$.  The reason for this nearly two orders of magnitude improvement is thought to be primarily due to reduced energy storage in lossy materials \cite{Shnirman2005, Martinis2005, OConnell2008, Frunzio2013}, though quasiparticle dynamics may also play a role \cite{Martinis2009, Catelani2011, Sun2012, Riste2012c}.  The dielectric quality of surfaces and interfaces of materials are suspected to be of poor quality compared to the bulk.  Relatively small features like those found in the planar architecture tend to focus a large fraction of stored energy near these surfaces, while large features will have a smaller participation ratio in that lossy material because the field lines will penetrate deeper into the bulk.  Thus, instead of attempting to improve the material properties as many believed was required, the 3D architecture bypasses the issue by reducing the qubit's susceptibility to the problem.  In the process of studying these qubits, it also became clear that proper thermalization and filtering was crucial to getting good performance and is responsible for another factor of two in coherence \cite{Sears2012, PetrenkoMM2013}.

These improvements come at some cost.  The qubit is suspended in the middle of a superconducting cavity, greatly limiting our ability to have direct control of it.  We cannot, for example, run wires for flux bias to the qubit nor interface these wires through the wall of the cavity without modifying the mode structure (and, likely the lifetime) of the cavity.  Since the cavity is typically superconducting, threading external flux is also not a possibility.  (We could instead use a copper cavity \cite{Rigetti2012}, but the cavity quality factor will be much lower, requiring a much larger qubit-cavity detuning to limit Purcell relaxation.)  Is this a significant disadvantage?  Though flux bias lines are extensively used in this thesis for the purpose of entangling gates, that is unlikely to be their most important application\footnotemark.   However, fast tunability would play a critical role in demonstrations of hardware-efficient quantum error correction \cite{Leghtas2012}, entanglement distillation between distant pairs of qubits \cite{Bennett1996, Bennett1996b, Bennett1996c}, and distribution of quantum information between subsystems \cite{Bennett1993}.  The coupling of a qubit to the cavity bus could be modulated to control the inherited nonlinearity and the dispersive shift of the oscillator in real time, useful for continuous-variable quantum information processing \cite{Yin2012, Leghtas2012, Leghtas2013}.  Controlling the interactions between individual qubits, particularly those coupled to more than one cavity, could be used to shuttle quantum information between distant subsystems.  Fine-tuning system parameters would ease the implementation of schemes that have strict parameter requirements \cite{Nigg2012}.  Individual QND qubit readout, reset, or drives also requires individual qubit addressability, and are crucial for quantum error correction and are some of the DiVincenzo criteria \cite{DiVincenzo2000} enumerated in \sref{subsec:divincenzo}.

\footnotetext{Recently, an all-RF entangling gate which requires no qubit tunability has been demonstrated  \cite{Paik2013}.  The qubits used in that experiment were single-junction devices and not susceptible to flux noise, therefore enjoying longer $T_2$ times.}

The capacity to address individual qubits is a desirable capability, but we would like to regain it without sacrificing the improvements of 3D cQED.  Fortunately, work on coupling a single 3D qubit to two different cavities \cite{Kirchmair2013} has suggested a solution.  As shown in \figref{fig:3doctomotivation}(b), a ``vertical transmon'' can be laid across two cavities, with antennas sticking into both, and yielding strong coupling\footnotemark.  (This has also been done previously in the planar architecture, as reported in Ref.~\citenum{Johnson2010}.)  As a result, the Josephson junction of the qubit is located between these two cavities in the bulk of the aluminum and it is no longer isolated.  We could potentially thread in a current-carrying wire from the side to apply magnetic flux, as shown by the dashed lines on the right of (b).  This is the principle concept behind the {\it tunable 3D cQED architecture}.  In this section, we first show how the tunable architecture is assembled which will give a broad overview of the design principles.   We then introduce the specific design of the tunable cavities themselves, the two versions of the vertical transmon qubits, and the flux bias lines.

\footnotetext{The vertical transmon is actually a middle-ground between the full 3D qubit shown \figref{fig:3doctomotivation}(a) and the planar geometry in the sense that its participation in surfaces is somewhat higher than the 3D geometry.  Though other differences may be responsible, this increased participation could be the reason that vertical transmons thus far seem to be limited to $T_1 \approx 20-40\us$.  So, for now at least, we have compromised some coherence for improved control.}

\subsection{Tunable 3D cQED assembly}
\label{subsec:tunableassembly}

\begin{figure}
	\centering
	\includegraphics{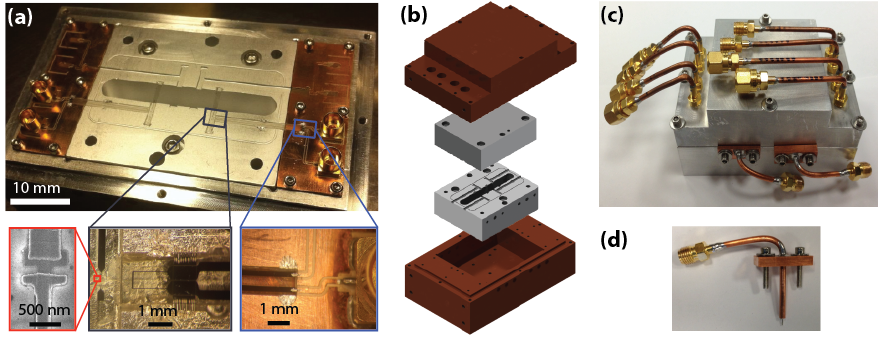}
	\mycaption{Single-cavity tunable 3D architecture assembly}{\capl{(a)} As described in the body of the text, the tunable 3D architecture employs vertical transmons located in trenches traversing the cavity that are coupled via  antenna that stick in.  These qubits are controlled with separately fabricated flux bias lines, which are able to access the qubit SQUID loop because it is located in the bulk of the material.  The FBLs attach to conventional PC boards located outside the cavity, on a sample holder that surrounds it.  \capl{(b)} Assembly diagram of the four components of the architect, showing the ``3D octobox'' sample holder in copper (though it most commonly made of aluminum), inside which the cavity is placed.  \capl{(c)} Picture of a fully-assembled device, showing the eight cables that can be used to control flux bias.  \capl{(d)} Picture of a coaxial RF coupler used to address the cavity, inserted through a hole in the bottom of the sample holder and into the cavity.}
	{\label{fig:3doctoassembly}}
\end{figure}

To implement the idea inspired by the two-cavity 3D device for the tunable 3D architecture, we place flux bias control wires adjacent to the vertical transmon qubits in the bulk of the cavity material.  Both single-cavity and two-cavity variants of the architecture exist.  As shown in \figref{fig:3doctoassembly}(a), the single-cavity device has pockets for as many as four tunable vertical transmon qubits.  Adjacent to each pocket are trenches for flux bias lines that are patterned on separate slivers of sapphire and abut against the qubit at the location of its junction (middle inset).  The slivers protrude from the side of the cavity and are received by two copper PC boards attached to the sample holder surrounding the cavity (right inset).  In a manner identical to the planar devices, these FBLs are wire-bonded to copper traces that are themselves soldered to a Rosenberger connector.  The qubits and the flux bias lines are held in place by flattening indium metal around them at several locations.  This is shown in the right inset for the case of the flux bias line and in the middle inset for the qubit (the latter case is more difficult to see due to the low contrast between aluminum and indium).  For additional shielding and to better connect the two cavity halves, indium wire can also be placed around the cavity in a rounded-rectangular groove.  

The architecture is assembled in four main sections, as shown in \figref{fig:3doctoassembly}(b).  The cavity is placed in the pocket of a sample holder box (copper-colored in the figure) and is secured with three screws.  To the sides of the cavity are placed the copper PC boards, which are similarly screwed to the sample holder.  In this configuration, the qubits and flux bias lines can be inserted and secured with indium, the FBLs wirebonded, and the indium seal placed.  The second half of the cavity is placed over the top, which must be carefully aligned by scrutinizing the matching flux bias line trenches on the top, and is secured with four more screws.  Finally, the sample holder lid which mates the PC board connectors to conventional cables is placed over the top and screwed to the bottom part.  Since this lid has eight flux bias line ports, it is known as a ``3D octobox.''  A picture of a fully-assembled device is shown in \figref{fig:3doctoassembly}(c), where the eight cables have been installed.  RF cavity control is achieved in a similar way as with conventional 3D cQED, with an RF coaxial cable sticking into the side of the cavity through a corresponding hole in the octobox.  As shown in \figref{fig:3doctoassembly}(d), the coupler soldered to a copper bar that is used for thermalization, shorting of ground planes, and securing it to the octobox.

\begin{figure}
	\centering
	\includegraphics{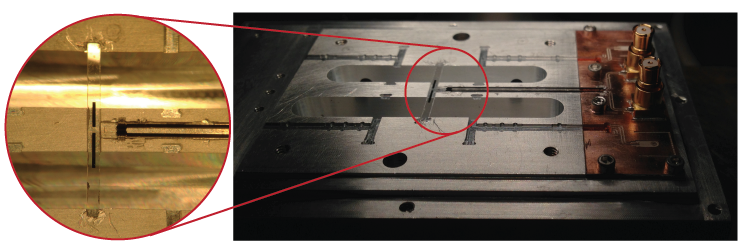}
	\mycaption{Two-cavity 3D octobox assembly}{The two-cavity version of the tunable 3D architecture can support as many as five tunable qubits, with a pair of qubits coupled to each cavity and a fifth qubit coupled to both.  It is compatible with the single-cavity sample holder, and so its assembly is essentially identical.  Here, a variant of the architecture is shown that merges the bottom cavity with the bottom of the sample holder in an attempt to cut down on RF crosstalk of the couplers via the small gap between the cavity and sample holder that otherwise arises.}
	{\label{fig:3dtwooctoassembly}}
\end{figure}

The two-cavity version of the tunable architecture is shown in \figref{fig:3dtwooctoassembly}.  It is designed and assembled much like the single-cavity version.  The primary difference is that the different locations for the qubits and flux bias lines require a distinct PC board design.  Each cavity supports two local tunable qubits, with a fifth tunable qubit located between the two cavities in the same geometry as the original two-cavity vertical transmon.  There is room for two RF couplers on either side of the octobox, enabling full RF control of both cavities.  Since the octobox has only four ports on either side, two of the flux bias lines on the side with the fifth qubit's FBL are shorted to the PC board instead of returning their current through a cable.  The two-cavity device also omits the indium trench for lack of space.

\subsection{Cavity design}
\label{subsec:tunablecavitydesign}

\begin{figure}
	\centering
	\includegraphics{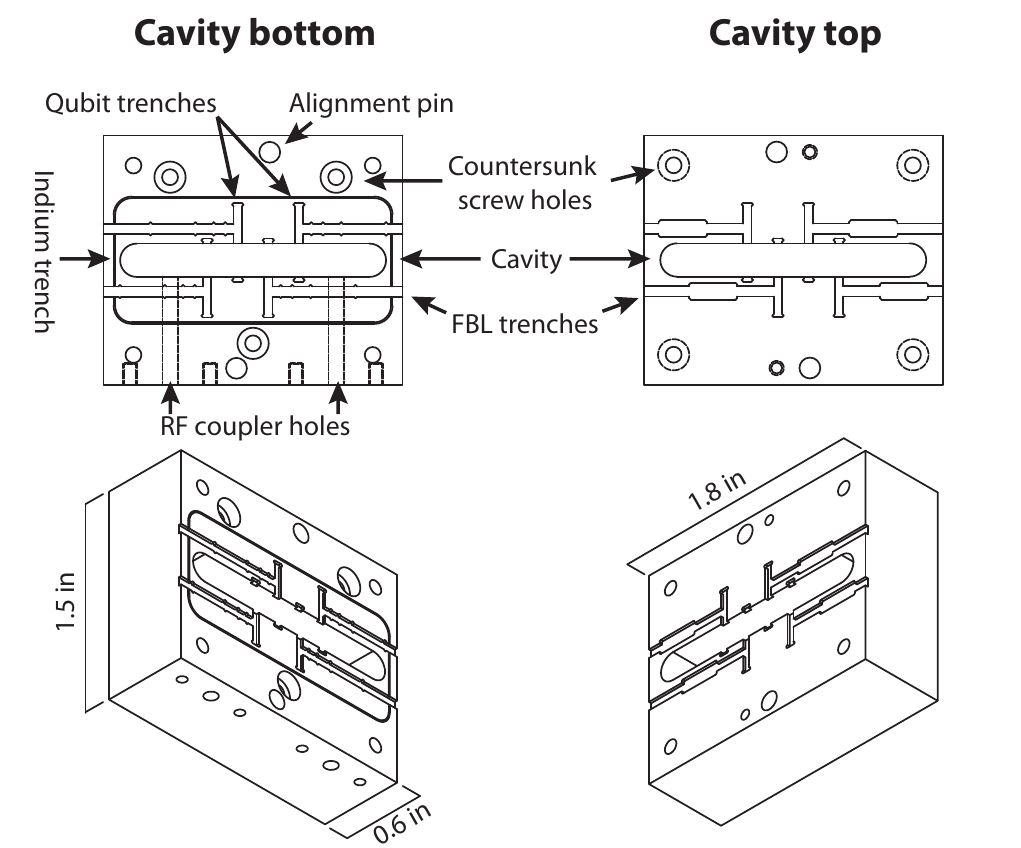}
	\mycaption{Tunable single cavity design}{The two halves of the cavity are shown on the left and right, with three-dimensional perspectives of each below.  The main features of the cavity design are highlighted, with the cavity itself located in the center and qubit trenches laying across.  Flux bias line trenches come in from the sides of the cavity and abut against the qubit trench at the location of the SQUID.  Countersunk screw holes hold the assembly together, with each mating to the screw pattern on the octobox bottom shown in \figref{fig:tunableoctoboxdesign}. }
	{\label{fig:tunableonecavitydesign}}
\end{figure}

Diagrams and 3D perspectives for the top and bottom sections of the tunable single-cavity device are shown in \figref{fig:tunableonecavitydesign}.  Both halves are 1.5" x 0.6" x 1.8" in size and typically are machined out of aluminum.  Focusing on the bottom half, we see at the center a single rounded-rectangular cavity pocket.  Across this cavity run four qubit trenches, with matching indentations on the opposite side of the cavity to support the qubit substrate.  Each qubit has a matching trench for a flux bias line that approaches from the side and abuts upon the location of the qubit SQUID loop; note that two different lengths of FBLs are required here.  Alongside the FBL trenches are several semi-circular ``ears'' into which indium is packed to secure the flux bias line.  Additional ears are located at the top and bottom of the qubit and serve the same purpose\footnotemark.  An indium wire can be placed in the trench surrounding the cavity and qubits to provide additional shielding.  This wire must be interrupted at the locations of the flux bias lines, and might not be necessary.  An alignment pin similar to the one shown in \figref{fig:3doctomotivation}(b) can also ease assembly.  There are three countersunk screw holes used to attach the cavity to the octobox sample holder; due to the countersinking, the head of the screw is below the cavity plane and does not interfere with assembly.  There are four clear holes on the corners of the device, through which we inserted the screws securing the top cavity.  Finally, there are two holes drilled from the outer wall of the device to the cavity for RF coupling (dashed lines).  Adjacent to both holes are tapped holes to which the RF couplers are fastened.

\footnotetext{In addition to making it possible to machine.  Since this trench will be milled out with a circular tool, if we did not have these ears, the edges of the trench would be rounded and a rectangular substrate would not fit.}

The top of the cavity is designed in much the same way.  There is a matching rounded-rectangular cavity pocket and matching trenches for the qubit and flux bias lines.  Those bias line trenches flare out to make room for wirebonding of the flux bias line filters to the body of the cavity. Without the extra room, the wirebonds would be sheared off by the wall of the trench during assembly.  There are four countersunk screw holes to hold the cavity together, with the countersinking on the back side surface (not shown in the perspective view).  Finally, next to the alignment pin holes are tapped screw holes that may be used to separate the two cavities in a controlled manner if they become stuck together.

\begin{figure}
	\centering
	\includegraphics{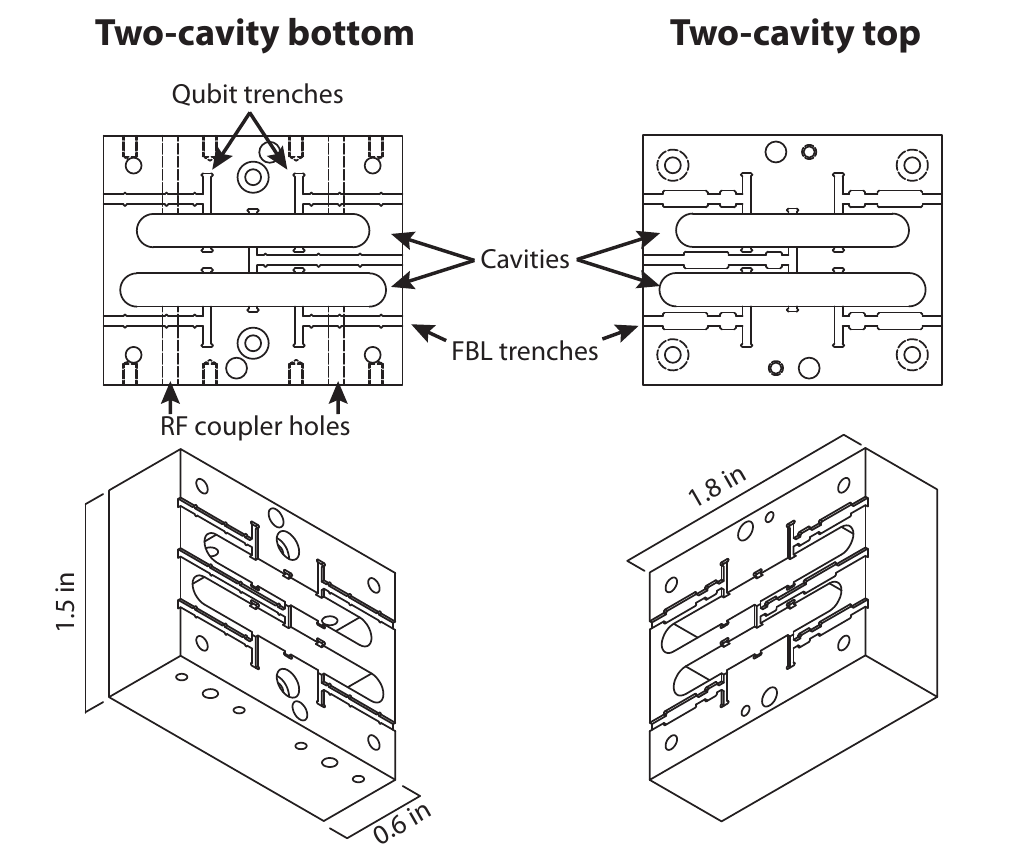}
	\mycaption{Tunable two-cavity design}{The features of the two-cavity version are much the same as the single-cavity.  The device supports as many as five tunable qubits, with four coupled to individual cavities and one coupled to both.  There are also four RF ports granting full control of both cavities.}
	{\label{fig:tunabletwocavitydesign}}
\end{figure}

We show the two-cavity device in \figref{fig:tunabletwocavitydesign}.  There, we see two cavities of slightly different size (with unloaded bare cavity frequencies of about $7.8\ghz$ and $9.2\ghz$).  Both the top and bottom cavities have qubit trenches traversing them, positioned at the same location as the single-cavity devices so that the same flux bias lines can be recycled.  A fifth tunable qubit is located between the two cavities at the exact center.  The FBL and qubit trenches also have the ``ears'' used to secure their contents with indium.  There are four RF coupling holes with matching tapped screw holes for two cavities.  Countersunk screw holes are located at the top and bottom of the devices, used for assembly, though the bottom device has two screws rather than three due to lack of space.  For the same reason, the indium seal trench has been omitted.  There are also alignment pins and corresponding tapped screw holes on the top of the device to ease assembly and disassembly.

\subsection{Octobox design}
\label{subsec:tunableoctodesign}

\nomdref{Asma}{SMA}{sub-miniature version A microwave connector}{subsec:tunableoctodesign}

\begin{figure}
	\centering
	\includegraphics{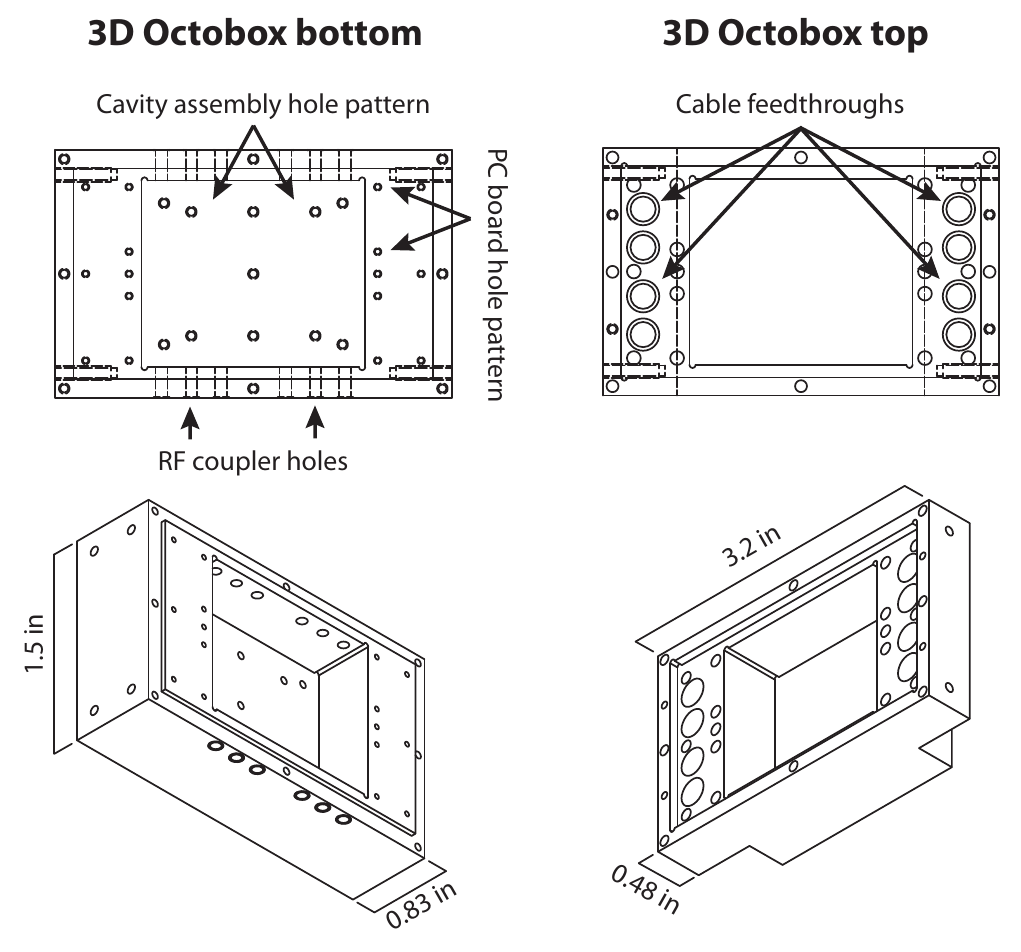}
	\mycaption{Octobox design}{The cavity is inserted into the pocket in the middle of the bottom half, alongside which PC boards are placed to interface with the flux bias lines.  The bottom of the octobox primarily serves as a platform for screwing together the rest of the assembly, with a screw pattern in the cavity pocket and on the FBL platform.  Clear holes for RF couplers are also located on the long sides of the box.  The top of the sample holder, which has a matching pocket for the top of the cavity, primarily functions to interface with the PC board connectors and convert them into SMA cables.  It also secures tightly to the octobox bottom, providing some additional magnetic shielding (if it is made out of a superconductor) and light shielding.  On both sides of the assembled octobox are six tapped screw holes, used to secure the sample holder to the fridge or to secure other experimental paraphernalia.  }
	{\label{fig:tunableoctoboxdesign}}
\end{figure}

A diagram of the 3D octobox sample holder is shown in \figref{fig:tunableoctoboxdesign}.  As before, we show separate diagrams for the top and bottom of the device, each with 3D perspective views below.  Focusing first on the bottom section, we see that the holder is large enough -- 1.5 in by 0.83 in by 3.2 in -- to fully enclose the cavities described in the previous section in a pocket at its center.  The floor of this pocket is patterned with tapped holes used for assembling the cavity, and is symmetric with respect to both cardinal axes of the device to simplify assembly.  Next to this pocket is an elevated platform for the PC boards with a matching screw hole pattern.  This pattern is again symmetric about the long axis of the octobox and supports both single-cavity and two-cavity variants.  The perimeter of the sample holder is lowered slightly to produce a lip that hopefully makes it more light-tight.  On this strip are tapped screw holes used to attach the top of the octobox.  There are twelve clear holes for the four RF couplers and tapped holes on the sides used for attaching the sample to the fridge, thermalization, and potentially attaching other samples to cool alongside.

The primary purpose of the top of the octobox is to interface with the Rosenberger connectors for the flux bias lines.  There are four holes on either side of the cavity used to capture the sample holder cables using a Fairview Microwave SMP-M 0.086" SR bulkhead connector (model number SC5161).  The length of this connector is specific; we pinch in the sides of the octobox directly above the PC board to form a ``top hat'' shape.  The large part of the box contains the pocket for the top half of the cavity.  There are additional shallow holes above the PC board screw pattern to make room for the heads of the screws.  There are also eight clear holes around the perimeter that match up with the tapped holes on the bottom for assembly.  An additional line of tapped holes are on the sides for thermalization.  The octobox is generally made out of aluminum to provide additional magnetic isolation, but could be made out of copper if thermalization is a larger concern.  By design, the coherent ``quantum'' systems are exclusively located inside the superconducting cavities.

\subsection{Qubit design}
\label{subsec:tunablequbitdesign}

\begin{figure}
	\centering
	\includegraphics{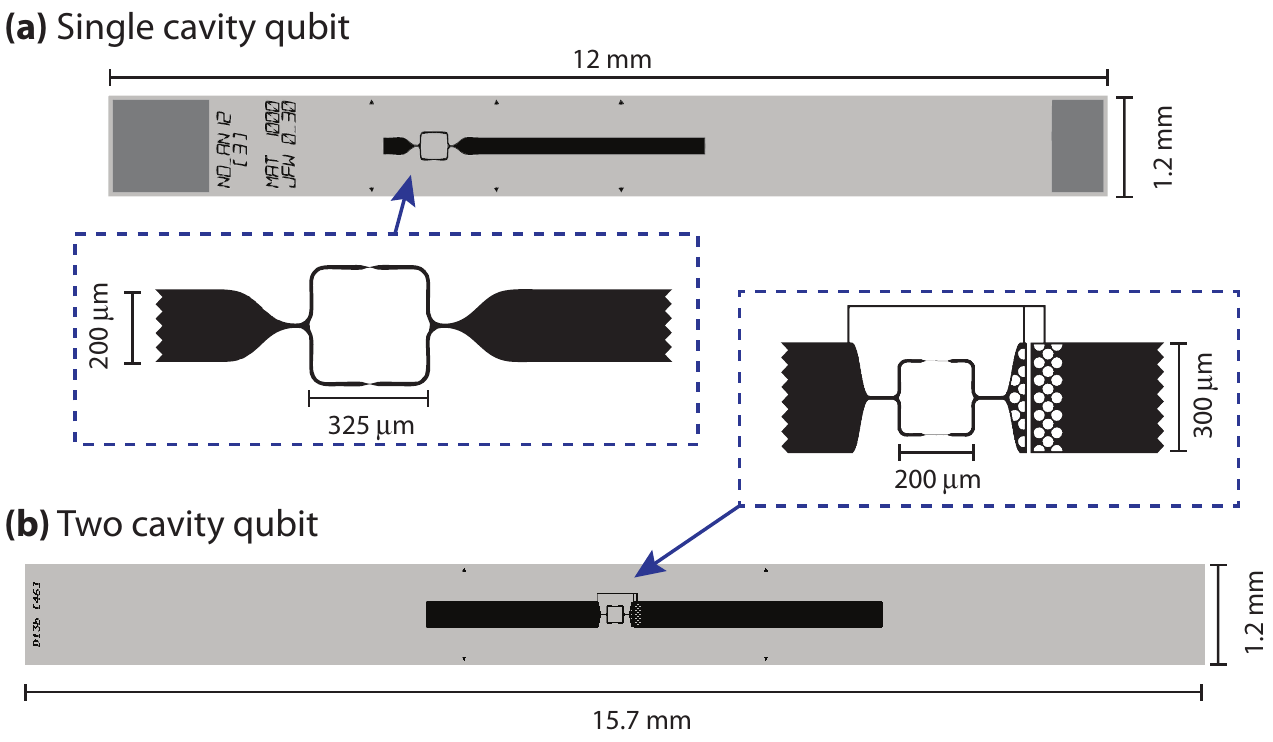}
	\mycaption{Tunable 3D qubit design}{Qubits are defined by four parameters: $E_J$, $E_C$, $g$, and flux sensitivity.  \capl{(a)} In the case of the single-cavity qubit, all four of these parameters can be chosen independently in the limit that one side of the antenna is much larger than the other.  \capl{(b)} For two cavities, we need both antennas to be long and couple to their respective cavities.  A capacitor between the junction and one branch of the antenna is required to keep the charging energy (and therefore, qubit anharmonicity) high enough.}
	{\label{fig:3dqubitdesign}}
\end{figure}

The tunable vertical transmon designs are shown in \figref{fig:3dqubitdesign}.  There are two variants: qubits designed to couple to a single cavity, or couple to two.  These qubits have four design parameters: the maximum Josephson energy $E_J$, the charging energy $E_C$, the coupling strength $g$ (or, for a fixed frequency, $\chi$) to the cavity or cavities, and the flux sensitivity (e.g. area of the loop, setting the magnetic field per flux quantum).  The Josephson energy is set by the area of the junctions and the amount of oxidation; the remaining three are set by the gross geometry that is visible in the figure.  The charging energy is approximately set by the smaller of the two island capacitances to ground, the coupling strength by the length of the antenna penetrating the cavity, and the flux sensitivity set by the size of the SQUID loop.

In the case of a single-cavity qubit, the design is straightforward because each of these parameters can be engineered essentially independently from one another.  We are able to make one side of the antenna as short as we like to set the charging energy.  Similarly, the length of the other side sets only the coupling because in the limit that this capacitance is much larger than the other side, the smaller one drops out\footnotemark.  The flux loop size is also mostly irrelevant to both parameters and can be changed without redesigning the entire qubit.  Thus, as we shown in \figref{fig:3dqubitdesign}(a), we can see that the left antenna is much smaller than the right, which sticks into the cavity by approximately $1.5\mm$.  (The location of the rightmost pair of arrows on the qubit denotes the wall of the cavity when it is installed.)  Though we have this design freedom, quantitatively predicting the qubit parameters requires numerical simulation of the geometry using a program like Ansoft HFSS.  This qubit also demonstrates an unusual feature: silver pads bookending the 12 mm by 1.2 mm sapphire substrate.  When shorted to the cavity with indium wire, this was intended to better thermalize the qubit and act as a quasiparticle drain.

\footnotetext{A corollary of this is when the two island capacitances are very different, the mode of the transmon is quite asymmetric as well.  It will therefore couple to both the common and differential modes of a flux bias line.  We calculated in \sref{subsec:fblrelaxation} that the contribution to $T_1$ of the common mode dominates that of the differential mode, which is unfortunate, since that is exactly the component we become more sensitive to as we utilize this trick.  One way to solve that problem is to use filtering, as we have done, but it may have been unnecessary if we were more clever about qubit design.}

Since we need both antennas to strongly couple to both cavities, we do not have the freedom of making one side of the qubit tiny for a two-cavity device.  In the first two-cavity 3D cQED device \cite{Kirchmair2013}, this trick was still viable given the possibility of spacing the cavities more closely (only 2 mm apart), placing the junction off-center, and making the antenna very thin.  In the case of the tunable architecture, however, we must have the cavities at least 4 mm apart to fit the flux bias line and require that the SQUID loop is near the center so that we can couple sufficient flux (see \figref{fig:3dtwooctoassembly}).  In order to circumvent this problem, we add a new sophistication: a capacitor that couples one side of the qubit island to its antenna.  As seen in the inset of (b), there is a very short piece of metal on the right side adjacent to a long antenna.  The holes surrounding the capacitive gap between these two features are present for technical reasons: to help avoid overdosing during electron beam lithography and shorting out the capacitor as a result.  The size of this capacitor and the length of the antenna then sets the size of the coupling to the right cavity and the position of the gap relative to the SQUID loop sets the charging energy.  We are free to choose the length of the opposite antenna to address other concerns.  The wires connecting the two sides of the junction and the auxiliary antenna are there to prevent the junctions from getting destroyed by static discharge during dicing, and are removed prior to installation.  Comparing the two qubit designs, we also see that the SQUID loop in the case of the single-cavity device is a great deal larger than the two-cavity device.  This is due to the fact that the single-cavity qubit was designed first, when we were not yet confident in our ability to thread flux.  As it became clear that flux coupling was not a problem, we chose to reduce the size of the loop in an attempt to reduce the magnitude of flux noise.  We also increased the pitch size of the antenna from $200\um$ to $300\um$ to reduce surface participation.

\subsection{Flux bias line design}
\label{subsec:tunablefbldesign}

\begin{figure}
	\centering
	\includegraphics{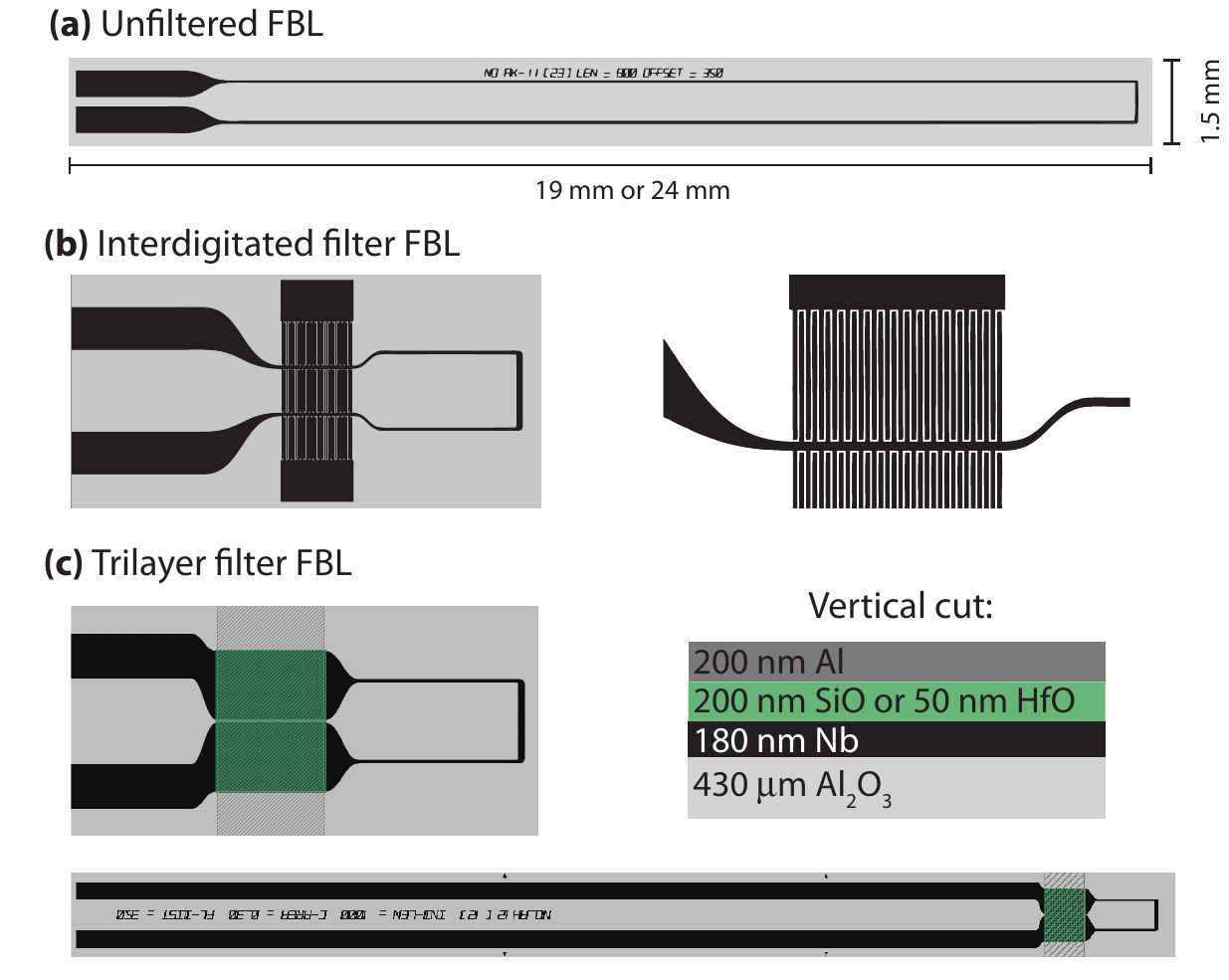}
	\mycaption{Flux bias line design}{The flux bias lines used in the tunable 3D cQED architecture are patterned on narrow sapphire substrates.  \capl{(a)} The prototypical FBL is a wire that approaches as close as possible to its qubit, with large pads on one end suitable for wire bonds.  This design unacceptably shortens qubit lifetime, however, necessitating filtering.  \capl{(b)} There are two approaches to implementing the FBL filter, depending solely on the capacitor.  The easier implementation uses an interdigitated finger capacitor between the two lines and the lines and some metal islands.  The islands are shorted out to the cavity.  \capl{(c)} The superior though more involved method uses a three-layer parallel plate capacitor design, which can attain much higher capacitances without self-resonance issues.  The constituent layers are shown to the right.  
	Note that in all these cases, the vast majority of the length of the FBL is simply a transmission line.  The length of this line is varied depending on its use, since there are three distinct FBL lengths with two for single-cavity qubits and one for the two-cavity versions.
	}
	{\label{fig:fbllitho}}
\end{figure}

The flux bias lines used in the tunable 3D architecture are on their own $1.5\mm$ by $\sim20\mm$ sapphire substrates.  The simplest form of these, shown in \figref{fig:fbllitho}(a), is a wire that passes from one of the PCB ports to the other, coming as close as possible to the qubit SQUID to maximize coupling.  As explained in \sref{subsec:fblrelaxation}, however, this type of bias line design is unacceptable because it will shorten the lifetime of the qubit via capacitive coupling.  The solution to this problem is to add a low-pass filter (\sref{subsec:fblfiltering}) to the line in the form of an inductor and capacitor.  In \figref{fig:fbllitho}(b-c), we show two versions of this filter.  In both cases, the inductance of the low-pass filter is produced by a length of wire $1\mm$ long, corresponding to approximately $1\nH$.  It is not possible to get substantially more than this from a wire at this frequency without running into problems of self-resonances.  As a result, the two bias line designs differ only in their capacitor implementation.  

The first-generation filter design, shown in \figref{fig:fbllitho}(b), uses an interdigitated capacitor.  The center pin of the FBL fans out with numerous fingers, which are interleaved with fingers of metal islands above and below the line.  There are also fingers between the two lines, granting extra capacitance to the differential mode.  During operation, the islands on the top and bottom are shorted to the electrical ground of the cavity using either compressed indium or wire bonds.  This filter helps somewhat by increasing the measured $T_1$ by approximately a factor of four.  It cannot, however, provide sufficient capacitance to solve the problem.  The surface length of the fingers sets the frequency scale of self-resonance; adding more fingers to get more capacitance would push this self-resonance frequency down, violating the lumped-element approximation, and making it inoperable as a capacitor at the relevant frequencies.  We knew this would be the case prior to measuring it, however, and only tested it because the fabrication was no more complicated than the unfiltered FBL found in (a).  We also enlarged the width of the transmission line used for the filtered FBLs to attempt a better impedance match.

A superior filter is shown in \figref{fig:fbllitho}(c), which uses a three-layer lithography process.  Instead of relying on the capacitance between two pieces of metal on the same plane (which is extremely inefficient) this approach uses the third dimension to produce a canonical parallel plate capacitor with a dielectric in between.  The lithography is straightforward: first, as shown in black in (c), the flux bias line is patterned in niobium.  It flares out considerably at the point where we wish to have our capacitor.  Next, a thin layer of dielectric -- we have tried $200\nm$ of silicon monoxide or $50\nm$ of hafnium oxide -- is deposited in the area designated in green.  Finally, a $200\nm$ thick strip of aluminum is deposited over the whole thing (denoted in gray), which constitutes the other side of the capacitor.  The aluminum is shorted to the bulk of the cavity with wire bonds.  (This type of filter is shown in both \figref{fig:3doctoassembly} and \figref{fig:3dtwooctoassembly}.)  This approach has the advantage of getting a capacitance many times higher before being concerned about self-resonances.  In practice, it seems to eliminate any relaxation through the FBL (\sref{sec:tunablequbitlifetime}).

\section{Dilution fridges and wiring diagram}
\label{sec:fridges}

\nomdref{Alo}{LO}{local oscillator}{sec:fridges}
\nomdref{Arf}{RF}{radio frequency}{sec:fridges}
\nomdref{Aiq}{IQ}{in-phase / quadrature}{sec:fridges}

Most of the experiments detailed in this thesis were performed in a Cryoconcept ``wet'' helium dilution refrigerator.  The fridge uses a bath of liquid helium to get to 4 Kelvin and a standard closed-circuit $^3\mathrm{He}$-$^4\mathrm{He}$ dilution unit to attain its base temperature of $T_b \sim 10-15\mK$.  The advertised cooling power at $100\mK$ is $200\uw$.  The helium bath must be filled approximately twice a week with $50~\mathrm{L}$ of liquid.  The level of this bath constantly changes, as does the fridge temperature profile, causing the attenuation of the cables to drift slightly in time.  More recently, the lab has shifted toward using Oxford Instruments brand ``dry'' fridges that use a high-performance pulse tube to reach $4~\mathrm{K}$.  These have the advantage of being much cheaper to run as they do not require liquid helium.  This provides both more experimental space and less overall bulk; it also requires much less work to cool (due to e.g. the lack of need for an indium seal) and operate.  Given the rapidly increasing price of helium in the context of these advantages, it is likely that many cryogenic labs will shift toward these dry fridges if experimentally possible\footnotemark.

\footnotetext{One example of where this is unfeasible is for experiments that are extremely sensitive to vibration.}

\begin{figure}
	\centering
	\includegraphics{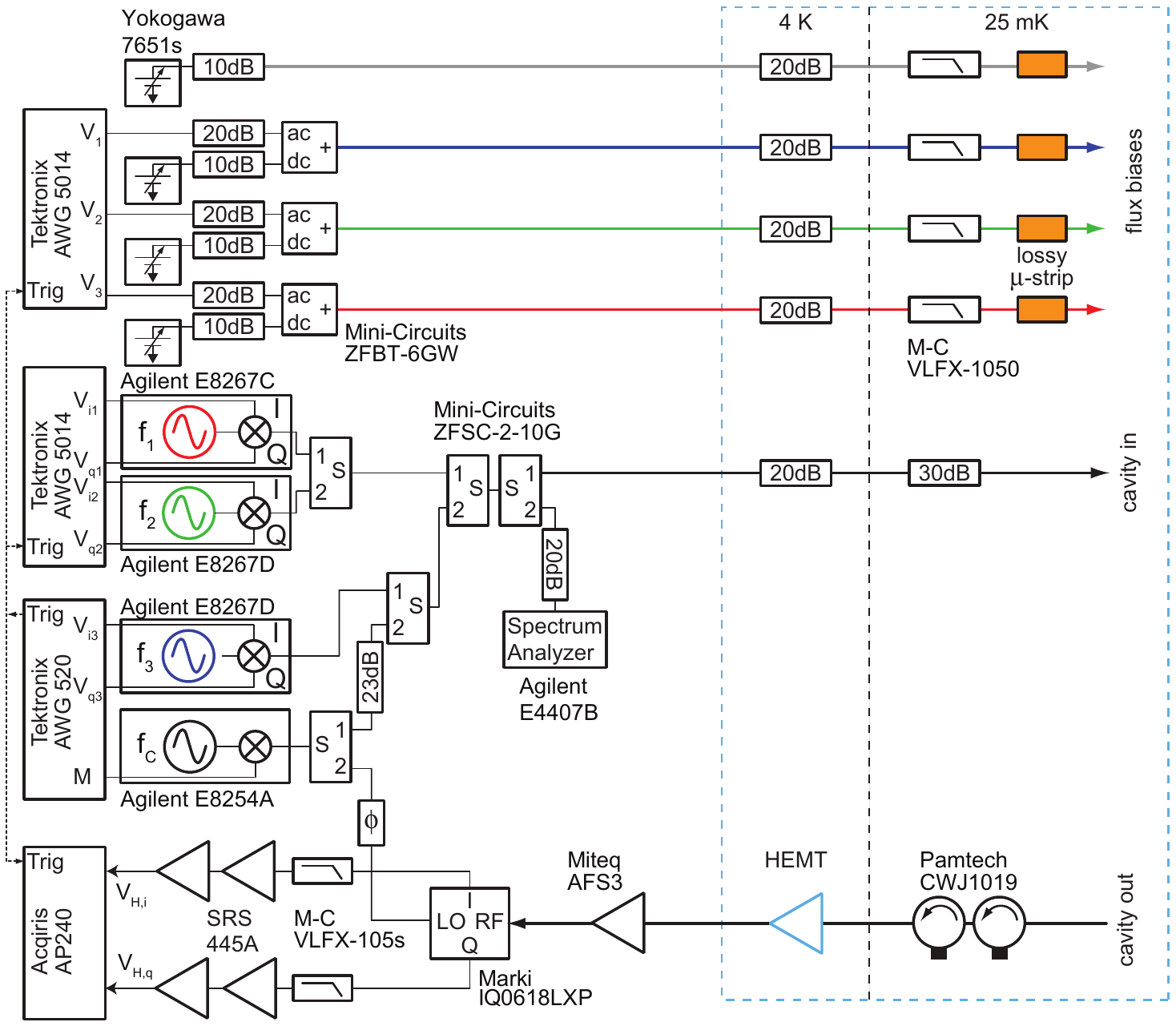}
	\mycaption{Diagram of room-temperature electronics and fridge cabling}{Cables going into the fridge must be attenuated at $4\K$ and base to prevent thermal noise from disrupting the experiment.  For flux bias line cables, attenuation at base would produce too much heat because of the large current required, so reactive filtering and lossy microstrips are used instead.  On the return side, we cannot attenuate our signal and instead use microwave circulators to reject room temperature thermal and amplifier back-action noise.  A HEMT amplifier is located at $4\K$, after the circulators, for the first stage of signal amplification; further amplification takes place at room temperature.  The overall amplification chain noise temperature is approximately $10\K$.  At room temperature, flux bias control signals are produced with Yokogawa voltage sources for DC biases and Tektronix AWGs for fast pulses.  A combination of vector and scalar generators controlled with additional AWGs produce single-qubit rotations and measurements.  The output of the amplification chain is mixed down to DC and digitized by an Acquiris analog to digital converter. \figthanks{DiCarlo2010}}
	{\label{fig:wiringdiagram}}
\end{figure}

The wiring diagram used to produce GHZ states, as described in \chref{ch:entanglement}, is shown in \figref{fig:wiringdiagram} and is representative of all the experiments.  Let us first explain what happens inside the fridge, as denoted by the dashed blue box.  Thermal Johnson noise coming down cables would constitute an unacceptable noise source to the sample, and must be ``filtered'' by lowering its effective temperature via cold attenuation.  In the low effective-temperature limit, voltage noise is linear in temperature $S_V(\omega) \approx 4 k_b T R$, so  we can attenuate at a cold stage by a factor $\alpha = T_{\mathrm{hot}} / T_{\mathrm{cold}}$.  Thus, cables carrying signals into the fridge are attenuated by $20\dB$ at 4 Kelvin to cut down room temperature $290\K$ noise.  The RF drive line used for measurement and single-qubit rotations (black) is further attenuated by $30\dB$ to go from $4\K$ to base.  The lossy stainless steel cables themselves (which are used for their minimal thermal conduction) constitute an additional $10\dB$ of attenuation, for a total of $\sim80\dB$.  Since resistively attenuating the large DC currents used for flux bias would generate too much heat at base, the flux bias line cables (gray, blue, green, red) use reactive low-pass and lossy strip-line filters.  They reject Johnson noise as well as prevent erroneous qubit coupling, while maintaining a $50\ohm$ impedance match at low frequency.

On the amplification (output) side, we do not have the luxury of attenuating our small signals, and instead use microwave circulators.  Circulators are three-port non-reciprocal devices by which microwaves going into one port are transmitted only to the port to the right \cite{Pozar, Kamal2011}.  Waves going to port 1 would come out port 2, those going into port 2 would go to port 3, and into 3, out 1.  We terminate one of these ports with a matched $50\ohm$ load to absorb the noise coming down from room temperature, but the signal coming up from the sample is transmitted.  This effective two-port device is also known as an isolator and can be purchased by itself.  Reverse-isolation is not perfect, so we typically use two or more in series.  In addition to rejecting noise from room temperature, circulators also block the back-action noise produced by the high electron mobility transistor (HEMT) amplifier.  This amplifier has about $40\dB$ of gain and a noise temperature of $5\K$.  If we include cable losses and reflections between the device and the amplifier, the total amplification chain noise temperature is closer to $10\K$.  We use additional room temperature amplification as well, bringing the total output gain to $\sim70\dB$.

Outside of the fridge, there are three sets of equipment.  Starting from the top of \figref{fig:wiringdiagram}, we have the four flux bias line channels.  This wiring diagram is for a cQED sample with four flux-tunable qubits.  In the GHZ experiment described in \chref{ch:entanglement}, only three of these qubits were used and the fourth tuned far away and left in its ground state.  The first FBL channel (in gray) provides a quiescent DC bias for this unused qubit, which needs no fast control.  It is attenuated at room temperature by $10\dB$ to reduce the noise coming from the Yokogawa 7651s voltage source.  The remaining three flux channels (in blue, green, and red) have a Yokogawa voltage source to provide DC offsets, but are also driven by a channel of a Tektronix 5014 arbitrary waveform generator (AWG) for fast control.  This AWG outputs four 14-bit channels at 1 GSample/s, with a voltage range of $\pm 2.25\V$.  It is attenuated by $20\dB$ and combined with the DC bias using a Mini-Circuits ZFBT-6GW bias-T.  In this experiment, the bias-T had a $100\KHz$ high-pass on the AC port, which limits the duration of fast flux pulses to a few microseconds.  For the purposes of the GHZ experiment, this was more than slow enough.  However, it may not be sufficient for longer-lived future experiments and could be replaced by a resistive splitter that has no intrinsic timescale.  It would be possible to use the AWG 5014 to provide both the AC and DC biasing of the qubits, though the Yokogawa sources are more stable and have higher resolution.  Separating the tasks allows for greater attenuation of the relatively noisy AWG outputs.

The middle set of equipment connected to the black RF input cable generates the microwave drives used in the experiments.  In \figref{fig:wiringdiagram}, qubit rotations are generated using wideband I-Q modulated vector generators (Agilent E8267C and 8267D).  The mixer control voltages are generated with a combination of a second Tektronix 5014 AWG and a Tektronix 520 AWG (which differs from the 5014 by only having two channels with 10 bits of vertical resolution each). The 520 AWG additionally generates a gate pulse for the Agilent E8254A RF scalar generator for measurement pulses and a clock to synchronize the AWGs which sets the repetition rate of the experiment.  The outputs of all four generators are combined with MiniCircuits ZFSC-2-10G splitters, the output of which is split to send half to the fridge and the other half to a diagnostic spectrum analyzer.  As we show below in \sref{subsec:mixercalibration}, the relatively expensive and exotic IQ generators can be readily replaced with much cheaper IQ mixers connected to RF generators that serve the same function.

The final set of equipment constitutes the measurement chain.  In this experiment, the output of the amplification chain was mixed down to DC by mixing it with a copy of the measurement pulse.  This copy, serving as the local oscillator (LO) of the mixer, is tuned with a phase shifter in order to isolate the readout signal to one quadrature.  The output voltages $I$ and $Q$ are then low-pass filtered and amplified again (with an SRS 445A preamplifier) before being digitized with an Acqiris AP240 digitizer.  This card has two input channels and can operate at a $1\ns$ sampling interval with 8 bits of resolution.  In all three sets of equipment, the AWGs, microwave synthesizers and acquisition card are clocked with a Rubidium $10\mhz$ frequency standard (SRS FS725, not shown).  Compared to the other two sets of equipment, the measurement chain varies most widely between experiments.  Often, a separate LO generator is used to mix down the measurement signal to some finite frequency to avoid low frequency noise.  The Acqiris digitizer card has increasingly been phased out in favor of a card made by AlazarTech that has a much faster connection to its host computer and can acquire and process data in real time.  The use of field programmable gate arrays (FPGAs) for real-time data acquisition, analysis, and conditional pulse sequencing has also recently become possible \cite{Riste2012a, Riste2012b}.

\section{Pulse generation at room temperature}
\label{sec:roomtemppulsegeneration}

\nomdref{Aawg}{AWG}{arbitrary waveform generator}{sec:roomtemppulsegeneration}
\nomdref{Aaps}{APS}{arbitrary pulse sequencer}{sec:roomtemppulsegeneration}
\nomdref{Acw}{CW}{continuous-wave}{sec:roomtemppulsegeneration}

Precise generation of microwave pulses at room temperature is required for high-fidelity single-qubit control.  Learning the most effective way to do this took considerable effort, but can now be easily reproduced with standard components.   Broadly speaking, microwave pulse generation is done by modulating the amplitude of a CW microwave tone in time with voltages provided by an AWG or APS (arbitrary pulse sequencer).  As shown above in \figref{fig:wiringdiagram}, one approach to applying this modulation is to use a vector signal generator such as the Agilent E8267D to both generate and shape a microwave tone.  The modulation is taken care of internally, and so one only needs to supply the modulation voltages from an AWG.  This approach works well: the instruments have built-in amplitude and pulse modulation, IQ leakage cancellation, and can control output power over fifteen decades, accurate to the hundredth of a dB.  Unfortunately, they are as expensive as they are capable.  They typically cost three times as much as an equivalent scalar generator (even without options, a new generator lists at \$101,192).  As we scale up both the number and complexity of our qubit experiments, it is impractical to continue relying on vector signal generators.
	
Fortunately, the relevant functionality of the vector generator can be replicated with much cheaper components.  A judicious combination of a microwave mixer, band-pass filters, a power amplifier, an arbitrary waveform generator, and a scalar generator, paired with a robust calibration scheme, can deliver performance essentially indistinguishable from a vector generator for a third of the price or less.  These components are also not a ``black box'' with layers of proprietary (and potentially unnecessary) Agilent technology between the experimentalist and the output, so more reliable and less noisy performance can also be expected.  The following section will serve to explain the various calibrations necessary and the reasoning behind them, as well as considerations to make when assembling the components.

\subsection{Calibrating the mixer}
\label{subsec:mixercalibration}

\nomdref{Assb}{SSB}{single sideband modulation or mixing}{subsec:mixercalibration}
\nomdref{Avswr}{VSWR}{voltage standing wave ratio}{subsec:mixercalibration}

The heart of our pulse modulation scheme is a microwave IQ mixer (here, the Marki IQ0618MXP).  These are four-port passive microwave devices which take a local oscillator (LO) microwave tone and convert it to an amplitude and phase-modulated radio frequency (RF) tone based on the applied voltage on the I (in-phase) and Q (out-of-phase) modulation ports.  Specifically, an ideal mixer's RF output should be proportional to the instantaneous low-frequency (e.g. DC to a few hundred MHz) voltage on the modulation ports (produced by an AWG), with the sine component of the RF given by I and the cosine by Q.  However, like all things in experimental physics, microwave mixers are not perfect.  We now introduce and explain the resolution to five important non-idealities.
	
\subsubsection{Nonlinearity}
\label{subsubsec:nonlinearity}

The simplest problem to fix is related to the nonlinearity of the mixer.  The proportionality between applied voltage and output power is only linear in the low-voltage limit.  The output power will compress (similar to the gain of a microwave amplifier when too much power is applied) if this voltage is too large, leading to distortions of the pulse.  For example, the voltage corresponding to a $\pi$ pulse will not be exactly twice that of a $\pi/2$ pulse.  The transfer function could be measured and inverted, but that is more trouble than it is worth.  A better solution is to ensure that you not apply very large voltages, staying well within the linear limit.  This comes at a cost of maximum output power.  Even when using a high-power mixer (sourced with a $16~\mathrm{dBm}$ LO), it is not possible to drive a fast $8\ns$ $\pi$ pulse for a typical experiment when limited to low, linear voltages.  For that reason, we use a power amplifier.

The power amplifier we typically use is the Mini-Circuits ZVE-3W-183+, which has a maximum output power of $3~\mathrm{W}$, a gain of $35\dB$, and bandwidth of $6-18\ghz$\footnotemark.  This amplifier requires a lot of power ($2~\mathrm{A}$ at $15~\mathrm{V}$) so a special power supply is necessary.  Its gain has proved sufficient to drive fast pulses on typical planar or 3D qubits while staying well within the linear regime of the mixer.  As with most amplifiers, however, its gain is not adjustable.  Since the input power to the mixer must be fixed (at e.g. $16~\mathrm{dBm}$), the course-grain tuning of the pulse power is done by adding attenuation to some combination of the output of the mixer, amplifier, or AWG.  Fine-tuning can be done by changing the amplitude of the AWG output.  This is somewhat inconvenient if you intend to move your qubit around in frequency significantly (correspondingly requiring large changes in spectroscopy power).  During those situations, it may be best to invest in a computer-controlled variable attenuator if a vector generator is unavailable.  Once the qubit's frequency is fixed for the purposes of an experiment, tuning up is straightforward.

\footnotetext{The ZVE-3W-183+ has a relatively high noise temperature and probably has more gain than is necessary so other amplifiers might be better suited to this task if a large output power is not required.}
	
One convenient way of ensuring that you do not apply too much voltage to the mixer is to add $20~\mathrm{dB}$ of attenuation to the output of the AWG.  Then, even when the AWG is at its maximum output voltage ($4.5~\mathrm{V}$ for the Tektronix 5014), the mixer is still guaranteed to be sufficiently linear.  This also has a significant advantage of increasing the effective DC voltage resolution seen by the mixer, which, as we will see, is crucial for effectively canceling the IQ leakage.  It may be preferable to split this into two $10\dB$ attenuators placed on either end of the cable going from the AWG to the mixer to minimize reflection delays.

\subsubsection{IQ leakage}
\label{subsubsec:iqleakage}

Finite LO leakage is one of the less trivial non-idealities of the mixer.  Even when there is zero applied voltage to both modulation ports, some LO will couple to the RF port of the mixer.  Uncorrected, this leakage is too large for our purposes -- on the order of $30\dB$ lower than the maximum output power -- which, if in resonance with the qubit, would drive a huge population.  Fortunately, this leakage can be significantly diminished by applying DC ``offset'' voltages to both of the mixer ports.
	
The optimal voltages are a function of LO frequency and power, and can drift by a small amount with time.  The easiest way to determine the voltages is with a spectrum analyzer.  With an AWG, you would run a sequence that applies zero nominal voltage to the mixer.  You then adjust the AWG's offset voltages to minimize the mixer output by monitoring the leakage on the spectrum analyzer.  The leakage drops off sharply as you approach the optimal offset voltages, so the efficacy of this cancellation is a strong function of available voltage resolution.  As mentioned in the previous section, the cancellation can be significantly improved by attenuating the output of the AWG by $20\dB$, which effectively increases the voltage resolution by a factor of ten.  Further attenuation does not seem to improve cancellation beyond that, possibly due to noise of the AWG output.  It may be possible that a less noisy voltage source could improve cancellation, though other considerations then become important as well.
	
\subsubsection{Pulse modulation}
\label{subsubsec:pulsemodulation}
	
While IQ leakage can be greatly reduced with applied offset voltages, even a small resonant leakage can cause problems.  This is especially true as the $T_2$, and therefore the Rabi decay time, of qubits increase.  When using a vector generator, we were able to avoid this by instantaneously turning the output of the generator off when not intentionally driving.  This instantaneous control is known as ``gating'' or ``pulse modulation'' and can be toggled on nanosecond timescales to provide another $\sim80\dB$ of isolation.  While it is still necessary to cancel IQ offsets with the vector generator, gating greatly reduces our sensitivity because the leakage only occurs during a tiny fraction of the experimental duty cycle when pulses are intentionally applied.  Unfortunately, we cannot use the internal modulation of the RF generator when using a mixer for pulse shaping.  The mixer's output is distorted immediately after such a pulse due to a turn-on transient.  While it may be possible to correct for these distortions, it is much easier to avoid the issue entirely by using an additional switch further down the chain.  

We have used the Hittite HMC547LP3 chip to provide pulse modulation.  These chips have excellent bandwidth (DC-20 GHz), low insertion loss ($< 2 \dB$), high isolation ($< 45 \dB$), extremely fast response time ($< 3 \ns$), and are relatively inexpensive when purchased as a single chip.  The only disadvantage is that the TTL signal used to switch them is an inconvenient voltage, namely $0$ to $-5\V$.  Ideally we would use one of the plentiful marker bit channels output by the Tektronix to do this modulation, though in practice we have to use two analog channels of the AWG because the marker bits can only source 2 V.  Fortunately, the $0$ to $-5\V$ specification is somewhat negotiable, and seems to work with the $0$ to $4.5\V$ provided by the 5014 analog channels, or even the $0$ to $2.5\V$ provided by the 520 analog channels.  (You have to be a little unconventional with the Tektronix 520 by using $0$ to $2.5\V$ on one channel and $-2.5$ to $0\V$ on the other.)  We hope to soon produce a board with some analog electronics to take a single $0$ to $2\V$ marker bit from the 5014 and convert it to two $0$ to $-5\V$ channels necessary to drive the board.  Hittite also sells connectorized switches (part number HMC-C019), but they cost nearly \$2000.

\subsubsection{Single sideband mixing}
\label{subsubsec:ssb}

If a Hittite switch or the necessary AWG channels are unavailable, there is an alternative technique called {\it single sideband mixing} (SSB) \cite{Carson1915}.  The idea is that by multiplying the pulse waveform a sine and cosine of some frequency on the I and Q channels respectively, the resulting pulse frequency will be offset from the local oscillator.  Typical SSB frequencies are $50$ to $200\mhz$, which means that the LO can be that far detuned from the qubit.  The maximum SSB frequency is set by the bandwidth of the Tektronix and the mixer, and was about $300\mhz$ for the equipment used in this thesis.  Additionally, it is possible to use a single mixer chain to pulse at more than one frequency.  This is very useful for pulsing multiple number-split peaks or two separate qubits that are close in frequency.  The mathematical details can be found in Blake Johnson's thesis \cite{JohnsonThesis}, and have been implemented in the Mathematica and Matlab pulse generation code used in the Schoelkopf lab.  Sideband modulation is also sensitive to two additional non-idealities of the mixer, which, as we will discuss in the next subsection, gives us a convenient method of tuning up their corrections.

\subsubsection{IQ skewness and amplitude imbalance}
\label{subsubsec:iqskew}
	
When single sideband mixing is working properly, the only tone going through the mixer should be at the intended frequency (e.g. the LO minus the SSB frequency).  However, there are two more tones: one at the LO frequency, and one at the opposite sideband (LO plus the SSB frequency).  The tone at the LO frequency is due to IQ leakage, and can be cancelled as described in the previous section.  The upper sideband, however, is due to two more sources of mixer imperfection: skewness and amplitude imbalance.  The presence of either of these errors will cause power to be delivered at this unwanted frequency.

Applied voltages on the I and Q ports should produce RF tones exactly out of phase from one another.  However, the two ports will not actually be perfectly orthogonal.  We describe this imperfection by imagining that the Q axis is rotated by some angle relative to where it should be, so there is some projection of Q onto I.  This {\it skewness syndrome} can be corrected by rotating Q back using a linear transformation.  (The mathematics of this have already been worked out and are included in the pulse generation notebook.)
	
The same voltages on the I and Q ports should produce the same amount of power at the RF port.  In reality, there is often a small {\it amplitude imbalance} between these ports.  If you are attenuating the output of the AWG, small differences in the size of the various attenuators can cause the same error.  This will result in a rotation around the $x$-axis being different than one around the $y$-axis for the same nominal voltage.  This error is corrected by multiplying one quadrature voltage by some factor.

How do we determine this rotation angle and scale factor?  It turns out to be easily done, as with IQ leakage, using a spectrum analyzer and Tektronix 5014 AWG.  First, set the AWG to play a sequence of a very long (e.g. $15\us$) single-sideband shifted square wave.  Then, using a spectrum analyzer, monitor the power coming out of the mixer at the unwanted sideband frequency.  We then want to adjust the scale factor and angle to minimize this leakage.  Changing the scale factor is accomplished by adjusting the amplitude of only one of your two AWG outputs.

Controlling the rotation angle is a bit more subtle.  The AWG 5014 has the capability of delaying the output of channels relative to one another.  This delay is set in multiples of $5~\mathrm{ps}$, up to a maximum of several nanoseconds.  In the limit of a long SSB-shifted tone, this delay turns out to be mathematically equivalent to a phase shift according to $\Delta\phi = 2\pi f_{ssb} \Delta t$.  (Both the Tektronix AWG 520 and the BBN Arbitrary Pulse Sequencers lack this feature, so some other method  of determining $\phi$ is necessary when using those instruments.)  Thus, you can extract the rotation angle and scale factor by iteratively adjusting the amplitude and channel delay to minimize the upper sideband leakage.  Once known, these parameters should be incorporated into the sequence file and not left enabled on the AWG.  (For short pulses, the AWG delay is {\it not} the same as a skewness rotation.)

\subsubsection{Filtering}
\label{subsubsec:filtering}
	
The final mixer correction is also the simplest.  The mixers we use (the Marki IQ0618MXP) have a very large RF output at twice the LO frequency.  This frequency is so high that it is not immediately obvious that it would cause problems, but is unsettling in the context of higher excited states and multi-photon transitions.  Fortunately, it is simple to remove: add a band-pass or low-pass filter to the output of either the mixer or power amplifier.  For qubits between $6$ and $8\ghz$, a pair of Mini-Circuits VBFZ-6260-S+ filters work well, but other filters would certainly be acceptable.  Where best to place these filters relative to the amplifier is not obvious.  It is beneficial to prevent the large tones from going into the amplifier to cut down on higher-order products, but it is also advantageous to have the filters after the amplifier to reduce the noise sent to the fridge.  In practice, we have simply had one filter before and one filter after the amplifier, but it is not clear that it matters either way.

\subsection{Assembly and reflections}
\label{subsec:assembly}	
	
\begin{figure}
	\centering
	\includegraphics[scale=1]{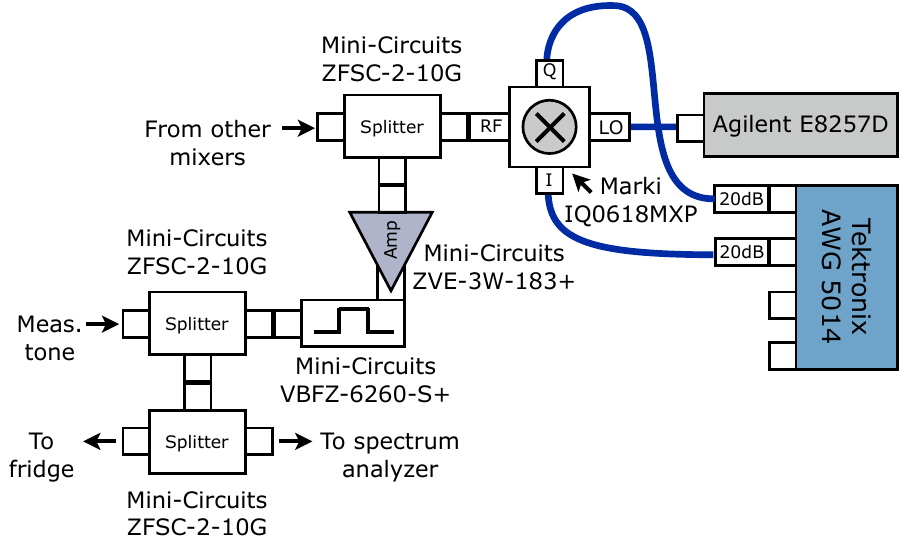}
	\mycaption{Typical mixer assembly scheme}{The components discussed so far are assembled as shown.  Multiple qubit channels can be made via the addition of more mixers to the first splitter.  To better control reflections, it may be better to split the $20\dB$ AWG attenuation between either end of the cable -- that is, $10\dB$ at the AWG and $10\dB$ at the mixer.  In this scheme, an extra splitter is added right before sending the output to the fridge for attaching a spectrum analyzer to.  It is important to terminate this port when not actively using the spectrum analyzer, as the additional cable length going to it can cause problems with reflections. 
	}
	{\label{fig:mixer_assembly}}
\end{figure}

A schematic of a typical setup is shown in \figref{fig:mixer_assembly}.  This scheme is scalable to many qubits -- simply duplicate everything to the right of the first splitter as necessary.  If only one spectroscopy tone is necessary, omit the first splitter and plug the mixer's RF port directly into the power amplifier.  It should be noted that additional savings (compared to a vector generator) can be had by using a cheaper scalar microwave generator known as a LabBrick.  It costs a tenth the price or less of a full-featured Agilent scalar generator -- \$3,000 vs. \$35,000 or more.  However, compared to Agilent generators, LabBricks have significantly worse phase noise, frequency resolution, power precision, and dynamic range.  Fortunately, these disadvantages are irrelevant for the single-qubit rotations, though likely are unacceptable for qubit measurements or biasing of quantum-limited amplifiers.

\begin{figure}
	\centering
	\includegraphics[scale=1.15]{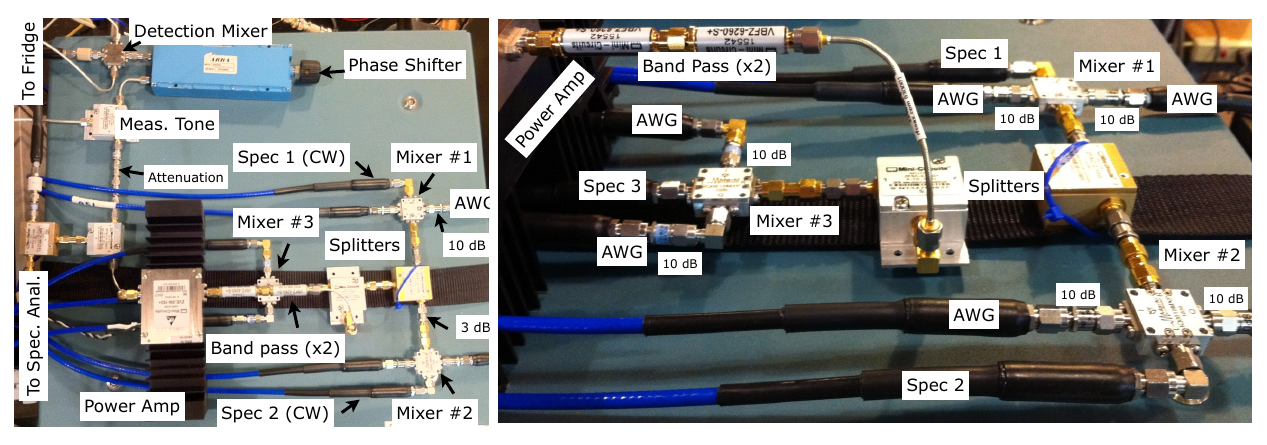}
	\mycaption{Photograph of mixer assembly}{\capl{(a)} An overhead view of a typical setup for generating pulses for three separate qubits.  \capl{(b)} A sideways view, showing the third mixer more clearly.  Here, we have $10\dB$ of attenuation at the mixer and another $10\dB$ at the AWG (not shown).  Two band-pass filters are also used to reject unwanted frequencies, and are placed before the amplifier.  In this picture, both were attached before the amplifier, but in later implementations one was put before and one after, in order to reduce amplifier noise power being sent to the fridge.  An ``autodyne'' measurement scheme is used here, where the RF and LO are the same generator.  The length of cable between the mixers and the splitter headed to the fridge is minimized wherever possible.  Note that the Hittite switch was not yet integrated into this scheme at the time of the picture, but would normally go after the power amplifier.	}
	{\label{fig:mixer_assembly_pictures}}
\end{figure}

In addition to knowing which components to use and effective ways of tuning them up individually, it is important to consider reflections when assembling the apparatus.  Any impedance mismatch between components can cause reflections of time-dependent pulses, which, when combined with significant lengths of cabling, will result in ``ghosts'' of your pulse arriving as much as tens of nanoseconds after the intended pulse.  This can cause significant rotation errors and can be very difficult to diagnose.  While it would be best to use only components with well-controlled impedances, they are not always available.  In particular, the mixers and splitters we use appear to have unfavorable VSWRs.  However, the nature of the reflection matters a great deal.  If the echo takes place only a few nanoseconds after your intended pulse, it will be calibrated away easily.  If it comes in a long time after, when you may be performing some subsequent pulse, that rotation can be significantly distorted.  (See \sref{subsec:allxy} for more on this.)  Therefore, we aim to minimize the reflection delay by limiting the length of the cable through which the pulses travel.  Note in particular that cable length only matters for \textit{time-dependent} pulses.  You are free to pipe the CW generator output as far as you like, so we prefer to have the mixers as close to the top of the fridge as possible.  A photograph of a typical setup for three qubits is shown in \figref{fig:mixer_assembly_pictures}.  

\section{Conclusions}

This chapter introduced the two specific architectures for realizing cQED.  The planar or 2D architecture, which has been studied extensively in this lab for many years, is the subject of much of the remaining content of this thesis.  We summarized its properties and showed how the devices are fabricated and packaged.  We also introduced a new tunable 3D cQED architecture, which has the useful flux control found in the 2D architecture but also enjoys the improved qubit lifetimes of 3D qubits.  Getting this to work involves re-engineering every constituent part, including the ``samples'' (of which there were two variants, with one or two cavities), the sample holders, the qubits, and the flux bias lines.  We also showed three generations of flux bias line designs, each implementing more sophisticated filtering than its predecessor.  Finally, we discussed how these samples are cooled in dilution fridges and operated with an overview of the room-temperature control equipment involved.  It is possible to use either a vector-modulated RF generator to generate our microwave control pulses, or, with proper calibrations, a much cheaper scalar generator and RF mixer.  In the next chapter, we will go into detail about the initial measurements to bring a new qubit experiment online, the best way to accurately tune-up single-qubit rotations, and will introduce the basic single-qubit experiments that we will use as tools in future chapters.

\setcounter{chapter}{4}
\chapter{Single Qubit Gates}
\thumb{Single Qubit Gates}
\lofchap{Single Qubit Gates}
\lotchap{Single Qubit Gates}
\label{ch:singlequbitgates}


\lettrine{A}{ccurate} generation of microwave pulses is an important experimental capability when working with superconducting qubits. This has already proven true with applications such as state tomography, which requires measuring the projection of a state along various axes of the Bloch sphere and is sensitive to errors in both the Bloch angles $\theta$ and $\phi$.  In the future, single qubit gates with verifiable fidelities in excess of 99.9\% will be crucial for practical error correction and other applications.  With recent advances in qubit coherence \cite{Paik2011, Rigetti2012, Chang2013}, the dominant source of gate infidelity is no longer necessarily $T_1$, but rather the technical details of pulse generation and non-ideal terms of the system Hamiltonian (such as higher levels or spurious couplings).

This chapter begins by explaining the conventional experiments that must be done with a new cQED device to bring it online.  This involves finding the cavity frequency by measuring transmission (at both high and low drive powers) and performing spectroscopy to find the qubit.  We will then examine the methods we have developed for tuning up and verifying pulse calibrations.  These range from simple Rabi oscillations to more sophisticated sequences such as ``AllXY'' which significantly increase the precision of the tune-up.  AllXY is both much more sensitive to certain kinds of errors than other sequences and detects many linearly-independent error syndromes at the same time.  Finally, we will discuss the kinds of techniques that will be necessary to further increase and accurately measure pulse fidelities.

\section{Experimental bring-up}
\label{sec:expbringup}

When a new cQED device is first cooled down and measured, there are several standard experiments that are always done to determine basic device parameters.  Through many repeated trials, we have gotten quite good at them; what might have taken weeks or months in the past has now been smoothed out or automated to the point of only taking hours or days.  This section introduces the standard suite of experiments to set up measurements and find and track the qubit as a function of applied magnetic field.  These experiments also serve to debug the experimental setup, as they will take advantage of many of the basic capabilities that will be required for more sophisticated experiments.

\subsection{Cavity transmission}
\label{subsec:transmission}
	
The first and most basic measurement that should be done in every cQED experiment is to find the resonance frequency of the cavity.  This is done by measuring transmission through the cavity ports as a function of frequency, and serves several purposes.  First and most obviously, knowing the cavity frequency is necessary for qubit measurements\footnotemark.  Transmission also verifies that the experiment is set up correctly (amplifiers powered, cables connected, generators on, etc).  It can be done with either a network analyzer or a computer-based data acquisition card and heterodyne detection \cite{SchusterThesis}.  While a network analyzer is the faster option (since this is exactly the kind of measurement they are designed to do), it is usually a good idea to try and make things work with the instruments that will be used for the qubit experiment in order to debug more of your setup.

\footnotetext{We saw in \sref{subsec:dispersivelimit} that the simplest readout mechanism involves measuring the transmission through the cavity at the frequency corresponding to the qubit in the ground state.  This requires knowing the hybridized cavity frequency, and uses a very low RF power.  As we will discuss in \chref{ch:qubitmeasurement}, there is another means of measurement that involves using a very high power and a different frequency -- the so-called {\it bare cavity frequency}.  Unlike the dispersive cavity location, this frequency is set entirely by the physical geometry of the device and does not depend on the coupling to or detuning of the qubit.  Both measurement schemes play a role in current experiments, and so the bring-up experiments depend slightly on which readout you use.}

\begin{figure*}
	\centering
	\includegraphics[scale=1]{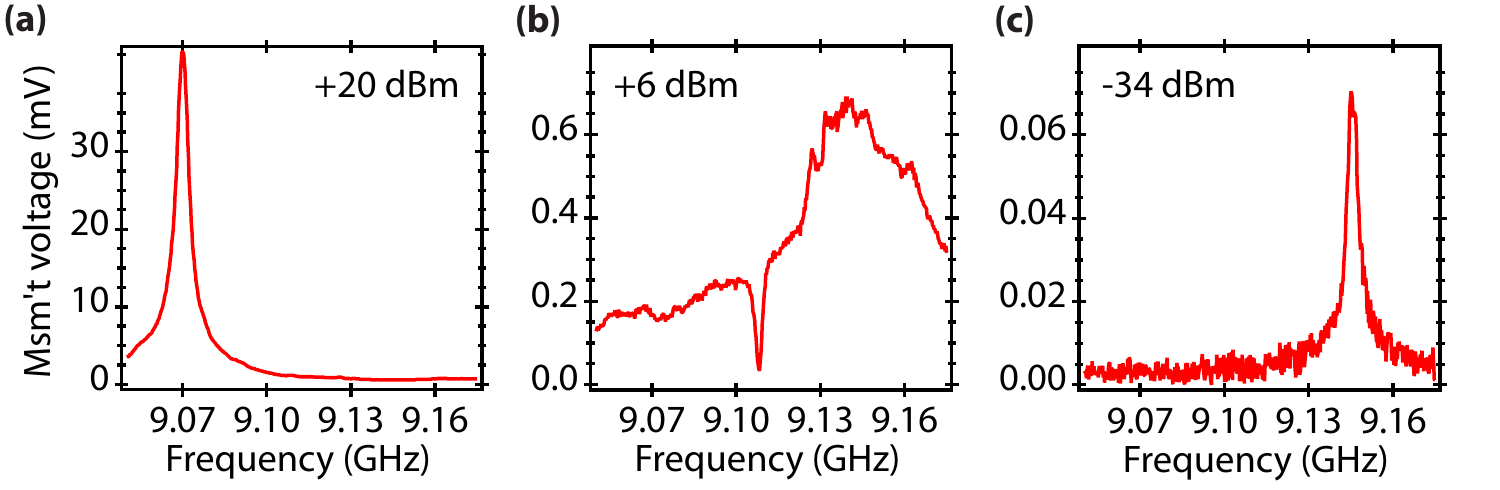}
	\mycaption{Cavity transmission for various drive strengths}{\capl{(a)} Transmission using a relatively large drive strength (+20 dBm), showing the ``bare cavity'' frequency and linear response of the resonator.  \capl{(b)} As we turn down power (to +6 dBm), the cavity reveals a strongly nonlinear response, indicating the presence of a qubit which is hybridizing with the cavity mode.  \capl{(c)} At sufficiently low drive strengths (-34 dBm), the cavity has a mean occupation of only one photon and returns to linear response at its ``low power'' frequency used for dispersive readout.  Notice the differing $y$-scales for each case.
	}
	{\label{fig:transmission}}
\end{figure*}
	
The power that should be used for this transmission experiment is actually rather important, because of the cavity anharmonicity inherited from the qubit.  The frequency and structure of cavity transmission depends very strongly on input power, which is the basis of the high-power readout scheme \cite{Reed2010b} described in \sref{sec:jcreadout}.  As shown in \figref{fig:transmission}(a-c), we see that, as we lower the drive power, the lineshape of transmission changes from a lorentzian response at the {\it bare cavity frequency} at high power, to a highly nonlinear response at medium power, and finally to a weak lorentzian response at the dispersively shifted frequency at very low ($\sim 1$ photon mean excitation) power.  This behavior is only present when a qubit is strongly coupled to the cavity, so seeing any non-linear power dependence of the cavity frequency and response is an easy way to detect the presence of qubits.  It makes sense to first use a relatively large amount of power to find the bare cavity frequency and then repeat the measurement with substantially less power to try and detect the presence of nonlinearity.
	
In experiments where a low-power dispersive readout mechanism is going to be used, measuring the frequency of the cavity at $\sim1$ photon mean occupation is necessary.  Since knowing the input power required to drive one photon is a function of line attenuation, cavity $Q$, and other parameters that may only be roughly known, a good tactic is to turn down the drive power until the apparent cavity frequency stops changing.  Especially in the case of high $Q$ cavities, this may become impractical if the signal becomes too small.  In that case, once the qubit is found, number splitting (\sref{subsec:dispersivelimit}) can be used to more efficiently measure the ultra-low power cavity frequency.  In experiments where the qubit frequency can be changed, the low-power cavity frequency will also move, and must be found at every magnetic flux.  This is in contrast to the high-power bare cavity frequency, which is independent of qubit detuning.  It is also worth noting that the difference between the high-power and low-power cavity frequencies and the approximate coupling strength of the qubit and cavity can be used to calculate the qubit detuning.  If these frequencies are almost the same, the qubit is weakly coupled and/or far detuned.
	
\begin{figure*}
	\centering
	\includegraphics[scale=1]{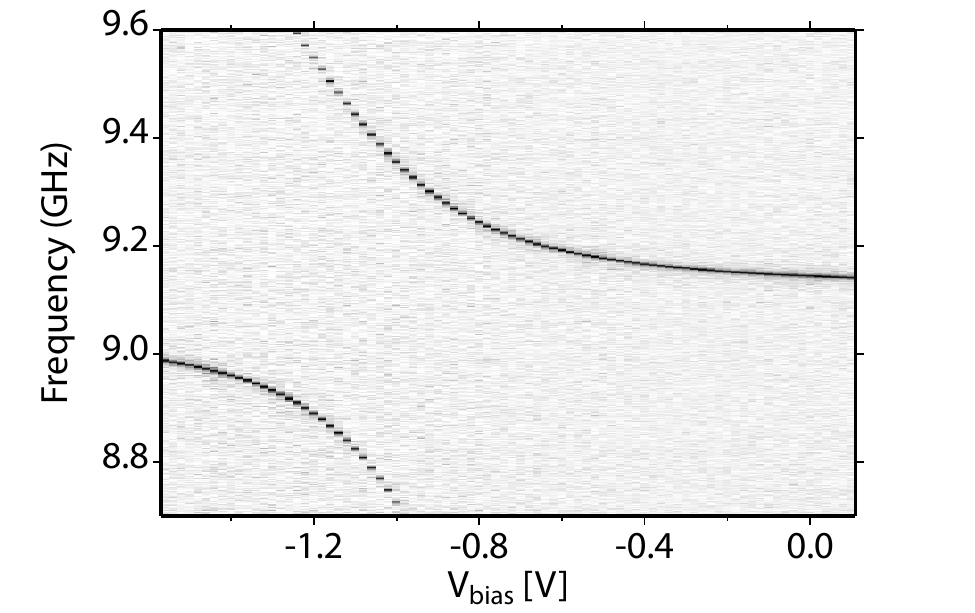}
	\mycaption{Transmission vs. applied qubit flux}{When the qubit frequency is tuned through the cavity, the two hybridize and make an avoided crossing.  This is known as a vacuum-Rabi splitting.  These data were taken using the four-qubit device described in \sref{sec:fourqubitdevice}.
	}
	{\label{fig:vacuumrabi}}
\end{figure*}
	
Measuring transmission as a function of qubit frequency can also inform you about device parameters.  If the qubit's maximum frequency (at the zero SQUID flux point) is above the cavity, tuning the qubit through the cavity will produce {\it vacuum-Rabi splitting} curve like the one shown in \figref{fig:vacuumrabi}.  There, as described in \sref{subsec:qubitcavitycoupling}, the cavity response splits into two peaks because of the avoided crossing between the cavity and qubit.  The size of this splitting indicates the coupling strength between the two.  Even if the qubit does not actually come into resonance, the cavity will still shift in frequency away from the qubit as it approaches.  This can be useful for finding the magnetic field corresponding to the maximum qubit frequency since it is usually offset from $0$ by ambient magnetic fields.  In either case of maximum qubit frequency, the conversion of applied control voltage to SQUID flux quanta can be measured by tuning over a full period of flux.

\subsection{Spectroscopy}
\label{subsec:spectroscopy}

\nomdref{Cfmax}{$f_{\mathrm{max}}$}{maximum transition frequency of a flux-tunable qubit}{subsec:spectroscopy}

\begin{figure*}
	\centering
	\includegraphics[scale=1]{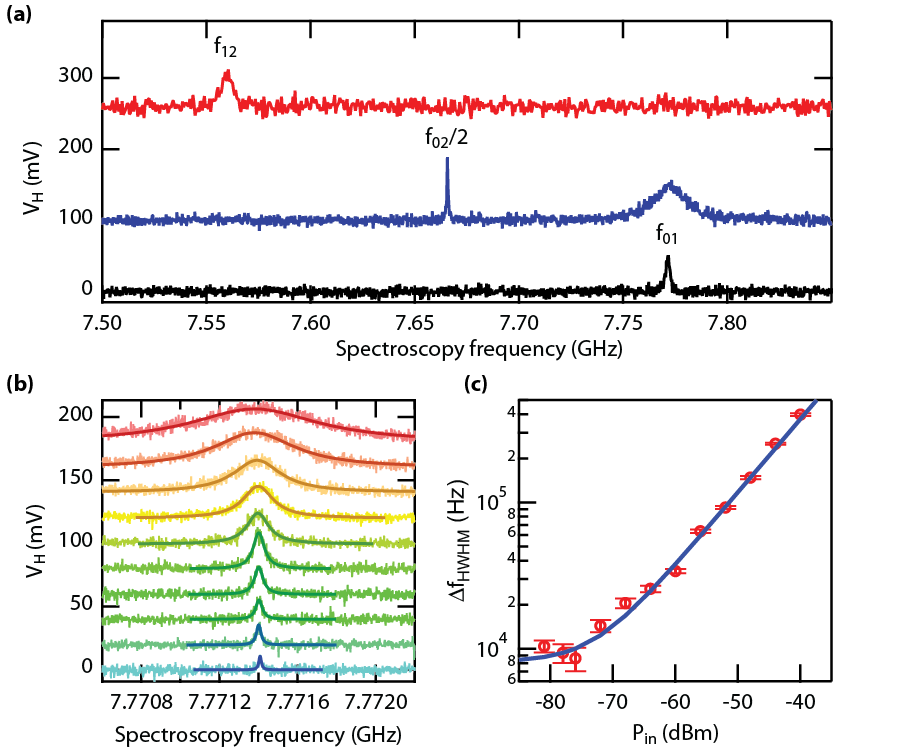}
	\mycaption{Qubit spectroscopy}{\capl{(a)} In black, spectroscopy of a 3D qubit is performed with a low drive power.  Blue shows the same trace with significantly more power, broadening the $0\rightarrow1$ qubit transition and turning on a two-photon transition directly between $|0\rangle$ and $|2\rangle$.  Knowing both of these frequencies tells you the qubit anharmonicity.  This is verified with a third experiment, shown in red, where the $0\rightarrow1$ transition is saturated with a CW tone and spectroscopy is done with a second tone, revealing the $1\rightarrow2$ transition.  The difference between these two frequencies is a direct measure of anharmonicity, with the $f_{02}/2$ transition located directly between.  
	\capl{(b)} The $f_{01}$ transition as a function of drive power, demonstrating power broadening.  
	\capl{(c)} The FWHM of the qubit transition as a function of spectroscopy drive power, fit to the expected power dependence.
	\figadapt{Paik2011}.  
	}
	{\label{fig:spectroscopy}}
\end{figure*}

The next experiment generally performed to bring-up an experiment is spectroscopy.  There are several variations commonly used in the field, but they all involve applying a microwave pulse at some frequency followed by a qubit measurement\footnotemark.  The frequency of the pulse is typically scanned and, when it drives transitions of the qubit, the subsequent measurement will indicate as such.  The simplest conceptual version of spectroscopy is known as {\it pulsed spectroscopy}, where the qubit pulse is turned off prior to a measurement being turned on.  The qubit pulse could either be some fast gaussian rotation, or more commonly, a long saturation tone.  A tone that is much longer than the qubit $T_2$ will drive some equilibrium incoherent population of the qubit, and is convenient because that population is relatively insensitive to the pulse power.  (A short, coherent drive would require carefully tuning the $\pi$ pulse power at each detuning to maximize fidelity, and could even eliminate signal if we inadvertently drive a $2n\pi$ pulse.  However, this approach can yield a larger signal if SNR is a challenge.)  The spectral width of a long pulse is much smaller than the faster gaussian, which allows for narrower line widths.  An example of saturation pulsed spectroscopy is shown in \figref{fig:spectroscopy}(a).

\footnotetext{In addition to pulsed and CW spectroscopy discussed here, we will later introduce ``swap spectroscopy'' as well as ``flux spectroscopy'' which add a fast-flux pulse to the mix.  They serve to measure avoided crossings in the time domain, which is vital to calibrating sudden entangling gates or to calibrate flux pulse amplitudes.  The second of these is a requirement when using fast flux to quantitatively measure a system.}

Prior to the high-power readout coming into common usage \cite{Reed2010b}, spectroscopy was more often done with both the qubit saturation and measurement tones on continuously.  One would step the frequency of a continuous-wave (CW) microwave tone while monitoring the transmission of second tone at the low-power dressed cavity frequency to detect qubit transitions \cite{ChowThesis}.  This is known as ``CW spectroscopy.''  Since this approach does not require time-domain control it is quite straightforward to set up.  It has the disadvantage of producing a more-complicated spectrum than pulsed spectroscopy, however.  There is an equilibrium population of photons in the cavity, so qubit transitions that depend on the number of photons will be visible.  The most concrete example of this is qubit number-splitting (\sref{subsec:dispersivelimit}) which will either split or broaden the qubit response, though higher-order transitions may also be visible.  This is in contrast to pulsed spectroscopy, where only one microwave pulse at a time is ever on and many of these complications are absent.  CW spectroscopy is also incompatible with the high-power readout mechanism, since the qubit state is scrambled by the measurement tone, so finding the dressed cavity frequency for each detuning is necessary.  The experimental simplicity of CW spectroscopy is often out-weighed by the convenience and high fidelity of the high-power readout, so CW spectroscopy is seldom performed anymore.

When using a long saturation tone for spectroscopy, the drive strength is an important consideration.  As described in Ref.~\citenum{Schuster2005}, at low drive power, the spectroscopic linewidth of the qubit is set by the pure dephasing time $T_2^*$.  This is because the inverse of this number can be taken as the intrinsic uncertainty of the qubit frequency.  As the power of the drive is increased, the observed linewidth will also grow, a phenomenon known as {\it power broadening}.  The expected power dependence is given by $2\pi \Delta f_{\mathrm{HWHM}} = \sqrt{1/T_2^2 + 4 \Omega_R^2 T_1/T_2}$, where $T_1$ and $T_2$ are the qubit lifetime and dephasing time and $\Omega_R$ is the drive Rabi rate \cite{Schuster2005}.  It can be understood as the drive causing stimulated emission of the qubit, shortening its lifetime and therefore $T_2^*$.  Power broadening is actually a useful effect, since recent 3D qubits can have intrinsic linewidths less than $10\khz$ \cite{Paik2011} and would be difficult to find without substantial broadening.  Moreover, as the applied power is increased, the average excited state qubit population initially grows linearly, and so its response gets stronger.  For larger drive strengths, the average qubit population approaches $1/2$ and saturates.  The power dependence of the qubit transition is shown in \figref{fig:spectroscopy}(b-c).

A large spectroscopy power can also drive higher-order qubit transitions.  For example, a two-photon transition directly from the ground state to the second excited state ($|0\rangle\rightarrow|2\rangle$) is allowed, but requires substantially more drive power than it takes to saturate $|0\rangle\rightarrow|1\rangle$.  This is shown in the blue trace of \figref{fig:spectroscopy}(a).  This effect is very useful in that it is the easiest way to directly measure qubit anharmonicity.  The transition frequencies are related by $\omega_{02/2} = \left(\omega_{01} + \omega_{12}\right)/2$, and $\omega_{12} = \omega_{01} + \alpha$, where $\alpha$ is defined as the qubit anharmonicity.  Finding both the $\omega_{01}$ and $\omega_{02/2}$ transitions, therefore, also determines $\alpha$.  The fact that this is the anharmonicity can be verified by saturating the $0\rightarrow1$ transition with a CW tone and then repeating spectroscopy, directly revealing the $1\rightarrow2$ transition.  Even higher-order qubit transitions, such as a three-photon transition directly from $|0\rangle\rightarrow|3\rangle$ are also allowed.  When two qubits are coupled to the same cavity, a two-photon transition between $|00\rangle\rightarrow|11\rangle$ is also possible at the exact average of the two qubit transition frequencies, and is sometimes referred to a ``Bell-Rabi'' transition because it can be used to directly prepare an entangled Bell state \cite{Poletto2012}.  Thus, if you see a spurious transition in your spectrum it is often a good idea to first try turning down the spectroscopy power to see if it goes away, meaning that it can be ascribed to some higher-order effect.

\begin{figure*}
	\centering
	\includegraphics[scale=1]{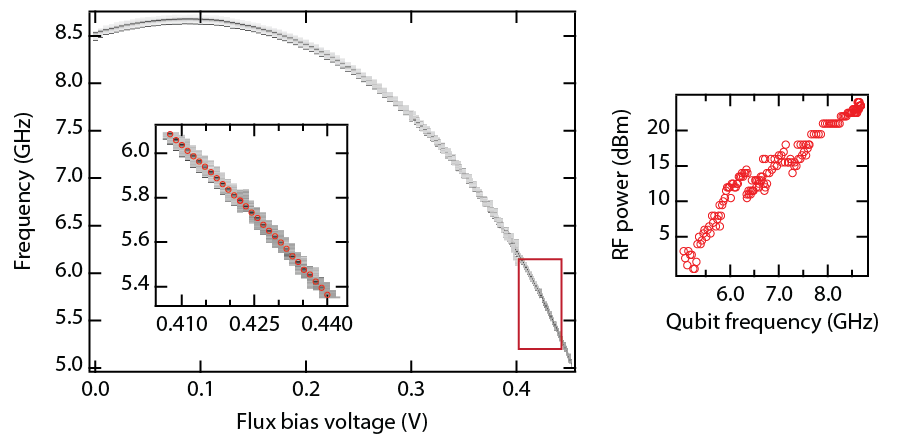}
	\mycaption{Spectroscopy as a function of applied magnetic flux}{The qubit frequency is tracked as a function of applied voltage on the qubit's flux bias line.  The qubit frequency tunes proportionate to $\sqrt{|\mathrm{cos}\left(\pi \Phi/\Phi_0\right)|}$ (\equref{eqn:transmonfluxtune}), where $\Phi$ is the flux threaded through the qubit SQUID loop.  In the inset, we see each individual trace with the qubit frequency denoted with a red circle.  The required measurement power changes as a function of qubit frequency, making it necessary to periodically re-tune as the qubit flux is stepped.
	}
	{\label{fig:spectroscopyvsflux}}
\end{figure*}

Performing spectroscopy as a function of applied magnetic field is another common experiment for tunable devices, but poses its own challenges.  At zero applied magnetic flux, the qubit should nominally be at its maximum frequency but in practice is offset due to finite environmental magnetic field.  Once a flux-tunable qubit is found at one position, then, the next step is to ``track'' it to its maximum frequency by doing pulsed spectroscopy as a function of applied magnetic field.  This tells you not only the qubit $f_{\mathrm{max}}$ but can also serve to verify that the qubit spectrum is clean of spurious avoided crossings.  Some typical data are shown in \figref{fig:spectroscopyvsflux}(a).  One complication of this process is due to the fact that both the optimal RF and spectroscopy drive power change as a function of detuning.  Periodically, the high-power readout contrast should be maximized by comparing readout contrast as a function of measurement power.  This involves making a measurement at a given power with and without the a tone applied to the qubit.  This is not actually the same thing as readout fidelity, but it is typically an excellent analog that is experimentally convenient and easily automated.  The ``optimal'' spectroscopy power is more ambiguous, but is typically tuned to keep the width of the qubit transition within some range (e.g. $3-5\mhz$).

\section{Single-qubit pulse tune-ups}
\label{sec:pulsetuneups}

Once the qubit has been located and readout has been set up, the next task is to calibrate single-qubit gates.  As we saw in \sref{sec:roomtemppulsegeneration}, much of the work to tune-up an IQ mixer can be done with room-temperature measurements.  However, there are several pulse parameters that must be measured with actual qubit experiments.  The simplest example is the attenuation and mixer voltage that corresponds to a $\pi$ rotation on a qubit.  This is set by, among other things, cavity $Q$, qubit detuning, coupling strength,  cable attenuation, and insertion loss -- all of which are only roughly known prior to measurement.  In this section we will introduce, in order of increasing complexity and precision, a series of experiments that are already in common use for calibrating pulses.  The section will culminate with the ``AllXY'' sequence, which tests the result of all combinations of two single-qubit gates and is quite sensitive to a variety of error syndromes.  

Despite the relative sophistication of this procedure, it is clear that in the future more sensitive techniques will become necessary.  We will conclude with a brief discussion of the shape those experiments might take.  Depending on the application, some or all of these experiments are not necessary.  The best pulses are only needed for experiments like state tomography that require high-precision control of the whole qubit Bloch sphere; measuring something like qubit $T_1$ requires little calibration.

\subsection{Rabi}
\label{subsec:rabi}

\begin{figure*}
	\centering
	\includegraphics[scale=1]{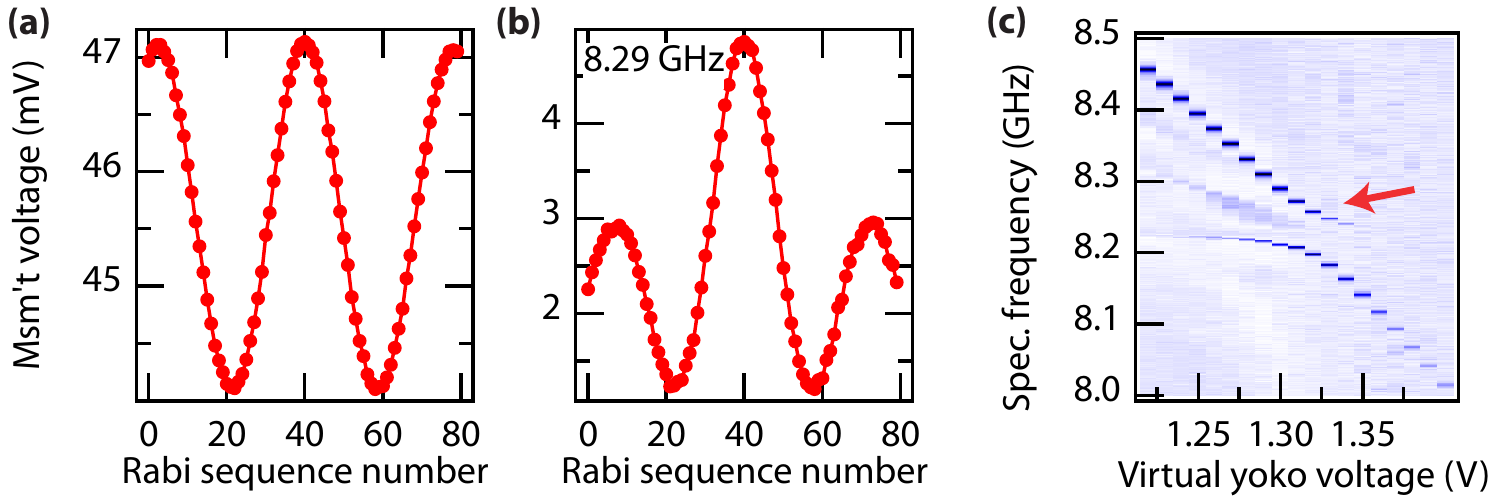}
	\mycaption{Rabi oscillations}{\capl{(a)} A qubit is rotated by some angle $\theta$, controlled by the amplitude of a resonant gaussian microwave pulse, and measured.  The $x$-axis of this plot is experiment number, where 41 has no rotation, experiments greater than 41 have positive rotation angles, and lesser have negative ones.  For example,  point 19 corresponds to a $-\pi$ rotation.  
	\capl{(b)} A ``bad'' Rabi oscillation showing that the qubit is hybridized with some other Hilbert space.  A $2\pi$ (e.g. sequence number 0 or 80) pulse does not return the qubit to the ground state, indicating leakage from the qubit computational space.  
	\capl{(c)} Spectroscopy in the vicinity of an avoided crossing.  The red arrow indicates the frequency and bias of the qubit in \capl{(b)}, showing that the system is indeed hybridized with some spurious degree of freedom.
	}
	{\label{fig:rabi}}
\end{figure*}

As discussed in \sref{subsec:rabi}, one of the simplest experiments one can do with a qubit is a Rabi oscillation.  There, the qubit is rotated by some angle about the $x$-axis of the Bloch sphere via the application of a resonant microwave tone.  As a function of that angle, the $z$ projection of the qubit will oscillate.  There are two common versions of the experiment.  First, a ``time Rabi,'' where a constant-power tone is applied to the qubit for a variable period of time.  The probability of being found in the excited state after this pulse is $\mathrm{sin}^2\left(\frac{\Omega_R}{2} \tau\right)$ where $\Omega_R$ is the Rabi rate and $\tau$ is the amount of time the pulse is applied.  The more useful realization for the present purpose is the ``power Rabi,'' where a fixed-length pulse with a variable amplitude is applied to the qubit\footnotemark.  This amplitude is swept, again yielding a sinusoidal response.  To tune-up the sequence, we adjust the Rabi rate so that a $\pi$ rotation happens at some desired voltage or DAC value.  All possible rotation angles can be interpolated from that calibration as long as the pulse amplitude modulation is linear in voltage.  An example of this is shown in \figref{fig:rabi}(a).
	
\footnotetext{A time Rabi is exactly equivalent to a power Rabi from the point of view of the physics, but it is experimentally convenient to impose a constant time for pulses and vary amplitudes.  This is true not only because it is easier to time things out when you have ``clock cycles,'' but also because the amplitude resolution of AWGs is typically much better than their time resolution.}

A Rabi experiment also serves as a general check of whether a qubit is behaving well.  The oscillation is only perfect in the limit of addressing a true two-level system, and so if this assumption is violated, the resulting curve will be distorted.  This may come about, for example, if the bandwidth of the qubit pulse is comparable to the anharmonicity of the qubit or if the qubit is hybridized with some other degree of freedom (such as a two-level system or a cavity mode).  An example of this second case is shown in \figref{fig:rabi}(b).  This qubit was near a spurious avoided crossing, shown in (c), distorting the oscillation.  A $2\pi$ pulse does not return the qubit to its ground state, indicating that there is leakage from the two-level qubit Hilbert space.  Another common distortion is an overall offset of the curve along the $x$-axis because of mixer IQ leakage (which drives an equilibrium qubit excited state population).  It is also not unusual to see that a negative and positive pulse amplitude drive different apparent rotations, which might be due to a mixer nonlinearity or imbalance.  Generally, a Rabi oscillation will not look as good as shown in \figref{fig:rabi}(a) unless both the mixer (see \sref{subsec:mixercalibration}) and pulse parameters (enumerated in \sref{subsec:allxy}) have been tuned carefully.

\subsection{Ramsey}
\label{subsec:ramsey}
	
As we described in \sref{subsec:spectroscopy}, the rough frequency of the qubit is first determined with pulsed spectroscopy.  However, because of power broadening and finite pulse bandwidth, this procedure typically enjoys a precision of only $\sim1 \mhz$.  That experiment typically will not use a mixer, and so the frequency measured there might also have a systematic offset from the one you will have in the ``real'' time-domain experiment.  This might be due to Stark shifts from IQ leakage or equilibrium cavity photon population due to a too-fast experimental repetition rate.  Thus, the precise qubit frequency must be measured using a {\it Ramsey experiment}.

A Ramsey oscillation is the gold-standard of frequency calibration for qubits as well as for precision instruments like atomic clocks \cite{Essen1955}.  The idea is to use the fact that a detuning $\Delta$ of the qubit from the RF generator frequency reference looks like a $z$-gate being continuously applied at a rate $\Delta$.  Thus, even tiny detunings can be measured by integrating this detuning as a function of time, limited only by the pure dephasing rate $T_2^*$.  The procedure begins by putting the qubit on the equator of the Bloch sphere with a $\pi/2$ pulse.  (Though errors in this rotation angle merely reduces the contrast of the Ramsey fringes.)  The qubit is then allowed to evolve for some period of time before a second $\pi/2$ rotation is applied.  For a finite detuning, the qubit will oscillate between being rotated back to the ground state and the excited state as a function of the delay time.  Because of dephasing, the amplitude of this oscillation will exponentially decay\footnotemark ~ at a rate $T_2^*$.  

\footnotetext{This is also the standard method of measuring the qubit decoherence time $T_2^*$.  Another common version of this procedure includes a $\pi$ pulse on the qubit exactly between the two $\pi/2$ pulses -- this is known as a Hahn echo.  The $\pi$ pulse flips the direction of evolution, thereby making the qubit insensitive to low-frequency noise \cite{Hahn1950}.  This will increase the effective coherence time, known as $T_2^{\mathrm{echo}}$, when the spectrum of the qubit noise is skewed toward lower frequencies, as it often is.}

In order to distinguish between the exponential decay due to dephasing and the oscillation due to detuning, it is desirable to have four or five full oscillations in the first two $1/e$ times.  Thus, you should either intentionally detune from where you think the qubit is or to sweep the phase angle of the second $\pi/2$ pulse to simulate a detuning.  (The second case might be preferable if you cannot source enough power to have a fast Rabi rate, and so have relatively slow, and therefore spectrally narrow, pulses.  This requires having full control of both quadratures of your mixer, however, which might not always be an experimental priority.)  The longer $T_2^*$ is, the smaller the detuning required to keep the number of oscillations constant.  Fitting these data, and subtracting the intentional (real or ``simulated'') detuning gives you a measurement of the unknown component of detuning.  The accuracy of this detuning is set by the quality of the fit (and so also the signal to noise of the data), but in general will be much more accurate than spectroscopy.  For example, the frequency of a 3D transmon was recently measured to better than $100 \hz$ \cite{Paik2011}.

\subsection{AllXY}
\label{subsec:allxy}

\nomdref{Adrag}{DRAG}{derivative removal by adiabatic gate}{subsec:allxy}

Rabi and Ramsey calibrations are adequate tune-ups for most basic single-qubit experiments.  However, higher quality rotations are often desirable for applications where a high quantitative value of fidelity is the goal, as for algorithms and state tomography.  For this, we need more sophisticated tune-up sequences.  While a Rabi oscillation is only second-order sensitive to the Rabi rate at its poles (though more sensitivity can be recovered by curve fitting the oscillation), concatenations of $\pi/2$ and $\pi$ pulses are, for example, first-order sensitive and errors can be intentionally amplified with additional $\pi$ pulse repetitions.  Other parameters like the orthogonality of rotations around the $x$ and $y$ axes, pulse corrections for higher excited transmon states, reflections, and the consequences of simultaneous pulses or coupling to other qubits are also not reliably measured by Rabi and Ramsey experiments.
	
In order to more accurately calibrate these parameters, we have developed a pulse sequence known as {\it AllXY}.  All combinations of one or two single-qubit rotations around the $x$- and $y$-axes by an angle of $\pi/2$ or $\pi$ are performed on a qubit that is then measured.  These are a member of the ``Clifford set''\footnotemark ~ of quantum gates \cite{Bravyi2005}.  In each case, the qubit should either end up on the north pole,  the equator, or the south pole of the Bloch sphere.  Other rotations like $\pi/4$ can be linearly interpolated from these two calibration points.  Each pulse combination is sensitive to various errors to varying degrees.  Different errors will then have a distinct fingerprint in the deviation from the ideal response, providing a quick way of diagnosing problems.  Moreover, the sequence was not designed with certain errors in mind, and is quite general.  Indeed, new types of pulse errors were discovered as a result of trying to make the sequence behave properly.

\footnotetext{The Clifford set is a finite subgroup of $U(2^N)$ (the space of operations on $N$ qubits) that is generated by the Hadamard $H$ and phase-shift $K$ gates on any single qubit and controlled-NOT gates $\Lambda(\sigma_x)$ between any two \cite{Bravyi2005}. Recall that these are given by 
$H=\frac{1}{\sqrt{2}}\left( \begin{smallmatrix} 1&1\\ 1&-1\end{smallmatrix} \right)$, 
$K=\left( \begin{smallmatrix} 1&0\\ 0&i\end{smallmatrix} \right)$, and
$\Lambda(\sigma_x)=\left( \begin{smallmatrix} I&0\\ 0&\sigma_x\end{smallmatrix} \right)$.  All of the gates used in AllXY can be made using only combinations of $H$ and $K$.}

The order of the 21 different pulse combinations was chosen to give each error syndrome an obvious signature.  The first tier of ordering, as seen below in \tref{table:allxy_sequence}, is determined by the place that the qubit should end up on the Bloch sphere.  First are those pulses which should return the qubit to the ground state, then the equator, and finally the excited state.  Pulses that end up on the north or south pole are often relatively insensitive to errors (though, as we will see, they uniquely indicate some errors and so are still important to perform), and so the most valuable information is primarily given by the pulses ending on the equator.

The ordering of the equator-bound pulses is first given by their sensitivity to over-rotations.  Each combination varies from being only second-order sensitive to the rotation angle to being several times as sensitive as a normal $\pi/2$ rotation.  That is, a $X(\pi/2)$ rotation followed by a $X(\pi)$ rotation would be three times as sensitive to over-rotations than just a $X(\pi/2)$ rotation.  Though both will end up on the equator (where a $z$ measurement is first-order sensitive), in the first case the over-rotations of both pulses add together, magnifying the error syndrome.  (The notation $N(\theta)$ denotes a rotation about the $\hat{n}$ axis by an angle $\theta$.)  Similarly, a $X(\pi/2)$ rotation followed by $Y(\pi)$ will be only as sensitive as a $X(\pi/2)$ pulse, because the second $Y$ pulse will not rotate the qubit to first order since the qubit will be in an eigenstate of that operation.  Rotations that end up on the north or south pole of the Bloch sphere are second-order sensitive because the expected value of $z$ is proportional to the cosine of the angle.  By ordering the pulses according to this sensitivity, too much or too little power yields a characteristic ``step'' pattern, shown below in \figref{fig:allxy_data_and_sim}(a), which can be distinguished with even a relatively poor signal to noise ratio.  For that reason, this approach is much more sensitive than a Rabi oscillation for tuning up power, even without curve fitting the data.  (However, if the power is significantly off, the errors are so large as to be uninterpretable -- you must first do a Rabi oscillation to get close.)  The remaining order is given by first $X$ rotations then $Y$ rotations in the first pulse position.  This is helpful because the two axes feel the opposite effect of detuning, giving a zig-zag pattern to both detuning (\figref{fig:allxy_data_and_sim}[b]) and, as we will see later, an additional pulse parameter used to compensate for phase errors due to the presence of higher excited-state levels being incorrect \cite{Motzoi2009}.  The leading-order dependence on small errors of amplitude and detuning are listed in the table (the code used to produce these equations is found in \aref{ap:mathematica}).

\begin{center}
\begin{table}\small
	\begin{tabular}{ | c | c | c | c | c | }
		\hline
	 	{\bf Ideal~$\langle z \rangle$} & {\bf First pulse} & {\bf Second pulse} & {\bf Power dependence} & {\bf Detuning dependence } \\ \hline
			
		1 	& Id 			& Id 			& $1$ 										& 1				\\ \hline
		1 	& $X(\pi)$ 		& $X(\pi)$ 		& $1-8\epsilon^2+\mathcal{O}(\epsilon^4)$ 	& $1-\frac{\pi^2\epsilon^4}{32}+\mathcal{O}(\epsilon^6)$ 			\\ \hline
		1 	& $Y(\pi)$ 		& $Y(\pi)$ 		& $1-8\epsilon^2+\mathcal{O}(\epsilon^4)$ 	& $1-\frac{\pi^2\epsilon^4}{32}+\mathcal{O}(\epsilon^6)$  			\\ \hline
		1 	& $X(\pi)$ 		& $Y(\pi)$ 		& $1-4\epsilon^2+\mathcal{O}(\epsilon^4)$ 	& $1-\epsilon^2+\mathcal{O}(\epsilon^3)$  	\\ \hline
		1 	& $Y(\pi)$ 		& $X(\pi)$ 		& $1-4\epsilon^2+\mathcal{O}(\epsilon^4)$ 	& $1-\epsilon^2-\mathcal{O}(\epsilon^3)$  	\\ \hline

		0	& $X(\pi/2)$ 	& Id 			& $-\epsilon+\mathcal{O}(\epsilon^3)$ 		& $(1-\frac{\pi}{2})\epsilon^2-\mathcal{O}(\epsilon^4)$ 		 	\\ \hline
		0	& $Y(\pi/2)$ 	& Id 			& $-\epsilon+\mathcal{O}(\epsilon^3)$	 	& $(1-\frac{\pi}{2})\epsilon^2-\mathcal{O}(\epsilon^4)$ 		 	\\ \hline
		0	& $X(\pi/2)$ 	& $Y(\pi/2)$ 	& $\epsilon^2-\mathcal{O}(\epsilon^4)$	 	& $-2\epsilon+\mathcal{O}(\epsilon^3)$	\\ \hline
		0	& $Y(\pi/2)$ 	& $X(\pi/2)$ 	& $\epsilon^2-\mathcal{O}(\epsilon^4)$ 		& $2\epsilon-\mathcal{O}(\epsilon^3)$	\\ \hline
		0	& $X(\pi/2)$ 	& $Y(\pi)$ 		& $\epsilon-\mathcal{O}(\epsilon^3)$ 		& $-\epsilon-\mathcal{O}(\epsilon^2)$	\\ \hline
		0	& $Y(\pi/2)$ 	& $X(\pi)$ 		& $\epsilon-\mathcal{O}(\epsilon^3)$ 		& $\epsilon-\mathcal{O}(\epsilon^2)$	\\ \hline
		0	& $X(\pi)$ 		& $Y(\pi/2)$ 	& $\epsilon-\mathcal{O}(\epsilon^3)$ 		& $-\epsilon-\mathcal{O}(\epsilon^2)$	\\ \hline
		0	& $Y(\pi)$ 		& $X(\pi/2)$ 	& $\epsilon-\mathcal{O}(\epsilon^3)$ 		& $\epsilon-\mathcal{O}(\epsilon^2)$	\\ \hline
		0	& $X(\pi/2)$ 	& $X(\pi)$ 		& $3\epsilon-\mathcal{O}(\epsilon^3)$ 		& $\frac{3\pi\epsilon^2}{8}+\mathcal{O}(\epsilon^4)$	\\ \hline
		0	& $X(\pi)$ 		& $X(\pi/2)$ 	& $3\epsilon-\mathcal{O}(\epsilon^3)$ 		& $\frac{3\pi\epsilon^2}{8}+\mathcal{O}(\epsilon^4)$	\\ \hline
		0	& $Y(\pi/2)$	& $Y(\pi)$ 		& $3\epsilon-\mathcal{O}(\epsilon^3)$ 		& $\frac{3\pi\epsilon^2}{8}+\mathcal{O}(\epsilon^4)$	\\ \hline
		0	& $Y(\pi)$ 		& $Y(\pi/2)$ 	& $3\epsilon-\mathcal{O}(\epsilon^3)$ 		& $\frac{3\pi\epsilon^2}{8}+\mathcal{O}(\epsilon^4)$	\\ \hline
			
		-1	& $X(\pi)$ 		& Id 			& $-1+2\epsilon^2+\mathcal{O}(\epsilon^4)$	& $-1 + \frac{\epsilon^2}{2}+\mathcal{O}(\epsilon^4)$  			\\ \hline
		-1	& $Y(\pi)$ 		& Id 			& $-1+2\epsilon^2+\mathcal{O}(\epsilon^4)$	& $-1 + \frac{\epsilon^2}{2}+\mathcal{O}(\epsilon^4)$  			\\ \hline
		-1	& $X(\pi/2)$ 	& $X(\pi/2)$ 	& $-1+2\epsilon^2+\mathcal{O}(\epsilon^4)$	& $-1+2\epsilon^2+\mathcal{O}(\epsilon^4)$ 	\\ \hline
		-1	& $Y(\pi/2)$ 	& $Y(\pi/2)$ 	& $-1+2\epsilon^2+\mathcal{O}(\epsilon^4)$	& $-1+2\epsilon^2+\mathcal{O}(\epsilon^4)$	\\ \hline
			
	\end{tabular}		
	\mycaption{AllXY pulse sequence}{The first and second pulse are listed and ordered according to where the qubit should ideally end up (on the north, equator, or south pole of the Bloch sphere).  The analytically calculated leading-order power and detuning error dependence of the qubit $z$ projection are shown.}
	{\label{table:allxy_sequence}}
\end{table}
\end{center}
	
\subsubsection{Single-qubit error syndromes}

As mentioned in the previous section, the AllXY sequence is set up so that the myriad pulse error types produce distinct error syndromes.  While many of these syndromes can be calculated with unitary matrix multiplication, others are due to either physical flaws in the experiment (e.g. reflections) or effects related to more complicated underlying physics (e.g. DRAG corrections) and are more difficult to model.  Nevertheless, we have identified numerous syndromes and, because they are linearly independent from one another (though not necessarily orthogonal), single-qubit pulse errors can be quickly diagnosed.  If you are interested in measuring only a single syndrome (e.g. amplitude deviations), there are other sequences that can be both faster and more sensitive; AllXY has the advantage of identifying a wide range of issues including problems that we were not previously classified.  One example of this is the syndrome associated with qubit-qubit coupling which was noticed in AllXY before it was understood or modeled.

\begin{figure*}
	\centering
	\includegraphics[scale=1]{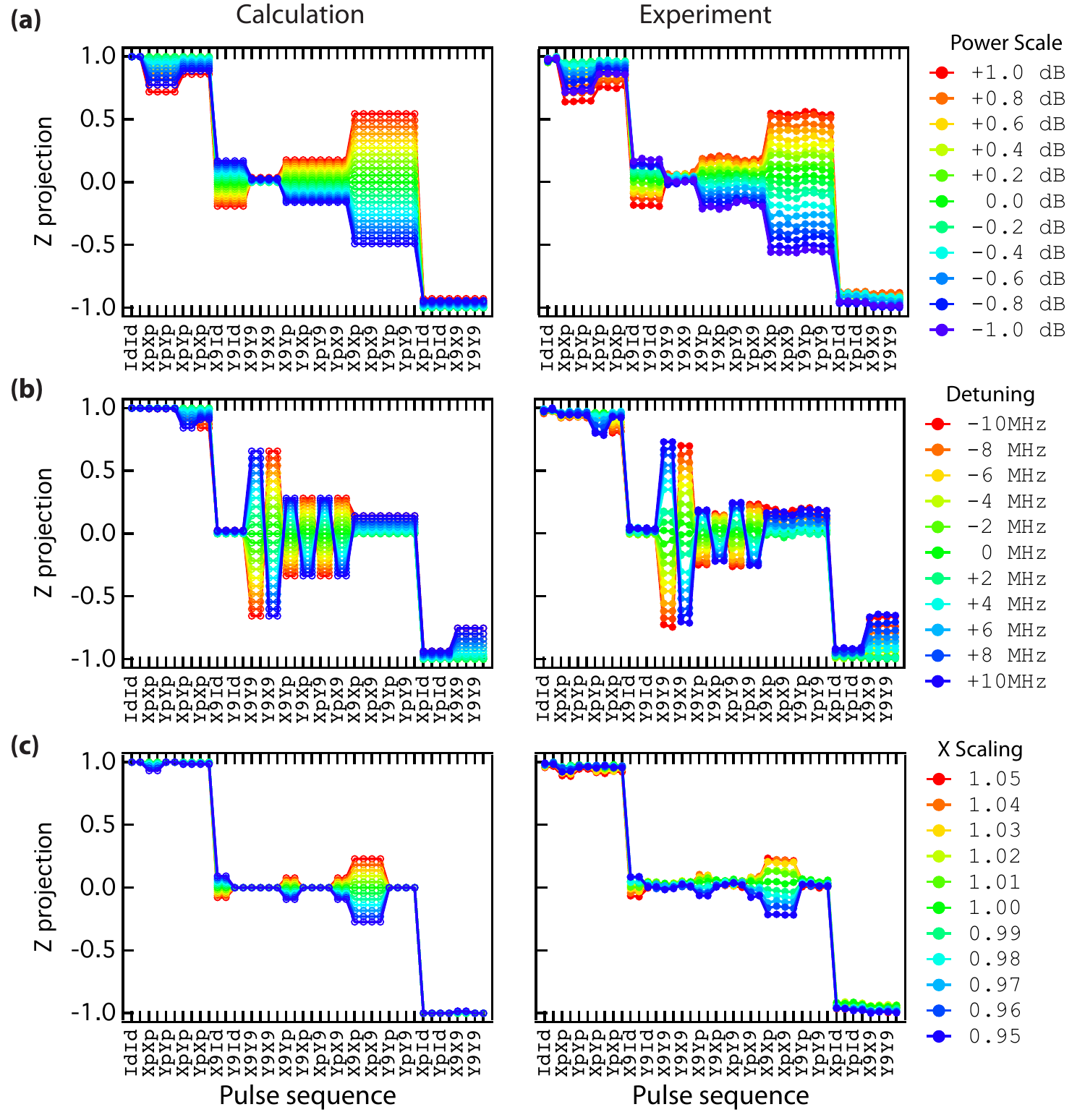}
	\mycaption{Experimental and simulated syndromes for amplitude, detuning, and skew-type errors}{On the left, a calculation using unitary matrix evolution for each type of error are shown.  The right shows the experimental reproduction of these same errors, showing excellent correspondence.  Each error signature is distinct, making it possible to detect several error syndromes simultaneously.
	}
	{\label{fig:allxy_data_and_sim}}
\end{figure*}
	
The matrix describing the rotation of a qubit by an angle $\theta$ about the axis $\hat{\sigma}$ is given by $\hat{U}(\theta,\hat{\sigma})=e^{-i\frac{\theta}{2}\hat{\sigma}}$.  An error in pulse power by $x$ dB (note the logarithmic power units) is correctly captured by this equation by scaling $\theta$ by a factor of $10^{(x/20)}$ (see \aref{ap:mathematica}).  As shown in the left column of \figref{fig:allxy_data_and_sim}, we can analytically calculate the expected $z$-projection following such errors.  We see experimental data on the right, showing excellent correspondence.  A detuning error is understood as an additional $z$ field to $\hat{\sigma}$, scaled by the ratio of the detuning to the Rabi rate.  A good approximation to this syndrome is given by $\hat{\sigma}\rightarrow\hat{\sigma}+\hat{\sigma}^\prime$, where $\hat{\sigma}^\prime=-2\pi \Delta \delta t \sigma_z$, $\Delta$ is the detuning, $\delta t$ is the single-qubit gate duration, and $\sigma_z$ is the $z$ Pauli matrix.  Other more esoteric syndromes, such as a scale factor between the $x$ and $y$ quadratures (that might happen if the IQ mixer is imbalanced), can also be modeled by scaling all $\sigma_x$ rotations by the appropriate factor.

\begin{figure*}
	\centering
	\includegraphics[scale=1]{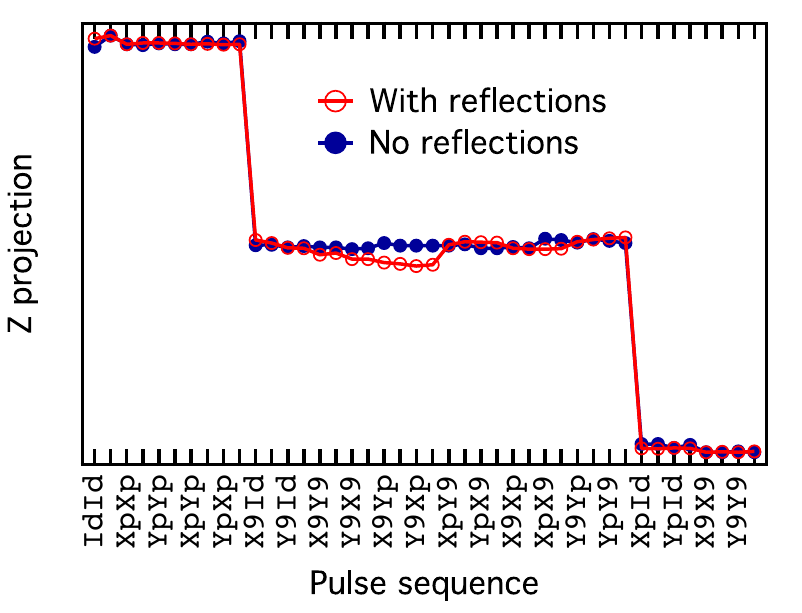}
	\mycaption{AllXY reflection error syndrome}{The first six equator pulses are insensitive to the action of the second pulse while the last six are first-order sensitive.  If reflections are an issue, the second pulse will be modified by the reflection of the first arriving at the qubit at the same time, and so will modify its behavior.  Here we see AllXY data with and without pulse reflections.  The drive power was tuned to minimize the overall error and so we actually see more of a problem with the ``insensitive'' pulses because we are using too little power.  Nevertheless, there is a clear difference between the first six and last six pulses due to this reflection.
	}
	{\label{fig:allxy_reflections}}
\end{figure*}
	
There are other syndromes that are not easily modeled, but nevertheless offer unambiguous syndromes.  The first example of this is that of reflections.  As described in \sref{subsec:assembly}, reflections are caused by impedance mismatches in the pulse conditioning chain.  Reflections of the first pulse which arrive at the sample after a delay long enough to collide with the second pulse will change the resulting qubit evolution.  That is, the effect of the first pulse will depend on what pulse happens after it, and second pulse on what happened before it.  In order to understand the syndrome associated with this problem, we observe that the second pulse of the first six sequences that end on the equator should have no effect.  For example, the fifth sequence is first $X(\pi/2)$ followed by $Y(\pi)$.  If the first pulse is successful, the qubit will end up parallel to the $y$-axis, and therefore rotations about that axis will do nothing.  For the last six sequences, however, the second pulse does rotate the qubit (e.g. $X(\pi/2)$ then $X(\pi)$).  We therefore expect that when reflections are an issue, the first six pulses on the equator should be approximately correct, while the last six should be distorted.  Indeed, this is what we experimentally observe, shown in \figref{fig:allxy_reflections}.

\begin{figure*}
	\centering
	\includegraphics[scale=1]{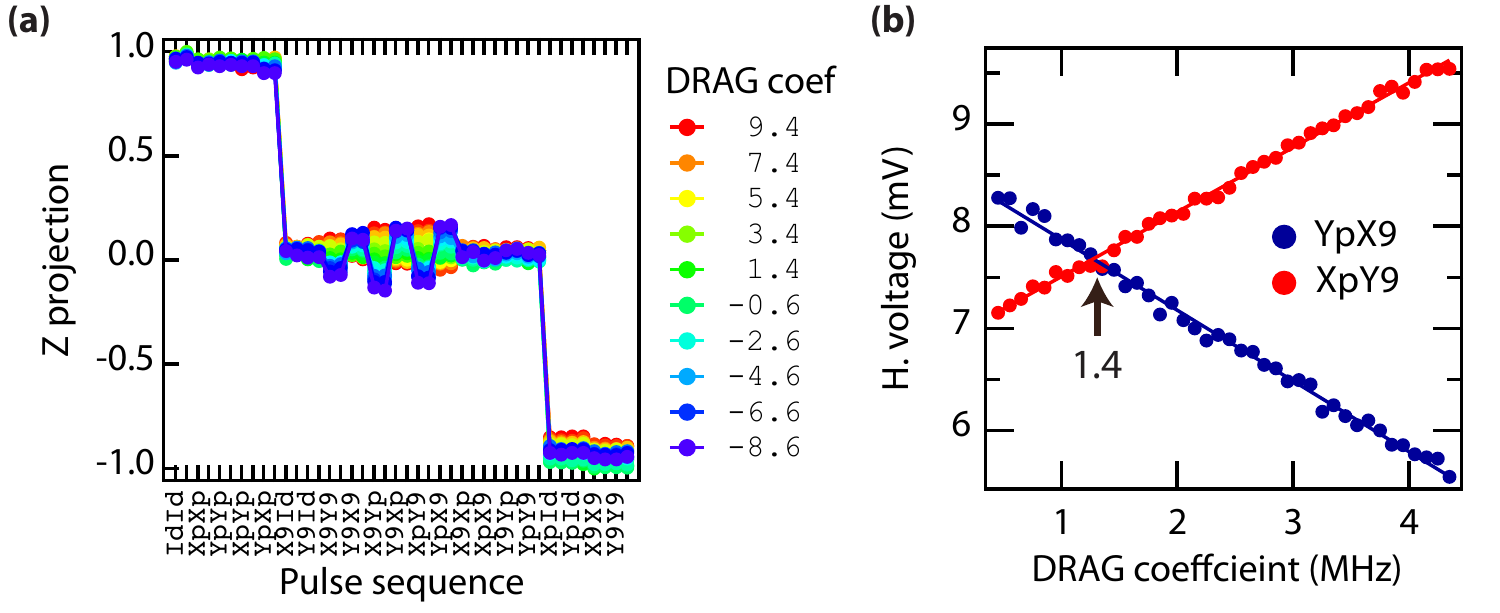}
	\mycaption{DRAG coefficient error syndrome}{The DRAG pulses used to cancel phase errors due to the presence of higher transmon levels are defined by one additional experimental parameter.  Here, that parameter is given by the number of megahertz that the qubit pulse should instantaneously be detuned by at a $\pi$ pulse amplitude.  (The pulse detuning will be scaled by the instantaneous pulse amplitude.)  The error syndromes associated with this parameter are challenging to calculate analytically, but are easily measured, as shown in \capl{(a)}.  An easy way to tune-up this parameter is to choose two of the AllXY pulse sequences whose errors come in with opposite sign, and measure them both with varying DRAG coefficient.  This will produce two lines that cross at the optimal DRAG coefficient value, as shown in \capl{(b)}.
	}
	{\label{fig:allxy_drag}}
\end{figure*}

Another syndrome not easily calculated but nevertheless crucial to tuning up pulses is associated with the technique of {\it derivative removal by adiabatic gate} or DRAG \cite{Motzoi2009}.  As described in section 4.2.3 of Jerry Chow's thesis \cite{ChowThesis}, this technique corrects for gate phase errors caused by the presence of higher excited states of the transmon qubit \cite{Chow2010}.  The lowest-order correction involves either continuously detuning the pulse as a function of its instantaneous amplitude or adding a copy of the derivative of the primary pulse to its orthogonal quadrature.  In both cases, there is a scale factor for this correction.  Though Motzoi, {\it et al.} calculate the value of this parameter as an analytical function of matrix elements and qubit anharmonicity in their original paper \cite{Motzoi2009}, the optimal value (and even its sign) differs dramatically due to filtering effect of the cavity.  We tune it up as a free parameter based on the observation of its syndrome in AllXY.  As shown in \figref{fig:allxy_drag}(a), the syndrome is reminiscent the one associated with detuning (consistent with the fact that this is principally an error in phase), though it is still distinct and distinguishable.  In order to efficiently tune-up the DRAG parameter, we take two of the AllXY pulses which exhibit the opposite sign of error -- $YpX9$ and $XpY9$ -- and perform each as a function of the parameter value.  This yields two lines that cross at the point where the parameter is optimal, as shown in \figref{fig:allxy_drag}(b).  The technique of choosing two AllXY pulses and sweeping a control parameter is very general.  It also lends itself to curve-fitting programs, which are better at determining the intersection of two lines than the naked eye.
	
\subsubsection{Two-qubit error syndromes}

\begin{figure}
	\centering
	\includegraphics[scale=1]{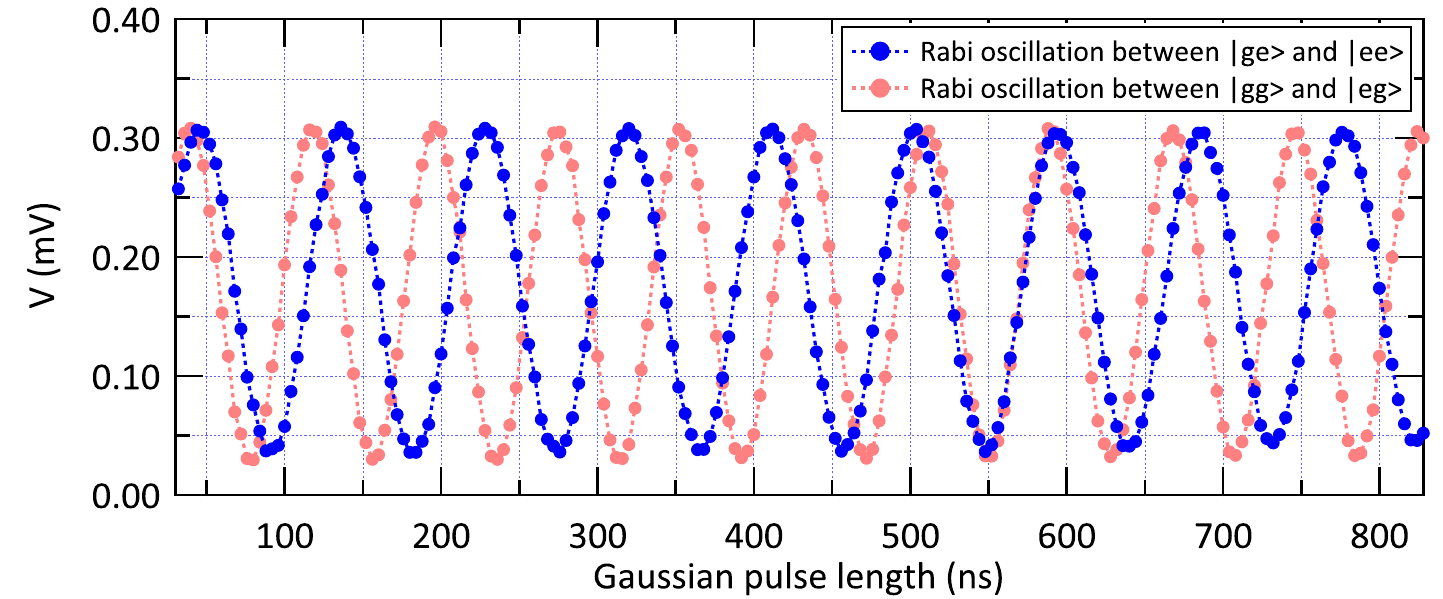}
	\mycaption{State-dependent Rabi rate}{A time-Rabi is performed on a qubit with and without another qubit in its excited state.  The Rabi rate depends strongly on the state of that adjacent qubit because of the hybridization of the qubit eigenstates.  This coherent error is equivalent to an unintentional two-qubit entangling gate and causes a new single-qubit error endemic to multi-qubit devices.  The ratio of the Rabi rates for the two cases is 1.17.
	}
	{\label{fig:statedependentrabi}}
\end{figure}

\begin{figure}
	\centering
	\includegraphics[scale=1]{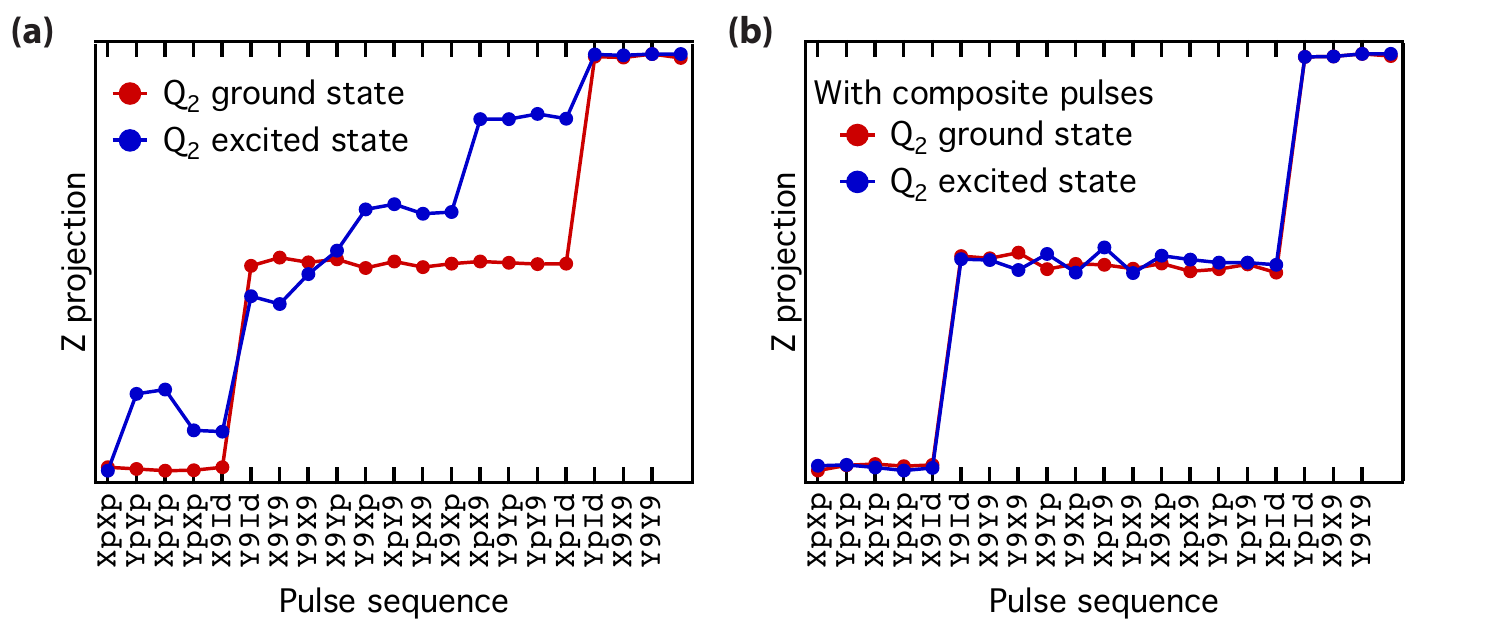}
	\mycaption{Two-qubit joint AllXY}{\capl{(a)} The AllXY sequence is run on a qubit with and without an adjacent qubit in its excited state.  The state-dependent Rabi rate and frequency shift associated with that excitation cause huge errors.  Note that ideally both of these traces should look identical and tuned up; deviations indicate that single-qubit pulses are actually multi-qubit ones.  Errors can be mitigated to a certain extent by splitting the difference between the two cases, but ultimately a new tactic must be taken.  
	\capl{(b)} Joint AllXY using BB1 composite pulses.  These pulses are designed to be insensitive to amplitude errors, and so significantly mitigate the multi-qubit syndrome \cite{Paik2013, Wimperis1994}.
	In both \capl{(a)} and \capl{(b)}, repeated pulse sequences are averaged together, giving a total of 21 measurements rather than the 42 shown in previous figures.
	}
	{\label{fig:jointallxy}}
\end{figure}

There are also error syndromes associated with coupling to additional qubits \cite{Gambetta2012}.  The hybridization of these qubits produces an effective $ZZ$ interaction and causes the matrix element coupling one qubit to the microwave drive to depend the state of the other qubit.  In one particularly extreme case \cite{Paik2013}, the Rabi rate of a qubit was observed to change as much as 17\% if the other qubit in the cavity flipped its state, as shown in \figref{fig:statedependentrabi}.   This causes big problems when trying to perform accurate pulses on both qubits, because by definition a single-qubit rotation cannot depend on the state of other qubits.  AllXY run for the case of the adjacent qubit being in the ground and excited state are shown in \figref{fig:jointallxy}(a).  When the qubit is in its ground state, AllXY looks essentially perfect, but by exciting that adjacent qubit the sequence is ruined.   Fortunately, by using the BB1 composite pulse sequence, which is designed to reduce sensitivity to amplitude errors, this issue can be largely mitigated as shown in \figref{fig:jointallxy}(b).  For more information on this composite pulsing, see Ref.~\citenum{Wimperis1994}.

\subsection{Future tune-up sequences}
\label{subsec:futuretuneups}

AllXY has the advantage of being simultaneously more sensitive to errors than conventional experiments like a Rabi oscillation, having unique signatures for various error types, and detecting many different error types.  However, if there is a particular error that you are interested in studying, it is possible to design experiments that are more sensitive.  For example, amplitude errors could be directly detected by applying a $\pi/2$ pulse followed by $N$ $\pi$ pulses.  Each additional $\pi$ pulse amplifies the error syndrome, and since the qubit is initially on the equator, it is first-order sensitive to these errors.  It is easy to imagine that equivalent pulse sequences can be made arbitrarily sensitive to particular errors, limited only by qubit decoherence.

Verifying that pulses are correct can also be a challenge.  It is possible to construct robust tests for specific kinds of errors, but that requires knowing what to look for.  For the purposes of quantum error correction, or simply to report gate fidelity to the community, we are interested in knowing the absolute difference between a gate and its ideal unitary.  {\it Randomized benchmarking} is one attractive protocol recently developed for this purpose \cite{Emerson2007, Knill2008, Chow2009, Gambetta2012, Magesan2012a, Magesan2012b, Gaebler2012, Corcoles2013}.  There, a large number of randomly chosen (but known) pulses are applied to a qubit prior to measurement.  The pulse sequence is arranged so that the qubit should end up in the ground state after the sequence.  It can be shown that {\it any} type of gate error will be mapped to a depolarization channel \cite{Gambetta2012}.  Then, any remaining excited state population, averaged over many sets of chosen gates, is a direct measure of gate infidelity.  This process tells you nothing about {\it what} is wrong with a gate, but does indicate its overall fidelity.

\section{Summary}

In this chapter, we have listed the experiments and procedures necessary to bring a new cQED qubit experiment from initial cool down to being well-calibrated.  We showed how to measure the cavity frequency with a transmission measurement, and how measuring transmission as a function of drive power can quickly detect the presence and approximate detuning of a qubit.  With the measurement set up, the qubit can be found by using spectroscopy.  Of the several variants, by far the most common and conceptually simple version is pulsed spectroscopy.  With the qubit found, we can then go about tuning up single-qubit rotations.  Gross calibrations are done with Rabi and Ramsey oscillations to get the pulse power and qubit frequency correct, respectively.  We have developed a more sophisticated sequence known as AllXY which is both more sensitive to detuning and pulse power and detects numerous other error syndromes that the earlier sequences cannot.  However, if you are interested in tuning up a particular pulse parameter, there are other protocols that can be made much more sensitive than AllXY by concatenating multiple rotations together.  Quantifying the fidelity of pulses is challenging, but one promising approach is to use randomized benchmarking.  In the next chapter, we will discuss in greater detail the various measurement schemes that we previously alluded to.

\setcounter{chapter}{5}
\chapter{Qubit Measurement}
\thumb{Qubit Measurement}
\lofchap{Qubit Measurement}
\label{ch:qubitmeasurement}


\lettrine{T}{he} measurement of qubits has been a major line of research for the last several years, both for the field in general and this thesis specifically.  Broadly speaking, quantum measurement involves entangling a qubit with some other degree of freedom which can be measured to infer the qubit state.  Measuring the {\it pointer state} of our ancillary degree of freedom projects the qubit along some axis, which is usually defined to be the qubit $z$-axis.  In cQED, the extra degree of freedom is the cavity and the pointer state is its displacement.  The fundamental mechanism for measurement in cQED is shown in the dispersive Jaynes-Cummings Hamiltonian (\sref{subsec:dispersivelimit}).  There, the cavity frequency depends on the state of the qubit, so driving the cavity entangles the cavity and qubit states since the cavity displacement depends on the detuning of the drive.  This physics is straightforward in principle, but we will see that optimizing readout, when faced with experimental realities, can be subtle and complicated.

There are three main sections to this chapter.  The first introduces the experimental use of the low-power dispersive measurement (see \sref{subsec:dispersivelimit}).  There, we apply a weak tone to the cavity and measure the resulting transmission.  Since the cavity displacement depends on the qubit due to the dispersive coupling, the amount of transmitted light similarly encodes the qubit state.  When used with a conventional microwave amplifier, however, and especially in experiments that do not explicitly optimize system parameters for measurement, the resulting fidelity can be low.  The situation has improved recently with the development of practical quantum-limited microwave amplifiers which dramatically reduce added noise at the cost of increased complexity.  This section will introduce experimental quantum measurement in cQED, and will provide much of the language and intuition used later.

The second section seeks to optimize the fidelity of the dispersive readout.  As we will see, under normal circumstances the qubit and cavity lifetimes are interrelated by the {\it Purcell effect}.  This inhibits our ability to increase measurement fidelity by independently optimizing those lifetimes.  In response, we introduce the {\it Purcell filter}, which breaks (or at least, engineers improvements in) the relationship between the qubit and cavity lifetime.  It can significantly increase qubit lifetime and therefore dispersive measurement fidelity, and has the advantage of being compatible with quantum-limited amplifiers.  As an ancillary benefit, we will discover that the Purcell filter can also enable efficient reset of the qubit by swapping long-lived qubit excitations into the short-lived cavity mode.

The third and final section concerns an unexpected behavior which was accidentally discovered when measuring the first Purcell filter device.  As a function of drive strength, cavity transmission initially looks linear and lorentzian at low drive strengths but rapidly develops anharmonic behavior as the inherited qubit nonlinearity comes into play with higher excitations.  It was long thought that these nonlinearities, which eliminate readout contrast, set the maximum drive strength that could be used to measure.  However, we find that if you drive even harder, the cavity eventually restores to a linear lorentzian response at a different frequency.  The power required to reach this {\it bright state} depends on the initial qubit state, providing what turns out to be a very high fidelity qubit measurement that is not affected by following amplifier noise.  While this mechanism works well for a wide range in qubit and cavity parameters, it scrambles the qubit and cavity state (e.g. is not quantum non-demolition) and is not applicable to some experiments.

\section{Dispersive readout}
\label{sec:dispersivereadout}

\nomdref{Ahemt}{HEMT}{high electron mobility transistor amplifier}{sec:dispersivereadout}
\nomdref{Asnr}{SNR}{signal-to-noise ratio}{sec:dispersivereadout}
\nomdref{Ajpc}{JPC}{Josephson parametric converter}{sec:dispersivereadout}
\nomdref{Ajba}{JBA}{Josephson bifurcation amplifier}{sec:dispersivereadout}

We saw in \sref{subsec:dispersivelimit} that, in the dispersive limit, the Jaynes-Cummings Hamiltonian simplifies to
\begin{equation}\label{eq:dispersivejc2}
	\hat{H}=\hbar \omega_r \hat{a}^\dagger \hat{a} + \hbar \frac{\omega_q}{2} \sigma_z + \hbar \chi \hat{a}^\dagger \hat{a} \sigma_z 
\end{equation}
where $\omega_r$ is the cavity frequency, $\omega_q$ the qubit frequency, and $\chi$ the dispersive shift.  	Recall that the first term is responsible for the cavity oscillator (with the zero-point energy included, but suppressed from now on), the second responsible for the two-level qubit\footnotemark, and the third responsible for the state-dependent frequency shift.  In the case of a true two-level qubit, $\chi=g^2/\Delta$, but for transmon qubits, this parameter becomes $\chi = \frac{g^2}{\Delta} \frac{\alpha}{\Delta - \alpha}$ \cite{Koch2007}.  We have already taken the rotating wave approximation, ignoring terms like $\hat{a}^\dagger \sigma^+$ and $\hat{a} \sigma^-$ which do not conserve energy, and assumed that the detuning $\Delta = |\omega_r - \omega_q| \gg g$, where $g$ is the vacuum-Rabi splitting.  

\footnotetext{If we include higher transmon levels, we find that each has a different dispersive shift to the cavity.  It is therefore possible, if you can measure both the phase and amplitude of the transmitted signal, to discriminate between several qubit states in a single shot \cite{LiuMM2013}.}

For the purposes of qubit readout, we group the dispersive term with the cavity number operator, giving us $\hat{H}_c = \hbar \hat{a}^\dagger \hat{a} \left( \omega_r + \chi \sigma_z \right)$ (\sref{subsec:dispersivelimit}).  This shows that cavity frequency depends on the qubit state; the frequency when the qubit is in the ground state is $2\chi$ higher than when it is in the excited state.  In the case of transmon qubits, $\omega_r$ is a convenient frequency with which to write the Hamiltonian and has no physical significance.  Both the ground and excited state peaks are shifted up in frequency relative to the bare cavity frequency, but the ground state shift is larger, by what we call $2\chi$.  (In the Cooper-pair box limit, the two $\chi$ shifts are actually equal and opposite, but this symmetry is broken by the higher transmon levels.)

\begin{figure}
	\centering
	\includegraphics[scale=1]{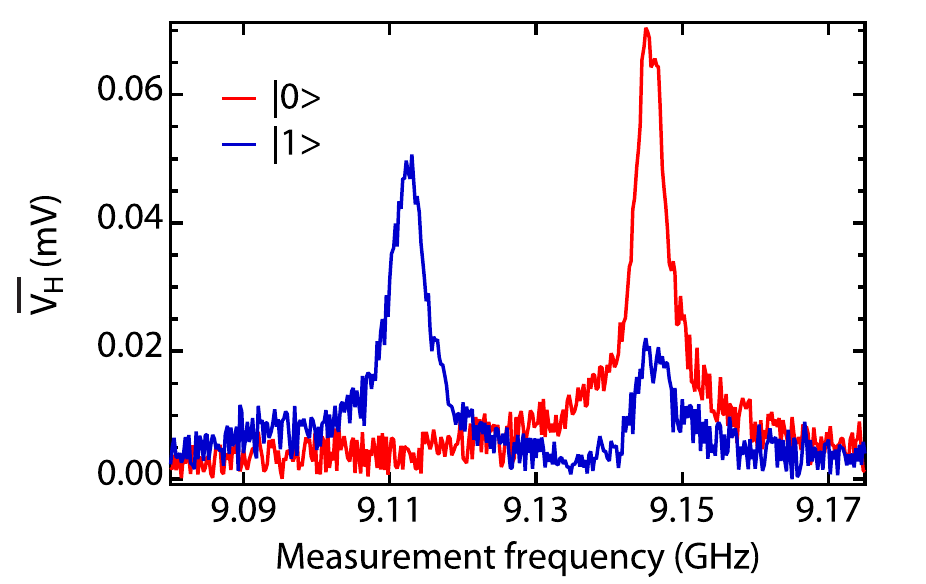}
	\mycaption{Dispersive cavity shift}{We measure transmission as a function of frequency when the qubit is in the ground and excited state by preparing the qubit then pulsing on the measurement tone for a time ($400\ns$) that is small compared to the qubit $T_1$.  We see two distinct cavity frequencies, different by the dispersive shift $2\chi$.  During the measurement, there is a finite chance for the qubit to decay, giving us an additional peak for the blue curve at the ground state cavity frequency of $\sim9.15\ghz$.  By measuring the transmission at e.g. $9.15\ghz$, we can infer the qubit state by mapping a high transmission to ground state and a low transmission to excited state.  These data were taken using the device described in \sref{sec:fourqubitdevice}.
		}
	{\label{fig:dispersivecavity}}
\end{figure}

We can demonstrate the dispersive cavity shift with a simple pulsed transmission experiment.  We first prepare the qubit in either its ground or excited state and then measure the transmission of a short RF tone, as shown in \figref{fig:dispersivecavity}.  There, the dispersive shift $2\chi$ is much larger than the cavity linewidth $\kappa$, so the two cavity frequencies are well-resolved.  The qubit has some chance of decaying during the measurement, so we see a small peak in the blue curve corresponding to the qubit being in the ground state.  This measurement was done with a drive power small enough that the mean cavity occupation is approximately one photon so as to avoid the nonlinear cavity effects discussed in \sref{subsec:cavnonlinearity}.  

In order to measure the qubit, we apply a tone at the ground-state cavity frequency and measure how much is transmitted.  If we get a relatively large transmission, we can infer that the qubit must be in the ground state, while a relatively small signal indicates that the qubit is excited.  That is, we take advantage of a {\it state-dependent cavity transmission}.  If we send a {\it single} photon of the relevant frequency through the cavity and see it come out the other end, we have in principle measured the qubit as long as the two cavity frequencies are well resolved.  (Alternatively, we could measure at the exact middle of these two frequencies -- at $\omega_r$ -- and see a large difference in the {\it phase} of the outgoing signal.  This is advantageous in some cases.)

Of course, in real life things are not so simple \cite{Gambetta2007}.  While it is true that a single photon can {\it project} the qubit into a $z$-eigenstate (e.g. collapse a superposition), detecting that photon is very difficult.  This is essentially a statement about the energy scale involved: microwave photons carry very little energy and cannot ionize atoms, so we do not (yet) have the luxury of a single-photon detector in this frequency range.  Instead, we use heterodyne detection to down-convert and digitize the signal directly.  This process, where the signal coming out of the fridge is mixed with a local oscillator $\sim25\mhz$ detuned and sent to a digitizer card, necessarily includes several stages of amplification that add noise.  Even when using the best commercially-available cryogenic high electron mobility transistor (HEMT) amplifiers, this noise power is typically twenty times larger than the signal power of a single photon.

One thing we could try is to send several photons through the cavity.  As long as the qubit does not decay, we are free to do this because the third term in \eref{eq:dispersivejc2} commutes with the the other terms.  (Formally, this is because the measurement is QND to the qubit state; when we measure the qubit, it remains in the same state in which we had measured it.)  How much signal could we gather with this process?  We must remember that we can only send photons through the cavity as long as the qubit has not decayed.  Since we cannot know when that has happened, we aim to integrate for a time $\Delta t$ which is on the order of the $T_1$ of the qubit.  (Exactly how long you should measure turns out to be a bit complicated, and is touched on in a footnote below.)  Next, the amount of information we extract per photon is not necessarily one full bit.  If the dispersive shift is smaller than the cavity linewidth, for example, we will gain an amount of information expressed as $\sin^2(\theta)$, where $\theta = \mathrm{tan}^{-1}\left(\frac{2 \chi}{\kappa}\right)$, which is bounded between 0 and 1 bit.  (We get a full bit when the state-dependent shift $\chi$ is much larger than the cavity bandwidth $\kappa$.)  Finally, each of these photons will give us $\hbar \omega_r$ of energy and the rate of photon collection is set by the bandwidth of the cavity $\kappa$.  If we have $\langle n \rangle$ equilibrium cavity photons, our information gain (in units of energy) during a measurement will be given by
\begin{equation}
		E_{\mathrm{signal}}= \langle n \rangle \hbar \omega_r \mathrm{sin}^2(\theta) T_1 \kappa.
\end{equation}
If we specify the noise of our measurement chain with an effective noise temperature $T_N$ our signal to noise ratio is given by $\mathrm{SNR} = E_{\mathrm{signal}} / k_B T_N$.

For most devices, this number turns out to be rather small.  Assuming fairly favorable numbers of a mean cavity occupation $\langle n \rangle = 10$ photons\footnotemark, an $8\ghz$ measurement cavity, an integration time of $3\us$, $\kappa=5\mhz$, and a $10~\mathrm{K}$ noise temperature (typical when using only a HEMT), we have $\mathrm{SNR} \sim 1$.  (In Cooper-pair box devices or more recent 3D transmon devices with longer lifetimes, this number can be more like $5-10$.)  As a result, for a conventional amplification chain, the fraction of the time you can correctly identify the qubit state with a dispersive measurement is bounded to about 70\%.  For devices with parameters less favorable for dispersive measurement (e.g. a longer cavity lifetime), fidelity values have been observed as low as $5-10\%$, though this is atypical.  Robust quantum error correction requires fidelities in excess of 99\%, so this performance must be improved.

\footnotetext{The reason we cannot just increase the number of intra-cavity photons will be discussed in \sref{sec:jcreadout}.  Ten photons is approximately where cavity nonlinearity and a reduction in qubit $T_1$ start to dominate.}

Fortunately, significant progress has recently been made in the development of specialized quantum-limited amplifiers \cite{Spietz2009, Castellanos2009, Bergeal2009}.  The idea is to put an extremely low-noise amplifier like a Josephson parametric converter (JPC) or Josephson bifurcation amplifier (JBA) between the cavity and the cryogenic commercial amplifier, so as to boost the signal and drown out the added HEMT noise.  This significantly reduces the effective amplification chain noise temperature, increasing SNR and measurement fidelity.  For example, if we reduce the noise temperature to three times the quantum limit, or $100~\mathrm{mK}$, and re-calculate the signal to noise ratio with the parameters as above, we find $\mathrm{SNR} \sim 50$ (or, with state-of-the-art lifetimes, several hundred).  As a result, dispersive measurements can now resolve individual quantum jumps of the qubit state in real time \cite{Vijay2011} and are approaching $99.5\%$ fidelity \cite{Hatridge2013}.  This comes at a cost of complexity.  The amplifiers have to be biased with a very stable RF source, require extra circulators, and have narrow amplification bandwidth that must be matched with the cavity.  Nevertheless, it is clear that as the engineering challenges are overcome, the use of these devices will become more routine (a process which is coinciding with the writing of this thesis), enabling high fidelity QND measurement and real-time feedback.  For the purposes of this chapter, however, we focus on improvements we can make without using these special devices. 

Another method to circumvent amplifier noise is to intentionally make the cavity extremely nonlinear with the addition of a Josephson junction.  The cavity will then exhibit a ``latching'' Kerr bistability between two long-lived and classically distinguishable states.  The qubit state can be mapped onto the choice of cavity state and read out over a relatively long period of time, yielding higher fidelity \cite{Siddiqi2004, Siddiqi2006, Boulant2007, Mallet2009}.  As with low-noise amplifiers, this approach also increases complexity of both fabrication and the physics of the device since the cavities require an additional junction.  Though we will not further discuss this approach, it is as an interesting context for \sref{sec:jcreadout}, in which we produce a qualitatively similar behavior without the need of an extra cavity junction.  

\section{The Purcell filter}
\label{sec:purcellfilter}

In considering how to optimize the fidelity of dispersive readout, we quickly reach an impasse.  On the one hand, we would like to have the fastest possible cavity (e.g. a large $\kappa$) to maximize the rate of measurement photon transmission; on the other hand, decreasing cavity lifetime can also reduce qubit lifetime due to the {\it Purcell effect} \cite{Purcell1946, Houck2008}.  For a fixed signal to noise ratio, the optimal measurement time is a fixed fraction of $T_1$.  Higher SNRs require shorter and shorter integration times to optimize the resulting measurement fidelity \cite{Gambetta2007}.  The reason for this is simple: the readout is QND but the qubit is free to decay during the measurement.  When this happens, all measurement integration following the decay event will actually {\it lower} the measurement faithfulness, because we are interested in what the qubit state was at the beginning of the measurement and not what it is currently\footnotemark. Thus, lowering qubit $T_1$ will lower measurement fidelity.

\footnotetext{
	The optimal length of integration is complicated by the possibility of both linear and nonlinear filtering.  One example of linear filtering is to multiply the measurement tone by an exponential decay.  The data at the beginning of the measurement will then count for more than that of later measurements, ameliorating the effect of finite qubit lifetime.  Indeed, such a filter can improve with low SNRs, though its effectiveness drops rapidly for larger SNRs \cite{Gambetta2007}.  
	
	Processing data in real time with a nonlinear filter can do even better if you have knowledge of the distribution of measurement outcomes as a function of integration time \cite{Hatridge2013}.  For example, if, shortly after the measurement is started, the integrated homodyne voltage is already located far down one of the tails of the ``S-curve'' distribution, then you already know with high confidence what the qubit state was and the measurement is complete.

	In the case of the simple boxcar integration of the measurement tone that is used in most experiments, the x\% of $T_1$ integration time for 1-x fidelity rule of thumb is a good approximation.  That is, in order to get 99\% measurement fidelity, you need high enough SNR to accurately read out the qubit within no more than 1\% of $T_1$. 
}

\subsection{The Purcell effect}
\label{subsec:purcelleffect}

The Purcell effect modifies the lifetime of any quantized system coupled to a resonant circuit or cavity \cite{Purcell1946}.  Depending on the detuning of the system transition frequency from the cavity resonance frequency and the lifetime of the cavity, the rate of decay of the quantum system can be strongly enhanced \cite{Purcell1946,Goy1983} or suppressed \cite{Kleppner1981,Hulet1985,Jhe1987} relative to the decay rate to the electromagnetic continuum.  In cQED, qubits are often sufficiently detuned to have suppressed decay rates compared with the continuum, but $T_{1}$ can still be limited by radiative decay through the cavity.  That is, the qubit decaying through the cavity is always a possible mode of relaxation, but it may or may not be dominant depending on the coupling strength and cavity lifetime.  When a significant fraction of qubit relaxation is decay through the cavity, we say that the qubit is ``Purcell-limited.''

\nomdref{Gcgammakappa}{$\gamma_\kappa$}{Purcell decay rate}{subsec:purcelleffect}

We can make a rough approximation of the Purcell lifetime with the single-mode Jaynes-Cummings Hamiltonian (\eref{eq:jch}).  In this case, $\hat{H} = \hbar \omega_r \left(\hat{a}^\dagger \hat{a} + \frac{1}{2} \right) + \frac{\hbar \omega_q}{2} \hat{\sigma}_z + \hbar g \left(\hat{a}^\dagger \sigma_- + \hat{a} \sigma_+\right)$, with $g$ the coupling strength, $\omega_r$ the cavity frequency, and $\omega_q$ the qubit frequency.  Using first-order non-degenerate perturbation theory (expanding in powers of $\frac{g}{\Delta}$), we find that the dressed qubit-like eigenstate of the system is given by $|\mathrm{dressed}\rangle \approx |e,0\rangle + \frac{g}{\Delta}|g,1\rangle$ (see Appendix B of David Schuster's thesis \cite{SchusterThesis}). There, $|e,0\rangle$ is the undressed eigenstate of an excited qubit and zero photons, $|g,1\rangle$ is the undressed qubit in the ground state and one photon, and $\Delta$ the qubit-cavity detuning (see \sref{subsec:qubitcavitycoupling}).  We can then say that the Purcell decay rate is given by the decay of the photon-like component of this state at the cavity decay rate $\kappa$, $\gamma_{\kappa} = \left(\frac{g}{\Delta}\right)^2 \kappa$.

\begin{figure}
	\centering
	\includegraphics{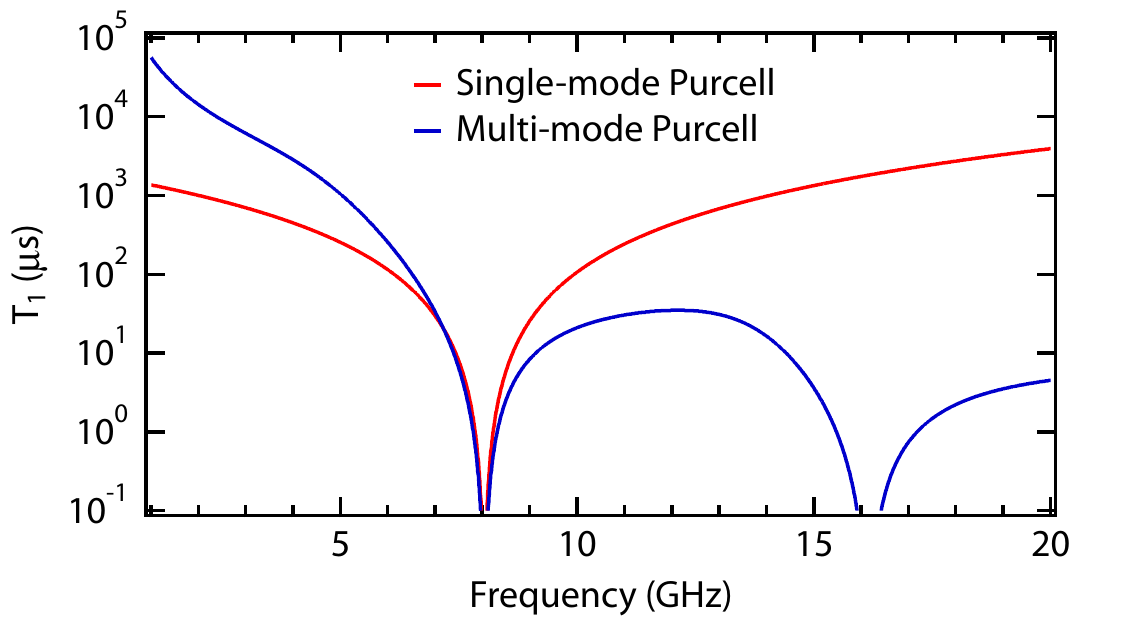}
	\mycaption{Multi-mode vs single-mode Purcell decay rates}{We calculate the Purcell lifetime for a single mode with $T_1^{-1} = \gamma_\kappa = (g/\Delta)^2 \kappa$ for a device with $g = 300\mhz$.  We consider a symmetric device where $C_{\mathrm{in}} = C_{\mathrm{out}} = 1\ff$.  We use the relation $q_i = \omega C_i Z_0$, $\kappa_i = \left(\frac{2}{\pi}\right) q_i^2 \omega$ and $\kappa = \kappa_{\mathrm{in}} + \kappa_{\mathrm{out}}$.  $Q_{\mathrm{tot}}= 124,000$.  We compare this to a multi-mode circuit model using a transmission line of length $l=7650\um$, an impedance $Z_0=50\ohm$, a capacitive qubit coupling $C_c = 20\fF$ and a qubit capacitance $C_q = 55\fF$.  In both cases, the resonance frequency of the cavity is $8.02\ghz$.  Note the significant difference between the two predictions when substantially detuned from the first cavity resonance.}
	{\label{fig:multimodedecay}}
\end{figure}

\nomdref{Ccq}{$C_q$}{qubit capacitance}{subsec:purcelleffect}
\nomdref{Cy}{$Y$}{circuit admittance}{subsec:purcelleffect}
\nomdref{Gomegaq}{$\omega_q$}{qubit transition frequency}{subsec:purcelleffect}

This is an excellent approximation for the case of a single-mode cavity.  However, we are coupling to a transmission line or 3D cavity that has many independent spatial modes which can also couple strongly to and induce decay of the qubit \cite{Houck2008}.  As a result, when the qubit is appreciably detuned from any particular mode, the single-mode approximation is very poor, as shown in \figref{fig:multimodedecay}.  To correctly calculate the Purcell effect, we must incorporate the full impedance environment seen by the qubit.  As explained in \sref{subsec:fblrelaxation}, the relationship between qubit $T_{1}$ due to spontaneous emission and admittance $Y$ of the coupled environment is  
\begin{equation}
	\label{eq:mmpurcell}
	T^{\mathrm{Purcell}}_{1}=\frac{C_{q}}{\mathrm{Re}[Y(\omega_{\mathrm{q}})]},
\end{equation} 
where $C_{\mathrm{q}}$ is the qubit capacitance [\figref{fig:purcelldesign}(a)] \cite{Esteve1986,Neeley2008}.  This equation, which is equivalent to Fermi's golden rule, has previously been used to accurately model $T^{\mathrm{Purcell}}_{1}$ when all modes of the cavity are taken into account in the calculation of $Y$ \cite{Houck2008} and the qubit anharmonicity is relatively small.  In that paper, $Y$ is calculated with an equivalent circuit model of the device, with a transmission line crucially playing the role of the cavity.  As of the writing of this thesis, calculating the Purcell lifetime in 3D cavities is an ongoing project.

\subsection{Optimizing dispersive fidelity}
\label{subsec:optimizedispersive}

Returning to the issue of measurement fidelity, we see that we have arrived at an impasse.  If we are Purcell-limited, the ratio of $\kappa$ and $\gamma_{\kappa}$ is fixed and the number of collected measurement photons per qubit lifetime cannot be adjusted by changing the cavity decay rate.  (Not being Purcell-limited is actually more detrimental because our $\kappa$ is smaller than it could be without harming $T_1$; we are losing information in a way that does not increase our measurement fidelity.)  We can therefore only improve the SNR by increasing the amount of information carried per photon, using a larger number of equilibrium measurement photons, or lowering the noise power of the amplifier chain.  In practice, however, the amount of information per photon is already close to one and the number of measurement photons is already at its maximum (which is found by driving the system as hard as possible without lowering qubit $T_1$ or measurement contrast).  Amplifier noise is ripe for improvement but, for now, we are looking for solutions that do not require modifying the measurement apparatus.  Similarly, increasing $\Delta$ would increase the Purcell lifetime but will the decrease information per photon to precisely balance it as long as $\chi$ is not (wastefully) larger than $\kappa$.  

These issues limited us to about 70\% measurement fidelity in conventional 2D cQED \cite{Johnson2010}.  The issue is normally more severe than this because experiments are rarely optimized for readout.  We generally prefer for $T_1$ to be as long as possible to attain high gate fidelities and therefore would not be Purcell-limited.  Similarly, parameters like $\Delta$ or $\chi$ may be otherwise constrained by engineering multi-qubit interactions \cite{Reed2012} or quantum optics experiments \cite{Johnson2010}.  For example, the dispersive readout in the device first used to make a three-qubit GHZ entangled state \cite{DiCarlo2010} had a single-shot dispersive measurement fidelity of less than 5\%.

We see that optimizing dispersive readout fidelity is complicated because our parameters are interrelated.  Specifically, we want $T_1$ to be long so we can maximize our measurement time, but we also have to maximize the cavity lifetime to accommodate for the Purcell effect.  Increasing the cavity lifetime means our measurement photons leak out more slowly, eliminating the ``transmitted photons per qubit lifetime'' gains we may have made by increasing the qubit $T_1$\footnotemark.  A better solution would improve qubit $T_{1}$ independent of the cavity lifetime, leaving its optimization to our discretion.

\footnotetext{In a ``normal'' cQED device, the maximum fidelity is given when $\chi \approx \kappa$ and the qubit $T_1$ is just barely Purcell-limited.  This strikes a good balance between coupling strength lowering $T_1$ and increasing information per photon.  The point of this section is to circumvent this ``normal'' bound.}

Fortunately, there is a way around this problem because there is no intrinsic reason that the Purcell decay rate cannot be modified, even with fixed $\kappa$ and coupling.  A key feature of \eref{eq:mmpurcell} is that the impedance is evaluated at the frequency of the qubit only.  We can modify the admittance at the qubit frequency without substantially affecting the cavity if the two are far detuned from one another.  In particular, if we make the admittance purely reactive at $\omega_{\mathrm{q}}$ (that is, make $Y$ imaginary-valued) then $T^{\mathrm{Purcell}}_{1}$ diverges and the Purcell decay channel is shut off.  This solution decouples the choice of cavity $Q$ from the Purcell decay rate, freeing us to optimize readout fidelity by increasing $\kappa$ while enjoying a long qubit $T_1$.  As we will see, this admittance engineering can be realized with conventional circuit elements placed in an experimentally convenient location.

\subsection{Purcell filter implementation}
\label{subsec:purcellfilterimp}

\begin{figure}
	\centering
	\includegraphics[scale=1]{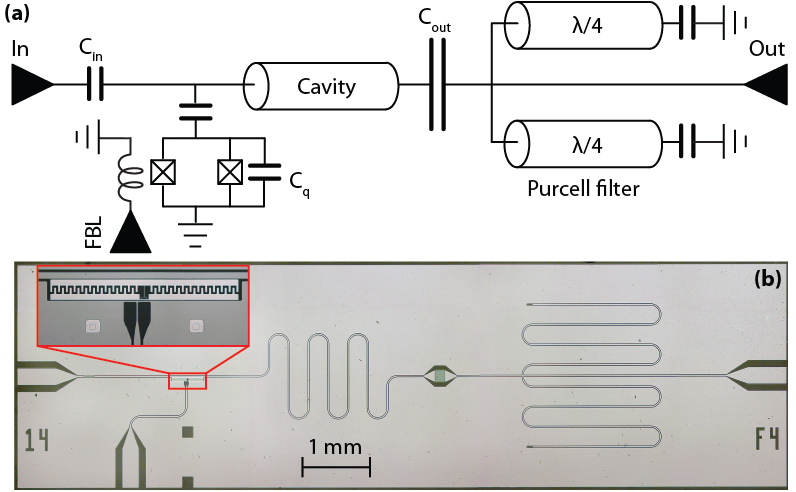}
	\mycaption{Design and realization of the Purcell filter}{\capl{(a)} Circuit model of the Purcell-filtered cavity design.  The Purcell filter, implemented with $\lambda/4$ open-circuited transmission-line stubs, inhibits decay through $C_{\mathrm{out}}$ near its resonance $\omega_{\mathrm{f}}$.  We use two identical stubs above and below the major axis of the chip to maximize the symmetry of the device, in an effort to suppress undesired on-chip resonances.  A flux bias line is included to tune the frequency of the qubit in-situ by changing the magnetic flux through the qubit SQUID loop.
	\capl{(b)} Optical micrograph of the device with inset zoom on transmon qubit.  Note the correspondence of the circuit elements directly above in \capl{(a)}.  The device is made with standard 2D cQED fabrication techniques: the resonator and filter structure are niobium patterned on a sapphire substrate, while the single transmon qubit (shown in the inset, top left) is fabricated using standard shadow masked double-angle aluminum deposition.  The device is 2 by 7 mm and is wire-bonded to a copper PC board and cooled in a standard octobox sample holder in a helium dilution fridge.  
	\figthanks{Reed2010}
	}
{\label{fig:purcelldesign}}
\end{figure}

We implemented the idea of shorting out the admittance at the qubit frequency with what we have deemed a ``Purcell filter.''  In this first experiment, the filter was realized with a transmission-line stub terminated in an open circuit placed outside the output capacitor of a low-$Q$ CPW cavity.  As shown in \figref{fig:purcelldesign}, the length of the stub is set to act as a $\lambda/4$ impedance transformer to short out the $50\ohm$ environment at its resonance frequency $\omega_{\mathrm{f}}$.  In this configuration, the Purcell filter only inhibits decay through the output capacitor, $C_{\mathrm{out}}$, so that capacitor is made much larger than the input capacitor $C_{\mathrm{in}}$.  We cannot filter both capacitors because that would eliminate (or at least severely inhibit) our ability to directly drive the qubit; however, the input capacitor is so small that the Purcell decay through that channel is negligible compared to non-radiative (e.g. dielectric) relaxation.  The total coupling capacitance in this device was approximately $80\ff$, giving a cavity bandwidth of $\kappa/2\pi = 20\mhz$.  The bare cavity frequency is $\omega_{\mathrm{c}}/2\pi=8.04\ghz$, the filter is at $\omega_{\mathrm{f}}/2\pi=6.33\ghz$, and a flux bias line is used to address a single transmon qubit with a maximum frequency of $9.8\ghz$, a charging energy $E_{\mathrm{C}}/2\pi \hbar$ of $350\mhz$, and a resonator coupling strength $g/2\pi$ of $270\mhz$.

It is worth emphasizing that the concept of a Purcell filter does not imply a particular implementation.  One might consider any structure that additionally modifies the impedance environment of a cavity for the purposes of reducing the Purcell decay rate of a subsystem to be a filter.  Indeed, a waveguide below cutoff has recently been constructed for use with a 3D cavity \cite{Hatridge2013b}. 

\begin{figure}
	\centering
	\includegraphics[scale=1]{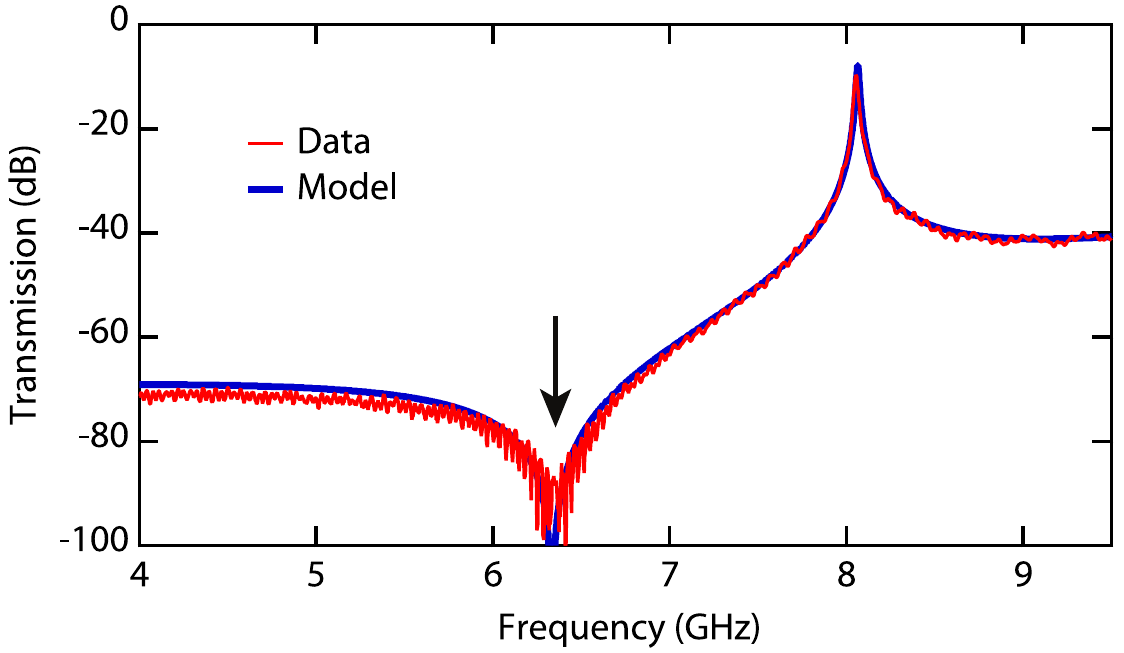}
	\mycaption{Diagnostic transmission data of the Purcell filter}{We measured transmission through the Purcell-filtered cavity at $4.2\K$.  At $\omega_{\mathrm{f}}$ (arrow) the Purcell filter shorts out the $50\ohm$ output environment, producing a $30\dB$ drop in transmission.  The high frequency oscillations visible are due to a small impedance mismatch along the measurement cables.  A circuit model incorporating only the experimental parameters $C_{\mathrm{in}}$, $C_{\mathrm{out}}$, $\omega_{\mathrm{c}}$, and $\omega_{\mathrm{f}}$ shows excellent correspondence.  This gives us confidence our design and understanding of the Purcell filter.
	\figthanks{Reed2010}
	}
{\label{fig:purcelltransmission}}
\end{figure}

We can verify that our device is working as expected by first measuring transmission at $4.2\K$ (below the superconducting transition temperature of the Niobium resonator) and comparing it with the circuit model used in design.  As seen in \figref{fig:purcelltransmission}, there is a dip corresponding to inhibited decay through $C_{\mathrm{out}}$ at $\omega_{\mathrm{f}}$.  A circuit model accurately predicts this dip and the overall structure of the transmission.  The model only misses the high frequency oscillations that are due to slight impedance mismatches (e.g. reflections) in the measurement chain.  The model used four input parameters: the input and output capacitance (which were simulated with the ANSYS Maxwell software package) and the frequencies of the cavity and filter (free parameters based on the measurement), lending confidence to our ability to predict and engineer the impedance of the structure.

\begin{figure}
	\centering
	\includegraphics[scale=1]{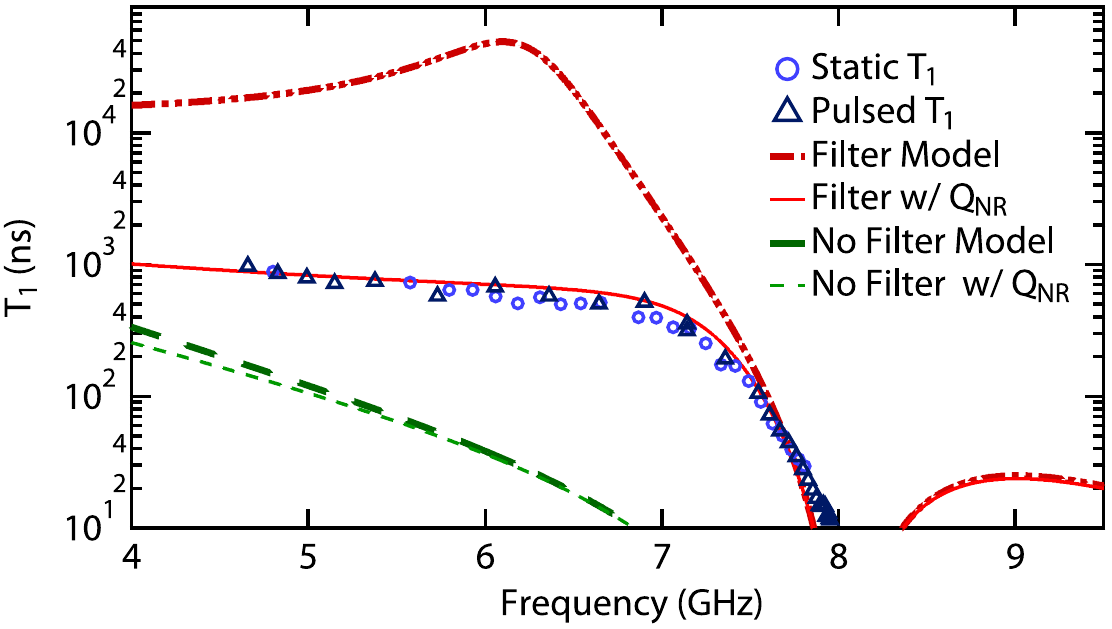}
	\mycaption{Purcell-filtered qubit $T_{1}$ and comparison to models}{We measure qubit $T_1$ using two methods.  The first is a static measurement (circles): the qubit is excited and measured after a wait time $\tau$. The second (triangles) is a dynamic measurement: the qubit frequency is tuned with a fast flux pulse to an interrogation frequency, excited, and allowed to decay for $\tau$, and then returned to its operating frequency of $5.16\ghz$ and measured.  This method allows for accurate measurement even when $T_{1}$ is extremely short.  When measuring near resonance with the pulsed method, coherent oscillations are noticeable as the qubit excitation swaps to the cavity and back (not shown).  Measurements using the two methods show excellent agreement. The top dashed curve is the predicted $T^{\mathrm{Purcell}}_{1}$, while solid curve additionally includes non-radiative internal loss with best-fit $Q_{\mathrm{NR}}=2\pi f T^{\mathrm{NR}}_{1}\approx27,000$.  The two lower curves correspond to an unfiltered device with the same $C_{\mathrm{in}}$, $C_{\mathrm{out}}$, and $\omega_{\mathrm{c}}$, with and without the internal loss.  In this case, the Purcell filter gives a $T_{1}$ improvement by up to a factor of $\sim$50 ($6.7\ghz$), but would be much higher in the absence of $Q_{\mathrm{NR}}$.
	\figthanks{Reed2010}
}
{\label{fig:lifetime}}
\end{figure}

Finally, to demonstrate the filtering effect, we measured the qubit $T_{1}$ as function of frequency.  As shown in \figref{fig:lifetime}, $T_{1}$ is accurately modeled by the sum of the Purcell rate predicted by our filtered circuit model and a non-radiative internal loss $Q_{\mathrm{NR}}\approx27,000$.  The source of this loss was not definitely known, though some possibilities proposed at the time were surface two-level systems \cite{Shnirman2005, Martinis2005, OConnell2008, Frunzio2013}, dielectric loss of the tunnel barrier oxide \cite{Martinis2005} or corundum substrate, non-equilibrium quasiparticles \cite{Martinis2009, Catelani2011, Sun2012, Riste2012c}, or effects due to IR radiation or inadequate sample thermalization (which may or may not be related to quasiparticle creation).  More recent advances in qubit coherence \cite{Paik2011, Chang2013} have strongly indicated surface dielectric loss as the primary culprit.  This model contains only the fit parameter $Q_{\mathrm{NR}}$ combined with the independently measured values of $g$, $E_{\mathrm{C}}$, $\omega_{\mathrm{c}}$, $\omega_{\mathrm{f}}$, $C_{\mathrm{in}}$, and $C_{\mathrm{out}}$.  

\begin{figure}
	\centering
	\includegraphics[scale=1]{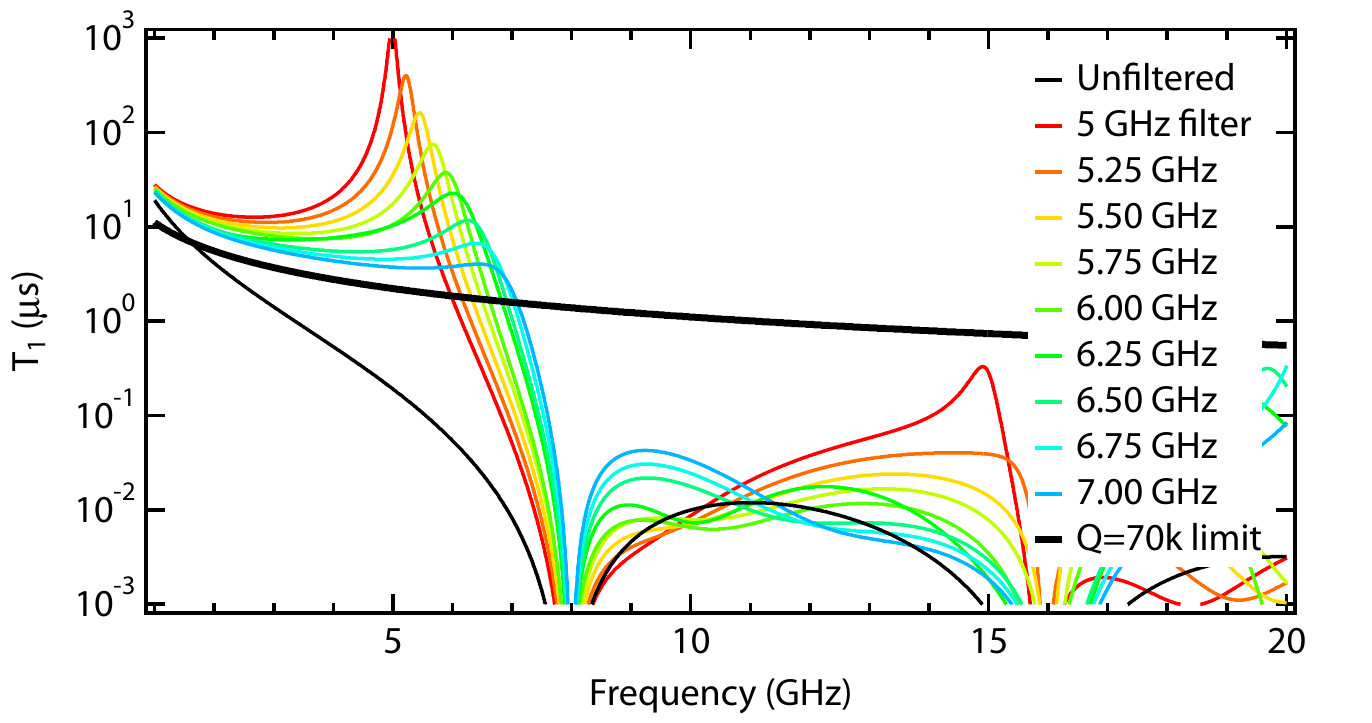}
	\mycaption{Calculated Purcell-filtered lifetimes for various filter frequencies}{We simulate the circuit shown in \figref{fig:purcelldesign}(a) for various lengths of Purcell filter, corresponding to different filter center frequencies.  The cavity $Q$ is a weak function of the cavity frequency, with $Q=\{60, 94, 146, 196, 268, 372, 445, 521, 605\}$ for filters centered on $f_f = 5\ghz$ to $7\ghz$, in steps of $250\mhz$.  This dependence comes from our having fixed the output coupler capacitance, but it would be possible to restore a constant-$Q$ by changing $C_{\mathrm{out}}$ along with $f_f$.  We also plot the unfiltered case and a constant $Q=70,000$ $T_1$ line that estimates the non-radiative planar $T_1$.
	}
{\label{fig:purcellfilterfreqs}}
\end{figure}

To calculate the improvement of $T_1$ due to the filter, we rely on our model to simulate what would occur in its absence.  The model, which calculates the Purcell decay rate for the case of an unfiltered circuit with the same $\kappa$ with and without the experimentally measured non-radiative decay is shown in the green dashed lines.  The improvement to $T_{1}$ due to the Purcell filter is found to be as much as a factor of $50$ at $6.7\ghz$, and would be much greater in the absence of $Q_{\mathrm{NR}}$ (comparing the red and green dashed lines).   We also calculated the predicted Purcell lifetime for various filter center frequencies, shown in \figref{fig:purcellfilterfreqs}. 

For small detunings, the qubit $T_1$ approaches the lifetime of the cavity itself, in the range of a few tens of nanoseconds.  A $T_{1}$ that brief can be a challenge to measure because dispersive readout (used in this experiment) will have a very low SNR.  This issue was avoided by using fast flux control \cite{DiCarlo2009}.  For measurements at small $\Delta$, the qubit is pulsed to the detuning under scrutiny, excited and allowed to decay, then pulsed to $5.16\ghz$ where measurement fidelity is higher, and interrogated.  This method has the advantage of being able to measure small lifetimes without sacrificing readout fidelity.  It would be easier to pulse the qubit at the home position and only then flux it because the qubit pulse frequency would be fixed, but the prescribed order also verifies that the qubit frequency at a given flux amplitude is truly known.

\subsection{Qubit reset}
\label{subsec:qubitreset}

One consequence of the Purcell filter is that the device exhibits an exceptionally large dynamic range in $T_1$.  We realize about a factor of $80$ between the longest and shortest times measured (a ratio that would be significantly larger if not for $Q_{\mathrm{NR}}$).  This can be viewed as a major ancillary benefit, one so substantial that it might be advertised as enabling the primary application of the Purcell filter.  That benefit is for qubit reset.  As we have already seen with the example of measurement-free \cite{Mermin2007} quantum error correction (\sref{subsec:quantumrepetitioncode}), there are many applications that benefit from or require on-demand reset of qubit state.  Experimental repetition rates can be greatly enhanced when they are otherwise limited by $T_{1}$.  At the end of a calculation, our qubit state will in general be in some unknown superposition of its ground and excited state.  If we want to do another calculation, we need to restore the qubit state unconditionally to some known state.  This is normally done by waiting for the qubit to decay back to the ground state at a rate $1/T_1$.  As qubit lifetimes increase \cite{Paik2011}, this can be inconveniently slow.  Resetting the qubit would greatly increase experimental bandwidth.  Similarly, experiments that use a qubit to make repeated measurements of a coupled system also require resetting the qubit between interrogations \cite{Johnson2010}.  

\begin{figure}
	\centering
	\includegraphics[scale=1]{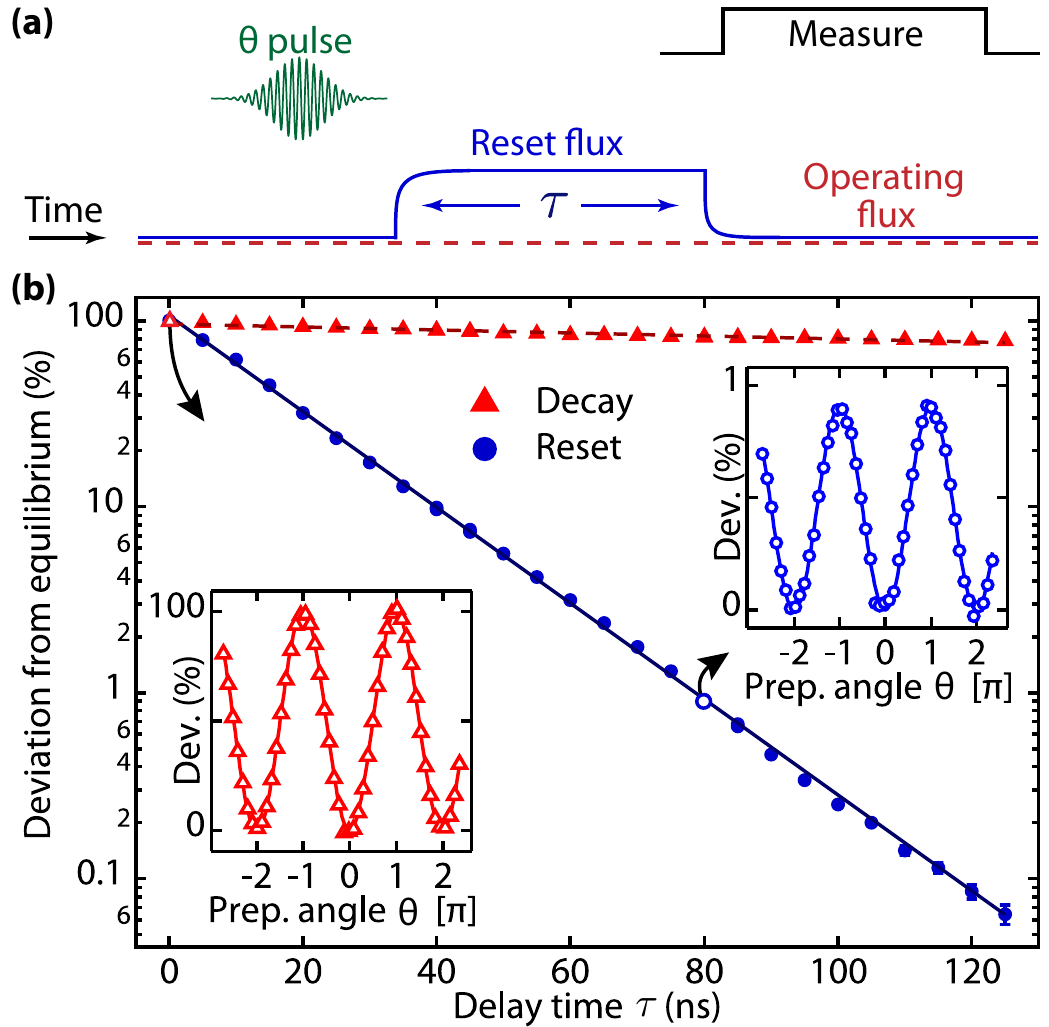}
	\mycaption{Fast qubit reset using Purcell filter}{\capl{(a)} Schematic of a pulse sequence used to realize a qubit reset and characterize its performance.  The fidelity of reset was quantified using a modified Rabi oscillation scheme.  The qubit is first rotated around the x-axis by an angle $\theta_{\mathrm{i}}$ at the operating frequency of $5.16\ghz$ and then pulsed into near resonance with the cavity (solid line) or left at the operating frequency (dashed line) for a time $\tau$.  The state of the qubit is measured as a function of $\theta_{\mathrm{i}}$ and $\tau$ after being pulsed back to $5.16\ghz$.  
	\capl{(b)} Remaining Rabi oscillation amplitude as a function of $\tau$, normalized to the amplitude for $\tau$=0.  This ratio gives the deviation of the qubit state from equilibrium.  Curves are fit to exponentials with decay constants of $16.9 \pm 0.1\ns$ and $540 \pm 20\ns$ respectively. Insets: Measured Rabi oscillations for $\tau=0$ (lower left) and $\tau= 80\ns$ (top right).  Note that the vertical scales differ by a factor of $100$.  This measurement technique was carefully chosen to be insensitive to thermal excitations of the qubit or cavity -- we are measuring the relaxation back to equilibrium, whatever that equilibrium may be.  A sinusoidal signal of known phase is also easy to fit accurately, leading to minimal uncertainties.  Fit uncertainties for the remaining equilibrium deviation are included here, but only visible for $\tau > 120\ns$.
	\figthanks{Reed2010}
	}
{\label{fig:reset}}
\end{figure}

With our Purcell filtered device, we can exploit the sudden drop in qubit lifetime in the vicinity of the cavity to perform reset.  We normally operate far detuned from the cavity, where the qubit $T_1$ is not Purcell-limited.  To reset, we can quickly move the qubit frequency near resonance with the cavity using a fast flux pulse.  The qubit excitation is dumped into the cavity and quickly decays to the environment\footnotemark.  Though this could work for any flux-tunable cQED device, the Purcell filter greatly improves the flexibility of this process.  It can exhibit a dramatically increased ``lifetime contrast'' -- the ratio of the longest to shortest qubit $T_1$ -- through the use of a low-$Q$ cavity that does not limit qubit $T_{1}$ at its operating frequency.  This concept has been used in other experiments, but by resetting through a spurious lossy two-level system that couples to the qubit rather than an engineered cavity mode \cite{Shalibo2010}.

\footnotetext{It is worth noting that reseting with the Purcell filter need not require fast flux tunability.  There are several methods of unconditionally swapping excitations from the qubit into the cavity \cite{Wallraff2007, Schliesser2008, Geerlings2013}, but they all benefit from a large dynamic range between cavity and qubit lifetimes.  We need some non-unitary evolution to reset the qubit, which can be granted by the cavity decay.  Another possible non-unitary initialization is to measure the qubit and use that classical information to feed back and set the state \cite{Riste2012a, Johnson2012}.  This has the advantage of being something that we will need to do anyway in a ``real'' quantum computer, giving us initialization essentially for free.}

The efficacy of reset in this device is readily quantified using a modified Rabi oscillation sequence.  As shown in \figref{fig:reset}(a), the qubit is rotated by some angle and then reset for some time before being measured.  The amplitude of the remaining sinusoidal Rabi oscillation indicates the degree to which the qubit is out of equilibrium after the reset time $\tau$.  The protocol was chosen so as to be insensitive to any equilibrium thermal population of the qubit.  The non-equilibrium population is found to exhibit pure exponential decay over three orders of magnitude.  The qubit can be reset to $99.9\%$ in $120\ns$ or any other fidelity depending on $\tau$ (e.g. $99\%$ for $80\ns$ or one ``9'' every $40\ns$).  The sequence is also performed with the qubit remaining at the operating frequency during the delay to demonstrate the large dynamic range of $T_{1}$ available in this system.  The dynamic range would be much higher in the absence of the non-radiative $Q_{\mathrm{NR}}$ limiting qubit lifetimes to well below the Purcell limit when far detuned.

One potential concern with using this technique in larger systems is what would happen were we to reset one qubit by transferring its population into the cavity, which is itself coupled to other qubits.  This photon would cause a $\chi$ shift on all the others, and since the decay of that excitation out of the cavity is stochastic (e.g. the time it takes to decay, and therefore the phase it evolves, is random), would result in dephasing of all the other qubits \cite{Sears2012}.  This could be avoided by either turning off the $\chi$s of the other cavities by detuning them or by using separate reset cavities for each qubit.  Having separate cavities might seem a bit extravagant, but it should not be any more costly (and it could potentially be much more useful) than a flux bias line per qubit.  This is especially attractive given recent methods of entangling subsets of qubits with all-microwave drives \cite{Chow2011, Poletto2012, Paik2013} which may eliminate the need for flux bias lines.  This extra cavity mode could also be implemented for individual qubit readout, a functionality that fulfills one of the DiVincenzo criteria (\sref{subsec:divincenzo}).  

\subsection{Purcell filter summary}

The Purcell filter seeks to solve the problem of radiative decay of the qubit through the cavity to which it is coupled.  We approached this problem in the context of measurement fidelity, but it should be understood as a general problem of maximizing qubit coherence when coupled to a lossy external impedance (\sref{subsec:fblrelaxation}).  One can always reduce coupling or increase cavity lifetime, but doing so will have unwelcome implications for measurement fidelity.  Fortunately, there is no reason that the qubit needs to decay through a coupled cavity at all.  As long as the frequencies of the respective modes are sufficiently detuned, the cavity and qubit sample distinct impedance environments.  We exploited this fact by creating a Purcell filter using a piece of shorted transmission line.  This object shorts out the coupled environment at the qubit frequency while minimally changing it at the cavity frequency.  The resulting device demonstrated that the Purcell effect was shut off in accordance with our expectations and also exhibited a very large dynamic range in qubit $T_1$.  Combining this resource with fast flux, we showed the ability to reset our qubit back to equilibrium very rapidly by dumping the excitation into the cavity.

You may have noticed that we introduced the design of the Purcell filter for increasing measurement fidelity, but we did not actually demonstrated that it helps.  This is due to the fact that we were never able to carefully study its effect.  During our exploration and optimization of measurement performance in the first Purcell device, we discovered a new readout mechanism.  It is found only at very high drive strengths and at an unusual cavity frequency, but can deliver very high measurement fidelity.  This phenomenon diverted our attention from the applications of the Purcell filter to dispersive readout and is the subject of the next section of this chapter.

\section{Jaynes-Cummings readout}
\label{sec:jcreadout}

\nomdref{Cfbare}{$f_{\mathrm{bare}}$}{bare cavity frequency}{sec:jcreadout}

One aspect of measurement that we have not yet explored is what happens as a function of measurement power.  It seems rather crucial -- why not simply increase the drive power, increasing the signal while keeping noise constant?  Unfortunately, as alluded to earlier in the chapter, there is an optimal number of measurement photons that you may use, beyond which the overall measurement fidelity will decrease.  There are several underlying causes of this fact, but the one we focus on in this section is inherited cavity nonlinearity\footnotemark.  For the purposes of qubit readout, we like to model the cavity as a perfectly harmonic oscillator whose transition frequency happens to depend on the qubit state (e.g. \eref{eq:dispersivejc2}), but the reality is more complicated.  The cavity mode hybridizes with the qubit itself, such that photons stored in the cavity mode drive currents not only in the coplanar strip line (or the walls of your 3D resonator) but across the Josephson junction of the qubit as well.  Thus the cavity inherits some qubit-like behavior -- nonlinearity -- that is not contained in the typical dispersive Hamiltonian.

\footnotetext{The other primary reason is that the qubit $T_1$ is observed to reduce as a function of incident drive in many systems.  The reason for this is not well understood.  Typical values of where the qubit lifetime is significantly affected is on the order of $\sim 10-20$ photons.  For dispersive measurement, the SNR is directly proportional to this number.  Fortunately, for the purposes of the JCR readout scheme described in this section, this effect seems to be irrelevant.}

This section will introduce what happens, experimentally, to the cavity transmission as a function of drive power.  We will find that the cavity initially becomes strongly nonlinear and loses state-selective readout contrast.  When we drive much harder, the cavity eventually reaches a high-transmission {\it bright state} where its nonlinearity has shut off and it is restored to a harmonic oscillator.  We observe that the power required to drive the cavity to this state depends strongly on the initial qubit state.  We use this fact to make a qubit measurement, which, because of the large number of equilibrium photons in the bright cavity state, has a very large signal-to-noise ratio and a fidelity not limited by amplifier noise.  We conclude with a discussion of the underlying physics, which until that point will have been described only phenomenologically, and offer an intuitive (if simplified) explanation of the dynamics.  All of the data in this section are the result of measurements done on the four-qubit device described in \sref{sec:fourqubitdevice}.

\subsection{Cavity nonlinearity}
\label{subsec:cavnonlinearity}

The framework we most often employ to describe the mechanism of qubit readout is that of a perfect harmonic oscillator cavity with a qubit state-dependent frequency.  This is a good approximation in the limit of few excitations (e.g. less than one mean photon), but it rapidly breaks down as the cavity occupation increases.  As mentioned in the introduction to this section, this is because the cavity mode is hybridized with the qubit and therefore ``inherits'' some nonlinearity.  There are two equivalent ways of viewing this hybridization: as either a quantum mechanical effect where the eigenmodes of the system are hybridizations of the undressed qubit-like and photon-like excitations, or as a classical effect where the EM mode of the cavity routes some finite current through the qubit's Josephson junction.  While to first-order the form of this nonlinearity is encapsulated by a simple $(\hat{a}^{\dagger}\hat{a})^2$ ``Kerr'' term\footnotemark, at higher powers, the cavity's behavior is quite exotic and unexpected.

\nomdref{Abbq}{BBQ}{black-box quantization or barbecue}{subsec:cavnonlinearity}

\footnotetext{Another common model for the cavity-qubit system that has recently gained popularity following the development of the black-box quantization (BBQ) model \cite{Nigg2011} is to treat both the qubit and cavity as a harmonic oscillator with some $(a^{\dagger}a)^2$ nonlinearity.  In the limit of low cavity occupations, this model again has great success.  For example it can correctly predict the single-photon Kerr effect \cite{Kirchmair2013} and cavity-cavity ``cross-Kerr'' coupling in devices with more than one cavity.  However, as we will see, even this more sophisticated model will not correctly predict the high-power behavior of the cavity.  The cavity nonlinearity is not constant, but rather reduces with excitation number.  Indeed, the form of the nonlinearity cannot be captured with a finite number of Taylor expansion terms of the Josephson potential.}

\nomdref{Cah}{$\overline{A_{\mathrm{H}}}$}{averaged homodyne amplitude}{subsec:cavnonlinearity}

\begin{figure}
	\centering
	\includegraphics[scale=1]{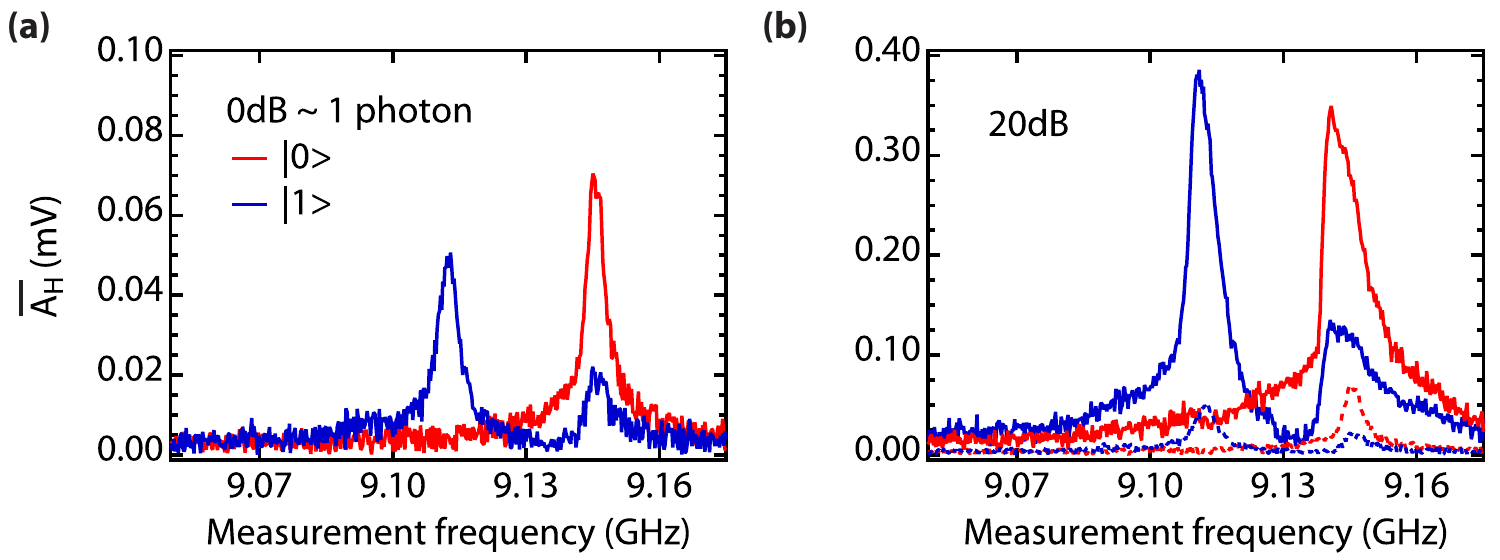}
	\mycaption{Cavity transmission for low drive strengths}{\capl{(a)} Dispersively shifted cavity response for excited (blue) and ground (red) states of the $8\ghz$ qubit with $\sim1$~photon mean cavity occupation.  We reference this power to $0\dB$.  For all plots in this figure and the next one, the qubit tuned to $8\ghz$ is prepared, a measurement tone is pulsed on, and the responding homodyne amplitude of the transmitted signal is averaged for $400\ns$ to yield $\overline{A_{\mathrm{H}}}$.  The mV scale used is arbitrary, but consistent to ease comparison.
		\capl{(b)} Cavity transmission for the same preparations and frequency range as \capl{(a)}, but when driving with 100 times (+20 dB) more incident power.  Transmission is inhibited by the cavity nonlinearity, which also distorts the line into an asymmetric shape, bending in the direction of the nonlinearity.  Data from \capl{(a)} is plotted in dashed lines for comparison.
		\figadapt{Reed2010b}
		}
	{\label{fig:transmission1}}
\end{figure}

Cavity transmission as a function of drive power reveals this nonlinearity.  Here, we measure the device detailed in \chref{ch:entanglement}, which is a two-dimensional cQED device with a cavity at $9.070\ghz$ and four qubits tuned to $6$, $7$, $8$, and $12.3\ghz$ (when making single-qubit measurements the $8\ghz$ device is addressed).  As shown in \figref{fig:transmission1}(a), at extremely low drive power the cavity response is Lorentzian and therefore harmonic.  This is because the nonlinearity has no consequence if we only sample the 0 or 1 photon states.  (The fact that two photons are not exactly twice the energy of one photon is not noticed if we never populate the second photon).  We also observe the dispersive shift, with transmission for the qubit prepared in the ground state in red and the excited state in blue.  If we wanted to make a conventional measurement, we would drive at this power at about $9.145\ghz$ (the red peak), inferring that the qubit was in the ground state if we got a relatively high transmission and vice versa.

When we drive much harder, we start to see evidence of the cavity being non-harmonic.  This is shown in \figref{fig:transmission1}(b), in which we have increased the drive power by a factor of $100$.  We refer to this power as being $+20\dB$, where $0\dB$ is defined by the power that produces a one-photon mean cavity occupation.  The cavity line shape is asymmetric, with the slope on the low-frequency side being much steeper than that of the high-frequency side.  This is the canonical behavior of a Kerr-Duffing oscillator, where the negative cavity anharmonicity pushes the resonance down in frequency as the occupation increases.  The peak transmitted homodyne amplitude would be a factor of $10$ higher if the cavity were linear but has only increased by a factor of $4$.  This is attributable to either pure cavity dephasing or a photon-blockading effect, where further excitation is inhibited when an additional photon shifts the transition frequency of the cavity out of resonance with the drive.  Photon blockading would be correctly predicted by an $(\hat{a}^\dagger \hat{a})^2$ anharmonicity.

\begin{figure}
	\centering
	\includegraphics[scale=1]{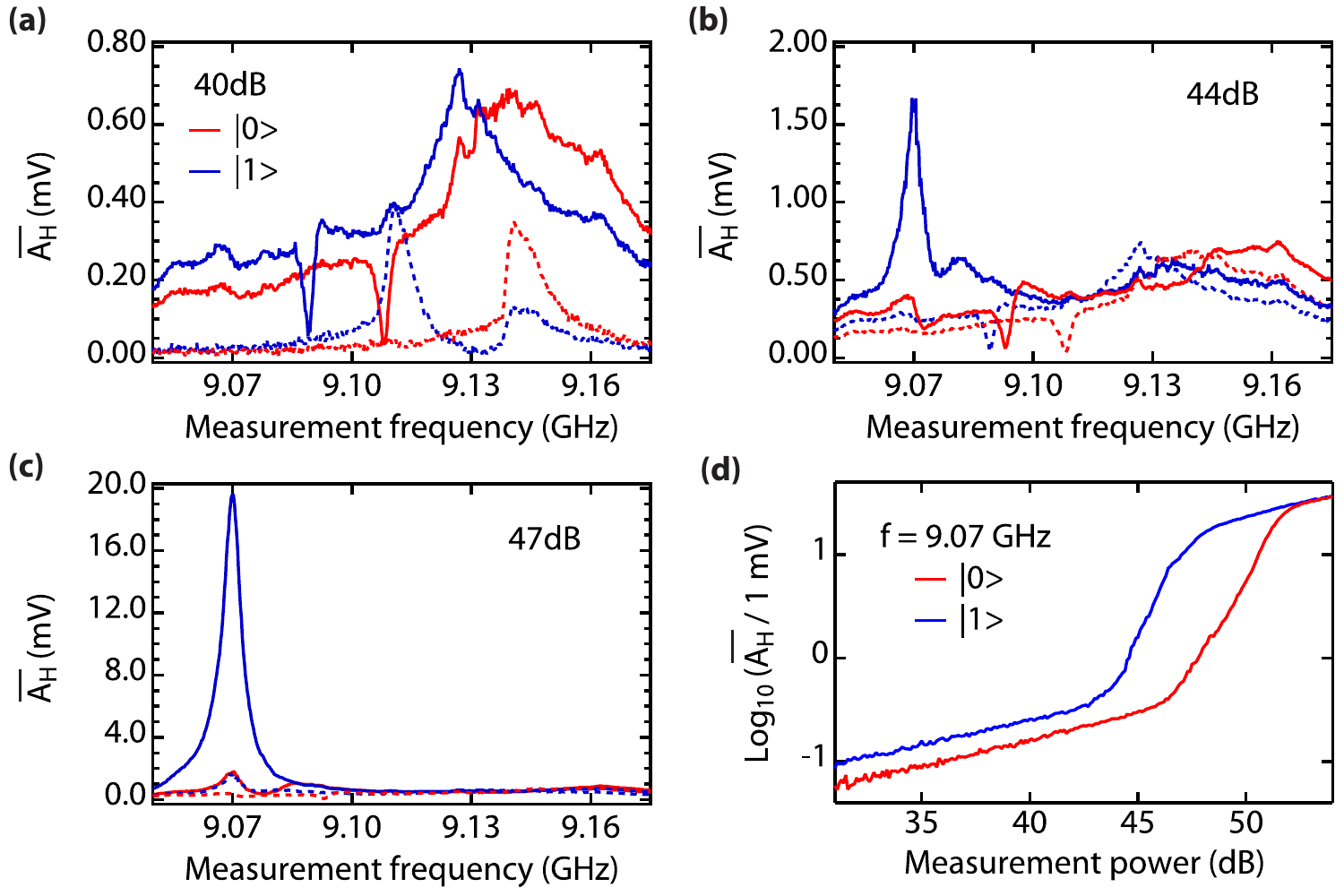}
	\mycaption{Cavity transmission at large drive strengths and bare cavity response vs. power}{\capl{(a)} Cavity response for increasing drive power, with data for previous power plotted with dashed lines.  Transmission is inhibited by the cavity's inherited nonlinearity, limiting dispersive measurement fidelity.  Note that increasing drive amplitude $10\dB$ from the dashed to solid lines increases $\overline{A_{\mathrm{H}}}$ by a factor of $\sim2$ and complicates the frequency dependance.  
		\capl{(b)} At slightly higher drive strengths than \capl{(a)}, a distinct resonance starts to emerge from the chaotic background, indicating that the physics driving this response is not simply a Kerr-Duffing nonlinearity.  Detuned from this peak, the transmission is not substantially changed from \capl{(a)}.
		\capl{(c)} At large drive power, the feature at $f_{\mathrm{bare}}$ grows to near-unity transmission, indicating that the cavity anharmonicity has shut off.  For this power the system only reaches its bright state when the qubit is excited due to the asymmetry of the dispersive cavity shift about $f_{\mathrm{bare}}$.  This asymmetry is characteristic to the transmon qubit \cite{Schreier2008} but might be possible to simulate for other designs \cite{Bishop2010}.
		\capl{(d)} Response at $f_{\mathrm{bare}}$ versus input power, showing a steep jump in transmission corresponding to the onset of the bright state at a qubit-state dependent power.  Though transmission state dependence exists elsewhere, the behavior here is especially amenable for use as a qubit readout because the transmission difference is large compared to amplifier noise.  On either side of the jump, the increase in power is a linear function of power (e.g. a straight line on a log plot, where the $x$-axis is in exponential units) because the cavity response is linear in both cases.  On the left, the cavity excitation is too low to sample the nonlinearity; on the right the nonlinearity has been turned off as the cavity is in its bright state.
		\figadapt{Reed2010b}
		}
	{\label{fig:transmission2}}
\end{figure}

The photon-blockade or dephasing effect is even more dramatically illustrated in \figref{fig:transmission2}(a).  There we have increased the drive power by another factor of $100$ but see an increase in homodyne voltage by less than a factor of two.  (Note the presence of a square root here, with the drive measured in power units but the transmission in voltage units; we should expect only ten times more voltage in linear response.)  The line shape no longer resembles that of a conventional resonance: it is quite broad and rough and the excited-state cavity transmission has shifted back toward the ground state, reducing the state-dependent transmission difference.  There is also a sharp dip at lower frequencies, possibly due to an interference between the multiple solutions of the now-bifurcated cavity \cite{Drummond1980}.

\begin{figure}
	\centering
	\includegraphics[scale=1]{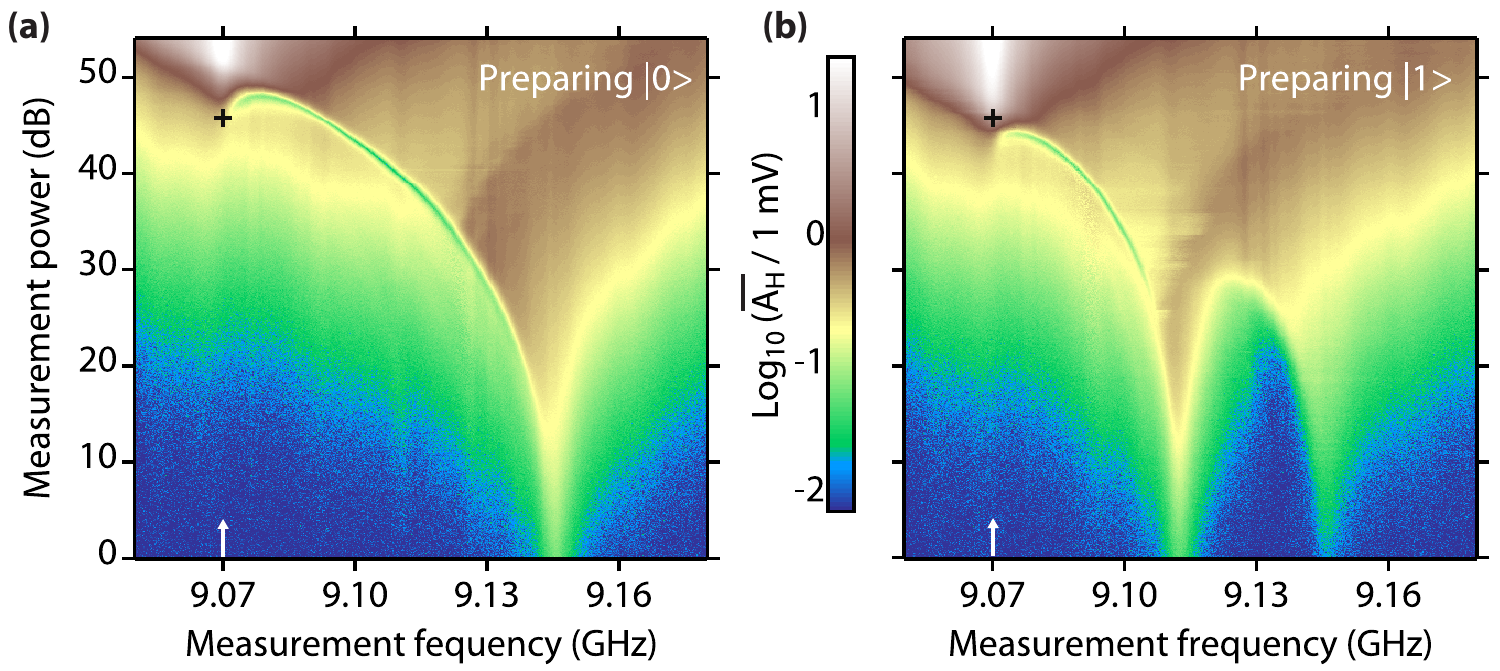}
	\mycaption{Cavity amplitude response}{We plot the log magnitude of cavity response as a function of drive power and frequency for \capl{(a)} qubit ground and \capl{(b)} excited state preparations.  The cavity continuously evolves from its low-power linear behavior through the anharmonic bistable region and to the bright state.  There are two peaks present in \capl{(b)} due to qubit relaxation during measurement.  The symbols (+) denote the optimal power and frequency for qubit readout, where the cavity response for the two qubit states is maximally different.  The dip that follows the cavity up to the bright state on its low-frequency side is likely an interference between the two bistable cavity solutions with differing phase.
	\figadapt{Reed2010b}
		}
	{\label{fig:transmissioncolor}}
\end{figure}

For a long time, it was believed that this bifurcation was the end of the cavity-transmission-vs-power story.  For the purposes of readout, increasing power would initially help, but not as much as it ``should'' if the cavity were linear because of photon blockading effects.  Beyond a certain point, the cavity would be so distorted that the state-dependent transmission difference would drop, and fidelity would suffer.  Indeed, this would be true if the cavity had a constant Kerr-type nonlinearity, which is well understood and even regularly used to create low-noise Josephson bifurcation amplifiers \cite{Spietz2009, Castellanos2009, Bergeal2009}.  It turns out, however, that the form of the cavity nonlinearity is not so simple.  As shown in \figref{fig:transmission2}(b), when we increase the power a further $4\dB$ as compared with (a), a distinct peak emerges from the chaotic background.  If we further increase the power, that peak grows to completely dominate the transmission, as shown in (c).  This behavior indicates that the cavity anharmonicity must shut off at a certain point, restoring the cavity to linear response.  (That is, the anharmonicity is a function of power and asymptotically decreases to zero with increasing occupation, as we will see in \sref{subsec:highpowertheory}.)  At this point, the cavity is referred to as being in its {\it bright state}, where transmission has dramatically increased (to near-unity) compared to lower drive strengths, and its line shape and frequency cease to evolve with power.  The frequency at which the cavity goes bright is always the same and is known as the {\it bare cavity frequency}.  This is the frequency at which the cavity would resonate in the absence of the qubit's Josephson junction.  As shown in \figref{fig:transmission2}(c), the transmission of the cavity in this state is huge compared to its response when we drive with only a few dBs lower power.  Transmission for the two qubit preparations is summarized in \figref{fig:transmissioncolor}.

The strongly-driven Jaynes-Cummings Hamiltonian has been previously studied for the case of the cavity and qubit in resonance \cite{Bishop2010}.  There, at drive strengths so high that they would be inaccessible with essentially any other system, higher-order photon number resonances with characteristic $\sqrt{n}$ spacing appear, in excellent agreement with theory.  The dispersive limit was not studied for the reasons described in the previous paragraph, but the story is similar.  We can reach a regime of the Hamiltonian where the drive strength (or, excitation level) is so high that qualitatively new physics emerges \cite{Reed2010b, Bishop2010, Boissonneault2010}.  In contrast to the resonant case and to our great fortune and surprise, this new physics has practical applications for the purposes of qubit measurement.

\subsection{Single and multi-qubit measurement}
\label{subsec:jcqubitmeasurement}

The ability to use this phenomenon to read out the qubit was demonstrated in \figref{fig:transmission2}(c).  There, we have driven the cavity to the bright state for the qubit in the excited state, but not for the ground state.  The transmission difference at the bare cavity frequency is therefore huge.  We get a small signal driving at $9.070\ghz$ with $47\dB$ if the qubit is in its ground state but a very large one if it is excited.  Qualitatively, this is the same mechanism we use for the dispersive readout.  The primary difference here is that we are dealing with much larger signals.  In the case of the dispersive readout, we have a mean cavity occupation of approximately ten photons; in this device, there are thousands of times more.  Thus, our measurement fidelity is not dominated by the signal-to-noise ratio, nor set by things like amplifier noise.  (Let us hasten to point out that a very large SNR for distinguishing the bright state from the dim state does not imply unity measurement fidelity; the mapping of the qubit state onto the semi-classical cavity pointer state can still be imperfect.)

The emergence of the high-power peak, we postulated, results from the fact that at a certain occupation the cavity anharmonicity reduces to zero (or at least to less than the cavity linewidth).  This explanation made no mention of the qubit state, however, so why do we see a state dependence of the critical power?  It originates from the same underlying physics as the conventional low-power readout: the qubit-state selective dispersive cavity shift.  In our prescription for measurement using this effect, we drive at the bare cavity frequency with a certain power.  This means that we are initially driving far off-resonant from the cavity transition frequency when it contains a few photons.  However, the amount by which we are off-resonant depends on the qubit state because of the dispersive shift.  If the qubit is excited, the cavity starts nearer to the bare cavity frequency (since $\chi$ is negative) and the measurement drive, making it somewhat easier to input photons.  Thus, the cavity will reach the occupation necessary for the bright state at a slightly lower drive strength compared to the ground-state case.  For a fixed drive strength, the cavity will be much more likely to excite to its bright state if it starts out closer to the drive.  This gives us our state-selective transmission and qubit measurement mechanism.

\subsubsection{Single-shot measurement fidelity}

\begin{figure}
	\centering
	\includegraphics[scale=1]{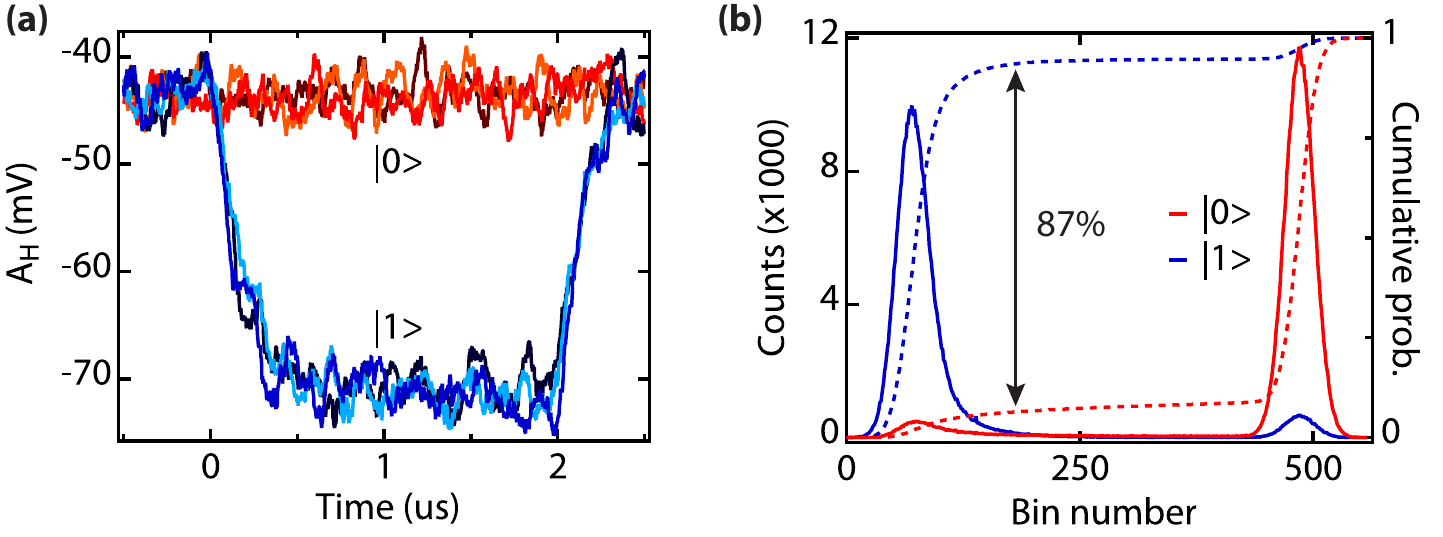}
	\mycaption{Measurement transients and histograms for single-shot measurement of one qubit}{\capl{(a)} Three measurements of single-shot system amplitude response $A_{\mathrm{H}}(t)$ to driving at $f_{\mathrm{bare}}$ with a power $P_{\mathrm{meas}}$ for prepared qubit ground and excited states, low-passed to the response time of the cavity $1/\kappa\approx 100\ns$.  They are reproducible and well-distinguished, demonstrating that the response is large compared to the measurement noise.  
	\capl{(b)} Histograms and S-curves quantifying measurement fidelity of the $8\ghz$ qubit.  An ensemble of single-shot responses are integrated for $500\ns$ and their distribution plotted. The two histogram peaks (solid lines) are well separated with few counts between them.  Integrating these yields ``S-curves'' (dashed lines), with their maximal difference indicating a single-shot fidelity of 87\%.
		\figadapt{Reed2010b}
		}
	{\label{fig:fidelity}}
\end{figure}

The state-dependent cavity critical power provides a mechanism for qubit readout, but how good is this measurement?  Specifically, we are interested in knowing what fraction of the time the measurement result is an accurate representation of the qubit state.  We can directly measure this fraction by repeatedly preparing the qubit in a known computational state ($|0\rangle$ or $|1\rangle$) and measuring, then comparing what we should have found with what we did find.  This is as simple as doing a $\pi$ pulse or no pulse on the qubit, immediately measuring it, and then recording each single-shot measurement result.  (One aspect which is atypical for this measurement is that we are interested in knowing individual measurement outcomes without ensemble averaging\footnotemark.)  In \figref{fig:fidelity}(a), we show several single-shot measurement transients of our system when the qubit is prepared in either the ground or excited state.  At time $t=0$, a measurement tone at the critical power and frequency is pulsed on, and at $t=2\us$ it is turned off.  When the qubit is in the ground state, the response is so minimal as to be obscured by the noise, but when the qubit is excited the response is large and easily discriminated.  After only about one hundred nanoseconds, the signals are entirely distinguishable, which supports our expectation of a very high signal-to-noise ratio.

\footnotetext{For the vast majority of experiments we currently perform, you may average ensembles of measurement records together on the digitizer card before transmitting the data to the computer.  When the communication link between the card and the computer is slow (e.g. with the Acqiris 240 data acquisition card we have long used), this can provide huge advantages in repetition rate by e.g. getting several thousand runs of the experiment per data transfer rather than only a few.  When histogramming measurement results, however, we do not have that luxury; we have to transfer each unmolested trace to the computer for processing.  This makes collecting measurement histograms thousands of times slower than it should be.  Fortunately, our lab has started to transition to a much faster solution, the AlazarTech digitizer, with which data can be acquired and transmitted to the computer in real-time with essentially no overhead.  This opens up the possibility of, for example, histogramming and thresholding every measurement, which would reduce our sensitivity to gain drifts and give some physical meaning to our measurement values.  For example, our $y$-axis could be understood to be ``the fraction of time the cavity went bright,'' rather than just ``digitizer millivolts.''  This and measurement integration time are discussed further below.}

Given these single-shot measurement transients, we can readily quantify our measurement fidelity by histogramming that data.  We integrate each trace for some fixed amount of time and plot the distribution of the resulting values in \figref{fig:fidelity}(b), with a separate histogram for either of the qubit preparations.  These ``measurement histograms'' show a bimodal and well-separated distribution in both cases, with very few counts between the two peaks.  While the excited state qubit almost always results in the cavity going bright (here, a small bin number) and vice versa, there are also counts for each case indicating the opposite response and corresponding to the qubit state being incorrectly mapped to the cavity.  This unfaithful mapping might trivially come from a faulty qubit state preparation (e.g. thermal qubit population or imperfect $\pi$ pulses) or decay of the qubit state prior to measurement.  It might also be indicative of the underlying physics: most obviously, a small difference between critical state powers.  The difference in critical power for the two qubit states seems to be influenced by the ratio of $\chi/\kappa$.  Experimentally, if this value is small (e.g. the dispersive shift is not large compared to the cavity linewidth), we observe relatively low measurement fidelity.  In addition to simple cavity overlap, this may have to do with our square measurement pulse having a large spectral bandwidth.  We have seen some evidence that pulse-shaping and frequency-chirping can result in qualitatively different behavior \cite{PrivateCommunicationLuyanSun}.  Classical and perhaps quantum noise would also tend to wash out the cavity transition, as would a finite cavity lifetime.

The fact that there are very few counts between the two histogram distributions indicates that when we start driving the cavity, it either immediately reaches its bright state or it never does.  The behavior of the cavity for very long time scales under a continuous-wave drive seems to refute this, with random jumps of the cavity up to and down from the bright state \cite{PrivateCommunicationLuyanSun}.  These up and down rates are strong functions of measurement power, however, and mostly play a role at relatively modest drive strengths.  For the short time scales and drive strengths optimal for measurement, the cavity state will only rarely relax back from the bright state once it is attained.  This is because we had to drive much harder than necessary to maintain the bright state when we initially off-resonantly populated photons.

By integrating the two distributions as a function of bin number, we can define a single measurement fidelity number at the point where the difference of these two ``S-curve'' functions is largest.  By drawing a line at this maximum-difference bin, we say that any measurement result larger than that point most likely indicates that the qubit is in its ground state, and vice versa.  The difference tells us our measurement fidelity, which for this case is 87\%.  That is, our measurement result agrees with our qubit state preparation an average of 93.5\% of the time\footnotemark.  This is a substantial improvement over what the conventional dispersive readout would yield in this particular device, measured to be $\sim5\%$.

\footnotetext{Looking at the histograms in \figref{fig:fidelity}, the measurement fidelity is defined as one minus sum of the probability that either state is misidentified.  For example, if your measurement correctly identifies the excited state 90\% of the time and the ground state 95\% of the time, the fidelity would be $1-0.10-0.05 = 0.85$.  This definition correctly classifies the case of a measurement that merely ``flips a coin'', successfully identifying either state 50\% of the time.  The fidelity of that measurement would be $1-0.5-0.5 = 0$.  Similarly, a measurement that {\it always} says the qubit is in the ground state will have  a fidelity $1-0-1=0$.}

\subsubsection{Fidelity as a function of integration time}

\begin{figure}
	\centering
	\includegraphics{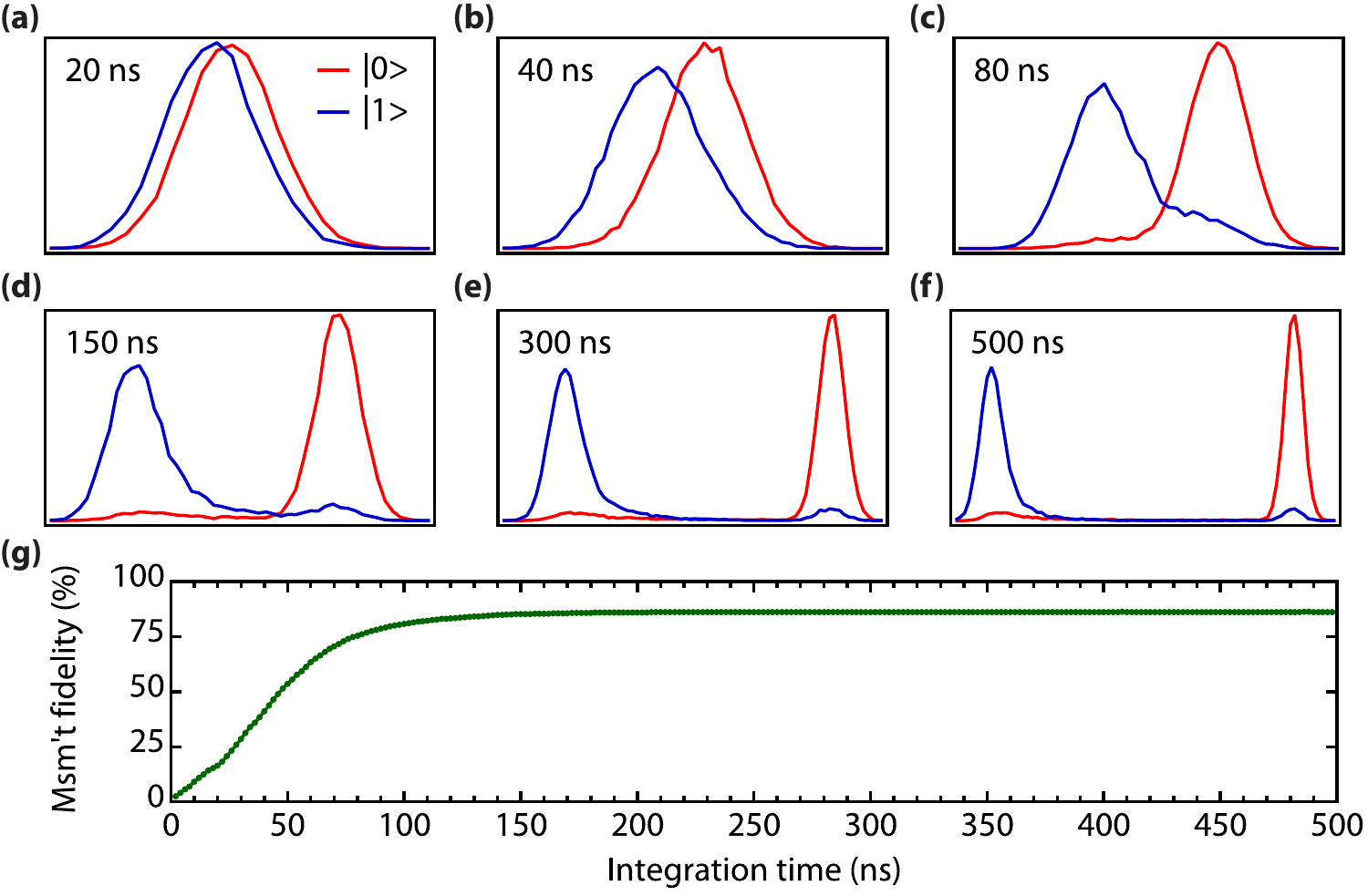}
	\mycaption{Measurement histograms and fidelity versus integration time}{\capl{(a-f)} Histograms of single-qubit measurements as a function of integration time.  Here, the same measurement transient dataset are processed and histogrammed, but with a variable integration as specified in each window.  At short times, the histograms are essentially overlapping, but for longer measurements they evolve apart and entirely separate.  
	\capl{(g)} Measurement fidelity as a function of integration time.  Integrating $V_{\mathrm{H}}(t)$ for $120\ns$ yields 80\% fidelity, while $240\ns$ yields 87\%.  After that point, the fidelity has saturated and increasing the SNR does not improve the result.  We are limited by the cavity pointer state incorrectly mapping the qubit state, rather than an inability to distinguish the pointer state.  Additional integration does increase the histogram separation, however, which may be useful if we are thresholding the data to convert a voltage to an estimated qubit state.
	\figadapt{Reed2010b}
	}
{\label{fig:integrationtimefidelity}}
\end{figure}

One thing we have not yet discussed is how long these measurements need to take.  In the case of the dispersive readout, the measurement time is of great importance because the qubit can decay, so further integration after this occurs lowers the fidelity.  In the case of the bright state readout, however, once the cavity has decided to go bright or not, it tends to stay that way -- that is, this is a {\it latching} readout.  As a result, we have the luxury of integrating as long as we like to get whatever SNR we desire.  In \figref{fig:integrationtimefidelity}, we show the resulting measurement fidelity for varying measurement integration times.  For this device, the measurement fidelity smoothly increases from 0 to 80\% in the first $100\ns$, and levels off at its final value of 87\% by $175\ns$.  This time is likely due to either the speed with which the system excites to the bright state, or simply how long it takes for the measurement chain noise to be drowned out by our growing signal.

Integrating for more than $240\ns$ does not increase fidelity in this device because we are limited by unfaithful qubit state mapping; it does, however, increase the signal-to-noise ratio.  We can see this increase by plotting the measurement histograms as a function of integration time.  After only $20\ns$ the distributions are virtually identical, but they split into our bimodal distribution as we integrate longer.  By $120\ns$, the fidelity has almost reached its final value, but the distributions remain close to one another.  Integrating for $300\ns$ or $500\ns$ does not substantially change the measurement fidelity but does increase the open space between the peaks.  This has one practical benefit.  If we were to use a fast digitizer to ``threshold'' the data at our optimal fidelity point, turning distributions of measurements into ones and zeros, having extra open space around the threshold value makes that process more robust.  Drifts in the gain or offsets of the measurement chain will not tend to change the thresholded measurement outcome for long integration times, while they might for smaller SNRs of the same nominal fidelity.

\subsubsection{Multi-qubit measurement}

We can extend this readout scheme to simultaneously measure multiple qubits.  The physics of multi-qubit readout is similar to the single-qubit case, but generalized for multiple basis states.  Each qubit has its own dispersive shift $\chi_i$, and since for e.g. three qubits there are eight separate basis states, there will be eight zero-power cavity frequencies.  For example, the state $|011\rangle$, where the second and third qubits are excited, will shift the cavity by $2\chi_2 + 2\chi_3$.  Since each basis state induces a different net dispersive cavity shift, there will be a hierarchy of drive strengths necessary to drive each cavity to the bright state.  Readout is done by choosing a measurement drive power above the onset power for all except the state or states one is trying to distinguish.  That way, the cavity should go bright for the states with critical powers below the selected drive power, and stay dim otherwise, again giving us a single classical bit of information.  This scheme is especially convenient for performing qubit state tomography \cite{Steffen2006}, which requires inferring multi-qubit correlations \cite{DiCarlo2009, Filipp2009, Chow2010b}.  As discussed in \sref{sec:statetomo}, the effective measurement operator for this scheme is $\ket{0^{\otimes N}}\bra{0^{\otimes N}}$ and contains all $Z$ correlations.  As we will see, this fact has been exploited to efficiently detect tripartite entanglement \cite{DiCarlo2010}.

\begin{figure}
	\centering
	\includegraphics{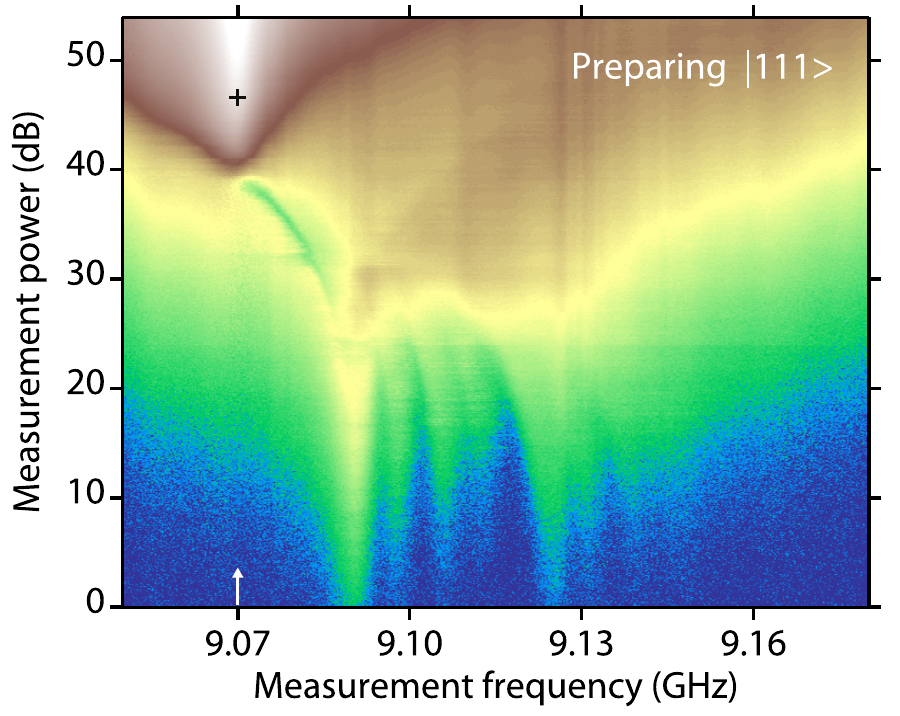}
	\mycaption{Pulsed cavity response $\overline{A_{\mathrm{H}}}$ for $\ket{111}$ state}{The eight ($2^3$) register states of three qubits induce a different dispersive cavity shift, each discernible at low power due to decay of the $\ket{111}$ state during measurement.  These frequencies were independently measured to be $( f_{\ket{000}}, f_{\ket{100}}, f_{\ket{010}}, f_{\ket{001}}, f_{\ket{110}}, f_{\ket{101}}, f_{\ket{011}}, f_{\ket{111}} ) = ( 9.145, 9.139, 9.131, 9.112, 9.124, 9.105, 9.097, 9.090 )~\mathrm{GHz}$, with $f_{\mathrm{bare}}=9.070\ghz$.  The most prominent secondary cavity position corresponds to the third qubit (at $8\ghz$) relaxing during measurement, consistent with its Purcell-limited \cite{Houck2008} $T_{\mathrm{1}}$ being the shortest of the three, with $T_{\mathrm{1}}^{\mathrm{Q}_{\mathrm{1}}}= 1.2\us$, $T_{\mathrm{1}}^{\mathrm{Q}_{\mathrm{2}}}= 1.0\us$, and $T_{\mathrm{1}}^{\mathrm{Q}_{\mathrm{3}}}= 0.6\us$.  The system excites to its bright state at lower power here than seen in \figref{fig:transmissioncolor} because of the smaller initial cavity detuning from $f_{\mathrm{bare}}$.  The color scale is identical to that in \figref{fig:transmissioncolor}.
	\figadapt{Reed2010b}
	}
	{\label{fig:transmissioncolor3q}}
\end{figure}

The device that we first demonstrated this effect with had a total of four qubits.  Three were tuned to $6$, $7$, and $8\ghz$, with the fourth at high frequency and unused (\sref{sec:fourqubitdevice}).  In \figref{fig:transmissioncolor3q}, we show transmission as a function of frequency and power when the $|111\rangle$ state (where all three qubits are in their first excited state) is prepared.  At low drive strengths, eight separate cavity peaks are visible because $|111\rangle$ has the possibility of decaying into each of the other seven basis states.  The cavity reaches its bright state at a substantially lower power -- about $41\dB$ -- than any other basis state, because it starts closest to the bare cavity frequency. 

\begin{figure}
	\centering
	\includegraphics[scale=1]{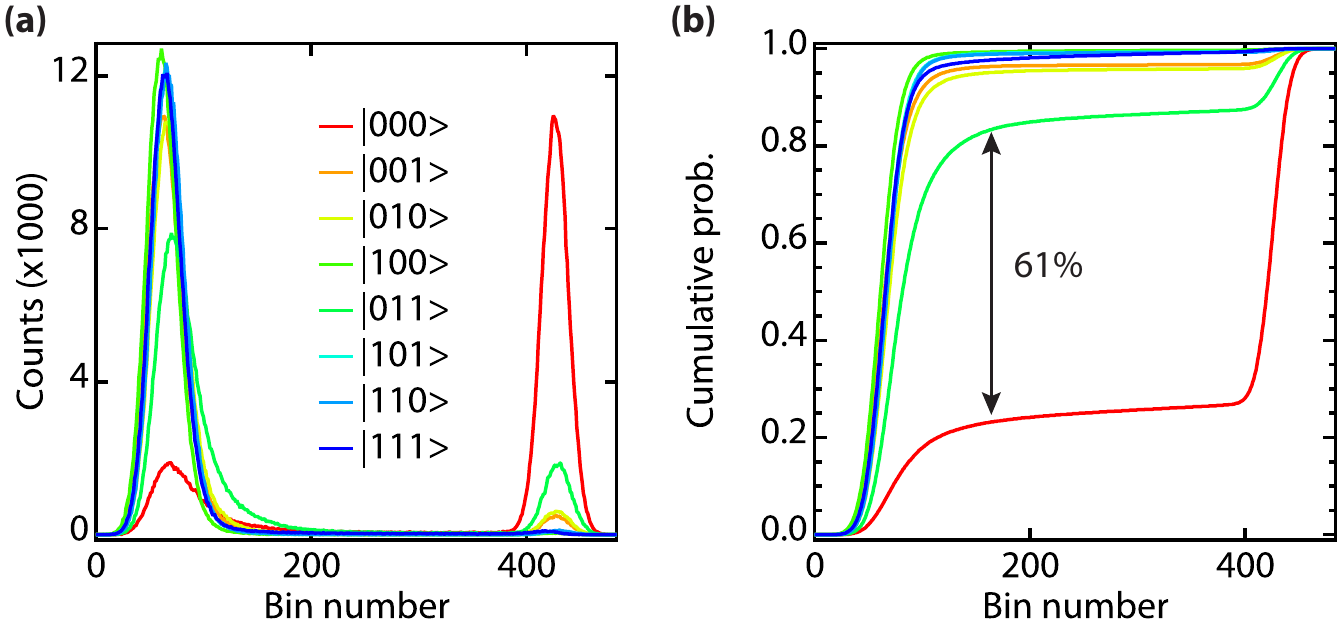}
	\mycaption{Joint qubit readout fidelity measurement}{\capl{(a)} Histograms for all eight basis states when driving with +49~dB at $f_{\mathrm{bare}}$.  Here we have tuned the drive strength to optimize the distinguishability of $|100\rangle$ from $|000\rangle$, which requires a large drive power and causes the cavity prepared with $|000\rangle$ to go bright about 25\% of the time.  We integrate long enough for the SNR to be large, separating the two peaks by a substantial margin.
		\capl{(b)} S-curves for joint qubit readout, indicating a minimum of 61\% fidelity for distinguishing the $\ket{000}$ from the least differentiated state.  If we add a $\pi$ pulse on the 1-2 transition of the $6\ghz$ qubit to increase its dispersive shift, the three-qubit fidelity can be increased to 80\%.
		\figadapt{Reed2010b}
		}
	{\label{fig:fidelity3q}}
\end{figure}

We typically want to distinguish the ground state $|000\rangle$ from all other basis states.  There is an ambiguity about how to optimally accomplish that task, however, because there are seven states against which we are potentially discriminating.  For example, we might choose our readout power to give the highest average fidelity over all states or perhaps the fewest spurious $|000\rangle$ bright-state counts; the exact tune-up might depend on precisely what you are interested in measuring.  Here we chose to optimize the discrimination of the least distinguishable state, $|100\rangle$ (where only the $6\ghz$ qubit is excited), and found that a drive strength $P_{\mathrm{joint}} \approx 49\dB$ was best for that purpose.  Histograms and S-curves for each of the seven pairs (discriminating each state from $|000\rangle$) are shown in \figref{fig:fidelity3q}.  Defining measurement fidelity is again ambiguous because this process gives us seven distinct numbers.  We report the most conservative metric, the single-state fidelity to the least distinguishable state (again, $|100\rangle$), and find a three-qubit measurement fidelity of 61\%.  That number could be substantially improved by either reducing detuning of that qubit or by performing a $\pi$ pulse on the 1-2 transition of that qubit prior to measurement to increase its effective dispersive shift.  When that pulse is included in the measurement process, the joint fidelity here is increased to about 80\%.  Note that optimizing against this state necessarily reduces the fidelity to the others.  That optimization leads us to drive rather strongly, giving a $\sim25\%$ chance of the cavity exciting for the ground state; all other fidelities are reduced by this factor.

\subsection{Theory of the Jaynes-Cummings nonlinearity}
\label{subsec:highpowertheory}

Now that we have presented the experimental observations of the high-power cavity response and its application to measurement, let us return to the discussion of the underlying physics.  This section will largely summarize the result of \cite{Bishop2010}, which qualitatively predicts the main features of the Jaynes-Cummings readout from first principles.  Bishop {\it et al.} consider the Jaynes-Cummings Hamiltonian of a two-level qubit coupled to a cavity under the influence of a drive:
	\begin{equation}\label{eq:transformedH}\hat{H} = \omega_r \hat{a}^\dagger \hat{a} + \frac{\omega_q}{2} \sigma_z + g \left(\hat{a} \sigma_+ + \hat{a}^\dagger \sigma_- \right) + \frac{\xi(t)}{\sqrt{2}}\left(\hat{a} + \hat{a}^\dagger\right)\end{equation} 
where $\omega_r/2\pi$ is the cavity frequency, $\omega_q/2\pi$ the qubit frequency, $g$ the coupling strength, and $\xi(t) = \xi \mathrm{cos}(\omega_d t)$ is the drive of amplitude $\xi$ and frequency $\omega_d$.  They then make the ``bad-cavity'' and strong-dispersive approximations, where the cavity lifetime is short compared to the qubit decay and dephasing rates ($\kappa > \gamma, \gamma_\phi$) and the qubit-cavity dispersive shift $\chi$ is greater than $\kappa$, respectively.  Dropping small terms, this yields the effective Hamiltonian
\begin{equation}
	\tilde{\hat{H}} = \omega_r \hat{a}^\dagger \hat{a} + \left( \omega_r - \Delta \right) \frac{\sigma_z}{2} + \frac{\xi}{\sqrt{2}}\left( \hat{a} + \hat{a}^\dagger \right) \mathrm{cos}(\omega_d t).
\end{equation}

This equation could be (and has been) integrated numerically with a quantum master equation \cite{Bishop2010}.  That is challenging here because there will be a large number of photons for relevant drive strengths and so the Hilbert space is huge.   Instead, if the cavity anharmonicity is smaller than its linewidth (e.g. the $N-1 \leftrightarrow N$ and $N \leftrightarrow N+1$ transition peaks overlap, or $N \gg g^4/\kappa \delta^3$), a semiclassical model will be a good approximation.  This is especially attractive given that the typical approach of perturbatively expanding \equref{eq:transformedH} in powers of $N/N_{\mathrm{crit}}$ will not converge when $N > N_{\mathrm{crit}}$.  There, the critical photon number $N_{\mathrm{crit}}= \delta^2/4g^2$ and $\delta$ is the cavity-qubit detuning.  For the parameters of this device, $N_{\mathrm{crit}} \sim 6$, so this is a very stringent limit.

Bishop rewrites \equref{eq:transformedH} using canonical variables $X = \sqrt{1/2}\left( \hat{a}^\dagger + \hat{a}\right)$ and $P=i \sqrt{1/2}\left(\hat{a}^\dagger - \hat{a}\right)$ to give
	\begin{equation}
		\tilde{H} = \frac{\omega_r}{2} \left(X^2 + P^2 + \sigma_z \right) + \xi X \mathrm{cos}(\omega_d t) - \frac{\sigma_z}{2} \sqrt{2g^2\left( X^2 + P^2 + \sigma_z \right) + \delta^2}.
	\end{equation}
Treating $X$ and $P$ as numbers (and not operators) in the semiclassical approximation, and incorporating cavity decay with an additional damping term, we solve for the steady state amplitude $A=\sqrt{X^2 + P^2}$ as
	\begin{equation}\label{eq:amplitude} 
		A^2 = \frac{\omega_r^2 \xi^2}{\{\omega_d^2 - [\omega_r - \chi(A)]^2\}^2 + \kappa^2 \omega_d^2}
	\end{equation}
with the cavity dispersive shift $\chi(A)$ depending on cavity amplitude as
	\begin{equation}
		\chi(A) = \sigma_z \frac{g^2}{\sqrt{2g^2\left(A^2+\sigma_z\right) + \delta^2}}.
	\end{equation}
This equation satisfies the normal approximation for the dispersive shift for a Cooper-pair box (e.g. two-level qubit) with $\chi(0) \approx \pm g^2/\delta$.  Crucially, under strong driving it also saturates with $\mathrm{lim}_{A\rightarrow \infty}\chi(A) = 0$.

With this result, we can start predicting the behavior of the cavity as a function of drive strength.  We expect the anharmonicity of the cavity (e.g. the difference in $\chi$ for $N$ and $N+1$ excitations) to be maximal at low power and continuously diminish as the drive strength is increased.  At the zero-power limit, we will see a linear response because we are only populating the 0 and 1 photon states and never sample the nonlinearity.  As we increase the drive strength, however, the anharmonicity will turn on and distort our line shape, consistent with \figref{fig:transmission1}.

At slightly higher drive strengths, based on our experience with Kerr nonlinearities, we expect the frequency response to bifurcate into two solutions.  The phases of the two solutions, one ``dim'' and one ``bright,'' will be almost opposite since they are effectively on alternate sides of the resonance.  This bifurcation is observed in the numerical simulations of \equref{eq:amplitude} done by Bishop, {\it et al}.  As the drive strength is turned up, rather than the bistable region growing forever with power as with a Kerr oscillator, the region shrinks and eventually vanishes, emphasizing that this cavity's nonlinearity is not constant.  This point corresponds exactly with the bare cavity frequency and critical power.  As a consequence, when driving at the bare cavity frequency, the system is never bistable (at least in this model's approximation).  Moreover, this bifurcation is responsible for the appearance of the dip in \figref{fig:transmissioncolor} that traces up to the high-power peak.  We expect destructive interference at the point where the populations of the two solutions are similar, weighted by the switching rates \cite{Drummond1980}.  This occurs where our heterodyne detection coherently averages over the ensemble of experiments.

Finally, for increasing drive strength at the bare cavity frequency, the cavity population increases and anharmonicity shrinks.  At some critical power, where $\chi(N) \lesssim \kappa$, we drive hard enough for the cavity to shift into resonance with the drive.  At this point, its population (and therefore transmission) rapidly grow while its anharmonicity shrinks -- thus reaching the {\it bright state}.  Note that this point (where we say $N=N_{\mathrm{bare}}$) necessarily happens at a much larger drive strength than $N_{\mathrm{crit}}$, so a perturbative expansion would fail to reveal this behavior.  It is unsurprising that this phenomenon was not theoretically predicted prior to its experimental discovery.  By looking at the numerical solution shown in the Bishop paper, we also see that the semiclassical model predicts the qualitative response as a function of drive power at the bare cavity frequency shown in \figref{fig:transmissioncolor}(a).  The success of this model at predicting essentially all of the behavior of the cavity is quite remarkable considering that it is a vast simplification relative to the actual system Hamiltonian.

One detail worth mentioning is that the model presented above does not predict that this would work as a readout mechanism.  The model assumes a two-level atom, implying that the dispersive cavity shift will be symmetric about the bare cavity frequency.  Without an asymmetry, the critical power for both cases would be essentially the same (broken only by qubit relaxation), eliminating our state-dependent transmission.  In order to have contrast, this symmetry must be broken.  For our system, this is accomplished by the fact that our qubits are actually weakly anharmonic transmon oscillators with higher excited states.  The repulsion of the states $|e, n+1\rangle$ and $|f, n\rangle$ (where $e$ is the first excited qubit state and $f$ the second) of the undressed Hamiltonian would, for example, be sufficient asymmetry.  Alternatively, if more than one two-level system were available, the extra shift from that qubit would give contrast to any single qubit, though not both jointly (since e.g. the $|00\rangle$ and $|11\rangle$ states would have the same shift).

\subsection{JCR summary}

Using the Jaynes-Cummings cavity nonlinearity to measure qubits has the significant advantage of delivering high fidelity without requiring any additional experimental hardware or change of sample design.  Essentially, we are using the qubit as its own amplifier.  Its fidelity is virtually independent of cavity Q as long as $\chi > \kappa$.  We have observed excellent measurement fidelities with cavity Q's ranging from a few hundred to several million.  (Note that this would {\it not} be true of a dispersive measurement, since its information collection rate is proportional to $\kappa$.)  The effect is seen in every cavity coupled to a transmon qubit.  As discussed in \sref{subsec:transmission}, it can even serve as a quick verification that qubits are present. We can measure the cavity at high power and then see if it either disappears abruptly or shifts in frequency as the power is lowered to detect the presence of a qubit, and estimate how strongly it is coupled or how far it is detuned using the difference in frequency between the high- and low-power cavity frequencies.

There are some practical aspects which are not fully understood.  Occasionally, the critical power seems to have a ``switchy'' behavior, seeming to randomly jump between two or more values.  The source of this tendency is unknown, but it is seemingly mitigated by either using slightly lower measurement power or by waiting for the system to ``calm down'' (it seems to happen more frequently shortly after the device is cooled).  When a qubit is positively detuned from the cavity, the readout is typically not as reliable because the cavity nonlinearity will have to change its sign.  The gain curve analogous to \figref{fig:transmission2}(d) can also exhibit a double-step behavior where the cavity seems to go bright in two distinct transitions when the qubit is far detuned or weakly coupled.  There, the understanding of an asymptotically decreasing anharmonicity is no longer true.  Though the physics are significantly more complex, it still seems to work as a reliable measurement mechanism.

Experimental repetition rates and qubit $T_1$s can also be affected by the high-power readout.  We did not initially notice any difference between the dispersive and high-power readout when using a relatively low-$Q$ cavity.  However, the 3D cQED architecture features photon lifetimes so long that the difference between waiting for a few dozen and a few thousand measurement photons to decay can be meaningful.  We have seen that if the repetition rate is not slow enough, the qubit $T_1$ and $T_2$ can be adversely affected.  The repetition rate required can also sometimes be much longer than one would expect from either the qubit or cavity decay time.  Since the 3D architecture generally operates in a much larger value of $\chi / \kappa$, the drive power needed to off-resonantly drive the cavity to its bright state can be extremely large and populate a huge number of photons if the cavity is allowed to come into equilibrium.  We therefore speculate that we might be driving the qubit junction to its normal state, and so are required to wait for quasiparticles to recombine.  One way of mitigating this effect would be to chirp the frequency of the drive tone from the low-power cavity down to the bare cavity frequency while its amplitude is increased \cite{PrivateCommunicationLuyanSun}.  This would lower the effective critical power as well as the total energy needed for measurement.

This readout mechanism is only appropriate for the end of an experiment because it is not QND to the qubit state.  In order to drive the cavity to its bright state, it is necessary to dump so much power into the cavity that the qubit state seems to be scrambled and likely left in a highly excited state.  This is likely due to some combination of dressed dephasing, qubit dressing \cite{Boissonneault2010}, or direct qubit transitions.  For experiments in which measurements must be QND, the Jaynes-Cummings readout is not an option.

\section{Conclusion}

Qubit measurement in cQED is accomplished by taking advantage of the dispersive shift of the cavity frequency.  Applying a displacement to the cavity therefore entangles the cavity together with the qubit state.  This displacement leaks out and is absorbed by our amplifier chain, projecting the qubit superposition along the $z$-axis.  If we have a high enough signal to noise ratio, we can detect this transmitted light to infer the state of the qubit.  With a conventional amplifier chain, this transmission is combined with a large amount of amplifier noise which obscures our ability to distinguish the signal.  The signal power is also limited by the cavity lifetime, which cannot be made too short because the Purcell effect will reduce qubit lifetime.  One solution to this is to use a so-called Purcell filter, which breaks the cavity-qubit lifetime relationship.  This apparatus has the ancillary benefit of increasing the dynamic range in qubit lifetime, which can be used for efficient qubit reset.

For conventional dispersive readout, the equilibrium number of cavity photons is limited by cavity anharmonicity.  Beyond a certain drive strength, the cavity's inherited nonlinearity will corrupt its response, reducing measurement fidelity as power is increased.  The nature of this nonlinearity is quite special since it turns off at extremely high drive strengths, giving the cavity a linear response at its bare cavity frequency.  Because the drive strength required to drive the cavity to this high-transmission state depends on the initial cavity detuning and thus on the qubit state, the effect can be used to readout the qubit state.  This mechanism, known as the {\it Jaynes-Cummings} or simply {\it high-power readout}\footnotemark, is extremely robust to device parameters.  Moreover, because the cavity is so highly excited and emits a large signal, its fidelity is not limited by amplifier chain noise and can be quite high.

\footnotetext{Calling it the ``Reedout'' has also been known to happen, as a play on words with my last name.}

The question of which readout mechanism is the ``best'' arises.  The choice often comes down to experimental requirements.  If a QND measurement is not required and fidelities of $\sim 90\%$ are acceptable, the high-power JC readout is extremely attractive.  It does not require any special hardware and will deliver relatively high fidelity across a wide range in qubit and cavity parameters.  Additionally, it always operates at the same cavity frequency independent of qubit detuning, and thus is easy to tune up.  Its main disadvantage is that it is not QND to the qubit state.  If you wish to make a qubit measurement and then manipulate the system as a result of that measurement (a function that will be extremely common in a ``real'' error-corrected quantum computer), the JCR will not work since the large number of cavity photons seem to scramble the qubit state.  There are other potential downsides to this scheme as well: the large amount of energy deposited in the cavity is suspected to heat up the system and create quasiparticles that may adversely affect qubit coherence.  The system also seems to require a long time to relax back to its ground state, reducing the experimental repetition rate that can be used.

The dispersive readout mechanism solves many of these issues, but introduces some of its own.  Since an extremely small amount of power is used, it can be QND to the qubit state and allow for repeated feed-forward measurements.  This small amount of energy decays quickly, which potentially speeds up experimental repetitions.  However, the small signal power is difficult to distinguish from conventional amplifier noise, requiring sophistications like the Purcell filter or ultra-low noise amplifiers to attain high measurement fidelity.  Fidelity is also a strong function of system parameters (cavity $Q$, qubit coupling and detuning, etc), which complicates sample design, especially for experiments that are not strictly about qubit measurement.  Nevertheless, it is clear that when implemented correctly, the dispersive mechanism can provide a well-understood, high-fidelity, and QND qubit measurement.  If a cQED quantum computer is ever built, it will certainly use some form of the dispersive readout mechanism.

\setcounter{chapter}{6}
\chapter{Tripartite Entanglement on Demand}
\thumb{Tripartite Entanglement on Demand}
\lofchap{Tripartite Entanglement on Demand}
\lotchap{Tripartite Entanglement on Demand}
\label{ch:entanglement}



\lettrine{W}{e} now turn our attention to generating and measuring entanglement in the cQED architecture.  This chapter serves two primary purposes.  First: to report our result of generating an entangled state of three qubits known as a Greenberger-Horne-Zeilinger or GHZ state.  As we saw in \sref{subsec:quantumrepetitioncode}, this state is of great interest because it is a prerequisite for demonstrating the simplest form of quantum error correction.  In showing how we generate a GHZ state, we will explain how to engineer and tune up entangling gates and introduce the distinction between adiabatic and sudden quantum trajectories.  The second purpose of this chapter is to introduce the idea of measuring quantum systems with tomography, which is useful for verifying the states and processes we claim to control.  Virtually everything discussed here will be revisited when we demonstrate quantum error correction in \chref{ch:qec}.  Many ideas presented here are also not specific to cQED, and can be applied to any quantum computing platform.

In the first section, we introduce the four-qubit device used to generate GHZ states and explain how the device is operated and calibrated.  Each qubit has its own flux bias line which is used to tune qubit frequencies in-situ.  We use them in two different limits: DC and fast-flux, respectively used to set the default frequency of each qubit and to tune their frequency rapidly to generate entanglement.  The flux lines must be calibrated for both tasks.  In the case of DC flux, there is substantial cross-coupling between each flux line and the neighboring qubits which can be measured and inverted to make virtual control knobs that address only one qubit.  For fast-flux, we need to apply a deconvolution kernel to compensate for ringing and finite bandwidth.  Using these linearized control knobs, we show pulsed spectroscopy of $Q_1$ moving up in frequency through avoided crossings with $Q_2$, $Q_3$, and the cavity.

Taking the qubit-qubit avoided crossings as an impetus, we then discuss how to engineer the controlled-phase gates used to generate entanglement between qubits.  We start by proposing a hypothetical ``J-swap'' gate that directly accesses the interaction zone between two qubits.  While this process would work to generate entanglement \cite{Quintana2012}, we find that it fails to be the desired gate in the case of $\ket{11}$ due to an interaction that was not revealed in single-tone pulsed spectroscopy.  Using two-tone spectroscopy technique, we measure the additional avoided crossing between the higher-excited $\ket{11}$ and $\ket{02}$ states, thus accounting for the failure of the J-swap.  We show how this crossing, rather than being an obstacle, can be used to produce two different kinds of controlled-phase entangling gates by moving into the crossing either adiabatically or suddenly.

Having established our ability to entangle qubits, we then describe how we measure and prove that the intended state was produced.  We reconstruct the density matrix of an ensemble of our states by repeatedly measuring certain correlations, a process known as {\it state tomography}.  The conventional method of obtaining these correlations involves measuring each of the three qubits individually and calculating the correlations with classical post-processing.  However, because we lack the ability to measure the qubits individually, we use an alternate method.  Our measurement operator is (approximately) $\hat{M}=\ket{0^{\otimes N}}\bra{0^{\otimes N}}$, which can be expressed as a sum of all combinations of $Z$ and $I$ Pauli operators.  By rotating the qubits immediately prior to measurement, we can effectively modify this measurement operator to contain all the Pauli correlations we seek.  Applying a full set of rotations, we produce sufficient linearly-independent measurement operators to infer every Pauli correlation and fully specify the experimental density matrix.  We demonstrate this capability with several simple test states.

We can extend this idea to perform tomography on a quantum {\it process}.  This technique tells us the full action of any quantum process by measuring state tomograms of a sufficient number of states to span the full Hilbert space.  These data can be converted to a $\chi$-matrix which maps input density matrices to output density matrices.  We demonstrate this technique by measuring the process matrix of a two-qubit controlled-phase gate.  The conventional $\chi$-matrix representation is difficult to interpret beyond its fidelity and is susceptible to systematic errors; we will discuss how these limitations have motivated researchers to suggest alternative formulations.

We complete the chapter by producing and measuring 2- and 3-qubit entanglement.  We introduce procedures that use the sudden cPhase gates introduced earlier to generate entangled Bell states of two qubits, and extend that idea to make a three-qubit entangled GHZ state.  We show that the so-called {\it Pauli bar representation} of a density matrix is especially useful for depicting entangled states.  This is in contrast that to the case of conventional ``cityscape'' plots, which are much more difficult to interpret when the state being shown is not easily written in the $z$ basis.  We conclude by evaluating various three-qubit entanglement witnesses and verifying that we have exceeded both biseparable and W-class bounds, unambiguously creating true three-qubit GHZ-class entanglement.

\section{Four-qubit cQED device}
\label{sec:fourqubitdevice}

\nomdref{Cq1q4}{$Q_1$-$Q_4$}{qubits in the four-qubit device}{sec:fourqubitdevice}

\begin{figure}
	\centering
	\includegraphics{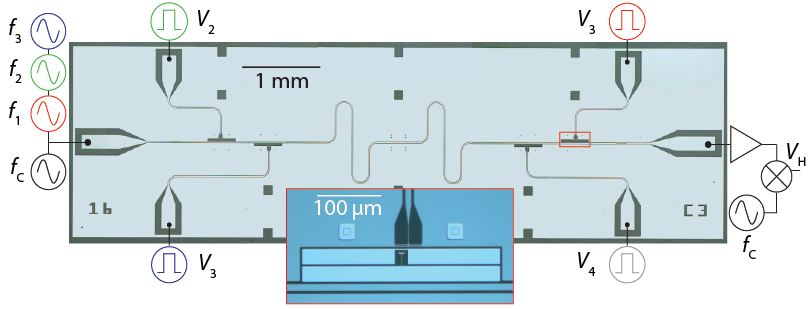}
	\mycaption{Optical micrograph of the four-qubit cQED device}
		{This device has a total of six ports, two for the input and output of the CPW cavity and four for the flux bias lines.  The bias lines allow us to tune the four qubit transition frequencies on nanosecond timescales with room-temperature voltages $V_i$.  We label each qubit $Q_1$ through $Q_4$.  In these experiments, we use the first three qubits and tune the fourth to its maximum frequency of $12.271\ghz$ to minimize its interactions with the others.  $8\ns$ microwave pulses resonant with the qubit transition frequencies $f_1$, $f_2$, and $f_3$ are used to drive single-qubit rotations around the $x$- and $y$-axis.  To measure the qubits, we pulse on a microwave tone at the bare cavity frequency $f_c=9.070\ghz$ and measure the resulting homodyne voltage $V_H$.
		\capl{(inset)} Optical micrograph of a typical transmon qubit in this device.  Note that the normal interdigitation between the two islands of the qubit was omitted in an attempt to reduce susceptibility to dielectric loss.  The termination of the FBL is also visible in this picture.
	\figthanks{DiCarlo2010}
	}
{\label{fig:fourqubitpic}}
\end{figure}

The device to which we will refer throughout this chapter is a cQED device with four qubits.  It employs the conventional planar cQED architecture to couple the qubits to a single CPW cavity, as shown in \figref{fig:fourqubitpic}.  The device serves as the context for the experimental capabilities we have as well as for the specific object that we used to produce the results reported in \chref{ch:entanglement} and \chref{ch:qec} (as well as large parts of \chref{ch:qubitmeasurement}).  We label the qubits $Q_1$ through $Q_4$, starting in the upper right corner and counting counterclockwise.  The transmission-line cavity has a bare frequency of $9.070\ghz$ and was employed for the high-power Jaynes Cummings readout mechanism \cite{Reed2010b} introduced in the previous chapter.

Each qubit is equipped with its own flux bias line, which is used to control the qubit frequencies in-situ by driving a current very close to a qubit's SQUID loop (\sref{sec:fluxbiaslines}).  We can send DC currents through these lines to set the ``default'' frequency of each qubit.  For the experiments reported here, we set the qubit frequencies to $(f_1,f_2,f_3) = (6.000,7.000,8.000)\ghz \pm 2\mhz$, with the fourth biased at its maximum frequency of $12.271\ghz$ and unused.  As we will see, these flux lines can also move the qubits very quickly with an RF pulse.  If we suddenly apply a voltage pulse to the line, the associated qubit will move to a new frequency within a few nanoseconds.  This timescale is fast when compared to the coherence time of the qubits, so we use this {\it fast-flux control} to manipulate the qubits during a given experiment.  As we will see starting in \sref{sec:fluxgates}, this manipulation can deterministically entangle qubits.

\subsection{Calibrating flux lines and spectroscopy}
\label{subsec:fluxlinecal}

\nomdref{Adac}{DAC}{digital to analog converter}{subsec:fluxlinecal}

\begin{figure}
	\centering
	\includegraphics{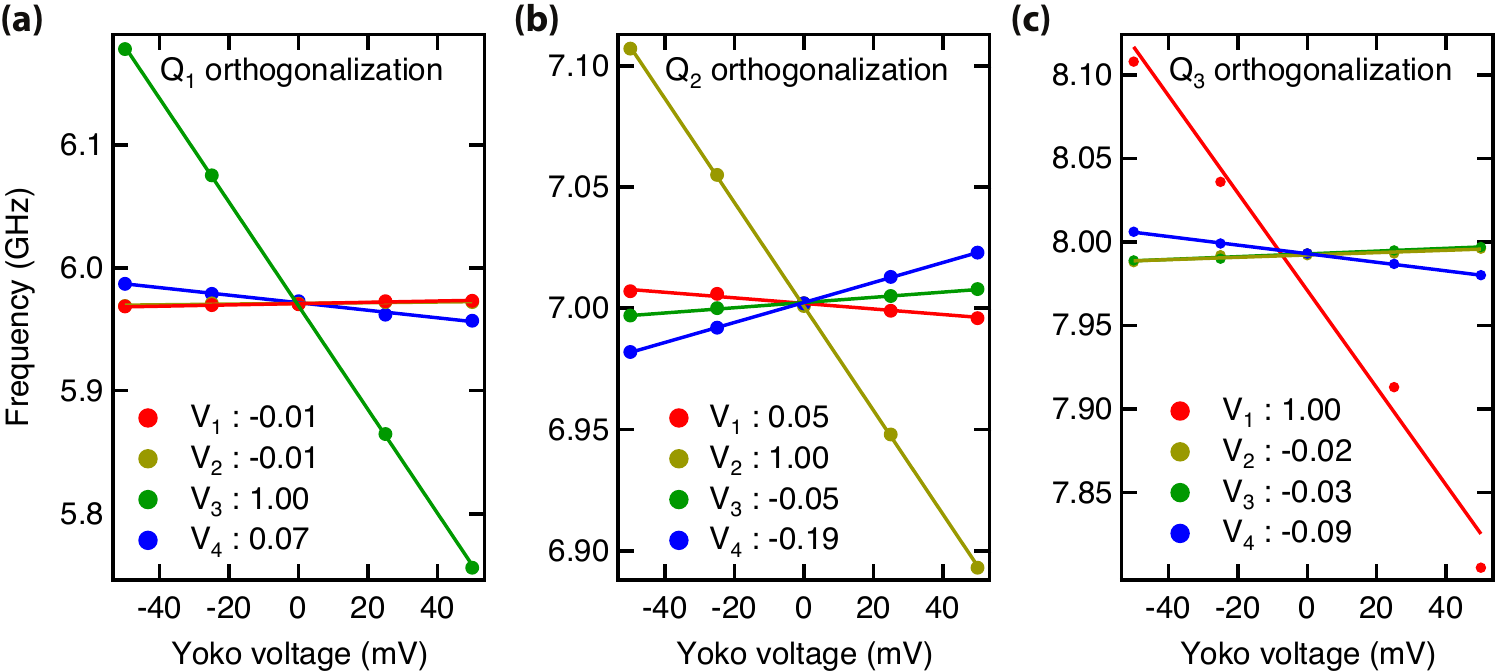}
	\mycaption{Flux bias line orthogonalization}
		{For each qubit, we measure its frequency as a function of the applied voltage on each flux bias line.  In each case, the strongest response is given by the flux line closest to that qubit, though we also observe that voltages applied to neighboring flux lines cause a frequency shift due to cross-coupling.  By measuring the slopes of these lines, we can orthogonalize the control scheme to produce ``virtual'' voltage sources that move one qubit at a time.  The fourth qubit was also measured, but is not shown.  We list the slopes of each line, normalized to the steepest response.
	}
	{\label{fig:fluxorth}}
\end{figure}

The flux bias lines require some calibration to work as desired.  Because the return path of the current on each line is not explicitly controlled, there is a large amount of DC cross talk between each line.  That is, changing the bias on any single qubit's flux line actually changes all of the qubit frequencies.  Fortunately, this effect can be measured and compensated for.  As shown in \figref{fig:fluxorth}, we measure the frequency of each qubit as a function of the voltage applied to each FBL.  While the bias line that is nominally intended to control the qubit has the most substantial effect, each of the other lines also change the qubit frequency.  For small voltages, the frequency dependence is approximately linear (e.g. we do not need to include the curvature of the qubit's flux dependence), so the ratio of the slope of each line accurately represents the relative magnitude of each coupling.  We insert each slope, combined with the analogous data for the fourth qubit (not shown here), into a matrix which we invert.  By multiplying our desired excursion by this matrix, we control ``virtual'' voltage sources that are orthogonalized to move only the desired qubit.  Properly done, this process can reduce cross-talk from 40\% to less than 1\%.

\begin{figure}
	\centering
	\includegraphics{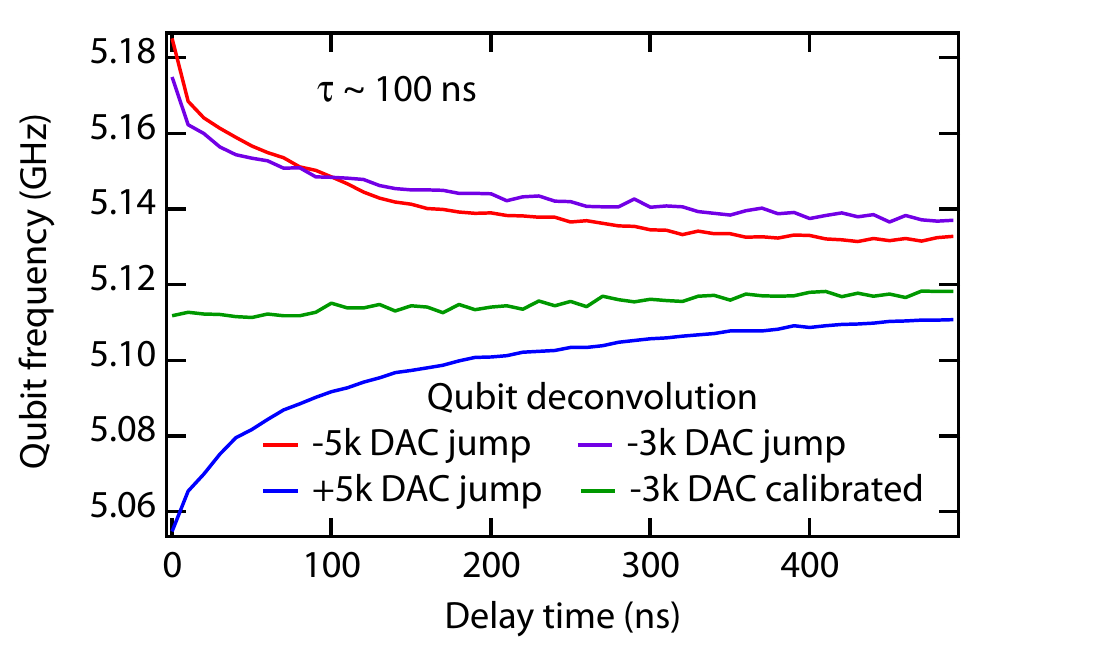}
	\mycaption{Fast-flux calibration}
		{When a square flux pulse is applied, the qubit will quickly move most of the way, but will slowly relax in $\sim100\ns$ to its final frequency.  We show this effect by applying a flux pulse to the qubit and then performing pulsed spectroscopy with gaussian pulses on the qubit as a function of frequency.  We then extract the peak frequencies from that data, and plot them as a function of time after and the size of the applied flux pulse (in units of DAC value).  We can apply a deconvolution kernel to the square flux pulse which shortens this effective response time, shown in green.
	}
	{\label{fig:fastfluxcal}}
\end{figure}

The second calibration necessary for our flux lines concerns fast time scales.  While we experimentally observe that the orthogonalization described in the previous chapter is not necessary for fast-flux pulses (e.g. less than $\sim1\us$), we do need to compensate for finite rise time and ringing.  Without any compensation, a square flux pulse can move a qubit several gigahertz in $2-4\ns$, but will take $10-100\ns$ to settle down to its final frequency, $\sim 10\mhz$ away,  as shown in \figref{fig:fastfluxcal}.  The waveform generator we typically use to generate these pulses, the Tektronix AWG 5014, also has $\sim 1\%$ transients for $\sim 1\us$ both before and after the flux pulse (e.g. it is acausal due to some internal signal processing).  Fortunately, again because of the linearity of the FBL, an integral transform can deconvolve away both of these effects.  The procedure for generating and using a deconvolution kernel, which describes the system's response to a delta-function impulse, is described in section 4.3 of Blake Johnson's thesis \cite{JohnsonThesis}.  In practice, we can settle to within $1 \mhz$ of the final frequency within $5-10\ns$ of the start of the flux pulse.

\begin{figure}
	\centering
	\includegraphics{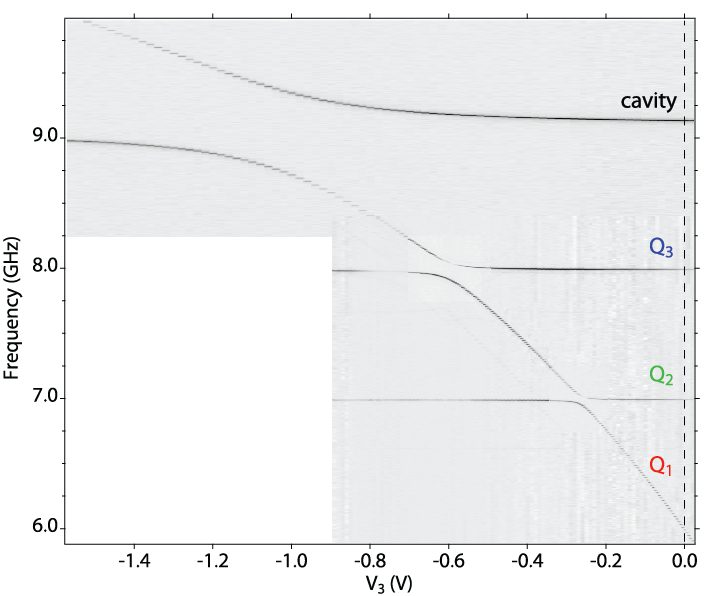}
	\mycaption{Spectroscopic characterization of the four-qubit device}
		{We perform pulsed spectroscopy on each qubit as a function of the orthogonalized flux bias on $Q_1$.  We find avoided crossings between $Q_1$ and $Q_2$, $Q_3$, and the cavity.  The size of these splittings indicates the strength of the coupling between the qubits and the cavity.  Above $8.2 \ghz$, we measured transmission instead of pulsed spectroscopy because the qubit is sufficiently hybridized with the cavity to produce a response.  The $x$-axis is an orthogonalized ``virtual'' voltage, with an offset which sets the qubits to their home positions.  The effectiveness of this calibration is reflected in the fact that $Q_2$ and $Q_3$ do not change frequency during this sweep.
	\figadapt{DiCarlo2010}
	}
	{\label{fig:fourqubitspec}}
\end{figure}

With the flux lines properly orthogonalized, we can demonstrate flux tuning of the qubits.  As shown in \figref{fig:fourqubitspec}, our qubits are at their home positions of $6$, $7$, and $8\ghz$ at $V_3=0$.  As we increase $V_3$, we move $Q_1$ up in frequency.  (The labels of the voltages and the qubits they control are not the same; one index is set by the qubit frequencies while the other is set by the geometry of the chip.  The voltage $V_3$ is an orthogonalized virtual voltage, with an offset which sets the qubits to their home positions at $V_i=0$ for $i=1..4$.  The extent to which $Q_2$ and $Q_3$ do not change their frequencies as we sweep $V_3$ demonstrates the efficacy of this calibration.)  As $Q_1$ tunes through $Q_2$ and $Q_3$, we see large J-type avoided crossings which are mediated by virtual interactions through the cavity (\sref{subsec:qubitqubitcoupling}).  When the qubits are in resonance, they hybridize into ``light'' and ``dark'' superposition states, named because the symmetries of (in this case) the upper state cause the coupling matrix element to cancel, turning off its dispersive shift and spectroscopic response \cite{Majer2007}.  At larger biases, the qubit comes into resonance with the cavity, demonstrating a vacuum-Rabi splitting (\sref{subsec:qubitcavitycoupling}).  (Data above $8.2\ghz$ are cavity transmission measurements rather than pulsed spectroscopy.)  The minimum separation of these peaks defines $2g \approx 620\mhz$.  To the resolution of all spectroscopy, there are no spurious avoided crossings which, as we will see, is a critical requirement for the pulsed qubit excursions used for multi-qubit flux gates.

\section{Two-qubit phase gates using fast-flux}
\label{sec:fluxgates}

With our qubits well characterized and controlled, we are now interested in generating and studying entanglement between them.  As discussed in \sref{subsec:multiplegates}, there are two classes of two-qubit gates: controlled-NOT and controlled-phase gates.  When the control qubit is in its excited state, the target qubit is flipped around the $x$-axis (e.g. a bit-flip) by a cNOT gate and is flipped around the $z$-axis (e.g. a phase-flip) by a cPhase gate.  The two gates are related to one another with single-qubit pulses -- for example, a cPhase is turned into a cNOT by bookending the target qubit with $\pi/2$ or Hadamard pulses.  The key concept is that they are {\it controlled}: something happens if and only if the qubits are in a particular state.

There are numerous ways of producing controlled operations in cQED, each with their own strengths and weaknesses.  The primary concern is the ultimate fidelity with which the operation can be performed\footnotemark.  In the case of the four-qubit device, our qubit lifetimes were only on the order of $1\us$, so fidelity of any operation is well approximated by the $T_1$ decay during its execution.  We are therefore interested in using interactions that are as fast (e.g. strong) as possible.  We aim to use FBLs to tune interactions terms of the system Hamiltonian on and off by moving the qubits together and apart from one another.  We will also only consider cPhase gates, which have fewer parameters to calibrate and more naturally generated in our architecture.

\footnotetext{Other concerns include how easy it is to calibrate the gate, whether it can be trivially repeated (e.g. the system has no memory of its application, outside of the qubit states), whether the gate requires extra hardware or engineering either at room temperature or in the device itself, how precisely you need to hit device parameters, and so on.  In this device, the coherence times were so short that essentially the only concern was gate time, but recently our qubits have become so coherent that we are afforded the luxury of optimizing for other considerations \cite{Paik2013}.  Gate fidelities also depend on parameters other than gate duration, though this is often only relevant when qubits are very coherent; this is explored in the cited paper.}

This section will explain two methods of generating controlled-phase gates and entanglement using fast-flux lines.  They exploit an avoided level crossing of non-computational states which is accessible only when both qubits are excited, thus providing the conditional nature of the gate.  The first method approaches this crossing adiabatically, which has the advantage of being less sensitive to parameters and to compensation of the flux bias line response function, but is relatively slow.  The second approaches the crossing suddenly, which saves time (and commensurately increases fidelity), but is more technically challenging.  Both of these techniques will ultimately be used in the next chapter to generate a three-qubit gate which we use to demonstrate basic quantum error correction.

\subsection{Adiabatic controlled-phase gate}
\label{subsec:adiabaticcphase}

To generate entanglement, broadly speaking, we require interactions between qubits\footnotemark.  Based on the spectroscopy of \figref{fig:fourqubitspec}, one obvious place where we would expect interactions is when they are near resonance with one another other, for example with $Q_1$ and $Q_2$ at $V_3=0.25~\mathrm{V}$.  Indeed, this is a viable way of preparing an entangled state.  Imagine that $Q_2$ is initialized in its excited state and $Q_1$ in its ground state (or vice versa) and we quickly move $Q_1$ into resonance with $Q_2$ with a fast-flux pulse.  If we are {\it sudden}, the state of our qubits will not change, but the eigenstates of the Hamiltonian will.  That is, though we begin in an energy eigenstate, by changing the Hamiltonian quickly with our fast flux bias, our wavefunction will no longer be an eigenstate, but rather a mixture of the newly hybridized eigenbasis.  Thus, as a function of time, the excitation that started in $Q_2$ will oscillate between qubits at the rate of the splitting between them.  Ideally, this interaction would produce a J-swap gate, with a unitary
\[ \left( \begin{array}{cccc}
1 & 0 & 0 & 0 \\
0 & \mathrm{cos}(Jt) & i\mathrm{sin}(Jt) & 0 \\
0 & i\mathrm{sin}(Jt) & \mathrm{cos}(Jt) & 0 \\
0 & 0 & 0 & 1 \end{array} \right)\]
where $J$ is again the interaction strength between the two qubits.  If we were to wait for half of the oscillation time $t=\pi/4J$, we would be left with the excitation half in $Q_1$ and half in $Q_2$, giving us a $\left(\ket{01} + \ket{10}\right)/\sqrt{2}$ maximally-entangled Bell state.  (Indeed, after this section was written, a paper was published that demonstrates the J-swap ``gate'' \cite{Quintana2012}, and it works exactly as we predicted.)  Unfortunately, as we will see, this interaction does not work if we start with both qubits simultaneously in their excited states (e.g. the $\ket{11}$ state); the lower right element of the J-swap unitary is {\it not} simply $1$.  This denigrates the operation from a true quantum gate (which must work for all inputs) to a method of preparing a particular Bell state.  It turns out that on the path of $Q_1$ toward $Q_2$, there is an extra interaction that only involves the $\ket{11}$ state, which causes this J-swap gate to fail because quantum amplitude leaks out of the computational Hilbert space.

\footnotetext{A measurement protocol could also be used, where, for example, the joint properties of the two qubits are interrogated in such a way that the state is projected onto the Bell basis \cite{Cabrillo1999, Kolli2009, Pfaff2013}.}

\begin{figure}
	\centering
	\includegraphics{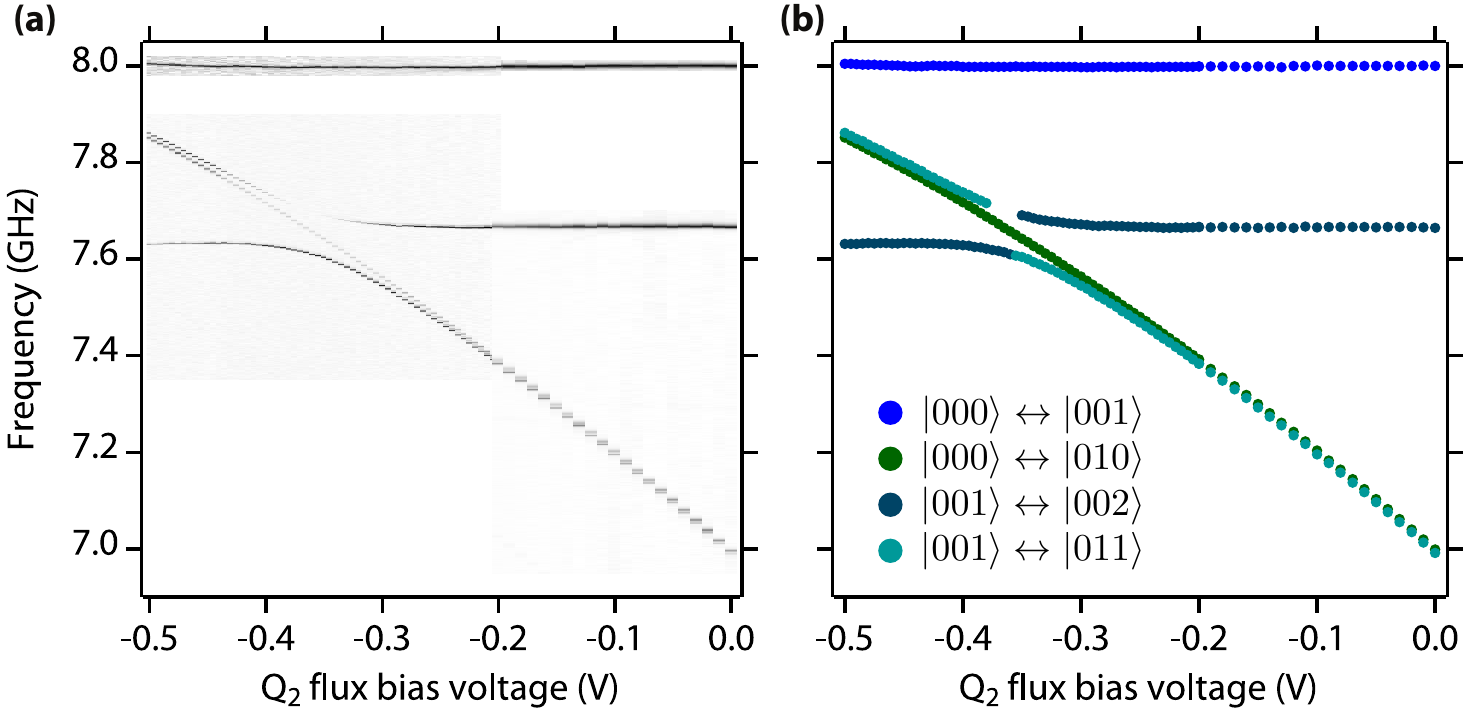}
	\mycaption{Two-tone spectroscopy of 011-020 avoided crossing}
	{\capl{(a)} For each vertical cut of these data, we first find the frequency of $Q_3$, which is near $8\ghz$.  Pulsing on a saturation tone $4\mhz$ detuned from this frequency, we then scan the frequency of a second pulsed tone, and measure the result.  Using this process, we drive transitions starting in either $\ket{000}$ or $\ket{001}$ because $\ket{001}$ is only virtually populated by our first, detuned tone.  As a result, we see both the normal transitions of $\ket{000} \leftrightarrow \ket{010}$ moving up in frequency as we tune the bias on $Q_2$ (analogous to \figref{fig:fourqubitspec}) and the two-excitation transitions of $\ket{001} \leftrightarrow \ket{002}$ and $\ket{001} \leftrightarrow \ket{011}$.  The states $\ket{011}$ and $\ket{002}$ have an avoided crossing which disrupts our J-swap gate but can be used to make a controlled-phase gate.  
	\capl{(b)} We locate each peak from the data in \capl{(a)} and label the transition our tones are driving.  Note that as we move through the avoided crossing between $\ket{011}$ and $\ket{002}$, the identities of the eigenstates swap.  This fact is reflected in their color.
	}
	{\label{fig:twotonespec}}
\end{figure}

We can investigate this extra interaction by performing two-tone spectroscopy.  Spectroscopy only detects transitions involving the prepared qubit state (typically the ground state).  In order to measure other transitions, we must prepare the qubit in a different state.  One method of doing so is to use {\it two-tone spectroscopy}, which, as the name suggests, involves applying two microwave tones simultaneously.  The process is iterative: at each given flux point, we first find the transition frequency of $Q_3$ (initially set at $8\ghz$, though it will change as the other qubit approaches) and turn on a saturation tone near that frequency to prepare an incoherent mixture of ground and excited states.  We then scan the frequency of a second tone and measure the resulting qubit populations, as with conventional pulsed spectroscopy.  In order to maintain readout contrast, the first tone is set to a frequency at a small fixed detuning (here, $-4\mhz$) from that qubit.  Transitions that involve populations of $Q_2$ will then be shifted by the opposite of that detuning in order to conserve energy -- that is, $Q_2$ will only be virtually populated until the energy deficit is paid by the second tone.  (If $Q_2$ were actually populated, we would not see these two-excitation transitions because the difference in measurement contrast between whichever final state and $\ket{010}$ would likely be small.  Our measurement operator is approximately $\ket{000}\bra{000}$, which does not discriminate between those states.)  As a result, we will also see transitions that start from the qubit in the ground state -- essentially, we are turning on sensitivity to extra transitions for free.

The result of this measurement is shown in \figref{fig:twotonespec}.  There, we move $Q_2$ up toward resonance with $Q_3$.  We have switched to examining the crossing between $Q_2$ and $Q_3$ instead of $Q_1$ and $Q_2$ because our data for the first case were cleaner; the physics is exactly analogous for either pair.  We observe the normal single-tone transitions of $\ket{000} \leftrightarrow \ket{010}$ (dark green) and $\ket{000} \leftrightarrow \ket{001}$ (blue), but also detect an extra avoided crossing at $7.62\ghz$ and $-0.33~V$.  For small voltages, we see a horizontal line $334\mhz$ below the $\ket{000} \leftrightarrow \ket{001}$ transition (dark blue).  This frequency is exactly the anharmonicity of $Q_3$ (modulo the $4\mhz$ offset), indicating that this transition is from the first excited state of $Q_3$ to the second excited state, $\ket{001} \leftrightarrow \ket{002}$.  We also observe an extra line that is initially parallel to $\ket{000} \leftrightarrow \ket{010}$ (teal), but peels off in a crossing with what we now understand to be $\ket{002}$.  This is the $\ket{001} \leftrightarrow \ket{011}$ transition, and so the avoided crossing is between $\ket{011}$ and $\ket{002}$.  We can now explain why our proposed J-swap gate will not work: because the qubits have negative anharmonicity, there will always be a large avoided crossing between $\ket{11x}$ and $\ket{02x}$ prior to the $\ket{01x} \leftrightarrow \ket{10x}$ swap.  (Recall that we estimated the size of this interaction in \sref{subsec:qubitqubitcoupling}.)  If we move through this transition with population in $\ket{11x}$, we cause quantum amplitude to leak into the non-computational state $\ket{02x}$.  As we will see below, the $\ket{11x}$ state will also evolve an additional phase difference during this process, further distorting it from the ideal J-swap.

\begin{figure}
	\centering
	\includegraphics{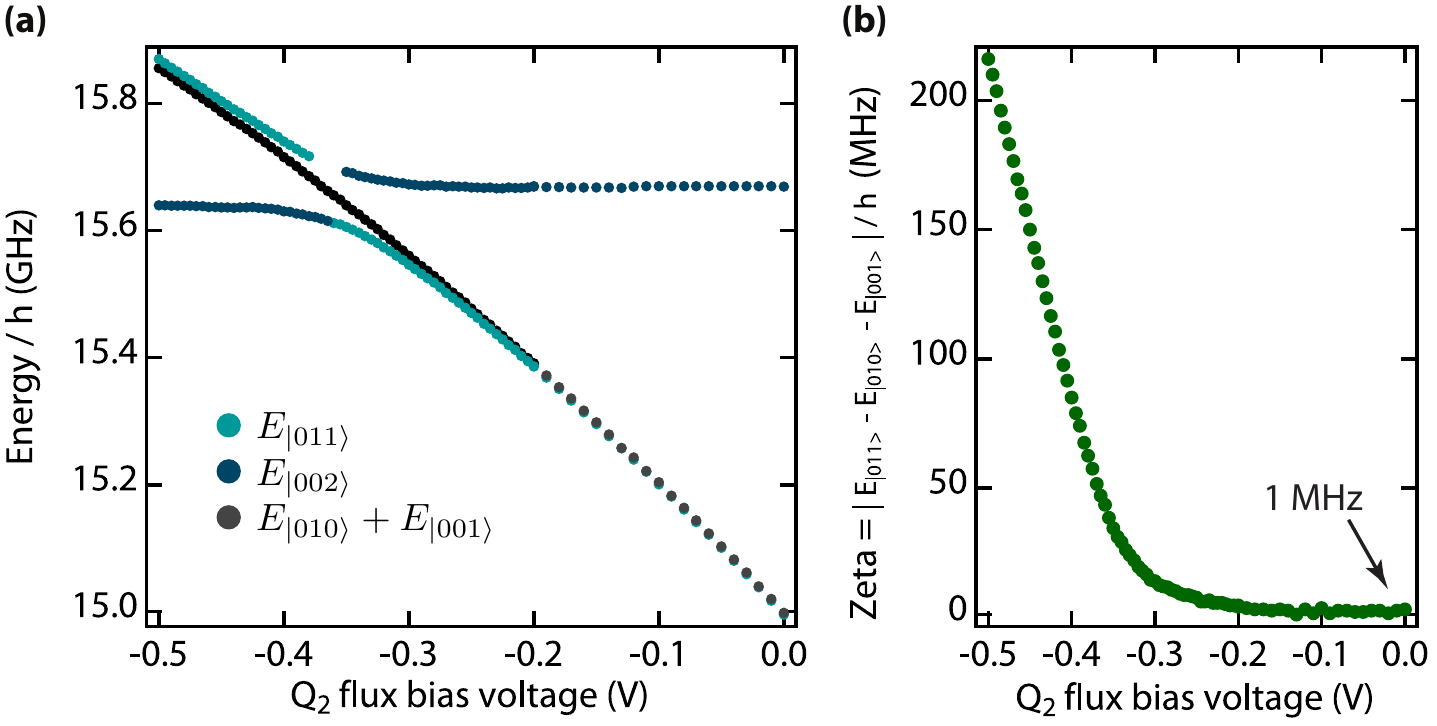}
	\mycaption{Conditional frequency shift of $\ket{011}$, $\zeta$}
	{\capl{(a)} Using the data from \figref{fig:twotonespec}(b), we calculate the energies of the $\ket{011}$ and $\ket{002}$ states and compare them to the sum of the energies of $\ket{010}$ and $\ket{001}$.  In the vicinity of the avoided crossing, there is a large difference between the energy of $\ket{011}$ and the sum of its constituents.  
	\capl{(b)} We define and plot $\zeta=f_{010} + f_{011} - f_{011}$, the difference between the energy of the $\ket{011}$ state and its constituents.  We consider the case of the crossing transversed adiabatically from the right, and so ignore the change in identity of $\ket{011}$ at $V=-0.35$.  As we approach this avoided crossing, we acquire conditional phase between $Q_2$ and $Q_3$ at a rate $\zeta$ because the avoided crossing is only accessible by the $\ket{011}$ or $\ket{111}$ computational states and is therefore conditional on both $Q_2$ and $Q_3$ being in their excited states.  At the home position of $Q_2$, $\zeta=1\mhz$, constituting an always-on $ZZ$ interaction between the qubits.
	\figadapt{DiCarlo2010}
	}
{\label{fig:zeta}}
\end{figure}

Far from being a nuisance as we have characterized it thus far, this extra avoided crossing is actually a valuable resource.  Consider the energies of the $\ket{011}$ state compared with the sum of its constituents, $\ket{010}$ and $\ket{001}$.  Initially, as shown in \figref{fig:zeta}(a), the two energies will be very similar because we are far from $\ket{002}$.  As we approach the avoided crossing, however, the energy of $\ket{011}$ will bend over and lag behind that of $\ket{010}$ and $\ket{001}$.  At this point, the phase evolution of $\ket{011}$ will be substantially different than its constituents, giving us a {\it conditional phase} -- evolution that only occurs when both qubits are excited.  We define the difference between the energies of the $\ket{011}$ state and its constituents to be a parameter $\zeta$, as shown in \figref{fig:zeta}(b).

To explain how we can use this interaction, it is helpful to first write out the unitary transformation of a general phase gate.  As the name implies, a phase gate does not change the excitation of the inputs, but does modify their phases.  Without loss of generality, we can write the action of any two-qubit phase gate as $\ket{00} \rightarrow \ket{00}$, $\ket{10} \rightarrow e^{i\phi_{10}} \ket{10}$, $\ket{01} \rightarrow e^{i\phi_{01}} \ket{01}$, and $\ket{11} \rightarrow e^{i(\phi_{10}+\phi_{01}+\phi_{11})} \ket{11}$ (note $\ket{11}$ acquires three different phases in this language).  The matrix describing this transformation is then the diagonal matrix:
\[ \left( \begin{array}{cccc}
1 & 0 & 0 & 0 \\
0 & e^{i\phi_{10}} & 0 & 0 \\
0 & 0 & e^{i\phi_{01}} & 0 \\
0 & 0 & 0 & e^{i(\phi_{10}+\phi_{01}+\phi_{11})} \end{array} \right).\]
This is a convenient way of parameterizing the phases of each state because it separates the origins of each phase.  The first two, $\phi_{10}$ and $\phi_{01}$, are known as {\it single-qubit phases}, and as we will see are due to the trivial phase evolution of any one qubit.  $\phi_{11}$, on the other hand, is a {\it two-qubit phase}, which is a phase evolution that occurs only when both qubits are exited and is associated with two-qubit interactions and entanglement.  This parametrization properly distributes phases; if we did a $z$-gate on a single qubit such that $\phi_{10}=\pi$, any state with the first qubit excited (that is, {\it both} $\ket{10}$ {\it and} $\ket{11}$) should have its phase flipped and we would have the unitary $\mathrm{diag}\{1, -1, 1, -1\}$\footnotemark.

\footnotetext{Note that our definition of single- and two-qubit phases is slightly different than the one found in Ref.~\citenum{DiCarlo2009}.}

Let us first consider how we would get a ``single-qubit'' phase.  The quantum amplitude of a state with energy $E=\hbar \omega_0$ evolves proportional to $e^{i \omega_0 t}$.  This evolution is normally eliminated because we work in the rotating frame of an RF generator tuned to the qubit transition frequency.  However, if the qubit were to be detuned from its reference frame with a flux pulse, the difference between the qubit's instantaneous energy and its reference would be responsible for a phase evolution.  As a function of time, the overall phase delay would be given by $\phi_{\mathrm{sq}} = \int_0^t{(\omega_0 - \omega(\tau))d\tau} = \int_0^t{\Delta(\tau) d\tau}$.  It is worth noting that to control a single-qubit phase, we do not actually need to change the associated qubit's frequency.  It is only defined relative to the rotating frame of its respective drive, so we need only advance or retard the phase of that oscillator to control it.  This is easily done in software by changing what defines the $x$- and $y$-axes for all pulses following a ``virtual'' $z$-gate, which comes down to changing the distribution of voltages on the I and Q ports of our mixer.  The fact that we are free to do this reveals the triviality of these single-qubit phases.  

In contrast, a two-qubit phase is a more subtle beast that can have real physical implications.  As you may have anticipated, one way of acquiring such a two-qubit phase involves the avoided crossing between $\ket{11}$ and $\ket{02}$.  As we have defined it above, the two-qubit phase is the difference in phase between $\ket{11}$ and its constituents.  Thus, we will acquire conditional phase at a rate proportional to the difference in energy between $\ket{11}$ and the sum of $\ket{10}$ and $\ket{01}$.  Having previously defined this quantity as $\zeta$, if we again tune $Q_2$'s frequency over some trajectory, the conditional phase acquired would be $\phi_{11} = \int_0^t{\zeta(\tau)d\tau}$.  Importantly, two-qubit phases are phase differences between {\it qubits} and cannot be generated in software\footnotemark.  Multi-qubit phase is very precious!

\footnotetext{We can make an analogy between qubit phases and potential energies.  Physically, the only thing that matters is the energy {\it difference} between a starting and ending state of some process.  We are free to renormalize an energy scale to whatever convenient zero we choose.  This is similar with phases: a single-qubit phase is like an energy that can be arbitrarily changed in our book-keeping, while a two-qubit phase is an energy difference that has physical implications.}

We have thus far neglected to emphasize that this mechanism relies on all of our flux pulses being {\it adiabatic} to the avoided crossing.  Adiabatic means that our qubit always remains in an instantaneous eigenstate of the time-dependent Hamiltonian.  This is only an issue when there are avoided level crossings, where eigenstates mix and change identity.  If we approach these splittings slowly enough, our wavefunction will evolve exactly with the changing eigenstate and stay on the same energy ``track.''  The timescale (or, energy change per unit of time) that we must be slow compared to is determined by the size of the splitting.  ``Slow'' depends on the details of the Hamiltonian, but in practice our approach must take a few times longer than the splitting period.  The larger the avoided crossing, the easier it is to be adiabatic to it.  The more closely you approach a crossing, the slower you must be to inhibit undesired tunneling.  Optimally acquiring two-qubit phase is a matter of finding the best trade-off between the magnitude of $\zeta$ and speed; the larger $\zeta$, the slower our trajectory must be.  The best solution is one in which you never stop moving the qubit, since any stationary time would be cut down by bringing the qubit closer to the crossing.

With our source of two-qubit phase established, we can now construct a cPhase gate.  The canonical conditional-phase gate is, as mentioned earlier, one which flips the phase of the qubit manifold if and only if both qubits are excited\footnotemark.  Thus, the unitary matrix should be $\mathrm{diag}\{1, 1, 1, -1\}$ and we can immediately see that we need $\phi_{10}=\phi_{01}=0$ and $\phi_{11}=\pi$.  To get the two-qubit phase, we apply a flux pulse that tunes the qubits close to one another (where $\zeta$ is large) while maintaining our adiabaticity to the avoided crossing.  We can fine-tune the time and amplitude of the pulse to satisfy $\phi_{11}=\int_0^t{\zeta(\tau) d\tau}=\pi$.  This conditional phase can be measured by applying a Ramsey sequence on one qubit while preparing the other in either its ground or excited state.  (This ``conditional Ramsey'' experiment is very useful whenever you are interested in measuring conditional phase evolution and will be discussed in more detail in \sref{subsec:toffolutuneup}.)  We satisfy our condition when the resulting oscillations are exactly $\pi$ out of phase from one another.  During the flux pulse, we will also get large single-qubit phases, since the qubits will be detuned from their phase references.  These are no problem, though, because we can simply measure them (using a procedure similar to that for two-qubit phases, where instead of toggling the excitation of one of the qubits we measure the qubit phase with and without the gate applied) and unwrap them in software.

\footnotetext{There is no sense of individual ``target'' or ``control'' qubits with a cPhase gate.  Each computational basis state has only one phase degree of freedom and only $\ket{11}$ gets the extra phase.  Each qubit therefore acts as both a control and a target.  This is in contrast to the controlled-NOT gate, where each qubit can individually flip and the gate can transmute one basis state to another.  Since a phase-flip gate can be easily converted to a bit-flip gate and vice versa, there is no substantive meaning underlying this asymmetry -- it is just an artifact of the basis we choose for computation.  This fact is the origin of the idea of a ``phase kickback'' and can complicate the interpretation of a quantum circuit diagram if you are not looking for it.}

There are four equivalent cPhase gates, differing only by which computational state picks up the $-1$ sign.  All the gates have $\pi$ conditional phase, but differ in their values of single-qubit phase.  For example, if we wished to put the $-1$ sign on the ground state, we would set $\phi_{10}=\phi_{01}=\pi$, in addition to $\phi_{11} = \pi$ (in this case we have the matrix $\mathrm{diag}\{1,-1,-1,-1\}$ which is equal to $\mathrm{diag}\{-1,1,1,1\}$ when we factor out a global phase).  Similarly, we can move the minus sign to $\ket{10}$ by setting $\phi_{10}=\pi$ and $\phi_{01}=0$ or put it on $\ket{01}$ with the opposite configuration.  These gates are not different evolutions since we can apply our $z$-gates in software.  Instead, you should consider them to be a means of keeping track of the single-qubit phase evolution that may be necessary to create a particular quantum state.

An adiabatic cPhase gate was used in the first demonstration of on-demand entanglement and quantum algorithms with superconducting qubits \cite{DiCarlo2009}.  It has the advantage of being straightforward to tune-up and insensitive both to pulse timings and flux trajectory.  However, it is relatively slow because we must be adiabatic to the qubit-qubit interaction we are using to generate conditional phase.  These gates typically take about $30-40\ns$, which is several times the $10-15\ns$ splitting period.  Since the coherence time of these devices was so low, this inefficiency is undesirable.  Fortunately, there is another version of this gate that operates at full speed which we describe in the next section.

\subsection{Sudden controlled-phase gate}
\label{subsec:suddencphase}

Now that we have explored what happens when approaching the $\ket{11} \leftrightarrow \ket{02}$ avoided crossing adiabatically, it is interesting to consider the case of moving {\it suddenly}.  This is the opposite limit, where we move so fast that our state wavefunction does not have a chance to evolve in response to the changing Hamiltonian.  In the vicinity of the avoided crossing, where eigenstates change their identities, our wavefunction will become a hybridization of the two states (e.g. $\ket{11} = \ket{+} + \ket{-}$ with $\ket{\pm} = (\ket{11} \pm \ket{02})/\sqrt{2}$ when we are in resonance with the crossing).  The two eigenstates have different energies because of the avoided crossing, and so their relative phase will evolve in time.  As this evolution proceeds, the projection of our wavefunction will oscillate between being primarily $\ket{11}$ and $\ket{02}$.  The amplitude of this oscillation will be given by the degree of hybridization at the location to which we jumped.  Moreover, because the oscillation is due to the presence of a higher level that only interacts when we are in $\ket{11}$, we will acquire conditional phase during this process.

\subsubsection{Toy model}

\begin{figure}
	\centering
	\includegraphics{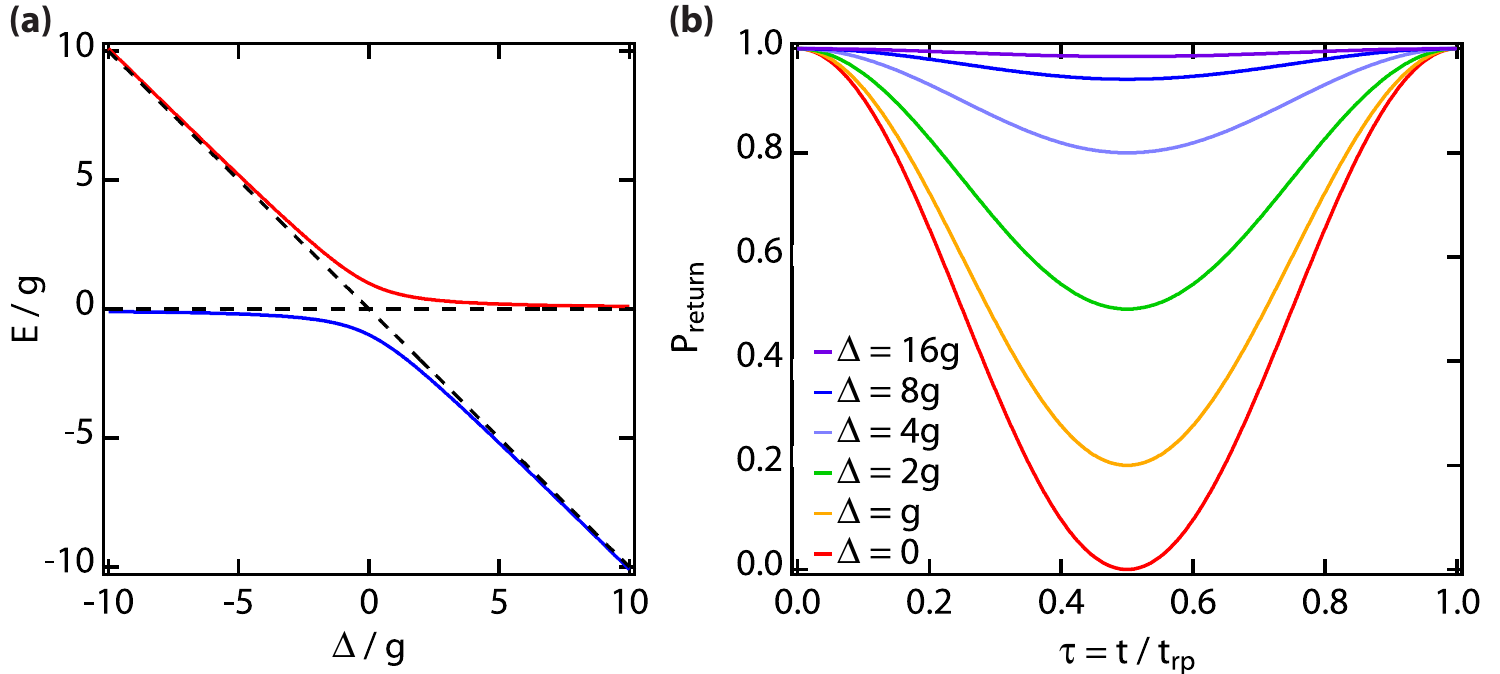}
	\mycaption{Model of a sudden approach to an avoided crossing}
	{\capl{(a)} We consider a toy model of the $\ket{11}\leftrightarrow\ket{02}$ avoided crossing with the two-state Hamiltonian $\hat{H} = \{\{0,g\},\{g, -\Delta\}\}$ and plot its eigenenergies as a function of $\Delta/g$.  This energy diagram closely resembles the avoided crossing found in \figref{fig:twotonespec}(a).  The dashed black lines show the eigenenergies in the absence of interaction.  
	\capl{(b)} We calculate the transition probability of a process where we start in the lower-energy undressed eigenstate, suddenly move to a certain detuning, wait for a period of time, then jump back and measure the projection to our original state.  Defining that time as $\tau=t/t_{\mathrm{rp}}$, where $t_{\mathrm{rp}}=1/\sqrt{(2g)^2+\Delta^2}$ is the rephasing time, we see that our projection oscillates as the phase between the dressed eigenstates evolves.  This oscillation is due to our wavefunction being a superposition of those dressed eigenstates.  The closer we move toward the avoided crossing, the more strongly hybridized our wavefunction, and therefore, the larger the oscillation.
	}
{\label{fig:acmodel}}
\end{figure}

We can see how this works with a toy model of an avoided crossing.  Consider the Hamiltonian $\hat{H} = \{\{0,g\},\{g, -\Delta\}\}$.  This matrix is diagonalized with the eigenenergies $E_\pm=\left(\pm\sqrt{(2g)^2 + \Delta^2} - \Delta\right)/2$, as plotted in \figref{fig:acmodel}(a), and has associated eigenvectors $v_\pm = \{\mathrm{cos}\left(\theta_m^\pm\right),\mathrm{sin}\left(\theta_m^\pm\right)\}$ with $\theta_m^\pm = \mathrm{cot}^{-1}\left(\frac{2g}{\Delta\pm\sqrt{(2g)^2+\Delta^2}}\right)$.  If we start far detuned from the crossing, with $\Delta \rightarrow \infty$, we have $v_- \approx \{0,1\}$ and $E_- \approx -\Delta$.  We are interested in what happens when we suddenly jump to a certain detuning and wait for some period of time.  The operator describing this evolution is given by
\begin{equation}U=\ket{v_+} \bra{v_+}e^{i E_+ t / h }+\ket{v_-}\bra{v_-}e^{i E_- t / h }\end{equation}
where we evaluate the eigenvectors and eigenenergies at the $\Delta$ to which we are jumping.  Starting in a pure state $\ket{v_0}=\{0,1\}$, we say the probability that we return to our initial state is given by $P_{\mathrm{return}}\left( t, \Delta, g \right) = |\langle v_0 | U | v_0 \rangle |^2$.  The ``rephasing'' time, defined by the first time that $P_{\mathrm{return}}\left( t_{\mathrm{rp}}, \Delta, g \right)=1$ and physically corresponding to one full period of oscillation, is given by $t_{\mathrm{rp}}=1/\sqrt{(2g)^2+\Delta^2}$.  Defining $\tau=t/t_{\mathrm{rp}}$, we can plot this return probability as a function of time for several values of $\Delta$, shown in \figref{fig:acmodel}(b).  As anticipated, the amplitude of this oscillation is given by the degree of hybridization; far from the avoided crossing, there is very little transfer of population, while in resonance the state is fully transferred to the other eigenstate and back.

\begin{figure}
	\centering
	\includegraphics{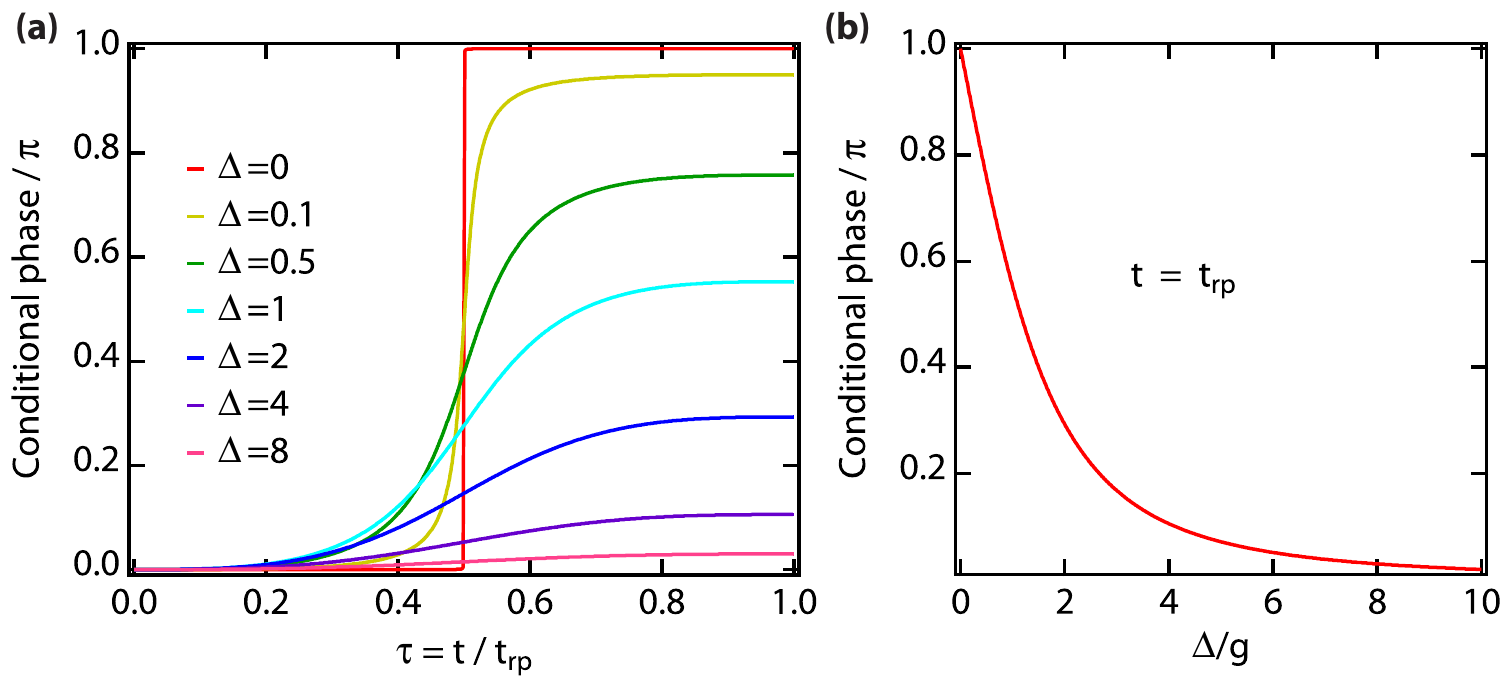}
	\mycaption{Evolution of conditional phase using sudden approach}
	{\capl{(a)} We plot the conditional phase acquired by the quantum amplitude of the original, lower-energy eigenstate as a function of the interaction time $\tau$ for several different values of $\Delta$. The slope of this curve and maximum conditional phase acquired increases with decreasing $\Delta$.  Note that we are examining only the phase of the projection on our original state; for non-integer values of $\tau$, our wavefunction will be a superposition of the two undressed eigenstates.
	\capl{(b)} Fixing $t = t_{\mathrm{rp}}$, we plot the conditional phase acquired as a function of the detuning to which we suddenly jump.  At $\Delta=0$, we acquire a full $\pi$ conditional phase shift, which decreases asymptotically to zero as $\Delta\rightarrow\infty$ with the functional form $\phi_c=\left( 1 - \frac{\Delta}{\sqrt{(2g)^2+\Delta^2}}\right)$.
	}
{\label{fig:acmodel_cp}}
\end{figure}

What is the conditional phase acquired during this process?  One thing that is initially confusing is that if we calculate the phase angle of $\langle v_0 | U | v_0 \rangle$ at $t=t_{\mathrm{rp}}$, we get $\phi_0=-\pi \left( 1 + \frac{\Delta}{\sqrt{(2g)^2+\Delta^2}}\right)$.  Does that mean we are getting more than $\pi$ of two-qubit phase when we wait off-resonance and so are maximally entangled at some intermediate time?  The answer is no -- we actually acquire less conditional phase for finite $\Delta$ -- but to see that we must consider what would happen if the avoided crossing were not there.  We have added dashed lines alongside the eigenenergies plotted in \figref{fig:acmodel}(a) to indicate the energies for the case of no interaction between the states.  This is our analog for the energies of the constituent computational states in the adiabatic case described in the previous section; it defines the single-qubit phase that we must subtract to determine our conditional phase.  The phase acquired by this state is given by $\phi_1 = E_-^{g=0} t = -2 \pi \Delta t$.  If we consider only the times that cause full oscillations, we have $\phi_1 = -2 \pi \Delta t_{\mathrm{rp}} =  \frac{-2\pi \Delta}{\sqrt{(2g)^2+\Delta^2}}$.  Our conditional phase is then $\phi_c = \phi_0 - \phi_1 = \pi \left( 1 - \frac{\Delta}{\sqrt{(2g)^2+\Delta^2}}\right)$.  This has the expected limits of $\phi_c \rightarrow 0$ as $\Delta \rightarrow -\infty$ and $\phi_c = \pi$ when $\Delta=0$.  We show the behavior of the two-qubit phase as a function of time and $\Delta$ (at $t=t_{\mathrm{rp}}$) in \figref{fig:acmodel_cp}.

\subsubsection{Swap spectroscopy and the sudden cPhase gate}

\begin{figure}
\centering
\includegraphics{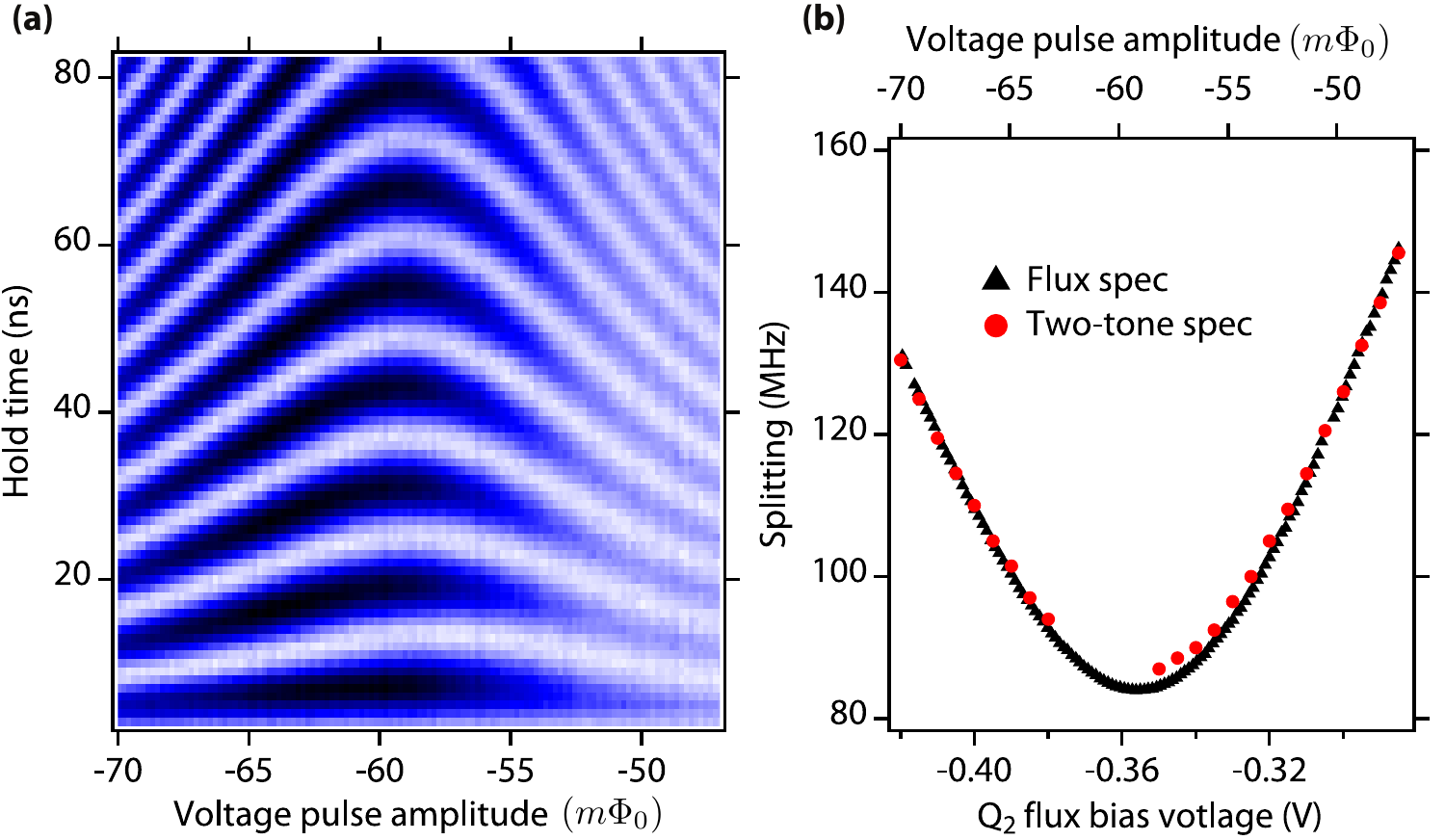}
	\mycaption{Swap spectroscopy of 011-002 avoided crossing}
	{\capl{(a)} We prepare the state $\ket{011}$ and suddenly move $Q_2$ such that the $\ket{011}$ eigenstate is near in energy to the $\ket{002}$ eigenstate.  We then wait for some amount of time, causing the quantum amplitude in our starting state to oscillate into the interacting state and back.  Finally, we return $Q_2$ to its home position, transfer the amplitude of $\ket{011}$ to $\ket{000}$ by repeating the initialization pulses, and measure.  We will get a low voltage (colored white) if, at the end of our waiting time, our population returned to $\ket{011}$ and a large voltage (blue) if the population was transferred to $\ket{002}$.  The data are asymmetric because our flux pulse is not perfectly sudden.  The data on the right correspond with smaller flux excursions, which, for fixed bandwidth, have slower velocities and are therefore less sudden.  
	\capl{(b)} We extract the frequency of oscillation for each vertical cut of the data in \capl{(a)} and plot it as a function of amplitude (top axis).  We compare this to the spectroscopic data from \figref{fig:twotonespec}(b) and find excellent correspondence. 
	}
{\label{fig:suddenchevron}}
\end{figure}

Now that we understand the underlying physical mechanism for a sudden cPhase gate, let us explore the details of its implementation.  In order to construct a gate with the $\ket{11} \leftrightarrow \ket{02}$ avoided crossing, we first use {\it swap spectroscopy} to measure where and how large it is.  The procedure is done by preparing some state of our qubit register, performing a sudden flux pulse on one of the qubits to a certain location, waiting for some time $\tau$, fluxing back, executing whatever qubit pulses would bring the prepared state back to the ground state, and finally, measuring.  The idea is that if we are near an avoided crossing at whatever flux location we jump to, waiting there will cause our population to oscillate between the prepared state and the state with which it is interacting.  We must have measurement contrast between the two potential states to detect whether or not this oscillation has occurred, so we jump back to our starting position and transfer whatever population remains in our prepared state back to its ground state.  Our readout will then discriminate between the qubit in the ground state (indicating that no leakage has occurred) and the qubit in some other state that it tunneled into during the waiting time.

In \figref{fig:suddenchevron}(a) we show the result of performing swap spectroscopy on the $\ket{110} \leftrightarrow \ket{020}$ avoided crossing when preparing $\ket{110}$.  We observe the characteristic ``chevron'' pattern of a suddenly-approached avoided crossing.  We plot the measurement result as a function of both flux pulse amplitude and the time we hold at each applied voltage.  Where we get a relatively small voltage, indicated in white, the qubit's wavefunction is unchanged; a dark pixel corresponds to a large voltage and population transfer to $\ket{020}$.  The frequency of this oscillation corresponds well with the prediction from our simple model.  The frequency of oscillation is minimal ($86\mhz$) when we move into resonance with the avoided crossing ($V=-55~\mathrm{m}\Phi_0$).  As we detune, the frequency increases according to $f = \sqrt{(2g)^2 + \Delta^2}$.  The amplitude of oscillation also decreases in a similar way.  In \figref{fig:suddenchevron}(a) we plot the oscillation frequency extracted from the chevron data (top and right axes) and overlay the spectroscopically measured splittings from \figref{fig:zeta} (bottom and left axes), showing excellent correspondence between the two methods.  Flux spectroscopy is much more precise since we have the luxury of fitting a sinusoid rather than two Lorentzians.

With this information, it is simple to construct our cPhase gate.  The amplitude of our flux pulse is given by the voltage corresponding to the slowest oscillation frequency we found in swap spectroscopy, $V\approx 55~\mathrm{m}\Phi_0$.  (Note that we have converted to units of $\Phi_0$ to have physical meaning, but experimentally we specify a voltage amplitude.)  We apply a square flux pulse with the fastest possible rise time to be maximally sudden for a time of $t=1/f\sim 12\ns$, where $f=86\mhz$ is the splitting of the avoided crossing.  As with the adiabatic case, we must also measure and correct single-qubit phases in software.  The resulting gate is of relatively high fidelity owing to its fast speed, as we will see in \sref{sec:processtomo}.

One potential concern is whether we are able to precisely time our excursion, since any error would cause some leakage into $\ket{020}$.  We can be neither perfectly sudden, nor can we exactly hit the timing of $11.62\ns$ corresponding to one oscillation.  While there are methods of detecting and compensating for these small timing errors (discussed in \sref{subsubsec:finitebandwidth}), for the purposes of a cPhase gate on qubits with relatively short $T_1$s, these errors are negligible enough to disregard.

\subsection{Other two-qubit gates}
\label{subsec:other2qgates}

The adiabatic and sudden fluxed cPhase gates that we have discussed are by no means the only methods of generating entanglement in cQED.  In fact, though they were the first practical gates demonstrated, they will almost certainly not represent the best approach going forward.  Our ideal gate is fast (as these are), but also has high ``intrinsic'' gate fidelity and does not require additional hardware in the device or at room temperature.  The {\it intrinsic} gate fidelity is a vague term which refers, in this context, to the infidelity of the gate unrelated to $T_1$ or $T_2$.  For example, non-adiabaticity or timing errors might cause leakage into the $\ket{02x}$ state, the size of which directly reduces the gate's fidelity.  These errors are also quite insidious, because they would not be corrected by most forms of quantum error correction \cite{FowlerMM2013}.  More importantly, these gates require fast flux lines.  In addition to requiring an extra cryogenic microwave line and room temperature electronics, making the qubits sensitive to flux inexorably decreases their $T_2$ coherence times because of flux noise.

One appealing gate that we have recently developed \cite{Paik2013} works by utilizing the dispersive shift of the cavity frequency -- exactly the same mechanism used in qubit readout.  We adiabatically apply a microwave pulse far detuned from the cavity such that the cavity state is instantaneously proportional to the drive state, and turn it on and off slowly so that the cavity returns to its ground state.  Each dispersively shifted cavity will take a different trajectory in Hilbert space and therefore will acquire a different phase (of both geometric and Stark-shift origins).  Decomposing these phases into single- and two-qubit types, we can tune-up a cPhase gate in the same way as normal.  This gate is interesting because it only requires microwave tones on the cavity -- something we automatically have for performing readout -- and has high intrinsic fidelity.  The two main sources -- measurement induced dephasing and non-adiabaticity -- can be made extremely small with appropriate choice of device parameters.  

There are other attractive microwave-only gates that have been developed \cite{Paraoanu2006, Rigetti2010, deGroot2010}, including the Bell-Rabi gate \cite{Poletto2012} that directly drives the $\ket{00} \leftrightarrow \ket{11}$ transition, and the cross-resonance gate \cite{Chow2011}, which relies on a state-dependent Rabi rate of the two qubits.  This is by no means an exhaustive list; indeed, superconducting qubits in cQED seem to love entangling with one another since they have large spatial wave functions.  Rather than being difficult to entangle, in fact, it seems that the challenge in moving toward high-fidelity gates will be to make sure the qubits do not {\it unintentionally} entangle with one another.

\section{State tomography}
\label{sec:statetomo}

\nomdref{Cmhat}{$\hat{M}$}{measurement operator}{sec:statetomo}
\nomdref{Cvh}{$V_H$}{homodyne measurement voltage}{sec:statetomo}

Now that we have the ability to generate entanglement, how can we prove that we have done so?  If we are given many copies of some unknown state, how can we determine what that state is?  The act of reconstructing the density matrix from an ensemble of states is called {\it state tomography}.  Constructing a state tomogram involves measuring some property of a quantum state; by averaging the result of that and other measurements over an ensemble of identically prepared states, we can infer every entry of the matrix.  We can express an $N$ qubit density matrix as a sum of Pauli correlations times their associated operator.  For a single qubit, we have $\rho = \frac{1}{2}\sum{ \mathrm{tr}\left(\rho I\right) +\mathrm{tr}\left( \rho X \right)X + \mathrm{tr}\left( \rho Y \right) Y +\mathrm{tr}\left( \rho Z \right)Z }$.  We define the quantity $\langle \hat{O} \rangle \equiv \mathrm{tr}\left( \rho \hat{O}\right)$.  This generalizes, for $N$ qubits, to $\rho = \frac{1}{2^N}\sum_{\hat{O}\in\{I^{\otimes N}..Z^{\otimes N}}\} \langle \hat{O}\rangle \hat{O}$ where the sum is taken over every $N$-qubit Pauli operator\footnotemark.  Thus, reconstructing the density matrix is the same as knowing each of the quantities $\langle \hat{O}\rangle$.

\footnotetext{For example, for two qubits there are 16 Pauli operators: $II$, $XI$, $YI$, $ZI$, $IX$, $IY$, $IZ$, $XX$, $YX$, $ZX$, $XY$, $YY$, $ZY$,$XZ$, $YZ$, and $ZZ$.  For three qubits there are 64, and for $N$, there are $4^N$. That is, the amount of information stored in a manifold of qubits grows exponentially with the number $N$.  The notation $I^{\otimes N}$ denotes the tensor product of $N$ Pauli $I$ operators, e.g. $I...I$.}

The most conceptually straightforward way to measure these Pauli correlations is with single-shot single-qubit measurements.  If we have the ability to measure $\langle \hat{O} \rangle$ for each qubit individually, we can calculate the correlations in software with classical processing.  For example, we would find $\langle ZZ\rangle$ by calculating the probability that, when we measured the $Z$ projection of both qubits at the same time, we found that both pointed in the same direction, and subtract from that the probability that they were both pointing in the opposite direction.  Similarly, calculating quantities such as $\langle XY \rangle$ would require measuring the $X$ projection of $Q_1$ simultaneous with the $Y$ projection of $Q_2$.  In our system, however, we do not have the capability of measuring each qubit individually.  Instead, our measurement operator (under normal circumstances) is one which projects the qubit manifold to the ground state: $\hat{M} = |0^{\otimes N}\rangle\langle 0^{\otimes N}|$.  We get a $0$ if all the qubits are in their ground state, and a 1 otherwise.  Taking the case of two qubits, we can express this operator in terms of Pauli operators as $\hat{M}_2 = \ket{00}\bra{00} = \left(II + ZI + IZ + ZZ\right)/4$, where $Z$ is the Pauli $z$ matrix\footnotemark.  We see in this representation that our measurement operator inexorably projects all of them.

\footnotetext{The operator $\ket{00}\bra{00} = \left(\begin{smallmatrix}1&0&0&0\\0&0&0&0\\0&0&0&0\\0&0&0&0\end{smallmatrix}\right)$ if we assume the qubit amplitude is confined to the computational subspace.  We have $II = \left(\begin{smallmatrix}1&0&0&0\\0&1&0&0\\0&0&1&0\\0&0&0&1\end{smallmatrix}\right)$, $ZI=\left(\begin{smallmatrix}1&0&0&0\\0&-1&0&0\\0&0&1&0\\0&0&0&-1\end{smallmatrix}\right)$, $IZ=\left(\begin{smallmatrix}1&0&0&0\\0&1&0&0\\0&0&-1&0\\0&0&0&-1\end{smallmatrix}\right)$, and $ZZ=\left(\begin{smallmatrix}1&0&0&0\\0&-1&0&0\\0&0&-1&0\\0&0&0&1\end{smallmatrix}\right)$.  Thus $\ket{00}\bra{00} = \left(II + IZ + ZI + ZZ\right)/4$.  Similarly, the measurement operator for three qubits $\hat{M}_3 = \ket{000}\bra{000}=III+ZII+IZI+IIZ+ZZI+ZIZ+IZZ+ZZZ$, and so on for more qubits.}

This measurement operator is actually ideal for tomography because it eliminates the need for single-shot measurements and for manually calculating correlations.  Consider what would happen if we performed qubit rotations -- say, a $\pi$ pulse on the first qubit -- just prior to measuring.  If we view that $\pi$ pulse as acting on the measurement operator rather than the qubit state, we would flip all of the $Z$ correlations of the second qubit, making $\hat{M}_2^{R^1_{\pi}} = II - ZI + IZ - ZZ$.  If we instead were to pulse only the second qubit, we would have $\hat{M}_2^{R^2_{\pi}} = II + ZI - IZ - ZZ$.  Taking the sum of the ensemble averaged values of these two cases, we would have $\langle \hat{M}_2^{R^1_{\pi}} \rangle + \langle \hat{M}_2^{R^2_{\pi}} \rangle = \langle II\rangle - \langle ZI\rangle + \langle IZ\rangle - \langle ZZ\rangle + \langle II\rangle + \langle ZI\rangle - \langle IZ\rangle - \langle ZZ\rangle = 2 \langle II \rangle - 2 \langle ZZ \rangle$.  Since $\langle II \rangle \equiv 1$, we can directly calculate $\langle ZZ \rangle = \frac{1}{2} \left( 2 - \langle \hat{M}_2^{R^1_{\pi}} \rangle - \langle \hat{M}_2^{R^2_{\pi}} \rangle\right)$.  Thus, by performing rotations on the ensemble immediately prior to measurement, we can calculate any single-qubit or two-qubit correlation with linear combinations of our measurement results.

The measurement operator is actually more complicated than $\ket{0^{\otimes N}}\bra{0^{\otimes N}}$ because of finite measurement fidelity and differing sensitivity to the various basis states.  For example, recalling the case of multi-qubit readout with the high-power Jaynes-Cummings readout mechanism described in \sref{subsec:jcqubitmeasurement}, we had $\sim67\%$ fidelity to the least-distinguishable $\ket{100}$ state, but in excess of $80\%$ fidelity to others.  This disparity could have been made even more extreme had we optimized for a state other than $\ket{100}$: the difference in fidelity could be two-to-one or more.  Taken in the ensemble, we will measure slightly different voltages for each basis state.  (Indeed, if we are in an eigenstate and given enough averaging, we can know not just whether or not we have the ground state, but also exactly what state we are in.)  Since our experiment gives us voltages and not measurement operator expectation values, we must compensate for this fact.

One simple way of calibrating the differing sensitivity to basis states is to modify our Pauli expansion of the measurement operator.  Rather than saying that $\hat{M}_2 = \left( II + ZI + IZ + ZZ \right)/2$, we add coefficients in front of each term of the operator and take the ensemble average such that our measurement voltage is given by $\hat{V}_2^M \equiv \beta_{II}\langle II\rangle + \beta_{ZI}\langle ZI\rangle + \beta_{IZ}\langle IZ\rangle + \beta_{ZZ}\langle ZZ\rangle$.  These $\beta$ coefficients have units of voltage, as does $\hat{V}_2^M$.  We can determine the values of these in a manner experimentally similar to our original construction of tomography.  We prepare each of the states $\ket{00}$, $\ket{10}$, $\ket{01}$, and $\ket{11}$ and measure them.  If our control is perfect, we should measure voltages of $\beta_{II}+\beta_{ZI}+\beta_{IZ}+\beta_{ZZ}$ when preparing $\ket{00}$, $\beta_{II}-\beta_{ZI}+\beta_{IZ}-\beta_{ZZ}$ for $\ket{10}$, and so on.  We can again make linear combinations to convert these results to $\beta$s.  Because their values tend to drift with time due to (for example) changes in helium level or local oscillator phase, we always measure them at the same time as the other tomographic measurements.

\begin{table} \small
	\centering
	\begin{tabular}{|c|c|c|c|}
		{\bf Measurement} 	& {\bf Rotation on $Q_1$} & {\bf Measurement operator}	\\ \hline
		1					& $I$					& $+\beta_{I}I+\beta_{Z}Z$	\\ 
		2					& $R_x^{\pi/2}$			& $+\beta_{I}I+\beta_{Z}Y$ 		\\ 
		3					& $R_y^{\pi/2}$			& $+\beta_{I}I-\beta_{Z}X$ 
	\end{tabular}
	\mycaption{Gate sequence for one-qubit state tomography}
	{The rotations on $Q_1$ effectively modify the measurement operator when we consider them to be part of the measurement process.  We measure the values of $\beta_{I}$ and $\beta_{Z}$ with separate experiments preparing $\ket{1}$ and $\ket{0}$, and perform them in series with the tomographic measurements to nullify slow drifts of measurement chain gain.  Combining that information with the result of the three operator measurements, we can extract the values of $\langle Z \rangle$, $\langle X \rangle$, and $\langle Y \rangle$, and therefore know the entire one-qubit density matrix.
	}
	\label{table:1qubittomosequence}
\end{table}

\begin{table} \small
	\centering
	\begin{tabular}{|c|c|c|c|}
		{\bf Measurement} 	& {\bf Rotation on $Q_1$} & {\bf Rotation on $Q_2$} & {\bf Measurement operator}				\\ \hline
		1					& $I$					& $I$						& $+\beta_{ZI}ZI+\beta_{IZ}IZ+\beta_{ZZ}ZZ$ \\ 
		2					& $R_x^{\pi}$			& $I$						& $-\beta_{ZI}ZI+\beta_{IZ}IZ-\beta_{ZZ}ZZ$ \\ 
		3					& $I$					& $R_x^{\pi}$				& $+\beta_{ZI}ZI-\beta_{IZ}IZ-\beta_{ZZ}ZZ$ \\ 
		4					& $R_x^{\pi/2}$			& $I$						& $+\beta_{ZI}YI+\beta_{IZ}IZ+\beta_{ZZ}YZ$ \\ 
		5					& $R_x^{\pi/2}$			& $R_x^{\pi/2}$				& $+\beta_{ZI}YI+\beta_{IZ}IY+\beta_{ZZ}YY$ \\ 
		6					& $R_x^{\pi/2}$			& $R_y^{\pi/2}$				& $+\beta_{ZI}YI-\beta_{IZ}IX-\beta_{ZZ}YX$ \\ 
		7					& $R_x^{\pi/2}$			& $R_x^{\pi}$				& $+\beta_{ZI}YI-\beta_{IZ}IZ-\beta_{ZZ}YZ$ \\ 
		8					& $R_y^{\pi/2}$			& $I$						& $-\beta_{ZI}XI+\beta_{IZ}IZ-\beta_{ZZ}XZ$ \\ 
		9					& $R_y^{\pi/2}$			& $R_x^{\pi/2}$				& $-\beta_{ZI}XI+\beta_{IZ}IY-\beta_{ZZ}XY$ \\ 
		10					& $R_y^{\pi/2}$			& $R_y^{\pi/2}$				& $-\beta_{ZI}XI-\beta_{IZ}IX+\beta_{ZZ}XX$ \\ 
		11					& $R_y^{\pi/2}$			& $R_x^{\pi}$				& $-\beta_{ZI}XI-\beta_{IZ}IZ+\beta_{ZZ}XZ$ \\ 
		12					& $I$					& $R_x^{\pi/2}$				& $+\beta_{ZI}ZI+\beta_{IZ}IY+\beta_{ZZ}ZY$ \\ 
		13					& $R_x^{\pi}$			& $R_x^{\pi/2}$				& $-\beta_{ZI}ZI+\beta_{IZ}IY-\beta_{ZZ}ZY$ \\ 
		14					& $I$					& $R_y^{\pi/2}$				& $+\beta_{ZI}ZI-\beta_{IZ}IX-\beta_{ZZ}ZX$ \\ 
		15					& $R_x^{\pi}$			& $R_y^{\pi/2}$				& $-\beta_{ZI}ZI-\beta_{IZ}IX+\beta_{ZZ}ZX$ 
	\end{tabular}
	\mycaption{Gate sequence for two-qubit state tomography}
	{We list the rotations for each of the two qubits to give 15 linearly independent measurement operators.  As with the single-qubit case, we measure the values of the $\beta$s by preparing all the computational basis states.  Combining these numbers with the 15 measurements is enough to infer the full two-qubit density matrix.
	}
	\label{table:2qubittomosequence}
\end{table}

Once we know the proper $\beta$ calibrations, we can infer every Pauli correlation by performing different rotations prior to measurement.  Just as doing a $\pi$ pulse converted $Z_i$ (the $z$ Pauli matrix acting on the $i$-th qubit) to $-Z_i$, $R_x^{\pi/2}$ maps $Z_i \rightarrow +Y_i$ and $R_y^{\pi/2}$ maps $Z_i \rightarrow -X_i$.  Doing all combinations of nothing, $\pi/2$ rotations around $x$ and $y$ and $\pi$ rotations around $x$ on both qubits will extract all the information of the density matrix.  The three pre-rotations necessary for single-qubit tomography are listed in \tref{table:1qubittomosequence}, and for two qubits, the required 15 pre-rotations can be found in \tref{table:2qubittomosequence}.  (For two qubits, we have suppressed the $\beta_{II} II$ term of the measurement operator, which is identical in all cases.)  There are $4^N-1$ entries since there are four gates and $N$ qubits plus one constraint ($\mathrm{tr}(\rho) = 1$).  Given this list of measurements, it is again a matter of converting those data into a density matrix.  It is also worth noting that these lists are not unique -- any combination that spans the Hilbert space would work, even a non-orthogonal one \cite{Merkel2012}.

\begin{table} \small
	\centering
	\begin{tabular}{|c|c|c|c|c|}
		{\bf \# } 	& {\bf $R_{Q_1}$} & {\bf $R_{Q_2}$} & {\bf $R_{Q_3}$} & {\bf Measurement operator}				\\ \hline
		1			& $I$			 & $I$			 & $I$			 & {$\scriptstyle+\beta_{1}ZII+\beta_{2}IZI+\beta_{3}IIZ+\beta_{4}ZZI+\beta_{5}ZIZ+\beta_{6}IZZ+\beta_{7}ZZZ$} \\ 

		2			 & $I$			 & $I$			 & $R_x^{\pi/2}$ & {$\scriptstyle+\beta_{1}ZII+\beta_{2}IZI+\beta_{3}IIY+\beta_{4}ZZI+\beta_{5}ZIY+\beta_{6}IZY+\beta_{7}ZZY$} \\ 

		3			 & $I$			 & $I$			 & $R_y^{\pi/2}$ & {$\scriptstyle+\beta_{1}ZII+\beta_{2}IZI-\beta_{3}IIX+\beta_{4}ZZI-\beta_{5}ZIX-\beta_{6}IZX-\beta_{7}ZZX$} \\ 

		4			 & $I$			 & $I$			 & $R_x^{\pi}$	 & {$\scriptstyle+\beta_{1}ZII+\beta_{2}IZI-\beta_{3}IIZ+\beta_{4}ZZI-\beta_{5}ZIZ-\beta_{6}IZZ-\beta_{7}ZZZ$} \\

		5			 & $I$			 & $R_x^{\pi/2}$ & $I$			 & {$\scriptstyle+\beta_{1}ZII+\beta_{2}IYI+\beta_{3}IIZ+\beta_{4}ZYI+\beta_{5}ZIZ+\beta_{6}IYZ+\beta_{7}ZYZ$} \\ 

		6			 & $I$			 & $R_x^{\pi/2}$ & $R_x^{\pi/2}$ & {$\scriptstyle+\beta_{1}ZII+\beta_{2}IYI+\beta_{3}IIY+\beta_{4}ZYI+\beta_{5}ZIY+\beta_{6}IYY+\beta_{7}ZYY$} \\ 

		7			 & $I$			 & $R_x^{\pi/2}$ & $R_y^{\pi/2}$ & {$\scriptstyle+\beta_{1}ZII+\beta_{2}IYI-\beta_{3}IIX+\beta_{4}ZYI-\beta_{5}ZIX-\beta_{6}IYX-\beta_{7}ZYX$} \\ 

		8			 & $I$			 & $R_x^{\pi/2}$ & $R_x^{\pi}$	 & {$\scriptstyle+\beta_{1}ZII+\beta_{2}IYI-\beta_{3}IIZ+\beta_{4}ZYI-\beta_{5}ZIZ-\beta_{6}IYZ-\beta_{7}ZYZ$} \\

		9			 & $I$			 & $R_y^{\pi/2}$ & $I$			 & {$\scriptstyle+\beta_{1}ZII-\beta_{2}IXI+\beta_{3}IIZ-\beta_{4}ZXI+\beta_{5}ZIZ-\beta_{6}IXZ-\beta_{7}ZXZ$} \\ 

		10			 & $I$			 & $R_y^{\pi/2}$ & $R_x^{\pi/2}$ & {$\scriptstyle+\beta_{1}ZII-\beta_{2}IXI+\beta_{3}IIY-\beta_{4}ZXI+\beta_{5}ZIY-\beta_{6}IXY-\beta_{7}ZXY$} \\ 

		11			 & $I$			 & $R_y^{\pi/2}$ & $R_y^{\pi/2}$ & {$\scriptstyle+\beta_{1}ZII-\beta_{2}IXI-\beta_{3}IIX-\beta_{4}ZXI-\beta_{5}ZIX+\beta_{6}IXX+\beta_{7}ZXX$} \\ 

		12			 & $I$			 & $R_y^{\pi/2}$ & $R_x^{\pi}$	 & {$\scriptstyle+\beta_{1}ZII-\beta_{2}IXI-\beta_{3}IIZ-\beta_{4}ZXI-\beta_{5}ZIZ+\beta_{6}IXZ+\beta_{7}ZXZ$} \\

		13			 & $I$			 & $R_x^{\pi}$	 & $I$			 & {$\scriptstyle+\beta_{1}ZII-\beta_{2}IZI+\beta_{3}IIZ-\beta_{4}ZZI+\beta_{5}ZIZ-\beta_{6}IZZ-\beta_{7}ZZZ$} \\ 

		14			 & $I$			 & $R_x^{\pi}$	 & $R_x^{\pi/2}$ & {$\scriptstyle+\beta_{1}ZII-\beta_{2}IZI+\beta_{3}IIY-\beta_{4}ZZI+\beta_{5}ZIY-\beta_{6}IZY-\beta_{7}ZZY$} \\ 

		15			 & $I$			 & $R_x^{\pi}$	 & $R_y^{\pi/2}$ & {$\scriptstyle+\beta_{1}ZII-\beta_{2}IZI-\beta_{3}IIX-\beta_{4}ZZI-\beta_{5}ZIX+\beta_{6}IZX+\beta_{7}ZZX$} \\ 

		16			 & $I$			 & $R_x^{\pi}$	 & $R_x^{\pi}$	 & {$\scriptstyle+\beta_{1}ZII-\beta_{2}IZI-\beta_{3}IIZ-\beta_{4}ZZI-\beta_{5}ZIZ+\beta_{6}IZZ+\beta_{7}ZZZ$} \\

		17			 & $R_x^{\pi/2}$ & $I$			 & $I$			 & {$\scriptstyle+\beta_{1}YII+\beta_{2}IZI+\beta_{3}IIZ+\beta_{4}YZI+\beta_{5}YIZ+\beta_{6}IZZ+\beta_{7}YZZ$} \\ 

		18			 & $R_x^{\pi/2}$ & $I$			 & $R_x^{\pi/2}$ & {$\scriptstyle+\beta_{1}YII+\beta_{2}IZI+\beta_{3}IIY+\beta_{4}YZI+\beta_{5}YIY+\beta_{6}IZY+\beta_{7}YZY$} \\ 

		19			 & $R_x^{\pi/2}$ & $I$			 & $R_y^{\pi/2}$ & {$\scriptstyle+\beta_{1}YII+\beta_{2}IZI-\beta_{3}IIX+\beta_{4}YZI-\beta_{5}YIX-\beta_{6}IZX-\beta_{7}YZX$} \\ 

		20			 & $R_x^{\pi/2}$ & $I$			 & $R_x^{\pi}$	 & {$\scriptstyle+\beta_{1}YII+\beta_{2}IZI-\beta_{3}IIZ+\beta_{4}YZI-\beta_{5}YIZ-\beta_{6}IZZ-\beta_{7}YZZ$} \\

		21			 & $R_x^{\pi/2}$ & $R_x^{\pi/2}$ & $I$			 & {$\scriptstyle+\beta_{1}YII+\beta_{2}IYI+\beta_{3}IIZ+\beta_{4}YYI+\beta_{5}YIZ+\beta_{6}IYZ+\beta_{7}YYZ$} \\ 

		22			 & $R_x^{\pi/2}$ & $R_x^{\pi/2}$ & $R_x^{\pi/2}$ & {$\scriptstyle+\beta_{1}YII+\beta_{2}IYI+\beta_{3}IIY+\beta_{4}YYI+\beta_{5}YIY+\beta_{6}IYY+\beta_{7}YYY$} \\ 

		23			 & $R_x^{\pi/2}$ & $R_x^{\pi/2}$ & $R_y^{\pi/2}$ & {$\scriptstyle+\beta_{1}YII+\beta_{2}IYI-\beta_{3}IIX+\beta_{4}YYI-\beta_{5}YIX-\beta_{6}IYX-\beta_{7}YYX$} \\ 

		24			 & $R_x^{\pi/2}$ & $R_x^{\pi/2}$ & $R_x^{\pi}$	 & {$\scriptstyle+\beta_{1}YII+\beta_{2}IYI-\beta_{3}IIZ+\beta_{4}YYI-\beta_{5}YIZ-\beta_{6}IYZ-\beta_{7}YYZ$} \\

		25			 & $R_x^{\pi/2}$ & $R_y^{\pi/2}$ & $I$			 & {$\scriptstyle+\beta_{1}YII-\beta_{2}IXI+\beta_{3}IIZ-\beta_{4}YXI+\beta_{5}YIZ-\beta_{6}IXZ-\beta_{7}YXZ$} \\ 

		26			 & $R_x^{\pi/2}$ & $R_y^{\pi/2}$ & $R_x^{\pi/2}$ & {$\scriptstyle+\beta_{1}YII-\beta_{2}IXI+\beta_{3}IIY-\beta_{4}YXI+\beta_{5}YIY-\beta_{6}IXY-\beta_{7}YXY$} \\ 

		27			 & $R_x^{\pi/2}$ & $R_y^{\pi/2}$ & $R_y^{\pi/2}$ & {$\scriptstyle+\beta_{1}YII-\beta_{2}IXI-\beta_{3}IIX-\beta_{4}YXI-\beta_{5}YIX+\beta_{6}IXX+\beta_{7}YXX$} \\ 

		28			 & $R_x^{\pi/2}$ & $R_y^{\pi/2}$ & $R_x^{\pi}$	 & {$\scriptstyle+\beta_{1}YII-\beta_{2}IXI-\beta_{3}IIZ-\beta_{4}YXI-\beta_{5}YIZ+\beta_{6}IXZ+\beta_{7}YXZ$} \\

		29			 & $R_x^{\pi/2}$ & $R_x^{\pi}$	 & $I$			 & {$\scriptstyle+\beta_{1}YII-\beta_{2}IZI+\beta_{3}IIZ-\beta_{4}YZI+\beta_{5}YIZ-\beta_{6}IZZ-\beta_{7}YZZ$} \\ 

		30			 & $R_x^{\pi/2}$ & $R_x^{\pi}$	 & $R_x^{\pi/2}$ & {$\scriptstyle+\beta_{1}YII-\beta_{2}IZI+\beta_{3}IIY-\beta_{4}YZI+\beta_{5}YIY-\beta_{6}IZY-\beta_{7}YZY$} \\ 

		31			 & $R_x^{\pi/2}$ & $R_x^{\pi}$	 & $R_y^{\pi/2}$ & {$\scriptstyle+\beta_{1}YII-\beta_{2}IZI-\beta_{3}IIX-\beta_{4}YZI-\beta_{5}YIX+\beta_{6}IZX+\beta_{7}YZX$} \\ 

		32			 & $R_x^{\pi/2}$ & $R_x^{\pi}$	 & $R_x^{\pi}$	 & {$\scriptstyle+\beta_{1}YII-\beta_{2}IZI-\beta_{3}IIZ-\beta_{4}YZI-\beta_{5}YIZ+\beta_{6}IZZ+\beta_{7}YZZ$} 
	\end{tabular}
	\mycaption{Gate sequence for three-qubit state tomography, part 1}
	{State tomography for three qubits is again analogous to the cases of one and two qubits, but needs 63 linearly independent measurement operators to infer all the Pauli correlators.  Only 32 fit on this page; the remaining 31 can be found on the following page.  As before, we measure the $\beta$s by preparing and measuring all 8 computational basis states.
	}
	\label{table:3qubittomosequence_a}
\end{table}

\begin{table} \small
	\centering
	\begin{tabular}{|c|c|c|c|c|}
		{\bf \# } & {\bf $R_{Q_1}$} & {\bf $R_{Q_2}$} & {\bf $R_{Q_3}$} & {\bf Measurement operator}																					 \\ \hline
		33			 & $R_y^{\pi/2}$ & $I$			 & $I$			 & {$\scriptstyle-\beta_{1}XII+\beta_{2}IZI+\beta_{3}IIZ-\beta_{4}XZI-\beta_{5}XIZ+\beta_{6}IZZ-\beta_{7}XZZ$} \\ 

		34			 & $R_y^{\pi/2}$ & $I$			 & $R_x^{\pi/2}$ & {$\scriptstyle-\beta_{1}XII+\beta_{2}IZI+\beta_{3}IIY-\beta_{4}XZI-\beta_{5}XIY+\beta_{6}IZY-\beta_{7}XZY$} \\ 

		35			 & $R_y^{\pi/2}$ & $I$			 & $R_y^{\pi/2}$ & {$\scriptstyle-\beta_{1}XII+\beta_{2}IZI-\beta_{3}IIX-\beta_{4}XZI+\beta_{5}XIX-\beta_{6}IZX+\beta_{7}XZX$} \\ 

		36			 & $R_y^{\pi/2}$ & $I$			 & $R_x^{\pi}$	 & {$\scriptstyle-\beta_{1}XII+\beta_{2}IZI-\beta_{3}IIZ-\beta_{4}XZI+\beta_{5}XIZ-\beta_{6}IZZ+\beta_{7}XZZ$} \\

		37			 & $R_y^{\pi/2}$ & $R_x^{\pi/2}$ & $I$			 & {$\scriptstyle-\beta_{1}XII+\beta_{2}IYI+\beta_{3}IIZ-\beta_{4}XYI-\beta_{5}XIZ+\beta_{6}IYZ-\beta_{7}XYZ$} \\ 

		38			 & $R_y^{\pi/2}$ & $R_x^{\pi/2}$ & $R_x^{\pi/2}$ & {$\scriptstyle-\beta_{1}XII+\beta_{2}IYI+\beta_{3}IIY-\beta_{4}XYI-\beta_{5}XIY+\beta_{6}IYY-\beta_{7}XYY$} \\ 

		39			 & $R_y^{\pi/2}$ & $R_x^{\pi/2}$ & $R_y^{\pi/2}$ & {$\scriptstyle-\beta_{1}XII+\beta_{2}IYI-\beta_{3}IIX-\beta_{4}XYI+\beta_{5}XIX-\beta_{6}IYX+\beta_{7}XYX$} \\ 

		40			 & $R_y^{\pi/2}$ & $R_x^{\pi/2}$ & $R_x^{\pi}$	 & {$\scriptstyle-\beta_{1}XII+\beta_{2}IYI-\beta_{3}IIZ-\beta_{4}XYI+\beta_{5}XIZ-\beta_{6}IYZ+\beta_{7}XYZ$} \\

		41			 & $R_y^{\pi/2}$ & $R_y^{\pi/2}$ & $I$			 & {$\scriptstyle-\beta_{1}XII-\beta_{2}IXI+\beta_{3}IIZ+\beta_{4}XXI-\beta_{5}XIZ-\beta_{6}IXZ+\beta_{7}XXZ$} \\ 

		42			 & $R_y^{\pi/2}$ & $R_y^{\pi/2}$ & $R_x^{\pi/2}$ & {$\scriptstyle-\beta_{1}XII-\beta_{2}IXI+\beta_{3}IIY+\beta_{4}XXI-\beta_{5}XIY-\beta_{6}IXY+\beta_{7}XXY$} \\ 

		43			 & $R_y^{\pi/2}$ & $R_y^{\pi/2}$ & $R_y^{\pi/2}$ & {$\scriptstyle-\beta_{1}XII-\beta_{2}IXI-\beta_{3}IIX+\beta_{4}XXI+\beta_{5}XIX+\beta_{6}IXX-\beta_{7}XXX$} \\ 

		44			 & $R_y^{\pi/2}$ & $R_y^{\pi/2}$ & $R_x^{\pi}$	 & {$\scriptstyle-\beta_{1}XII-\beta_{2}IXI-\beta_{3}IIZ+\beta_{4}XXI+\beta_{5}XIZ+\beta_{6}IXZ-\beta_{7}XXZ$} \\

		45			 & $R_y^{\pi/2}$ & $R_x^{\pi}$	 & $I$			 & {$\scriptstyle-\beta_{1}XII-\beta_{2}IZI+\beta_{3}IIZ+\beta_{4}XZI-\beta_{5}XIZ-\beta_{6}IZZ+\beta_{7}XZZ$} \\ 

		46			 & $R_y^{\pi/2}$ & $R_x^{\pi}$	 & $R_x^{\pi/2}$ & {$\scriptstyle-\beta_{1}XII-\beta_{2}IZI+\beta_{3}IIY+\beta_{4}XZI-\beta_{5}XIY-\beta_{6}IZY+\beta_{7}XZY$} \\ 

		47			 & $R_y^{\pi/2}$ & $R_x^{\pi}$	 & $R_y^{\pi/2}$ & {$\scriptstyle-\beta_{1}XII-\beta_{2}IZI-\beta_{3}IIX+\beta_{4}XZI+\beta_{5}XIX+\beta_{6}IZX-\beta_{7}XZX$} \\ 

		48			 & $R_y^{\pi/2}$ & $R_x^{\pi}$	 & $R_x^{\pi}$	 & {$\scriptstyle-\beta_{1}XII-\beta_{2}IZI-\beta_{3}IIZ+\beta_{4}XZI+\beta_{5}XIZ+\beta_{6}IZZ-\beta_{7}XZZ$} \\

		49			 & $R_x^{\pi}$   & $I$			 & $I$			 & {$\scriptstyle-\beta_{1}ZII+\beta_{2}IZI+\beta_{3}IIZ-\beta_{4}ZZI-\beta_{5}ZIZ+\beta_{6}IZZ-\beta_{7}ZZZ$} \\ 

		50			 & $R_x^{\pi}$   & $I$			 & $R_x^{\pi/2}$ & {$\scriptstyle-\beta_{1}ZII+\beta_{2}IZI+\beta_{3}IIY-\beta_{4}ZZI-\beta_{5}ZIY+\beta_{6}IZY-\beta_{7}ZZY$} \\ 

		51			 & $R_x^{\pi}$   & $I$			 & $R_y^{\pi/2}$ & {$\scriptstyle-\beta_{1}ZII+\beta_{2}IZI-\beta_{3}IIX-\beta_{4}ZZI+\beta_{5}ZIX-\beta_{6}IZX+\beta_{7}ZZX$} \\ 

		52			 & $R_x^{\pi}$   & $I$			 & $R_x^{\pi}$	 & {$\scriptstyle-\beta_{1}ZII+\beta_{2}IZI-\beta_{3}IIZ-\beta_{4}ZZI+\beta_{5}ZIZ-\beta_{6}IZZ+\beta_{7}ZZZ$} \\

		53			 & $R_x^{\pi}$   & $R_x^{\pi/2}$ & $I$			 & {$\scriptstyle-\beta_{1}ZII+\beta_{2}IYI+\beta_{3}IIZ-\beta_{4}ZYI-\beta_{5}ZIZ+\beta_{6}IYZ-\beta_{7}ZYZ$} \\ 

		54			 & $R_x^{\pi}$   & $R_x^{\pi/2}$ & $R_x^{\pi/2}$ & {$\scriptstyle-\beta_{1}ZII+\beta_{2}IYI+\beta_{3}IIY-\beta_{4}ZYI-\beta_{5}ZIY+\beta_{6}IYY-\beta_{7}ZYY$} \\ 

		55			 & $R_x^{\pi}$   & $R_x^{\pi/2}$ & $R_y^{\pi/2}$ & {$\scriptstyle-\beta_{1}ZII+\beta_{2}IYI-\beta_{3}IIX-\beta_{4}ZYI+\beta_{5}ZIX-\beta_{6}IYX+\beta_{7}ZYX$} \\ 

		56			 & $R_x^{\pi}$   & $R_x^{\pi/2}$ & $R_x^{\pi}$	 & {$\scriptstyle-\beta_{1}ZII+\beta_{2}IYI-\beta_{3}IIZ-\beta_{4}ZYI+\beta_{5}ZIZ-\beta_{6}IYZ+\beta_{7}ZYZ$} \\

		57			 & $R_x^{\pi}$   & $R_y^{\pi/2}$ & $I$			 & {$\scriptstyle-\beta_{1}ZII-\beta_{2}IXI+\beta_{3}IIZ+\beta_{4}ZXI-\beta_{5}ZIZ-\beta_{6}IXZ+\beta_{7}ZXZ$} \\ 

		58			 & $R_x^{\pi}$   & $R_y^{\pi/2}$ & $R_x^{\pi/2}$ & {$\scriptstyle-\beta_{1}ZII-\beta_{2}IXI+\beta_{3}IIY+\beta_{4}ZXI-\beta_{5}ZIY-\beta_{6}IXY+\beta_{7}ZXY$} \\ 

		59			 & $R_x^{\pi}$   & $R_y^{\pi/2}$ & $R_y^{\pi/2}$ & {$\scriptstyle-\beta_{1}ZII-\beta_{2}IXI-\beta_{3}IIX+\beta_{4}ZXI+\beta_{5}ZIX+\beta_{6}IXX-\beta_{7}ZXX$} \\ 

		60			 & $R_x^{\pi}$   & $R_y^{\pi/2}$ & $R_x^{\pi}$	 & {$\scriptstyle-\beta_{1}ZII-\beta_{2}IXI-\beta_{3}IIZ+\beta_{4}ZXI+\beta_{5}ZIZ+\beta_{6}IXZ-\beta_{7}ZXZ$} \\

		61			 & $R_x^{\pi}$   & $R_x^{\pi}$	 & $I$			 & {$\scriptstyle-\beta_{1}ZII-\beta_{2}IZI+\beta_{3}IIZ+\beta_{4}ZZI-\beta_{5}ZIZ-\beta_{6}IZZ+\beta_{7}ZZZ$} \\ 

		62			 & $R_x^{\pi}$   & $R_x^{\pi}$	 & $R_x^{\pi/2}$ & {$\scriptstyle-\beta_{1}ZII-\beta_{2}IZI+\beta_{3}IIY+\beta_{4}ZZI-\beta_{5}ZIY-\beta_{6}IZY+\beta_{7}ZZY$} \\ 

		63			 & $R_x^{\pi}$   & $R_x^{\pi}$	 & $R_y^{\pi/2}$ & {$\scriptstyle-\beta_{1}ZII-\beta_{2}IZI-\beta_{3}IIX+\beta_{4}ZZI+\beta_{5}ZIX+\beta_{6}IZX-\beta_{7}ZZX$}

	\end{tabular}
	\mycaption{Gate sequence for three-qubit state tomography, part 2}
	{These are the remaining 31 pulse sequences continued from the previous page.
	}
	\label{table:3qubittomosequence_b}
\end{table}

This method works just as well for more than two qubits.  For the case of three, we say that the measurement produces a voltage given by $\hat{V}_3^M \equiv \beta_{0}\langle III\rangle + \beta_{1}\langle ZII\rangle + \beta_{2}\langle IZI\rangle + \beta_{3}\langle IIZ\rangle + \beta_{4}\langle ZZI\rangle + \beta_{5}\langle ZIZ\rangle + \beta_{6}\langle IZZ\rangle+\beta_{7}\langle ZZZ\rangle$ (here, we have switched to enumerating rather than labeling the $\beta$ coefficients for the sake of brevity).  We can again calibrate the coefficients by creating and measuring all the computational basis states ($\ket{000}$ through $\ket{111}$), and, by pre-pending the measurement with all combinations of rotations, infer each Pauli correlation.  The only difference is that there are a {\it lot} more correlations to worry about.  While with two qubits we only had $4^2-1 = 15$ numbers to infer, with three we have $4^3-1 = 63$ numbers.  The list of tomographic pre-rotations necessary to measure all of these numbers is listed in \tref{table:3qubittomosequence_a} and \tref{table:3qubittomosequence_b}.  While the math for one qubit is trivial and two qubits isn't too taxing, it is wise to use a formal matrix inversion to convert our measurements to Pauli correlations when dealing with three qubits or more.  Listing the rotations required for four qubits or more is left as an exercise for the reader, for the sake of saving (an exponentially-increasing amount of) paper.

\begin{figure}
	\centering
	\includegraphics{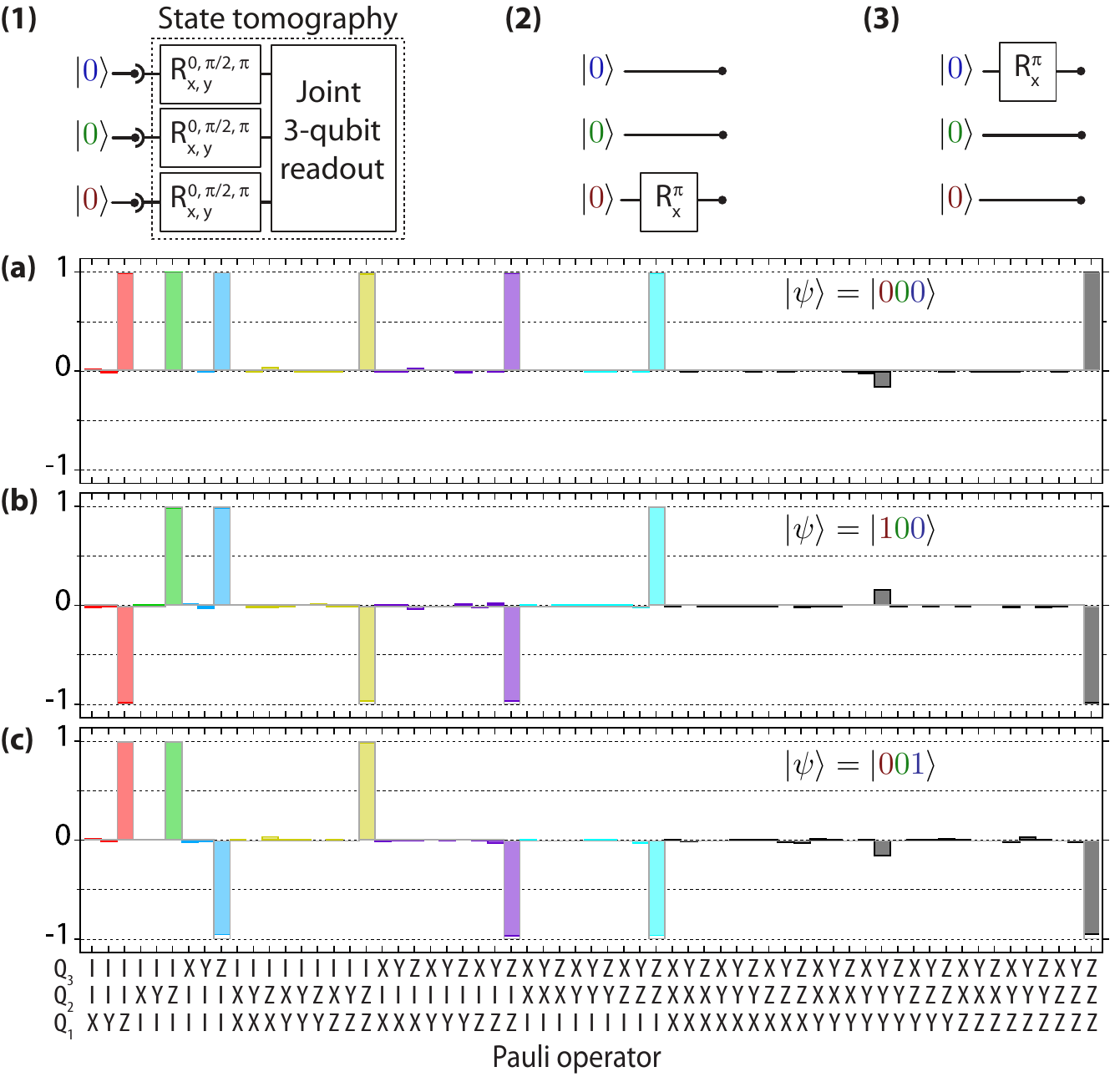}
	\mycaption{State tomography of separable states}
	{We prepare three separable states as shown in the circuit diagrams of \capl{(1)} $\ket{000}$, \capl{(2)} $\ket{100}$, and \capl{(3)} $\ket{001}$.  
	As shown in (1) and implied for the other two cases, immediately following state preparation we perform three-qubit state tomography.  The resulting tomogram is shown in \capl{(a-c)} for the three states in the Pauli bars representation.  We plot the expected values of the 63 Pauli operators for the state, grouping them with single-qubit operators (red, green, blue for operators on $Q_1$, $Q_2$, and $Q_3$ respectively), two-qubit operators (yellow, purple, and teal for $Q_1$ and $Q_2$, $Q_1$ and $Q_3$, and $Q_3$ and $Q_3$), and three-qubit operators (black).
	\figadapt{DiCarlo2010}
	}
{\label{fig:separable3qstatetomo}}
\end{figure}

We demonstrate three-qubit state tomography in \figref{fig:separable3qstatetomo}.  In (1-3) we show a circuit diagram for performing state tomography on several simple states.  Assuming that the qubits start in $\ket{000}$, we prepare some quantum state of the register and then perform tomography by applying the appropriate single-qubit rotations on all three qubits and measuring with our joint readout.  We show the resulting reconstructed density matrix in (a) using the {\it Pauli bar representation}, in which we plot the value of each Pauli correlation.  The types of correlations are segregated by color: red, green, and blue for single-qubit correlations of $Q_1$, $Q_2$, and $Q_3$ respectively, followed by yellow, purple, and teal for two-qubit correlations between $Q_1$ and $Q_2$, $Q_1$ and $Q_3$, and $Q_2$ and $Q_3$; finally, in black, we have three-qubit correlations.  As we will show in \sref{section:entanglementondemand}, in contrast to the conventional cityscape density matrix plot, this representation makes the presence of entanglement unambiguous.

We first show the resulting tomographic data for measuring the ground state.  As shown in \figref{fig:separable3qstatetomo}(a), we have seven unit-height bars for the $\langle ZII \rangle$, $\langle IZI \rangle$, $\langle IIZ \rangle$, $\langle ZZI \rangle$, $\langle ZIZ \rangle$, $\langle IZZ \rangle$, and $\langle ZZZ \rangle$ correlations.  The value of $\langle III \rangle$ is defined to be 1 and is suppressed in all plots, though it should be included when calculating state fidelity.  It is easy to see why these bars are all $1$.  For example, we can break the $ZIZ$ correlator up into $\langle 000 | ZIZ | 000 \rangle = \langle 0 | Z | 0 \rangle \langle 0 | I | 0 \rangle \langle 1 | Z | 1 \rangle$ because the operators commute through qubit states that they do not address.  Then, because $\langle 0 | Z | 0\rangle = 1$ and $\langle 0 | 0 \rangle = 1$, we have $\bra{000} ZIZ \ket{000}  = (+1)(+1)(+1) = +1$.  This same argument applies to all of the Pauli correlations involving only $I$ and $Z$.  Other correlations, for example $IIX$, are zero because the ground state is orthogonal to the eigenstates of the Pauli $\sigma_x$ operator.  In \figref{fig:separable3qstatetomo}(b) and (c) we show tomograms for the states $\ket{100}$ and $\ket{001}$.  These are similar to the ground state, except that the correlations involving the $Z$ operator for the flipped qubit are now negative because $\langle 1 | Z | 1 \rangle = -1$.

Looking at how much information is stored in only three qubits, you can imagine how complicated this gets as the number continues to increase.  Indeed, the exponential growth of the amount of information stored in a manifold of qubits is a reflection of the sample complexity of quantum mechanics that we're trying to leverage in building our computer.  The addition of each qubit multiplies the amount of information by about a factor of four.  We were fortunate that a single-state tomogram required only 5 minutes to measure in this device (a reflection of high readout fidelity and a fast repetition rate), but adding another qubit would translate that to a 20 minute task and one more would require more than an hour.  And as we will see in the next section, this scaling worsens dramatically when characterizing a full quantum process involving $N$ qubits.  Going forward, it will become more and more necessary to employ {\it compressed sensing} \cite{Donoho2006, Kosut2009, Gross2010} to measure our system, wherein we extract only the most relevant properties, chosen based on external knowledge or assumptions.

\subsection{Errors and physicality of state tomograms}
\label{subsec:tomophysicality}

\begin{figure}
	\centering
	\includegraphics{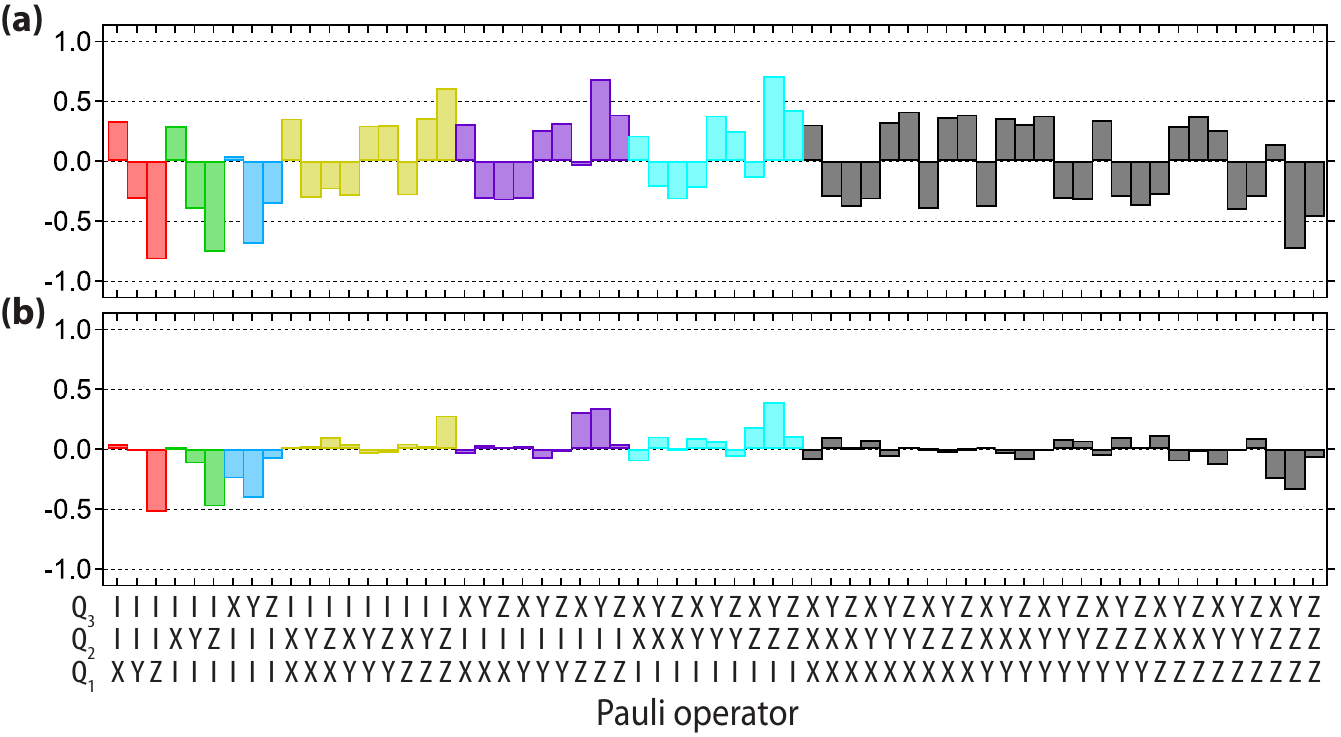}
	\mycaption{State tomography with population outside the Hilbert space}
	{\capl{(a)} We intentionally put population in the $\ket{102}$ state by preparing $\ket{111}$, moving suddenly into resonance with the $\ket{111} \leftrightarrow \ket{102}$ avoided crossing, waiting for half the splitting time, and then moving suddenly back.  The tomogram shows significant unphysical correlations because the tomographic pre-rotations do not address the non-computational population.  Experimentally, this causes a constant voltage offset in all of our measurements.  
	\capl{(b)} We manually subtract a constant voltage from all the measurements to reduce this effect, chosen by minimizing the squared amplitude of the density matrix (e.g. assuming that it is maximally mixed).  This tomogram is a more accurate representation of the computational space population. }
	{\label{fig:tomoextrahilbertpop}}
\end{figure}

There are several conditions that are important to keep in mind when using this method of performing state tomography.  This process assumes that the entire qubit density matrix occupies only the computational space; that is, the Hilbert space that our pre-rotations address.  To be more complete, our translation of the measurement operator $\ket{0^{\otimes N}}\bra{0^{\otimes N}}$ should also include operators that address higher excited states of the transmon, since a transmon in its second excited state will also give us a positive measurement outcome.  As shown in \figref{fig:tomoextrahilbertpop}(a), this can have dramatic effects on our reconstructed density matrix because any population outside the addressed space will always be present.  As shown in \figref{fig:tomoextrahilbertpop}(b), it will cause an offset voltage, as if our value for $\beta_0$ is incorrect.  (In Ref.~\citenum{DiCarlo2010}, this was explicitly excised with the addition of a best-fit parameter to $\beta_0$ that was chosen to minimize the sum of the magnitude of the bars.  This strategy was not used in Ref.~\citenum{Reed2012}.)  We also assume that our $\beta$ values are calibrated with perfectly prepared test states.  If there is some thermal equilibrium qubit population, for example, we will only address the ``pure'' quantum amplitude given by $P(\ket{0}) - P(\ket{1})$.  Thus, the fidelities we claim from tomography do {\it not} take into account the impurity of our ground state.  Finally, if our measurement pre-rotations are imperfect, our inferred values will also be flawed \cite{Merkel2012,Paik2013}.

All of these issues contribute to the un-physicality of our reconstructed density matrix.  Our density matrix automatically has the required property $\mathrm{tr}\left(\rho\right)=1$ since $\langle I^{\otimes N}\rangle \equiv 1$.  However, a physical matrix must additionally have all positive eigenvalues, which is not guaranteed here.  For example, nothing prevents the inferred value of a Pauli correlation from being larger than $\pm1$, be it due to noise or systematic errors.  This adds ambiguity to the definition of things like state fidelities, which involve taking a dot product of the measured Pauli correlations with their ideal values.  This is an especially large problem when fidelities approach $100\%$, since there is a significant difference in a state fidelity of $99.9\%$ and $99.0\%$ from the point of view of fault tolerant error correction.  Estimating the size of both systematic and random errors presents a challenge.

There are several strategies to deal with these sources of error.  You want your state tomograms to be as reliable as possible.  This involves calibrating single-qubit rotations as well as you can, both when addressing a single qubit and when simultaneously rotating multiple qubits.  Running sequences like AllXY with and without auxiliary $\pi$ pulses, as described in \sref{subsec:allxy}, can help, as can its more sensitive derivatives.  In a recent paper \cite{Paik2013}, it was also necessary to use NMR-inspired composite rotations that are less susceptible to certain types of errors.

It has become common to perform a {\it maximum likelihood estimation} of the density matrix \cite{James2001, Jezek2003}.  This process applies Lagrange multipliers to the inferred Pauli correlation values to enforce the physicality.  As described in section 2.5.2 of Jerry Chow's thesis \cite{ChowThesis}, this involves evaluating polynomial functions of measurement results and is therefore nonlinear.  This complicates the estimation of uncertainties.  A maximum likelihood estimation makes the smallest possible changes to the matrix to make it physical, which is most justifiable when random errors are the dominant source of un-physicality \cite{Riebe2006}.  These changes are arbitrary, however.  If we have systematic rather than random errors, there is no particular reason to trust the maximum-likelihood matrix more than the original one.  Nevertheless, many research groups \cite{Riebe2006, Chow2010b, Fedorov2011} use this process and argue that without enforcing physicality, metrics like state fidelity have no meaning.

Finally, there has been a recent proposal to perform tomography in a {\it self-consistent} way by oversampling the action of our measurement pre-rotations \cite{Merkel2012}.  The idea is to relax the assumption that the unitary rotations we perform prior to measurement are perfect, and instead to infer their effect by extensively sampling measurement results following their application.  While this approach should mitigate one source of systematic error, it still requires nonlinear convex optimization.  Though there exists a linear map of measurement results to the process matrix of each of our rotations, measurements have finite noise which will be magnified in the state tomogram unless it is mitigated.  In order to make this optimization tractable, the authors linearize the involved equations.  This has the effect of simplifying the math, but also requires that the rotations used are fairly close to the intended rotations around which the equations were expanded.  Nevertheless, this method promises to reduce sensitivity to errors and enforce physicality in a more realistic and thoughtful way than a relatively crude maximum likelihood estimation.

\section{Process tomography}
\label{sec:processtomo}

\nomdref{Gxchimn}{$\chi_{mn}$}{$\chi$-matrix representation of the quantum process matrix}{sec:processtomo}

What if we are interested not in the density matrix of a state, but rather, in the unitary evolution of a {\it process}.  For example, suppose someone gives us a black box which takes as an input $N$ qubits and gives the same number as an output, but with a modified state.  How can we determine the action of the box?  To begin, we might insert the ground state ($\ket{00}$, supposing for now that $N=2$) and measure the density matrix of the output state.  We must perform many measurements to perform state tomography, so we prepare an ensemble of our test ground states and apply the process to each.  This will tell us only a tiny amount about what the box does: we will learn its affect on the ground state.  For two qubits, that accounts for only one point in 4-dimensional Hilbert space with 16 orthogonal basis vectors.  We send another state through, selecting this time the state in which the first qubit is pointing along the $x$-axis and the other is in its ground state ($\left(\ket{00} + \ket{10}\right)/\sqrt{2}$).  This is actually a huge improvement over having only one point because quantum mechanics is linear.  Since we know the output density matrix for both states, we also know what happens to any {\it superposition} of those states: the result of {\it any} rotation about the $y$-axis on the first qubit alone.  If we know how the process affects each of the $4^N$ basis states for $N$ qubits, in principle, we know everything about the process.

In practice, we wish to express our process data in a more convenient way.  Specifically, following \eref{eqn:krausrep}, we want a set of operators ${E_i}$ for the process $\mathcal{E}(\rho) = \sum_i E_i \rho E_i^\dagger$\footnotemark.  We express these operators $E_i$ in some fixed basis of operators $\tilde{E}_i$ such that $E_i = \sum_m e_{im} \tilde{E}_m$, which we may plug into the expression for $\mathcal{E}$ above to give
\begin{equation}
	\mathcal{E}(\rho) = \sum_{mn} \tilde{E}_m \rho \tilde{E}^\dagger_n \chi_{mn}
\end{equation}
with $\chi_{mn} = \sum_i e_{im} e^*_{im}$.  The basis ${\tilde{E}_i}$ is arbitrary, but the $N$-qubit Pauli matrices are most often chosen.  The matrix $\chi$ is the ``chi-matrix representation'' of our process, and measuring it is our goal.  It will contain $d^4-d^2$ numbers, where $d$ is the dimension of the system (e.g. $d=2$ for one qubit).  From state tomography, we know $\mathcal{E}(\rho_j) = \sum_k \lambda_{jk} \rho_k$, where $\rho_k$ is our basis of states and $\lambda_{jk}$ are coefficients.  These $\lambda_{jk}$ are given by $\lambda_{jk} = \mathrm{tr}(\rho_k \rho)$, which we identify as the Pauli tomography bars (introduced in the previous section).   We relate our choice of operator basis to our state basis with a matrix $\beta$, defined by
\begin{equation}
	\tilde{E}_m \rho_j \tilde{E}_n^\dagger = \sum_k \beta_{jk}^{mn} \rho_k.
\end{equation}
There, {$\tilde{E}_i$} is again the operator basis in which we are expressing the process, $\rho_j$ are the basis states of our Hilbert space that we inserted into the black box and performed tomography on, and $\rho_k$ is the eigenbasis with which we express the resulting experimental density matrix.  Plugging this expression into the equation above, we have
\begin{equation}
	\sum_k \sum_{mn} \chi_{mn} \beta_{jk}^{mn} \rho_k = \sum_k \lambda_{jk} \rho_k.
\end{equation}
Since $\rho_k$ is a linearly independent basis, it follows that $\sum_{mn} \chi_{mn} \beta_{jk}^{mn} = \lambda_{jk}$.  Inverting $\beta$ with $\kappa = \beta^{-1}$, we arrive at an expression relating our tomographic measurement record $\lambda$ to the process matrix representation $\chi$ with
\begin{equation}
	\chi_{mn} = \sum_{jk} \kappa_{jk}^{mn} \lambda_{jk}
\end{equation}
where, again, the index $j$ labels the state we test with, $k$ specifies the basis with which we express that state, and $m$ and $n$ are the indices of the $\chi$-matrix and specify the operator basis of $E_i$.  Code which implements this math can be found in \aref{ap:mathematica}.

\footnotetext{This paragraph follows pages 390-392 of {\it Quantum Computation and Quantum Information} by Michael Nielsen and Isaac Chuang \cite{Nielsen2000}.}

\begin{figure}
	\centering
	\includegraphics{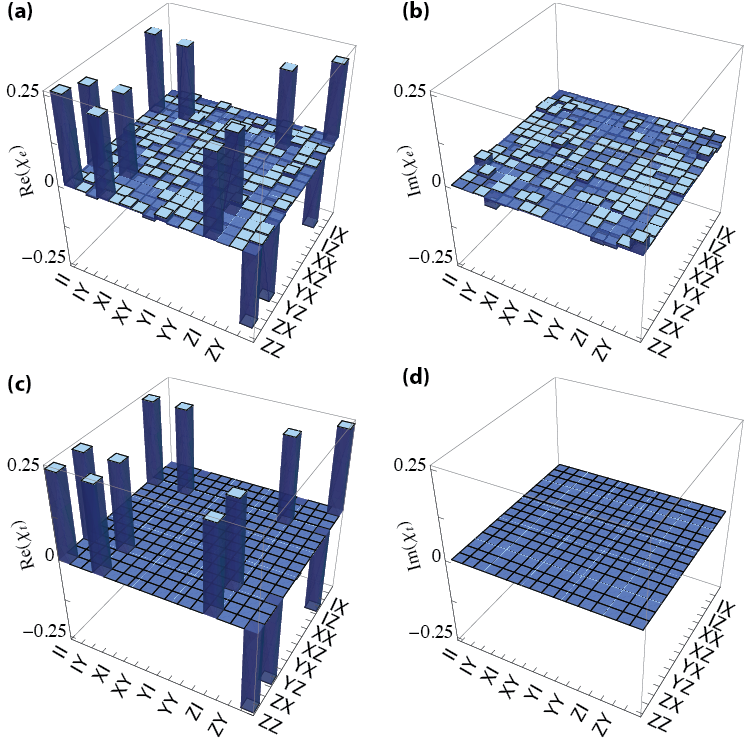}
	\mycaption{Process tomography of sudden two-qubit controlled-phase gate}
	{The plots in \capl{(a-b)} show the real and imaginary part of the experimentally measured process matrix $\chi_{e}$, while \capl{(c-d)} show the theoretically calculated $\chi_t$.  The operator basis used is II, IX, IY, IZ, XI, XX, XY, XZ, YI, YX, YY, YZ, ZI, ZX, ZY, and ZZ.  The fidelity of this operation is given by the dot product $\mathrm{tr}(\chi_e \chi_t) = 91.6\%.$ } 
	{\label{fig:suddenproctomo}}
\end{figure}

The previous paragraph is necessarily rather abstract, but do not lose sight of the basic idea.  We are simply applying some unknown process to a series of states that span the Hilbert space, and then, using the fact that quantum mechanics is linear, inferring what would happen to superpositions of those states.  Measuring process tomography is a matter of measuring {\it state} tomography many times.  In \figref{fig:suddenproctomo}, we show the experimental and theoretical process matrix for a sudden two-qubit controlled-phase gate.  The matrix $\chi$ can be complex-valued, depending on the operator basis used, and so we plot both the real and imaginary parts.  Just as with state tomography, process tomography can also suffer from unphysical errors.  For this reason, significant effort has been made recently in developing more robust and efficient ways to characterize processes \cite{deSilva2011, Fedorov2011, Steffen2012, Moussa2012, Flammia2012, Corcoles2013}.

\section{Entanglement on demand}
\label{section:entanglementondemand}

\begin{figure}
	\centering
	\includegraphics{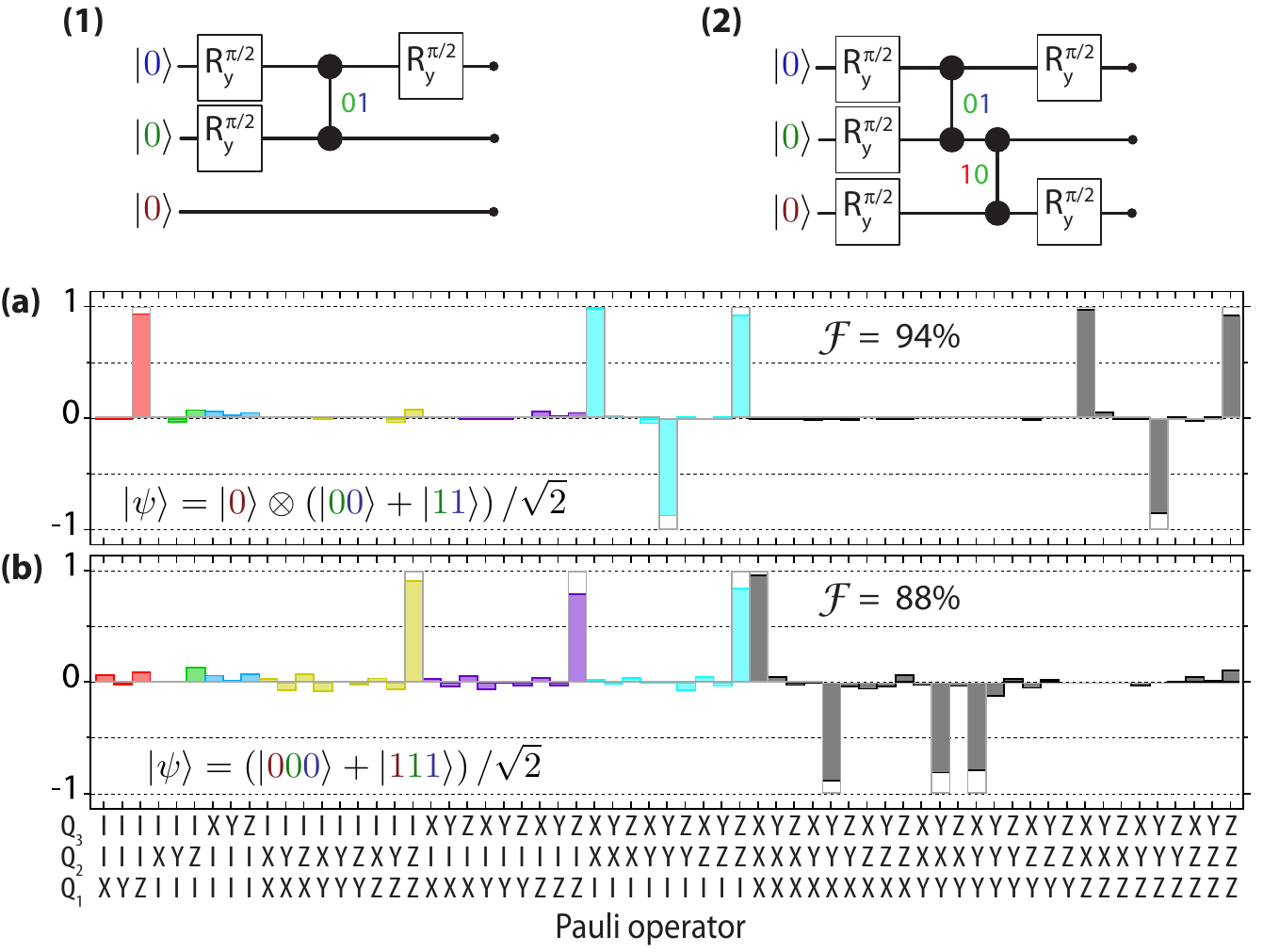}
	\mycaption{Circuit model and state tomography of Bell and GHZ states} 
	{In \capl{(1)} we show the circuit for preparing a Bell state between $Q_2$ and $Q_3$.  Lines between qubit tracks terminating in circles indicates a controlled-phase gate, with the state receiving the $-1$ sign indicated adjacent to the line.  
	In \capl{(a)} we show the Pauli bar tomogram of the resulting state.  $Q_1$ still has a single-qubit $Z$ correlation of $+1$ because it is in the ground state, but $Q_2$ and $Q_3$ have no single-qubit correlation since they are entangled.  They do have strong two-qubit correlations, however, which are echoed in the three-qubit correlations.  The fidelity of this state to its target, calculated by taking a dot product of the Pauli bar vector with the ideal one (including the value of +1 for $\langle III \rangle$), is $94\%$.  
	In \capl{(2)} we show a circuit to extend the entanglement to all three qubits, and plot the measured tomogram in \capl{(b)}.  The lack of single-qubit correlations and presence of strong two- and three-qubit correlations indicates that this is a three-qubit entangled GHZ state.  The resulting state fidelity is $88\%$.
	\figadapt{DiCarlo2010}
	}
{\label{fig:ghztomo}}
\end{figure}

Now that we have the tools of entangling gates and efficient state tomography, we can demonstrate entanglement.  We begin with the simple case of generating entanglement between only two qubits.  The gate sequence for generating a Bell state between $Q_2$ and $Q_3$ is shown in \figref{fig:ghztomo}(a).  We start by putting both qubits in a superposition of $\ket{0}$ and $\ket{1}$ with $\pi/2$ rotations about the $y$-axis of the Bloch sphere, with $Q_1$ unused.  This transfers $\ket{000} \rightarrow \ket{0} \otimes \left( \ket{00} + \ket{10} + \ket{01} + \ket{11} \right) / 2$.  We then apply a cPhase gate (in this case, we use the sudden version described above) with the $-1$ sign set to the $\ket{01}$ state, giving us $\ket{\psi} = \ket{0} \otimes \left( \ket{00} + \ket{10} - \ket{01} + \ket{11} \right) / 2$.  At this point, we are in a maximally entangled Bell state of the top two qubits, though it is not yet in an easily recognizable configuration.  To achieve that, we finish with a second $\pi/2$ rotation about $y$ on $Q_3$ to give us $\ket{\mathrm{Bell}_{23}}=\ket{0} \otimes \left(\ket{00} + \ket{11} \right)/\sqrt{2}$ -- the canonical Bell state.

We perform state tomography on the result of this procedure, shown in \figref{fig:ghztomo}(b).  The order of the Pauli correlations emphasizes entanglement.  As previously mentioned, the first three sets, denoted in red, green, and blue, show single-qubit correlations.  Note that while there is a strong (ideally, $+1$) $Z$ correlation for $Q_1$, there are no nonzero single-qubit correlations for either $Q_2$ or $Q_3$.  This might be an indication that both of those qubits are in a fully mixed state; however, looking further down the tomogram, we find that we have strong two-qubit correlations between these qubits (in $IXX$, $IYY$, and $IZZ$).  This is an unambiguous sign of entanglement between those two qubits: neither qubit has any individual character, but considered simultaneously with its pair, we observe strong correlations.  We also see that the two-qubit correlations involving $Q_1$ and either $Q_2$ or $Q_3$ are zero, again reinforcing the idea that $Q_2$ and $Q_3$ cannot be described separately.  Finally, we have three-qubit correlations which are a trivial extension of our two-qubit correlations plus $Q_1$ in its ground state.  The fidelity of this state to the intended state is $94\%$, calculated by taking the dot product of the ideal state vector with the measured values (including the $III$ bar) and dividing by 8.

To entangle all three qubits, we can extend this procedure with an additional cPhase gate.  As shown in \figref{fig:ghztomo}(2), we first put all three qubits on the equator of their Bloch sphere.  The state at this point will be an equal superposition of all computational states, given by 
\begin{equation*}
	\ket{\psi} = \left(\ket{000} + \ket{100} + \ket{010} + \ket{001} + \ket{110} + \ket{101} + \ket{011} + \ket{111} \right)/2\sqrt{2}.
\end{equation*}
Performing conditional phase gates between $Q_2$ and $Q_3$ with the conditional phase on $01$ flips the phase of $\ket{001}$ and $\ket{101}$ (since the state of $Q_1$ is irrelevant for that gate), and the second conditional phase gate between $Q_1$ and $Q_2$ with the phase on $10$ flips the phase of $\ket{100}$ and $\ket{101}$.  The concatenation of these two gates will thus flip $\ket{001}$ and $\ket{100}$ only, since $\ket{101}$ is flipped twice, yielding
\begin{equation*}
	 \ket{\psi} = \left(\ket{000} - \ket{100} + \ket{010} - \ket{001} + \ket{110} + \ket{101} + \ket{011} + \ket{111} \right)/2\sqrt{2}.
\end{equation*}
As in the two-qubit case, this state is already a maximally entangled three-qubit state, though in this basis it is difficult to see that.  Thus, to finish, we apply $\pi/2$ rotations about $y$ on $Q_1$ and $Q_3$, which give us the simple and unambiguously entangled state $\ket{GHZ} = \left(\ket{000} + \ket{111}\right)/\sqrt{2}$.  Like the Bell state, this state also has a special name: a three-qubit Greenberger-Horne-Zeilinger or {\it GHZ state} \cite{Greenberger1989} (\sref{subsec:entanglement}).

The presence of three-qubit entanglement is clear when looking at the measured state tomogram depicted in \figref{fig:ghztomo}(d).  Following the argument we made for the case of the two-qubit Bell state, we see here that there are {\it no} nonzero single-qubit correlations.  There are, however, strong two-qubit $ZZ$ correlations of any pair of qubits.  This alone tells us that all three qubits are entangled with one another.  Though they do not point in any particular direction by themselves, when measured as a group they have a definite direction.  There are also nontrivial three-qubit correlations of $XXX$, $YYX$, $YXY$, and $XYY$, which, as we will see later, are sufficient to witness three-qubit entanglement.  The fidelity of this state to the state we intended to create is $88\%$.

\subsubsection{An aside on the Pauli correlation representation}

\begin{figure}
	\centering
	\includegraphics{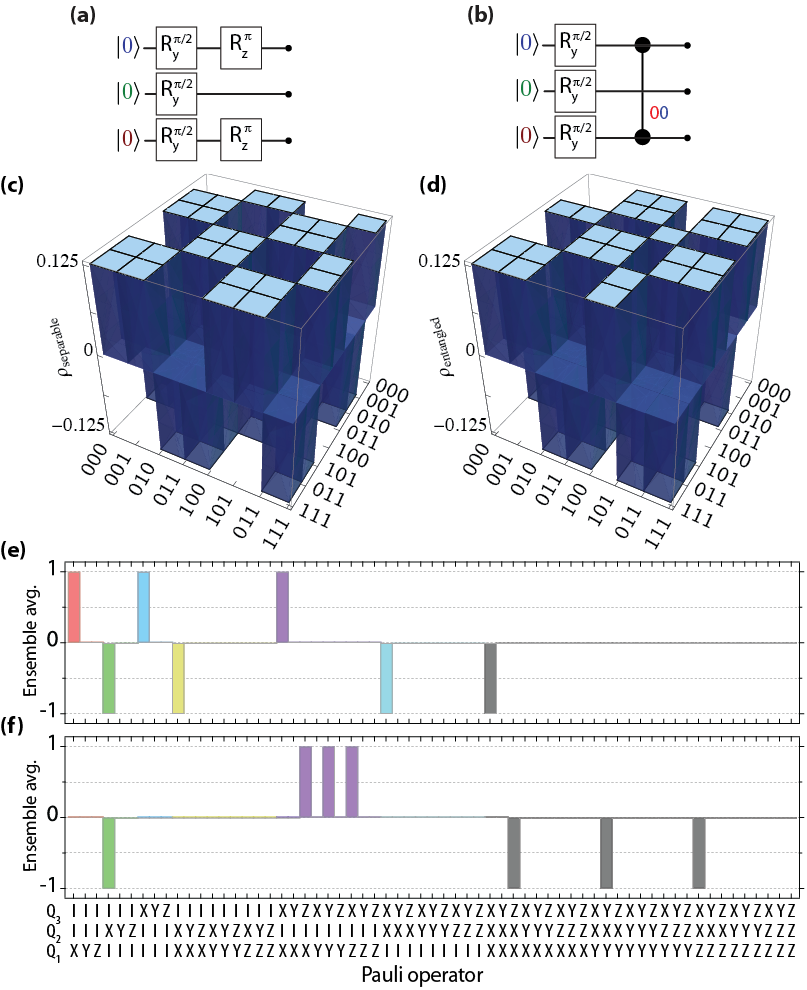}
	\mycaption{Calculated separable and entangled state tomograms}
	{We calculate the density matrices of the states produced by the circuits shown in \capl{(a)} and \capl{(b)} and plot their cityscape density matrices in \capl{(c)} and \capl{(d)}.  The state shown in \capl{(c)} is separable and the state in \capl{(d)} is a maximally-entangled Bell state between qubits 1 and 3, but it is very difficult to distinguish them in this representation (here $\rho$ is real).  The Pauli representation of the same states, shown in \capl{(e)} and \capl{(f)} unambiguously communicates that the first state is separable and the second entangled because the first has strong single-qubit correlations and few two-qubit ones, while the second is the opposite.   This demonstrates the advantage of the Pauli representation.
	}
{\label{fig:simcityscape}}
\end{figure}

A big advantage of the Pauli bar representation of a density matrix is the ability to read off the fact that these states are entangled.  We can compare this to the conventional ``cityscape'' density matrix representation, where each element of the matrix is displayed in a 3D bar plot\footnotemark.  In \figref{fig:simcityscape}(a-b), we depict circuit diagrams to create two states, one separable and the other fully-entangled.  The first case is the result of applying $\pi/2$ rotations around the $y$-axis to all three qubits and then flipping the phase of the top and bottom qubits with $\pi$ rotations around the $z$-axis.  The resulting separable state is $\ket{\psi} = \ket{000} - \ket{100} + \ket{010} - \ket{001} - \ket{110} + \ket{101} - \ket{011} + \ket{111}$.  We make the second state starting with the same initial single-qubit rotations, but then apply a controlled-phase gate between the top and bottom qubits conditioned on $\ket{00}$.  In the two-qubit subspace, this gate has the phases $\phi_{10} = \phi_{01} = \phi_{11} = \pi$.  In other words, we still apply the same single-qubit $z$ gates, but add an additional two-qubit rotation around $ZZ$.

\footnotetext{Both representations contain the same information since $\rho = \sum_{\hat{O} \in \{III...ZZZ\}} \langle \hat{O} \rangle \hat{O}$.  To convert in the opposite direction, we have $\langle \hat{0} \rangle = \mathrm{tr}\left(\rho\hat{O}\right)$.}

We calculate the density matrix resulting from both of these processes.  In \figref{fig:simcityscape}(c-d) we show the cityscape representation, plotting each of the elements of $\rho_{\mathrm{separable}}$ and $\rho_{\mathrm{entangled}}$, the density matrices for the separable and entangled states respectively.  As you can see, though the two matrices differ slightly, without labels and a lot of careful thought there would be no way of distinguishing that one matrix was fully entangled while the other is a separable state.  This reflects the fact that this state is pointing along the $x$-axis, and is quite complicated when expressed in the $z$ basis.  Contrast this representation to the Pauli bar plots shown in \figref{fig:simcityscape}(e-f) for the separable and entangled states, which clearly illustrates the difference.  In the first case, all three qubits are pointing in cardinal Bloch sphere directions (along $+x$, $-x$, and $+x$, respectively); in the second case the qubits $Q_1$ and $Q_3$ have no single-qubit correlations but have strong two-qubit correlations, unambiguously identifying the tomogram as of a fully-entangled Bell state of qubits 1 and 3.

\begin{figure}
	\centering
	\includegraphics{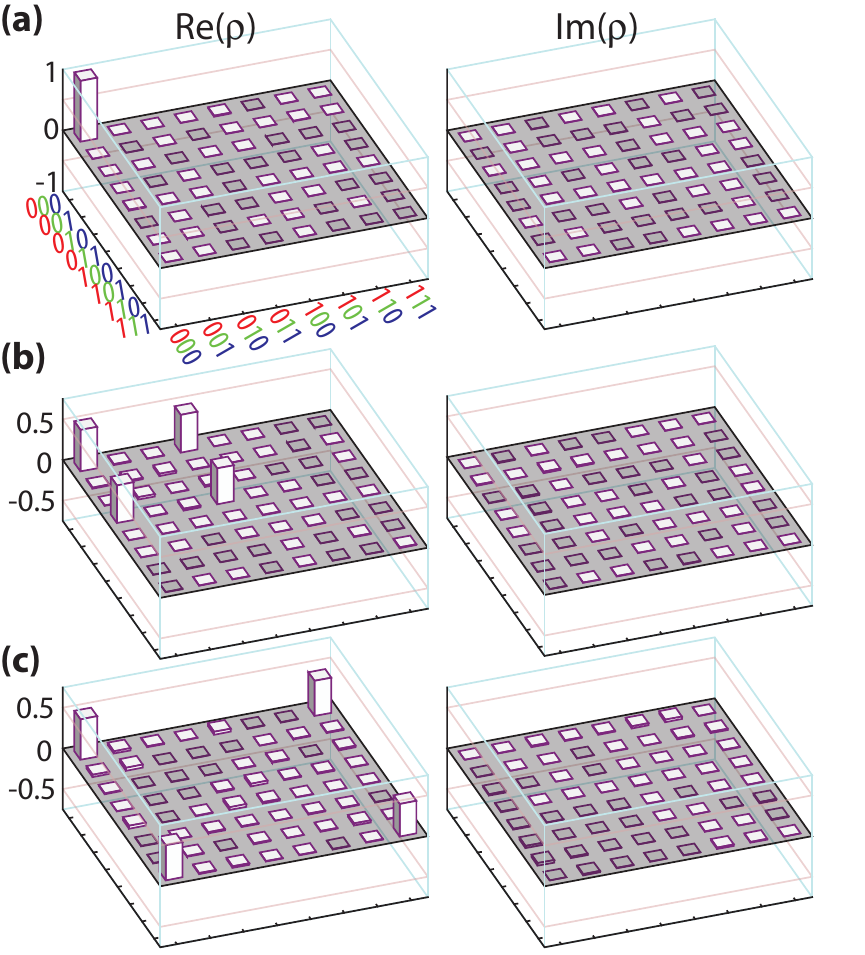}
	\mycaption{Cityscape density matrix representations of experimentally created states}
	{These are the same states shown in the Pauli representation in \figref{fig:separable3qstatetomo}(a) and \figref{fig:ghztomo}(a,b).  Entangled states of two $z$ eigenstates show the characteristic four corners of a square pattern.  These density matrices are  simple because their states are easily written in the $z$ basis. 
	\figthanks{DiCarlo2010}
	}
{\label{fig:expcityscape}}
\end{figure}

This problem is not as extreme when the state can be efficiently written in the $z$ basis.  In \figref{fig:expcityscape}, we show the density matrices corresponding to the Pauli bar data for the ground state (from \figref{fig:separable3qstatetomo}[a]), the Bell state between qubits 2 and 3 (\figref{fig:ghztomo}[a]), and the GHZ state (\figref{fig:ghztomo}[b]).  Since these states are easily written in the $z$ basis as $\ket{000}$, $\ket{0} \otimes \left( \ket{00} + \ket{11}\right)/\sqrt{2}$, and $\left(\ket{000} + \ket{111} \right)/\sqrt{2}$, the cityscape density matrices are not terribly complicated.  Nevertheless, because this symmetry is not necessarily present for any given quantum state, we strongly favor the Pauli bar representation.  This representation has also been extended to more naturally express process matrices \cite{Chow2012}, though that practice has not yet been adopted by Schoelkopf lab.

\subsubsection{Why entangle three qubits?}

Why is three-qubit entanglement interesting?  There are several reasons, but the primarily one is the prospect of doing error correction.  As we saw in \sref{subsec:quantumrepetitioncode}, the most basic quantum error correcting code requires three qubits.  This code starts identically to the procedure we have just demonstrated, encoding a quantum state in a three-qubit GHZ state.  If we change the rotation axis of the $\pi/2$ rotation on $Q_2$, we will create a different GHZ state.  Defining the angle $\phi$ relative to the $x$-axis of the Bloch sphere (so that $\phi=\pi/2$ is the $y$-axis), preparing $Q_2$ with a $\pi/2$ rotation around the axis $\hat{n}(\phi)$ will create the state $\ket{GHZ_\phi} = (\ket{000} - i e^{i\phi}\ket{111})/\sqrt{2}$.  The three-qubit codes, of which there are two distinct types, can protect this encoded state from either bit-flips or phase-flips, and demonstrate the first principles of quantum error correction.  The next chapter in this thesis will experimentally demonstrate both codes.

With three qubits, we can see the proliferation of distinct classes of entanglement for the first time, as mentioned in \sref{subsubsec:generatingentanglement}.  With two qubits, any increase of entanglement meant moving toward the manifold of Bell states.  With three qubits, however, there are two distinct classes of entanglement.  The first we have already seen, the GHZ-like states, whose canonical state is $\ket{GHZ} = \left(\ket{000} + \ket{111}\right)/\sqrt{2}$ but span the space given by any combination of single-qubit rotations on that state.  The second class, known as ``W''-like states, are most commonly expressed by the state given by $\ket{W} = \left(\ket{100} + \ket{010} + \ket{001}\right)/\sqrt{3}$, but again span the space of single-qubit rotations on that state.  The GHZ state is considered to be a higher form of three-qubit entanglement.  It exhibits stronger multi-qubit correlations (the $W$ state has some nonzero single-qubit correlations, for example), measuring any single qubit would project the entire state, and it is the kind of entanglement necessary for quantum error correction.  The $W$ state is nevertheless distinct, and cannot be transmuted to $GHZ$ class or back with local operations.

\subsection{Entanglement witnesses}
\label{sec:entanglementwitnesses}

\nomdref{Achsh}{CHSH}{Clauser-Horne-Shimony-Holt}{sec:entanglementwitnesses}
\nomdref{Alhv}{LHV}{local hidden variable}{sec:entanglementwitnesses}

We can make statements about the type and quality of entanglement we have generated without needing to know the entire density matrix.  These fall under the broad category of ``entanglement witnesses,'' the most famous of which are the CHSH (Clauser-Horne-Shimony-Holt) operators for two qubits \cite{Clauser1969, Chow2009}.  They were originally suggested in the context of fundamental tests of quantum mechanics like the EPR (Einstein-Podolsky-Rosen) \cite{Einstein1935} paradox and Bell's theorem.  Witnesses are operators whose expected values will never exceed a certain threshold for a separable state, but might if the ensemble contains entanglement.  To be concrete, one of the CHSH operators is given by $\widehat{CHSH} = \langle XX \rangle - \langle XZ \rangle + \langle ZX \rangle + \langle ZZ\rangle$, which has the property that any non-entangled state satisfies $\widehat{CHSH}<2$, while entangled states can reach values as high as $2\sqrt{2}$.  Another witness is given by $\hat{W} = \left(II-XX+YY-ZZ\right)/4$, such that $\mathrm{tr}\left(\rho \hat{W}\right)<0$ guarantees entanglement \cite{Chow2010b, Eisert2007, Clauser1969}.  Being expressed as operators, they have the advantage over other metrics of being linear.

How can we extend witnesses to three-qubit states?  The simplest is the state fidelity to a GHZ-class state.  The maximum fidelity of a {\it biseparable state} (a state with no more than two qubits entangled) to any GHZ state is $50\%$.  Any fidelity greater than that value thus guarantees the presence of some degree of three-qubit entanglement \cite{Acin2001}.  This metric also distinguishes between the types of three-qubit entanglement.  While either $W$ or $GHZ$-class states will exceed a GHZ fidelity of $50\%$, only states with $GHZ$-class entanglement, useful for quantum error correction, can have a fidelity of greater than $75\%$.  Thus, our state fidelity of $88\%$ definitively exceeds both these biseparable and GHZ bounds.  Note that we are free to choose {\it any} GHZ-class state for this purpose: we should exhaustively evaluate the fidelity of our test state to any state on the GHZ manifold and choose to compare against the largest value we find.

Though the fidelity metric is definitive, obtaining it requires reconstructing the entire density matrix.  (In principle, this is not necessary -- perhaps we could engineer our measurement operator to perform exactly $|GHZ\rangle\langle GHZ|$ -- but, for us, that would mean repeating the GHZ-creation step twice over.  This would be a challenge to tune-up reliably and independently.)  It is possible to detect three-qubit entanglement with linear witnesses, which necessitates fewer Pauli measurements than knowing the full density matrix.  For example, the Mermin sums $M_{S1}=\langle XXX \rangle - \langle YYX \rangle - \langle YXY\rangle - \langle XYY\rangle$ and $M_{S2}=-\langle YYY \rangle + \langle XXY \rangle + \langle XYX\rangle + \langle XYY\rangle$ satisfy $|M_{S1,2}| \leq 2$ for all biseparable states \cite{Toth2005}.  Notice that the first of these involves the nonzero three-qubit correlations we found in our GHZ state tomogram; they contain the information in which we are interested.  The second sum contains those bars that are maximal for another GHZ state, $\ket{000}+i\ket{111}$.

\begin{figure}
	\centering
	\includegraphics{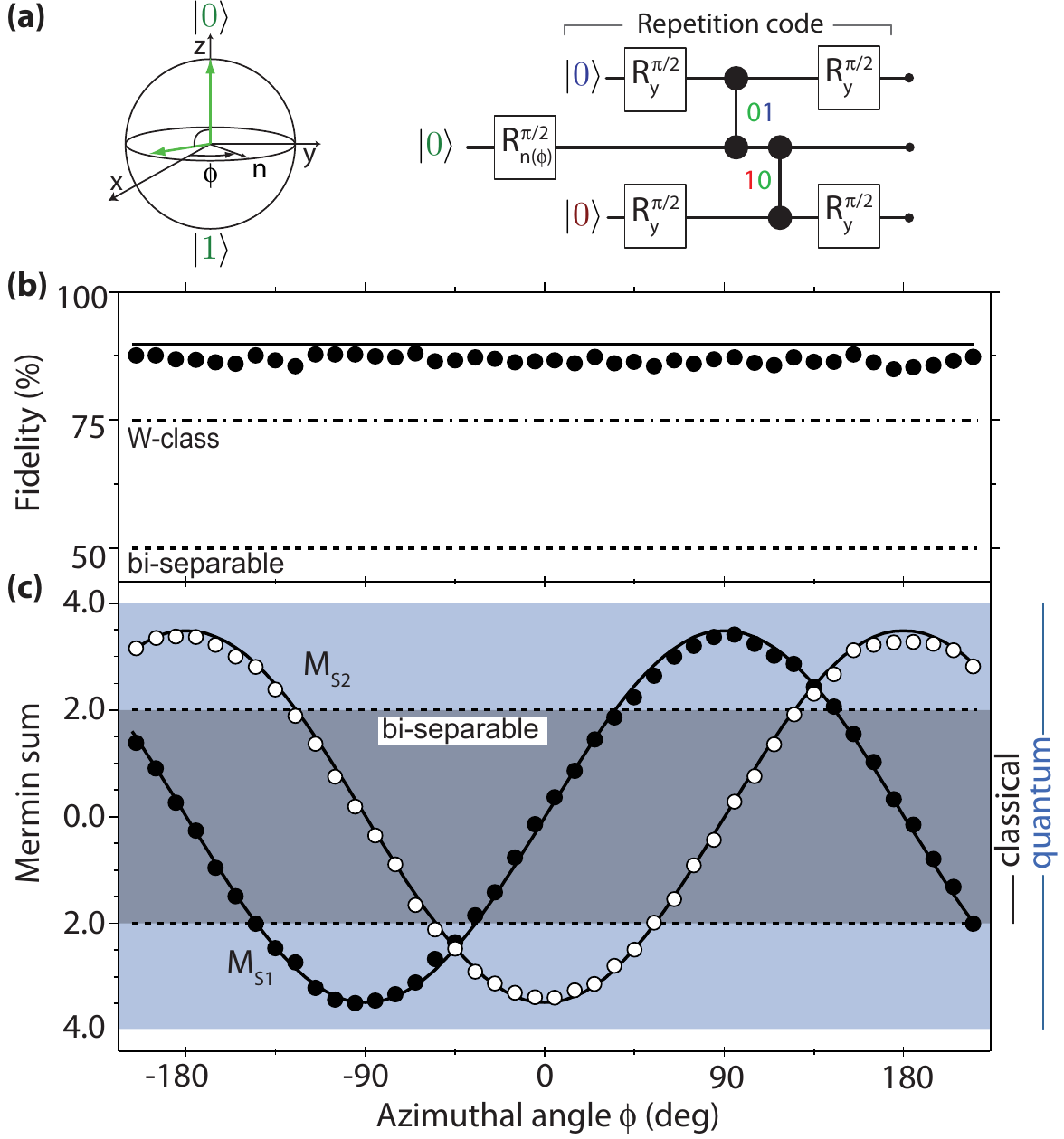}
	\mycaption{Witnessing entanglement with fidelity and Mermin sum inequalities}
	{\capl{(a)} We map $\left(\ket{0} - i e^{i\phi} \ket{1} \right)/\sqrt{2}$ superpositions of $Q_2$ into GHZ-class states $\ket{GHZ_{\phi}} = \left(\ket{000} - i e^{i\phi} \ket{111} \right)/\sqrt{2}$ using the circuit shown.  The indicated section constitutes the encoding step of the three-qubit quantum error correcting ``repetition'' code that will be discussed in the following chapter.  
	The fidelity of these states $F=\bra{GHZ_{\phi}} \rho \ket{GHZ_{\phi}}$ as a function of the azimuthal angle $\phi$ of the initial $\pi/2$ rotation of $Q_2$ is plotted in \capl{(b)} and has an average value of $87\%$.  This fidelity far exceeds biseparable and W-class bounds, witnessing stringent GHZ-class three-qubit entanglement.  
	\capl{(c)} Evolution of Mermin sums $M_{S1}=\langle XXX \rangle - \langle YYX \rangle - \langle YXY\rangle - \langle XYY\rangle$ and $M_{S2}=-\langle YYY \rangle + \langle XXY \rangle + \langle XYX\rangle + \langle XYY\rangle$ as a function of $\phi$.  Allowed values for quantum and biseparable states are labeled.  For all values of $\phi$, at least one of the sums exceeds the biseparable bound.  For \capl{(b)} and \capl{(c)}, the solid lines result from a master equation simulation including only qubit decay during the $81\ns$ pulse sequence.  The fact that it predicts a similar fidelity to our experimental result suggests that pure dephasing and other errors are small contributions to infidelity.  
	\figthanks{DiCarlo2010}
	}
{\label{fig:merminsum}}
\end{figure}

In order to elaborate on the behavior of these sums, we create a variety of GHZ-class states.  As previously mentioned, we take a cut along the manifold of GHZ states by changing the initial axis of rotation of the first pulse on $Q_2$.  Rotating about the axis $\hat{n}(\phi)$ where $\phi=0$ is the $x$-axis and $\phi=\pi$ is the $y$-axis, we create the state $\ket{GHZ_\phi} = (\ket{000} - i e^{i\phi}\ket{111})/\sqrt{2}$.  This is a maximally entangled GHZ state for all values of $\phi$, but the values of the three-qubit Pauli correlations change with this angle.  In \figref{fig:merminsum}(a) we show the circuit diagram for creating this state, and in (b) show the fidelity of the resulting state as a function of $\phi$.  For all values, the biseparable and W-class bounds are amply exceeded.  We also evaluate the Mermin sums $M_{S1}$ and $M_{S2}$ for all angles, shown in (c), and exceed the biseparable bound of at least one of the sums for all $\phi$.  It should be clear that although we defined only two Mermin sums, it is possible to create a $M_{S\phi}$, which is maximally violated for each value of $\phi$.  $|M_{S1,2}| \leq 2$ also defines a {\it local-hidden-variable} (LHV) bound in the same way as the Bell test, which our extremal value of $3.4\pm0.1$ violates by 14 standard deviations \cite{Mermin1990}.  However, we cannot refute local realism because we are not free of detection loopholes \cite{Kwiat1994}. 

\begin{figure}
	\centering
	\includegraphics{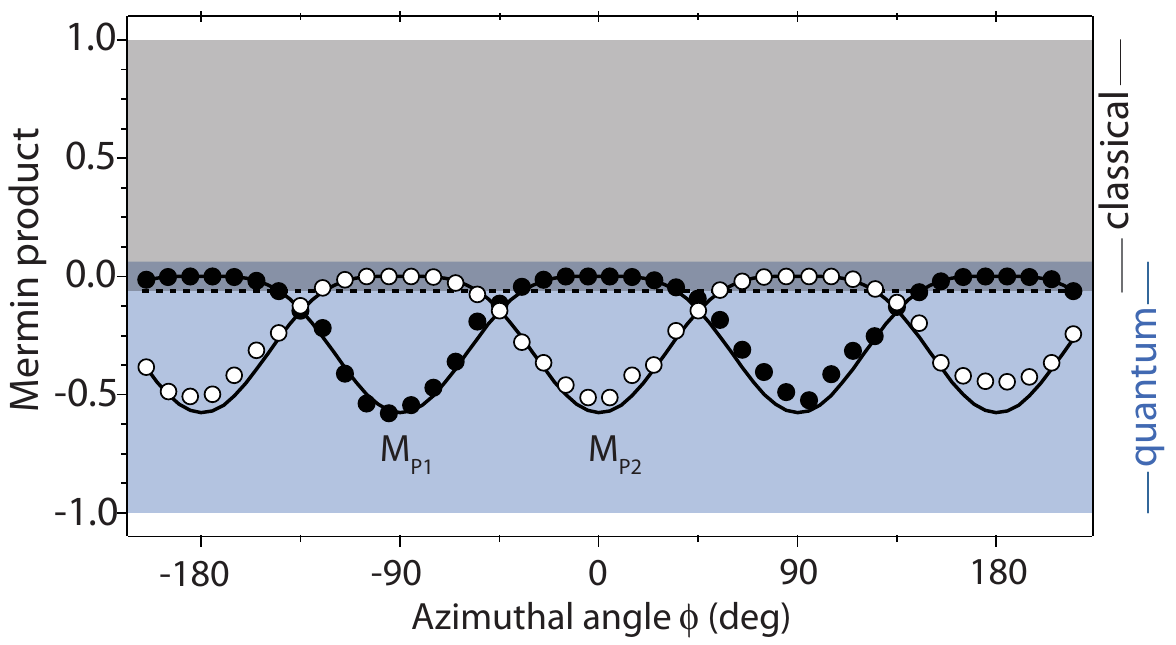}
	\mycaption{Witnessing three-qubit entanglement with Mermin product inequalities}
	{The values of the Mermin products $M_{P1}=\langle XXX\rangle\langle YYX\rangle\langle YXY\rangle\langle XYY\rangle$ and $M_{P2}=\langle YYY\rangle\langle XXY\rangle\langle XYX\rangle\langle YXX\rangle$ are plotted as a function of the angle $\phi$ of the prepared state $\ket{GHZ_{\phi}} = \left(\ket{000} - i e^{i\phi} \ket{111} \right)/\sqrt{2}$.  The most negative value measured, $-0.52\pm0.05$, violates the biseparable and LHV bound of $-1/16$ by $830\pm80\%$.  The solid lines are the result of a master equation simulation that includes qubit relaxation during the pulse sequence. 
	\figthanks{DiCarlo2010}
	}
{\label{fig:merminproduct}}
\end{figure}  

\Fref{fig:merminsum}(c) clearly demonstrates one downside of the Mermin sums: for most of its allowed range of values, a sum will not distinguish between classical behavior and quantum behavior.  Formally, this indicates that for three qubits, the biseparable range significantly overlaps with the quantum range.  We can compress this overlap by using a nonlinear metric, amplifying three-qubit correlations that are due to entanglement.  We define the Mermin products $M_{P1}=\langle XXX\rangle\langle YYX\rangle\langle YXY\rangle\langle XYY\rangle$ and $M_{P2}=\langle YYY\rangle\langle XXY\rangle\langle XYX\rangle\langle YXX\rangle$, which are a function of the same correlations as the Mermin sums.  For these metrics, we found through numerical exploration that biseparable states obey $-1/16 \leq M_{P1,2} \leq 1/64$, which is small compared to the range spanned by three-qubit quantum states, $-1 \leq M_{P1,2} \leq 1/16$ \cite{DiCarlo2010}.  The LHV range for the Mermin products is $-1/16 \leq M_{P1,2} \leq 1$.  Thus we see the advantage of the product over the sum: whereas for the sum the range for LHV states was fully inside the range for quantum states, the product values only overlap for $|M_{P1,2}| \leq 1/16$.  \Fref{fig:merminproduct} shows the result of evaluating this product.  The most negative value we find is $-0.52 \pm 0.05$, which is distinguished from the biseparable and LHV bounds by $830\pm80\%$.  As with the Mermin sum, at least one of the products falls outside the compatibility region for all values of $\phi$. 

\section{Conclusion}
\label{section:entanglementconclusions} 

This chapter focused on methods to create and measure entanglement.  We introduced a four-qubit cQED device that we will continue to study in the next chapter.  That device features flux bias lines to tune qubit frequencies in-situ, but they require calibration to work effectively.  We used this flux control to implement two versions of a controlled-phase two-qubit entangling gate.  These employ an avoided crossing with a non-computational state, which is approached either adiabatically or suddenly.  We then introduced how to perform state and process tomography efficiently by taking advantage of the joint measurement operator to extract multi-qubit correlations.  Finally, we demonstrated both two- and three-qubit entanglement and calculated entanglement metrics to detect the presence of high-quality entanglement.  As we will see in \chref{ch:qec}, all of these tools and techniques are directly extendible to implementing three-qubit quantum error correction.

\setcounter{chapter}{7}
\chapter{Quantum Error Correction with cQED}
\thumb{Quantum Error Correction with cQED}
\lofchap{Quantum Error Correction with cQED}
\label{ch:qec}


\lettrine{T}{his} chapter explains how we have implemented the most basic form of quantum error correction in a cQED device.  Though this demonstration has been done before in NMR \cite{Cory1998, Knill2001, Boulant2005, Moussa2011} and trapped ions \cite{Chiaverini2004, Schindler2011}, this was the first realization in the solid state.  We perform the quantum repetition code that was introduced in \sref{subsec:quantumrepetitioncode}, implementing the measurement-free version because we do not yet have the ability to individually measure and feed-forward ancilla states.  As we saw, the key to this code is the three-qubit ccNOT or Toffoli gate, which coherently reverses an error on the primary qubit after decoding.  Though this gate can be constructed with four of our existing cPhase gates, we instead implement a faster version using the higher level structure of the transmon qubits.  Our gate uses both adiabatic and sudden interactions, leveraging our understanding of the two cPhase gates that we have discussed previously.  In the first half of this chapter, we will detail how our gate works and is tuned up before verifying its action with both state and process tomography.

We combine this gate with our ability to create GHZ-like states to implement quantum error correction in the second half of the chapter.  After encoding an arbitrary quantum state into the three-qubit manifold using the GHZ circuit, we apply intentional bit-rotation errors to one of the qubits.  We then reverse the encoding step, correct the errors with the Toffoli gate, and perform state tomography on the output to verify our success.  We also demonstrate phase-flip error correction, which only requires a small change to the bit-flip code.  To more realistically evaluate its performance, we apply errors on all three qubits simultaneously with some effective probability.  The signature of success is a quadratic dependence of the fidelity of the final quantum state on this error probability, which unambiguously verifies that errors are being corrected as expected.

\section{Toffoli gate}

\begin{figure}
	\centering
	\includegraphics{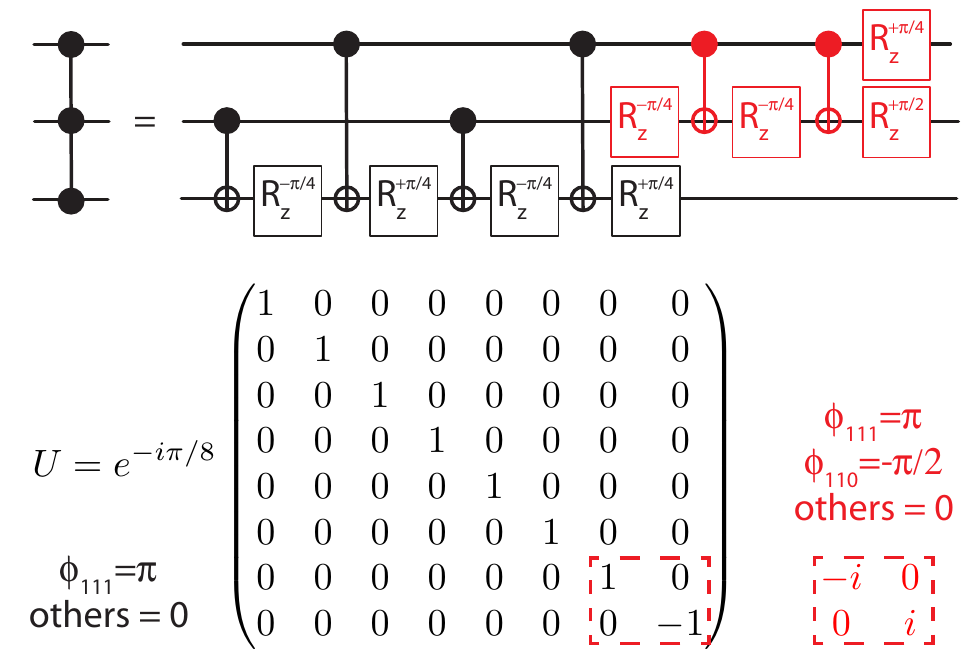}
	\mycaption{Construction of the ccNOT gate with cNOT gates}
	{The three-qubit Toffoli gate necessary for autonomous quantum error correction can be constructed out of two-qubit cNOT gates.  The full construction requires at least six cNOT gates and eight single-qubit rotations \cite{Shende2009}.  This can be reduced to four of each by utilizing the fact that we have quantum error correction in mind, so spurious two-qubit phase entanglement between the ancilla qubits is inconsequential.  The resulting unitary matrix is shown for the two cases, with a red dashed area that indicates the result if we employ our QEC optimization.  In our system, the cNOT gates would be implemented with cPhase gates, but that does not change the cost of the gate since the single-qubit rotations associated with mapping cPhase to cNOT can be compiled with the existing ones.}
	{\label{fig:toffoli_deconstruction}}
\end{figure}

As shown in \sref{subsec:quantumrepetitioncode}, the key ingredient to the autonomous three-qubit QEC code is the Toffoli gate which implements the correction.  This operation, which is also known as the {\it controlled-controlled-NOT} or ccNOT gate, acts on a manifold of three qubits simultaneously and is therefore a three-qubit gate.  In addition to QEC, this gate is important for a variety of applications such as Shor's factoring algorithm \cite{Shor1995}.  For that reason, it has attracted significant experimental interest with recent implementations in linear optics \cite{Lanyon2008}, trapped ions \cite{Monz2009}, and superconducting circuits \cite{Fedorov2011, Mariantoni2011}.  Though we have not yet encountered such a gate in this thesis, since the combination of the cPhase gate and single-qubit rotations is universal, we know that we can directly construct a Toffoli gate from techniques already established.  One possible construction is shown in \figref{fig:toffoli_deconstruction}, comprised of six cNOTs.  (We could equivalently use cPhase gates here, with appropriate changes to the single-qubit rotations.)  We can optimize this gate for the purpose of autonomous error correction by taking advantage of the fact that the code calls for two of the qubits to be measured or reset at the end of the operation.  Thus, any spurious entanglement between the ancillas is harmless and need not be corrected.  This freedom enables us to make a smaller construction using only four cNOTs, also in \figref{fig:toffoli_deconstruction} by omitting the red gates (which involve only the top two ancilla qubits) and swapping the bottom corner of the resulting unitary matrix with $\mathrm{diag}\{-i,i\}$.

Even this optimized pseudo-Toffoli takes quite a long time, however.  If we assume the cPhase gates take $18\ns$ and the single-qubit rotations $12\ns$ (including $2\ns$ of padding for each), it requires at least $130\ns$.  This is quite intimidating, given that the coherence times for these qubits were less than $1 \us$ and including the cost of making and unmaking a GHZ state to perform the error correction code.  Can we do better?  While it has been proven that a six-gate Toffoli is the shortest possible construction using cNOT gates \cite{Shende2009}, that fortunately only applies to two-level qubits.  We can potentially use the higher level structure of the transmon qubits to engineer a more efficient gate.

How might we expect such a gate to work?  We found in the last chapter that a conditional interaction is produced by an avoided crossing between exactly one of the computation states ($\ket{11}$) and a non-computational higher-excited state ($\ket{02}$).  Energy conservation and selection rules dictate that this interaction is very strong when both qubits were excited but non-existent for any other basis state.  The same reasoning applies for a three-qubit gate: an interaction that affects a single basis state of the three-qubit manifold is conditional on all three qubit states.  Moreover, we know that this interacting state must be triply-excited.  If we interacted $\ket{111}$ with a lesser state like $\ket{102}$, that interaction would be exactly symmetrical with $\ket{011}$ and $\ket{002}$, and would merely yield a two-qubit interaction.

\subsection{Efficient Toffoli using higher levels}
\label{subsec:toffoliwithhigherlevels}

Here we will show how we used exactly that interaction of $\ket{111}$ with $\ket{003}$ to build an efficient Toffoli-sign gate.  This gate, also known as the {\it controlled-controlled-Phase} or ccPhase, is related to the Toffoli by single-qubit rotations, just as the cPhase is to the cNOT.  We must discuss two major complications in getting it to work.  The first complication is that the direct interaction of these two states is extremely weak.  There is no coupling of $\ket{111}$ to any state that changes its excitation by more than one to first order (e.g. there are strong interactions with states like $\ket{201}$ but not with $\ket{300}$).  Higher-order terms do enable some coupling, but will be less than $1\mhz$ and too slow to be useful.  In \sref{subsec:threequbitphase}, we will see the signature of this interaction and will verify that it is indeed tiny.  To avoid this, we begin by swapping the quantum amplitude of the state $\ket{111}$ into $\ket{102}$, and, because the state of $Q_1$ is irrelevant, $\ket{011}$ to $\ket{002}$.  The $\ket{102}$ state has a large avoided crossing with $\ket{003}$ that we use to generate our three-qubit conditional phase.  We then reverse the swap to transfer back to the computational space.

The second complication is that a three-qubit phase gate has seven independent parameters, four of which are due to nontrivial multi-qubit interactions.  We can see this by generalizing the notation that we introduced for the case of a two-qubit phase gate.  We distinguish between types of phases, again mapping singly-excited states like $\ket{100}$ to $e^{i\phi_{100}}\ket{100}$ and doubly-excited states like $\ket{110}$ to $e^{i\left(\phi_{100}+\phi_{010}+\phi_{110}\right)}\ket{110}$, the latter of which has the conditional phase $\phi_{110}$, as seen with two-qubit gates.  One difference here, however, is that there are {\it three} independent two-qubit phases between each pair of qubits ($\phi_{110}$, $\phi_{101}$, and $\phi_{011}$), so $\ket{101}$ maps to $e^{i\left(\phi_{100}+\phi_{001}+\phi_{101}\right)}\ket{101}$, with a phase linearly independent from the others.  Finally, there is also a {\it three-qubit} phase $\phi_{111}$ that is conditional on all three qubits being excited, so that the state $\ket{111}$ maps to $e^{i\left(\phi_{100}+\phi_{010}+\phi_{001}+\phi_{110}+\phi_{101}+\phi_{011}+\phi_{111}\right)}\ket{111}$.  We can summarize this by saying that any three-qubit phase gate can be written in the form
{\begin{equation*} 
	\begin{pmatrix}1&0&0&0&0&0&0&0\\
	0&e^{i\phi_{100}}&0&0&0&0&0&0\\
	0&0&e^{i\phi_{010}}&0&0&0&0&0\\
	0&0&0&e^{i\phi_{001}}&0&0&0&0\\
	0&0&0&0&e^{i\left(\phi_{100}+\phi_{010}+\phi_{110}\right)}&0&0&0\\
	0&0&0&0&0&e^{i\left(\phi_{100}+\phi_{001}+\phi_{101}\right)}&0&0\\
	0&0&0&0&0&0&e^{i\left(\phi_{010}+\phi_{001}+\phi_{011}\right)}&0\\
	0&0&0&0&0&0&0& e^{i\sum\phi}
	\end{pmatrix},
\end{equation*}}
where we have again factored out the global phase, order of states as $\ket{000}$, $\ket{100}$, $\ket{010}$, $\ket{001}$, $\ket{110}$, $\ket{101}$, $\ket{011}$, and $\ket{111}$, and $\sum\phi = \phi_{100}+\phi_{010}+\phi_{001}+\phi_{110}+\phi_{101}+\phi_{011}+\phi_{111}$.  Thus, in order to fully tune-up our gate, we need to engineer an interaction that will provide us with the correct three-qubit phase in addition to allowing us to control two-qubit phases as necessary.

This section will explain each component of implementing a three-qubit Toffoli-sign gate.  We start by discussing how the three-qubit phase is generated by first transferring population from $\ket{111}$ into $\ket{102}$.  We perform this transfer suddenly, which requires certain tricks to compensate for technical limitations.  We next explain how the three-qubit phase is adiabatically acquired by bringing $\ket{102}$ near in energy to $\ket{003}$.  We then reverse the swap and return the population of $\ket{102}$ back to $\ket{111}$, after which we show how to control two of the two-qubit phases (that being enough since we can choose to ignore the remaining two-qubit ancilla phase).  We report the resulting pulse sequence and discuss how to measure and verify each of the phases.  Finally, state and process tomography is shown, proving that the gate works as intended.

\subsection{\texorpdfstring{$\ket{111} \rightarrow \ket{102}$ transfer}{111 -> 102 transfer}}

\begin{figure}
	\centering
	\includegraphics{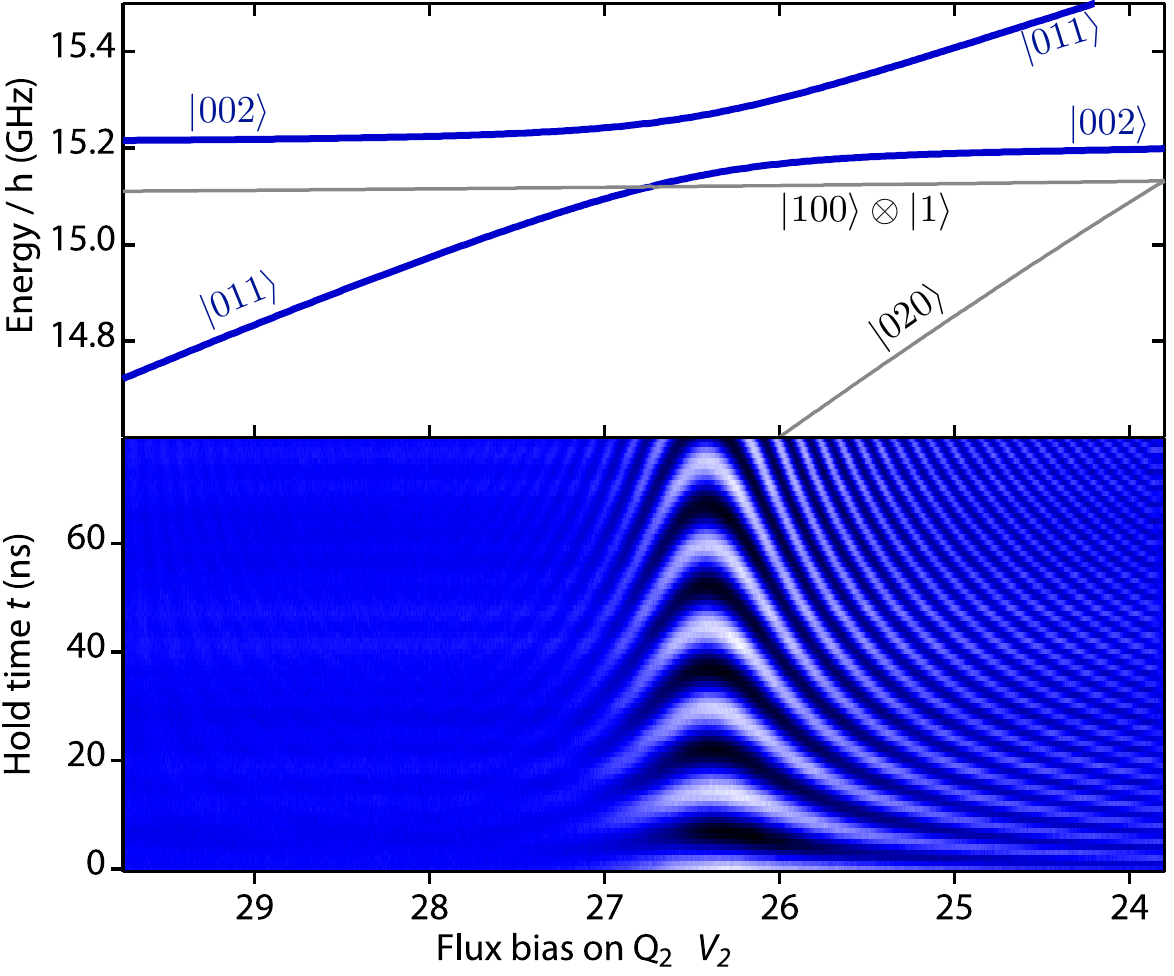}
	\mycaption{Calculated spectrum and time-domain measurements of the $\ket{111} \leftrightarrow \ket{102}$ avoided crossing}
	{The energy spectrum of doubly-excited states in the vicinity of the avoided crossing between $\ket{011}$ and $\ket{002}$ is shown with both \capl{(top)} a numerical diagonalization of the system Hamiltonian and \capl{(bottom)} a time-domain measurement as a function of the applied magnetic flux on $Q_2$.  \capl{(top)} The frequencies for the involved eigenstates are blue and the non-interacting eigenstates of similar energy are gray.  The notation $\ket{abc} \otimes \ket{d}$ indicates respectively the excitation level of each qubit and the cavity photon number.  When omitted, $d=0$.  \capl{(bottom)}  The state $\ket{011}$ is prepared and a square flux pulse of duration $t$ and amplitude $V_2$ is applied.  Coherent oscillations produce a chevron pattern, with darker colors corresponding to population remaining in $\ket{002}$.  This crossing is identical to that between $\ket{111}$ and $\ket{102}$, aside from a $6\ghz$ offset since $Q_1$ is not involved in the interaction.
	\figthanks{Reed2012}
	}
	{\label{fig:chev23}}
\end{figure}

We aim to use the interaction of $\ket{111}$ with $\ket{003}$ to generate a three-qubit phase.  But, because of selection rules, the direct coupling between these states is first-order prohibited and is therefore extremely slow.  Instead, we first swap the quantum amplitude of $\ket{111}$ into the non-computational state $\ket{102}$, which interacts strongly with $\ket{003}$.  Since this transfer does not involve the state of $Q_1$, it will also swap $\ket{011}$ into $\ket{002}$.  In \figref{fig:chev23}(a), we show the calculated energy levels in the vicinity of the avoided crossing of these two states as we move the frequency of $Q_2$ up toward $Q_3$.  Using Mathematica, we numerically diagonalize the Hamiltonian given by \equref{eq:multitransmoncqed}, where we include six quantum states of the three transmons and the cavity.  The parameters of the Hamiltonian were set with independent experiments.  The two states we are concerned with -- $\ket{111}$ and $\ket{102}$ -- are colored in blue; states which are nearly degenerate but irrelevant because they do not couple to any populated level are colored in gray.  

This spectrum suggests two methods for transferring our population.  Since the identity of $\ket{111}$ transmutes into $\ket{102}$, we could potentially transfer our population with a full adiabatic passage.  However, the splitting size of these states is relatively small at only $67\mhz$, which means that the adiabatic transfer would take $\sim 3 / (67 \mhz) \approx 50\ns$ each way, for a total of $100\ns$ round trip.  As this is only the process by which to {\it begin} our gate, and given that we are competing against the four cNOT construction which takes approximately $130\ns$, clearly this approach will not yield the sought after improvement.

Fortunately, there is a second method for transferring population: the sudden approach.  In \figref{fig:chev23}(b) we show swap spectroscopy in the vicinity of this crossing.  As described in the previous chapter, we first initialize in the state $\ket{011}$ and then suddenly tune its energy to a certain value by adjusting the frequency of $Q_2$.  Because this move is made suddenly, $\ket{011}$ is no longer in an eigenstate and our state's projection oscillates between the undressed eigenstates.  After waiting some time $t$, we suddenly move back to our home position, pulse $\ket{011}\rightarrow\ket{000}$, and measure.  If we had transferred some population to $\ket{002}$ during our waiting time, our measurement will give a large value (black pixels); if not, we will be in the ground state and get a small value (white pixels).  (Note that we subtract the average of each vertical cut, so the color scale is not an absolute indication of state.  Near the avoided crossing, the black/white $\ket{002}$/$\ket{000}$ mapping is correct, but away from it blue corresponds to the ground state.)  If we jump in, wait until our population oscillates fully into $\ket{002}$, then jump away in either direction, we will have swapped our entire population from $\ket{011}$ into $\ket{002}$.  (For the purposes of this gate, we want to move further up in frequency to get $Q_2$ out of the way of $Q_1$, which will need to be increased in frequency to turn on our $\ket{102} \circlearrowright \ket{003}$ interaction.)  Moreover, this transfer is very fast, taking $\frac{1}{2} \frac{1}{67\mhz} \approx 7\ns$.  This is a huge improvement over the adiabatic case and preserves the possibility of making gate which is faster than the constructed one.

\subsubsection{Correction for finite bandwidth}
\label{subsubsec:finitebandwidth}

\begin{figure}
	\centering
	\includegraphics{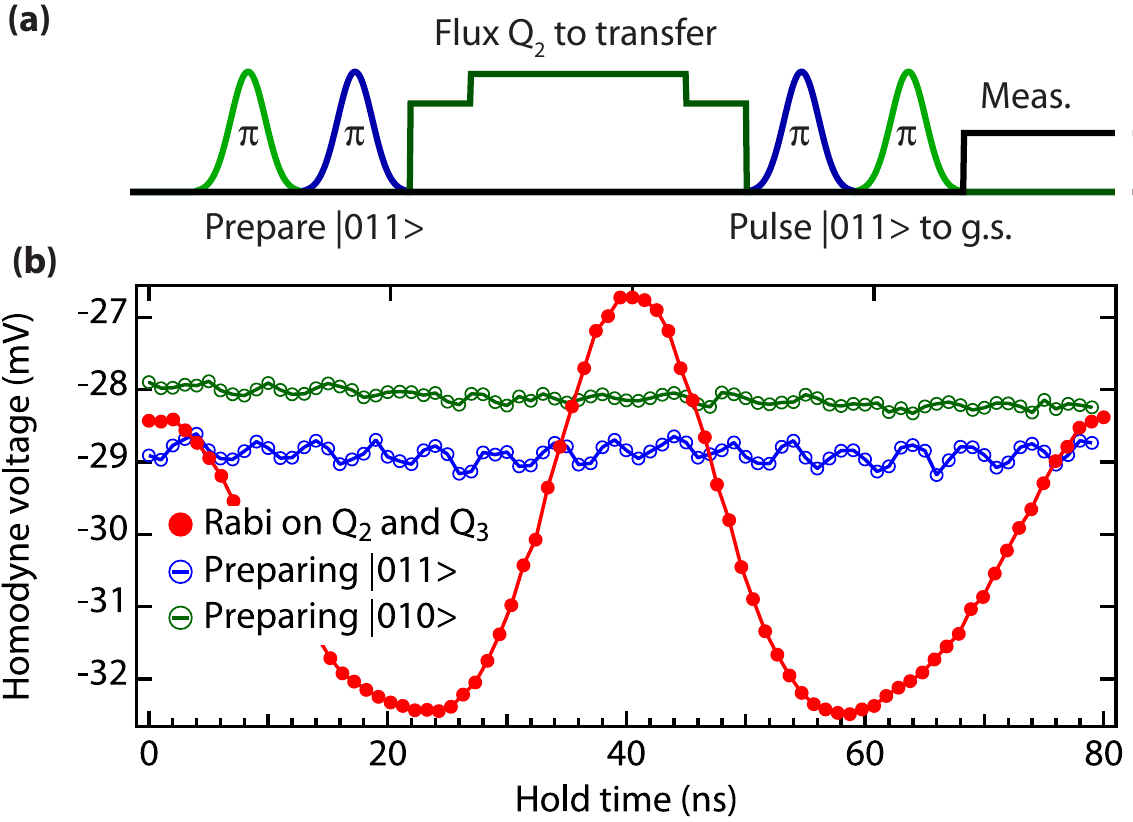}
	\mycaption{Eigenstate transfer verification test}
	{\capl{(a)} A sequence to test the efficacy of the transfer of $\ket{011}\rightarrow\ket{002}$ using a sudden flux pulse is shown.  The state $\ket{011}$ is first initialized with $\pi$ pulses on $Q_2$ and $Q_3$.  The frequency of $Q_2$ is then tuned with a square flux pulse which moves the state directly into resonance with $\ket{002}$ for $7\ns$.  This transfers the projection of the quantum state from $\ket{011}$ to $\ket{002}$, which we lock in by moving $Q_2$ further up in frequency, away from the interaction.  We wait for some time at this higher frequency and then reverse the swap with a symmetrical flux pulse, rotate $\ket{011}$ back to the ground state with two $\pi$ pulses, and measure.
	\capl{(b)} If the transfer was successful, this process should fully return the qubit manifold to its ground state.  However, if the transfer was incomplete and we were really in a superposition of states during our hold time, the relative phase will evolve and we will see oscillations as a function of time.  We see that, when preparing the state $\ket{011}$, there is a substantial oscillation due to incomplete transfer.  We compare this to the case of a Rabi oscillation on $Q_1$ and $Q_2$ simultaneously (both qubits are being pulsed, which is why the oscillation is not sinusoidal) to calibrate the $y$-axis.  There is an oscillation of approximately $5-10\%$ of the full measurement amplitude, which indicates that the transfer is imperfect.  We also plot the case of preparing $\ket{010}$, which also exhibits oscillations because it is non-adiabatic to the J crossing between $Q_2$ and $Q_3$.}
	{\label{fig:02transfer}}
\end{figure}

Our intent to transfer in $7\ns$ does warrant some cause for alarm.  The Tektronix AWG we use to generate our flux pulses has a bandwidth of about $300\mhz$, so we cannot expect the rise time of any flux pulse to be much faster than $2-3\ns$.  The flux pulse which we intend to be square will in fact have rounded edges and will not be perfectly sudden.  This error will cause us to leave behind population in $\ket{011}$, which we can test for by performing a modified version of swap spectroscopy.  As shown in \figref{fig:02transfer}(a), we first perform the transfer of $\ket{011}$ to $\ket{002}$ and leave $Q_2$ positively detuned from the avoided crossing, then wait for some varying amount of time.  We next undo the transfer with a flux pulse that exactly mirrors the first, and finally, we apply pulses to map $\ket{011}\rightarrow\ket{000}$ and measure.  If we are fully transferred to our target state, we will be in an eigenstate at our waiting point, and thus should see no evolution as a function of the waiting time.  However, if our transfer is incomplete and we are in a superposition, the phase between the two states will evolve, modifying the reverse process and yielding oscillations.  This is shown in \figref{fig:02transfer}(b).  There, we also show the case of preparing the state $\ket{010}$, which itself approaches an avoided crossing with $\ket{001}$ to which we must be adiabatic.  We compare it to the contrast of a $\ket{000} \circlearrowright \ket{011}$ Rabi oscillation and find that the size of the oscillation is fairly substantial -- almost 10\% -- which we must mitigate.

Fortunately, a simple theoretical procedure exists to deal with a finite transfer speed.  Consider a test Hamiltonian $\hat{H}=\frac{1}{2}\left( \Delta \sigma_z + g \sigma_x \right)$.  Our goal is to transfer the population of the lower eigenstate to the upper eigenstate using a finite-length pulse, starting from a large negative detuning and ending at some finite positive detuning.  We begin by adiabatically moving to a position at which we are symmetric about the avoided crossing relative to our final position.  That is, if we call our final position $F$, which has energy $E_F$, we want to move to an intermediate position $I$, such that $E_I = -E_F$ ($E=0$ for $\Delta=0$).  Note that while we must make this first movement adiabatically, if $F$ is sufficiently distant from $\Delta=0$, we can be very fast and still satisfy that condition (given that our speed limit there is set by $\Delta$, not $g$).  Second, we move as quickly as possible to exact resonance, taking time $t_s$.  (Our fast movement saved time, though this is not a necessity.)  At this point, we visualize the two states of the basis at $I$ on the Bloch sphere, and have slightly rotated from the north pole in some direction due to our non-sudden trajectory.  This small rotation is directly accountable for the error that we are trying to correct.

At this point, our goal is to wait for a period of time $\tau$ that will bring the qubit to the equator of this pseudo-Bloch sphere.  How much and in what direction our non-sudden trajectory has rotated the qubit determines the amount of time we need to wait.  In resonance, the Hamiltonian acts as $\sigma_x$ but our erroneous initial rotation could go in any direction.  In fact, during the sudden transfer we would expect extremely rapid changes in the axis of rotation as the eigenstates change and phase is acquired between them.  Therefore, $\tau$ is a sensitive function of the qubit's exact trajectory into the avoided crossing.  If it happened to rotate about $\sigma_y$, $\tau$ will be given by exactly $1/g$, but will be reduced by any component of the rotation around $\sigma_x$\footnotemark.

\footnotetext{If we moved into the avoided crossing adiabatically we would have moved to a location on the Bloch sphere corresponding to being an eigenstate of the $\sigma_x$ operator, and therefore would not move from that point during the waiting time.  But if we are able to move in adiabatically, we can simply repeat that path on the other side and accomplish the transfer.  Our goal is to be at an eigenstate at the end of the transfer; being one at every point during the process is a sufficient but not necessary condition.}

Assuming that we have calculated or can experimentally determine $\tau$, we want to wait for $2\tau$ to bring the qubit to a point mirrored across the $x-y$ plane of the Bloch sphere.  We then move as fast as we can to the final position $F$, which will perfectly complete our population transfer.  The reason this should be perfect is because of the symmetry that we have enforced at $t=t_s+\tau$, where we have moved to the equator of the Bloch sphere.  Since the avoided crossing is exactly symmetrical with respect to $\Delta=0$, if we start in an eigenstate at point $I$ and get to the equator of the Bloch sphere halfway through our transfer, then by symmetry we know that we will also be at an eigenstate at point $F$, regardless of the specific trajectory we took to get there.  The total transfer time is $2\left(t_s + \tau\right)$.

\begin{figure}
	\centering
	\includegraphics{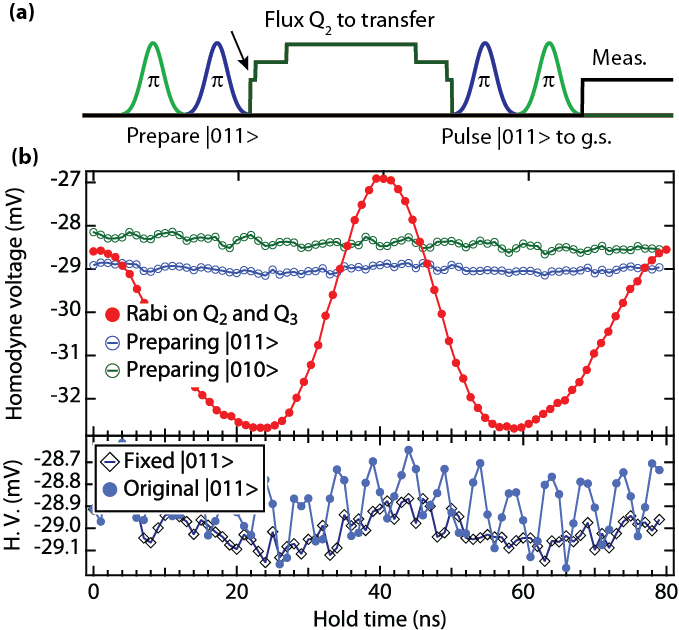}
	\mycaption{Eigenstate transfer verification test with pulse shaping}
	{\capl{(a)} To mitigate the imperfect state transfer of $\ket{011}\rightarrow\ket{002}$, we fine-tune the trajectory we take approaching this avoided crossing.  We leave the first DAC value of the pulse as a free parameter which we sweep to minimize the amplitude of oscillation following a repetition of the experiment, as described in \figref{fig:02transfer}(a).  The flux pulse is symmetric, and uses the same fine-tuned parameter for both directions of transfer.
	\capl{(b)} We show the resulting oscillation amplitude for preparations of $\ket{011}$ and $\ket{010}$, just as with \figref{fig:02transfer}(b).  Below, we zoom in on the oscillation of $\ket{011}$ and compare it to the data from \figref{fig:02transfer}(b), showing a substantial reduction of the oscillation amplitude.}
	{\label{fig:02transfertuned}}
\end{figure}

In the experiment, we lack the time resolution necessary to fine-tune the wait duration as prescribed above.  However, we can consider this to be theoretical guidance for searching for a more easily-implemented mitigation scheme.  We have found above that the specific trajectory we take through the avoided crossing can have significant consequences for the final state.  We thus suggested finding a specific time to match our trajectory.  However, in principle, we could choose a trajectory to match our waiting time just as well.  Thus, we hold constant the amount of time we sit in the crossing, but slightly tweak the path through which we go.  Specifically, while above we jumped directly into the crossing for exactly $7\ns$, let us instead jump to some different position relative to the crossing for $1\ns$, then jump into resonance for the remaining $6\ns$.  That position for the first nanosecond is a free parameter that we can adjust to minimize the oscillation we see in our modified swap spectroscopy sequence.  We show this exact result in \figref{fig:02transfertuned}, thus reducing the amplitude of spurious oscillation by a factor of three with this simple modification.  Additionally, this change does not worsen the non-adiabaticity of the $\ket{010}$ state approaching $\ket{001}$.

\subsection{\texorpdfstring{$\ket{102} \circlearrowright \ket{003}$ interaction}{102 - 003 interaction}}
\label{subsec:threequbitphase}

\begin{figure}
	\centering
	\includegraphics{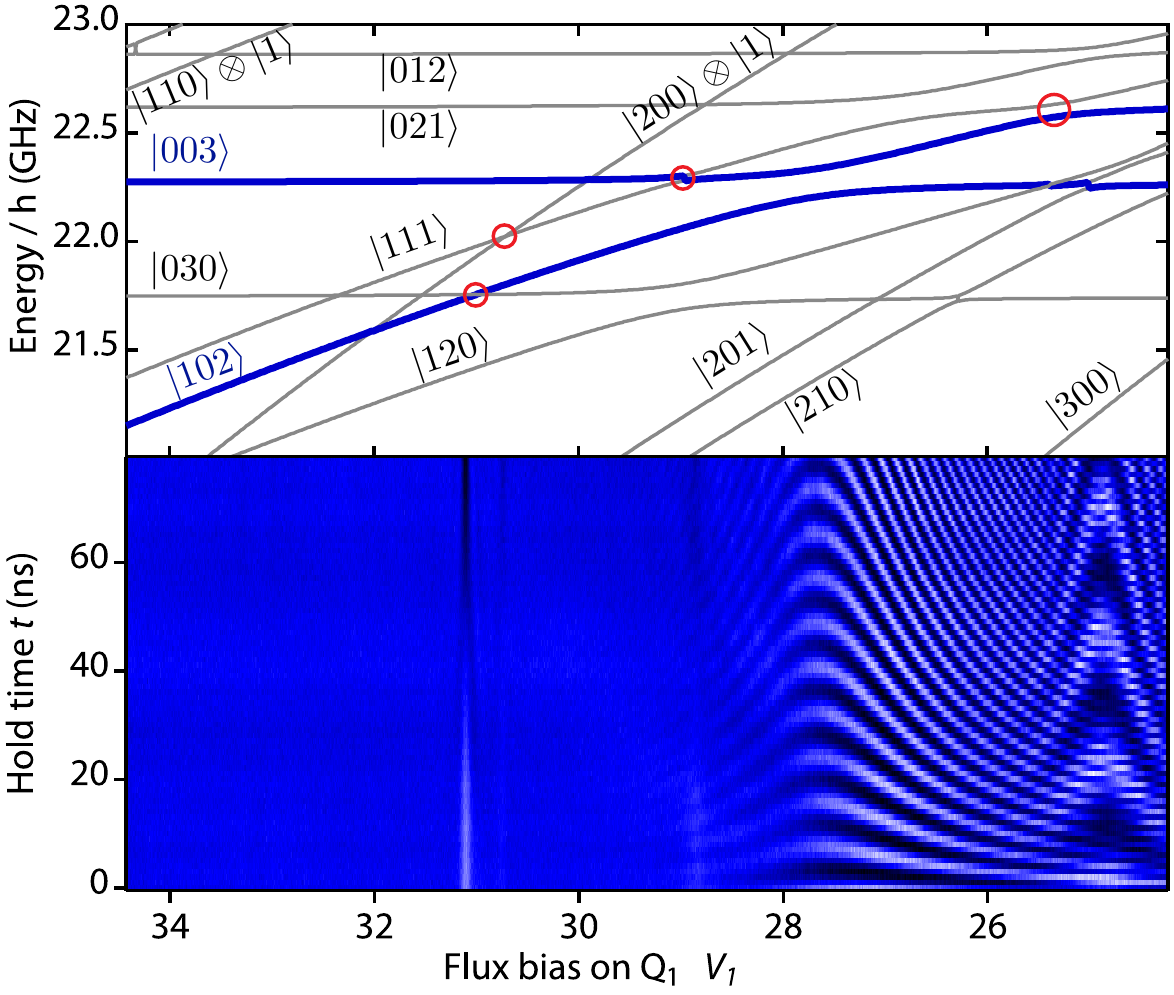}
	\mycaption{Calculated spectrum and time-domain measurements of the $\ket{102}\leftrightarrow\ket{003}$ avoided crossing}
	{The spectrum of triply-excited states showing the avoided crossing between $\ket{102}$ and $\ket{003}$ as a function of the flux bias on $Q_1$ is characterized in the same way as in \figref{fig:chev23}.   $\ket{102}$ is prepared by first making $\ket{111}$ and then performing the swap as described in \figref{fig:02transfertuned}.  As before, the relevant eigenstates are highlighted in blue.  Many additional eigenstates (shown in gray) are close in energy but are irrelevant because they do not interact with the populated states.  The large avoided crossing between $\ket{102}$ and $\ket{003}$ that is used to produce an adiabatic three-qubit interaction happens near $28~\mathrm{m}\Phi_0$.  Extra lines near $31~\mathrm{m}\Phi_0$ and $29~\mathrm{m}\Phi_0$ are due to higher-order interactions predicted by the Hamiltonian ($\ket{102}$ with $\ket{030}$ and $\ket{003}$ with $\ket{111}$), as is the larger first-order interaction at $25~\mathrm{m}\Phi_0$ ($\ket{102}$ with a hybridization of $\ket{021}$ and $\ket{111}$), but their impact on the gate protocol is negligible.
	\figthanks{Reed2012}}
	{\label{fig:chev003}}
\end{figure}

Now that we are able to efficiently transfer $\ket{111}$ into $\ket{102}$, we can turn on the interaction with $\ket{003}$ to acquire three-qubit phase.  We show the energy levels in the vicinity of the $\ket{102}\leftrightarrow\ket{003}$ avoided crossing in \figref{fig:chev003}(a), calculated, as before, with a numerical diagonalization of the system Hamiltonian.  We highlight the relevant states, $\ket{102}$ and $\ket{003}$, in blue.  In contrast to the previous plot which showed states with only two excitations, there is now a vast proliferation of nearly-degenerate states, shown in gray.  We label each state.  Note how the slope of each line is approximately proportional to the excitation of $Q_1$ (the qubit being tuned).  Several of these states have finite coupling to our populated states, though the coupling is high-order and therefore very weak.

We can demonstrate the coupling, as before, with swap spectroscopy.  We prepare the state $\ket{102}$ following the procedure we worked out in the previous section.  After we have moved to the final position $F$ of $Q_2$, we suddenly move $Q_1$ up in frequency, wait some time, flux back, undo the preparation of $\ket{102}$, and measure.  If our state leaks into some other eigenstate during the wait time, our process mapping $\ket{102}\rightarrow\ket{000}$ at the end will not address that population and our measurement will indicate that the qubit manifold is excited.  In \figref{fig:chev003}(b) we show the result of performing this measurement over the same range as (a).  The primary feature is a large avoided crossing at $27.5~\mathrm{m}\Phi_0$, which corresponds precisely with the location of where we expect $\ket{102}$ and $\ket{003}$ to be in resonance.  The splitting is extremely large, at $121\mhz$.  The size of this crossing is also reflected in the asymmetry of the coloring: the contrast on the lower-frequency side is smaller than on the larger-frequency side because we are more sudden the further we jump.

We see several other features in the swap spectrogram.  At $25~\mathrm{m}\Phi_0$ we have another fairly large ($\sim35\mhz$) crossing, corresponding to an interaction between the state $\ket{102}$ and a hybridization of $\ket{021}$ and $\ket{111}$.  This region is clearly very complicated, given the simultaneous interaction of several states.  Moreover, because we are not completely sudden to the crossing of $\ket{102}$ and $\ket{003}$, $\ket{102}$ is likely not fully populated at this point.  The contrast of that chevron will be reduced by the fraction of population not in $\ket{102}$, though the frequency and location of the oscillation will be unchanged.  We also see three more faint lines, with two near $31~\mathrm{m}\Phi_0$ and the third at $29~\mathrm{m}\Phi_0$.  These correspond to interactions of $\ket{102}$ with $\ket{030}$, the residual population of $\ket{111}$ with $\ket{200}\otimes\ket{1}$ (the second excited state of $Q_1$ plus one photon in the cavity), and $\ket{003}$ with $\ket{111}$.

The fact that these last two interactions are visible is a bit surprising.  In the first case, this is  present only because some population must have been left in $\ket{111}$.  This indicates that our transfer to $\ket{102}$ was not perfect, though we already knew this from the finite contrast of the oscillation in \figref{fig:02transfertuned}.  That contrast is only a few percent of the full readout voltage, which shows that our swap spectroscopy sequence is extremely sensitive and indicates that any place where we do {\it not} see a line is truly free from interactions.  The second oscillation, $\ket{003}$ with $\ket{111}$ tells us something interesting as well.  For one, we are only sensitive to this crossing because of the small population that we have erroneously left in $\ket{111}$.  Secondly, the smallness of this crossing confirms our original contention that the direct interaction of $\ket{111}$ and $\ket{003}$ is weak and that the intermediate swap to $\ket{102}$ is required.  Note that though it appears as if we have approximately one-half oscillation in $70\ns$, this is again merely an artifact of the way we normalize these data (subtracting the average from each horizontal cut).  While a half-oscillation would indicate the splitting is approximately $7\mhz$, in reality the coupling strength is a great deal less than that.

Getting good correspondence between the numerical diagonalization and swap spectroscopy data requires some work.  The relationship between qubit frequency and flux pulse amplitude must be calibrated.  Qubit frequency is not directly proportional to flux because the qubits have a finite maximum frequency; $\omega_{01}$ is proportional to $\sqrt{|\mathrm{cos}\left(\Phi / \Phi_0\right)|}$.  The numerical diagonalization's $x$-axis must be transformed by the inverse of this function.  We accomplish this by using ``flux spectroscopy,'' whereby we move the qubit to some location with fast-flux and then send in a short spectroscopy pulse at some frequency, and then measure.  The qubit will be populated when our pulse is on resonance with the transient qubit frequency, which allows us to map out the voltage-frequency relation.  Fitting those data to a $\sqrt{|\mathrm{cos}\left(V / V_0 + \phi\right)|}$, we can invert that function and transform the $x$-axis.  This process is also used in \figref{fig:chev23}, but is less noticeable because there is only one avoided crossing.

We also need to include accurate values of all the Hamiltonian parameters in our model.  In particular, the frequency of the cavity and all the qubits' maximum frequencies, the qubit frequencies during the gate (e.g. where $Q_2$ and $Q_3$ end up after the $\ket{102}$ transfer), and all of the qubit anharmonicities and cavity coupling strengths are required.  We independently measured these values with conventional and flux spectroscopy and input them into the model.  We see excellent agreement with the first four avoided crossings, though there is some small aberration with the one at $25~\mathrm{m}\Phi_0$.  This may be due to small errors in our system parameters or due to the presence of higher modes of our cavity which were not included in the model.  We could fine-tune the parameters to correctly fit all the locations, but that process is a bit unscientific.  Nevertheless, the quality of correspondence indicates that our simple Hamiltonian accurately models the system even for these relatively highly excited states.

Now that we have mapped out the flux spectrum of the $\ket{102}$ state, it is straightforward to acquire three-qubit phase.  We adiabatically move $Q_1$ up in frequency to move $\ket{102}$ close to resonance with the $\ket{003}$ state.  As in the case of the adiabatic two-qubit gate, $\ket{102}$ experiences a frequency shift relative to its constituents ($\ket{100}$ plus $\ket{002}$) due to the avoided crossing, thus yielding three-qubit phase.  In principle this phase could also be acquired suddenly, but the size of the avoided crossing is too large for our room temperature electronics to approach.  Moreover, the large size makes an adiabatic approach both easy and fast.  We can acquire $\pi$ three-qubit phase in only $20 \ns$, tuning with the location of the closest approach.  We might also be concerned that being adiabatic would give us problems with the three spurious avoided crossings we found along the way.  Fortunately, however, these are so small that even while we are adiabatic to $\ket{102}$-$\ket{003}$, we will be sudden to these other interactions, making them wholly irrelevant.

\subsection{Two-qubit phases}

We have now seen how the combination of a sudden transfer to $\ket{102}$ and an adiabatic interaction with $\ket{003}$ can give us full control over the three-qubit phase $\phi_{111}$.  But, as we discussed in \sref{subsec:toffoliwithhigherlevels}, there are three more two-qubit phases that describe a three-qubit phase gate.  These phases, $\phi_{101}$, $\phi_{011}$, and $\phi_{110}$, must be dealt with as well.  Fortunately, because we reset or measure the ancilla qubits at the end of the QEC code, conditional phase between those qubits does not matter.  Moreover, the code is entirely symmetric with respect to the qubits until the correction step (and with the exception of the very first single-qubit pulses), so we can use whichever two qubits are most convenient for ancillas.  For reasons that will soon become clear, we choose $Q_1$ and $Q_3$.  Thus, the value of $\phi_{101}$ is inconsequential and we need only to control $\phi_{011}$ and $\phi_{110}$.

In order to tune-up the remaining two phases, we need to find a location where we will very rapidly acquire the relevant two-qubit phase so we can control it by changing some parameter by a small amount.  It is important that this small adjustment does not change the other two- or three-qubit phases significantly.  Otherwise, it would be difficult to find a location where all of our conditions are simultaneously satisfied.  That is, we might be able to get $\phi_{110}=0$, but then find that $\phi_{111}\ne \pi$, so we tune that back up, but then $\phi_{110}$ might be wrong again, and so on.  While this process may eventually converge with sufficient iterations, it would be time consuming and laborious.

To our great fortune, there is automatically such a point in this gate where the phase $\phi_{011}$ is acquired rapidly.  After we have transferred the population of $\ket{011}$ to $\ket{002}$ and $\ket{111}$ to $\ket{102}$, we move to a point positively detuned from that crossing, where the transition frequency of $Q_2$ is even closer to $Q_3$.  At that point, the state $\ket{x11}$ is detuned from $\ket{x02}$ ($x=0,1$) by approximately $100\mhz$, and so the phase $\phi_{011}$ is wrapping extremely rapidly.  While we do not necessarily have much freedom to control the amount of time we wait at this location (the minimum time is set by making $\phi_{111}=\pi$ and we would not want to wait any longer than necessary), we do have a large amount of freedom in choosing the exact frequency to which we tune $Q_2$ (using the language of the previous subsection on finite bandwidth, we are choosing the location of point $F$).  By finely adjusting this, we change the difference in the energy of $\ket{x11}$ and $\ket{x02}$ and so control the conditional phase.  Moreover,  since we acquire $\phi_{111}$ adiabatically, this change doesn't do much to the other two conditional phases.  Therefore, it is fairly insensitive to the exact location of the $\ket{102}-\ket{003}$ avoided crossing; $Q_1$ need not get very close to $Q_2$ during the process and so will not interact strongly.

We are not quite as lucky in regard to the remaining phase $\phi_{110}$.  The qubits $Q_1$ and $Q_2$ are only close together when we are acquiring three-qubit phase, so we cannot fine tune their interaction there independently.  Instead, we must append an extra controlled interaction between these qubits to control their mutual phase.  In principle, we could do this correction suddenly by going on a revolution at the exact detuning from the avoided crossing that cancels the spurious phase (as described in \sref{subsec:suddencphase}); in practice, however, this is too much trouble since we would need to choose a different pulse detuning and duration every time $\phi_{110}$ changes, which it does frequently during the tune-up process.  We instead perform this interaction adiabatically so we can adjust the phase by simply changing the minimum detuning between $Q_1$ and $Q_2$.  Since the uncorrected phase $\phi_{110}$ turns out to be small, we can save time by performing a $\pi$ pulse on $Q_2$ prior to the adiabatic interaction, so the phase evolves in the opposite direction and we need not phase-wrap.  At this point, it is clear why we chose to ignore the $\phi_{101}$ phase.  Those two qubits are not nearest-neighbors so bringing them close to each other to control their conditional phase would inexorably cause large changes in the other conditional phases.

\subsection{Pulse sequence and the cRamsey phase tune-up procedure}
\label{subsec:toffolutuneup}

\begin{figure}
	\centering
	\includegraphics{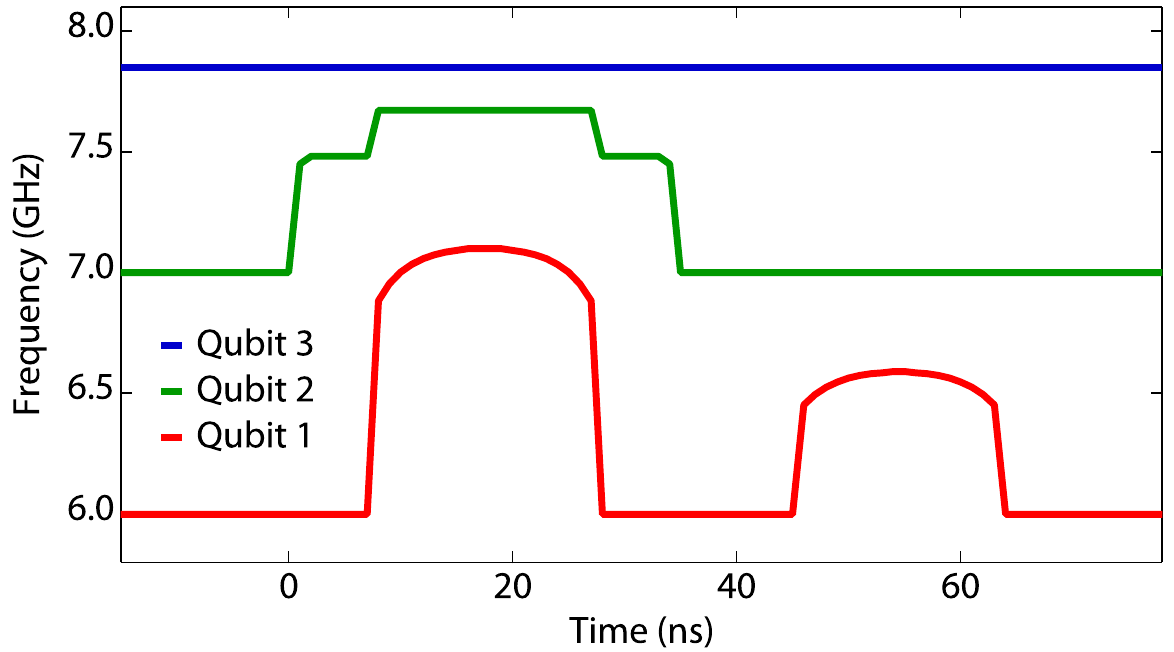}
	\mycaption{Toffoli-sign pulse sequence}
	{The frequency of each qubit as a function of time during our efficient Toffoli-sign gate is shown.  $Q_2$ is first moved up to transfer the quantum amplitudes of $\ket{111}\rightarrow\ket{102}$ and $\ket{011}\rightarrow\ket{002}$.  The final waiting position is chosen to tune-up the two-qubit phase between $Q_2$ and $Q_3$.  $Q_1$ is then moved up in frequency to adiabatically approach the $\ket{102}\leftrightarrow\ket{003}$ avoided crossing.  At that point, we acquire three-qubit phase.  Afterwards, the swap is reversed, returning all quantum amplitudes to the computational basis.  Finally, we apply a $\pi$ pulse to $Q_2$ (not shown, happening at $t=36\ns$) and perform an adiabatic cPhase gate between $Q_1$ and $Q_2$ to correct their two-qubit phase.
	\figadapt{Reed2012}
	}
{\label{fig:toffolipulse}}
\end{figure}

We now have all the ingredients of the efficient Toffoli-sign gate.  Before we continue, let us briefly recap our decisions thus far.  The gate is initiated by suddenly transferring $\ket{011}$ to $\ket{002}$ and $\ket{111}$ to $\ket{102}$ with a flux pulse on $Q_2$.  The first nanosecond of the transfer is fine-tuned to compensate for our finite control bandwidth and time resolution.  We then wait in the state $\ket{102}$, and choose the location of $Q_2$ during this time to set the first of our two-qubit phases $\phi_{011}=0$.  Simultaneously, we move $Q_1$ up in frequency to bring the state $\ket{102}$ close to an avoided crossing with $\ket{003}$.  This causes the energy of $\ket{102}$ to diverge from the sum of its components, $\ket{100}$ and $\ket{002}$, giving us a three-qubit conditional phase.  We tune this to equal $\pi$ by adjusting the time and the detuning of this pulse.  Moving $Q_1$ back to its home position, we then reverse the swap, dumping the amplitude of $\ket{102}$ back into $\ket{111}$ (and $\ket{002}$ back to $\ket{011}$).  Finally, we perform a $\pi$ pulse on $Q_2$ and adiabatically move $Q_1$ up to bring the states $\ket{11x}$ close to resonance with $\ket{02x}$, thus fixing the $\phi_{110}$ conditional phase.  A cartoon depicting the resulting pulse sequence is shown in \figref{fig:toffolipulse}.

Looking at the pulse sequence, it is clear that there are quite a few parameters that we need to experimentally measure in order to perform this gate.  First of all, we need to know precisely what voltage we must apply to our flux line to bring the $\ket{111}$ state into resonance with $\ket{102}$ as well as approximately how long we need to wait to efficiently transfer our quantum amplitude.  Swap spectroscopy is the best way of doing this, as was previously shown in \figref{fig:chev23}.  We also need to fine-tune the transfer to $\ket{102}$, which is done by repeatedly measuring the oscillation in population while waiting at position $F$, as shown in \figref{fig:02transfertuned}.  We also measure swap spectrograms for all other computational state preparations as a function of the frequency of $Q_2$ to make sure there aren't any unknown interactions that are approaching.  We also need to approximate the distance we need to move $Q_1$ up to acquire three-qubit phase, and verify the absence of any extra crossings that affect our other basis states.  To do this, we again perform swap spectroscopy for the $\ket{111}$ state, as shown in \figref{fig:chev003}.  In principle, we have now proven that we can acquire three-qubit phase, though we still need to work out how to go about measuring it and other conditional phases.

\nomdref{Acramsey}{cRamsey}{Conditional Ramsey sequence, used for measuring conditional phases.}{subsec:toffolutuneup}

How do we measure these conditional phases?  In the previous chapter, we mentioned that a modified Ramsey oscillation can be used to measure the conditional phase between two qubits.  In this sequence, we prepare one of our qubits along the equator of the Bloch sphere and the other qubit in either its ground or excited state.  Next, we apply the gate, do a second $\pi/2$ pulse about a varying angle on the first qubit, return the second qubit to the ground state, and measure.  We get two sinusoidal oscillations that will be offset from one another by the exact two-qubit conditional phase $\phi_{11}$.  How do we see that?  First, consider the Ramsey oscillation when the second qubit is in the ground state.  As a function of the angle of the final $\pi/2$ pulse, we oscillate between $\ket{00}$ and $\ket{10}$.  We define the action of our gate as mapping $\ket{00}\rightarrow\ket{00}$ and $\ket{10}\rightarrow e^{i\phi_{10}}\ket{10}$.  The rotation axis that will move the qubit up to the excited state will thus be rotated by $\phi_{10}$ from the axis of our first $\pi/2$ pulse, which we can call $x$ without loss of generality.  Following this same logic, when the second qubit is excited, we oscillate between $\ket{10}$ and $\ket{11}$ and $\ket{11}\rightarrow e^{i\left(\phi_{10}+\phi_{01}+\phi_{11}\right)}\ket{10}$.  We are only sensitive to phases involving the state of the first qubit because our $\pi/2$ pulse addresses only that qubit.  The phase-lag of the resulting oscillation is given by $\phi_{10}+\phi_{11}$ -- that is, it does not depend on the value of $\phi_{01}$.  Finally, if we examine the phase difference between these curves, we get $\left(\phi_{10} + \phi_{11}\right) - \left(\phi_{10}\right) = \phi_{11}$, the two-qubit conditional phase, as promised.  This {\it conditional-Ramsey technique} (cRamsey) measures the difference in phase acquired by a single qubit depending on the state of the other qubit during some process.

\begin{figure}
	\centering
	\includegraphics{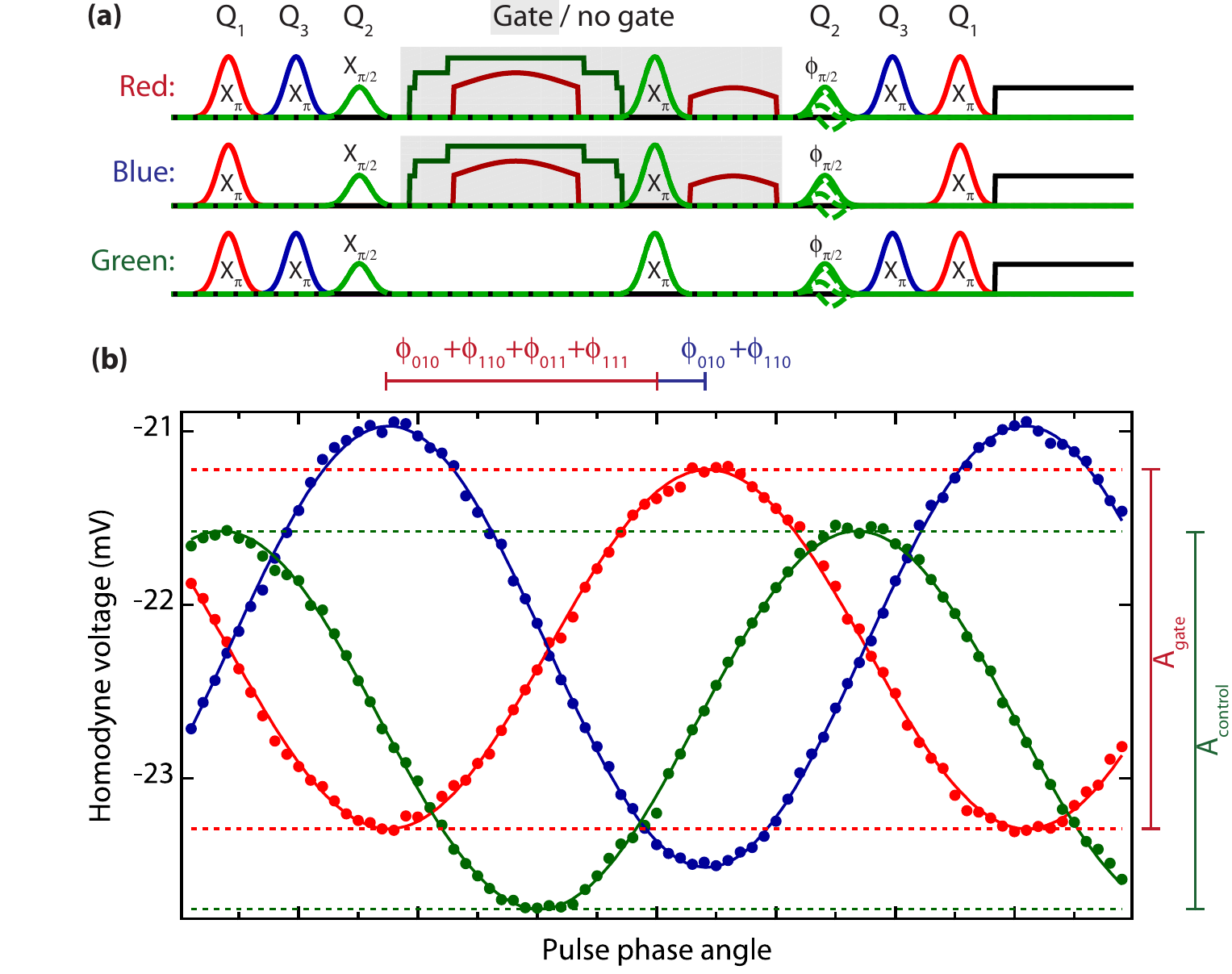}
	\mycaption{Conditional-Ramsey phase tune-up sequence for measuring $\phi_{111}+\phi_{011}$}
	{\capl{(a)} The pulse sequence for three Ramsey experiments is shown.  For red and green, a superposition of the states $\ket{101}$ and $\ket{111}$ is prepared; a superposition of $\ket{100}$ and $\ket{110}$ is made for blue.  The Toffoli gate we are testing is applied to the red and blue cases.  After each, a $\pi/2$ pulse around the $\phi$-axis is applied to $Q_2$, and, before being measured, the remaining qubits are pulsed back to their ground states.  This procedure tells us the phase difference between the two prepared states.  In red, we measure the phase $\phi_{010} + \phi_{110} + \phi_{011} + \phi_{111}$; in blue, we measure $\phi_{010}+\phi_{110}$.  Subtracting the two will give us the value of $\phi_{111} + \phi_{011}$, which, combined with similar experiments with different qubit preparations,  fully specifies the ccPhase gate.  We are not interested in the phase of the green experiment in this case, but the amplitude of oscillation indicates any loss in fidelity.
	\capl{(b)} The result of performing the experiment described in \capl{(a)}.  We indicate the phase delays of the red and blue experiments which we fit and subtract from one another to extract $\phi_{111} + \phi_{011}$.   We also highlight the amplitudes of the red and green experiments.  Ideally, these amplitudes should be identical, which would indicate that the gate is working properly.  Additional decoherence because of being highly excited and imperfect state transfer slightly reduces the amplitude of red by about 4\%.
}
{\label{fig:toffoliphasetuneup}}
\end{figure}

What happens when we apply cRamsey to three qubits?  If we only excite two qubits, the above description remains unchanged.  We can therefore use this technique without modification to measure and tune-up the two two-qubit phases of the gate.  Measuring the three-qubit phase is more complicated, however.  Consider what happens if we perform a Ramsey sequence on $Q_2$ with $Q_1$ excited and $Q_3$ excited or not, as described in \figref{fig:toffoliphasetuneup}.  The first oscillation will be between $\ket{100}$ and $\ket{110}$, and tells us $\phi_{010}+\phi_{110}$.  The second oscillation is between $\ket{101}$ and $\ket{111}$, giving us $\phi_{010}+\phi_{110}+\phi_{011}+\phi_{111}$.  Subtracting these two, we see the phase delay is given by $\phi_{011} + \phi_{111}$.  This indicates that we cannot only measure the three-qubit phase in which we are interested.  Unfortunately, the only means to circumvent this is to independently measure $\phi_{011}$.

The cRamsey technique provides another important piece of information about your gate: its approximate fidelity.  The contrast of the Ramsey curves after a gate, normalized to a Ramsey where nothing happens between the two $\pi/2$ pulses, is given by the remaining coherence of the target qubit.  Contrast can be reduced if quantum amplitude leaks out of the Hilbert space or for any other reason that the control qubits are not returned to their ground state.  Almost nothing except technical problems with the measurement apparatus will {\it increase} contrast.  This is a good first test; measuring the ratio of Ramsey contrast is much faster than the more formal methods determining gate fidelity (such as quantum process tomography), so verifying with this method before proceeding with those techniques can prevent a lot of wasted effort.

With the two techniques of swap spectroscopy and cRamsey, we can fully tune-up the Toffoli-sign gate.  After extracting the parameters from swap spectroscopy as described above, we tune-up the phases $\phi_{011}$ by adjusting the waiting position of $Q_2$, represented by the flat portion of the green curve in \figref{fig:toffolipulse}.  Since we acquire phase so rapidly at that position, there are usually several waiting locations that will satisfy $\phi_{011}=0$, which we choose between experimentally to maximize the Ramsey contrast of the other basis states.  We set $\phi_{111}$ with the height of the first pulse on $Q_1$ (the red curve).  Finally, we adjust $\phi_{110}$ with the height of the second pulse on $Q_1$.  This tune-up is iterated several times until all the phase values are stable.  Given that, we can also measure the single-qubit phases for unwrapping in software.

\subsection{Toffoli-sign gate tomography}
\label{subsec:toffolitomography}

\begin{figure}
	\centering
	\includegraphics{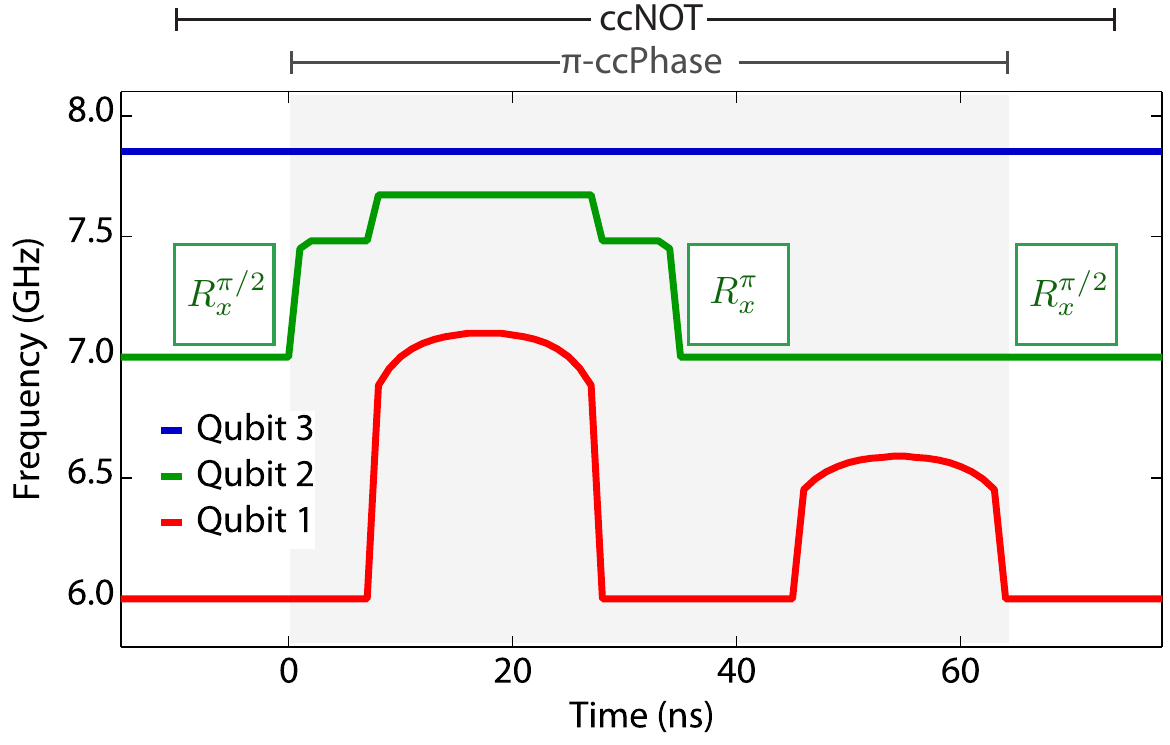}
	\mycaption{ccNOT implemented with Toffoli-sign gate}
	{The full pulse sequence for generating a controlled-controlled-NOT gate with our efficient Toffoli-sign gate is shown.  $Q_2$ is pulsed by $\pi/2$ around the $x$-axis before the Toffoli-sign gate is applied.  Because of our $\pi$ pulse on $Q_2$ to save time on the unwrapping of the $\phi_{011}$ phase, we refer to this operation as a $\pi$-ccPhase.  Afterwards, another $\pi/2$ gate is applied to $Q_2$.  If the ccPhase flipped the sign of the manifold, this gate will combine with the other rotations on $Q_2$ to flip its state.  If not, it will rotate $Q_2$ back to its original value.
	\figthanks{Reed2012}}
	{\label{fig:toffoliccnot}}
\end{figure}

We first demonstrate our tuned-up gate by measuring its classical action.  The ccPhase gate, which maps $\ket{111}$ to $-\ket{111}$, has no effect on pure computational states because the extra phase acts globally.  Instead, we can implement a ccNOT gate by appending $\pm\pi/2$ pulses on $Q_2$ both before and after a ccPhase gate, in the same way that we have previously mapped a cPhase to a cNOT.  For the two input states where both ancillas are excited ($\ket{101}$, $\ket{111}$, becoming $\ket{101}\pm\ket{111}$ after the pulse), $Q_2$ acquires a $\pi$ phase shift which flips the phase of the second pulse, and so the two pulses combine for a full rotation.  (The uncorrected $\phi_{101}$ phase is irrelevant because it does not affect $Q_2$.)  Other computational states do not see the phase of $Q_2$ shifted, so the pulses cancel.  We show the actual phase in the shaded region of \figref{fig:toffoliccnot}, which differs slightly from a pure ccPhase because of the $\pi$ pulse used to speed the unwrapping of $\phi_{110}$.  A ccNOT is thus constructed by making both the pre- and post-rotations a positive $\pi/2$ rotation.   

\begin{figure}
	\centering
	\includegraphics{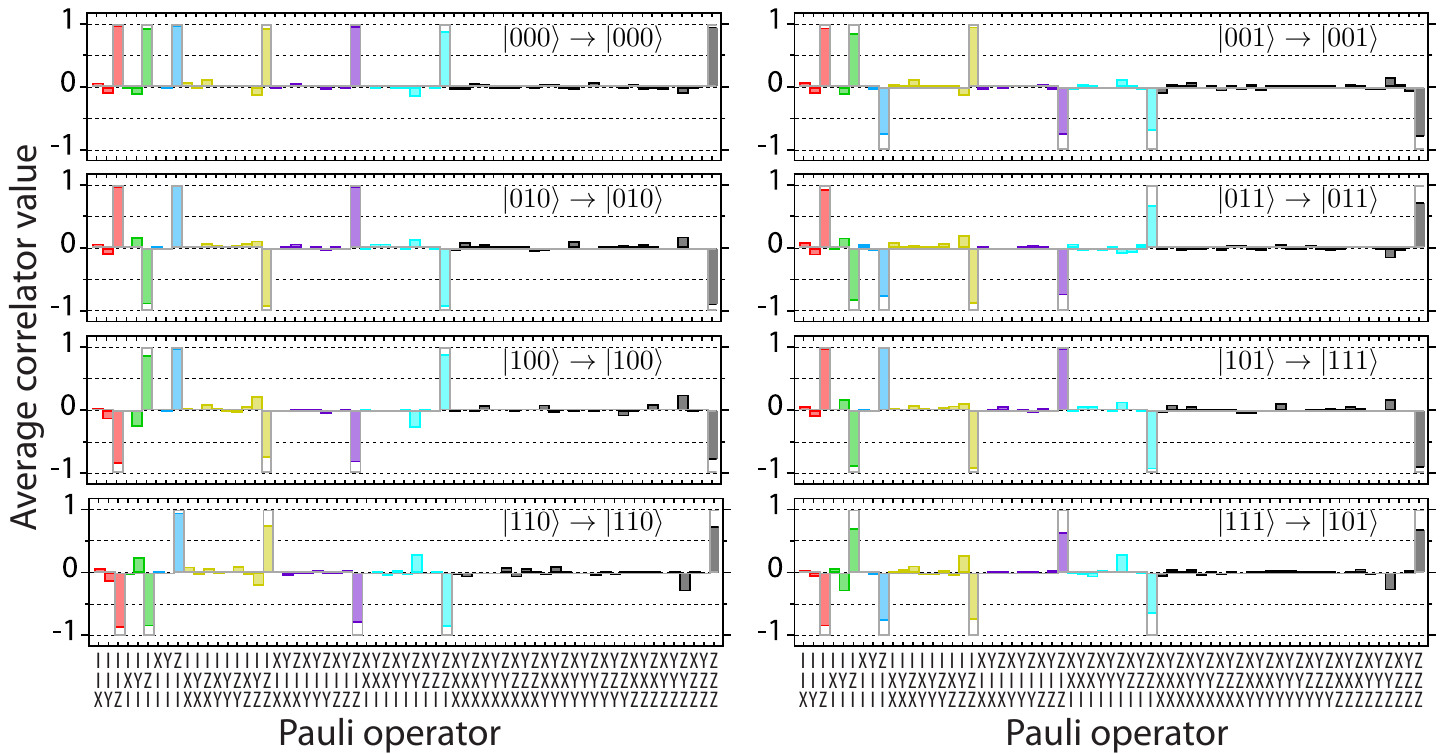}
	\mycaption{ccNOT classical verification state tomograms}
	{We can measure the classical action of our ccNOT gate by preparing the eight computational states, applying the gate to them, and measuring the result with state tomography.  We show each tomogram in the Pauli-bar representation, with gray bars indicating the ideal values of each correlation.  Each tomogram is labeled with the state that was prepared and its ideal final state.  Since the gate is conditional on the states of $Q_1$ and $Q_3$, $Q_2$ is flipped if and only if both those qubits are excited.
	\figthanks{Reed2012}}
	{\label{fig:truthtabletomo}}
\end{figure}

A ccNOT gate ideally swaps the states $\ket{101}$ and $\ket{111}$ but does not affect the remaining computational states.  To verify this, we prepare the eight computational states, apply the gate, and measure the resulting output with three-qubit state tomography \cite{DiCarlo2010}.  We report each tomogram in \figref{fig:truthtabletomo}.  We label each tomogram with the state that was prepared and what it should ideally be mapped to, and use gray to outline the bars of the ideal target state.  Generally speaking, the more highly excited the input state, the less faithful the mapping.  Note that the lack of erroneous two- and three-qubit correlations indicates that there is no significant loss of population from the computational subspace (\sref{subsec:tomophysicality}).  

\begin{figure}
	\centering
	\includegraphics{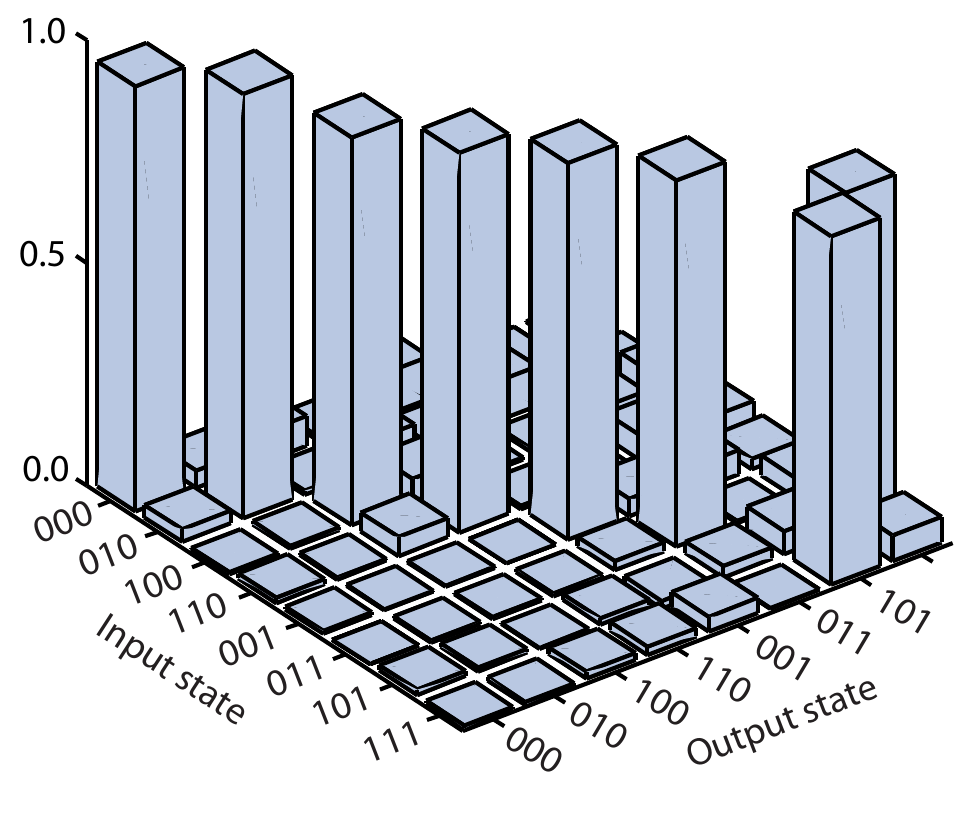}
	\mycaption{ccNOT classical truth table}
	{We take the projection of each tomogram shown in \figref{fig:truthtabletomo} onto the eight computational states and plot the result as a function of the prepared state.  This is the classical ``truth table'' of the ccNOT gate.  Note that the matrix is diagonal except for the bottom corner, where the qubit is flipped if and only if the two control bits, $Q_1$ and $Q_3$, are excited.  The correct state is reached with an average of $85\pm1\%$ fidelity.
	\figthanks{Reed2012}}
	{\label{fig:truthtable}}
\end{figure}

We can measure the fidelity to each target state by taking a dot product of the measured values with the target state.  The states $\ket{000}$, $\ket{010}$, $\ket{100}$, $\ket{110}$, $\ket{001}$, $\ket{011}$, $\ket{101}$, and $\ket{111}$ map to their ideal targets with fidelities of 95.4, 95.2, 87.1, 85.2, 84.4, 82.2, 78.1, and 75.7\%, respectively.  The unconventional order of states is commensurate with the order of fidelity.  States that differ only by the excitation of $Q_2$ are similar because that qubit is immediately pulsed onto the equator so its state will not change the net excitation of the system during the gate nor modify the effect of $T_1$.  We then order according to the knowledge that $Q_1$ has a longer $T_1$ than does $Q_3$.  By taking the projection of each tomogram to the remaining seven computational states, we generate the classical truth table, shown in \figref{fig:truthtable}.  The intended state is reached with $85 \pm 1\%$ fidelity on average.

\begin{figure}
	\centering
	\includegraphics{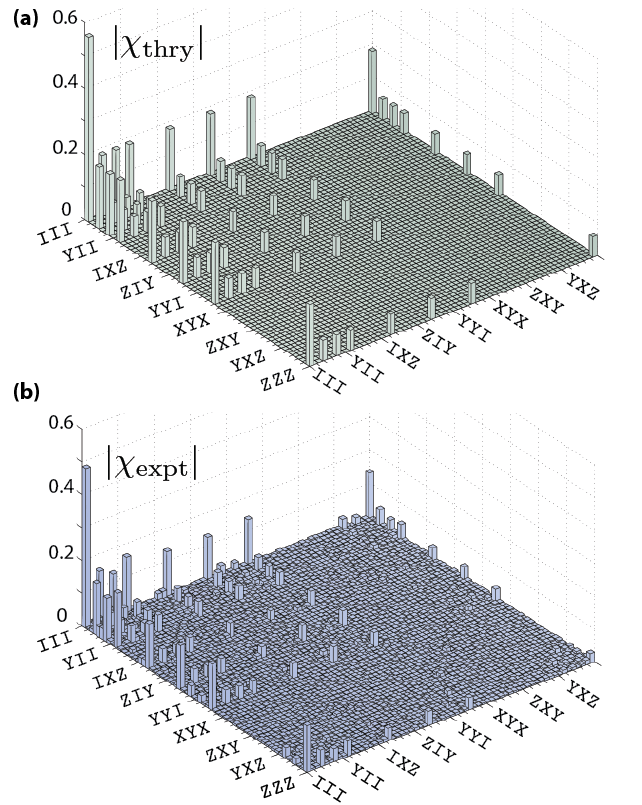}
	\mycaption{Toffoli-sign gate process tomography}
	{\capl{(a)} A theoretical prediction of the $\chi$-matrix of the gate, including the spurious phase $\phi_{101}$ set to its experimentally-measured value of $57$ degrees.  
	\capl{(b)} Full quantum process tomography of the Toffoli-sign gate, acquired with the procedure described in \sref{sec:processtomo}.  In both cases, the absolute value of $\chi$ is shown.  The fidelity of the experimental gate to its theoretical ideal is $78\pm1\%$.  The order of operators here are as follows: III, IIX, IIY, IIZ, IXI, IYI, IZI, XII, YII, ZII, IXX, IYX, IZX, IXY, IYY, IZY, IXZ, IYZ, IZZ, XIX, YIX, ZIX, XIY, YIY, ZIY, XIZ, YIZ, ZIZ, XXI, YXI, ZXI, XYI, YYI, ZYI, XZI, YZI, ZZI, XXX, YXX, ZXX, XYX, YYX, ZYX, XZX, YZX, ZZX, XXY, YXY, ZXY, XYY, YYY, ZYY, XZY, YZY, ZZY, XXZ, YXZ, ZXZ, XYZ, YYZ, ZYZ, XZZ, YZZ, and ZZZ.  We do not make use of the maximum-likelihood estimator commonly used to require the physicality of the $\chi$-matrix so that the reported elements of $\chi$ and the fidelity are linearly related to the raw measurements \cite{Chow2010b}.
	\figthanks{Reed2012}
	}
{\label{fig:toffoliprocesstomo}}
\end{figure}

Since we only input computational states, the truth table is only sensitive to classical action.  We therefore complete our verification by performing full quantum process tomography (QPT) on the ccPhase gate, which detects the evolution of quantum superpositions of computational states.  Instead of only eight, we prepare 64 input states which span the computational Hilbert space and then perform state tomography on the result of the gate's action on each state (see \sref{sec:processtomo}).  The resulting $\chi$-matrix is shown with both the theoretical prediction and experimental result in \figref{fig:toffoliprocesstomo}.  The fidelity is found to be $78 \pm 1$\% to a process in which the spurious two-qubit phase between $Q_1$ and $Q_3$ is set to the independently measured value of 57 degrees.  Though this phase does not affect the operation of error correction, it is detected by QPT.  Due to this extraneous phase, the gate is most accurately described as a cc-$e^{i\phi}Z$ gate.  The infidelity is consistent with the expected energy relaxation of the three qubits during the $85 \ns$ procedure, with some remaining error owing to qubit transition frequency drift during the 90 minutes required to collect the full dataset.  This completes our construction and verification of our efficient Toffoli-sign gate.  

\section{Realizing the bit-flip code}

\begin{figure}
	\centering
	\includegraphics{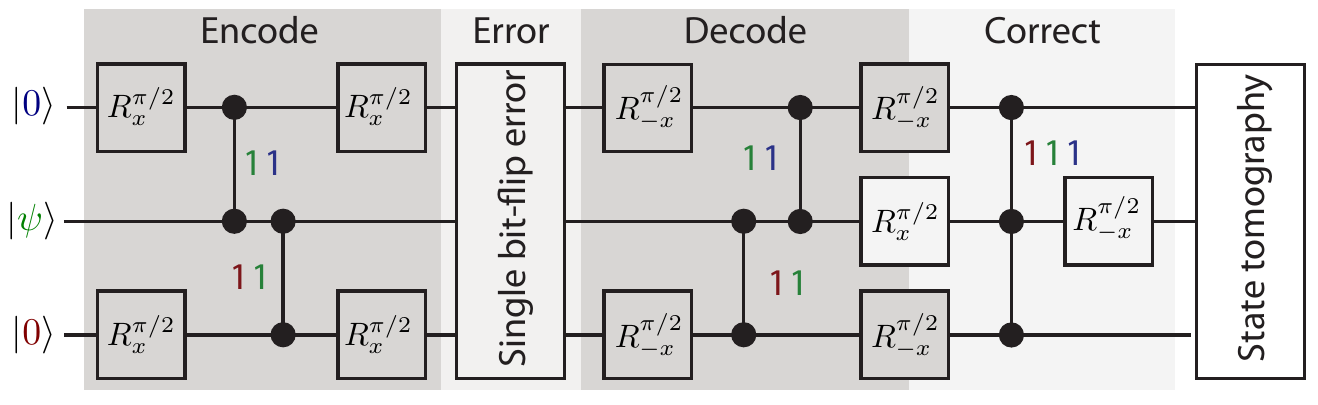}
	\mycaption{Bit-flip QEC circuit}
	{The error correction protocol starts by encoding the quantum state to be protected in a three-qubit GHZ-like state by entangling the two ancilla qubits $Q_1$ and $Q_3$ with $Q_2$ through the use of $\pi/2$ rotations and two cPhase gates (vertical lines terminating in solid circles).  The number adjacent to each cPhase indicates which state receives a $\pi$ phase shift (e.g. for each the single-qubit phases $\phi_{01}=\phi_{10}=0$, with $\phi_{11} = \pi$).  A single $y$-rotation of a known angle $\theta$ is then performed on a single qubit to simulate a bit-flip occurring with probability $p=\sin^2(\theta/2)$.  The state is then decoded, leaving the ancillas in a product state indicating which single-qubit error occurred.  For finite rotations, the ancillas will be in a superposition of the error occurring or not, with each tensor-multiplied by the associated single-qubit state of $Q_2$.  If an error has occurred on $Q_2$, the ccNOT gate implemented with our ccPhase gate (represented by three solid circles linked by a vertical line) at the end of the code will correct it.  We then perform three-qubit state tomography to verify the result.  Note that in the actual experiment, we compile consecutive single-qubit rotations together to save time and increase fidelity.  Specifically, the error $y$-rotation on the ancillas would normally be placed between a positive and negative $\pi/2~x$-rotation associated with turning the cPhase gate into a cNOT, and so we compile these three single-qubit gates into one $z$-gate.
	\figthanks{Reed2012}}
	{\label{fig:bitflipcircuit}}
\end{figure}

With our Toffoli gate in hand, and having leveraged our knowledge of how to create GHZ states (as introduced in \chref{ch:entanglement}), we can now demonstrate the quantum repetition code.  We first examine the bit-flip code, which is shown in \figref{fig:bitflipcircuit}.  To review, the code begins by encoding some quantum state in a three-qubit entangled state through the use of cPhase gates.  This is an exact replica of the code we used to create GHZ states in \chref{ch:entanglement}, but now modified so the state of $Q_2$ is arbitrary.  (When $Q_2$ is located on the equator, the resulting state is a maximally-entangled GHZ state \cite{Greenberger1989, DiCarlo2010, Neeley2010}; for any other state, the resulting state may or may not exhibit some level of entanglement.)  As before, we use the sudden cPhase gates that employ the $\ket{11} \leftrightarrow \ket{02}$ avoided crossing between the involved qubits.  The state $\alpha\ket{0}+\beta\ket{1}$ is thus encoded as $\alpha\ket{000}+\beta\ket{111}$.

Recall that this state has the property that any two-qubit $ZZ$ product is $+1$ regardless of the values of $\alpha$ and $\beta$.  Moreover, if any single qubit is flipped, one or both of these $ZZ$ products will flip this sign as well.  For example, if $Q_3$ were flipped, we will have the state $\alpha\ket{001}+\beta\ket{110}$, whose $Z_1 Z_2$ expectation value is $+1$ but whose $Z_2 Z_3$ product is $-1$.  As seen in \tref{table:ghzerroreigenvalues}, if no more than one qubit flips, knowledge of the two $ZZ$ products uniquely specifies which error occurred.  Since we do not have extra ancilla qubits to extract these error syndromes in a fault-tolerant way, we instead reverse the encoding, leaving $Q_2$ in a state that may or may not have been flipped and $Q_1$ and $Q_3$ containing the values of the $Z_1 Z_2$ and $Z_2 Z_3$ products (as explained in \sref{subsubsec:msmtbasedqrc}).  Specifically, if $Q_2$ were flipped, both of the ancillas will be left in their excited state.  As discussed in \sref{subsubsec:autonomousec}, we can then use these qubits as the control bits of our three-qubit Toffoli gate to rotate $Q_2$ back to its original state conditional on the error.  At this point, the ancillas store the entropy associated with the error.  If we wished to apply another round of QEC, we would need to reset them to return the qubit register to its original state, perhaps by using the method described in \sref{subsec:qubitreset}.  Thus, we see that this code can correct for full bit-flips.

\subsection{Finite rotation errors}

With classical bits, the only errors that may occur are full bit-flips, mapping $0\leftrightarrow 1$.  With quantum bits, however, {\it finite} rotations may occur because superpositions of $0$ and $1$ are allowed.  Luckily, the code we just introduced works for finite errors as well.  To show this, we will step through the evolution of the quantum register at each point in the code.  Prior to encoding, $Q_2$ is in an unknown state $\alpha\ket{0}+\beta\ket{1}$ and the ancilla are in their ground state, so the system state is $\ket{\psi}= (\alpha\ket{0}+\beta\ket{1})\otimes\ket{00}$, where we list $Q_2$ followed by the two-qubit ancilla state of $Q_1$ and $Q_3$.  After the encoding step, we are in the canonical state $\ket{\psi}=\alpha\ket{000}+\beta\ket{111}$, which is sensitive to errors.  Suppose we now perform a rotation on $Q_2$, not of $\pi$, but some {\it finite} angle $\theta$.  Though we implement this as a coherent rotation, in the context of the error correcting code this can be viewed as a bit-flip that occurs with probability $p=\mathrm{sin}^2(\theta/2)$.  The wavefunction will now be a superposition of the two possibilities, $\ket{\psi} = \sqrt{1-p}(\alpha\ket{000}+\beta\ket{111}) + \sqrt{p}(\alpha\ket{010}+\beta\ket{101})$\footnotemark.  Decoding, we will be left with $\ket{\psi} = \sqrt{1-p}(\alpha\ket{0}+\beta\ket{1})\otimes\ket{00} + \sqrt{p}(\beta\ket{0}+\alpha\ket{1})\otimes\ket{11}$: that is, the state will be a superposition of $Q_2$ in the correct state with the ancillas indicating no error plus $Q_2$ flipped with the ancillas indicating as such.  Finally, we apply the ccNOT gate on this wavefunction, which coherently flips $Q_2$ if and only if both $Q_1$ and $Q_3$ are both in their excited state.  This results in the wavefunction $\ket{\psi} = \sqrt{1-p}(\alpha\ket{0}+\beta\ket{1})\otimes\ket{00} + \sqrt{p}(\alpha\ket{0}+\beta\ket{1})\otimes\ket{11} = (\alpha\ket{0} + \beta\ket{1})\otimes(\sqrt{1-p}\ket{00} + \sqrt{p}\ket{11})$, where in the last step we have rewritten $\psi$ to highlight the fact that the wavefunction of $Q_2$ is now correct regardless of $p$.  Thus, we have shown that this code works even for finite rotations because the ccNOT gate operates only on the subspace where a full bit-flip has occurred and otherwise does nothing.

\footnotetext{If we did not know that the rotation was coherent, we could also write the state of the system at this point as a density matrix $\rho = (1-p)(\alpha\ket{000}+\beta\ket{111})(\alpha^*\bra{000}+\beta^*\bra{111}) + p(\alpha\ket{010}+\beta\ket{101})(\alpha^*\bra{010}+\beta^*\bra{101})$.}

\begin{figure}
	\centering
	\includegraphics{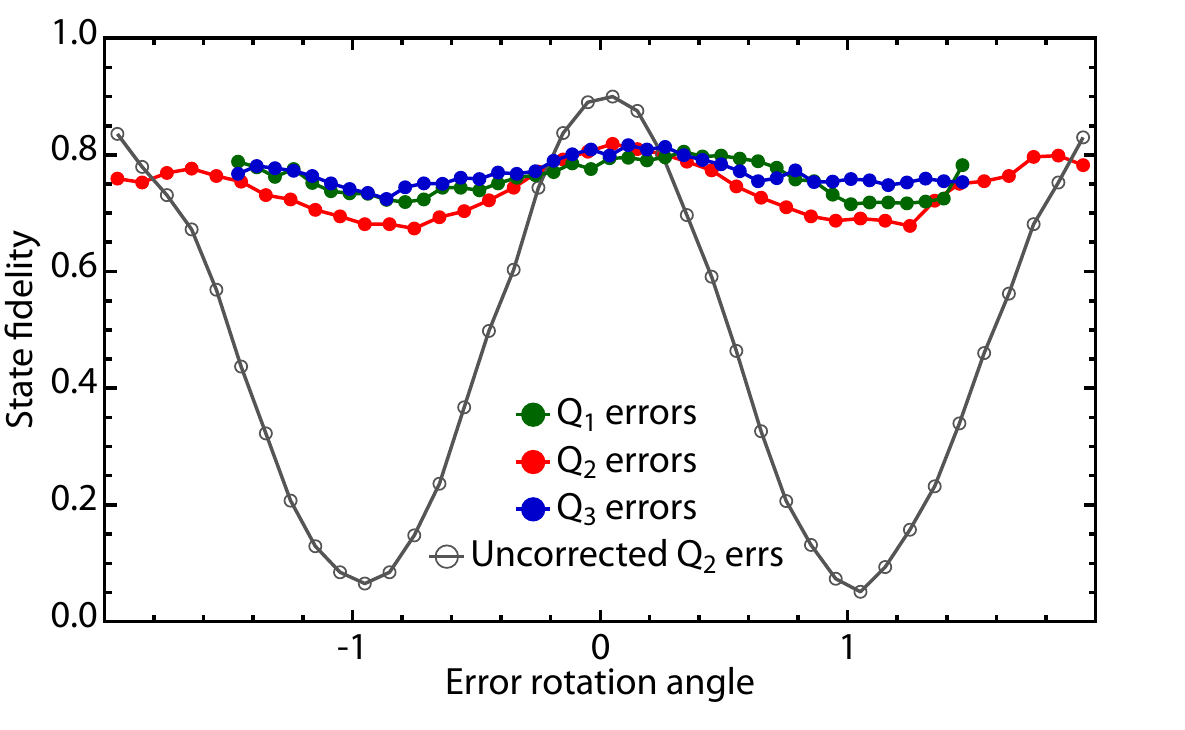}
	\mycaption{Bit-flip quantum error correction with finite error rotations}
	{State fidelity to the created state $\ket{\psi}=\ket{\mathrm{+X}}$ after applying an error on one of the qubits, with and without error correction.  Ideally, the curves would be flat lines at unit fidelity.  Finite excited-state lifetimes cause oscillations and displacement downward because errors change the excitation level of the system.
	\figthanks{Reed2012}}
{\label{fig:bitflipqec}}
\end{figure}

We demonstrate the process just described by implementing the circuit shown in \figref{fig:bitflipcircuit}.  We choose the initial state $\ket{\psi}=\ket{\mathrm{+X}}$ to encode in the three-qubit state and perform single deterministic rotations around the $y$-axis by some angle to simulate errors.  We then un-encode and correct the error, before performing state tomography to measure the fidelity of $Q_2$ to its original state.  As shown in \figref{fig:bitflipqec}, we plot this fidelity for the case of error correction both performed or not performed with errors applied to $Q_2$.  For the case of no error correction, we still wait for an equivalent period of time but do not involve $Q_1$ or $Q_3$, so as to indicate the loss of fidelity due to the decoherence of $Q_2$.  We also demonstrate the case of errors on the two ancilla qubits individually, because when doing error correction, their state undergoing an error cannot be allowed to corrupt the encoded state.  The error correcting code should perfectly correct any single rotation, so we should expect unit fidelity for all angles when we apply it.  However, because of qubit decay and the varying excitation levels of the qubits depending on the error performed (e.g. when $Q_2$ is flipped, $Q_1$ and $Q_3$ are excited and thus can decay), the curves show a small oscillation centered at 75.7\% fidelity.  Nevertheless, they demonstrate a significantly reduced sensitivity to $\theta$ as compared to the uncorrected case, showing that our error correction successfully ameliorates the errors.

\subsection{Verifying ancilla states}

\begin{figure}
	\centering
	\includegraphics{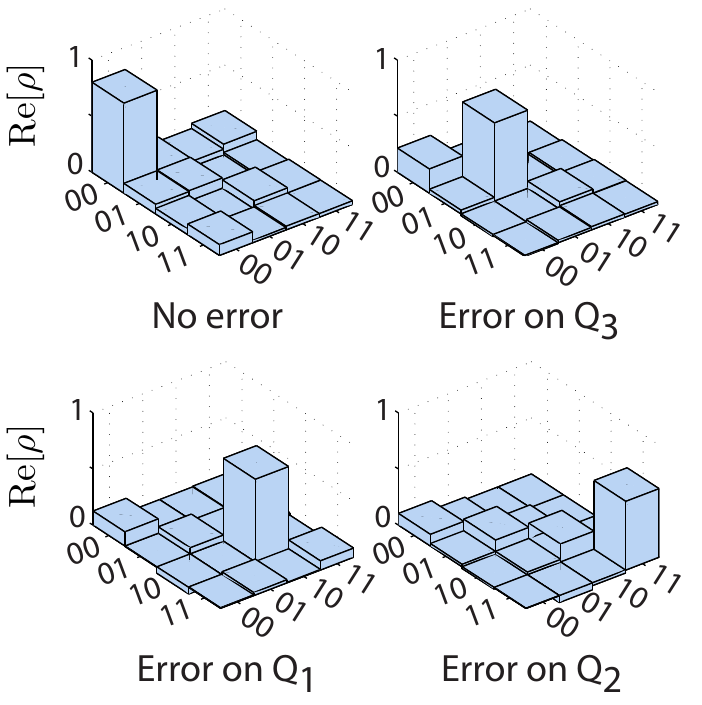}
	\mycaption{Ancilla states after bit-flip quantum error correction}
	{Two-qubit density matrices of the ancillas after each of the four possible full bit-flip errors have occurred.  The fidelities of each of these states to the ideal error syndromes, which are respectively $\ket{00}$, $\ket{01}$, $\ket{10}$, and $\ket{11}$, are $(81.3\%, 69.7\%, 73.1\%, 61.2\%)$
	\figthanks{Reed2012}}
	{\label{fig:bitflipancilla}}
\end{figure}

Our demonstration of a reduced oscillation contrast is encouraging, but in some sense reflects a null result.  We do something and nothing happens.  How can we show that the code truly works as expected?  One approach is to measure the two-qubit density matrices of the ancilla qubits.  After the four possible full-flip errors (no error or a flip of one of the three qubits), they should be in a computational product state that indicates which error occurred.  As shown in \figref{fig:bitflipancilla}, this is indeed the case, allowing for finite fidelity due to decoherence.  These ``debugging data'' demonstrate, at least in the case of full flips, that the code operates as expected.  We could also examine the case of finite rotations mapping to the correct states, where the ancilla should be left in some superposition state of error and no error.  In practice, however, it is easier and more impressive to look for the signature of error correction when we apply simultaneous errors.  We will show this in \sref{subsec:simultaneouserrors}.

\section{Realizing the phase-flip code}
\label{sec:realizingphaseflipcode}

\begin{figure}
	\centering
	\includegraphics{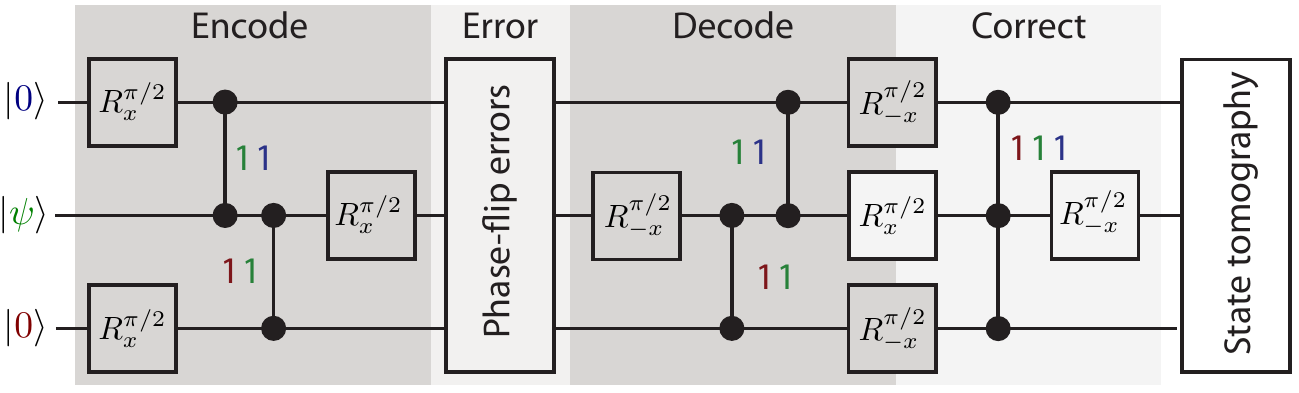}
	\mycaption{Phase-flip quantum error correction circuit}
	{The phase-flip error correction protocol differs from the bit-flip one described in \figref{fig:bitflipcircuit} in only the single-qubit rotations.  Those gates effectively rotate the coordinate system, mapping phase-flips to bit-flips and vice versa, so the remainder of the procedure is exactly the same as for the bit-flip case \cite{Nielsen2000}.  We perform errors on all three qubits simultaneously with $z$-gates of known rotation angle, which is equivalent to phase-flip errors with probability $p=\mathrm{sin}^2(\theta/2)$.  These $z$-gates are implemented ``in software'' by rotating the frame of reference of subsequent $x$- and $y$-rotations.  As with the bit-flip code, if a single error has occurred on the primary qubit, the ccNOT gate at the end of the code will undo it.
	\figthanks{Reed2012}}
	{\label{fig:phaseflipcircuit}}
\end{figure}

The quantum repetition code can correct for one type of error: bit-flips or phase-flips.  Thus far, we have discussed the case of bit-flip errors.  These have the advantage of being intuitive because of their direct classical analog.  However, they are in some sense a less important error in our system.  Bit errors only occur on their own if we have incorrectly tuned-up pulses (or, perhaps, some spurious mixer leakage, etc).  They also happen during spontaneous emission, but only in combination with phase-flip errors.  (That is, the $T_1$ process is {\it not} a bit-flip and it can only be described as a combination of bit- and phase-errors; see \sref{subsubsec:superoperators}.)  Phase-flips, on the other hand, are a much more common kind of error associated with {\it inhomogeneous dephasing} ($T_\phi$ dephasing not due to $T_1$).  Therefore, we are interested in implementing the phase-flip error correcting code.  Fortunately, as shown in \figref{fig:phaseflipcircuit}, the phase-flip code is a very straightforward modification of the bit-flip code.  The only difference is in the rotations performed after the entanglement.  Rather than rotating the primary qubit as in phase-flip correction, the ancillas are $\pi/2$ pulsed.  This can be understood as a change in coordinate system, converting phase-flips to bit-flips and vice versa; the remainder of the code is exactly the same as in the previous case \cite{Nielsen2000, Tornberg2008, Schindler2011}.

\subsection{Simultaneous errors}
\label{subsec:simultaneouserrors}

We originally introduced the quantum repetition code in the context of a classical binary symmetric channel (\sref{subsec:classicalrepetitioncode}).  We wished to transmit a bit through a noisy communication channel that had a probability $p$ of flipping it.  We found that we could reduce our sensitivity to this error if we communicated each bit three times.  The resulting error rate was a quadratic function of $p$, given by $3p^2-2p^3$.  Critically, as long as $p$ was small -- less than 1/2, which is not a very strict requirement, since a channel with error probability 1/2 has zero information carrying capacity -- then the overall error rate of this approach was smaller than sending only one bit.  The reason $p$ must be small is that by transmitting more bits of information, we are also adding ways for the information to be corrupted.

This is also true for quantum error correction.  When we add extra qubits to encode our quantum state, we also add extra ways by which the system could be corrupted.  The qubits we use to encode are no less likely to experience an error than the qubit we are trying to protect.  This idea shows that our demonstration of error correction is quite artificial because we intentionally apply exactly one error at a time.  A more realistic error model would account for the fact that in real physical systems, errors occur at approximately the same rate on all qubits rather than on one at a time.  Our error correction scheme can only succeed if no more than error happens, and will fail if there are two or three coincident errors.  As in the classical case, if we have a single-qubit error rate $p$, the probability of two or more errors is given by $3p^2-2p^3$ and so the fidelity of the error correction process should be given by $1 - 3p^2 + 2 p^3$.  For a scheme with gate fidelity limited by decoherence, these coefficients will be smaller, but crucially, any linear dependence on $p$ will be strongly suppressed.  If the qubits experience errors at different rates, these coefficients would again be modified, but the linear dependence would remain abated.  This suppression of linear dependence on $p$ is the unambiguous signature of quantum error correction.

\subsection{Process tomography}
\label{subsec:qecproctomo}

\begin{figure}
	\centering
	\includegraphics{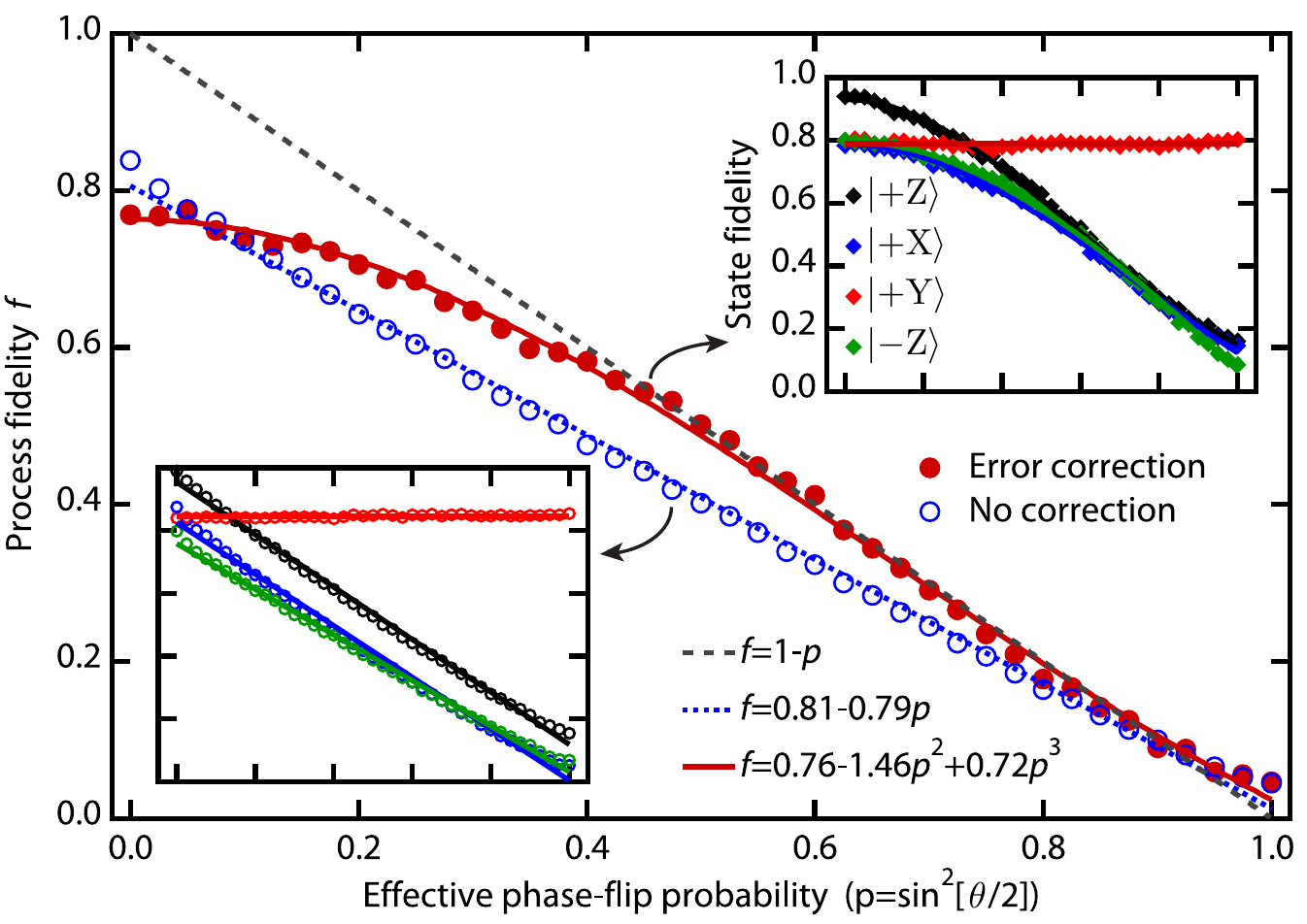}
	\mycaption{Process tomography of phase-flip quantum error correcting circuit}
	{We perform quantum process tomography on the phase-flip QEC circuit shown in \figref{fig:phaseflipcircuit} as a function of the effective error probability.  The fidelity of the process matrix of the protected qubit to the identity operation is plotted as a function of $p$.  As the code corrects only single-qubit errors, it will fail on the three-qubit subspace where more than one has occurred, which happens with a probability $3p^2 - 2p^3$.  These coefficients are reduced for processes with finite fidelity.  The process fidelity is fit with $f = (0.760 \pm 0.005) - (1.46 \pm 0.03) p^2 + (0.72 \pm 0.03) p^3$.  If a linear term is allowed, its best-fit coefficient is $0.03 \pm 0.06$.  We compare this to the case of no error correction to simulate the improvement that occurs when the decoherence of $Q_2$ is normalized away (blue symbols).  We also plot the simulated fidelity of a decoherence-free but non-corrected system (black dashed line), indicating that we do not show a net improvement for any rate of artificial errors.
	\capl{(insets)} The constituent state fidelities of the four basis states used to produce the process fidelity data for the case of \capl{(right)} error correction and \capl{(left)} no correction.  The $x$-axes of both plots are the same as the main panel, and they share the same $y$-axis.  The state $\ket{\mathrm{+}Y}$ is immune to errors because its encoded state is an eigenstate of two-qubit phase-flips.
	\figthanks{Reed2012}
}
{\label{fig:phaseflipqpt}}
\end{figure}

We demonstrate the procedure shown in \figref{fig:phaseflipcircuit} by creating some state $\ket{\psi}$ of $Q_2$, applying the protocol, and then measuring the state of $Q_2$ using state tomography.  In contrast to the bit-flip case, we are now going to do something slightly more sophisticated.  Instead of testing some specific chosen state $\ket{\psi}$ that could be more or less susceptible to the errors that are being applied, we instead perform full quantum process tomography on the protocol (\sref{sec:processtomo}).  To do so, we create four states ($\ket{0}$, $\ket{1}$, $\ket{+X}$ and $\ket{+Y}$) that span the single-qubit Hilbert space of $Q_2$.  We apply the QEC circuit to each of these states and measure the result with state tomography.  The resulting state fidelities for the case of performing or not performing error correction are plotted in the insets of \figref{fig:phaseflipqpt}.  For the case of no error correction, we apply identical single-qubit rotations to $Q_2$ without involving the ancillas.  We insert appropriate delays to have the same total procedure duration to indicate the infidelity due to the decoherence of $Q_2$, then combine each of these state tomograms into a single process matrix and calculate the fidelity of that process matrix to the identity.  The two process fidelities are plotted and fit to a polynomial in \figref{fig:phaseflipqpt}.  

For no error correction, we find a purely linear dependence on error probability given by $f=0.81-0.79p$.  However, when we apply the correction, we suppress all linear dependence, finding that the data are well-fit by $f=0.76 - 1.46p^2 + 0.72p^3$.  If we include a linear term in this function, its best-fit value is found to be consistent with zero.  The suppression of linear dependence on $p$ when performing error correction was precisely the signature that we had hoped to find, and proves that the circuit works as expected.  We anticipated that the fidelity of the process would only be improved by error correction if the error probability was less than $1/2$, below which it should be reduced.  However, in \figref{fig:phaseflipqpt}, we see that the error corrected case is always better than the non-error corrected case.  This is the result of an accident: below 50\% fidelity, decoherence of our state actually {\it improves} our state fidelity since a fully-mixed state has 50\% fidelity to any state.  (Its $II$ Pauli correlation is always defined to be $+1$, and is included in the fidelity.)  Moreover, when we are performing quantum error correction, we are much more susceptible to decoherence since three qubits are involved rather than one.  Thus, we get a boost in fidelity that pushes us above the uncorrected case.  Nevertheless, if the code was perfect and decoherence was nonexistent, when $p>1/2$, the protocol should indeed reduce our net fidelity.

In \figref{fig:phaseflipqpt}, we also plot the fidelity of a process without error correction in the absence of any overhead.  This simulates a process in which we perfectly prepare some state and immediately apply an error before measuring the resulting fidelity.  As you can see, this line is always above the experimental lines, which indicates that our error correction protocol never actually improves effective error rates.  Put another way: while the protocol corrects errors, the associated overhead is so large that the net fidelity of some process will never be improved.  This is not the fault of the code, but rather of the system itself.  The qubits we used in this planar structure are simply too error-prone.  Their coherence times were on the order of $1\us$, which is on the order of the time it takes to run the correction protocol.  In order for error correction to make sense, the overhead needs to be small compared to the improvement associated with going from a linear dependence on $p$ to a quadratic one.  Thus, in order to show an actual improvement (and perform error correction for a reason other than its own sake), we need substantially better qubits.

\section{Conclusion}

In this chapter, we have demonstrated the first realization of the most basic form of quantum error correction -- the quantum repetition code -- in a superconducting circuit.  We realized both bit- and phase-flip error correction, testing both major conceptual components of the nine-qubit Shor code \cite{Shor1995b} which can protect from arbitrary single-qubit errors by concatenating the two codes.  The key to our implementation was an efficient three-qubit Toffoli-sign gate, which is required for the autonomous repetition code.  Our gate made use of an avoided crossing between the $\ket{102}$ and $\ket{003}$ non-computational states to acquire three-qubit conditional phase.  Accessing this avoided crossing required intentionally leaving the computational space with a sudden swap operation.  We characterized our gate with both quantum state and quantum process tomography, and found fidelities approaching $80\%$.  The gate requires approximately half the time of an equivalent construction with one- and two-qubit gates. 

Combining this gate with a sudden cPhase gate, we implemented the bit-flip QEC code.  We began by encoding the qubit state that we wished to protect in a three-qubit GHZ-like state and then applied a deterministic rotation on one of the qubits to simulate a bit-flip error.  We showed how the code works with finite error rotations and that the error syndromes are being correctly encoded in the ancilla qubits.  We also implemented the phase-flip code, but used a more sophisticated error model that simulated simultaneous errors on all three qubits.  We performed process tomography on this circuit, and found the expected suppression of linear dependence of process fidelity on error probability.  Though this proved that the code was working as expected, the fidelity never exceeded the case of no error correction without overhead, which indicates that the improvement of the fidelity of a real physical process will require significant advances in both gate fidelity and device complexity.

\setcounter{chapter}{8}
\chapter{The Tunable 3D cQED Architecture}
\thumb{The Tunable 3D cQED Architecture}
\lofchap{The Tunable 3D cQED Architecture}
\label{ch:tunable}


\lettrine{I}{n} the previous chapter, we showed that we have built up the sophistication to perform basic quantum error correction with superconducting qubits.  We concluded by noting that although the circuit worked as anticipated, it did not improve error rates because our qubits were too short-lived.  We need much more coherent qubits that are as controllable as those we have already used.  Our answer to this requirement is the tunable 3D cQED architecture, first introduced in \sref{sec:tunable3darchitecture}.  That architecture combines the extremely high coherence times of 3D cQED with the ability to tune the system in-situ with fast flux bias lines.  In this chapter, we will summarize the experimental results attained thus far with tunable 3D cQED.  Though the results are unpublished, we aim to give a flavor of the possibilities enabled by this architecture.

We begin by showing that our flux bias lines constitute a significant relaxation channel.  We measure qubit $T_1$ as a function of frequency.  Both the magnitude and structure of these data provide strong support for our hypothesis that the qubit relaxes through the FBL because of capacitive coupling.  We confirm this by adding a low-pass filter to the FBL, and demonstrate that the qubit's relaxation is significantly ameliorated when we fabricate a big enough filter capacitor.  We also use flux spectroscopy to confirm that this low-pass filter does not adversely affect the response time of our fast flux bias lines.

In the second part of this chapter, we introduce several experiments to measure the system Hamiltonian parameters as a function of qubit frequency.  In the device we are describing, the cavity inherits enough anharmonicity from the qubit when it is at its maximum frequency that we can prepare photon number states.  By combining this ability with fast flux control, we show how to measure cavity lifetime, coherence time, nonlinearity, and dispersive shift as a function of qubit frequency.  While the results are of interest in and of themselves, this section primarily serves as an instruction manual for building up and understanding sophisticated experiments using fast flux.  In particular, when measuring the cavity dispersive shift $\chi$, we choose to perform a sequence that is much more challenging than necessary, in order to demonstrate the technical capabilities of the architecture.

We focus the last third of the chapter on using the cavity as a quantum resource.  Our cavity, being a harmonic oscillator,  has an infinitely large Hilbert space that makes efficient measurement of its quantum state challenging.  We introduce the Husimi $Q$ quasi-probability distribution, known more simply as the $Q$ function, which more naturally represents an oscillator density matrix.  We use the $Q$ function to measure the unusual evolution of a coherent state subject to the Kerr-type nonlinearity our cavity inherits from the qubit.  After a particular duration of this evolution, our coherent state is mapped to a superposition of two coherent states known as a {\it Schr\"{o}dinger cat state}.  By tuning our qubit away, we show that we can stop this Kerr evolution and freeze the cat state.  This demonstrates our ability to control the cavity Hamiltonian on demand by tuning it continuously between qubit-like and cavity-like limits.  It also highlights the possibility of using the cavity as an efficient quantum memory.  We conclude the chapter by noting that these capabilities enable a host of new experiments.  These will be the topic of the thesis's final chapter.

\section{Qubit lifetimes and FBL filtering}
\label{sec:tunablequbitlifetime}

\begin{figure}
	\centering
	\includegraphics[scale=1]{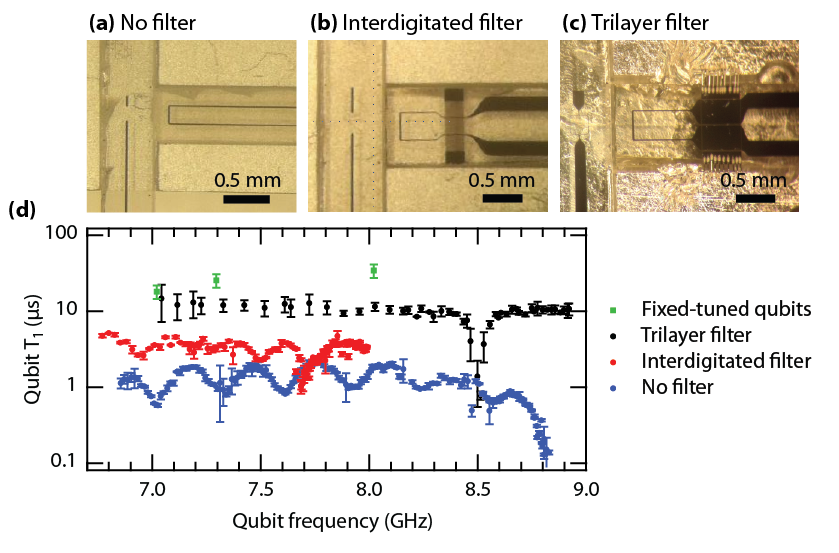}
	\mycaption{FBL filter performance}  
		{Optical micrographs of the flux bias lines with \capl{(a)} no filtering, \capl{(b)} an interdigitated capacitor filter, and \capl{(c)} a three-layer quasi-parallel plate capacitor.  The detailed designs of each are shown in \figref{fig:fbllitho}.  
		\capl{(d)} Qubit lifetime as a function of frequency for the three FBL designs, and for three qubits that do not have FBLs.  In the absence of filtering, the qubit lifetime is only about $2\us$ and exhibits oscillations that we ascribe to an impedance mismatch with a filter several feet away (blue dots).  Qubits addressed by the interdigitated filter have a slightly longer lifetime ($3-5\us$), but still show oscillations (red dots).  If we use a trilayer filter that has a sufficiently large filter capacitor, those oscillations disappear and we observe qubit lifetimes between $8$ and $20\us$ (black dots).  We compare this to identically-designed qubits that do not have flux bias lines, which we measure to have lifetimes as high as $40\us$ (green dots).  Spurious avoided crossings at $8.8\ghz$ for the blue points and $8.5\ghz$ for the black cause dips in qubit $T_1$ but are unrelated to the flux bias line.  For each frequency, the value is the mean and the error bars are the standard deviation of repeated measurements.}
	{\label{fig:filteredT1s}}
\end{figure}

The central feature of the tunable 3D cQED architecture is the flux bias line.  In \sref{subsec:fblrelaxation}, we showed that in the absence of additional filtering, the capacitive coupling of the qubit to the FBL will create an unacceptable channel for relaxation.  We measured the qubit lifetime when controlled with such an unfiltered FBL to confirm our expectations.  In \figref{fig:filteredT1s}(b), we see that the unfiltered lifetime $T_1$ is approximately $2\us$, which is consistent with the predictions of our circuit model.  There is also a prominent oscillation with a period of about $300\mhz$.  We associate this oscillation with an impedance mismatch some distance $d=n c / f \approx 2~\mathrm{m}$ from the qubit.  This indicates that the qubit is affected by the environment all the way to the lossy low-pass FBL filters located outside the sample box.  Moreover, we have measured several qubits without flux bias lines (both with and without SQUID loops) and found their lifetimes to be in the range of $20-40\us$.  As expected, the qubit seems to be limited by the FBL.

To fix this problem, we implemented two generations of FBL filters, which are shown in \figref{fig:filteredT1s}.  They both seek to solve the issue of FBL relaxation by low-pass filtering the line described in \sref{subsec:fblfiltering}.  The key to this filter is the size of the capacitance of each line to ground.  The difference between the two filters is how this capacitor is fabricated (\sref{subsec:tunablefbldesign}).  In the first generation, we implemented an interdigitated finger capacitor that requires only one layer of fabrication.  The size of this capacitor was limited by its self-resonance frequency, and so could not provide enough capacitance to shut off the relaxation channel.  The resulting $T_1$ data are shown in \figref{fig:filteredT1s}(b), where we see that the lifetimes are improved by some factor, but remain considerably lower than qubits without FBLs.  Moreover, the oscillations as a function of frequency are still present, which indicates that the qubit is still sampling the impedance environment outside of the sample box.

The second generation FBL implemented a much larger filter capacitor by using a three-layer lithography process.  This easily produces capacitances much larger than should be necessary to eliminate FBL relaxation.  This three-layer structure is analogous to a parallel-plate capacitor, where we have deposited either silicon monoxide or hafnium oxide between two metal pads.  The resulting qubit lifetime data are significantly improved, with $T_1$s approaching $20\us$, and showing no evidence of oscillation.  We have also measured $T_2$ times in several devices and found a flux noise density five to ten times higher than typical values in planar devices (data not shown).  The source of this increased noise is not yet known, but may have something to do with the increased size of the SQUID loop or technical problems with the grounding of the fridge and measurement setup.

Though the qubit lifetimes in the filtered device are much closer to those we measured without FBLs, it is not yet the same.  What can account for this persistent degradation of performance?  One potential concern is the dielectric loss of the material used in the filter capacitors.  They have recently been measured to have an internal $Q_i \approx 500-2000$ \cite{PrivateCommunicationKevinChou}, which is rather alarming.  We can estimate this loss, again, with a circuit argument.  The conductance across the capacitor will be given by $G=\omega C \tan \delta$.  As long as this value is small compared to $1/(50 \ohm)$, then the dielectric loss should be negligible.  Plugging in $C=10\pF$, $\omega / 2\pi = 9 \ghz$, and even assuming $\tan\delta=1$, we have $1/G = 1/(2 \ohm)$, indicating that the dielectric loss of this capacitor is irrelevant.  This treatment does not account for direct hybridization of the qubit mode with this dielectric, however.  If the qubit stores as little as 0.1\% of its energy there, it could constitute the dominant source of relaxation.  Whether or not this is a prominent effect remains an open question as of the writing of this thesis.

\subsection{Fast flux performance}

\begin{figure}
	\centering
	\includegraphics[scale=1]{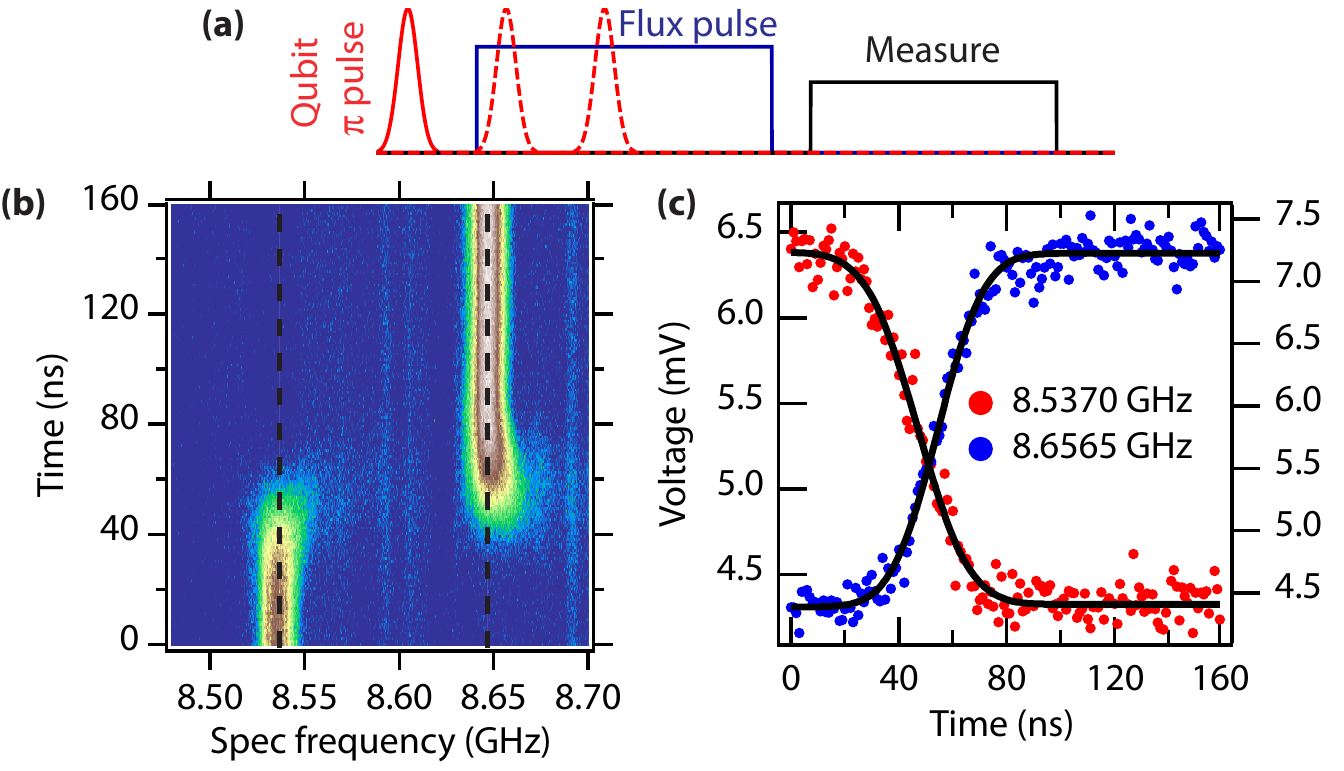}
	\mycaption{Fast flux characterization with flux spectroscopy}  
		{\capl{(a)} To perform flux spectroscopy, we suddenly detune the qubit from its starting position with a square flux pulse, depicted in blue.  After holding for some time, the qubit is returned to its home position and measured.  During this time, we apply a gaussian spectroscopy pulse at some frequency and delay relative to the flux pulse.  Here, the width of that gaussian is $\sigma=20\ns$ and its total duration is $4\sigma$.
		\capl{(b)} The ensemble-averaged measurement is plotted as a function of the delay and frequency of the spectroscopy pulse.  We see features for small times at $8.537\ghz$ and for large times at $8.657\ghz$, indicating that the qubit initially starts at the first frequency but jumps to the second upon the application of the flux pulse.
		\capl{(c)} Taking cuts along the two relevant frequencies, we see that these features are rounded in time.  The rounding is due to the temporal extent of the gaussian pulses.  Fitting them to error functions, we find that the offset between the centers of the two curves is only a few nanoseconds.  This indicates that the qubit is moving between the two positions in that time.  Note that the two y-scales are different for the two plots because the spectroscopy pulse drives a slightly different rotation at the two frequencies because of the cavity filtering.
	}
	{\label{fig:fastflux}}
\end{figure}

The addition of the low-pass filter to the flux bias line potentially slows down its response time.  The characteristic frequency of the filter is set by the size of the capacitor, which we aim to control with lithography.  What if we accidentally make a much larger capacitor than we intend?  Evidence would not appear in the $T_1$ data nor in the static flux tuning; it would only become apparent in a slowing of the rise time of an applied flux pulse.  We can directly measure this time with flux spectroscopy.  As shown in the cartoon of \figref{fig:fastflux}(a), we perform pulsed spectroscopy simultaneous with a flux pulse.  Some fixed time after the experiment begins, we move the qubit from its home position to another frequency as fast as possible with a square flux pulse.  We wait there for a few microseconds, flux back, and measure the qubit.  During this time, we apply a gaussian $\pi$ pulse of some frequency at some delay relative to the flux pulse.  We then ensemble average as a function of both pulse frequency and delay.

The result of this experiment is shown in \figref{fig:fastflux}(b).  We see that the qubit initially starts at $8.537\ghz$ and, at $t\approx50\ns$, jumps to a frequency of $8.657\ghz$.  We take cuts along both frequencies and plot them, shown in (b).  The observed rounding is due to the temporal width of the qubit spectroscopy pulse used to interrogate the system.  To see this, suppose that the qubit moves suddenly away during the application of the pulse, at $t_f$.  If its final position is far detuned from the spectroscopy frequency, the rotation that the qubit experiences will be given by the integral of the pulse from $t=0$ to $t_f$.  Since the pulse is gaussian and the rotation angle is given by the integral of the applied amplitude, the rotation angle is specified by the integral of that gaussian: an error function.  We fit both sets of data to that function with time constants of $t=20\ns$.  The difference between the centers of the two fits gives the transit time of the qubit, here being $5\ns$.

\begin{figure}
	\centering
	\includegraphics[scale=1]{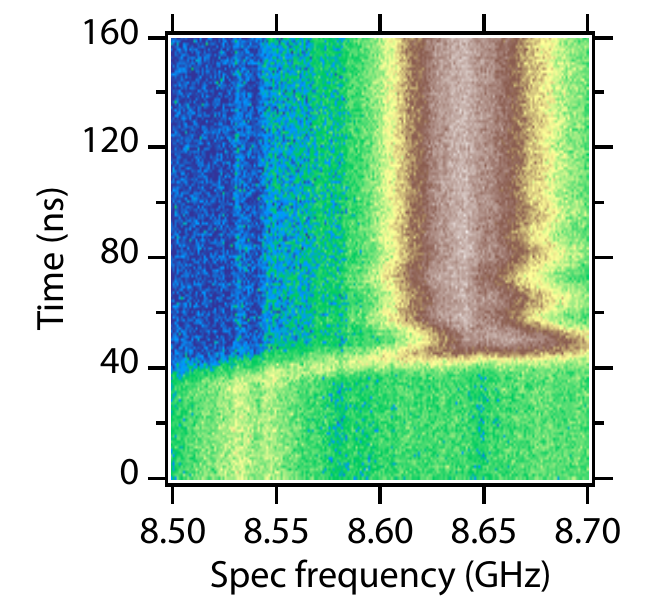}
	\mycaption{Flux spectroscopy with $5\ns$ gaussian pulse width}
	{We repeat the experiment shown in \figref{fig:fastflux} with $5\ns$ gaussian pulses instead of $20\ns$.  The spectral width of the qubit response is significantly broadened, but the temporal resolution is commensurately higher.  As a result, we can resolve ringing of the qubit frequency once it reaches its final frequency. These dynamics can be mitigated with linear deconvolution (\sref{subsec:fluxlinecal}). 
	}
	{\label{fig:fluxspec2}}
\end{figure}

This experiment is a bit unusual in that it aims to be sensitive to both frequency and time.  Since we use relatively long $20\ns$ pulses in \figref{fig:fastflux}(b), the data is relatively smooth in time and spectrally narrow.  However, as we show in \figref{fig:fluxspec2}, by using faster $5\ns$ pulses, we substantially broaden the qubit response.  This increases our temporal resolution, revealing fast ringing of the qubit frequency once it has neared its final frequency.  As described in \sref{subsec:fluxlinecal}, this ringing is due to the response function of the Tektronix AWG and the flux line itself, and can be mitigated with linear deconvolution.

\section{Device characterization with Fock states}
\label{sec:devicewithfock}

\nomdref{Ck}{$K$}{cavity Kerr anharmonicity}{sec:devicewithfock}

Since our flux bias line filtering is working as desired, we now turn our attention to performing experiments with a particular device.  This section reports on a single-cavity 3D device with a single trilayer-filtered tunable qubit.  The Hamiltonian of this system is well-approximated by
\begin{equation}
	\hat{H}/\hbar = \omega_q \hat{b}^\dagger \hat{b} - \frac{\alpha}{2} \hat{b}^\dagger \hat{b}^\dagger \hat{b} \hat{b} + \omega_c \hat{a}^\dagger \hat{a} - \frac{K}{2} \hat{a}^\dagger \hat{a}^\dagger \hat{a} \hat{a} - \chi \hat{a}^\dagger \hat{a} \hat{b}^\dagger \hat{b},
\end{equation}
where $\hat{b}^\dagger / \hat{b}$ are the qubit raising and lowering operators, $\hat{a}^\dagger / \hat{a}$ are the cavity raising and lowering operators, $\alpha$ is the qubit anharmonicity, $K$ is the cavity anharmonicity (also known as the self-Kerr term), and $\chi$ is the state-dependent dispersive shift \cite{Nigg2011, Bourassa2012}.  In this treatment, the cavity and qubit are governed by the same dynamics, differentiated only by the magnitude of $\alpha$ and $K$.  Note that this is a slightly different definition of $\chi$ than the one used in previous chapters.  Here we adopt the more modern convention where number split peaks are separated by $\chi$ \cite{Nigg2011, Kirchmair2013}.  In previous chapters, including \sref{subsec:dispersivelimit}, we had defined that separation to be $2\chi$.

The qubit in this device has a maximum frequency that is very close to vacuum-Rabi.  The cavity therefore inherits enough $K$ anharmonicity that we can directly pulse it to create photon number Fock states (\sref{subsec:harmonicoscillators}).  As we will show, this capability introduces the potential for a host of experiments that make it straightforward to measure system parameters like cavity coherence, Kerr anharmonicity, and the dispersive $\chi$ shift as a function of qubit frequency.  These experiments also demonstrate the power and flexibility of fast flux tuning.  Combined with the long cavity and qubit coherence times, a variety of interesting experiments are operable.

\subsection{System spectrum}

\begin{figure}
	\centering
	\includegraphics[scale=1]{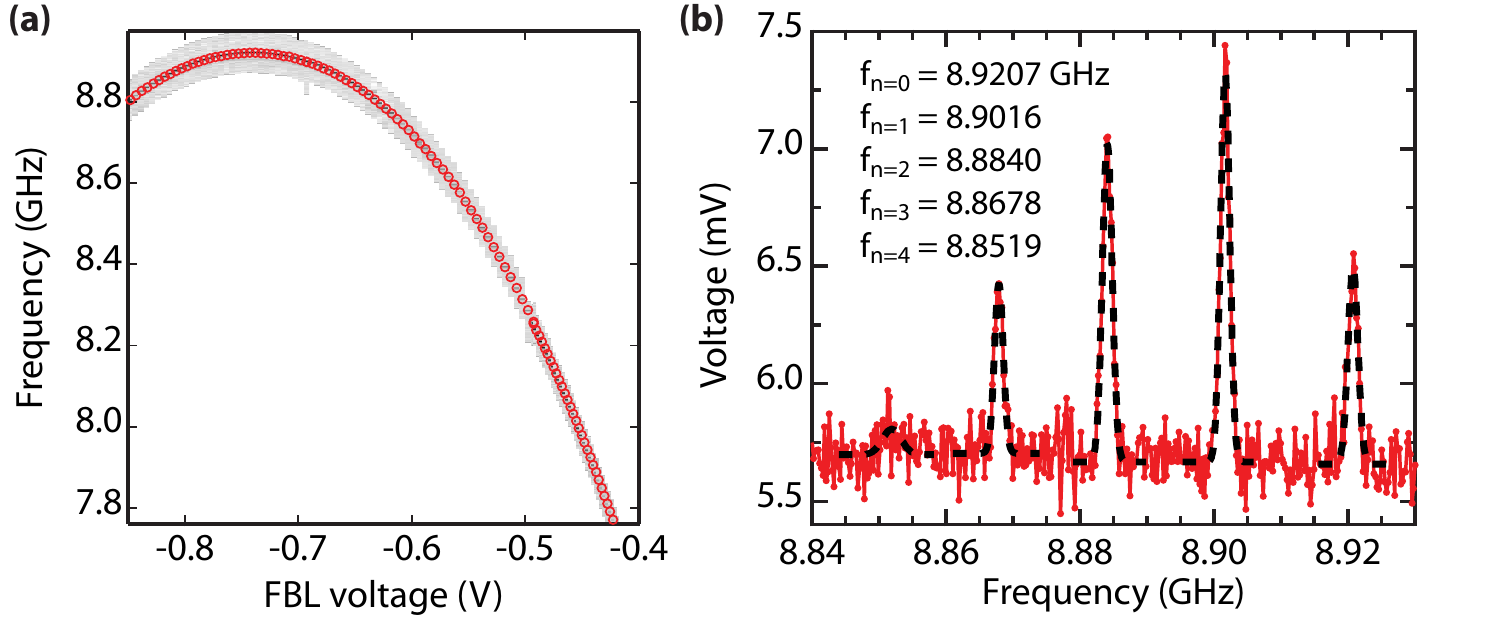}
	\mycaption{Qubit spectroscopy as a function of frequency and qubit number splitting at $f_{\mathrm{max}}$}  
		{\capl{(a)} Qubit frequency as a function of applied flux bias line voltage.  We measure with pulsed spectroscopy, as described in \sref{subsec:spectroscopy}.  Red circles indicate the qubit frequency at each voltage point.  Here, the bare cavity frequency is $9.169\ghz$ and the qubit maximum frequency is $8.921\ghz$.  
		\capl{(b)} At $f_{\mathrm{max}}$, the qubit is strongly hybridized with the cavity.  We apply a weak tone at the cavity frequency to set up some equilibrium photon population.  The qubit transition frequency splits according to the number of photons in the cavity, as described in \sref{subsec:dispersivelimit}.  Normally, these peaks would be evenly spaced, but because the qubit is so close to the cavity non-dispersive corrections become important.  As a result, $\chi_{01} = 19.1\mhz$, $\chi_{12} = 17.6\mhz$, $\chi_{23}=16.2\mhz$, and $\chi_{34}=15.9\mhz$.}
	{\label{fig:fmaxspec}}
\end{figure}

Recall from \sref{sec:expbringup} that one of the first characterization experiments is to measure qubit frequency as a function of applied flux bias voltage.  We see from that measurement that this device is a little unusual, as shown in \figref{fig:fmaxspec}(a).  Though the qubit does not cross the cavity and vacuum-Rabi split, it has an extremely close approach, with a qubit maximum frequency of $8.921\ghz$ and a bare cavity frequency of $9.169\ghz$.  Their minimum detuning of only $250\mhz$ is comparable to the coupling strength $g\approx150\mhz$, making the qubit strongly non-dispersive at $f_{\mathrm{max}}$.  This fact is illustrated clearly when we measure qubit number splitting, shown in \figref{fig:fmaxspec}(b).  As described in \sref{subsec:dispersivelimit}, normal number-split peaks are evenly spaced by $\chi$.  However, here the splitting is not evenly spaced, with, for example, the difference between the qubit transition frequency with 0 and 1 photons of $19.1\mhz$, but only $17.6\mhz$ between 1 and 2 photons.  This is an indication that the qubit and cavity are so strongly hybridized at $f_{\mathrm{max}}$ that the dispersive approximation has broken down.

\begin{figure}
	\centering
	\includegraphics[scale=1]{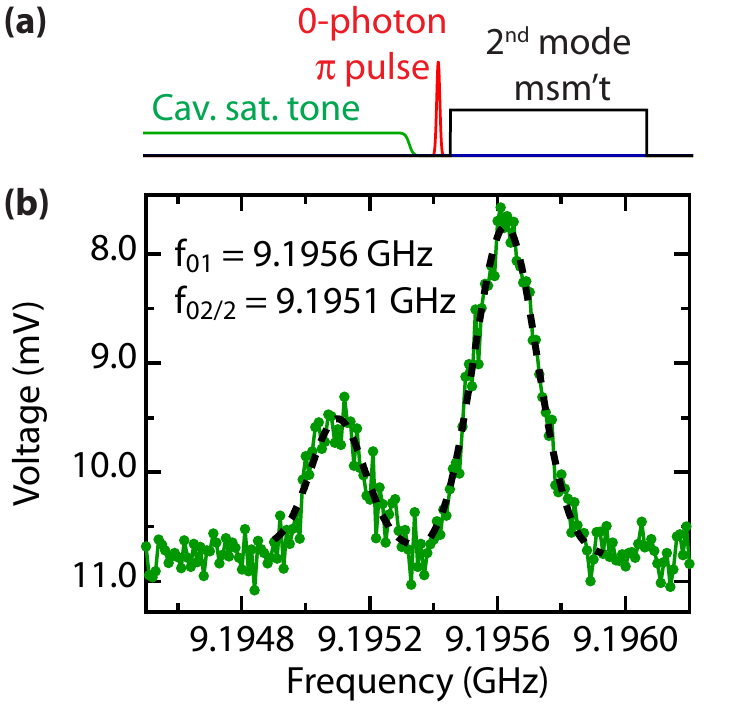}
	\mycaption{Pulsed spectroscopy of anharmonic cavity}
	{\capl{(a)} In order to measure the spectrum of the cavity, we must slightly modify our conventional pulsed spectroscopy sequence.  We begin with a long cavity saturation tone, as with normal spectroscopy.  In order to be sensitive to the resulting cavity state, we map its population onto the qubit by applying a $\pi$ pulse on the $0$-photon number-split peak at $8.9207\ghz$.  To detect the state of the qubit, we measure with the second spatial mode of the cavity at $11.244\ghz$.  We do this to avoid our qubit measurement contrast being sensitive to the number of photons stored in the first mode.  The qubit is also strongly coupled to the second mode because it is physically offset from the center of the cavity, a symmetry point usually used to eliminate the coupling strength.  We will continue using this trick to measure the cavity state throughout this chapter.
	\capl{(b)} The resulting cavity spectrum reveals resolved $f_{01}$ and $f_{02/2}$ transitions, just as we see with a strongly-driven qubit (\sref{subsec:spectroscopy}).  This indicates that the cavity has inherited an anharmonicity of about $1\mhz$ due to its hybridization with the qubit.  Note that the $y$-axis is inverted because we get a large voltage when the qubit pulse conditional on zero photons succeeds.  When our spectroscopy tone is in resonance with a cavity transition, we populate the cavity, causing the qubit pulse to fail and giving us a dip in measurement voltage.}
	{\label{fig:cavpspec}}
\end{figure}

We observe another indication of this non-dispersive behavior in the spectrum of the cavity with the qubit at its maximum frequency.  The procedure to perform cavity spectroscopy is shown in \figref{fig:cavpspec}(a), where we make use of qubit number splitting as well as the fact that the qubit couples to more than one spatial mode of the cavity.  The resulting data are shown in (b), demonstrating that we can resolve the individual $f_{01}$ and $f_{02/2}$ transitions.  The cavity is so strongly hybridized with the qubit that it inherits appreciable anharmonicity; in essence, the cavity behaves like a transmon qubit.  At $f_{\mathrm{max}}$, this anharmonicity is approximately $1\mhz$, which is substantially larger than the cavity linewidth.  This is known as a {\it Kerr nonlinearity}, and, as we will see in \sref{sec:cavitykerrexpts}, it has interesting consequences for the dynamics of the cavity \cite{Yin2012, Hoffman2011, Kirchmair2013}.

\subsection{Cavity Rabi oscillation}
\label{subsec:cavityrabi}

\begin{figure}
	\centering
	\includegraphics[scale=1]{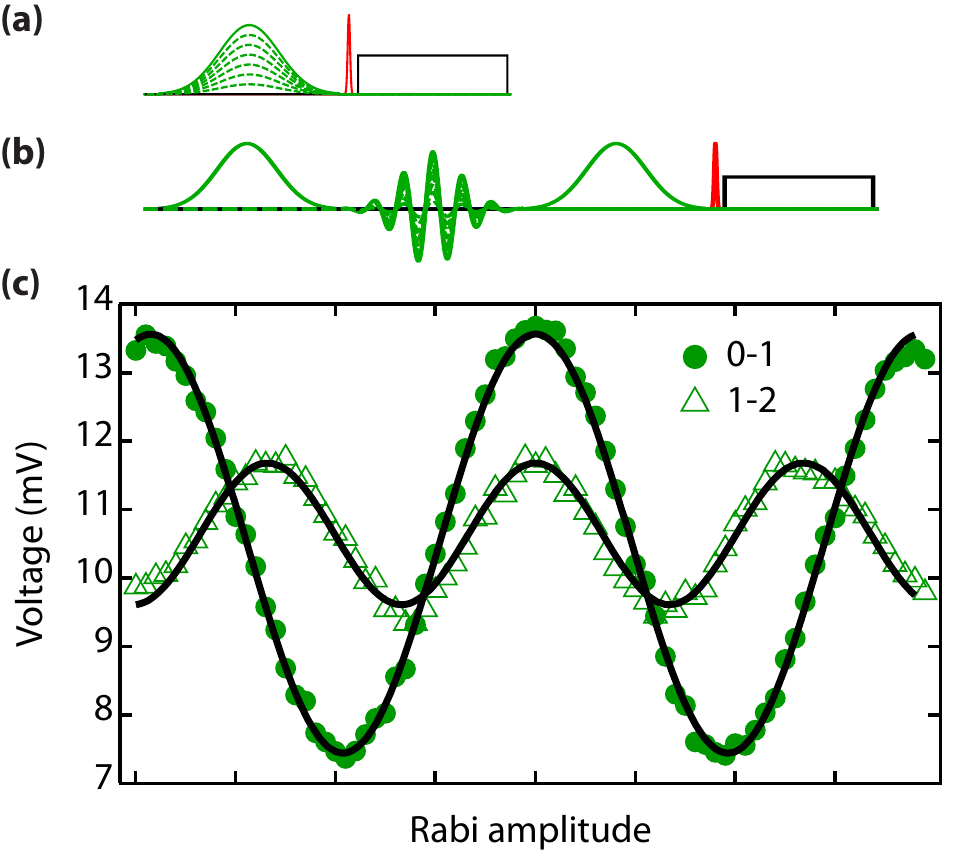}
	\mycaption{Cavity Rabi oscillation on 0-1 and 1-2 Fock number states}
		{\capl{(a)} To perform a Rabi oscillation on the cavity, we first apply a pulse of some varied amplitude on resonance with the 0-1 transition.  This is identical to rotating a qubit, but, because of the small cavity anharmonicity, we use an extremely long $\sigma=1\us$ gaussian pulse that is spectrally narrow enough to address only the desired transition.  In order to measure the resulting cavity state, we map it to the qubit with a number-selective $\pi$ pulse and measure with the second cavity mode, as described in \figref{fig:cavpspec}.  
		\capl{(b)} We can extend this procedure to perform a Rabi oscillation on the 1-2 transition of the cavity.  We first prepare a 1-photon Fock state with a $\pi$ pulse, then do a second pulse of varying amplitude at the 1-2 frequency.  This pulse is depicted as oscillating in the cartoon so as to indicate frequency modulation (\sref{subsubsec:ssb}).   To have contrast between the 1- and 2-photon Fock states, we pulse $1\rightarrow0$ with a second $\pi$ pulse prior to measurement.
		\capl{(c)} The results of the two experiments are plotted as a function of the amplitude of the Rabi pulse.  As expected, the oscillation frequency of the 0-1 oscillation is slower than the 1-2 oscillation by $\sqrt{2}$.  The amplitude of the 1-2 oscillation is also reduced because the sequence duration is a substantial fraction of the cavity lifetime.
		}
	{\label{fig:cavrabi}}
\end{figure}

The substantial anharmonicity of the cavity makes it possible to directly populate Fock number states.  Just as with a qubit, we apply a gaussian pulse to the cavity resonance frequency that is sufficiently spectrally narrow to address only the desired transition.  The difference here is that the anharmonicity is so small that we must use extremely slow pulses, here a gaussian $\sigma = 1 \us$.  The procedure to do a Rabi oscillation on the cavity is shown in \figref{fig:cavrabi}(a).  After this slow gaussian pulse, we again map the resulting cavity state to the qubit by performing a $\pi$ pulse on the 0-photon peak and measuring its state with the second cavity mode, done previously with cavity spectroscopy.  As shown in (b), we can extend this procedure by applying a rotation on the 1-2 transition of the cavity as well.  The resulting oscillations shown in (c) are consistent with our expectations, with the 1-2 oscillation sped up by $\sqrt{2}$ and its amplitude reduced due to the longer pulse sequence length.

\subsection{Cavity lifetime}

\begin{figure}
	\centering
	\includegraphics[scale=1]{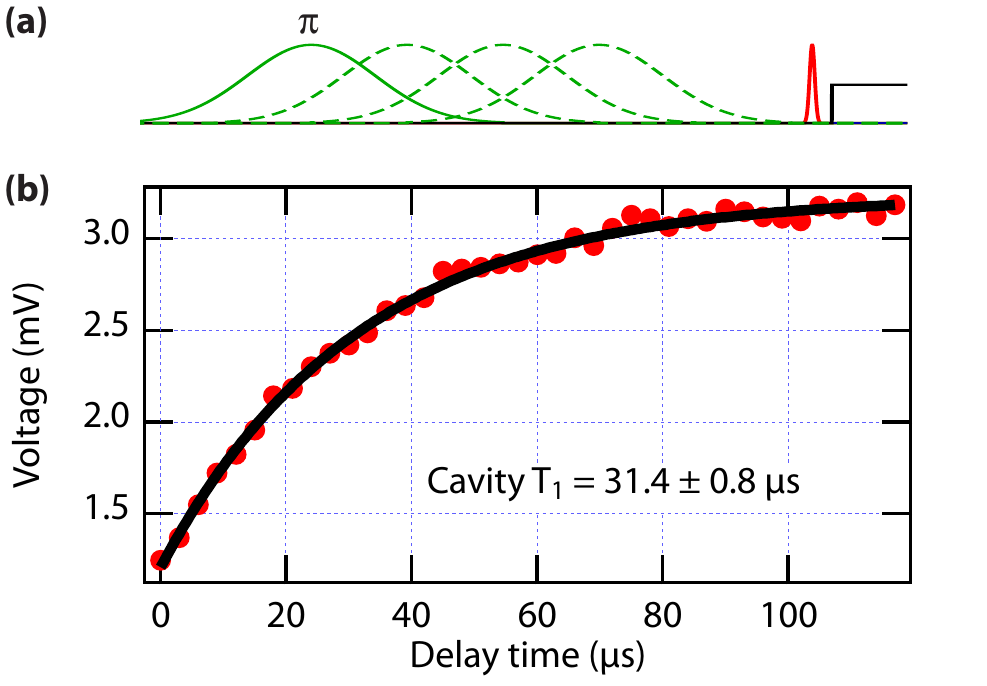}
	\mycaption{Cavity $T_1$ measured with Fock state}
		{\capl{(a)} To measure the cavity lifetime, we first prepare a 1-photon Fock state with a $\pi$ pulse on the cavity and wait for some amount of time.  We then measure the resulting cavity state with a number-selective $\pi$ pulse and second-mode measurement, as before.  
		\capl{(b)} We plot the average remaining cavity population as a function of the delay time.  The cavity lifetime is found to be $31.4\pm0.8\us$ with the qubit at $f_{\mathrm{max}}$.}
	{\label{fig:cavt1}}
\end{figure}

The ability to create single Fock states in the cavity is extremely useful for characterizing its properties.  The simplest example experiment is to measure the cavity lifetime.  Due to cavity anharmonicity and the glassy physics that sometimes dominates its loss, the apparent lifetime of the cavity can be a strong function of the magnitude of its excitation \cite{OConnell2008}.  Since we are normally interested in the low-power lifetime, we can only displace with an extremely weak tone that populates at most $\sim1$ photon.  We can then either watch this energy leak out in the time domain or measure the cavity spectral width in the frequency domain with transmission.  (The second tactic actually measures some combination of $T_1$ and $T_2$.)  Both approaches suffer from a low SNR since the signal we are looking for is very small.  It can also be difficult to ascertain that you are driving with a small enough power.  A superior method for measuring cavity lifetime is shown in \figref{fig:cavt1}(a).  There, we replicate the normal sequence for measuring qubit $T_1$ by applying a $\pi$ pulse to the cavity with our slow gaussian and delaying for a variable length of time prior to measurement.  As before, we measure by mapping the cavity state to the qubit with a number-selective $\pi$ pulse.  The result is shown in (b), where we find a cavity lifetime of about $30\us$ when the qubit is at its maximum frequency.

\begin{figure}
	\centering
	\includegraphics[scale=1]{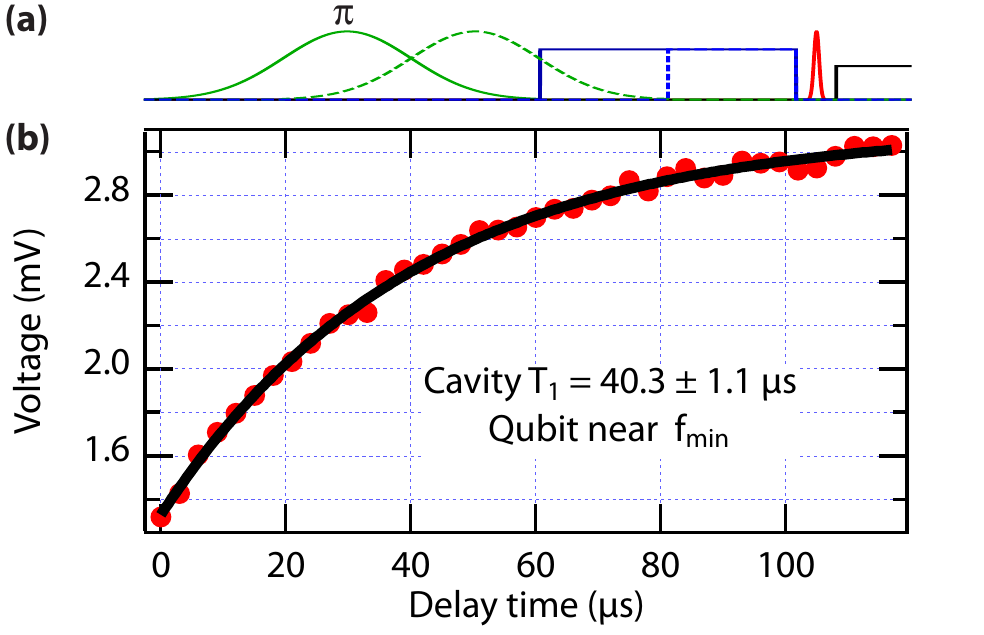}
	\mycaption{Cavity $T_1$ with qubit moved to its minimum frequency}
		{\capl{(a)} We measure the cavity lifetime with the qubit far detuned by adding a qubit flux pulse during the waiting time of the procedure described in \figref{fig:cavt1}.
		\capl{(b)} The cavity lifetime when the qubit is near its minimum frequency is found to be $40.3\pm1.1\us$.  This is an improvement of more than 20\%, indicating that the cavity was relaxing through the qubit due to hybridization.}
	{\label{fig:cavt1flux}}
\end{figure}

We can extend this protocol to measure the cavity lifetime as its hybridization with the qubit is turned off.  If the qubit is lossier than the cavity, the dispersive coupling may limit the cavity's lifetime.  The procedure is shown in \figref{fig:cavt1flux}(a).  The sequence is identical to the one described in \figref{fig:cavt1}, except that during the waiting time, the qubit is moved away to some other detuning with a flux pulse.  In \figref{fig:cavt1flux}(b), we show the resulting measurement when the flux pulse amplitude is chosen to move the qubit near its minimum frequency, estimated to be $2-3\ghz$.  The cavity lifetime is found to be $40\us$, an improvement of over 20\% as compared to when the qubit is at $f_{\mathrm{max}}$.  This indicates that the cavity hybridization caused it to decay faster than it would have otherwise.

\begin{figure}
	\centering
	\includegraphics[scale=1]{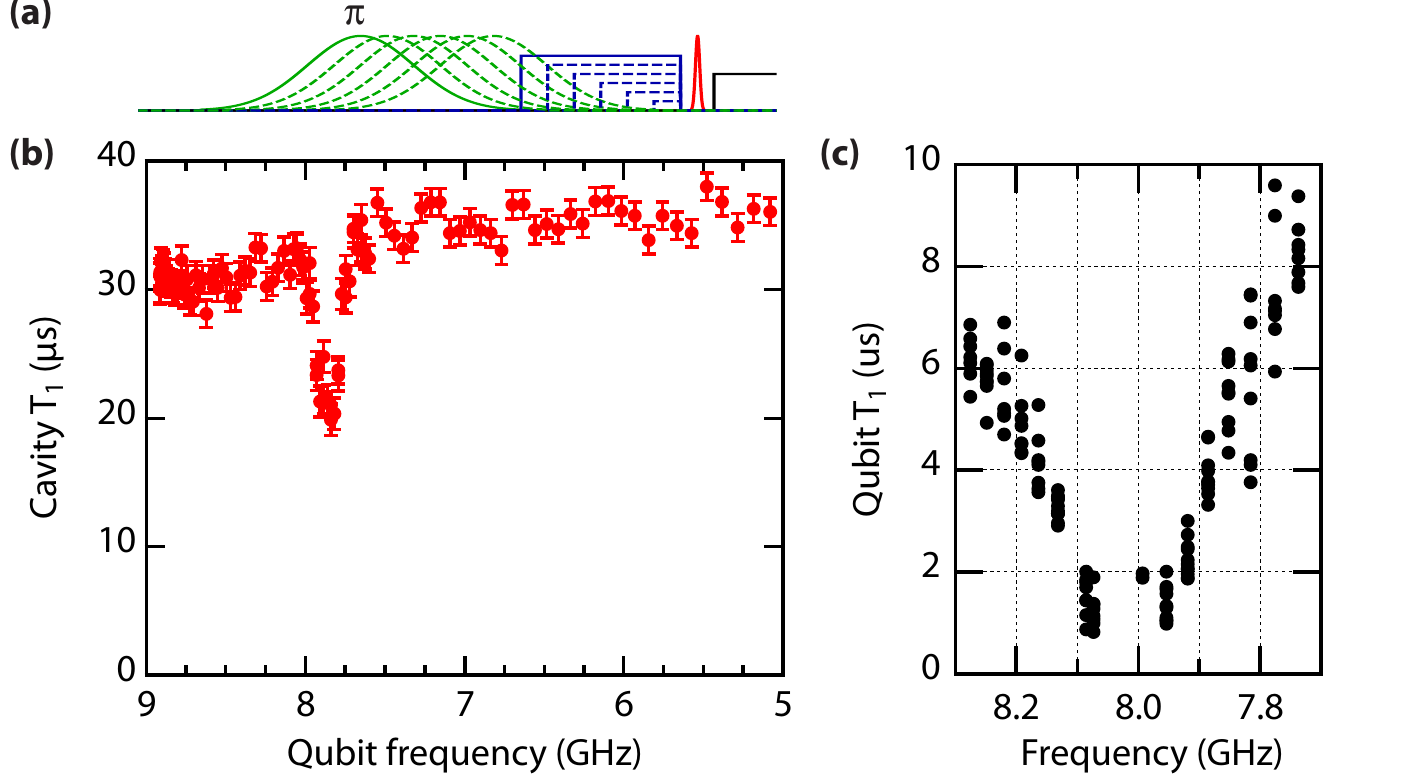}
	\mycaption{Cavity $T_1$ as a function of qubit frequency}
		{\capl{(a)} To measure cavity lifetime as a function of qubit frequency, we repeat the experiment described in \figref{fig:cavt1flux} as a function of the flux pulse amplitude.
		\capl{(b)} Cavity lifetime as a function of qubit frequency shows a general trend upward as the qubit is decoupled from the cavity.  We relate the amplitude of the applied flux pulse to the frequency of the qubit by measuring flux spectroscopy (see \figref{fig:fastflux}) for several points and fitting the resulting data to a $\sqrt{\cos}$ curve that we invert (not shown).  The measured cavity lifetime drops out near $8\ghz$.  
		\capl{(c)} This drop-out is associated with a spurious avoided crossing in the qubit spectrum that causes the qubit lifetime to become very short.  Since the cavity mode is hybridized with the qubit, the cavity lifetime also suffers.  Essentially, this is the reverse of the Purcell effect (described in \sref{subsec:purcelleffect}).}
	{\label{fig:cavt1fluxed}}
\end{figure}

We can measure the evolution of cavity $T_1$ by repeating the fluxed-cavity-lifetime experiment as a function of the flux pulse amplitude, as shown in \figref{fig:cavt1fluxed}(a).  The resulting data, shown in (b), reveal a general trend upward as the qubit is detuned and the cavity becomes less hybridized.  At $8\ghz$, however, there is a feature showing a dramatic drop-out in cavity lifetime.  Looking at the measured qubit lifetime in this device, we see that there is a drop-out at the same frequency corresponding to a spurious two-level system.  This additional loss is shared with the cavity, and causes its lifetime to diminish due to this ``reverse'' Purcell effect (\sref{subsec:purcelleffect}).

\subsection{Cavity coherence}
\label{subsec:cavcoherence}

\nomdref{Ct2echo}{$T_2^{\mathrm{echo}}$}{dephasing time when using a Hahn echo sequence}{subsec:cavcoherence}
\nomdref{Ct2star}{$T_2^*$}{dephasing time when performing a Ramsey measurement}{subsec:cavcoherence}

\begin{figure}
	\centering
	\includegraphics[scale=1]{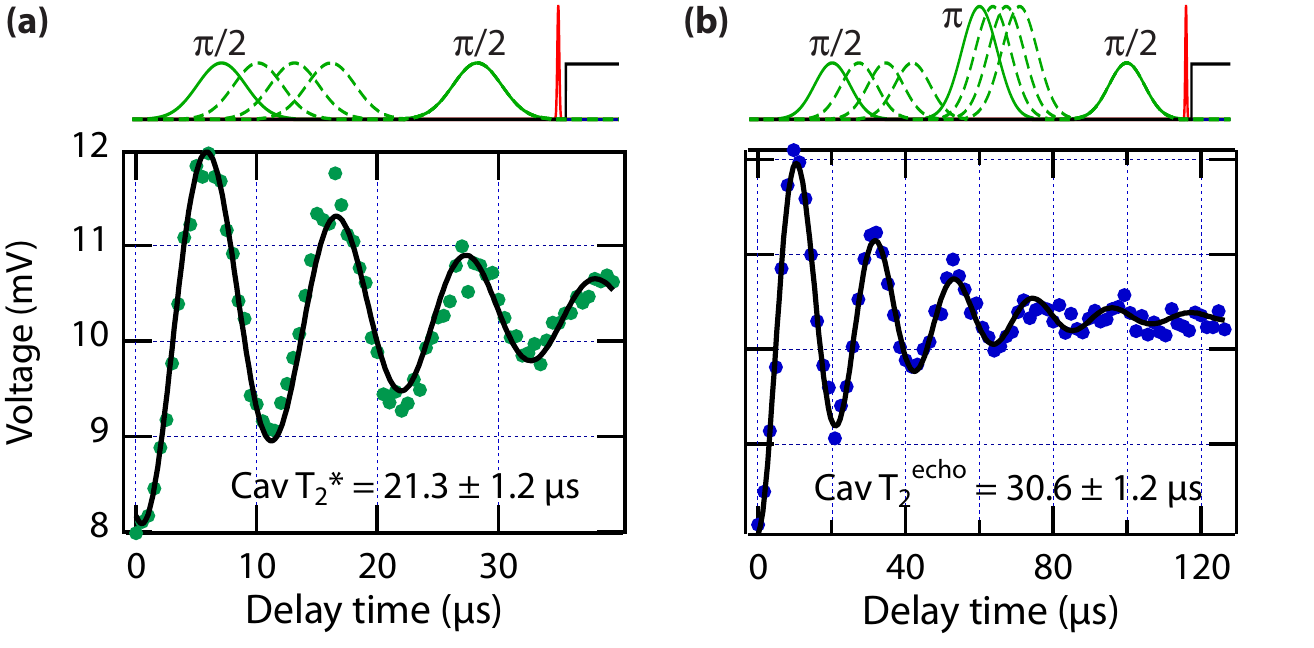}
	\mycaption{Cavity $T_2^*$ and $T_2^{\mathrm{echo}}$}
		{\capl{(a)} The cavity coherence time $T_2^*$ is measured in the same way as with a qubit.  We prepare a superposition of 0 and 1 photons by applying a $\pi/2$ pulse to the cavity.  We then delay for some period of time before applying a second $\pi/2$ pulse and measure by mapping the cavity state to the qubit with a number-selective qubit rotation.  The cavity pulses are slightly detuned so as to see an oscillation.  The resulting dephasing time is found to be $21.3\pm1.2 \us$, with the uncertainty given by the fit.  This corresponds to a pure dephasing time of approximately $30 \us$. 
		\capl{(b)} We can also perform a Hahn echo experiment with the addition of a $\pi$ pulse on the cavity exactly half way through the experiment.  The dephasing time increases to $30.6\pm1.2 \us$.  Here, we sweep the angle of the final $\pi/2$ pulse to see an oscillation, though that fact is not indicated in the cartoon for clarity.}
	{\label{fig:cavt2}}
\end{figure}

We can also use Fock states to measure the cavity coherence time $T_2$.  As shown in \figref{fig:cavt2}, the procedure for measuring cavity $T_2^*$ and $T_2^{\mathrm{echo}}$ is identical to the procedure for qubit coherence, excepting for the need to map the cavity state to the qubit for measurement.  The measured coherence time reveals that the cavity experiences some inhomogenous dephasing, with $T_{\phi} = \left(\frac{1}{T_2} - \frac{1}{2 T_1}\right)^{-1} \approx 30 \us$.  Applying an echo improves the decay time somewhat.  We are left to wonder: what is causing this dephasing?  While there are possible explanations such as vibration or heating to consider, the most likely culprit is the qubit itself.  Just as we saw with the cavity lifetime, the cavity mode is strongly hybridized with the qubit and so may inherit some of its dephasing.

\begin{figure}
	\centering
	\includegraphics[scale=1]{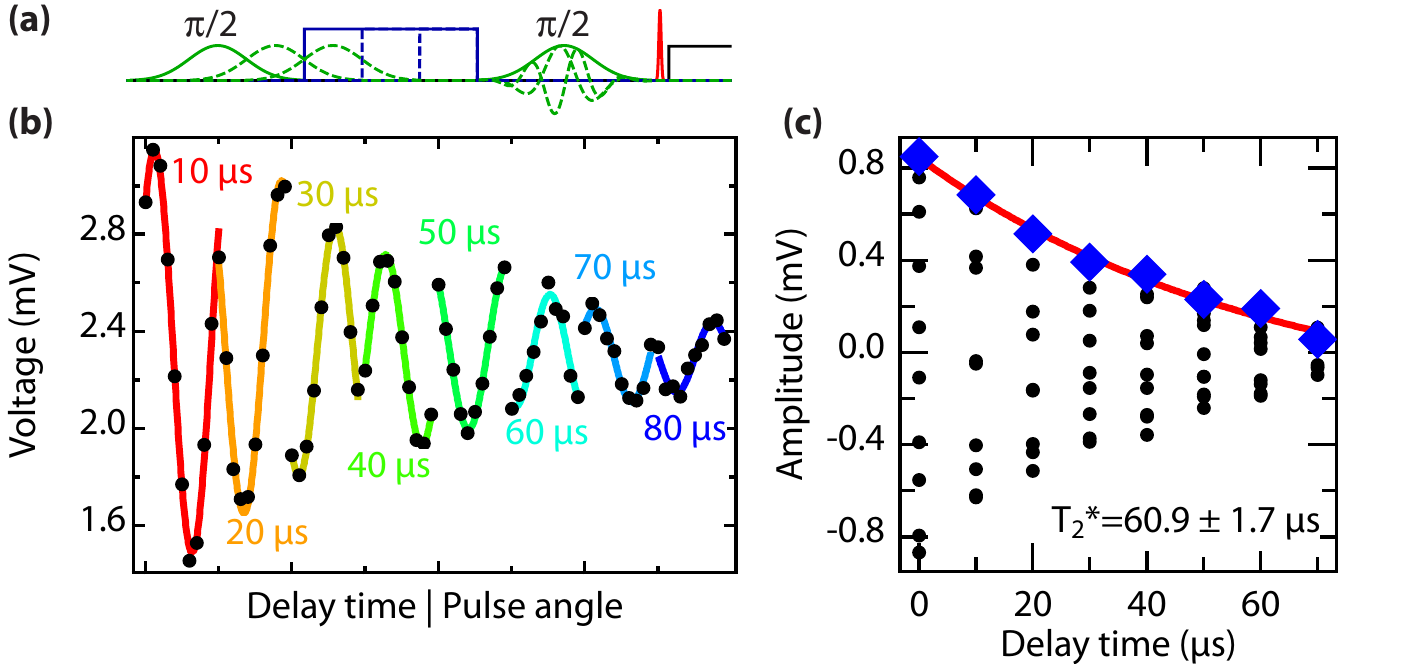}
	\mycaption{Procedure for measuring cavity $T_2$ with the qubit detuned}
		{\capl{(a)} Detuning the qubit with a flux pulse changes the cavity frequency by as much as $20\MHz$.  In order to measure the cavity dephasing time, we therefore must use a sequence that is insensitive to its absolute phase evolution.  We first prepare a superposition of 0 and 1 photons with a $\pi/2$ cavity pulse, as before.  We then move the qubit away for some time, do another $\pi/2$ pulse on the cavity, and measure its state in the usual way.  We repeat each time with several angles of the second $\pi/2$ pulse.
		\capl{(b)} The resulting raw data of this procedure show eight different sinusoids for delay times of $10-80\us$.  At each time, we get a sinusoidal oscillation with some phase.  This phase is set by the fast dynamics of the cavity, which seems random from our point of view.  The amplitude of the oscillation, which we extract by fitting the data, is proportional to the amount of phase coherence of the cavity at the specified time.
		\capl{(c)} We plot the extracted amplitudes as a function of the delay time (shown in blue).  We also plot the raw data (in black).  Fitting the amplitudes to an exponential, we find that the cavity dephasing time has improved to $60.9\pm1.7 \us$ when we have moved the qubit approximately $2\ghz$ away from its maximum frequency.  Given that the cavity lifetime here is approximately $35\us$, this corresponds to a pure dephasing time in excess of $400\us$.
	}
	{\label{fig:cavt2_fluxproc}}
\end{figure}

To test this theory, we would like to measure the $T_2$ of the cavity as a function of the qubit frequency.  Unfortunately, doing so is not as simple to generalize because of dynamical phases.  The cavity frequency changes by as much as $20\mhz$ when the qubit is detuned from its home position, causing the resulting cavity Ramsey oscillation to be extremely fast.  In order to resolve this phase evolution and get an accurate fit to the exponential decay, we would need to sample at a rate fast compared to $1/(20\mhz) = 50\ns$.  Considering that we expect dephasing times on the order of $30-80 \us$ and we would want to measure for several time constants, the number of required samples approaches $10,000$, which is quite unfeasible.  Alternatively, we can use the method shown in \figref{fig:cavt2_fluxproc}(a).  There, we choose several fixed times and repeat the Ramsey experiment for each while sweeping the angle of the final $\pi/2$ pulse.  At each time we get a sinusoid with a phase that, from our point of view, is random.  The amplitude of this oscillation tells us the amount of phase coherence remaining at that time.  As shown in (b), we can plot the amplitudes for each time and fit them to an exponential decay.  We see that when we have detuned the qubit by about $2\ghz$, the cavity coherence time has improved to more than $60\us$.  This corresponds to a pure dephasing time of $T_{\phi} = \left(\frac{1}{60 \us} - \frac{1}{2 \cdot 35 \us}\right)^{-1} \approx 400 \us$.  Clearly, the inherited dephasing from the qubit was dominating the cavity coherence.

\begin{figure}
	\centering
	\includegraphics[scale=1]{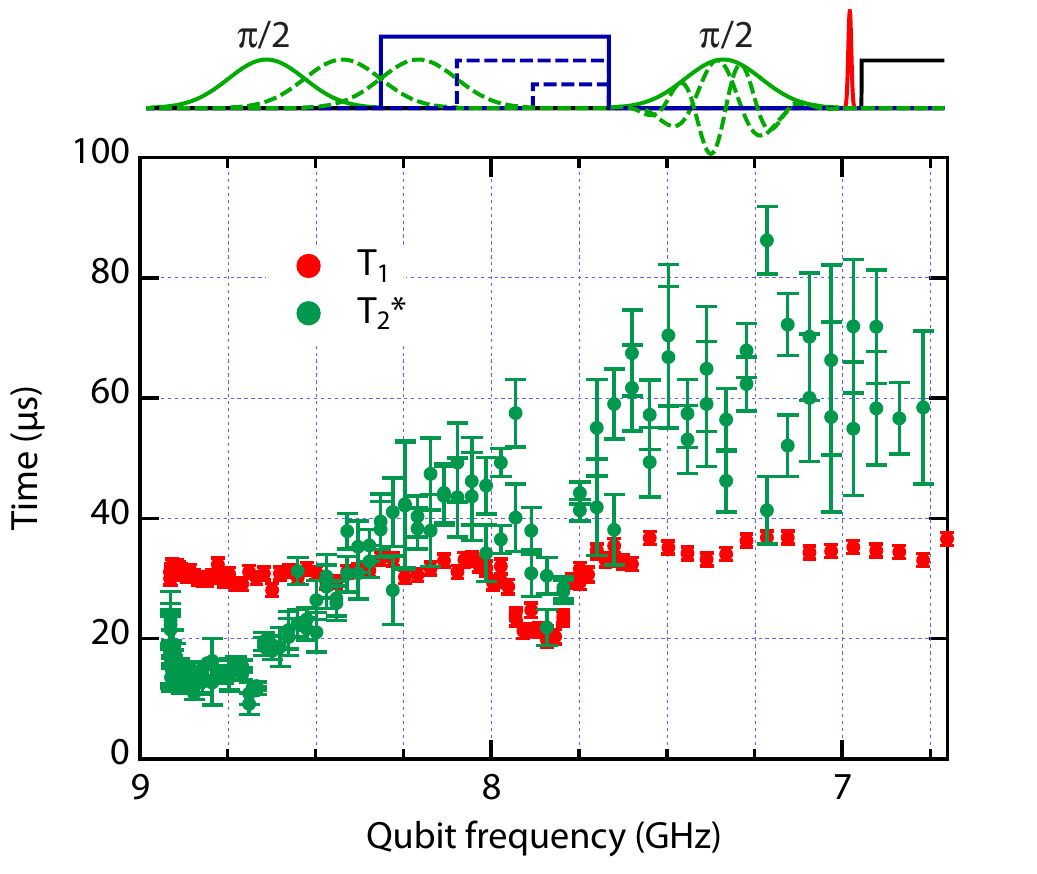}
	\mycaption{Cavity $T_1$ and $T_2$ as a function of qubit frequency} 
		{We perform the experiment described in \figref{fig:cavt2_fluxproc} as a function of qubit frequency by varying the amplitude of the qubit flux pulse.  We plot the measured cavity dephasing time, revealing that it initially decreases as the qubit lifetime rapidly degrades away from $f_{\mathrm{max}}$.  When the qubit is at $8.914\ghz$, the cavity $T_2$ is $22\us$, but reduces to only $12\us$ with the qubit at $8.846\ghz$.  For larger detunings, however, the dephasing time increases as the cavity becomes less hybridized and therefore becomes less susceptible to qubit dephasing.  Error bars are the standard deviation of several repeated measurements.  Uncertainties are larger for big detunings because we are unable to measure for larger delay times than $80\us$ due to the memory limitations of our AWG.  Therefore, we do not sample far on the exponential curve when the cavity coherence time gets long.  As before, we relate the applied flux pulse amplitude to qubit frequency with flux spectroscopy.  We also plot the previously-measured cavity $T_1$ for comparison, which shows that the cavity approaches $T_2=2T_1$ for large detunings.} 
	{\label{fig:cavt2_fluxed}}
\end{figure}

As before, we can reveal the evolution of this decoupling by measuring $T_2$ as a function of qubit frequency.  As shown in \figref{fig:cavt2_fluxed}, this is done by repeating the procedure from \figref{fig:cavt2_fluxproc} for a variety of flux pulse amplitudes.  The resulting data show two general trends.  At first, as we detune the qubit from its maximum frequency, the cavity coherence time is reduced because the qubit dephasing time is shrinking faster than the cavity is becoming decoupled.  (It is somewhat difficult to see this with the chosen $y$-scale, but the lifetime degrades from $22\us$ with the qubit at $8.914\ghz$ to $12\us$ with the qubit at $8.846\ghz$.)  As the qubit is further detuned, however, the cavity becomes less and less hybridized and thus less susceptible to the inherited dephasing.  The dephasing time therefore increases steadily to $\sim 70\us$, corresponding to nearly pure dephasing.  We conclude that, in the absence of the qubit, the cavity experiences essentially no inhomogenous dephasing.

\subsection{Cavity nonlinearity}
\label{subsec:measurecavnonlinearity}

We can continue building up procedure complexity and measure the cavity anharmonicity $K$.  The Kerr effect is due to hybridization of the cavity with the qubit that gives the cavity some nonlinearity.  As described in \sref{subsec:cavnonlinearity}, this can be seen as the first-order Taylor expansion of the Josephson potential, giving the cavity Hamiltonian a $\frac{K}{2}(\hat{a}^{\dagger}\hat{a})^2$ term.  While many materials exhibit some degree of Kerr nonlinearity, they are generally so lossy that the effect requires very high excitation to be observed.  The combination of large Josephson nonlinearities and extremely low cavity losses found in 3D cQED has recently enabled the study of the ``single-photon Kerr effect,'' where the quantization of photon number plays an important role in the physics \cite{Kirchmair2013}.  One way of quantifying this importance is to define the phase shift per photon $\phi=K/\kappa$, where $\kappa$ is the cavity bandwidth.  This phase shift is normally not resolvable when $\kappa > K$.

\begin{figure}
	\centering
	\includegraphics[scale=1]{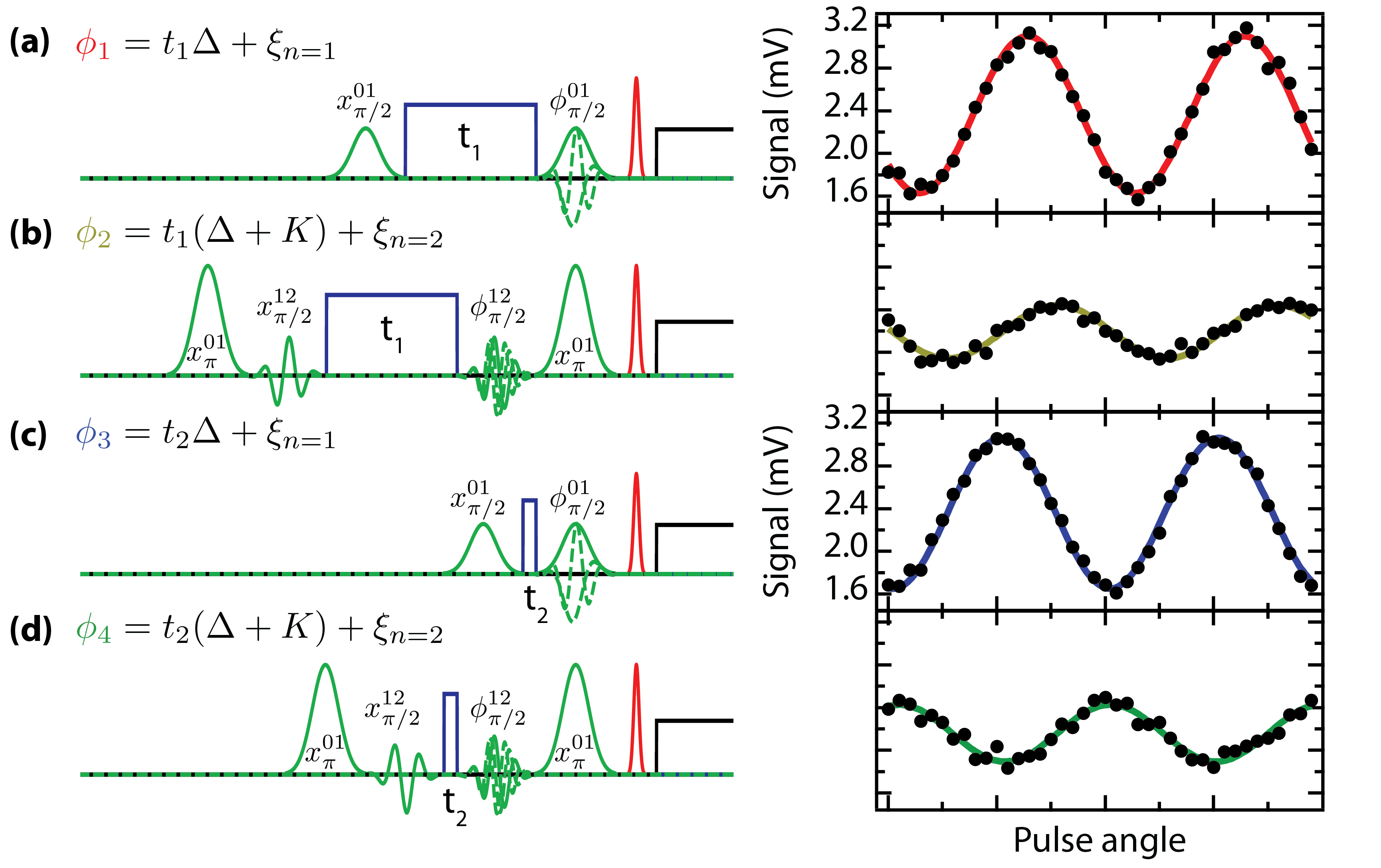} 
	\mycaption{Cavity Kerr measurement procedure}
	{To measure the cavity anharmonicity, we leverage our ability to create Fock state superpositions in the cavity.  Due to dynamical phases, we require four independent experiments.
	\capl{(a)} In the first experiment, we prepare a superposition of 0 and 1 photons with a $\pi/2$ rotation on the cavity.  This rotation is considered to be about the $x$ axis and addresses the $0\leftrightarrow 1$ transition, and so is labeled $x_{\pi/2}^{01}$.  We then move the qubit away for some time $t_1$ and then apply a $\pi/2$ cavity rotation around the $\phi$ axis, labeled as $\phi_{\pi/2}^{01}$, and measure.  We repeat this experiment as a function of $\phi$ to get a sinusoidal oscillation.  The phase of this curve will be given by $\phi_1 = t_1 \Delta + \xi_{n=1}$, where $\Delta$ is the amount by which the cavity frequency changes due to the qubit moving and $\xi_{n=1}$ is a dynamical phase associated with one photon undergoing that trajectory in time.
	\capl{(b)} The second experiment repeats the first except using a superposition of 1 and 2 photons.  We prepare that state by applying a $\pi$ pulse on the $0\leftrightarrow 1$ transition and then a $\pi/2$ rotation on the $1\leftrightarrow 2$ transition using single sideband modulation.  After waiting for $t_1$, we apply a second $\pi/2$ cavity rotation around the $\phi$ axis followed by a second $\pi$ pulse to map $1\rightarrow0$ to restore contrast between 1 and 2 photons for our subsequent measurement.  The phase of the resulting oscillation is be given by $\phi_2 = t_1 \left(\Delta + K \right) + \xi_{n=2}$ where $K$ is the Kerr anharmonicity and $\xi_{n=2}$ is the dynamical phase acquired by the 2-photon Fock state.  Note that the oscillation amplitude is significantly reduced compared to \capl{(a)} because the sequence takes a substantial fraction of the cavity lifetime due to the slow cavity pulses.  
	\capl{(c-d)} The third and fourth experiments simply repeat the first and second, but with a different waiting time $t_2$.  We therefore have $\phi_3 = t_2 \Delta +\xi_{n=1}$ and $\phi_4 = t_2 \left(\Delta + K \right) + \xi_{n=2}$, noting that the dynamical phases are identical to the first two cases so long as $t_2$ is longer than any qubit dynamics.  We thus have four equations and four unknowns, and can solve for $K/2\pi=\frac{\left(\phi_1 - \phi_2\right) - \left(\phi_3 - \phi_4 \right)}{2 \pi \left( t_2 - t_1\right)}$.}
	{\label{fig:measurekerr}}
\end{figure}

We want to directly measure the cavity Kerr anharmonicity using Fock states.  The essential fact that we are going to employ is that the energy of two photons is slightly less than twice the energy of one photon due to $K$.  A procedure for measuring $K$ as a function of qubit frequency is shown in \figref{fig:measurekerr}, where we measure the phase evolution of a superposition of $0$ and $1$ or $1$ and $2$ photons after moving the qubit away for some time.  The $0-1$ superposition acquires a phase due to the cavity frequency shifting in response to the qubit being detuned, while the $1-2$ gets both phase as well as another due to being lower in energy by $K$.  Both also acquire a dynamical phase due to the finite transit time of the qubit.  We perform two more experiments with different waiting times to calibrate those away.  We can solve the resulting system of four equations to extract $K$.

\begin{figure}
	\centering
	\includegraphics[scale=1]{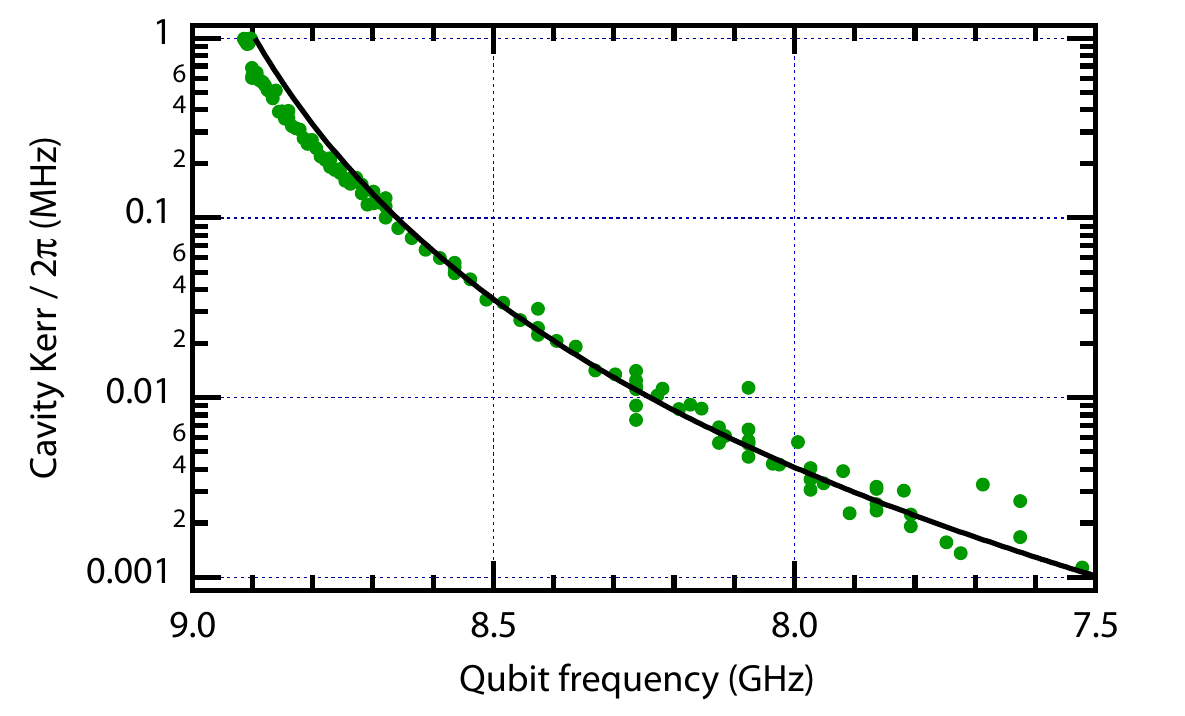}
	\mycaption{Cavity Kerr as a function of qubit frequency}
	{We perform the experiment described in \figref{fig:measurekerr} as a function of the qubit flux pulse amplitude.  As before, we relate the flux amplitude to the resulting qubit frequency with flux spectroscopy.  We plot the measured Kerr value as a function of qubit frequency.  Note that the $y$-axis is logarithmic, and that the Kerr parameter drops three orders of magnitude when we detune the qubit by only $1.5\ghz$.  We plot a $\Delta^{-4}$ curve fit to data starting at $8.5\ghz$ where the dispersive approximation should hold.  The data is well-fit by this power law, indicating that the Kerr effect can be reduced arbitrarily low with further detuning.}
	{\label{fig:cavitykerr}}
\end{figure}

We can immediately perform this experiment as a function of qubit flux pulse amplitude (as shown in \figref{fig:cavitykerr}).  We find that the Kerr reduces from $\sim1\MHz$ to $\sim 1\KHz$ when we detune the qubit by only $1.5\ghz$.  This dramatic reduction indicates that we have the ability to completely turn off the Kerr effect, in the sense that $\kappa \gg K$.  Indeed, the reason we were not able to measure much lower than $1\khz$ was because this device had a $\kappa = 1/(2\pi\cdot 40\us) = 4 \khz$ and our superposition state decayed faster than it acquired phase.  For larger detunings where the dispersive approximation applies \cite{Boissonneault2009}, the data conform to a $\Delta^{-4}$ curve.  Even if our cavity was substantially longer lived \cite{Reagor2013}, this indicates that we could still turn off the Kerr nonlinearity by detuning slightly further.

\subsection{Cavity dispersive shift}

\begin{figure}
	\centering
	\includegraphics[scale=1]{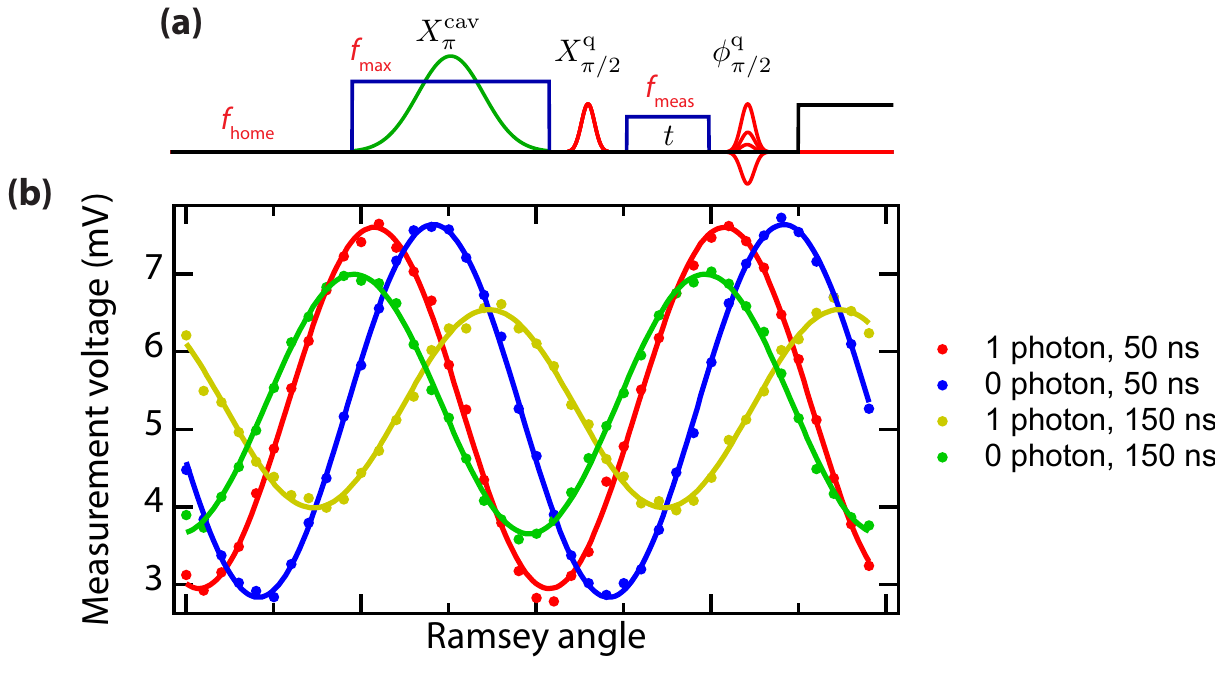}
	\mycaption{Sequence to measure cavity $\chi$}
		{\capl{(a)} The prototype of a set of experiments to measure the cavity dispersive shift $\chi$ as a function of qubit position is shown.  The qubit starts at some home position where $\chi$ is relatively small.  Here we choose $f_{\mathrm{home}}\approx8.5\ghz$ where $\chi= 3\MHz$.  We then move the qubit up to its maximum frequency $f_{\mathrm{max}}$, where the cavity inherits enough anharmonicity to be $\pi$ pulsed.  We prepare a one-photon Fock state before moving the qubit back to its home position.  By applying a $\pi/2$ pulse, we then prepare a superposition of ground and excited qubit states.  We perform this rotation as fast as possible so as to be maximally insensitive to the cavity state.  We then move the qubit to some frequency $f_{\mathrm{meas}}$ where we wish to measure the $\chi$-shift (which can be either above or below $f_{\mathrm{home}}$) and wait for some time $t$.  Finally, we flux the qubit back to $f_{\mathrm{home}}$ and perform a Ramsey experiment by varying the angle of a final $\pi/2$ pulse and measure with the second cavity mode.
		\capl{(b)} We perform the experiment described above for preparations of both 0 and 1 cavity photons by either performing or omitting the cavity $\pi$ pulse.  (In either case the flux pulse sequence is unchanged.)  As with the sequence for measuring the cavity Kerr (described in \figref{fig:measurekerr}), here we also have dynamical phases associated with a finite flux pulse rise time that we must measure and cancel out.  This is done, as before, by repeating the two experiments with a different integration time, here chosen to be either $50\ns$ or $150\ns$.  The four phases we will measure are then given by $\phi_1 = t_1 \Delta + \xi_{0}$, $\phi_2 = t_1 \left(\Delta + \chi\right) + \xi_{1}$, $\phi_3 = t_2 \Delta + \xi_{0}$, and $\phi_4 = t_2 \left(\Delta + \chi\right) + \xi_{1}$, where $\Delta/2\pi = f_{\mathrm{home}} - f_{\mathrm{meas}}$ and $\xi_i$ is the dynamical phase for $i=0,1$ photons.  We solve this system of equations, giving $\chi/2\pi=\frac{\left(\phi_2 - \phi_1\right) - \left(\phi_4 - \phi_3 \right)}{2 \pi \left( t_2 - t_1\right)}$.
		}
	{\label{fig:chimeas}}
\end{figure}

We can measure the cavity $\chi$-shift as a function of qubit frequency.  As described in \figref{fig:chimeas}, the sequence is quite similar to the one we used to measure the cavity Kerr.  We prepare either 0 or 1 photon, put the qubit on its equator, flux the qubit to the frequency at which we want to know $\chi$, wait for some time, then move the qubit back home and interrogate its phase with a Ramsey experiment.  Just as was necessary with Kerr, we have to perform four experiments to cancel out dynamical phases.

\begin{figure}
	\centering
	\includegraphics[scale=1]{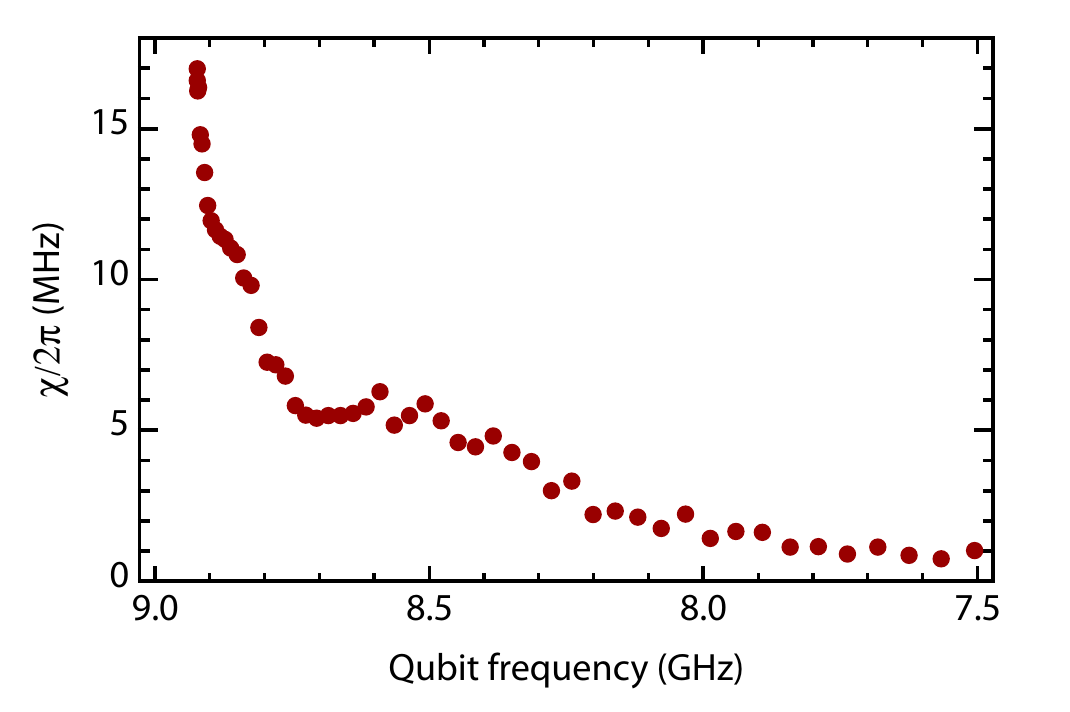}
	\mycaption{Cavity-qubit dispersive shift $\chi$ as a function of qubit frequency}
		{We measured cavity $\chi$ as a function of qubit frequency using the sequence described in \figref{fig:chimeas}.  Note that each frequency point required re-writing the sequence file, since we operated at three distinct frequencies and could not simply sweep the flux pulse channel voltage.  The measured $\chi$-shift is non-asymptotic as a function of detuning, indicating that there is something unusual going on.  One theory is that because the qubit is initially so close to the cavity, there may be some non-dispersive effects or avoided crossings that add complication.  Another possibility is that, because the qubit is so large and distributed, non-lumped-element effects may come into play.  Nevertheless, the point of this experiment was to prove that it could be measured, rather than understand in detail the parameters of the device.}
	{\label{fig:cavitychi}}
\end{figure}

This sequence is somewhat more complicated than is necessary.  It calls for performing $\pi/2$ pulses on the qubit at a frequency $f_{\mathrm{home}}$ that was distinct from $f_{\mathrm{max}}$.  By choosing $f_{\mathrm{home}}$ to be far from the cavity, the qubit number splitting can be smaller than our qubit pulse bandwidth.  That way we can successfully apply a number {\it insensitive} $\pi/2$ pulse regardless of the number of cavity photons.  This requires high-quality flux deconvolution (see \sref{subsec:fluxlinecal}) because we perform a qubit pulse shortly after the qubit has moved down from $f_{\mathrm{max}}$.  Operating at $f_{\mathrm{home}}$ is a requirement if, for example, we wanted to interrogate the cavity parity by performing a flux excursion such that $\int n\chi dt = n \pi$ \cite{Leghtas2012, Vlastakis2013}.  However, for the purposes of this experiment, the cavity is always in an energy eigenstate of either 0 or 1 photons.  Therefore, we could apply our qubit rotation at two different frequencies depending on the preparation, and thus apply all of our rotations at $f_{\mathrm{max}}$.  This would be a huge technical simplification, since any residual flux bias line ringing would be strongly damped as a result of the qubit being near its maximum frequency.  Nevertheless, as shown in \figref{fig:cavitychi}, we were able to perform the prescribed more complicated sequence.  This proves that we can successfully deconvolve away the ringing, enabling more sophisticated experiments that require operating the system at several distinct frequencies.

\section{Cavity dynamics}
\label{sec:cavitykerrexpts}

Our lab has recently become interested in using the cavity itself as a quantum resource.  Cavities may be longer-lived than even the best 3D qubits \cite{Reagor2013}, have a infinitely-large Hilbert space that can store huge amounts of quantum information \cite{Gottesman2001, Maitre1997, Mariantoni2011, Kirchmair2013, Vlastakis2013, Leghtas2013}, and suffer from only simple error processes that could potentially be monitored and reversed in real time \cite{Leghtas2012, Sun2013}.  These properties are very attractive for using the cavity as a quantum memory, which may be error correctable \cite{Leghtas2012} and fault-tolerant.  This section introduces some of the basic building blocks to store and interrogate quantum information in a cavity.  We measure as an example the Kerr evolution of a coherent state stored in the cavity, and show how we can shut off this evolution using fast flux.  As with the last section, the control and techniques we demonstrate here generalize directly to future experiments.

\subsection{The Husimi Q function}
\label{subsec:qfunction}

\nomdref{Cqalpha}{$Q(\alpha)$}{Husimi quasi-probability distribution}{subsec:qfunction}

How do we measure the quantum state of a cavity?  Since its Hilbert space is infinitely large, coming up with an efficient and physically-motivated representation is crucial to making good use of this resource.  As we saw in \sref{subsec:harmonicoscillators}, one possible route is to use the photon number Fock basis \cite{Johnson2010}.  This is most analogous to how we think of qubit quantum states, but is not a very natural choice for cavities.  The canonical cavity state is a ``coherent state,'' which has a complicated representation in the Fock basis.  Since we are likely to use coherent states frequently, it makes sense to use a representation that acknowledges that fact.  We therefore seek to use coherent states as the fundamental basis, rather than Fock states.  One consequence of representing our infinite Hilbert space with those states is that we must now explicitly represent our state using {\it continuous}, rather than discrete, variables \cite{Cahill1969, Braunstein2005}.  This reflects the physical reality that two coherent states always have finite overlap, with $\langle \alpha | \beta \rangle = e^{-|\alpha-\beta|^2}\ne \delta\left(\alpha-\beta\right)$.

The simplest ``quasiprobability'' distribution for representing quantum states of light is known as the {\it Husimi Q function} \cite{Husimi1940}.  It is defined\footnotemark~ as the modulus squared of the resonator state $|\psi\rangle$ with a coherent state $|\alpha\rangle$, $Q(\alpha) = \frac{1}{\pi} | \langle \alpha | \psi \rangle |^2$.  The coefficient $1/\pi$ is included for normalization, such that $\int_{\alpha \in \mathbb{C}} Q(\alpha)d\alpha^2 = 1$.  That is, $Q(\alpha)$ is proportional to the probability that $|\psi\rangle$ is the coherent state $|\alpha\rangle$.  As expected, if our cavity contains a coherent state $|\psi\rangle = |\beta\rangle$, then $Q(\alpha) = \frac{1}{\pi}e^{-|\alpha-\beta|^2}$.  We can also represent $Q$ using the displacement operator $\hat{D}(\alpha) = e^{\alpha \hat{a}^\dagger - \alpha^* \hat{a}}$.  Recall from \sref{subsec:harmonicoscillators} that the displacement operator represents the action of a coherent tone on a harmonic oscillator, and creates coherent states as $\ket{\alpha} = \hat{D}(\alpha)|0\rangle$.  Using $\hat{D}^\dagger(\alpha) = \hat{D}^\dagger(-\alpha)$, we can write $Q(\alpha) = \frac{1}{\pi} |\bra{0} \hat{D}(-\alpha) \ket{\psi}|^2$.  

\footnotetext{This can be generalized to a cavity density matrix as $Q(\alpha) = \frac{1}{\pi} \langle \alpha | \hat{\rho} | \alpha \rangle$, the normalized trace of the density matrix over the coherent state basis.}

\begin{figure}
	\centering
	\includegraphics[scale=1]{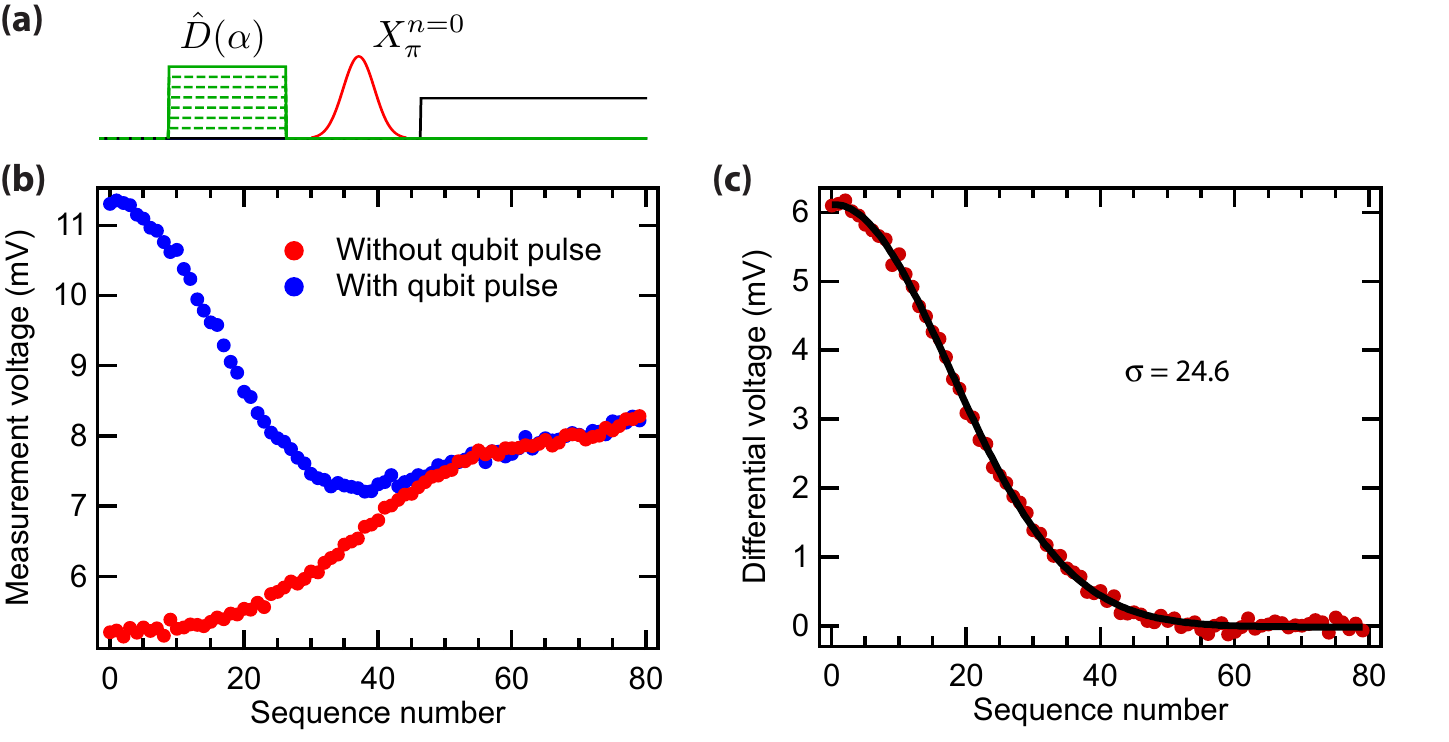}
	\mycaption{Q function calibration}
		{\capl{(a)} We apply a real-valued displacement of some magnitude $\alpha$ before interrogating the cavity with a qubit $\pi$ pulse conditioned on the $n=0$ photon number state.  We measure the qubit state with the second cavity mode.
		\capl{(b)} Due to the cross-Kerr effect, the measurement voltage will depend on the size of $\alpha$.  In red, we show the result of the experiment described in \capl{(a)} when the qubit $\pi$ pulse is omitted, as a function of the displacement.  If the measurement was truly only sensitive to the qubit state as we desire, this line should be flat.  At this point we have not yet calibrated the our displacement magnitude, so we instead plot as a function of sequence number.  In blue, we show the result when we include the qubit pulse.
		\capl{(c)} We plot the difference of the red and blue lines from \capl{(b)} and fit the result to a gaussian distribution.  The quality of this fit indicates that we can successfully subtract away the cross-Kerr error.  This is because, at least in the low-photon limit, our measurement operator is approximately linear in the total number of excitations.  If we were to displace substantially further, or pulse higher number states, we would start to see an compression in measurement contrast that we cannot subtract away.  Physically, this compression originates with the cavity bright-state probability approaching unity.  The characteristic width $\sigma$ of the gaussian fit also calibrates our displacement scale.  The probability of having $n$ photons given a displacement $\alpha$ is given by $P(|\alpha|) = |\alpha|^{2n} e^{-|\alpha|^2}/n!$.  For $n=0$, $P$ is a gaussian, and when $P=1/e$, $\alpha=1$.  Thus, the best-fit value of $\sigma=24.6$ indicates that, at the voltage corresponding to that sequence number, we are doing a displacement of unit magnitude.
		}
	{\label{fig:cal_qfuncs}}
\end{figure}

\begin{figure}
	\centering
	\includegraphics[scale=1]{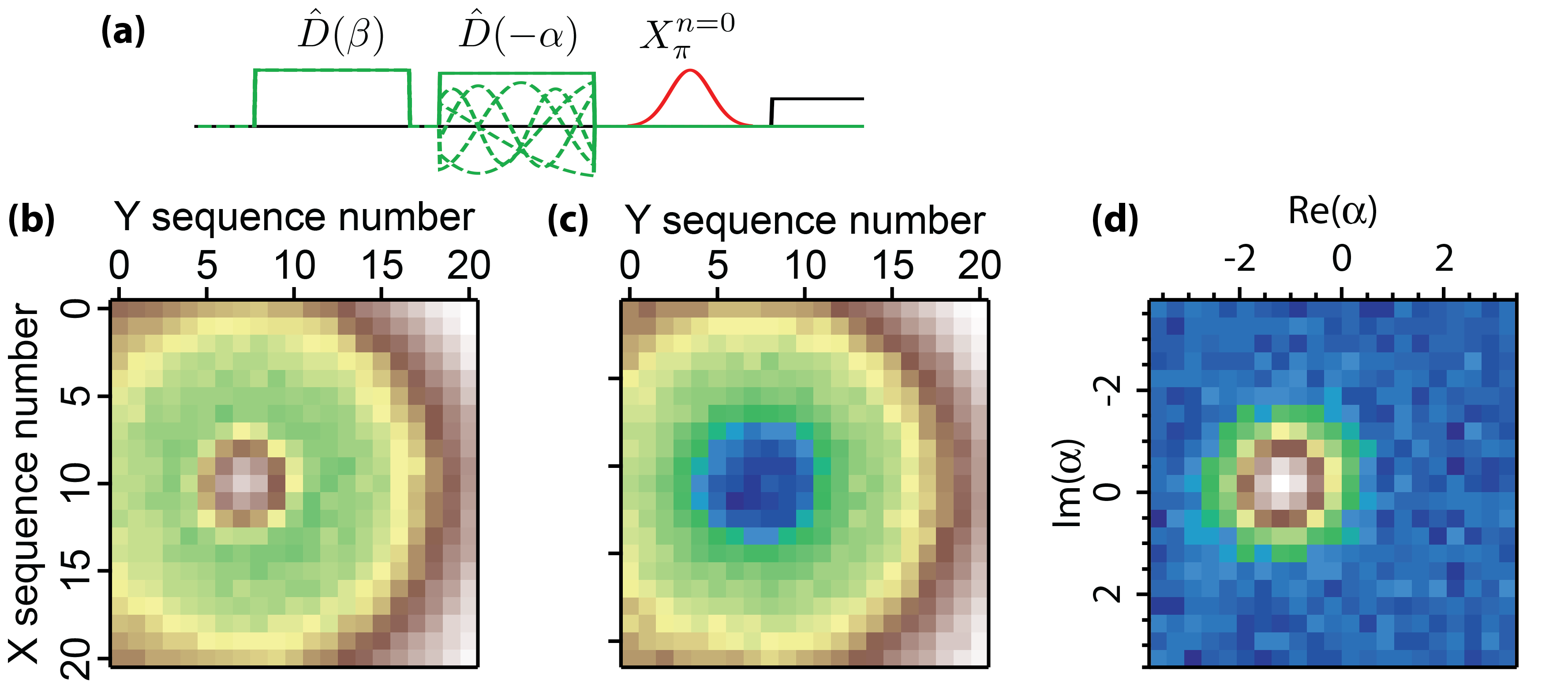}
	\mycaption{Procedure for measuring the Q function and example data}
		{\capl{(a)} To measure $Q$, we prepare a test state with a displacement $\hat{D}(\beta=-1.5)$.  We then measure $|\bra{0} \hat{D}(-\alpha) \ket{\psi}|^2$ by applying a second displacement $\hat{D}(-\alpha)$, a qubit $\pi$ pulse conditioned on 0 photons, and finally, a measurement through the second cavity mode.
		\capl{(b)} If we omit the qubit pulse (as described in \figref{fig:cal_qfuncs}) we still get measurement contrast as a function of $\alpha$ due to the cross-Kerr effect.  This measurement is used to calibrate away this effect.  Here, the $x$ and $y$ axes are proportional to the real and imaginary parts of $\alpha$.  We measure for 121 separate values of $\alpha$.
		\capl{(c)} Including the qubit pulse, we see a large peak slightly offset from the origin due to our prepared state $\ket{\beta}$.  As expected, however, we also see an overall background that must be removed.
		\capl{(d)} By subtracting \capl{(b)} from \capl{(c)}, we arrive at our $Q$ function.  It is ideally given by a gaussian centered at $\beta$.  Here, we calibrate the $x$- and $y$-axes using the method described in \figref{fig:cal_qfuncs}.}
	{\label{fig:meas_qfuncs}}
\end{figure}

This expression for $Q$ is highly suggestive of an experimental procedure.  To measure $Q(\alpha)$, we could displace the cavity by $-\alpha$ and measure the probability that there are zero photons using a number-selective qubit $\pi$ pulse conditioned on $n=0$.  As before, we use the first cavity mode as storage for the quantum state and use the second cavity mode to measure the qubit.  This plan will work, though we must acknowledge a complication \footnotemark.  Due to the cross-Kerr effect, there is a direct dispersive coupling between the first and second cavity mode.  Our second-mode measurement voltage will therefore depend on the number of photons in the first cavity mode.  This dependence will corrupt our measurement because the answer to whether or not the cavity has zero photons should not change if it has one photon or one hundred.  For relatively small displacements at least, the linearity of the measurement can be used to compensate for this effect (demonstrated in \figref{fig:cal_qfuncs}).  For each measurement of $Q$, we measure both with and without the qubit $\pi$ pulse and subtract the two measurements, which effectively cancels the cross-Kerr component.  We use this strategy to measure the $Q$ function of a coherent state in \figref{fig:meas_qfuncs}.

\footnotetext{Note that we are omitting from our discussion some other details of measuring a $Q$ functions that are necessary for a quantitative analysis.  These complications include the finite bandwidth and selectivity of our qubit $\pi$ pulse, thermal population of either the qubit or cavity, the linearity of the cavity displacement, and others.  Discussion of these issues can be found in references~\citenum{Kirchmair2013} and \citenum{Vlastakis2013}.}

\subsection{Kerr evolution}

The $Q$ function is not limited to measuring coherent states.  How might we create some other quantum state of light?  As we mentioned in \sref{subsec:harmonicoscillators}, any coherent tone applied to the cavity can only cause a displacement.  We require some source of nonlinearity to create a more interesting non-classical state.  Interesting non-classical states of light have been made before by building up Fock states \cite{Hofheinz2008}, but this is rather cumbersome.  Fortunately, there is an easier way.  As we have discussed in \sref{subsec:cavnonlinearity} and \sref{subsec:measurecavnonlinearity}, the cavity automatically has some nonlinearity due to its hybridization with the qubit.  The phase evolution of the $n$th Fock state is therefore delayed, with $\ket{n} \rightarrow e^{i\frac{K}{2} n^2 t} \ket{n}$.  Coherent states are thus no longer eigenstates, and evolve according to 
\begin{equation}
	\label{eq:kerrvolution}
	|\psi(t)\rangle = e^{i \frac{K}{2}(\hat{a}^\dagger \hat{a})^2 t} |\beta\rangle = e^{-|\beta|^2/2}\sum_n \frac{\beta^n}{\sqrt{n!}} e^{i\frac{K}{2} n^2 t} \ket{n} \ne \ket{\beta'}.
\end{equation}
By displacing our cavity and waiting, we can create interesting non-classical states.  

\begin{figure}
	\centering
	\includegraphics[scale=1]{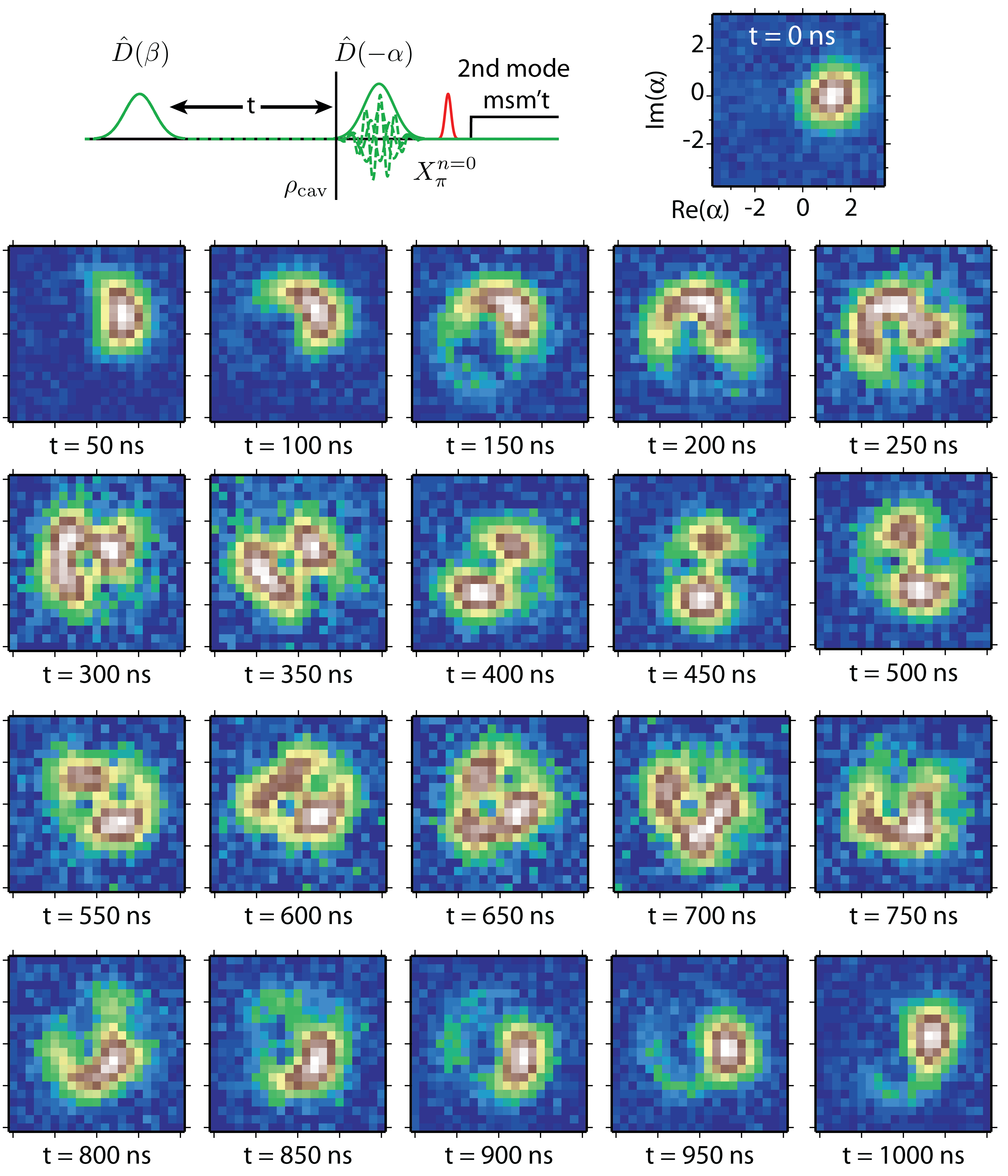}
	\mycaption{Evolution of a coherent state under the Kerr nonlinearity}
		{The cartoon in the upper left corner shows the procedure for producing and measuring these states, where we initially displace the cavity, wait some time, and then measure the $Q$ function.  The resulting evolution is stroboscopically probed every $50\ns$ until the state eventually re-coheres and the evolution repeats.  Note that we were in the $\frac{K}{2} \hat{a}^\dagger \hat{a}^\dagger  \hat{a} \hat{a}$ rotating frame of the cavity in these data rather than the $\frac{K}{2} (\hat{a}^\dagger a)^2$ frame, giving us a net $\pi$ phase shift in $1\us$.}
	{\label{fig:kerrevolution}}
\end{figure}

As first observed with two physical cavities in Ref.~\citenum{Kirchmair2013}, we demonstrate this Kerr evolution using the procedure shown in \figref{fig:kerrevolution}.  There, we initially displace our oscillator by a magnitude $\beta=1.5$.  This takes only $5\ns$ and, since its bandwidth is much greater than $K$, is completely insensitive to the cavity nonlinearity.  We then wait for some time $t$, during which the state evolves as written in \equref{eq:kerrvolution}.  Finally, we measure $Q$.  The measured distributions are also shown in \figref{fig:kerrevolution}.  For short times, the Kerr evolution is closely approximated by a rotation in phase space by an angle $\phi_{\mathrm{Kerr}} = K t (|\beta|^2 + 1/2)$ \cite{Kirchmair2013}.  (This fact was used to measure even smaller values of $K$ by compensating with larger $\alpha$, but in practice it is considerably more cumbersome than the procedure described in \sref{subsec:measurecavnonlinearity}.)  For longer times, the state spreads angularly because amplitude components further from the origin rotate with a larger velocity due to the $n^2$ dependence of the Kerr effect.  After a time $T_{\mathrm{rev}} = \frac{2\pi}{K}\approx 1\us$, the state revives to a coherent state, but with a $-1$ phase since $e^{i\frac{K}{2} n^2 t} = (-1)^n$ for $t=T_{\mathrm{rev}}$ \cite{Yurke1988}.  There a small tail is visible due to inhomogenous cavity dephasing inherited from the qubit.

The state appears to be much more complicated between this early spreading out and the eventual revival.  For example, we see at $t\approx450\ns$ that our state is two distinct blobs.  This is a superposition of $\ket{\beta}$ and $\ket{-\beta}$, and is known as a ``Schr\"{o}dinger cat'' state.  The name refers to the fact that it is a superposition of ``alive'' and ``dead'' macroscopic quantum states.  Although it is difficult to resolve with this relatively small initial displacement $\beta$, the Kerr evolution in general produces a superposition of an arbitrarily large number of evenly-spaced coherent states.  For integer fractions $q$ of $T_{\mathrm{rev}}$, we can write the quantum state as \cite{Haroche2006, Kirchmair2013}
\begin{equation}
	\ket{\psi\left(\frac{T_{\mathrm{rev}}}{q}\right)} = \frac{1}{2q} \sum_{p=0}^{2q-1} \sum_{k=0}^{2q-1} e^{ik(k-p)\frac{\pi}{q}} \ket{\beta e^{i p \frac{\pi}{q}}}.
\end{equation}
This is a ``multi-component Schr\"{o}dinger cat'' state and is made up of $q$ coherent states.  Note that the cat shown at $t=450\ns$ should be exactly symmetric, but is not.  We attribute this to the cavity having substantial higher-order nonlinearity due to its extremely strong coupling with the qubit.  If we move the qubit slightly away to slow down the Kerr evolution, the cats appear to be more symmetric (data not shown).

\subsection{Freezing Schr\"{o}dinger cats}

\begin{figure}
	\centering
	\includegraphics[scale=1]{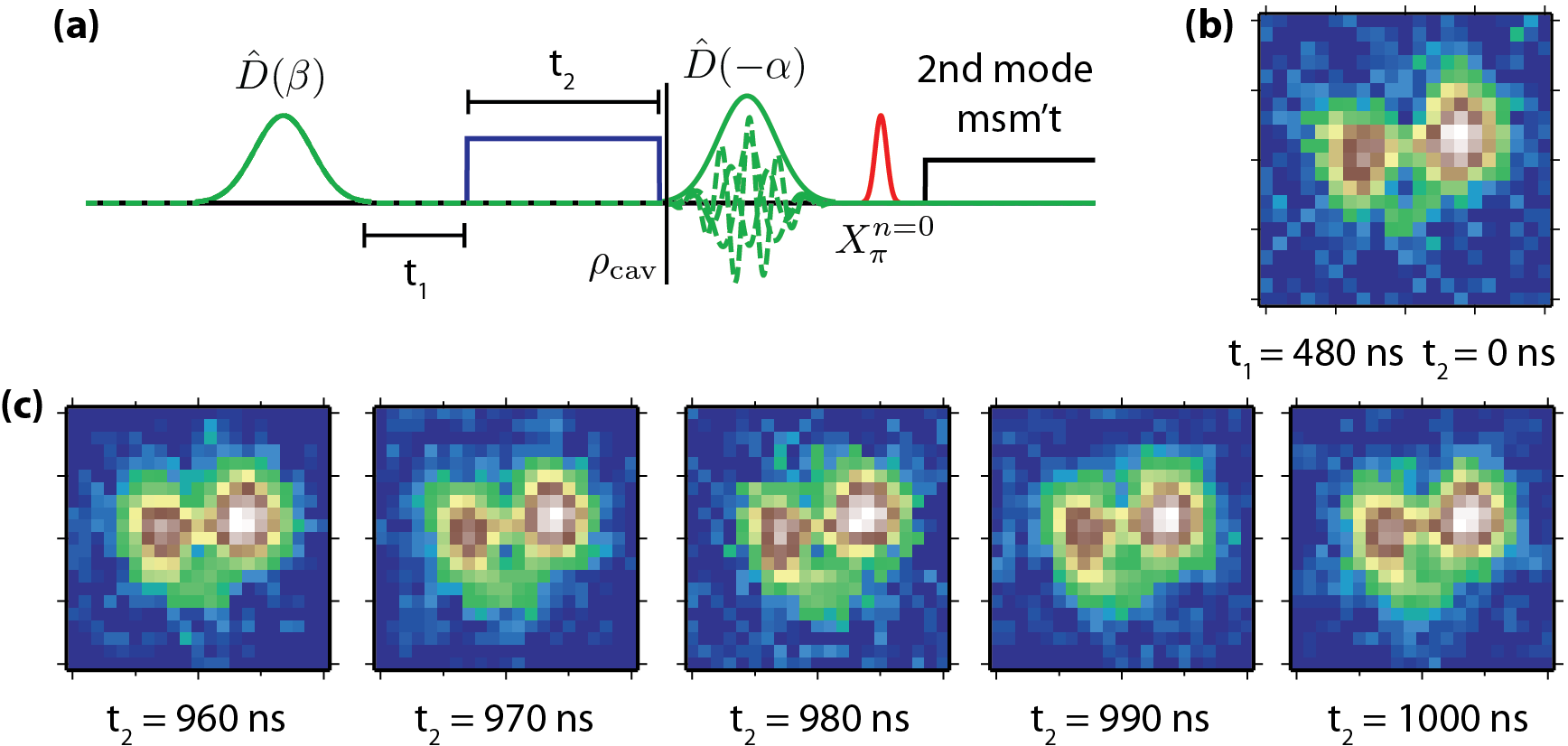}
	\mycaption{Freezing Schr\"{o}dinger cat states with fast flux}
		{\capl{(a)} We create a Schr\"{o}dinger cat state by displacing the cavity with $\beta=1.5$ and waiting for $t_1=480\ns$.  To freeze this state, we move the qubit far away, making $K\rightarrow 0$, and wait for some time $t_2$.  We then measure the $Q$ function.
		\capl{(b)} The $Q$ function of state we are aiming to freeze, with $t_1=480\ns$ and $t_2=0\ns$ is plotted.  This state does exhibit some asymmetry due to higher-order cavity nonlinearities, but for the purposes of this experiment is deemed to be the ideal state we wish to protect. 
		\capl{(c)} As described in \capl{(a)}, after creating our Schr\"{o}dinger cat, we move the qubit suddenly away for some period of time.  We show the resulting $Q$ function for several values of $t_2$, demonstrating that the evolution has completely shut off.  During the $40\ns$ of evolution shown, the cat would become noticeably deformed if $K$ had not been reduced.  Note that these data have been rotated in software to compensate for the $22\mhz$ detuning that the cavity has from our LO when the qubit is far-detuned. }
	{\label{fig:frozencats}}
\end{figure}

The Kerr evolution therefore gives us the ability to create interesting non-classical states of light.  However, the evolution shown in \figref{fig:kerrevolution} loops around forever, or at least until the state decays to 0 photons and becomes an eigenstate of the Kerr Hamiltonian.  If we stored some quantum information in this cavity, it would only be accessible every $T_{\mathrm{rev}}$.  Fortunately, we can eliminate this restriction by using our fast flux control.  As we showed in \figref{fig:cavitykerr}, we can turn $K$ down to essentially zero by detuning the qubit.  We can therefore start with the qubit near $f_{\mathrm{max}}$ to create some quantum state with the Kerr evolution or direct cavity pulses and then move it away to freeze the evolution of that state, as shown in \figref{fig:frozencats}.  This demonstrates that we are able to dynamically tune the Hamiltonian of the cavity from being qubit-like to cavity-like on demand to suit our needs.

\section{Conclusions}

This chapter demonstrated some of the basic qualities and functionality of the tunable 3D cQED architecture.  We showed that if we do not filter the flux bias lines, they will be an unacceptable qubit decay channel.  Consistent with our expectations, this channel can be turned off with proper low-pass filtering.  Moreover, this filtering does not appear to substantially degrade the fast performance of our flux bias lines since the associated time scales of the qubit and FBL are sufficiently different ($\sim 500\mhz$ vs $\sim 5-10\ghz$). 

We then turned our attention to a particular device with a single qubit and cavity.  The qubit maximum frequency was very close to resonance, making it possible to directly Rabi oscillate the cavity and create photon number Fock states.  Combined with fast flux control, this enabled us to perform a variety of characterization experiments.  We showed how we can directly measure the lifetime and coherence of the cavity as a function of qubit frequency, demonstrating that the cavity inherits substantial dephasing from the qubit.  We were able to directly measure and instantaneously control the cavity Kerr anharmonicity and the dispersive $\chi$ shift.  These experiments primarily serve as an explanation of the kinds of techniques and tools that are available with fast flux control, though also provide a rich dataset for validating theoretical predictions.

Finally, motivated by the promising qualities of the 3D cavities, we introduced some basic tools for manipulating and measuring continuous-variable quantum states.  In particular, we showed how to measure the Husimi $Q$ quasi-probability distribution using the tools we have already established.  Using the $Q$ distribution, we measured the Kerr evolution of a coherent state stored in our nonlinear cavity.  At a particular time, this evolution maps us to a superposition of two coherent states known as a Schr\"{o}dinger cat state.  Using fast flux, we froze the evolution of this state, demonstrating our ability to dynamically change the cavity between an anharmonic qubit-like system to a linear memory.  In the next and final chapter of this thesis, we will further explore this idea to motivate future applications of the tunable 3D architecture.

\setcounter{chapter}{9}
\chapter{Conclusions and Future Work}
\thumb{Conclusions and Future Work}
\label{ch:conclusion}


\lettrine{D}{uring} the last several years, superconducting circuits have made dramatic progress in both complexity and coherence time.  We have highlighted these trends in this thesis, culminating respectively with the demonstration of the most basic form of quantum error correction and the development of the tunable 3D cQED architecture.  While some breakthroughs such as the high-power readout mechanism were unexpected, other progressive trends were more purposeful.  Increasing from two entangled qubits to three was primarily a matter of improved engineering.  Similarly, the three-qubit Toffoli gate required for error correction was a logical progression from the two-qubit gates pioneered in earlier work.  (The improved coherence of 3D cQED lies somewhere in between these two extremes, as it was guided by principle but still exploratory.)  The fact that these devices are so well understood that they can be expanded and built upon speaks to the advantages of a solid-state quantum architecture.  Better understanding of physics and development of techniques can be leveraged to do progressively more sophisticated experiments.  Indeed, many high-profile results have become standard techniques for later experiments.

The experimental path thus far has largely followed that of earlier systems like NMR and trapped ions \cite{NistRoadmap}.  We have now accomplished many of the ``proof-of-principle'' experiments where any level of functionality can evidence success, but future quantum information goals are qualitatively different.  New landmarks will require combining increased device complexity with higher gate and measurement fidelity.  One such goal is to demonstrate a logical qubit with a lowered effective error rate using quantum error correction.  This has never been accomplished in any system despite the fact that the whole future of quantum computing is predicated on its success.  While there are a variety of proposals \cite{Nielsen2000, Shor1995b, Steane1996, Kitaev1997, Bravyi1998, Fowler2012, Leghtas2012}, a great deal of theoretical interest and, more recently, experimental ambition \cite{Chow2012, Barends2013} has concerned the ``surface code'' \cite{Kitaev1997, Bravyi1998, Fowler2012}.  This code has advantages of a high threshold ($\sim1\%$) and simple hardware requirements.  However, the signature of an operational surface code -- an improved logical error rate -- only emerges from the collective behavior of a large number of qubits.  Even assuming gate fidelities continue to increase, it is hard to imagine fewer than one hundred qubits being used in an initial demonstration.  And if merely controlling one hundred qubits is not challenging enough, each of them must be endowed with record-breaking gate fidelity and high fidelity in-situ readout and reset.  Progress is being made on each requirement \cite{Reed2010, Vijay2011, Chow2012, Riste2012a, Riste2012c, Geerlings2013, Hatridge2013,  Paik2013}, but combining all those advances with a large number of coupled qubits is a much bigger challenge.  

Fortunately, there are more direct paths to interesting scientific results.  One example is the recent proposal of Leghtas, {\it et al.} to error correct a quantum state stored in the Hilbert space of a harmonic oscillator \cite{Leghtas2013}.  It is not yet known to be fault tolerant and does not yet enjoy the considerable theoretical support of the surface code, but it is much simpler to implement.  None of the experimental requirements seem beyond the reach of an academic group in the coming years, and several would constitute a high-profile result in their own right.  Moreover, for plausible experimental parameters, this ``cat code'' could realize an improved effective cavity coherence time.  There is also ongoing theoretical development of the idea, which might lead to a fault-tolerant implementation that eliminates the need for a more technical approach like the surface code. 

In this chapter, we propose a series of experiments to expand on the work presented in this thesis.  We will start with a list of technical improvements that could be incorporated into the tunable 3D cQED architecture (\sref{sec:3ddev}).  Some of these are straightforward extensions of what has already been demonstrated, while others are more exploratory and aspirational.  We will then summarize a few exciting qubit experiments that could be achieved with an improved 3D cQED or a similar architecture (\sref{sec:qubitexpts}) before concluding with a general outlook on superconducting quantum computing.

\section{Tunable 3D cQED development}
\label{sec:3ddev}

A considerable amount of development of the 3D cQED architecture has already occurred, but there is room for more.  One promising example is to replace the flux bias lines with microwave resonators for individual qubit readout.  We already have high-bandwidth RF connectors on the PC board for flux control, so this would be a simple matter of a new lithographic design of the FBL.  It may also be possible to design a circuit that could be used for both flux bias and microwave cavity coupling because the relevant frequencies are so different.  Additionally, because the FBL wafer is so large, it would be easy to integrate complications like a Purcell filter to the output of the resonator.  This would enable efficient qubit reset, as we saw in \sref{subsec:qubitreset}.  When combined with a quantum-limited amplifier \cite{Siddiqi2004, Boulant2007, Spietz2009, Castellanos2009, Bergeal2009, Vijay2009, Abdo2011, Hatridge2013}, a Purcell-filtered resonator would also enable high-fidelity and QND dispersive measurements.

A resonant structure like the Purcell filter may also represent a better way to design 3D FBL lines.  In addition to being easier to fabricate, it would eliminate the need for the dielectric capacitors that might currently be limiting coherence.  We could also investigate the dependence of qubit lifetime on physical dimensions.  We have already seen that moving from a $50\um$ to a $300\um$ antenna pitch improves qubit lifetime considerably; there may be other simple and effective changes available to us that have not been tried.  Similarly, we could measure flux noise density as a function of the size of our SQUID loop to see if progress could be made on that front.  Recently, there has also been promising work on using surface treatments to cut down flux noise \cite{Anton2012}.  We have not ruled out the possibility that our flux noise is the result of a technical problem with the experimental setup, so testing a tunable 3D cQED device in another fridge would be informative.  Improved cryogenic filtering might also ameliorate qubit lifetimes, and validating our setup against others that are known to be of high quality would be reassuring.

Larger architectural design changes are also worth consideration.  There have been some issues with RF coupler cross-talk in multi-cavity designs that could be solved by re-designing the octobox sample holder.  We could retrofit the design to support a waveguide Purcell filter \cite{Hatridge2013b} if we wished to measure through a low-Q 3D resonator. A re-design could also render more robust the use of high-purity aluminum cavities.  The required etching of these cavities increases physical gaps, which can cause problems when we are relying on tight fits to confine microwaves in the normal design.  An RF choke \cite{PrivateCommunicationTeresaBrecht, Pozar} may improve shorting the two sides of the cavity together.  Particularly when sapphire wafers are present, the quality of the connection between the two sides could dominate the cavity quality factor.  Finally, we could scale the architecture to support more cavities and qubits as experimental requirements dictate.

\section{Qubit experiments}
\label{sec:qubitexpts}

Though many of the straightforward quantum information experiments have been done, there are a wealth of future directions for superconducting qubits.  As ever, we want to improve qubit gate and measurement fidelity, inventing more robust or scalable multi-qubit gates, and investing in technical capabilities like FPGAs for real-time measurement processing \cite{Riste2012a, Riste2012b, Johnson2012}.  

Thinking about the architecture of a large-scale quantum computer has become increasingly necessary.  One issue that must soon be addressed is that superconducting qubits tend to interact with one another, particularly when they mutually coupled to the same cavity.  Implementing the surface on a large-scale quantum computer will require restricting the length scale of these interactions.  As we discuss below, one approach is to make ``modules'' that interact only through an explicit measurement.  This has the advantages of eliminating any spurious interactions and of being scalable, though it also somewhat increases physical overhead. 

We will also provide a few examples of interesting experiments that can be done with current (or soon-to-be current) multi-cavity devices.  Having multiple cavities enables conceptually new experiments like entanglement distillation and syndrome extraction, both of which are necessary for fault-tolerant error correction.

\subsection{Modules}

One challenge with scaling to larger superconducting cQED systems is that qubits tend to interact with one another over relatively large length scales.  While this opens up a variety of ways to generate multi-qubit gates \cite{DiCarlo2009, DiCarlo2010, Neeley2010, Fedorov2011, Reed2012, Chow2011, Poletto2012, Paik2013}, it presents a challenge for single-qubit rotations \cite{Paik2013, Chow2012, Gambetta2012, Johnson2012}.  As we saw in \sref{subsec:allxy}, even two qubits in a single cavity can have dramatic implications for maintaining control over each qubit individually.  To a certain extent, composite pulses and understanding the details of these interactions can mitigate the control problem, but these long-range interactions also tend to propagate errors to other qubits, creating a nightmare for fault tolerance (\sref{subsec:faulttolerance}).  Though it would be possible to fabricate a device with a large number of qubits, accurately controlling and error-correcting such a system will require limiting the range of these interactions.

One approach to controlling those interactions is to physically break the connection between qubits.  Suppose we have some relatively small system of a few qubits and several cavities.  If we understood the details of all of the (finite number of) interactions well enough, we might control each qubit with extremely high fidelity.  We could then network a pair of these smaller devices together via a dispersive measurement that bounces off both devices prior to being detected \cite{Cabrillo1999, Kolli2009, Pfaff2013}.  These modules may be separated by relatively large distances (the length of an SMA cable) and would only interact via the intentional measurements.  Entanglement between devices is heralded by this measurement, which might be repeated or ``distilled'' if the interactions between modules prove faulty.  This approach is advantageous in that it is the only currently known way to robustly scale superconducting qubits, and because it has a straightforward experimental path.

What are the initial steps along that path?  First, we need to develop better tools for predicting the higher-order terms of a Hamiltonian and the Purcell effect.  Black-box quantization \cite{Nigg2011} has been shown to work well for one qubit and cavity, but more work is necessary to take into account multiple qubits and cavities, especially when $E_C$ is not small.  Validating these theoretical calculations will require a detailed measurement of all the couplings in a system, and performing those measurements as a function of qubit frequency via flux bias line or external-coil tuning would be even better.  

Developing the technology to measure multiple physical cavities is also important.  One approach is making two systems with the same cavity frequency and $\chi$-shift, and connect them with circulators and cables that are as low-loss as possible.  As of the writing of this thesis, the first results of this experiment were just announced out of Irfan Siddiqui's lab in Berkeley \cite{RochMM2013}.  Another approach is to connect the cavities to the signal and idler ports of a JPC amplifier \cite{Abdo2011, Hatridge2013}, which would eliminate the requirement for circulators prior to amplification and for the cavities to be at the same frequency \cite{DevoretMLS130422}.

\subsection{Multi-cavity experiments}

There are also a variety of interesting qubit experiments which take advantage of having more than one cavity which are currently possible or will be soon.  Most multi-qubit experiments conducted so far have involved several qubits in the same cavity.  Even if we do not use the module approach, we clearly need to limit the number of qubits that directly couple to one another in order to scale larger.  Fortunately, we already have a prototypical system with which to study this idea: the two-cavity 3D cQED device (\sref{subsec:tunablecavitydesign}).  There, we have a single qubit coupled to both cavities, and can have several coupled to each individual cavity.  This enables two conceptual classes of experiments: entanglement distillation and syndrome extraction.

\subsubsection{Entanglement distillation}

Entanglement distillation is a process by which we take many imperfect Bell pairs shared between two locations and convert them into a smaller number of higher-fidelity pairs \cite{Bennett1996, Bennett1996b, Bennett1996c, Nielsen2000}.  Suppose that we can generate entanglement between distant qubits, perhaps by shuttling excitations through a qubit that is coupled to both.  Entanglement generated in this way will likely be lower fidelity than operations limited to inside a given cavity.  For example, the coherence time of the shuttle qubit will be inferior to the single-cavity qubits if the shuttler is flux sensitive and the others fixed-tuned.  By repeatedly measuring quantities about the two pairs, we can increase the fidelity of the pairs (conditioned on the ``favorable'' outcomes of random measurements).

The most basic type of entanglement distillation would require five qubits in cQED: one ancilla qubit and one communication qubit per cavity, and one shuttle qubit coupled to both cavities\footnotemark.  We also require the ability to measure both ancilla qubits individually.  That protocol \cite{Campbell2007, Campbell2008} begins by entangling two communication qubits by some process that is assumed to be relatively low fidelity.  The communication qubits are then entangled with the ancilla qubits and subsequently measured along some basis.  If the results of both measurements are favorable and the fidelity of both the ancilla-communication entanglement and the measurements are high, then the resulting entanglement fidelity between the communication qubits is increased.  If the measurement outcome is unfavorable, however, the entanglement may either have been destroyed or reduced in fidelity, depending on the protocol.  Fortunately, with at least some schemes \cite{Campbell2007}, the probability of finding subsequent favorable measurements increases with each successful repetition.  

\footnotetext{The fifth qubit is not required, but some mechanism to controllably entangle the two distant sets is necessary.}

This general scheme -- in which ancillas are measured and a ``good'' result increases fidelity -- is also used in other QI protocols.  For example, there is no known way to fault-tolerantly implement non-Clifford gates like $\hat{T}=e^{-i\pi \hat{Z}/8}$ in certain error correction protocols like the surface code \cite{Nielsen2000}.  This gate is needed for universality, so another approach is crucial.  One method is known as ``magic state distillation'' \cite{Bravyi2004, Bravyi2012}.  There, $k$ imperfect copies of the state $\ket{A} = \hat{T}\ket{+} \sim \ket{0} + e^{i\pi/4}\ket{1}$ are generated.  These states are then entangled with one another and various Clifford projections are measured.  If all the measurements are found to have the value $+1$, we will have $k$ higher-fidelity copies of $\ket{A}$.  This protocol can be repeated to enhance the fidelity of $\ket{A}$ to the effective logical error rate of our computer, and then teleported into the circuit to implement $\hat{T}$ as necessary.

\subsubsection{Syndrome extraction}

All error correction protocols rely on extracting error syndromes from encoded logical qubits.  As we saw in \sref{subsec:quantumrepetitioncode}, this information can either be fed into a quantum gate to implement a correction or it can be measured, classically processed, and fed back.  It turns out that measuring the error syndromes is a much more effective method, since classical logic operations can be essentially perfect.  This dramatically lowers the overhead associated with fault-tolerant error correction.  However, extracting these syndromes is a challenge itself.  In cQED, measurement of transmission through a cavity projects every qubit coupled to that cavity.  If we intend to measure only one ancilla qubit that contains some error syndrome, we would have to use an individual cavity coupled solely to that qubit.  This is one of the motivations for replacing or supplementing FBLs with cavities, as we have previously discussed.

A method that avoids the need for individual cavities for syndrome extraction was recently proposed \cite{Nigg2012}.  There, we have a manifold of $N$ qubits coupled to the same cavity.  By using the dispersive $\chi$-shift and a cavity displacement, we can map the value of any Pauli correlation to the quantum state of the cavity.  The cavity will either be in the state $\ket{+\alpha}$ or $\ket{-\alpha}$ depending on whether the value of some observable is $+1$ or $-1$.  This observable can be chosen arbitrarily by performing rotations on the qubits and by echoing away the $\chi$-shifts of qubits that should not be measured.  The state of the cavity is then read out with some ancilla that is coupled to both the primary cavity and a secondary cavity.  It is convenient for that qubit to be flux-tunable to control its dispersive coupling to the storage cavity.  Since this qubit will be entangled with the value of the syndrome, measuring it will project the system as desired in QEC.  Moreover, with recent developments of fast FPGA logic \cite{Riste2012a, Riste2012b} and low-noise amplifiers \cite{Vijay2011, Hatridge2013}, a high-fidelity QND measurement could be repeated several times.  Based on the outcomes of several repetitions, one could, for example, prepare some known entangled Bell or GHZ state.

\section{Outlook on a superconducting quantum computer}

Despite the inevitable challenges, there are many reasons to be optimistic about the future of superconducting qubits.  Only fourteen years have passed since the first Rabi oscillations of a Cooper pair box were demonstrated at NEC \cite{Nakamura1999}.  Since then, there have been vast improvements to all aspects of the technology.  One of the most important advances has been in coherence time.  Superconducting qubit coherence has improved nearly five orders of magnitude since 1999, from $2\ns$ to $\sim200\us$.  This is of key importance because coherence had long been the most pressing open question in evaluating the prospects of a superconducting quantum computer.  In contrast to systems like quantum dots, the source of this decoherence is not known, and it was feared to be an intrinsic limit of Josephson junctions.  Recent results \cite{Paik2013, Rigetti2012, Riste2012b, Chang2013, Geerlings2013b} have shown that these junctions are perfect, at least to the limits of our ability to measure them.  Dielectric losses and thermalization were likely the sources of the previous $\sim \us$ ceiling.  Moreover, since the original demonstration of 3D cQED, transmon coherence has improved by another factor of four \cite{PetrenkoMM2013} and further advances are hopefully on their way.  Two-dimensional cQED has also enjoyed impressive advances in coherence, with recent devices exhibiting $20-60\us$ $T_2$ times \cite{Barends2013, Chang2013}.  It is fair to say that coherence time no longer represents the primary obstacle to a superconducting quantum computer, though there always remains room for improvement.

In contrast to the problem of coherence time for which the origin and solution were unknown, many of the current and upcoming challenges are better defined.  We need to continue making advances with gate fidelity and figure out how to scale to larger systems without sacrificing that progress.  There is a significant difference between demonstrating fault-tolerant fidelities in a system with only two qubits and demonstrating the same in a system with a hundred.  This creates a conflict: how can we learn to improve gate fidelity in a system that does not yet exist?  Or rather, how can we design an architecture that takes full advantage of the tricks and improvements we have gleaned from smaller devices?  One short-term tactic would be to separate the problems of fidelity and scaling into two research avenues.  We can continue studying how to improve the coherence and fidelity of relatively small systems while simultaneously thinking about ways to scale to large ones \cite{Mariantoni2011, Barends2013}.  Investigating how to mitigate spurious long-distance communication is an important concern, but has ample room for innovation and creativity.

Fortunately, the challenge of scaling plays directly into the strengths of superconducting qubits.  Unlike previous quantum architectures, cQED devices are solid-state systems fabricated in a clean room and controlled with electronic signals.  (This statement disregards the fact that 3D cavities are made in a machine shop, since there is no reason to believe that is a requirement.)  These properties of superconducting qubits are reminiscent of the silicon transistor, the most replicated human-made object in history.  It seems likely that methods of inhibiting qubit cross-talk, constructing high-fidelity gates, and implementing room-temperature control will all be soon in coming.  These will clear the way to scaling to very large superconducting devices, and potentially, a true quantum computer.

\begin{center}\decofourright \decofourleft\end{center}

\begin{singlespace}
\renewcommand*{\bibpreamble}{\thumb{Bibliography}}
\bibliographystyle{utphysMDR} 
\newcommand{\maybebackrefprint}{ \backrefprint}
\bibliography{thesis}

\end{singlespace}


\begin{appendices}
\cleardoublepage
\phantomsection
\addcontentsline{toc}{part}{Appendices}


\chapter{Current-Flux Coupling}
\nothumb
\label{ap:fluxcoupling}

\newcommand{\Cl}{C_{_L}}
\newcommand{\Cr}{C_{_R}}
\newcommand{\Ci}{C_{_I}}

\newcommand{\Ql}{Q_{_L}}
\newcommand{\Qr}{Q_{_R}}
\newcommand{\Qi}{Q_{_I}}

\newcommand{\Qa}{{\color{red}Q_{_A}}}
\newcommand{\Qb}{{\color{blue}Q_{_B}}}

\newcommand{\Vo}{V^{^O}}
\newcommand{\Ve}{V^{^E}}

\lettrine{I}{n} this appendix we reproduce the calculations of Nissim Ofek and Kevin Chou for the coupling of a flux bias line to a loop in a superconducting box.\\
~\\
\hrule
~\\

Here I try to estimate the flux coupling to a current segment (carrying current I) located between two SC plates. The distance between the plates is $w$ and I want to calculate the field at distance $d$.

Without the plates, the field goes down like $1/d$. The effect of the plates is to attenuate it by some factor.

So the current segment if located at $(0,0)$ and the plates are at $y=w/2$ and $y=-w/2$. Zero perpendicular field at the surface of the plates can be attained by current images located at $(0,n\cdot w)$ for $n=\pm1,2,3,\dots$ with alternating currents $I_n = (-1)^nI$.

The expression I want to get is the ratio $B_y(d,0)/B_y^0(d,0)$. Where $B_y^0(d,0)$ is the field at $(d,0)$ without the plates, which is proportional to $1/d$. And $B_y(d,0)$ is the field with the presence of the plates, in terms of the ratio $r=w/d$.

\begin{eqnarray}
G(d,0)\equiv\frac{B_y(d,0)}{B_y^0(d,0)}&=&\sum_{n=-\infty}^{\infty}(-1)^n\frac{1}{(n\cdot r)^2+1}
\end{eqnarray}

This expression can be analytically solve for the two extreme cases, $r\gg1$ (very close to the source) and $r\ll1$ (very far from the source). 

Case I, $r\gg1$:

\begin{eqnarray}
\sum_{n=-\infty}^{\infty}(-1)^n\frac{1}{(n\cdot r)^2+1}&\approx&1+\sum_{n\neq0}(-1)^n\frac{1}{(n\cdot r)^2} \\
&=&1+\frac2{r^2}\sum_{n=1}^{\infty}\frac{(-1)^n}{n^2}\\
&=&1-\frac{\zeta(2)}{r^2}\approx1-\frac{\pi^2}{6r^2}
\end{eqnarray}

So the effect of the plates is negligible near the source (of course ... ).

Case II, $r\ll1$:

Now $n\cdot r$ increments in small steps. I will rewrite the sum as follows:

\begin{eqnarray}
\sum_{n=-\infty}^{\infty}(-1)^n\frac{1}{(n\cdot r)^2+1}&=&1-2\sum_{m=1}^\infty\left[\frac{1}{[(2m-1)\cdot r]^2+1}-\frac{1}{[2m\cdot r]^2+1}\right]\\
&=&1-2\sum_{m=1}^\infty\left[\frac{1}{x^2+1}-\frac{1}{(x+r)^2+1}\right]_{x=(2m-1)\cdot r}\\
&=&\divideontimes
\end{eqnarray}

I define: 
$$f(x)=\frac{1}{x^2+1}-\frac{1}{(x+r)^2+1},$$
so
\begin{equation}
\divideontimes=1-2\sum_{m=1}^\infty f\left[(m-\frac12)\cdot 2r\right]\equiv1-2\cdot S.
\end{equation}

It is quite obvious that $\lim_{r\rightarrow0}S=\frac12$ and so the whole expression goes to zero. Still, we are interested to see at what rate this occurs.

\begin{equation}
	\begin{split}
	\int_{x=0}^\infty f(x)dx&=\sum_{k=0}^{\infty}\left[\sum_{m=1}^{\infty}\frac{2r^{2k+1}f^{(2k)}[(m-\frac12)\cdot 2r]}{(2k+1)!}\right]\\
	&=\sum_{m=1}^\infty2r\cdot f\left[(m-\frac12)\cdot 2r\right]+\sum_{k=1}^{\infty}\left[\sum_{m=1}^{\infty}\frac{2r^{2k+1}f^{(2k)}[(m-\frac12)\cdot 2r]}{(2k+1)!}\right]\\
	&=2r\cdot S+\sum_{k=1}^{\infty}\left[\sum_{m=1}^{\infty}\frac{2r^{2k+1}f^{(2k)}[(m-\frac12)\cdot 2r]}{(2k+1)!}\right]
	\end{split}
\end{equation}

Now, 
\begin{equation}
\int_{x=0}^\infty f(x)dx=\left.\left[\arctan x-\arctan(x+r)\right]\right|^{\infty}_0=\arctan r
\end{equation}

And so,
\begin{equation}
S = \frac{\arctan r}{2r}-\sum_{k=1}^{\infty}\left[\sum_{m=1}^{\infty}\frac{r^{2k}f^{(2k)}[(m-\frac12)\cdot 2r]}{(2k+1)!}\right]
\end{equation}

For $r<<1$ we can write:
\begin{eqnarray}
S=\frac{\arctan r}{2r}&\approx&\frac{r-\frac13r^3}{2r}=\frac12\cdot\left[1-\frac{r^2}{3}\right]
\end{eqnarray}

At the beginning I plugged it into get:
\begin{equation}
1-2\cdot S=\frac{r^2}{3}.
\end{equation}

Obviously, I was sloppy. I should also calculate the big sum at least to order $r^2$. This is quite easy in fact:

\begin{equation}\begin{split}
\sum_{m=1}^{\infty}\frac{r^{2k}f^{(2k)}[(m-\frac12)\cdot 2r]}{(2k+1)!}&=\left.\frac{r^{2k-1}}{2(2k+1)!}\cdot f^{(2k-1)}\right|^\infty_0\\
&-\sum_{l=1}^\infty\left[\frac{r^{2(l+k)}}{(2k+1)!(2l+1)!}\sum_{m=1}^\infty f^{2(l+k)}[(m-\frac12)\cdot2r]\right]
\end{split}\end{equation}

So, calculating $S$ to the second order in $r$ gives:

\begin{eqnarray}
S&=&\frac12\cdot\left[1-\frac{r^2}3\right]+\frac{r}{2\cdot3!}\cdot\frac{2r}{(r^2+1)^2}\\
&=&\frac12\cdot\left[1-\frac{r^2}{3}\right]+\frac12\cdot\frac{r^2}{3}\cdot\left[1-2r^2+3r^4+\dots\right]\\
&=&\frac12-\frac16\cdot\left[2r^4-3r^6-\dots\right]
\end{eqnarray}

So the field decays at least as $r^4$!

It is quite easy to continue to the next order, but it happens that $G(d,0)$ can be matched quite accurately by:
\begin{equation}
G(d,0)=\frac{1}{\cosh\left(\frac{d}{2w\cdot\sigma}\right)^2}\qquad; \qquad\sigma=0.955\cdot\frac{\pi^2}{24}
\end{equation}

\newpage

This is OK, but not exactly what I need. I want to see what is the coupling due to a finite segment of current. To do so, I will calculate the flux threading a square of width $w$ and height $h$ positioned $d$ away from a current segment of length $L$. I will assume that the segment lies on the $x$-axis.

\begin{center}
\includegraphics[scale=0.8]{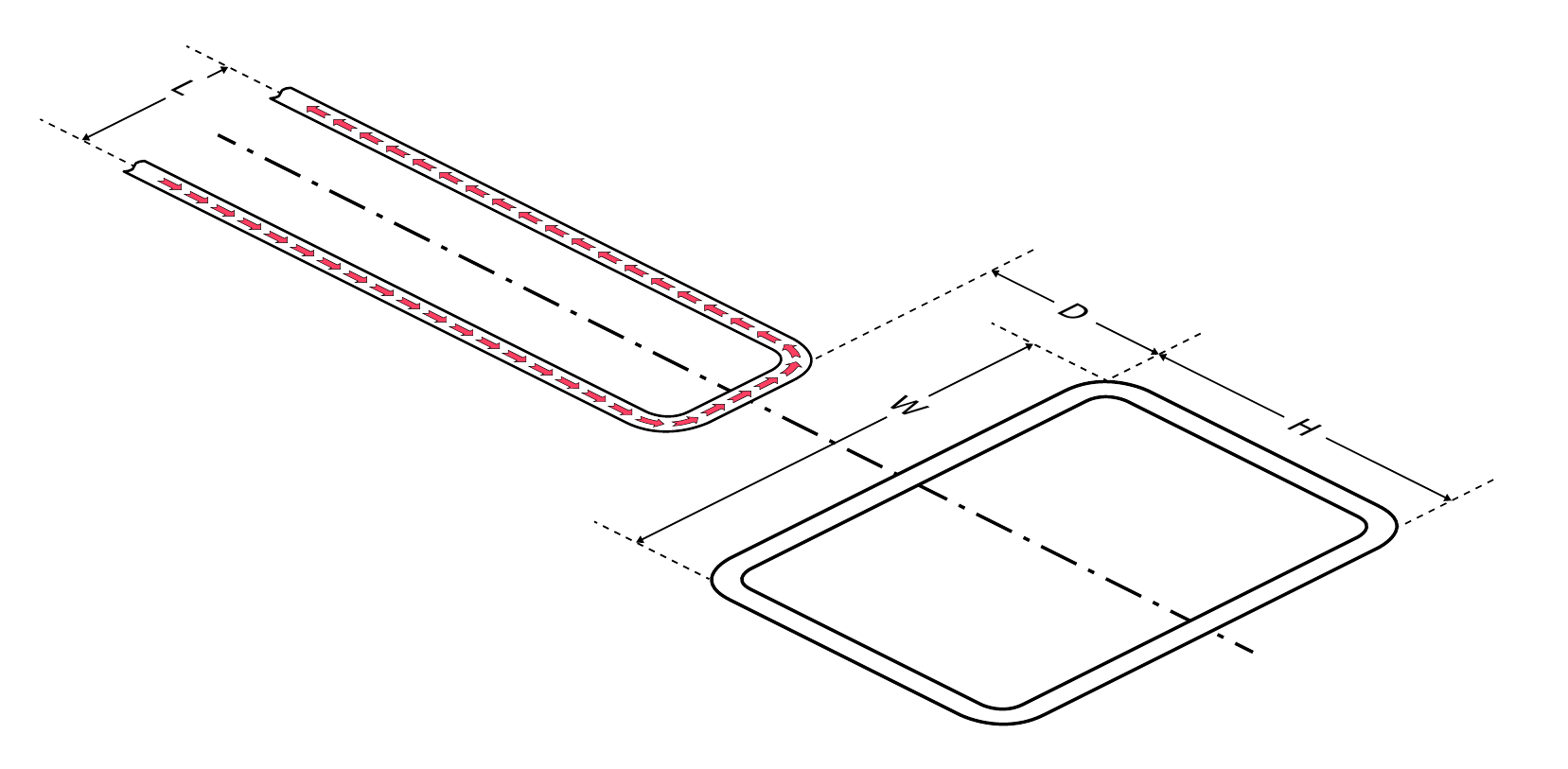}
\end{center}

I start from Biot Savart Lew:
$$B = \int \frac{\mu_0I}{4\pi}\frac{\vec{dl}\times\vec{r}}{|r|^3}$$

First, I calculate the field at position $(x,y)$ due to the current segment:

\begin{eqnarray}
B(x,y)&=&\frac{\mu_0I}{4\pi}\mathop{\int}_{-L/2}^{L/2}\frac{\vec{dl}\times\vec{r}}{|r|^3}\\
&=&\frac{\mu_0I\hat{z}}{4\pi}\mathop{\int}_{-L/2}^{L/2}\frac{y}{\left((x-l)^2+y^2\right)^{3/2}}dl=\otimes
\end{eqnarray}
$\vec{r}$ stands for $(x,y)-(l,0)=(x-l,y)$, and $\vec{dl}$ is simply $(dl,0)$. Hence:

By using $x-l=y\tan\theta$, this integral is transformed to be:
\begin{eqnarray}
\otimes&=&\frac{\mu_0I\hat{z}}{4\pi y}\mathop{\int}_{\arctan\left(\frac{2x-L}y\right)}^{\arctan\left(\frac{2x+L}y\right)}\cos\theta d\theta\\
&=&\frac{\mu_0I\hat{z}}{4\pi y}\left[\sin\left(\arctan\left(\frac{2x+L}{y}\right)\right)-\sin\left(\arctan\left(\frac{2x-L}{y}\right)\right)\right]\\
&=&\frac{\mu_0I\hat{z}}{4\pi y}\left[\frac{\frac{L}2-x}{\sqrt{\left(\frac{L}2-x\right)^2+y^2}}+\frac{\frac{L}2+x}{\sqrt{\left(\frac{L}2+x\right)^2+y^2}}\right]
\end{eqnarray}

For a point very close to the segment, $y\ll\frac{L}2-x, \frac{L}2+x$, this expression reduces to:
$$\frac{\mu_0I\hat{z}}{2\pi y}$$

Which is the field due to infinite current line.

The next step is to integrate this field within the area of the squid:

\begin{eqnarray}
\Phi&=&\mathop{\int}_D^{D+H}dy\mathop{\int}_{-\frac{W}2}^{\frac{W}2}dx\,B(x,y)\\
&=&\frac{\mu_0I}{4\pi}\mathop{\int}_D^{D+H}\frac{dy}{y}\mathop{\int}_{-\frac{W}2}^{\frac{W}2}\left[\frac{\frac{L}2-x}{\sqrt{\left(\frac{L}2-x\right)^2+y^2}}+\frac{\frac{L}2+x}{\sqrt{\left(\frac{L}2+x\right)^2+y^2}}\right]dx\\
&=&\frac{\mu_0I}{2\pi}\mathop{\int}_D^{D+H}\frac{dy}{y}\mathop{\int}_{-\frac{W}2}^{\frac{W}2}\frac{\frac{L}2+x}{\sqrt{\left(\frac{L}2+x\right)^2+y^2}}dx\\
&=&\frac{\mu_0I}{2\pi}\mathop{\int}_D^{D+H}\frac{dy}{y}   \left.\left[\sqrt{\left(\frac{L}2+x\right)^2+y^2}\right]\right|_{-\frac{W}2}^{\frac{W}2}\\
&=&\frac{\mu_0I}{2\pi}\mathop{\int}_D^{D+H}\frac{dy}{y}\left[\sqrt{\left(\frac{L+W}{2}\right)^2+y^2}-\sqrt{\left(\frac{L-W}{2}\right)^2+y^2}\right]=\oplus
\end{eqnarray}

As a test, we can check for $H\ll D$ and $L\gg W,D$, and see that we get the area, $H\cdot W$ times the field due to infinite length current line.

We have here two integral of the form:
$$\int\frac{dy}{y}\sqrt{u^2+y^2}=\sqrt{u^2+y^2}+u\log y -u\log\left\{u^2+u\sqrt{u^2+y^2}\right\}+C$$

It should be mentioned that this function is symmetric in $u$. It should be, looking at what we actually integrate. It is less obvious from the integral itself.

So the whole expression can be written in the following way:

\begin{eqnarray}
\oplus&=&\frac{\mu_0I}{2\pi}\left.\left.\left[\sqrt{u^2+y^2}+u\log y -u\log\left\{u^2+u\sqrt{u^2+y^2}\right\}\right]\right|_{y=D}^{D+H}\right|_{u=\frac{L-W}{2}}^{\frac{L+W}{2}}
\end{eqnarray}

So this is the analytic expression for the coupling as a function of the current, segment length, frame distance, width and height: $f(L,D,W,H)$.

If the segment is not in the same plane of the frame, say the frame is positioned at lower altitude, $A$, then the same formula can be used be slight changes:

\begin{center}
\includegraphics[scale=0.8]{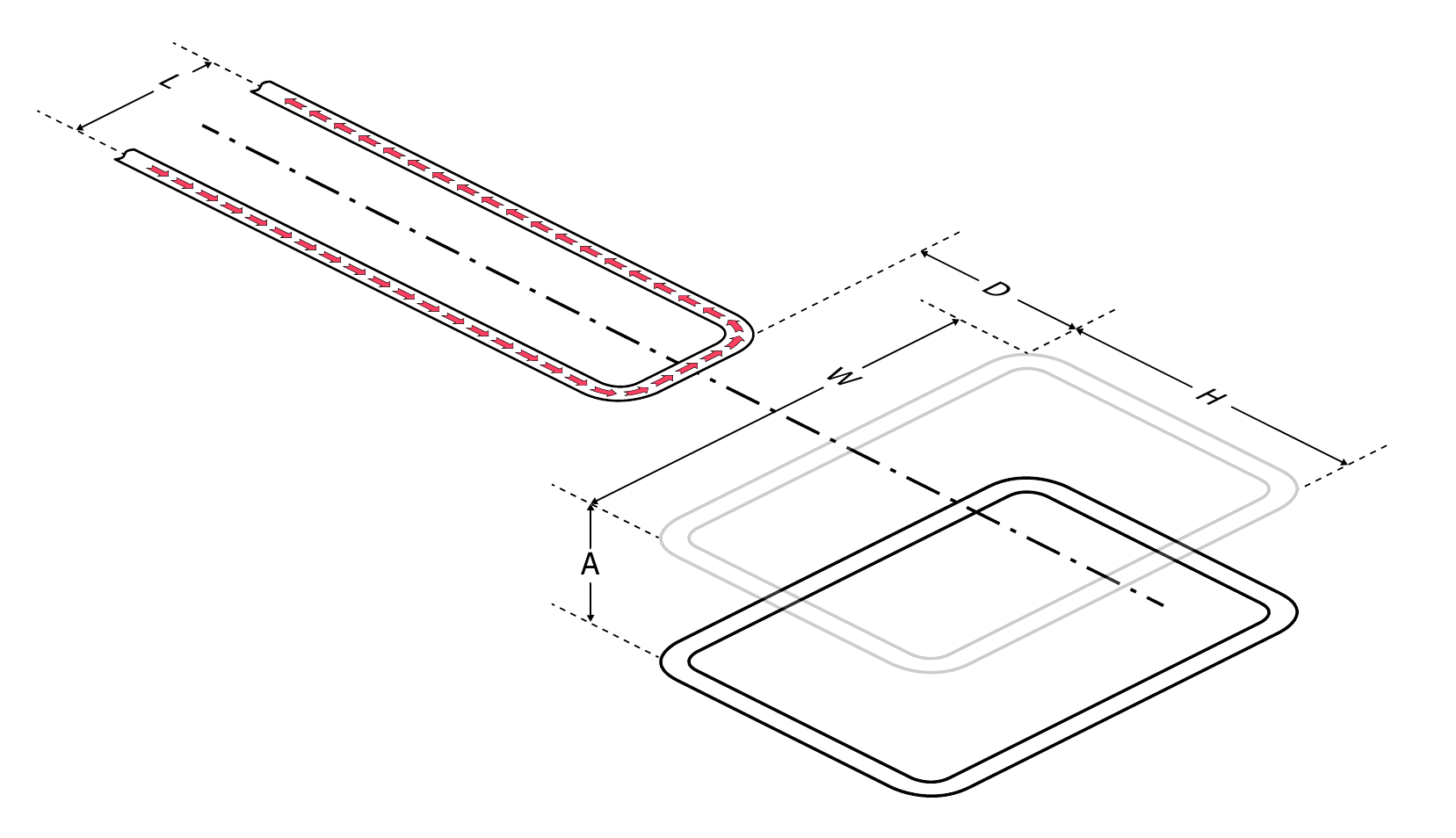}
\end{center}

$$g(L,D,W,H,A)=f(L,\sqrt{D^2+A^2}, W, \sqrt{(D+H)^2+A^2} - \sqrt{D^2+A^2})$$

We can also calculate the flux coupling when the frame is not sitting symmetric relative to the current segment, say with some offset $O$:

\begin{center}
\includegraphics[scale=0.8]{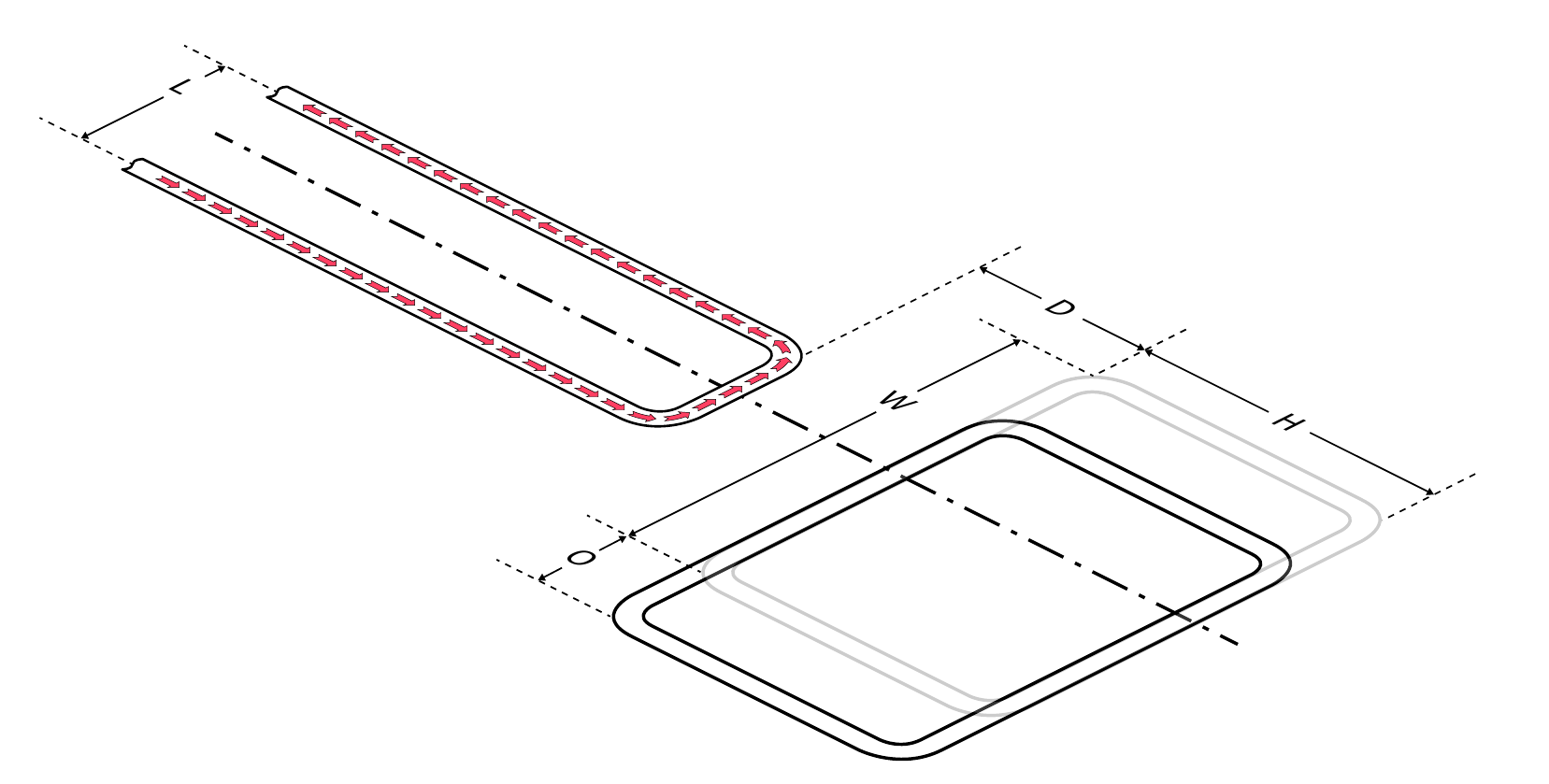}
\end{center}

If $W<2*O$, then:
$$h(L,D,W,H,O)=\frac12\left[f(L,D,W+2O,H)+f(L,D,W-2O,H)\right]$$
Otherwise:
$$h(L,D,W,H,O)=\frac12\left[f(L,D,W+2O,H)-f(L,D,2O-W,H)\right]$$.

And, of course, we can combine the offset and the altitude together.

This will be useful if we want to take into account the image currents cause by the SC box. 
As long as the dimensions are of the order of the width and height of the cavity, we can take the main term alone. For example, a frame of 100$\um$ by 100$\um$ located 400$\um$ away from a 500$\um$ current segment carrying 1mA has one flux quantum threading it. For the same parameters, the first image caused by the top SC plane, assuming the cavity is 1mm high, gives 8\% of the flux. We have to of these, so taking the first term only has an error no more than 16\% (we have two such images). The error due to the images cause by the side SC planes of the order of 9\%, so the total error is of the order of 35\%. We can still get half a flux out of the same current.

\chapter{Mathematica Code}
\label{ap:mathematica}%
\chaptermark{Mathematica Code}%
\nothumb

\lettrine{H}{ere} we reproduce the Mathematica code used for state and process tomography and simulating AllXY syndromes.  This code can be used to convert experimental state tomograms into a $\chi$ process matrix, or to simulate the matrix of an ideal process.  This code works only for two qubits (\sref{sec:processtomo}), but can be easily generalized for any number.  However, in practice, Mathematica is too slow to operate on more than two qubits, so the code was re-written in Matlab for the process tomography of \sref{subsec:toffolitomography}.  That code is not included here, but is functionally identical.

We also calculate the error syndromes of AllXY both analytically and in comparison with data.  Since this code was written, a much better method has been developed using the QuTiP package for Python\footnotemark.  We include this code so that the plots from \sref{subsec:allxy} can be easily reproduced, but going forward it would be wise to port the code over to Python.

\footnotetext{\url{https://code.google.com/p/qutip/}}


\edef\listopt{[pages=1-last, pagecommand={}, offset=\xoff 0in, trim=\xtrim 0.6in \xtrim 0.6in,clip,quiet,thread]}
\expandafter
\includepdf
\listopt{./Appendices/MathematicaCodeAppendix\pngRes.pdf}

\end{appendices}

\backmatter
\chapter{Copyright Permissions\label{ch:copyperm}}%
\nothumb

\begin{bibunit}[utphysMDR2]
\def\bibliography#1{\putbib[#1]}

\begingroup
  \makeatletter
  \let\BR@bibitem\BRatbibitem
  \nobibliography{thesis}



\newcommand{\hiliteLSB}[1]{#1}
\renewcommand{\newblock}{\\*}

\begin{singlespace}
\begin{itemize}

	\item Figures \ref{fig:fourqubitdevice}, \ref{fig:wiringdiagram}, \ref{fig:fourqubitpic}, \ref{fig:fourqubitspec}, \ref{fig:zeta}, \ref{fig:separable3qstatetomo}, \ref{fig:ghztomo}, \ref{fig:expcityscape}, \ref{fig:merminsum}, and \ref{fig:merminproduct} reproduced or adapted with permission from: \\* \bibentry{DiCarlo2010}.

	\item Figures \ref{fig:purcelldesign}, \ref{fig:purcelltransmission}, \ref{fig:lifetime}, and \ref{fig:reset} reproduced with permission from: \\* \bibentry{Reed2010}.

	\item Figures \ref{fig:transmission1}, \ref{fig:transmission2}, \ref{fig:transmissioncolor}, \ref{fig:fidelity}, \ref{fig:integrationtimefidelity}, \ref{fig:transmissioncolor3q}, and \ref{fig:fidelity3q} adapted with permission from: \\* \bibentry{Reed2010b}.

 	\item Figure \ref{fig:spectroscopy} adapted with permission from: \\* \bibentry{Paik2011}.

	\item Figures \ref{fig:chev23}, \ref{fig:chev003}, \ref{fig:toffolipulse}, \ref{fig:toffoliccnot}, \ref{fig:truthtabletomo}, \ref{fig:truthtable}, \ref{fig:toffoliprocesstomo}, \ref{fig:bitflipcircuit}, \ref{fig:bitflipqec}, \ref{fig:bitflipancilla}, \ref{fig:phaseflipcircuit}, and \ref{fig:phaseflipqpt} reproduced or adapted with permission from: \\* \bibentry{Reed2012}.

\end{itemize}
\end{singlespace}

\endgroup

\end{bibunit}

\cleardoublepage

\end{document}